# Evaluación y modelado del rendimiento de los sistemas informáticos

**Xavier Molero**
*Departament d'Informàtica de Sistemes i Computadors*
*Universitat Politècnica de València*

**Carlos Juiz**
*Departament de Ciències Matemàtiques i Informàtica*
*Universitat de les Illes Balears*

**Miguel Jesús Rodeño**
*Departamento de Ciencias de la Computación*
*Universidad de Alcalá*

# Prólogo

El responsable de un sistema informático se suele encontrar, habitualmente, en la difícil tarea de responder a preguntas como, por ejemplo: ¿soportará mi sistema la carga de trabajo del año próximo?, y si no aguanta, ¿cómo he de ampliar mi sistema para que la soporte? O bien: de las ofertas que me presentan los suministradores, ¿cuál de ellas responderá adecuadamente a la carga futura? A estas cuestiones, y otras similares, se puede responder fiándose solamente de la inspiración de cada uno, procedimiento siempre peligroso por poco fiable, o utilizar técnicas cuantitativas que arrojen algo de luz para responderlas adecuadamente.

El presente libro de Molero, Juiz y Rodeño, *Evaluación y modelado del rendimiento de los sistemas informáticos*, representa un magnífico resumen de técnicas cuantitativas simples que ayudan a dar respuesta a las cuestiones anteriores. Su enfoque no es el de la teoría por la teoría, sino que, en cada capítulo, después de un pequeño repaso teórico, se entra de lleno en numerosos problemas agrupados en tres categorías: los que presentan la resolución completa, aquellos de los que sólo se da la solución y, finalmente, los que se deja su resolución a la discreción del lector. Aun cuando puede achacarse a algunos de los problemas resueltos una dimensión y complejidad puramente académicas, éstos no son de despreciar, pues ponen de manifiesto, a escala reducida, el proceso que debe seguirse —con ayuda de las herramientas adecuadas— en la solución de problemas equivalentes del mundo real y de dimensión industrial.

Revisando el índice de este libro encontramos, en el Capítulo 1, una introducción al concepto de evaluación del rendimiento, los distintos criterios que hay que considerar en esta evaluación (técnicos y económicos), y la relación, a veces engañosa, entre las prestaciones y el coste de un equipo informático. No siempre lo más caro es lo mejor y, además, hay que tener en cuenta que los sistemas informáticos no son buenos o malos *per se*, sino adecuados, o no, a una carga concreta de trabajo.

A partir del principio de que "aquello que no se puede medir no se puede controlar", el Capítulo 2 trata sobre la monitorización de sistemas y programas. Dicho de otra forma, está dedicado a obtener información tanto de lo que está sucediendo en un sistema, como del comportamiento de un programa (ligado a los datos que procesa), no sólo desde el punto de vista externo, sino también desde el interno; es decir, se observa el sistema y el programa como si se fuera el responsable de la instalación. Hay que tener en cuenta que, en la actualidad, la mayor parte de los sistemas informáticos trabajan dando servicio a usuarios, normalmente remotos, que creen tener todo el sistema a su disposición. En definitiva, se podría decir que los usuarios no son más que un conjunto de "depredadores" de recursos del sistema informático, de los que no se sabe exactamente en qué momento harán uso del mismo. Por lo tanto, *a priori*, no podemos saber cómo se consumirán los recursos que ofrece. Sin embargo, al menos *a posteriori*, hay que saber quién los ha consumido y en qué cantidad. Para ello es necesario conectar al sistema elementos de tipo software, aunque también, para ciertas variables, pueden ser útiles elementos físicos (hardware) de medida. El capítulo ofrece una visión general de un conjunto de herramientas empleadas con asiduidad en los sistemas más extendidos, de la forma en que se han de interpretar los resultados que ofrecen estas herramientas y, adicionalmente, del uso que se puede hacer de ellas.

El Capítulo 3 está dedicado a técnicas para la comparación de las prestaciones de diversos sistemas. El tema es sumamente interesante y pone de manifiesto cuán frágiles son las técnicas utilizadas en la actualidad para efectuar dichas comparaciones, que pueden, actuando de mala fe, demostrar una hipótesis y su contraria. En todo caso, la conclusión a extraer del mismo no es tanto la inutilidad de los ratios para efectuar comparaciones como la necesidad perentoria de definir, de forma simple y transparente, los objetivos del estudio de comparación, y la relación entre variables que permita poner de manifiesto, con la mayor claridad posible, las ventajas de unos sistemas respecto de otros.

Los Capítulos 4 y 5 exponen el análisis operacional y sus aplicaciones. Se trata de una técnica que relaciona las variables que describen las prestaciones y las que describen la carga. Estas relaciones se fundamentan sólo en hipótesis comprobables experimentalmente, sin recurrir a otras difícilmente demostrables, como las empleadas con el mismo objetivo en la teoría de colas de espera. Por tanto, el análisis operacional constituye una primera aproximación al modelado de sistemas informáticos, la cual puede emplearse cuando se pretendan estimar las prestaciones de un sistema que no existe, ya sea total o parcialmente (por ejemplo, en el caso de ampliación de los recursos de un sistema o de la carga que debe tratar). Parece increíble la cantidad de interesantes resultados que se pueden encontrar partiendo de simples medidas elementales del comportamiento del sistema informático.

En todos los problemas de evaluación de las prestaciones se toma como dato de entrada la carga del sistema. Pero, ¿qué se entiende por la carga de un sistema? ¿Cómo podemos representarla de forma compacta sin recurrir a todos los programas que se ejecutan en él? La respuesta a estas cuestiones está tratada en el Capítulo 6, de forma simple en cuanto a la dimensión de los problemas propuestos para poder resolverlos, sin recurrir a herramientas estadísticas poderosas. Sin embargo, la base de esas herramientas está claramente expuesta en el repaso teórico, y utilizada en los problemas resueltos. Ello ha de permitir al lector buscar las herramientas adecuadas cuando haya de resolver un problema similar.

Uno de los problemas con que se enfrenta el responsable de un sistema informático es el de la planificación de su capacidad, es decir, el de habilitar los recursos para poder dar servicio a los usuarios del mismo respetando sus exigencias de comportamiento, tanto si estos usuarios están en un entorno cerrado (usuarios internos de una empresa, por ejemplo) como abierto (usuarios de un servidor web). Las técnicas para tratar este problema están expuestas de forma sucinta en el Capítulo 7. Evidentemente, esta tarea requiere muchos de los conocimientos expuestos en los capítulos anteriores, y lo que se propone en este capítulo es cómo coordinarlos adecuadamente para el fin propuesto.

Finalmente, el Capítulo 8 está dedicado al tratamiento de un problema próximo a la realidad, aunque reducido a una escala que permite su tratamiento manual. Esta simplifi- cación sirve para poner de manifiesto cómo se usan las herramientas y técnicas propuestas y descritas en los siete primeros capítulos para resolver un problema de estudio muy con- creto: el análisis de las prestaciones del entorno de un servidor web. Se trata de un capítulo magnífico que actúa como síntesis de los conocimientos tratados en el libro, poniendo de manifiesto la coordinación que debe existir entre todas aquellas técnicas y herramientas.

En conclusión, el material que se encuentra este texto, aunque orientado principalmente a estudiantes universitarios, constituye la base idónea para cualquier técnico informático que deba hacer frente a problemas de evaluación de prestaciones, tanto para mejorar un sistema existente como para predecir el comportamiento de un sistema en el que alguno, o algunos, de sus componentes —del hardware o del software—, no estén disponibles y, por lo tanto, sus prestaciones no puedan evaluarse a partir de la medición directa.

Todas aquellas personas que lean o usen este libro deberían agradecer a Xavier Molero, Carlos Juiz, y Miguel J. Rodeño la ímproba tarea que representa la colección de informa- ción, su síntesis y exposición final mediante los problemas desarrollados, haciendo amena la lectura de temas que, de otra forma, resultarían absolutamente indigestos.

Ramon Puigjaner
*Catedrático de Universidad de Arquitectura y Tecnología de Computadores*
*Universitat de les Illes Balears*

Palma, abril de 2004

# Índice general

# Presentación

Las nuevas tecnologías empleadas en la computación distribuida y el desarrollo de Internet, no sólo como instrumentos de difusión y entretenimiento sino como herramienta de negocio, han cambiado la fisonomía de la informática en la última década. Con el auge en paralelo de los entornos cliente/servidor y de las comunicaciones, se ha aproximado la tecnología telemática a los usuarios finales, de tal modo que casi no se puede imaginar un ambiente de negocio que no considere la informática como una herramienta básica para su desarrollo. Sin embargo, tras una fase de descubrimiento de las funcionalidades y del potencial de de- sarrollo de estos sistemas, se ha pasado a la lógica preocupación por aspectos cuantitativos y no funcionales que ya se planteaban en arquitecturas centralizadas. Así, por ejemplo, la seguridad, la disponibilidad, la eficiencia o el rendimiento son algunas de las características que normalmente se evalúan a la hora de explotar un sistema informático. En particular, dado que el aplicativo de negocio descansa sobre sistemas cada vez más distribuidos, las consideraciones acerca del rendimiento de éstos produce consecuencias que se acusan en los resultados del propio negocio.

A nadie se le escapa que la imbricación entre una compañía y el soporte informático que lo sustenta ha tenido siempre un fiel reflejo en los costes o la inversión. Pero, defi- nitivamente, la proliferación de Internet y de las aplicaciones basadas en servidores web ha hecho posible una nueva forma de conseguir beneficio de la propia gestión de la información. Por ello, cada vez más, es necesario conocer con qué rapidez puede atender el sistema informático a las necesidades de los clientes, que no sólo usuarios, de los servicios ofrecidos telemáticamente. Aún más, el sistema puede ser la visión que los clientes perciben del propio negocio, con lo que uno de los aspectos esenciales en la explotación de éste, tal como lo fue en el pasado reciente, será conocer cómo evaluar su rendimiento, o, como se denomina en Latinoamérica, su desempeño. Este libro utiliza los conceptos clásicos de rendimiento para el análisis cuantitativo de los sistemas centralizados o distribuidos.

Las técnicas y modelos que se presentan, proveen del conocimiento básico para deter- minar el ámbito y magnitudes generales que sirven para representar el rendimiento de un sistema informático. No se ha pretendido cubrir todos los aspectos relacionados con la eva- luación del rendimiento, ni tampoco se ha profundizado en las particularidades de todas las situaciones, pero el texto permite descubrir aquellas nociones y técnicas fundamentales. El libro se ha enfocado a un plano eminentemente práctico; otros textos anteriores en castella- no han realizado un importante esfuerzo de condensación teórica que aquí se ven en parte actualizados y en parte extendidos con ejercicios, actividades propuestas y problemas.

El texto que el lector tiene en sus manos va dirigido principalmente a estudiantes y profesores universitarios que traten de materias de primeros cursos relacionadas con el aná- lisis cuantitativo de los sistemas informáticos y telemáticos. En este sentido, este trabajo consolida la experiencia de los autores en la impartición de varios cursos académicos sobre evaluación, explotación, modelado o análisis del rendimiento de los sistemas informáticos en universidades con estudios de informática. En consecuencia, su contenido se puede utilizar como herramienta docente aislada, o bien combinada con material más específico sobre las técnicas básicas aquí presentadas. Sin embargo, también puede ser utilizado por los profesionales de las tecnologías de la información, especialmente aquellos que han de garantizar cierta calidad de servicio del sistema que gestionan, ya que su enfoque es especialmente práctico. Aquí encontrarán los fundamentos que luego habrán de perfeccionar y focalizar en su entorno de explotación.

El libro se estructura en ocho capítulos que comparten una estructura similar (a excep- ción del último): tras una serie de apartados teóricos, se resuelven problemas siguiendo una orientación didáctica; después se plantean problemas adicionales de los que sólo se indica la solución, y a continuación se exponen una serie de problemas sin solución; el capítulo acaba con una propuesta de actividades complementarias relacionadas con la materia tra- tada. Respecto del contenido teórico hay que aclarar que su extensión es la mínima posible teniendo en cuenta que se ha intentado ofrecer un libro que sea autocontenido; esto es, con la información referida en los primeros apartados de cada capítulo el lector debe ser capaz de abordar, de manera satisfactoria, la resolución de los problemas planteados en él.

Respecto de la materia tratada en este texto, a continuación se refiere de manera sucinta el contenido de cada capítulo. El Capítulo 1 introduce el concepto del rendimiento, su relación con el coste y presenta la ley de Amdahl como medio básico de análisis. El Capítulo 2, se centra en la monitorización como herramienta indispensable para conocer el rendimiento de los sistemas informáticos en régimen de explotación. Dada la tecnología distribuida en la que estamos inmersos, se descubren los parámetros que los monitores basados en Unix (y Linux) pueden aportar en el análisis cuantitativo de un sistema. En el Capítulo 3 se presentan diversas técnicas para comparar el rendimiento de varios sistemas informáticos entre sí, poniendo especial énfasis en algunas de las unidades empleadas por programas de prueba (*benchmarks*) establecidos *de facto* en la comunidad informática.

Los Capítulos 4 y 5 se centran en el análisis operacional como herramienta clásica para resolver modelos simples de sistemas basados en redes de colas de espera. Se presentan también aplicaciones típicas de las leyes operacionales, como la determinación de los límites del rendimiento o del cuello de botella del sistema.

En el Capítulo 6 se introduce al lector en las técnicas para caracterizar la carga de trabajo a la que se somete a un sistema, incidiendo en la técnica de agrupamiento por clases. El Capítulo 7 muestra la utilidad de planificar la capacidad del sistema actual y predecir, aunque sea de forma sencilla, la capacidad futura a través de técnicas estadísticas simples.

Por último, el Capítulo 8 presenta el caso de estudio del rendimiento de un servidor web en una Intranet corporativa. Se han simplificado las técnicas y métodos para poder seguir con facilidad muchos de los conceptos y problemas resueltos en los capítulos precedentes, pero abarca lo suficiente para reconocer las tareas a las que se enfrenta un analista de prestaciones.

Los autores agradecen a todas aquellas personas que, de una u otra manera, han co- laborado en la elaboración de este material. Con su ayuda han contribuido a la difusión, en castellano, de este conocimiento clásico pero nunca tan actual como la evaluación y modelado del rendimiento de los sistemas informáticos.

Esperamos y deseamos que este trabajo resulte útil para el lector, porque éste era nues- tro principal objetivo, y le motive en el conocimiento de una materia que, particularmente a nosotros, nos apasiona. Finalmente, dada la profusión de problemas, y a pesar de que se han revisado a conciencia todos los capítulos, somos conscientes de que, todavía, algún error se haya podido deslizar en ellos. Confiamos en la comprensión del lector y le pedimos disculpas de antemano, toda vez que quedamos a su disposición para incorporar, en el futuro, las sugerencias que tiendan a mejorar este texto.

# Capítulo 1

## Introducción a la evaluación de rendimiento

Una manera sencilla —y a la vez intuitiva— de comparar sin subterfugios el rendimiento de diversos sistemas informáticos es utilizar como medida de prestaciones el tiempo de eje- cución de un programa o, más habitualmente, un conjunto de programas. Estos programas representan la carga de prueba en la que se basarán los resultados del estudio comparativo. Desde esta perspectiva, el computador más rápido será aquel que ejecute los programas en el menor tiempo. Sin embargo, huelga decir que la conclusión de cualquier estudio depende enteramente de los programas utilizados, y que cargas distintas pueden proporcionar, en muchos casos, conclusiones diferentes.

Por otro lado, todos los computadores tienen un precio que depende tanto de los costes de diseño como de los de fabricación y comercialización. Por ello resulta muy interesante relacionar el rendimiento de los sistemas informáticos con el coste. Por ejemplo, un com- prador potencial podría pensar ante la disyuntiva de elegir entre dos computadores: este computador cuesta el doble que otro, pero. . . ¿es el doble de rápido? Un diseñador, por su parte, también tiene que decidir si vale la pena tomar decisiones de diseño que incre- mentan el precio del computador a costa de mejorar el rendimiento del producto final. En consecuencia, será necesario establecer relaciones entre el precio y el coste para poder elegir entre varios productos o alternativas. Sin embargo, y como veremos más adelante, esta relación suele ser difícil de establecer, por lo que se suelen adoptar soluciones bastante elementales. La relación más utilizada es del tipo rendimiento/coste, esto es, consiste en dividir la medida de rendimiento entre el coste.

En lo que a nomenclatura se refiere, se emplearán los términos *rendimiento* y *presta- ciones* de forma equivalente, traducciones ambas del término anglosajón *performance*. A esta pareja de términos podemos añadir un tercero, *desempeño*, que es la variante más empleada en el ámbito hispanoamericano.

## 1.1. Relación entre rendimientos

En ocasiones resulta interesante hablar de rendimiento de un computador como el inverso del tiempo que tarda en ejecutar un programa. De esta manera, cuanto más rápido ejecute el programa, más alto será su rendimiento. En este apartado presentamos una manera muy simple de comparar las prestaciones de dos sistemas informáticos a partir de la ejecución de un programa. Consideremos dos computadores X e Y, los cuales tardan $T_X$ y $T_Y$ unidades de tiempo, respectivamente, en ejecutar este programa. Si $T_X = T_Y$ diremos que el rendimiento de ambas máquinas es igual o equivalente, ya que en ambas obtenemos el mismo tiempo de ejecución.

Imaginemos, por el contrario, que $T_X < T_Y$, esto es, el computador X tarda menos tiempo en ejecutar el programa. Esta relación nos permite afirmar que "X es más rápido que Y". Sin embargo, nuestro objetivo es cuantificar esta relación y decir que "X es *tantas veces más* rápido que Y". El valor numérico al que nos estamos refiriendo recibe el nombre de aceleración (*speedup*) y se puede calcular como la relación entre el tiempo de ejecución más grande y el más pequeño:

$$\text{Aceleración} = \frac{T_Y}{T_X}$$

Por lo tanto, la aceleración representa el incremento de rendimiento de una máquina respecto de la otra.

La manera de expresar esta aceleración en palabras adquiere múltiples formas. Por ejemplo, se puede hablar de mejora y decir que "X es tantas veces mejor que Y", pero en este caso hay que ser más preciso, porque el término *mejor* es una gradación del adjetivo *bueno*, cuyo significado depende de cada situación particular y no se restringe únicamente al rendimiento. Finalmente, hay ocasiones en que esta aceleración se expresa en términos porcentuales, esto es, "X es un $n$ % más rápido que Y", en cuyo caso la relación anterior se expresa:

$$\text{Aceleración} = \frac{T_Y}{T_X} = 1 + \frac{n}{100}$$

Imaginemos, por ejemplo, que $T_X = 36$ y $T_Y = 40$ segundos. En consecuencia, podemos afirmar que X es $40/36 = 1{,}11$ veces más rápido que Y o, alternativamente, que X es un 11 % más rápido que Y.

Respecto a la notación, y dependiendo del contexto, la aceleración o incremento de rendimiento suele representarse por $A$, $A_r$ o $\Delta A$.

## 1.2. El coste también cuenta

La comparación de precios entre computadores se puede llevar a cabo de la misma manera que la empleada para el rendimiento. Por ejemplo, si los costes de los computadores X e Y son $C_X$ y $C_Y$, respectivamente, el incremento (o también aceleración) del coste de una opción respecto de la otra se puede expresar dividiendo el coste más elevado entre el más bajo. Si suponemos que $C_X > C_Y$, entonces podemos escribir:

$$\text{Incremento} = \frac{C_X}{C_Y} = 1 + \frac{n}{100}$$

En consecuencia, esta expresión nos permitirá decir que "X es tantas veces más caro que Y", o que "X es un $n$ % más caro que Y". Por ejemplo, si $C_X = 625$ € y $C_Y = 550$ €, entonces se puede decir que X es $625/550 = 1{,}14$ veces más caro o, alternativamente, que X es un 14 % más caro que Y.

Respecto a la notación utilizada, y para diferenciar esta aceleración o incremento de coste con su homólogo de rendimiento, el primero se suele expresar mediante las variables $A_c$ o $\Delta C$. En el contexto de costes, sin embargo, resulta más intuitivo hablar de incremento de coste que de aceleración.

## 1.3. Relación entre prestaciones y coste

Sigamos con el supuesto de comparación del rendimiento de dos computadores X e Y mediante el tiempo de ejecución de un programa. Las expresiones empleadas hasta ahora nos han permitido cuantificar, de manera aislada, la relación entre sus prestaciones y la relación entre sus costes. Para realizar un análisis conjunto de precio y prestaciones no queda más remedio que establecer algún tipo de conexión entre ambas. Por ejemplo, se puede dividir el rendimiento de cada computador entre su coste y comparar ambas cantidades:

$$\frac{\text{Rendimiento}_X}{\text{Coste}_X} \quad \text{vs.} \quad \frac{\text{Rendimiento}_Y}{\text{Coste}_Y}$$

Las cantidades anteriores nos pueden ayudar a conocer qué opción de las dos, en su conjunto, ofrece una mejor relación entre el rendimiento obtenido y el precio que se va a pagar por él. Nótese que resultará mejor aquel sistema que obtenga el valor más elevado. Si aplicamos la expresión anterior para el ejemplo que estamos analizando obtenemos los siguientes valores:

$$\frac{\text{Rendimiento}_X}{\text{Coste}_X} = \frac{1}{T_X \times \text{Coste}_X} = \frac{1}{36 \times 625} = 4{,}44 \times 10^{-5}$$

$$\frac{\text{Rendimiento}_Y}{\text{Coste}_Y} = \frac{1}{T_Y \times \text{Coste}_Y} = \frac{1}{40 \times 550} = 4{,}55 \times 10^{-5}$$

Atendiendo a la relación entre prestaciones y coste podemos observar que la diferencia existente entre los dos computadores es muy pequeña. En particular, y dado que interesa maximizar este índice, el computador Y resulta ligeramente superior al computador X. Finalmente, nótese que en las expresiones anteriores hemos expresado el rendimiento del computador como el inverso del tiempo de ejecución de un programa; sin embargo, en general, el índice que se utiliza para expresar el rendimiento depende de cada estudio en particular.

Otro posible contexto, diferente al anterior, viene dado cuando se trata de analizar el efecto de una determinada mejora en un sistema informático. Por ejemplo, imaginemos que disponemos de un computador comprado hace cierto tiempo por un precio de 1.200 € al cual queremos añadir un nuevo disco duro que cuesta 350 €. ¿Cuál es el incremento de coste $\Delta C$ que supondrá hacer la actualización? El valor de este incremento se puede calcular dividiendo el coste del equipo con el componente añadido entre el coste del computador original:

$$\Delta C = \frac{\text{Nuevo coste}}{\text{Coste original}} = \frac{1.200 + 350}{1.200} = 1{,}29$$

En caso de que, en vez de añadir, se reemplazaran componentes, como por ejemplo un procesador, el cálculo del incremento de coste se puede hacer de otra manera, ya que el elemento que se reemplaza deja de formar parte del sistema. Imaginemos que el procesa- dor original costó 325 € y el nuevo cuesta 475 €. Una primera aproximación consiste en considerar que el coste del elemento a reemplazar hay que descontarlo del sistema, puesto que ya no forma parte de él. En este caso el incremento de coste será:

$$\Delta C = \frac{\text{Nuevo coste}}{\text{Coste original}} = \frac{1.200 - 325 + 475}{1.200} = 1{,}125$$

Una segunda aproximación supone que el coste del sistema, una vez hecho el reemplazo, incluye el precio que se pagó por el componente reemplazado:

$$\Delta C = \frac{\text{Nuevo coste}}{\text{Coste original}} = \frac{1.200 + 475}{1.200} = 1{,}4$$

Sin embargo, e independientemente de que haya reemplazo o no de componentes, cuan- do se trata de comparar entre sí diversas alternativas para la actualización de un sistema, el enfoque anterior adolece de que el incremento de coste del sistema global puede estar muy influenciado por el coste original. Esto es así, sobre todo, para aquellos componentes con mucha influencia en el rendimiento pero con poca repercusión en el coste de todo el sistema. Para evitar este problema procederemos como se indica a continuación.

Imaginemos un computador al que se le va a reemplazar el procesador, y en el que un programa de prueba tarda un tiempo $T$ en ejecutarse. Para efectuar el reemplazo existen dos alternativas posibles de costes: $C_1$ y $C_2$, respectivamente. Empleando cada una de las alternativas se obtienen tiempos de ejecución $T_1$ y $T_2$, respectivamente, con $T_1 < T$ y $T_2 < T$ (en caso contrario no habría mejora de rendimiento). Si hay que elegir el proce- sador que obtiene una mejor relación entre prestaciones y coste, entonces la comparación de rendimiento se hará basándose en el tiempo de ejecución del programa en el sistema completo (variables $T_1$ y $T_2$). Sin embargo, la comparación de costes se hará teniendo en cuenta únicamente el de cada alternativa (variables $C_1$ y $C_2$), dejando de lado el coste del sistema completo.

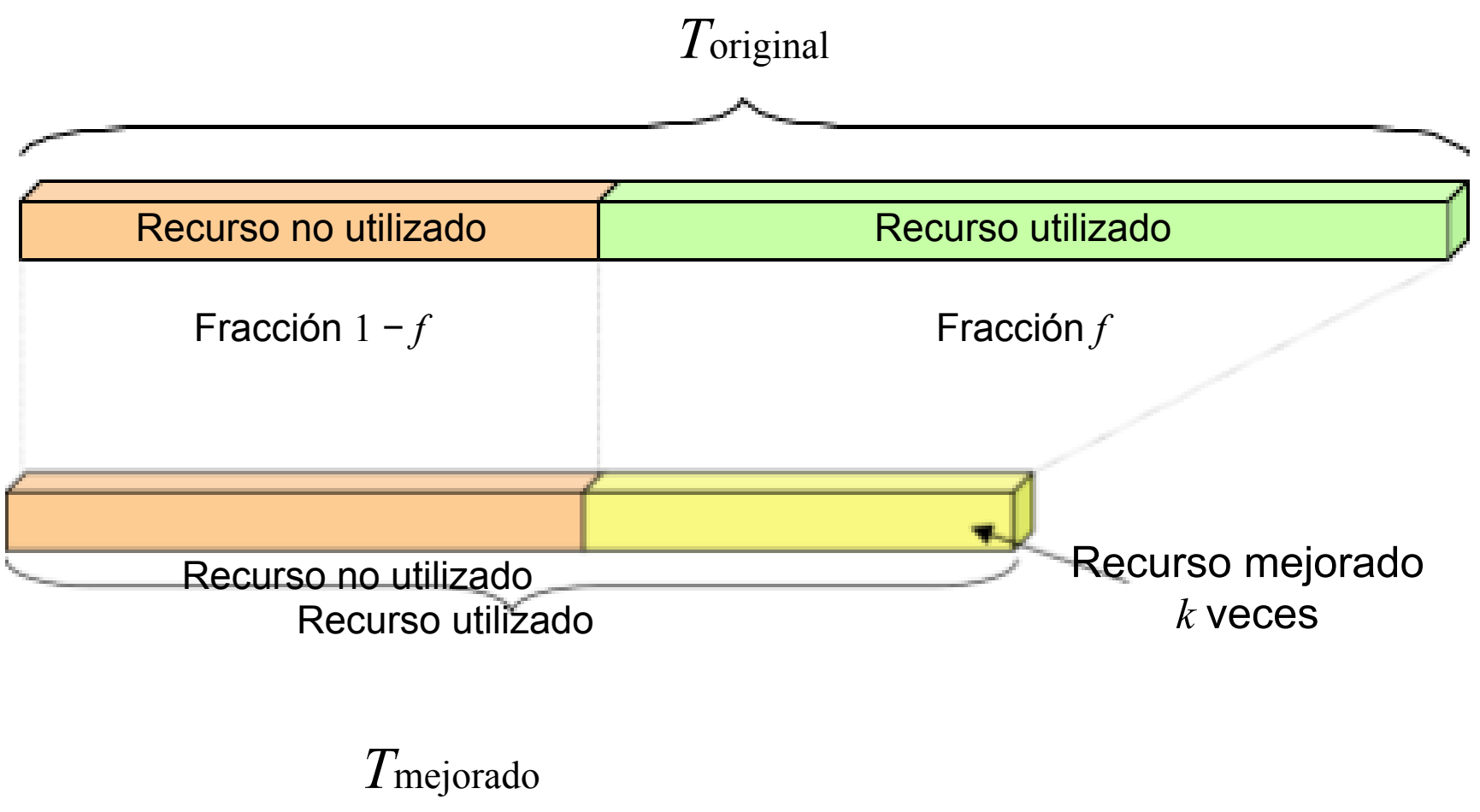

**Figura 1.1:** Tiempo original versus tiempo mejorado.

## 1.4. La ley de Amdahl

La ley que presentamos en este apartado acota, de una manera muy sencilla, el incremento de prestaciones obtenido en un sistema como consecuencia de la mejora de una o varias partes del mismo. Esta mejora, representada como incremento de rendimiento o velocidad, va a depender tanto de la calidad de las mejoras efectuadas como del tiempo en que éstas se utilicen. De forma abreviada, podemos decir que este incremento de prestaciones dará la medida de cómo un computador rinde, en relación con un rendimiento previo, después de efectuar en él una o varias mejoras.

Consideremos, para simplificar nuestro planteamiento, un computador que tarda un tiempo $T_{original}$ en ejecutar una determinada aplicación, y que nuestro objetivo es reducir este tiempo de ejecución mejorando una de las partes del computador. Supongamos que durante una fracción $f$ del tiempo original el programa hace uso exclusivo de un recurso del sistema (por ejemplo, el procesador). En consecuencia, podemos expresar $T_{original}$ como la suma de dos componentes disjuntos: uno en el que no se utiliza este componente más otro en el que sí se utiliza:

$$T_{original} = T_{original} \times (1 - f) + T_{original} \times f$$

La Figura 1.1 muestra gráficamente la relación entre el tiempo original y el tiempo obtenido tras mejorar el recurso en un factor de $k$. El nuevo tiempo de ejecución $T_{mejorado}$ que se obtendrá después de mejorar $k$ veces el recurso afectado puede calcularse teniendo
en cuenta que el segundo sumando de la expresión anterior se reducirá en un factor de $k$. Por tanto, podemos escribir:

$$T_{mejorado} = T_{original} \times (1 - f) + \frac{T_{original} \times f}{k}$$

Esta nueva expresión pone de manifiesto que el incremento de prestaciones conseguido con la mejora del recurso depende de la fracción de tiempo en que se emplea. Si dividimos ahora el tiempo original entre el tiempo mejorado obtendremos la cuantificación de la mejora de prestaciones global $A$ obtenida, referida también de manera abreviada como aceleración (*speedup*):

$$A = \frac{1}{1 - f + \frac{f}{k}}$$

La expresión anterior recibe el nombre de ley de Amdahl. Nótese que podemos conside- rar dos casos particulares. Si $f = 0$, entonces $A = 1$; es decir, cuando la fracción de tiempo en que se utiliza el componente mejorado es nula, no se consigue ninguna aceleración en el sistema global. Por contra, si $f = 1$, entonces $A = k$; esto es, la aceleración obtenida en el sistema global será equivalente al factor de mejora del componente si éste se utiliza durante todo el tiempo.

Resulta fácil examinar qué pasa cuando el factor de mejora se hace muy grande, esto es, $k \to \infty$. En este caso tendremos:

$$\lim_{k\to\infty} A = \lim_{k\to\infty} \frac{1}{1 - f + \frac{f}{k}} = \frac{1}{1 - f}$$

Este resultado pone de manifiesto que, independientemente de la mejora llevada a cabo en un sistema, el incremento de prestaciones global está limitado intrínsecamente por las operaciones que no están afectadas por esta mejora. Por ejemplo, imaginemos un programa que utiliza el procesador durante el 95 % del tiempo, mientras que el 5 % restante lo emplea en acceder a los dispositivos de entrada/salida. En este supuesto, la aceleración global más alta que se puede conseguir en la ejecución del programa reemplazando el procesador por uno nuevo más rápido resulta de 1/0,05 = 20. En particular, si este último fuera 2 veces más rápido que el original, la aceleración global sería de 1/(0,05+0,95/2) = 1,91, valor muy cercano a 2, que es la mejora neta del procesador; sin embargo, si el nuevo procesador fuese 50 veces más rápido, la aceleración global sería solamente de 1/(0,05 + 0,95/50) = 14,49, valor muy alejado de la mejora neta del procesador que es 50.

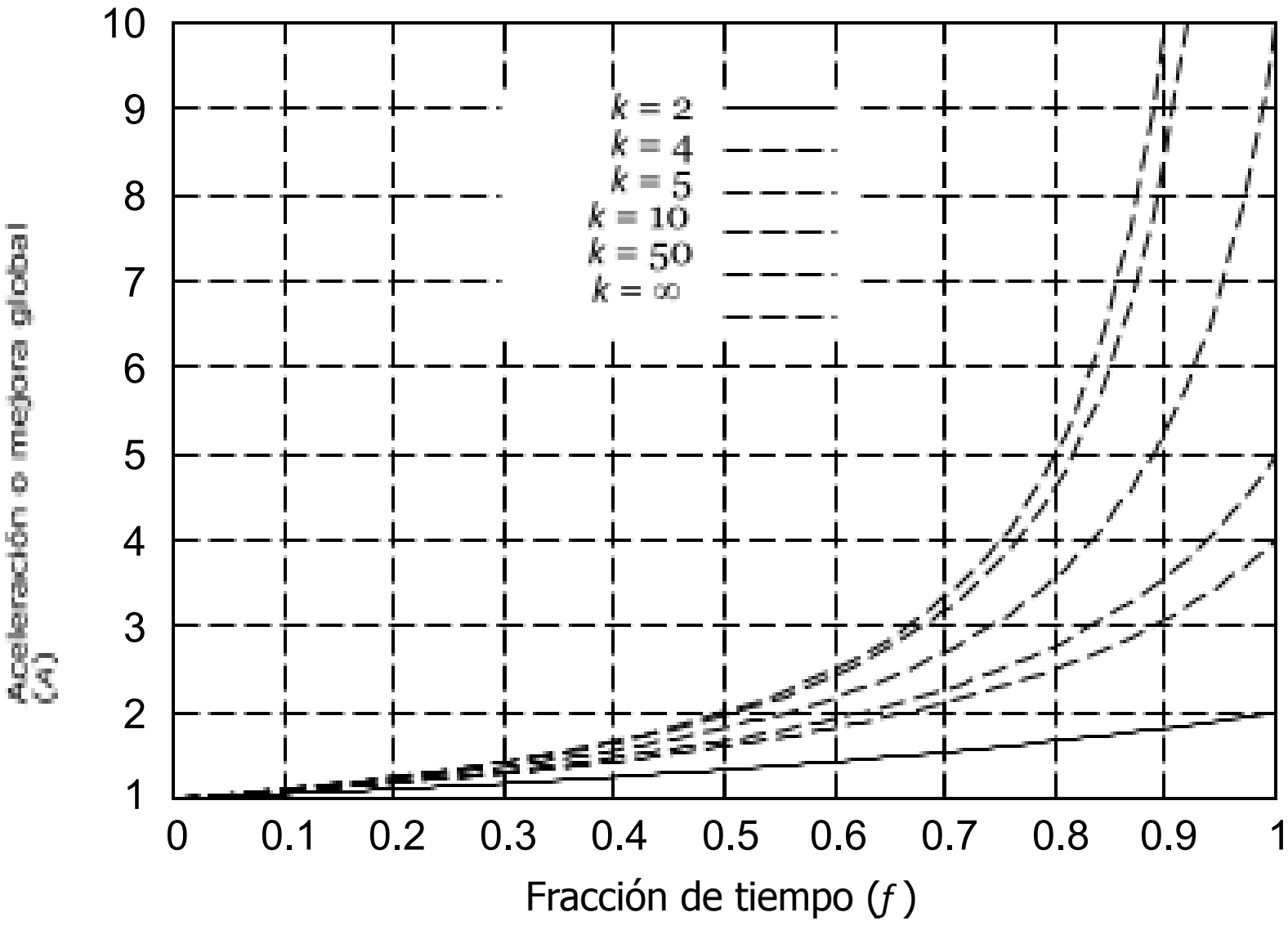

**Figura 1.2:** Relación entre la aceleración global $A$ y el factor de mejora $k$.

A veces también resulta interesante, conocidas $k$ y $A$, calcular la fracción de tiempo $f$. La expresión que relaciona estas variables, obtenida de la anterior, es la siguiente:

$$f = \frac{k \times (A-1)}{A \times (k-1)}$$

La Figura 1.2 muestra gráficamente, para diferentes valores del factor de mejora $f$, la relación entre la aceleración conseguida $A$ por el sistema global en función de la fracción de tiempo de uso del recurso mejorado. Tal como se aprecia en esta figura, las diferencias entre las curvas representadas comienzan a ser importantes a partir de fracciones de utilización superiores al 50 %, lo que indica que la traslación efectiva de factores altos de mejora en la aceleración global es significativa para utilizaciones altas del recurso mejorado.

La ley de Amdahl puede generalizarse fácilmente al caso en que se lleven a cabo mejoras sobre más de un recurso. En efecto, si se mejoran $n$ recursos del sistema en factores $k_1$, $k_2$, . . ., $k_n$, y cada uno de ellos se utiliza de manera exclusiva durante las fracciones de tiempo $f_1, f_2, \ldots, f_n$, respectivamente, la mejora global o aceleración obtenida se puede expresar:

$$A = \frac{1}{f_0 + \sum_{i=1}^{n} \frac{f_i}{k_i}}, \text{ con } f_0 = 1 - \sum_{i=1}^{n} f_i$$

La Figura 1.3 muestra gráficamente la evolución de la aceleración global $A$ en función de dos factores de mejora $k_1$ y $k_2$, con valores máximos de 10 y 5, respectivamente. La primera mejora se aplica durante el 60 % del tiempo, mientras que la segunda se usa en el 30 %. Nótese que el valor máximo de $A$ es 4,545 y se alcanza cuando el factor de ambas mejoras es el más alto.

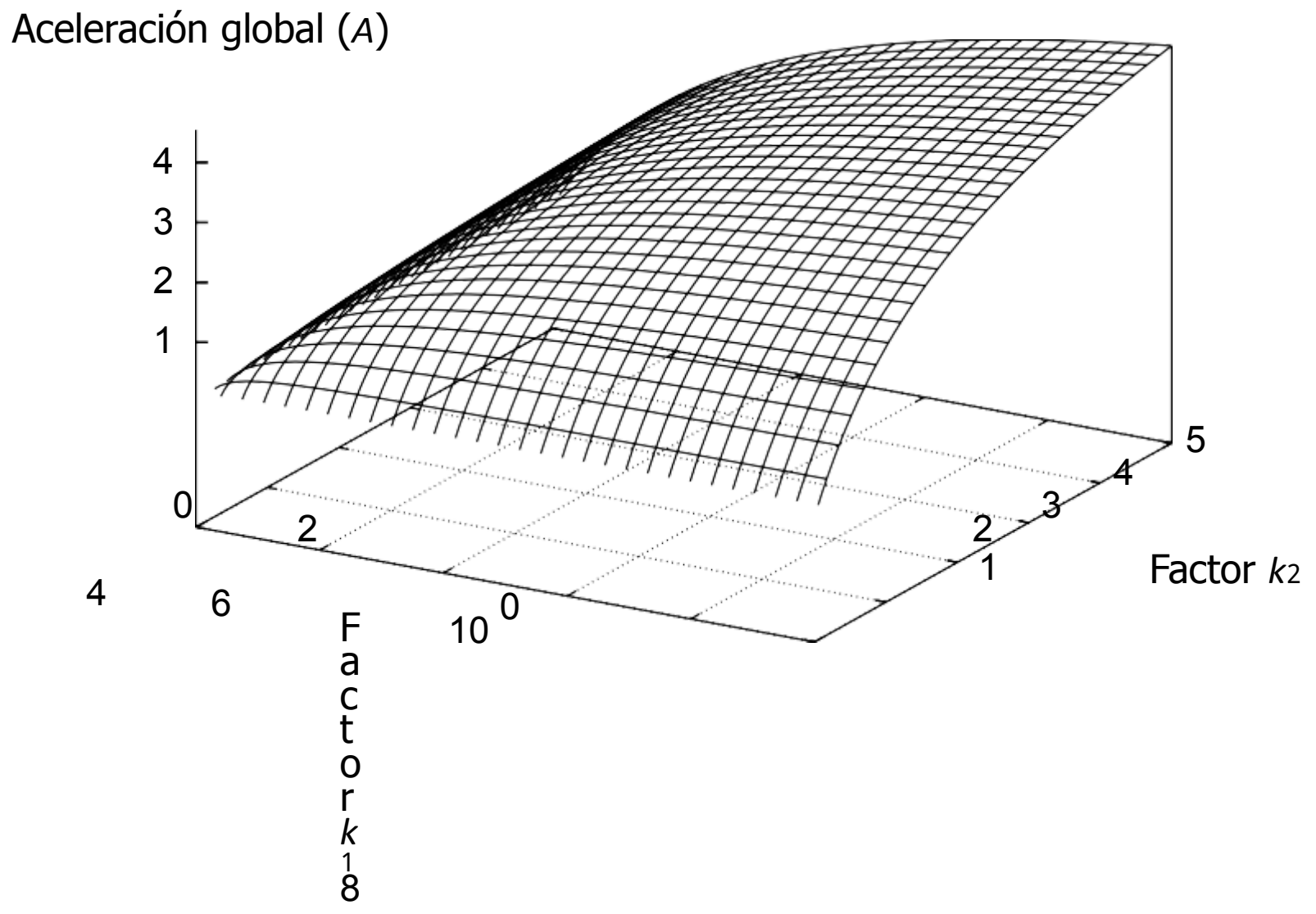

**Figura 1.3:** Aceleración global $A$ para dos factores de mejora $k_1$ y $k_2$.

Las ecuaciones anteriores muestran de manera directa que una mejora es más efectiva cuanto mayor es la fracción de tiempo en que se aplica. Así, para mejorar un sistema complejo, como por ejemplo una unidad central de proceso, es necesario optimizar aquellos elementos que se utilizan durante la mayor parte del tiempo. Esta estrategia se resume en dirigir todos los esfuerzos de optimización hacia el caso más común. Por ejemplo, la ruta de datos dentro del procesador, o las instrucciones más frecuentes en el diseño del repertorio de instrucciones.

Finalmente, no hay que perder de vista que la hipótesis de partida de la ley presupone la utilización de forma exclusiva de los componentes mejorados. Por tanto, hay que tener en cuenta el cumplimiento de esta condición para establecer la validez de los resultados obtenidos mediante esta ley, lo que en algunos casos prácticos limita su aplicabilidad.

## 1.5. Problemas resueltos

**PROBLEMA 1.1** Dos computadores A y B ejecutan el programa ESCACS en 35 y 87 segundos, respectivamente. Si sus costes son de 710 y 650 €, respectivamente, calcúlese cuál de ellos presenta una mejor relación entre prestaciones y coste.

**SOLuCión:** El computador A es más rápido que B, ya que tarda menos tiempo en ejecutar el programa que sirve de base al análisis de rendimiento. Esta mejora se calcula como:

$$A_r = \frac{87}{35} = 2{,}49$$

Por tanto, A es 2,49 veces más rápido que B. Sin embargo, el primero resulta más caro que el segundo. En particular,

$$\Delta C = \frac{710}{650} = 1{,}09$$

El computador A es un 1,09 veces más caro que B. Si comparamos el incremento de rendimiento de A respecto a B frente al incremento de coste, podemos ver que el primero es bastante más alto que el segundo (2,49 frente a 1,09). Si dividimos el rendimiento entre el coste obtendremos:

$$\frac{\text{Rendimiento}_A}{\text{Coste}_A} = \frac{1}{35 \times 710} = 4{,}024 \times 10^{-}$$

$$\frac{\text{Rendimiento}_B}{\text{Coste}_B} = \frac{1}{87 \times 650} = 1{,}768 \times 10^{-}$$

En consecuencia, la relación entre prestaciones y coste resulta mucho más eficiente para el computador A que para el B. Si dividimos estas dos cantidades entre sí, se puede comprobar que esta eficiencia es más de dos veces superior en A que en B. ■

---

**PROBLEMA 1.2** La utilización del procesador de un sistema informático es del 83 %. Calcúlese la mejora de rendimiento que se conseguirá si éste se sustituye por uno nuevo 2,5 veces más rápido.

**SOLUCIÓN:** La mejora introducida por el nuevo procesador será efectiva en una fracción de tiempo de 0,83, periodo durante el cual el tiempo de procesamiento se reducirá en un factor de 2,5. La mejora global de prestaciones se calcula aplicando la ley de Amdahl:

$$A = \frac{1}{1 - f + \frac{f}{k}} = \frac{1}{1 - 0{,}83 + \frac{0{,}83}{2{,}5}} = 1{,}99$$

En consecuencia, la actualización del sistema consigue una mejora de 1,99 respecto del computador original. ■

---

**PROBLEMA 1.3** Se quiere mejorar el rendimiento de un computador mediante la introducción de una unidad de coma flotante (FPU, *floating point unit*). Esta unidad permite reducir a la mitad el tiempo dedicado a la realización de las operaciones aritméticas. Calcúlese la mejora de rendimiento que se conseguirá si la aplicación ejecutada dedica el 60 % del tiempo a hacer cálculo aritmético. Si el programa tardaba 12 segundos en ejecutarse sin unidad de coma flotante, ¿cuánto tiempo tardará el computador en ejecutarlo después de añadir esta unidad?

**SOLuCIón:** La aceleración conseguida con la adición de esta nueva unidad se puede calcular fácilmente aplicando la ley de Amdahl:

$$A = \frac{1}{1 - f + \frac{f}{k}} = \frac{1}{0{,}4 + \frac{0{,}6}{2}} = 1{,}43$$

Por lo tanto, la mejora en el tiempo de ejecución es de 1,43 o del 43 %. El nuevo tiempo de ejecución puede calcularse directamente a partir del resultado anterior dividiendo el tiempo original entre el factor de mejora:

$$T_{mejorado} = \frac{T_{original}}{A} = \frac{12}{1{,}43} = 8{,}4 \text{ s}$$

o también, de manera directa, calculando el nuevo tiempo a partir de la fracción del mismo donde se aplica la mejora:

$$T_{mejorado} = T_{original} \times (1 - f) + \frac{T_{original} \times f}{k} = 12 \; 0{,}4 + \frac{12 \times 0{,}6}{2} = 8{,}4 \text{ s}$$

■

---

**PROBLEMA 1.4** Con el objetivo de mejorar el rendimiento de un computador se dispone de dos opciones diferentes:

1. Ampliar la memoria principal (250 €), con lo que se consigue que el 50 % de los programas se ejecuten tres veces más rápidamente.
2. Cambiar la placa base (150 €), con lo que el 70 % de los programas se pueden ejecutar en la mitad de tiempo.

Calcúlese qué opción de la dos presenta la mejor relación entre prestaciones y coste.

**SOLuCIón:** En primer lugar hay que calcular el incremento de rendimiento obtenido por cada una de las alternativas. Si etiquetamos con 1 y 2 a la primera y segunda alternativa, respectivamente, entonces podemos escribir aplicando la ley de Amdahl:

$$A_1 = \frac{1}{0{,}5 + \frac{0{,}5}{3}}$$

$$A_2 = \frac{1}{0{,}3 + \frac{0{,}7}{2}}$$

= 1,50

= 1,54

La segunda opción presenta una aceleración ligeramente superior a la primera. Si atende- mos a los costes, podremos observar que la primera opción resulta más cara. En concreto, el incremento de coste de la primera opción respecto de la segunda es:

$$\Delta C = \frac{250}{150} = 1{,}67$$

En consecuencia, la mejor opción es la segunda, ya que permite conseguir un rendimiento ligeramente superior a la primera y además resulta menos costosa. ■

---

**PROBLEMA 1.5** Un sistema informático actualiza su unidad de disco por una nueva 5 veces más rápida que la original.

1. ¿Qué utilización ha de tener esta unidad de disco si se quiere conseguir una aceleración global de rendimiento de 2?
2. ¿Y si se quiere conseguir una aceleración global de 6?

**SOLuCıón:**

1. En este caso basta con aplicar la ley de Amdahl y calcular el valor de la fracción de tiempo que hay que mejorar:

   $$f = \frac{k \times (A-1)}{A \times (k-1)} = \frac{5 \times (2-1)}{2 \times (5-1)} = 0{,}625$$

   Así, la unidad de disco ha de tener una utilización del 62,5 % para que el rendimiento del sistema experimente una mejora de 2.
2. Es imposible obtener una aceleración global de 6, ya que el componente mejorado lo hace en un factor menor, en concreto, de 5. El valor máximo de la mejora, por tanto, será de 5 siempre que la nueva unidad de disco se utilice durante el 100 % del tiempo. ■

---

**PROBLEMA 1.6** Un programa tarda en ejecutarse un total de 124 segundos. Durante este tiempo el procesador está ejecutando tres tipos diferentes de instrucciones: aritméticas de enteros, salto y coma flotante. La proporción del tiempo de ejecución en que se emplea cada tipo es del 28, 40 y 32 %, respectivamente. Se pide:

1. Calcular el incremento de prestaciones si se mejoran un 15 y un 45 % las instrucciones de aritmética entera y de salto, respectivamente.
2. Determinar cuánto se tienen que mejorar las operaciones de coma flotante si queremos rebajar el tiempo de ejecución original hasta los 95 segundos (con solo esta mejora).

**SOLUCIÓN:**

1. El tiempo de ejecución original se puede desglosar en el dedicado a cada tipo de instruc- ción:

$$124 = 124 \times (0{,}28 + 0{,}40 + 0{,}32) = 34{,}72 + 49{,}60 + 39{,}68$$

Las mejoras introducidas en los dos tipos de instrucciones afectan a los dos primeros sumandos, los cuales se rebajarán de acuerdo con el valor de estas mejoras. En particular, si las mejoras para las instrucciones de aritmética entera y de salto son del 15 y el 45 %, respectivamente, el nuevo tiempo de ejecución será:

$$T_{\text{mejorado}} = \frac{34{,}72}{1{,}15} + \frac{49{,}60}{1{,}45} + 39{,}68 = 30{,}19 + 34{,}21 + 39{,}68 = 104{,}08 \text{ s}$$

Por lo tanto, el incremento de rendimiento experimentado en el tiempo de ejecución del programa una vez introducidas las mejoras es:

$$A = \frac{T_{\text{original}}}{T_{\text{mejorado}}} = \frac{124}{104{,}08} = 1{,}19$$

Esta aceleración se podría haber calculado, de una manera más directa, utilizando la ley de Amdahl de la siguiente manera:

$$A = \frac{1}{0{,}32 + \frac{0{,}28}{1{,}15} + \frac{0{,}40}{1{,}45}} = 1{,}19$$

2. Si se rebaja el tiempo de ejecución original de 124 a 95 segundos, el incremento de rendimiento conseguido es de $124/95 = 1{,}31$. Esta aceleración se alcanza mejorando las instrucciones de coma flotante en una cantidad $k$ que desconocemos, pero que podemos calcular a partir de la siguiente relación:

$$\frac{1}{0{,}28 + 0{,}40 + \frac{0{,}32}{k}}$$

La resolución de esta ecuación establece un valor $k = 3{,}72$, esto es, las instrucciones de coma flotante han de mejorar 3,72 veces para conseguir un incremento de rendimiento de 1,31 en la ejecución del programa completo. ■

**PROBLEMA 1.7** Un computador sin memoria cache ejecuta un programa en 180 segun- dos. Si se incorpora una memoria de este tipo que funciona 15 veces más rápidamente que la memoria principal, calcúlese el porcentaje de accesos a memoria que deben satisfacerse por la memoria cache para que el programa consiga ejecutarse en 80 segundos.

**SOLuCIón:** Si se quiere reducir el tiempo original de ejecución de 180 a 80 segundos, el incremento de prestaciones a conseguir es de 180/80 = 2,25. Teniendo en cuenta que el factor de mejora en la memoria es $k = 15$, el valor de $f$ que queremos calcular será:

$$f = \frac{k \times (A - 1)}{A \times (k - 1)} = \frac{15 \times (2{,}25 - 1)}{2{,}25 \times (15 - 1)} = 0{,}60$$

Por tanto, para ejecutar el programa en 80 segundos habría que conseguir que al menos el 60 % de los accesos a memoria principal fueran satisfechos por la memoria cache. Evidentemen- te, esto no es más que una aproximación, puesto que el resultado final depende de la política de reemplazo que siga la memoria cache. ■

---

**PROBLEMA 1.8** El 70 % de las tareas de una aplicación informática son susceptibles de ser paralelizadas para su ejecución en un sistema multiprocesador. Si esta aplicación tarda 120 segundos en ejecutarse en una máquina secuencial, se pide:

1. ¿Qué aceleración experimentará la ejecución del programa en un sistema con 16 procesadores? ¿En cuánto tiempo se ejecutará el programa?
2. ¿En cuánto tiempo se ejecutará el programa si se dispusiera de infinitos procesadores?

**SOLuCIón:**

1. La mejora global conseguida en la ejecución del programa con 16 procesadores afecta a una fracción del 70 % del tiempo y su valor se calcula como:

$$\frac{1}{1 - f + \frac{f}{k}} = \frac{1}{0{,}3 + \frac{0{,}7}{16}} = 2{,}91$$

Nótese que esta aceleración se conseguirá siempre que seamos capaces de dividir el pro- grama secuencial en 16 subprogramas independientes (paralelización perfecta). Teniendo en cuenta esta hipótesis, con 16 procesadores ejecutando tareas de la aplicación, la me- jora conseguida es de 2,91. El tiempo de ejecución se puede calcular dividiendo el tiempo original entre el factor de mejora:

$$T_{\text{mejorado}} = \frac{T_{\text{original}}}{A} = \frac{120}{2{,}91} = 41{,}24 \text{ s}$$

2. Con infinitos procesadores, y suponiendo que las tareas pueden ser de un tamaño ar- bitrariamente pequeño, la parte paralela del tiempo de ejecución se reduciría hasta ser despreciable. Quedaría solamente, por tanto, la componente puramente secuencial de este tiempo:

$$T_{mejorado} = T_{original} \times 0{,}3 + \frac{T_{original} \times 0{,}7}{\infty} = 36 \text{ s}$$

La aceleración global que se conseguiría en el rendimiento en este caso vendría dada por la relación entre el tiempo de ejecución original y el tiempo mejorado que acabamos de calcular:

$$A = \frac{T_{original}}{T_{mejorado}} = \frac{120}{36} = 3{,}33$$

■

---

**PROBLEMA 1.9** Durante la fase de diseño de un computador se ha introducido una me- jora que incrementa en un factor $k = 10$ el rendimiento local de un cierto recurso. Esta mejora ha permitido reducir el tiempo de ejecución original de un programa. Una vez in- troducida la mejora se ha podido estimar que ésta se emplea durante el 50 % del tiempo mejorado.

1. ¿Cuál es el porcentaje del tiempo original de ejecución del programa afectado por la mejora?
2. ¿Cuál es la aceleración global conseguida?

**SOLuCıón:**

1. En primer lugar hay que señalar que la fracción del 50 % hace referencia al tiempo de ejecución después de introducir la mejora, y no se puede emplear directamente en la ley de Amdahl porque ésta utiliza la fracción del tiempo de ejecución antes de usar la mejora.

   Partimos de la siguiente expresión para el tiempo de ejecución mejorado:

$$T_{mejorado} = (1 - f) \times T_{original} + \frac{f}{k} \times T_{original}$$

   Como la mejora se usa durante el 50 % del tiempo de ejecución mejorado, entonces podemos asegurar que este último se divide en dos partes iguales, esto es:

$$(1 - f) \times T_{original} = \frac{f}{k} \times T_{original}$$

Resolviendo esta ecuación obtenemos:

$$f = \frac{10}{10+1} = 0{,}909$$

En consecuencia, la fracción del tiempo original durante el cual se utiliza la mejora es del 90,9 %.

2. La aceleración global se puede calcular aplicando directamente la ley de Amdahl y em- pleando la fracción de tiempo calculada en el apartado anterior:

$$A = \frac{1}{1 - f + \frac{f}{k}} = \frac{1}{0{,}091 + \frac{0{,}909}{10}} = 5{,}498$$

■

**PROBLEMA 1.10** Consideremos un computador dedicado exclusivamente a la gestión de reserva de billetes de un aeropuerto regional donde el procesador actual, que funciona a 66 MHz, es el cuello de botella, con una utilización media del 84 % (el 16 % restante se usa en tareas de entrada y salida). Se pide determinar cuál de las dos opciones siguientes para reemplazar el procesador presenta la mejor relación entre prestaciones y coste:

1. Procesador Bart a 550 MHz, con un precio de 480 €.
2. Procesador Homer a 500 MHz, con un precio de 270 €.

En ambos casos se supone que el computador admite los dos procesadores y que no hace falta recompilar la aplicación de reserva de billetes.

**SOLuCIón:** La solución de este problema requiere calcular la mejora en el rendimiento a partir de la frecuencia de funcionamiento del procesador, que es el único dato relativo a prestaciones de que disponemos, y compararla respecto al coste de cada opción.

La mejora de rendimiento para cada uno de los procesadores se calculará teniendo en cuenta que el factor de mejora local $k$ es la relación entre las frecuencias de reloj. En el caso de Bart esta mejora es $k = 550/66 = 8{,}33$, mientras que para Homer tendremos $k = 500/66 = 7{,}58$. La repercusión del incremento en la frecuencia de reloj depende de la fracción de tiempo en que el procesador se esté utilizando, y se calculará mediante la ley de Amdahl. Así, podremos escribir:

$$A_{Bart} = \frac{1}{\quad 66} = 3{,}83$$

$$0,16 + 0,84 \times 550$$

$$A_{Homer} = \frac{1}{0,16 + 0,84 \times \frac{66}{500}} = 3,69$$

En consecuencia, el incremento de rendimiento debido a la frecuencia de reloj en la ejecución de la aplicación de reserva de billetes es de 3,83 y 3,69, para los procesadores Bart y Homer, respectivamente.

La relación entre los costes de las dos alternativas se calculará dividiendo el coste de la opción más cara entre la más barata:

$$\Delta C = \frac{C_{Bart}}{C_{Homer}} = \frac{480\text{ €}}{270\text{ €}} = 1,78$$

A la vista de los resultados se comprueba que el procesador Bart proporciona unas prestacio- nes ligeramente mejores que Homer, pero a costa de un coste mayor. Ahora falta por cuantificar si este coste adicional justifica, al menos desde un punto de vista teórico, la ganancia en ren- dimiento. Esta comparación puede hacerse de manera muy simple calculando el cociente entre la aceleración y el coste ($A/C$) para cada procesador, que resulta:

$$\frac{A_{Bart}}{C_{Bart}} = \frac{3,83}{480} = 7,979 \times 10^{-3}$$

$$\frac{A_{Homer}}{C_{Homer}} = \frac{3,69}{270} = 13,337 \times 10^{-3}$$

En consecuencia, el procesador Homer obtiene una mejor relación entre el rendimiento conseguido y el coste que hay que pagar por él, y por tanto representa la opción a elegir. ■

---

**PROBLEMA 1.11** Un computador ejecuta una aplicación de gestión de base de datos. Cada transacción con esta base de datos tarda un total de 12 segundos. Un análisis llevado a cabo mediante un monitor de ejecución de programas ha permitido averiguar que el 78 % del tiempo que tarda cada transacción se debe al acceso al subsistema de discos, mientras que el resto del tiempo se emplea en operaciones del procesador.

1. Calcúlese el nuevo tiempo de respuesta de una transacción si el subsistema de discos se sustituye por uno nuevo cuatro veces más rápido.
2. Repítase el apartado anterior suponiendo que el subsistema de discos se sustituye por uno nuevo seis veces más rápido.
3. Determínese la mejora del rendimiento obtenida en cada una de las actualizaciones anteriores.

**SOLUCIÓN:**

1. Dado que la mejora sobre el sistema informático afecta únicamente al subsistema de discos, la fracción del tiempo de cada transacción que se verá afectada será el 78 % del tiempo total. El nuevo tiempo de respuesta se calcula como:

$$T_{4x} = 0{,}22 \times 12 + 0{,}78 \times \frac{12}{4} = 12 \times \left(0{,}22 + \frac{0{,}78}{4}\right) = 4{,}98 \text{ s}$$

2. En este caso podemos aplicar el mismo razonamiento que en el apartado anterior, pero ahora incrementando el factor de mejora:

$$T_{6x} = 0{,}22 \times 12 + 0{,}78 \times \frac{12}{6} = 12 \times \left(0{,}22 + \frac{0{,}78}{6}\right) = 4{,}20 \text{ s}$$

3. La mejora de rendimiento que obtenemos en los casos anteriores viene dada por la ace- leración en el tiempo de ejecución del programa. Esta aceleración se puede calcular, bien dividiendo el tiempo original entre el nuevo:

$$A_{4x} = \frac{12}{4{,}98} = 2{,}410$$

$$A_{6x} = \frac{12}{4{,}20} = 2{,}857$$

o bien aplicando directamente la ley de Amdahl:

$$A_{4x} = \frac{1}{0{,}22 + \frac{0{,}78}{4}} = 2{,}410$$

$$A_{6x} = \frac{1}{0{,}22 + \frac{0{,}78}{6}} = 2{,}857$$

La mejora del rendimiento experimentada por las transacciones que sirve la aplicación de base de datos cuando se emplea un subsistema de discos seis veces más rápido respecto de uno cuatro veces más rápido será:

$$\frac{A_{6x}}{A_{4x}} = \frac{T_{4x}}{T_{6x}} = \frac{4{,}98}{4{,}20} = 1{,}186$$

Se puede observar que el aumento de velocidad del subsistema de discos del sistema original de cuatro a seis veces se traduce en una mejora del tiempo de respuesta de las transacciones del 18,6 %. ■

**PROBLEMA 1.12** Se sabe que el tiempo de respuesta de una petición a un servidor web es de 23 segundos, y que el 72 % de este tiempo se emplea en acceder al subsistema de discos magnéticos. El coste del subsistema de discos de este servidor fue de 3.500 €.

Los responsables del funcionamiento del servidor web quieren mejorar el tiempo de respuesta experimentado por los usuarios. Con este objetivo, están estudiando la posibilidad de sustituir el subsistema de discos actual por uno nuevo tres veces más rápido con un coste de 4.800 €.

1. Calcúlese el nuevo tiempo de respuesta del servidor web que se conseguiría con la actualización del subsistema de discos.
2. Compruébese si vale la pena hacer esta sustitución teniendo en cuenta la relación entre el rendimiento y el coste.

**SOLUCIÓN:**

1. El tiempo de respuesta mejorado se puede calcular como:

$$T = 0{,}28 \times 23 + \frac{0{,}72}{3} \times 23 = 11{,}96 \text{ s}$$

Por tanto, la mejora de rendimiento o aceleración que se conseguiría con esta actualización se calcula dividiendo el tiempo de respuesta original entre el tiempo mejorado: 23/11,96 = 1,923.

2. La relación de costes entre el nuevo subsistema de discos y el original es la siguiente:

$$\Delta C = \frac{4.800\ €}{3.500\ €} = 1{,}371$$

Dado que la mejora de rendimiento (1,923) es superior al incremento del coste (1,371), podemos concluir que valdría la pena hacer la sustitución del subsistema de discos. ■

**PROBLEMA 1.13** Una aplicación informática integra un conjunto de tareas diferentes. La ejecución de esta aplicación sobre un computador monoprocesador tarda un total de 2 minutos y medio.

Un estudio llevado a término por un monitor de ejecución de programas ha revelado que el 65 % de las tareas que componen la aplicación informática se podrían ejecutar en paralelo. Esto quiere decir que, en un sistema con varios procesadores, las tareas se podrían ejecutar al mismo tiempo, y no una después de la otra, como ocurre en un computador convencional basado en una única unidad central de procesamiento. Se pide:

1. Expresar la aceleración $A(p)$ que se conseguiría ejecutando la aplicación en un sistema multiprocesador con $p$ procesadores.

2. Expresar el tiempo de ejecución $T(p)$ de la aplicación si se ejecuta en un sistema multiprocesador con $p$ procesadores.

3. ¿Qué aceleración experimentará el programa si se ejecuta en un sistema con 2 procesadores? ¿En cuánto tiempo se ejecutaría el programa?

4. ¿En cuánto tiempo se ejecutará el programam si el sistema dispone de 8 procesadores?

5. Estimar el tiempo de ejecución de la aplicación si se ejecuta en un hipotético sistema con infinitos procesadores.

**SOLuCIón:**

1. La aceleración del tiempo de respuesta de la aplicación informática $A(p)$ se puede expresar en función del número de procesadores $p$ por medio de la ley de Amdahl. Teniendo en cuenta que la paralelización afectará al 65 % de las tareas podremos escribir:

$$A(p) = \frac{1{,}0}{0{,}35 + \frac{0{,}65}{p}}$$

2. El tiempo de ejecución $T(p)$ se puede expresar directamente mediante la siguiente expre- sión:

$$T(p) = T(1) \times 0{,}35 + \frac{T(1)}{p} \times 0{,}65 = T(1) \times 0{,}35 + \frac{0{,}65}{p}$$

o también haciendo uso de la aceleración $A(p)$:

$$T(p) = \frac{T(1)}{A(p)}$$

3. La aceleración para 2 procesadores se obtiene sustituyendo el valor de la variable $p$ en la ecuación anterior:

$$A(2) = \frac{1}{0{,}35 + \frac{0{,}65}{2}} = 1{,}481$$

Por tanto, el tiempo de ejecución para 2 procesadores será:

$$T(2) = 150 \times 0{,}35 + \frac{0{,}65}{2} = 101{,}25 \text{ s}$$

4. Los cálculos para 8 procesadores son similares a los que hemos hecho en el apartado anterior:

$$A(8) = \frac{1}{0,35 + \frac{0,65}{8}} = 2,319$$

$$T(8) = 150 \times 0,35 + \frac{0,65}{8} = 64,688 \text{ s}$$

5. Si hubiera infinitos procesadores el tiempo de ejecución de la aplicación quedaría reducido a la parte puramente secuencial. Si calculamos el límite de la expresión de $T(p)$ cuando $p$ se hace muy grande tendremos:

$$\lim_{p\to\infty} T(p) = \lim_{p\to\infty} T(1) \times 0,35 + \frac{0,65}{p} = T(1) \times 0,35 = 52,5 \text{ s}$$

Nótese que se está suponiendo implícitamente que el número de tareas en que se puede descomponer la aplicación también tiende a infinito y su tiempo de ejecución se puede hacer tan pequeño como se quiera. En los sistemas reales este número de tareas y su tiempo de ejecución asociado estará acotado, y por lo tanto, la mejora vendrá afectada por un límite superior. Por ejemplo, si las tareas del programa que se pueden ejecutar en paralelo son 16, entonces disponer de más procesadores no afectará para nada al tiempo de ejecución. ■

**PROBLEMA 1.14** Un programa tarda 32 minutos en ejecutarse en un computador con un único procesador ($p = 1$). El 80 % del tiempo de ejecución del programa es susceptible de ejecutarse en paralelo por varios procesadores ($p \geq 2$). El 20 % restante de este tiempo se divide en dos partes iguales: una corresponde a la componente puramente secuencial del programa y la otra se dedica a las operaciones de entrada/salida. Se pide calcular:

1. La aceleración que se conseguirá si la unidad de entrada/salida se hace tres veces más rápida.
2. La aceleración máxima alcanzable por el programa suponiendo que se dispone de un número suficientemente grande de procesadores.
3. El número de procesadores que son necesarios para conseguir una aceleración $A = 3$ en la ejecución del programa.
4. Ídem que el caso anterior pero suponiendo que la componente paralela representa el 85 % del tiempo de ejecución en detrimento de la parte secuencial, que pasa a ser del 5 % (las operaciones de entrada/salida permanecen inalterables).

**SOLuCIóN:** En primer lugar, y para tener una idea más clara sobre cómo se distribuyen los 32 minutos del tiempo de ejecución del programa, podemos calcular el tiempo destinado a cada componente. De estos 32 minutos, 32 0,10 = 3,2 minutos se emplean en operaciones de entrada/salida, y la misma cantidad de tiempo corresponde a la componente secuencial del programa. El 80 % restante, esto es, 32 0,80 = 25,6 minutos, representa la parte paralela del programa.

1. Si la unidad de entrada/salida se acelera tres veces, el nuevo tiempo de ejecución se puede calcular sumando las diferentes partes que lo componen y dividiendo el tiempo de la entrada/salida entre 3:

   $$3{,}2 + \frac{3{,}2}{3} + 25{,}6 = 29{,}87 \text{ minutos}$$

   La aceleración conseguida en el tiempo de ejecución se calcula dividiendo el tiempo original entre el que se acaba de calcular: $A = 32/29{,}87 = 1{,}071$. De manera similar, esta aceleración se puede obtener también aplicando la ley de Amdahl:

   $$A = \frac{1}{0{,}10 + \frac{0{,}1}{3} + 0{,}8} = 1{,}071$$

2. La máxima aceleración que se podría conseguir en la ejecución del programa viene de- terminada por la expresión $1/(1 - f)$, donde $f$ representa la fracción de tiempo que se puede mejorar (en realidad se trata de la expresión obtenida a partir de la ley de Amdahl cuando el valor de $k$ tiende a infinito). En el caso que nos ocupa esta fracción $f$ incluye la parte paralela del programa (se excluye la dedicada a tareas de entrada/salida y la puramente secuencial). Por lo tanto, el valor máximo de la aceleración será:

   $$= \frac{1}{1 - 0{,}8} = \frac{1}{0{,}2} = 5$$

   En este caso extremo el tiempo de ejecución del programa sería de $32/5 = 6{,}4$ minutos, es decir, el tiempo de ejecución de la parte secuencial más el dedicado a operaciones de entrada/salida.

3. La expresión de la aceleración en función del número de procesadores $p$ del sistema utilizando la ley de Amdahl es la siguiente:

   $$A = \frac{1}{0{,}10 + 0{,}1 + \frac{0{,}8}{p}}$$

   Si se hace $A = 3$, el valor de $p$ que satisface esta restricción resulta $p = 6$. Así pues, con 6 procesadores la aceleración en la ejecución del programa que se consigue es 3, esto es, el programa se ejecuta en $32/3 = 10{,}67$ minutos.

4. En este caso bastará con calcular el valor de $p$ en la ecuación:

$$A = \frac{1}{0{,}05 + 0{,}1 + \frac{0{,}85}{p}}$$

que satisface $A = 3$. En este caso la solución viene dada por $p = 4{,}64$. Nótese que como el número de procesadores ha de ser un número entero, en este caso no podemos obtener un valor exacto de 3 en la aceleración $A$. Así, podemos elegir $p = 5$, con lo que la aceleración real que se conseguiría es:

$$\frac{1}{0{,}05 + 0{,}1 + \frac{0{,}85}{5}} = 3{,}125$$

o bien considerar 4 procesadores, con lo que la aceleración será:

$$\frac{1}{0{,}05 + 0{,}1 + \frac{0{,}85}{4}} = 2{,}759$$

■

## 1.6. Problemas con solución

**PROBLEMA 1.15** Si se sustituye el disco de un sistema informático por otro cuatro veces más rápido que el actual, calcúlese el uso que se debería hacer de este dispositivo si se quiere conseguir que el sistema funcione tres veces más rápido.

**SOLUCIÓN:** La utilización del disco será 0,889. ■

**PROBLEMA 1.16** El procesador de un sistema informático costó 900 € y es utilizado por una aplicación durante un 65 % del tiempo de ejecución. Determínese, teniendo en cuenta la relación entre prestaciones y coste, si valdría la pena sustituir este procesador por uno tres veces más rápido que costara el doble.

**SOLUCIÓN:** La ejecución de la aplicación informática experimentaría una aceleración de 1,77, mientras que el coste se incrementaría en un factor de 2. Suponiendo que la relación entre prestaciones y coste es lineal, no valdría la pena hacer esta sustitución■

**PROBLEMA 1.17** Se dispone de dos opciones diferentes para mejorar el rendimiento de un computador:

1. Ampliar la memoria principal, con lo que el 80 % de los programas se ejecuta 1,75 veces más rápidamente.
2. Añadir un nuevo disco duro con más capacidad y velocidad que el existente, con lo que se consigue que el 60 % de los programas se ejecuten en la tercera parte del tiempo original.

Suponiendo que el precio de las dos alternativas es el mismo, ¿cuál de ellas permite obtener mejores prestaciones?

**SOLUCIÓN:** La primera opción permite conseguir una aceleración de 1,52, mientras que la segunda obtiene 1,67. Por tanto, resulta preferible la segunda opción. ■

---

**PROBLEMA 1.18** Un computador ejecuta un programa en 130 segundos. Para incremen- tar el rendimiento se tienen dos opciones: mejorar en 3 la ejecución de las instrucciones de acceso a memoria o en 5 la ejecución de las instrucciones aritméticas. Un monitor ha des- velado que el primer tipo de instrucciones se utiliza durante el 40 % del tiempo, mientras el segundo lo hace durante el 30 %. Calcúlese la aceleración conseguida en los tres casos siguientes: (a) sólo se mejoran las instrucciones de acceso a memoria, (b) sólo se mejoran las aritméticas, y (c) se mejoran los dos tipos de instrucciones.

**SOLUCIÓN:** Las aceleraciones son (a) 1,36, (b) 1,32 y (c) 2,03. ■

---

**PROBLEMA 1.19** A partir de la ley de Amdahl, dedúzcase una expresión para $k$ en función de $A$ y $f$.

**SOLUCIÓN:** La expresión para $k$ queda:

$$k = \frac{A \times f}{1 - A + A \times f}$$

■

---

**PROBLEMA 1.20** Sean dos factores de mejora sobre un sistema informático $k_1 = 5$ y $k_2 = 3$. Suponiendo que ambas mejoras no se pueden utilizar de forma simultánea, calcúlese la relación entre las fracciones de tiempo $f_1$ y $f_2$ durante las cuales se han de emplear para que ambas obtengan la misma aceleración global.

**SOLUCIÓN:** La fracción $f_2$ debe ser 1,2 veces superior a $f_1$, esto es, la relación es $f_2 = 1{,}2 \times f_1$. Nótese que, dado que $k_2 < k_1$, la segunda mejora se tiene que utilizar durante más tiempo que la primera para conseguir un rendimiento equivalente. ■

---

## 1.7. Problemas sin resolver

**PROBLEMA 1.21** Un programa se ejecuta en 52 segundos en un computador personal. Se quiere reducir este tiempo mediante la sustitución del procesador, para lo cual existen dos alternativas:

1. Sixtium (330 €), que permite reducir el tiempo de ejecución hasta los 32 segundos.
2. Septium (295 €), que permite reducir el tiempo de ejecución hasta los 40 segundos.

¿Cuál de ellas permite obtener una mejor relación entre prestaciones y coste?

---

**PROBLEMA 1.22** Un computador que actúa de servidor web dispone de un disco duro con un tiempo medio de servicio de 40 ms. Un estudio de monitorización ha desvelado que este disco es el cuello de botella del sistema y tiene una utilización media del 91 %. Si este disco costó 260 €, discútase cuál de las siguientes opciones es la mejor para reemplazar el disco teniendo en cuenta la relación entre prestaciones y coste:

1. Disco con un tiempo medio de acceso de 34 ms y 130 €.
2. Disco con un tiempo medio de acceso de 29 ms y 140 €.
3. Disco con un tiempo medio de acceso de 17 ms y 220 €.

---

**PROBLEMA 1.23** Un programa se ejecuta en 112 segundos. Si las operaciones de multiplicación consumen el 84 % de este tiempo, ¿en cuánto se tendría que mejorar la velocidad de estas operaciones si queremos que el programa se ejecute seis veces más rápidamente?

---

**PROBLEMA 1.24** Un equipo de arquitectos de computadores está tratando de diseñar una unidad aritmético-lógica que minimice el tiempo de ejecución de un programa de prue- ba. Las instrucciones de aritmética entera consumen el 60 % del tiempo total de ejecución de este programa. De éstas, las de suma son empleadas durante el 30 % del tiempo total. Determínese cuál de las dos alternativas siguientes consigue obtener un mejor rendimiento en la ejecución del programa de prueba:

1. Triplicar la velocidad de ejecución de todas las instrucciones de aritmética entera.
2. Hacer que la instrucción de suma se ejecute 15 veces más rápidamente.

---

**PROBLEMA 1.25** Existen tres propuestas distintas de mejora para reducir el tiempo de ejecución de un cierto programa en un computador. La mejora A consigue una aceleración local $k = 25$, la mejora B obtiene $k = 15$ y, por último, la mejora C permite conseguir $k = 35$. Supóngase que no se pueden utilizar estas mejoras de manera combinada, esto es, en un instante dado solamente puede emplearse una de ellas.

1. Si las mejoras A y B se emplean cada una de ellas durante el 30 % del tiempo, ¿durante qué fracción de tiempo se debe utilizar la mejora C para que el incremento de rendimiento del computador sea 10?

2. Supóngase que las mejoras A, B y C se aplican durante las fracciones 40, 25 y 10 % del tiempo original de ejecución (sin mejoras), respectivamente. ¿Durante qué fracción del tiempo mejorado no se usa ninguna de las tres mejoras?

---

**PROBLEMA 1.26** Después de mejorar en un factor $k$ = 16 un determinado recurso de un computador, un monitor ha desvelado que este recurso se emplea durante el 75 % del tiempo de ejecución de un programa.

1. Calcúlese el porcentaje del tiempo de ejecución original afectado por la mejora introducida.

2. Determínese el incremento de rendimiento en la ejecución del programa.

---

## 1.8. Actividades propuestas

**ACTIVIDAD 1.1** Se han introducido dos mejoras A y B en el diseño de un computador. Las fracciones de tiempo en que se utilizan estas mejoras son del 50 y el 30 %, respectiva- mente. Dibújese un gráfico tridimensional representando la aceleración global $A$ conseguida en función de los factores de mejora, sabiendo que el máximo de ambos es $k_A = k_B = 20$.

**ACTIVIDAD 1.2** Se quiere estimar el incremento de rendimiento que supone utilizar el disco duro frente al disco flexible en operaciones de escritura. Para ello diséñese un pro- grama que se limite a escribir un fichero con un tamaño similar a la capacidad de un disco flexible. Ejecútese el programa desde el disco duro y desde el disco flexible. Repítase diez veces esta operación y calcúlese la media aritmética del tiempo de ejecución en cada caso, ya que los tiempos de ejecución pueden presentar variaciones.

**ACTIVIDAD 1.3** Cuantifíquese la mejora conseguida por el uso del coprocesador mate- mático de un computador en la ejecución de operaciones en coma flotante.

# Capítulo 2

# Monitorización de sistemas y programas

La monitorización es una técnica de uso generalizado para supervisar, analizar y evaluar el comportamiento y el rendimiento de los sistemas informáticos que están en funciona- miento. En este capítulo nos hemos centrado tanto en la monitorización de la actividad de los computadores como en el análisis del comportamiento de los programas. El entorno escogido es el sistema operativo Unix (o Linux), ya que la cantidad de herramientas de monitorización en este sistema es relativamente elevada y su uso está muy extendido entre los administradores de sistemas.

Revisaremos brevemente aquellos programas que se han venido utilizando de manera más extendida. Por su importancia, dedicaremos un apartado específico al monitor `sar`, y finalmente hablaremos del monitor para analizar la ejecución de programas `gprof`. La des- cripción de todas estas herramientas requeriría mucho más espacio del que aquí dedicamos, por lo que en última instancia el lector habrá de remitirse a la documentación aportada por los propios monitores.

## 2.1. Medida y monitor

En general, la monitorización de sistemas informáticos hace referencia a todo lo relativo a la extracción de información que permita conocer qué está sucediendo en ellos. Esto conlleva una serie de problemas que a menudo no son fáciles de resolver. Entre otros, hay que decidir qué datos hace falta recoger, hay que saber dónde se encuentran estos datos, y por último, es necesario acceder a ellos de manera que perturben lo mínimo el funcionamiento del sistema, para grabarlos con vistas a su posterior análisis.

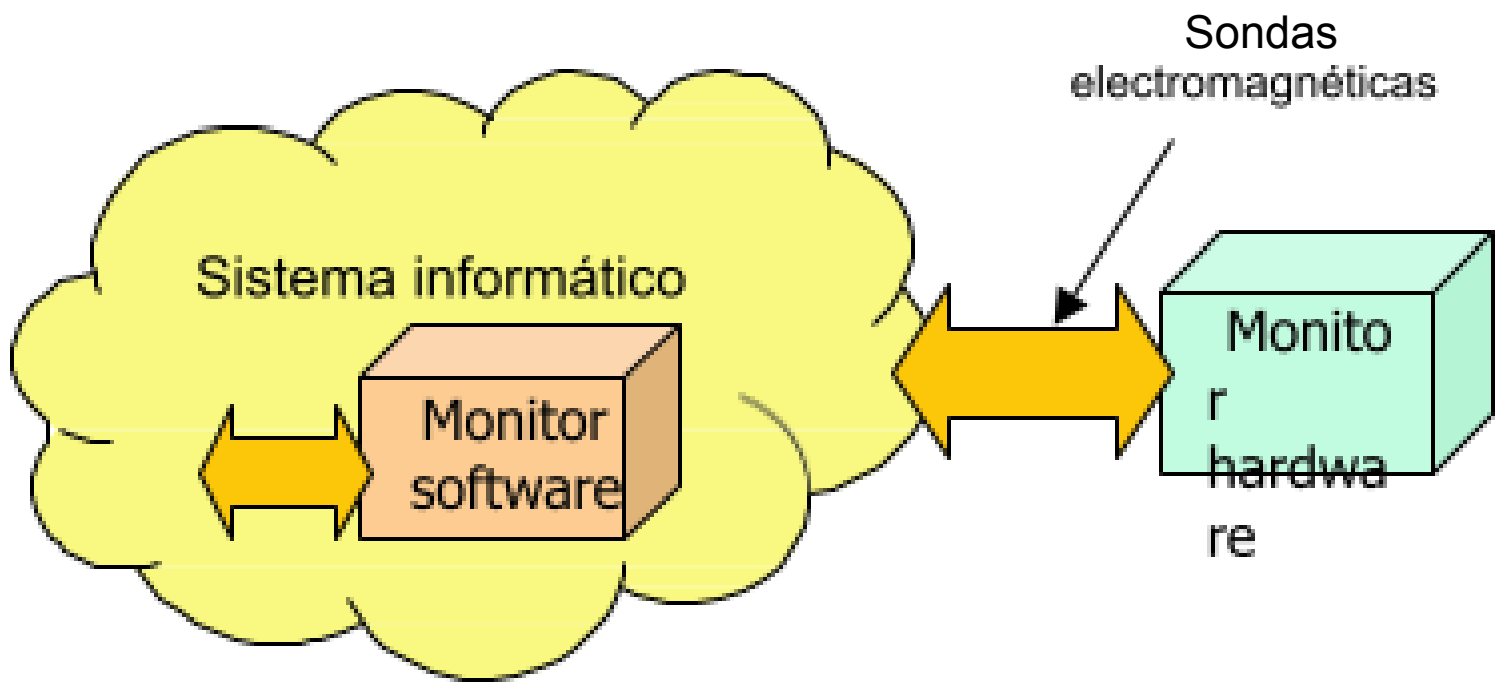

**Figura 2.1:** Situación de los monitores según su tipo.

La información que aporta la monitorización de la actividad de un sistema puede ser aprovechada de numerosas maneras. Por ejemplo, un analista podría servirse de la monitorización para construir modelos de la carga real. A su vez, un administrador puede conocer el consumo de recursos y detectar posibles cuellos de botella, o bien cambiar el valor de algunos parámetros del sistema operativo con el objetivo de ajustar mejor su funcionamiento. Incluso el mismo sistema operativo podría aprovechar los datos recogidos para adaptarse dinámicamente a la carga que soporta.

La toma de medidas en un sistema informático se puede hacer mediante dos posibles técnicas. La primera de ellas consiste en detectar todas las ocurrencias de aquello que queremos conocer. La segunda, a diferencia de la anterior, implica la toma de muestras a intervalos regulares de tiempo. La mayoría de las herramientas que monitorizan la acti- vidad de los sistemas informáticos emplean una mezcla de las dos: se utilizan contadores de sucesos que se muestren de forma periódica. Este muestreo periódico permite un aná- lisis estadístico sencillo, toda vez que el volumen de información recogida y su precisión dependerán de la cadencia con que se tomen las muestras. Mayor dificultad tiene la imple- mentación de un muestreo aleatorio, aunque de este modo se evitaría el sincronismo entre los estados por los que pasa el sistema y los intervalos fijos de monitorización.

Un monitor es una herramienta diseñada para observar el comportamiento de un siste- ma informático. Desde el punto de vista de su implementación, los monitores se clasifican en dos grupos principales: hardware y software (véase la Figura 2.1). Un monitor hardware es un dispositivo físico, independiente del sistema a examinar, que se conecta al mismo mediante un conjunto de sondas electrónicas. Este tipo de monitores no usan recursos del sistema monitorizado y, por tanto, su grado de intrusión es nulo. Sin embargo, presentan un difícil proceso de instalación, y se enfrentan a dos graves problemas. El primero de ellos es que gran parte de la información relevante del sistema es difícil de conseguir mediante sondas, ya que éstas sólo acceden a aquella información que se refleje en posiciones fijas de memoria. El segundo, quizás más importante incluso, es que el diseño actual de la mayoría de los equipos informáticos, que emplea una escala de integración muy elevada, no facilita la conexión de este tipo de dispositivos de medida.

Las razones anteriores determinan que la mayoría de los monitores disponibles sean programas instalados en el computador, ligados de manera más o menos íntima al sistema operativo que lo gestiona. El uso de programas aporta una gran flexibilidad en el proceso de monitorización, aunque se paga el precio de un grado de intrusión más elevado que en el caso de los monitores hardware. Dado que la toma de medidas por un monitor software implicará la ejecución de un programa, y para ello se usará el procesador del computador, el simple hecho de monitorizar implicará la perturbación del sistema sobre el que se realiza la medida. El grado de distorsión en la medida suele estimarse mediante una variable denominada sobrecarga del monitor (*overhead* ), que viene definida de la siguiente manera:

$$\text{Sobrecarga} = \frac{\text{Tiempo de ejecución del monitor}}{\text{Intervalo de medida}}$$

De la expresión anterior se deduce fácilmente que la sobrecarga del monitor se puede minimizar haciendo más grande el intervalo entre activaciones o disminuyendo su tiempo de ejecución.

## 2.2. Herramientas más comunes en Unix

La mayoría de los programas de monitorización hacen referencia a tres elementos principales del computador: procesador, disco y memoria. Sin embargo, hay otros monitores con no menos importancia, como los que aportan información sobre la carga del sistema, número de clientes conectados al computador o tiempo de ejecución de un programa. Muchos de los monitores que vamos a citar emplean ficheros con información estadística que el propio sistema operativo se encarga de mantener, como la contenida en el directorio /proc. Aquí podemos encontrar, por ejemplo, información sobre el procesador (fichero cpuinfo), interrupciones (fichero interrupts) o memoria (fichero meminfo).

La primera orden que vamos a considerar es uname (*system information*), que aporta información básica sobre el sistema operativo y la máquina. La opción -a imprime toda la información disponible:

```
$ uname -a
Linux osito 2.4.18-4GB #1 Wed May 29 15:47:24 UTC 2002 i686 unknown
```

El ejemplo refleja que la máquina se denomina osito y la versión del núcleo del sistema operativo es 2.4.18, compilada en mayo del año 2002.

### Carga del sistema

El sistema operativo Unix define la carga del sistema (*system load average*) como el número medio de procesos en la cola del núcleo. El valor de la carga se puede ver mediante la orden uptime:

```
$ uptime
 10:09am    up 10 days, 20:12, 1 user, load average: 1.20, 0.80, 0.10
```

La información mostrada refleja la hora del sistema (10:09 AM), el tiempo que lleva el sistema en marcha (10 días, 20 horas y 12 minutos), el número de usuarios que hay conectados (uno) y el valor medio de la carga en el último minuto, últimos 5 minutos y últimos 15 minutos. Algunos autores indican que un sistema en operación normal mostrará valores de la carga iguales o inferiores a 3, mientras que valores entre 4 y 7 representan cargas muy altas. En cualquier caso, estos valores son orientativos porque un mismo nivel de carga se tolera de forma diferente según las distintas configuraciones del sistema y según lo que ejecuten los programas.

## Tiempo de ejecución de un programa

La medición del tiempo de ejecución de un programa se lleva a cabo con la orden time. Esta variable es de especial importancia porque refleja, de forma directa, la percepción que el usuario tiene de las prestaciones del sistema informático. Por ejemplo, para saber cuánto tiempo necesitará la ejecución del programa quicksort, ejecutaremos la orden:

```
$ time quicksort real
          0m55.036s
user      0m51.580s
sys       0m1.300s
```

El resultado de la monitorización pone de manifiesto que el tiempo total transcurrido desde el inicio de la ejecución del programa hasta su conclusión es de 55,036 segundos (pa- rámetro real). De este tiempo total, 51,580 segundos se han empleado en la ejecución de código del programa por parte del procesador en modo usuario (parámetro user), mientras que durante 1,300 segundos el procesador ha ejecutado código en modo supervisor (pará- metro sys) como resultado de llamadas al sistema operativo efectuadas por el programa. Esto significa que la diferencia entre el tiempo total y el tiempo que realmente se está ejecutando código, 55,0356 51,580 1,300 = 2,156 segundos, representa la espera que el programa ha sufrido debido a la ejecución de otros programas que estaban en ejecución en el sistema y que competían por los mismos recursos, o a la espera por operaciones de entrada y salida.

## Actividad de los procesos

Una de las herramientas más utilizadas por los administradores de sistemas para saber qué procesos hay en ejecución y cuánta memoria consumen es top (*top CPU processes*). Este programa muestra, de manera dinámica, los procesos que están consumiendo tiempo de procesador ordenados de acuerdo con este consumo. La información se actualiza cada 5 segundos, aunque este valor es configurable. A continuación se muestra el resultado obtenido tras ejecutar esta orden:

```
$ top
  8:48am    up 70 days, 21:36,          1 user, load average: 0.28, 0.06, 0.02
47 processes: 44 sleeping, 3 running, 0 zombie, 0 stopped
CPU states: 99.6% user,          0.3% system,       0.0% nice,       0.0% idle
Mem:     256464K av,       234008K used,       22456K free, 0K shrd, 13784K buff Swap:
136512K av,                     4356K used, 132156K free                       5240K cached
```

| PID | USER | PRI | NI | SIZE | RSS | SHARE | STAT | LC | %CPU | %MEM | TIME | COMMAND |
|---|---|---|---|---|---|---|---|---|---|---|---|---|
| 9826 | carlos | 0 | 0 | 388 | 388 | 308 | R | 0 | 99.6 | 0.1 | 0:22 | simulador |
| 9831 | miguel | 19 | 0 | 976 | 976 | 776 | R | 0 | 0.3 | 0.3 | 0:00 | top |
| 1 | root | 20 | 0 | 76 | 64 | 44 | S | 0 | 0.0 | 0.0 | 0:03 | init |
| 2 | root | 20 | 0 | 0 | 0 | 0 | SW | 0 | 0.0 | 0.0 | 0:00 | keventd |
| 3 | root | 20 | 0 | 0 | 0 | 0 | SW | 0 | 0.0 | 0.0 | 0:00 | kapmd |
| 4 | root | 20 | 19 | 0 | 0 | 0 | SWN | 0 | 0.0 | 0.0 | 0:00 | ksoftiq |
| 5 | root | 20 | 0 | 0 | 0 | 0 | SW | 0 | 0.0 | 0.0 | 0:13 | kswapd |
| 6 | root | 2 | 0 | 0 | 0 | 0 | SW | 0 | 0.0 | 0.0 | 0:00 | bdflush |
| 7 | root | 20 | 0 | 0 | 0 | 0 | SW | 0 | 0.0 | 0.0 | 0:10 | kdated |
| 8 | root | 20 | 0 | 0 | 0 | 0 | SW | 0 | 0.0 | 0.0 | 0:01 | kinoded |
| 11 | root | 0 | -20 | 0 | 0 | 0 | SW< | 0 | 0.0 | 0.0 | 0:00 | recoved |
| 361 | root | 20 | 0 | 356 | 336 | 244 | S | 0 | 0.0 | 0.1 | 0:00 | syslogd |

La información aportada abarca diversos aspectos. La carga media del sistema se mues- tra en la primera línea, y equivale al resultado de ejecutar la orden uptime. A continuación se muestra el número de procesos en el sistema desde la última actualización. Estos procesos se clasifican, en virtud de su estado, en cuatro grupos: durmiendo (*sleeping*), en ejecución (*running*), zombis (*zombies*) y parados (*stopped* ). La utilización media del procesador des- de la última actualización también es un dato de gran importancia. Este uso se clasifica en cuatro apartados, según se emplee en ejecutar código de usuario, de sistema operativo, de procesos con baja prioridad (*nice*) o bien el procesador esté ocioso (*idle*). Finalmente, se muestra información acerca de la memoria principal y estadísticas sobre la memoria de intercambio (*swapping*). La Tabla 2.1 explica el significado del resto de los datos aportados por este monitor.

En este ejemplo se puede comprobar que el proceso activo que consume prácticamente la totalidad del tiempo del procesador es simulador. Este programa es responsable de más del 99 % del uso del procesador, y tan sólo ocupa 388 KB de memoria física, lo que representa un 0,1 % de la capacidad total.

Otra herramienta muy popular en relación con la ejecución de procesos, y comple- mentaria de la anterior, es la denominada ps (*process status*). Generalmente se emplea para aislar un proceso en particular, y dispone de un amplísimo abanico de opciones. Si se ejecuta la orden sin ningún argumento se muestra información relativa a la sesión del usuario:

| Campo | Descripción |
|---|---|
| PID | Identificador del proceso |
| USER | Nombre del usuario propietario del proceso |
| PRI | Prioridad del proceso |
| NI | Valor del parámetro *nice* del proceso |
| SIZE | Memoria total en KB que ocupa (datos, código y pila) |
| RSS | Memoria física en KB ocupada |
| SHARE | Memoria compartida en KB |
| STAT | Estado del proceso: R (*running*), S (*sleeping*), Z (*zombie*), D (*uninterruptible sleep*), T (*stopped* )<br>Modificadores: W (*swapped out* ), N (*running niced* ), >(*memory soft limit exceeded* ), <(*high niced level* ) |
| LC | Identificador del último procesador usado por el proceso |
| %CPU | Uso del procesador desde la última actualización |
| %MEM | Uso de la memoria física desde la última actualización TIME |
| | Tiempo de CPU que ha utilizado el proceso desde su inicio |
| COMMAND | Nombre del proceso (línea de órdenes) |

**Tabla 2.1:** Información aportada por top.

```
$ ps
  PID TTY          TIME CMD
29806 pts/0    00:00:00 bash
29939 pts/0    00:00:00 ps
```

Las diferentes y variadas opciones de ps se suelen emplean para obtener información más elaborada o bien relativa a otros procesos. Por ejemplo, para ver únicamente aquellos procesos que están en ejecución, con un formato de salida orientada a la legibilidad por parte del usuario, se puede utilizar la siguiente orden:

```
$ ps aur
USER       PID %CPU %MEM  VSZ  RSS TTY   STAT START TIME COMMAND
miguel   29951 55.9  0.1 1448  384 pts/0 R    09:16 0:11 tetris
carlos   29968 50.6  0.1 1448  384 pts/0 R    09:32 0:05 tetris
xavier   30023  0.0  0.5 2464 1492 pts/0 R    09:27 0:00 ps aur
```

Tal como se aprecia, la información aportada por este monitor es similar a la que proporciona la orden top, aunque la ventaja de este último radica en la comodidad de la actualización automática. En los datos anteriores se puede comprobar que hay tres procesos en ejecución, uno de los cuales es el propio monitor (véase la última línea), ordenados según el consumo que hacen del procesador. Téngase en cuenta que en este caso la orden se ha ejecutado en un sistema con dos procesadores.

## Actividad de la memoria

El monitor vmstat (*virtual memory statistics*) muestra información relativa al sistema de memoria, incluyendo datos sobre la memoria física y virtual. Además, ofrece datos relativos a la actividad de intercambio entre memoria y disco (*swapping*), transferencias con el disco, interrupciones, cambios de contexto y utilización del procesador. La sintaxis de esta orden es vmstat t n, donde t indica el tiempo transcurrido (en segundos, normalmente) entre dos muestras consecutivas, y n es el número de muestras. Por ejemplo:

```
$ vmstat 2 6
   procs                          memory      swap        io     system          cpu
 r  b  w     swpd    free    buff  cache       si   so     bi  bo     in   cs      us  sy id
 7  0  0     4356   21152 13848  46384       0    0      1    1      1    1       1   0 99
 3  0  0     4356   21152 13848  46384       0    0      0    0    102   38     100   0   0
 3  0  0     4356   21152 13848  46384       0    0      0    0    103   38     100   0   0
 3  0  0     4356   21152 13848  46384       0    0      0    0    103   35     100   0   0
 3  0  0     4356   21152 13848  46384       0    0      0    0    106   36     100   0   0
 3  0  0     4356   21152 13848  46384       0    0      0   49    113   38     100   0   0
```

Los datos de la primera línea se calculan desde el momento de la puesta en marcha de la máquina hasta el instante actual, por lo que no representa información de utilidad. Por otro lado, esta línea se muestra tan pronto como comienza a ejecutarse la orden; en consecuencia, se obtienen 5 muestras con información válida, y el periodo de medida total dura $5 \times 2 = 10$ segundos. Finalmente, si se quiere representar gráficamente la salida de esta orden es conveniente utilizar el parámetro -n para evitar la impresión de la cabecera con los nombres de las variables mostrados cada 20 muestras. La Tabla 2.2 refleja una breve descripción de los datos aportados por este monitor.

La visualización de los datos aportados por vmstat es muy sencilla si se utiliza, por ejemplo, el programa gnuplot. En la Figura 2.2 se muestra el resultado obtenido para las variables r, free, in, us, sy e id, tras un periodo de medida de 40 segundos, con una duración de un segundo entre dos muestras consecutivas.

La orden free (*free and used memory in the system*) sirve para tener de una manera rápida información sobre el estado de la memoria del sistema:

```
$ free
                     total       used       free     shared      buffers      cached
Mem:                256464     235464      21000          0        13904       46700
-/+ buffers/cache:             174860      81604
Swap:               136512       4356     132156
```

La primera línea refleja el estado de la memoria física, y la última el estado de la memoria de intercambio (*swapping*). Así mismo, también se dan datos sobre el uso de los buffers utilizados por el núcleo. Esta orden se puede ejecutar de manera periódica como free -s t, donde t indica la duración del intervalo entre muestreos consecutivos.

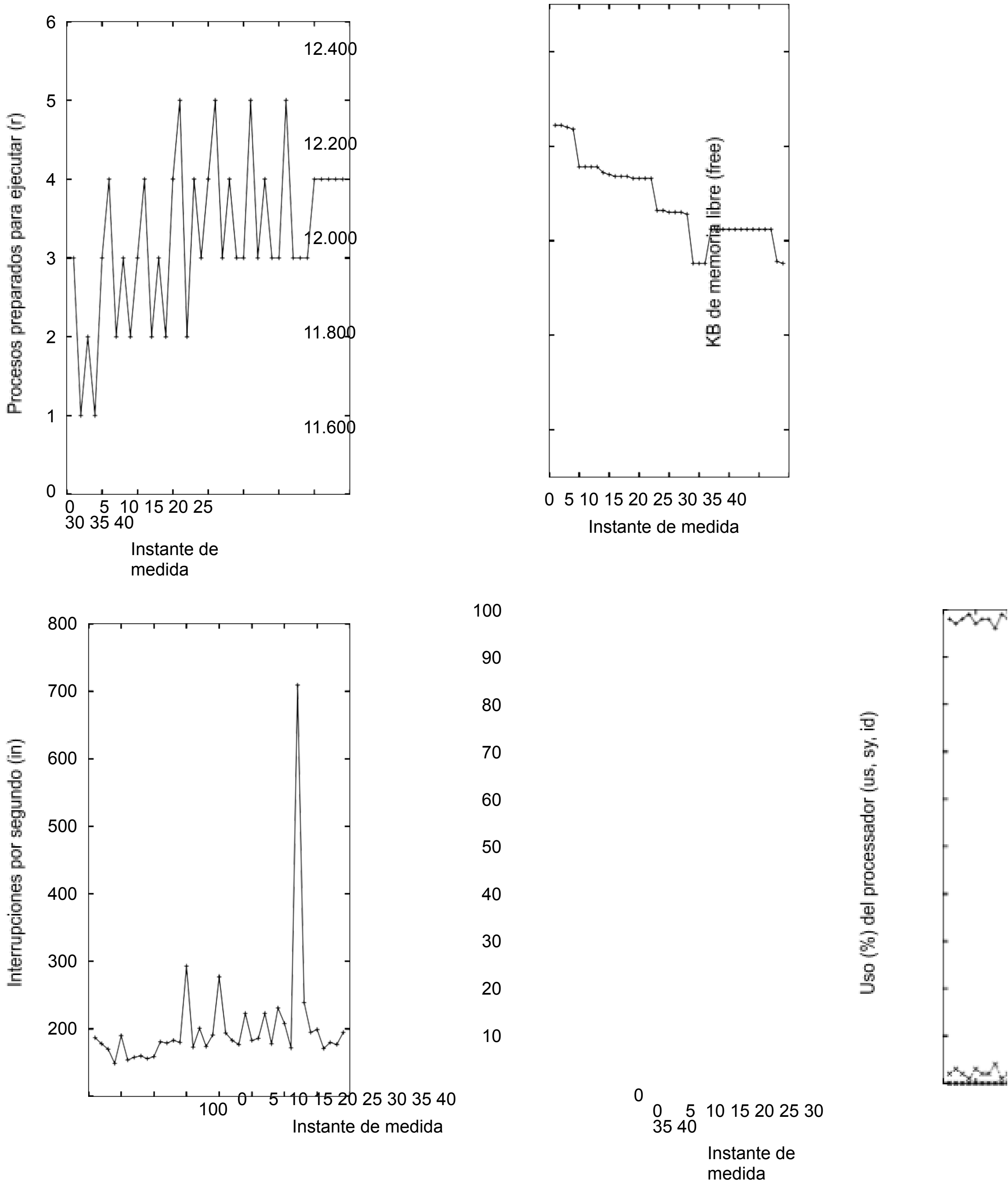

**Figura 2.2:** Representación gráfica de la información dada por vmstat.

| Campo | Descripción |
|---|---|
| r | Procesos esperando a ser ejecutados (*runnable*) |
| b | Procesos durmiendo ininterrumpidamente (*I/O blocked* ) |
| w | Procesos intercambiados (*swapped out* ) |
| swpd | Memoria virtual en uso (KB) |
| free | Memoria física libre (KB) |
| buff | Memoria usada como buffer (KB) |
| cache | Memoria usada como cache (KB) |
| si | Memoria intercambiada (KB/s) desde disco |
| so | Memoria intercambiada (KB/s) hacia disco |
| bi | Bloques de memoria por segundo enviados a disco |
| bo | Bloques de memoria por segundo recibidos desde disco |
| in | Interrupciones por segundo (incluyen las de reloj) |
| cs | Cambios de contexto por segundo |
| us | Uso del procesador ejecutando código de usuario |
| sy | Uso del procesador ejecutando código de sistema operativo |
| id | Porcentaje de tiempo con el procesador ocioso |

**Tabla 2.2:** Información aportada por vmstat.

### Actividad de los discos

El uso del sistema de ficheros (almacenamiento en disco) se puede examinar mediante la orden df (*filesystem disk space usage*):

```
$ df
Filesystem              1k-blocks      Used    Available Use%  Mounted on
/dev/hda2                 9606112   3017324    6100816    34%  /
/dev/hdb1                12775180   9236405    3140445    75%  /home
```

Los datos mostrados revelan que el sistema de ficheros se encuentra organizado en dos discos físicos (/dev/hda y /dev/hdb). En la partición del primer disco se encuentra montada la raíz del árbol de directorios (/), mientras que en la partición del segundo disco se ha montado el directorio /home, utilizado habitualmente para albergar los ficheros de los usuarios de la máquina. La ocupación de la capacidad del primer disco se sitúa en el 34 %, mientras que la del segundo llega hasta el 75 %.

También se puede averiguar la capacidad ocupada por un directorio concreto del sistema de ficheros mediante la orden du (*file space usage*):

```
$ du doc
160       doc/cartas
432       doc
```

El resultado de la orden indica que el directorio doc ocupa un total de 432 KB, de los cuales el subdirectorio cartas ocupa 160 KB.

Dentro de los monitores que dan información sobre la actividad de los discos del computador podemos destacar la herramienta hdparm, diseñada específicamente para discos con interfaz IDE. Esta herramienta permite tanto conocer los parámetros más importantes del mismo como cambiar algunos de sus valores de configuración (operación esta última harto peligrosa para los datos contenidos en la unidad). Tiene un gran número de parámetros de entrada, entre los que podemos destacar -g para obtener la geometría del disco, así como
-t y -T para obtener medidas de rendimiento del disco. Sirva el siguiente resultado como ejemplo de uso para averiguar la geometría del disco:

```
$ hdparm -g /dev/hda
/dev/hda:
  geometry        = 790/255/63, sectors = 12706470, start = 0
```

La información anterior muestra la geometría del disco mediante la tripleta cilindros, cabezales y sectores, así como el número total de sectores en el dispositivo y el desplazamiento (también en sectores) desde el principio del disco. Nótese que el producto de los paráme- tros geométricos anteriores no ofrece el mismo número de sectores mostrado por el monitor: 790 255 63 × 12.691.350, esto es, hay una diferencia de 12.706.470 12.691.350 = 15.120 sectores. Aunque desconocemos la razón última de esta discrepancia, sospechamos que se debe a que los discos actuales emplean la técnica ZBR (*zone bit recording*), esto es, orga- nizan su información en diversas zonas, donde cada una de ellas tiene su propio número de sectores por pista. Por ello una geometría homogénea para todo el disco, como la mostrada por el monitor, no deja de ser una aproximación, aunque, eso sí, muy precisa (el error en el número de sectores ronda el 0,1 %).

Finalmente, mostramos el siguiente resultado que hace referencia a la velocidad de lectura del disco: la primera línea muestra la velocidad de lectura de la memoria cache de entrada/salida del sistema operativo (en realidad no hay acceso físico al disco), mientras que la segunda muestra la velocidad de lectura (sectores secuenciales) que el disco es capaz de mantener sin tener en cuenta la sobrecarga del sistema de ficheros:

```
$ hdparm -tT /dev/hda
  Timing buffer-cache reads:     128 MB in   1.15 seconds =111.30 MB/sec
  Timing buffered disk reads:     64 MB in   6.04 seconds = 10.60 MB/sec
```

## Usuarios del sistema

Finalmente, la cantidad de usuarios conectados a una máquina también puede ser una buena estimación de la carga que soporta el sistema. En este sentido cabe resaltar la orden w (*who is logged on and what they are doing*) que sirve para obtener información acerca de qué usuarios están conectados a la máquina y qué están haciendo:

```
$ w
10:04am up 70 days, 22:52, 2 users, load average: 0.31, 0.07, 0.02
USER     TTY     FROM                LOGIN@   IDLE   JCPU    PCPU   WHAT
xavier   pts/0   songoku.disca.up    8:47am   0.00s  8.25s   0.02s  w
carlos   pts/1   osito.disca.upv.    10:02am  1:58   0.07s   0.07s  -bash
```

La orden muestra que hay dos usuarios conectados. El parámetro JCPU incluye el tiempo total de procesador usado por todos los procesos dependientes del terminal especificado; este tiempo también incluye los trabajos en segundo plano (*background* ). El parámetro PCPU indica el tiempo empleado por el proceso especificado en el campo WHAT.

## 2.3. El monitor sar

La orden sar (*system activity reporter* ), por sí sola, es una de las herramientas más po- tentes disponibles para los administradores de sistemas a fin monitorizar la actividad en un computador. Su historia arranca prácticamente desde el desarrollo de los primeros sis- temas Unix, y en la actualidad existen varios ejemplos de implementación para Linux. En nuestro caso utilizaremos la versión mantenida por Sebastien Godard que se encuentra disponible en el sitio http://perso.wanadoo.fr/sebastien.godard y que pertenece a un conjunto de utilidades denominado sysstat. En este paquete también pueden encontrar- se otras herramientas útiles para monitorización como mpstat e iostat, que describimos brevemente al final de este apartado.

Otra versión para Linux de este monitor clásico, denominada atsar, con una instala- ción y un manejo similar al anterior, se encuentra disponible en la dirección de Internet ftp.atcomputing.nl/pub/tools/linux. En esta misma dirección pueden obtenerse otras utilidades como atop, una versión bastante mejorada de top, o akins, que muestra infor- mación relacionada con variables y tablas del núcleo del sistema operativo.

La utilidad principal de sar es la detección de cuellos de botella en el sistema. Esta orden no solamente ofrece la posibilidad de mostrar lo que está ocurriendo en el sistema cuando se ejecuta de manera interactiva, sino que también se puede utilizar para guar- dar información sobre la carga y el estado del mismo en ficheros históricos con el fin de examinarla con posterioridad. En general, los sistemas que tienen instalado este monitor suelen activarlo a intervalos regulares de unos pocos minutos. Ésta es la información que, de manera automática, se guardará en ficheros históricos diariamente y podrá recuperarse posteriormente para analizar el comportamiento del sistema.

En realidad, esta herramienta consta de dos órdenes complementarias. La primera de ellas es sadc (*system activity data collector* ), que es la que realmente recoge todos los datos relacionados con la actividad del sistema y construye con ellos un registro en formato binario. La orden sar, que da nombre al monitor, lee los datos binarios que recoge sadc y los traduce a formato de texto legible por nosotros. La relación entre ambas se muestra en la Figura 2.3. En esta figura se aprecia cómo sadc recolecta información de contadores mantenidos por el sistema operativo. El registro elaborado se puede almacenar en un fichero histórico para que sea leído posteriormente por sar, o bien pasarlo directamente a sar en tiempo de ejecución para la confección del informe en modo texto.

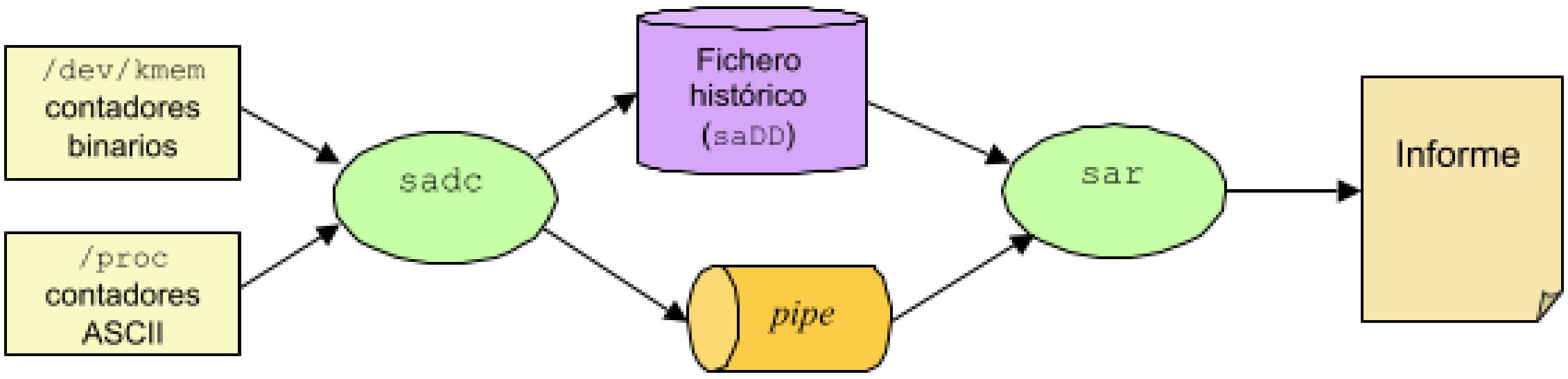

**Figura 2.3:** Funcionamiento del monitor sar.

| **Parámetro** | **Actividad que detalla** |
|---|---|
| u | Utilización del procesador |
| B | Paginación de la memoria virtual |
| c | Creación de procesos |
| b | Transferencias con la entrada/salida |
| d | Transferencias para cada disco |
| I | Sistema de interrupciones |
| n | Conexión de red |
| q | Carga media del sistema |
| r | Sistema de memoria |
| w | Cambios de contexto |
| W | Intercambio (*swapping*) |
| -x PID | Estadísticas sobre un proceso |

**Tabla 2.3:** Información aportada por sar.

La selección de la información a incluir en el informe se hará mediante los parámetros de entrada a la orden sar. Por ejemplo, el informe contendrá toda la información almacenada por sadc utilizando el parámetro -A. En la Tabla 2.3 se detallan algunos de los diferentes parámetros que se pueden utilizar y el tipo de información que aportan (la lista no es exhaustiva, y en todo caso depende de la versión del monitor utilizada).

La ejecución de sar sin ningún parámetro ofrece información sobre la utilización del procesador. En este caso, se puede observar que la máquina dispone de dos procesadores instalados (sistema biprocesador), por lo que los datos mostrados son un promedio de ambos:

```
$ sar
00:00:00        CPU     %user     %nice   %system     %idle
00:05:00        all      0.09      0.00      0.08     99.83
```

```
00:10:00            all      0.01      0.00      0.01      99.98
...
11:15:00            all      0.02      0.00      0.02      99.96
11:20:00            all      0.44      0.00      0.20      99.36
11:25:00            all      0.05      0.00      0.02      99.92
```

El formato de la salida es simple: la primera columna refleja el instante de tiempo en que se tomó la medida, mientras que las columnas restantes hacen referencia a una variable en particular. En este ejemplo se puede apreciar que el monitor toma muestras cada 5 minutos. Si quisiéramos obtener estadísticas para un procesador concreto (el 0 o el 1), hay que especificarlo en la orden:

```
$ sar -U 0
00:00:00            CPU     %user     %nice   %system     %idle
00:05:00              0      0.04      0.00      0.02     99.94
00:10:00              0      0.02      0.00      0.00     99.98
...
11:15:00              0      0.02      0.00      0.02     99.96
11:20:00              0      0.40      0.00      0.24     99.35
11:25:00              0      0.09      0.00      0.04     99.87
```

Nótese que las muestras comienzan en la medianoche. Esto es así porque sar guarda la información histórica por días. Las muestras tomadas a lo largo de un día se van alma- cenando en un fichero histórico con el nombre saDD, en formato binario, donde DD indica el día del mes. Si no se indica en la orden, el monitor muestra la información relativa al día actual. En caso contrario, hay que especificar el día concreto en cuestión. Por ejemplo, para mostrar la utilización del procesador 0 en el día 21 del mes actual debemos indicar dónde se encuentra el fichero histórico correspondiente:

```
$ sar -U 0 -f /var/log/sa/sa21
00:00:00              0      0.61      0.00      0.19      99.21
00:05:00              0      0.14      0.00      0.06      99.80
...
23:50:00              0      0.02      0.00      0.01      99.97
23:55:00              0      0.76      0.00      0.33      98.91
```

También se pueden seleccionar las medidas tomadas durante un periodo concreto del día. Por ejemplo, la siguiente orden se puede emplear para mostrar los datos relativos a la actividad del sistema de disco (se especifica el parámetro -d) desde las 12:00 hasta las 12:15 del día 20 del mes:

```
$ sar -d -s 12:00:00 -e 12:15:00 -f /var/log/sa/sa20
12:00:00            DEV       tps    sect/s
12:05:00         dev3-0      0.14      1.33
12:10:00         dev3-0      0.08      0.85
12:15:00         dev3-0      0.10      1.09
```

La columna tps indica el número de transferencias por segundo enviadas al disposi- tivo, y sect/s muestra el número de sectores transferidos por segundo desde o hacia el dispositivo. El tamaño de cada sector es de 512 bytes.

A continuación se puede ver una muestra de la información mostrada mediante el parámetro -b, que sirve para consultar la actividad de la unidad de entrada/salida en su conjunto. En este caso la información aportada distingue entre las transacciones por segundo de lectura y las de escritura. Así mismo, esta misma información se muestra como bloques leídos o escritos por segundo.

```
$ sar -b
00:00:00        tps     rtps     wtps   bread/s   bwrtn/s
00:05:00       0.74     0.39     0.35      7.96      3.27
00:10:01       0.09     0.00     0.09      0.00      0.91
00:15:00       0.15     0.00     0.14      0.03      1.36
00:20:00      65.12    59.96     5.16    631.62    162.64
```

Como se ha indicado, la orden sar se puede emplear de manera interactiva, y en este caso se utiliza cuando se desea conocer algún aspecto del funcionamiento del sistema a intervalos de tiempo menores que los especificados en la recolección de la información histórica. En este sentido el programa se ejecuta especificando el número de muestras a tomar y el periodo entre cada muestra. Por ejemplo, la orden sar 2 30 recoge un total de 30 muestras, con un intervalo de 2 segundos entre dos muestras consecutivas.

Como ejemplo de visualización de resultados, la Figura 2.4 muestra la evolución de diversos aspectos de la actividad del sistema. En concreto, se ha representado el número de bloques transmitidos desde el disco debido a paginación, el número de páginas activas, el número de procesos preparados para ejecutarse, y la utilización del procesador.

Finalmente, referiremos aquí algunos detalles de los monitores mpstat y iostat, am- bos incluidos en el paquete sysstat. El monitor mpstat (*processors related statistics*) da información sobre la utilización de los procesadores de un computador. Por ejemplo, a continuación se visualiza el uso de uno de los procesadores en un sistema biprocesador durante 5 muestras consecutivas, espaciadas tres segundos:

```
$ mpstat -P    1 3 5
12:07:03       CPU    %user    %nice  %system    %idle    intr/s
12:07:06         1   100.00     0.00     0.00     0.00     63.00
12:07:09         1   100.00     0.00     0.00     0.00     66.00
12:07:12         1   100.00     0.00     0.00     0.00     44.00
12:07:15         1   100.00     0.00     0.00     0.00     74.00
12:07:18         1   100.00     0.00     0.00     0.00     50.00
Average:         1   100.00     0.00     0.00     0.00     59.40
```

Por otro lado, la orden iostat proporciona información sobre la actividad relacionada con los dispositivos de bloques y particiones, e incluye también datos sobre la utilización del procesador. La sintaxis es similar a la anterior, incluyendo el número de muestras a tomar y el intervalo entre ellas. Si no se utilizan parámetros se comporta igual que vmstat, mostrando las estadísticas desde que el computador se puso en marcha, y por tanto, no aporta información útil:

```
$ iostat
cpu-avg:   %user   %nice   %sys   %idle
            3.70    0.02   0.48   95.81

Device:      tps  Blq_read/s  Blq_wrtn/s  Blq_read  Blq_wrtn
dev2-0      0.00        0.00        0.00       133         0
dev3-0      0.55        4.53        6.62  11726226  17108122
dev3-1      0.01        0.00        0.61      2698   1590072
```

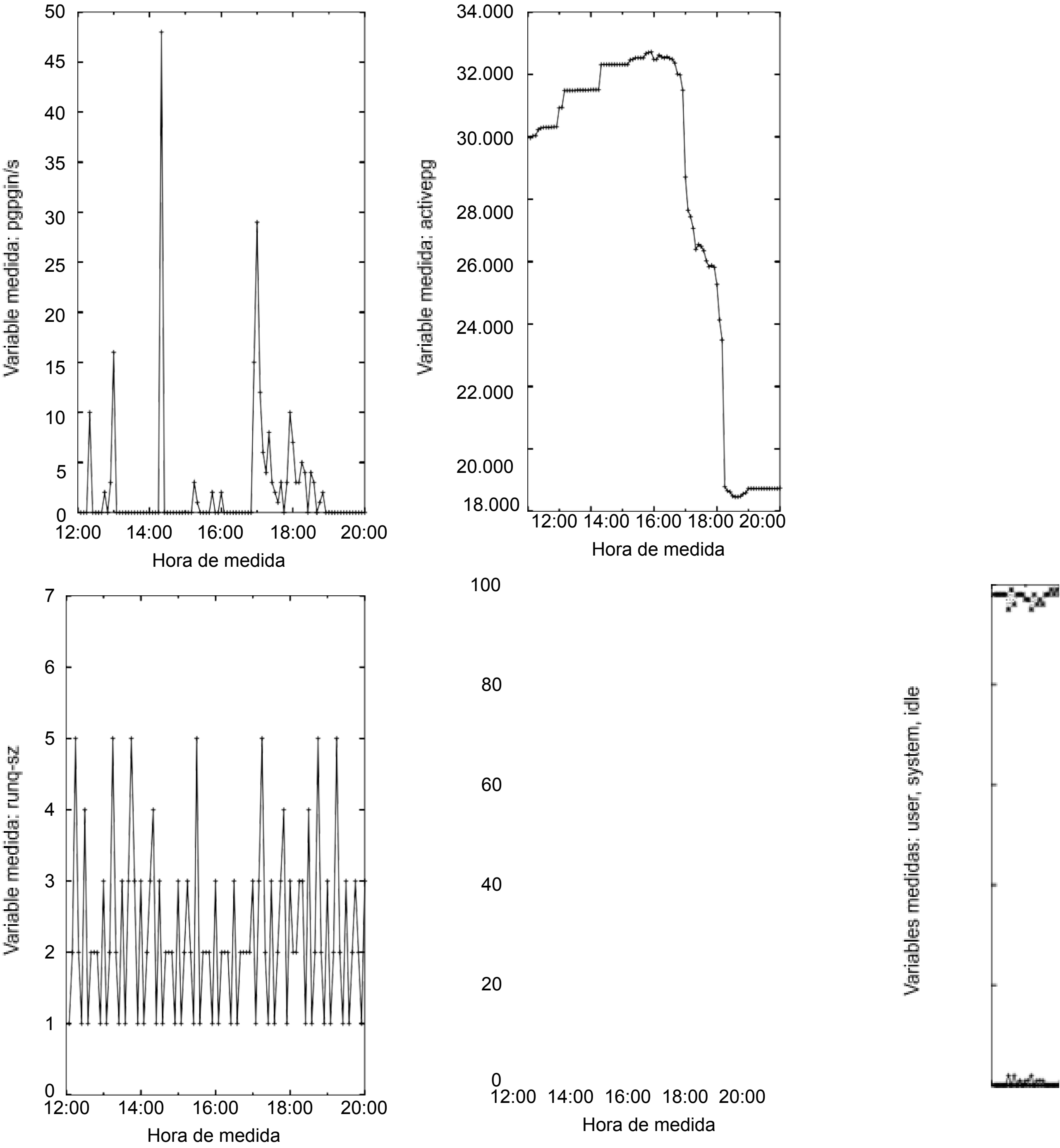

**Figura 2.4:** Representación gráfica de la información dada por sar.

## 2.4. Monitorización de programas

La monitorización o análisis de programas (*profiling*) es una técnica utilizada para obtener información sobre la ejecución de los mismos. Empleamos esta técnica, por ejemplo, cuando se quiere conocer qué parte del código de un programa es la que más tiempo de ejecución consume, o cuál es la secuencia de llamadas entre procedimientos. El análisis de programas se lleva a cabo mediante tres etapas principales, esquematizadas en la Figura 2.5:

1. El código fuente del programa a estudiar se debe compilar y enlazar especificando las opciones necesarias para la monitorización. Esta fase también se denomina instrumentación, ya que dotamos al código ejecutable de los instrumentos necesarios para recoger información relativa a su ejecución.
2. El programa compilado con las opciones de análisis correspondientes, esto es, el programa instrumentado, se ejecuta para poder recoger los datos de la monitorización en uno o varios ficheros (*data profiles*). Debido al proceso de monitorización, la ejecución del programa instrumentado se puede ralentizar ligeramente respecto a la ejecución del programa original.
3. En esta última fase, se ejecuta la herramienta adecuada para leer la información recogida durante la ejecución del programa instrumentado.

La herramienta más popular en Unix para estudiar el comportamiento de los programas es gprof. Esta orden se emplea principalmente cuando se necesita información sobre los procedimientos y funciones declarados en un programa. Si suponemos que nuestro programa se denomina prog.c, las etapas a seguir para analizar su comportamiento son:

```
$ gcc prog.c -o prog -pg -g -a
$ prog
$ gprof prog > prog.gprof
```

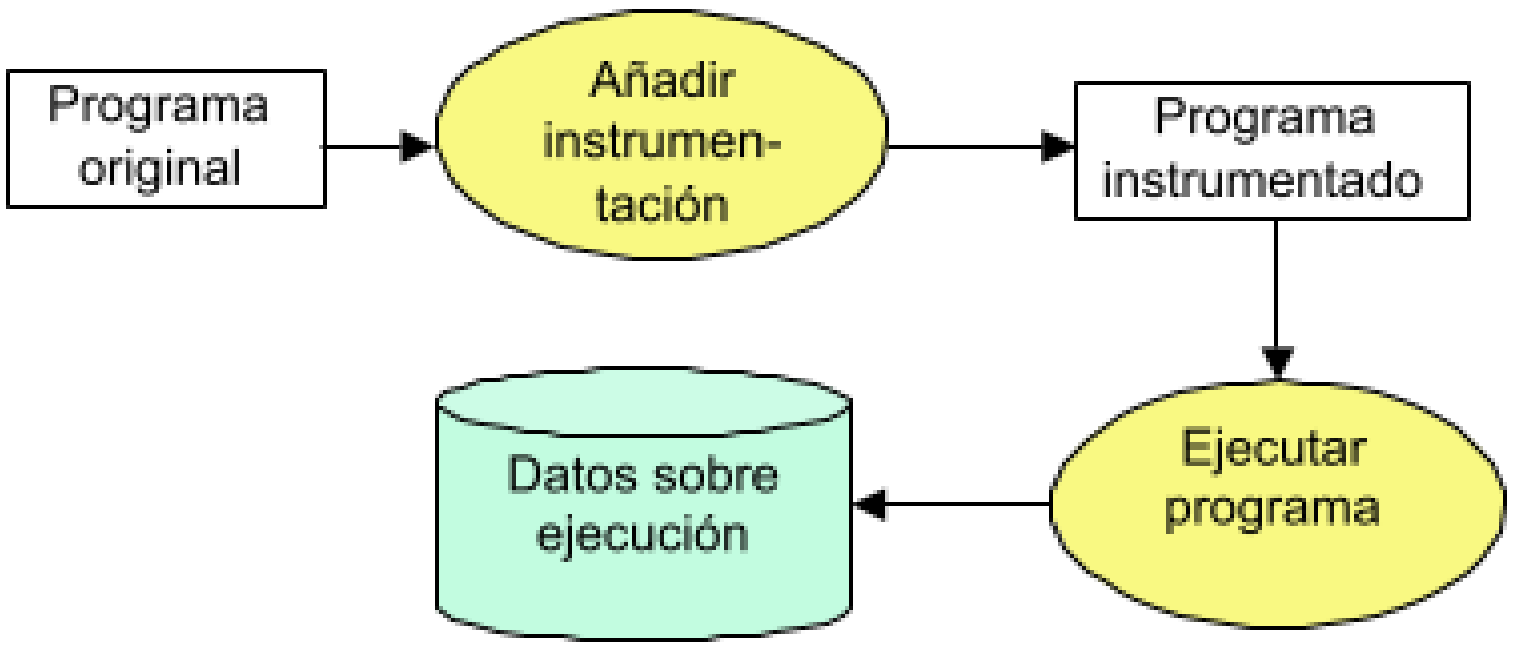

**Figura 2.5:** Etapas en la monitorización de programas.

La primera línea instrumenta completamente el programa mediante las opciones de compilación -pg, -g y -a. Estas opciones indican la recogida de datos relativos a las funciones, a las líneas del código fuente y a bloques de instrucciones, respectivamente. La segunda línea ejecuta el programa instrumentado, el cual genera un fichero de datos en formato binario con la información recogida denominado gmon.out. Finalmente, la tercera línea ejecuta la orden gprof para analizar los datos recogidos y guarda el resultado de este análisis en el fichero prog.gprof en formato texto, con el objetivo de que sea legible por el usuario.

Así mismo, la orden gprof también dispone de opciones según la información que nos interese conocer. Entre otras, las más importantes son las predeterminadas: -p y -q. La primera visualiza la distribución del tiempo de ejecución entre los procedimientos del programa (*flat profile*), mientras que la segunda muestra información relativa al grafo de dependencias entre los procedimientos (*call profile*).

Supongamos que queremos analizar el siguiente programa denominado bucles.c que tiene el siguiente código:

```
#include <math.h>
double a=3.14,b=6.34,c=-3.03; long i;

void main(){
    producto(); producto(); producto(); división();
    división(); atangente();}

producto(){
  for (i=0; i<50000000; i++)
    c=a*b;}

división(){
  for (i=0; i<30000000; i++)
    c=a/b;}
```

```
atangente(){
  for (i=0; i<30000000; i++)
    c=atan(a);}
```

Este programa consta de tres procedimientos distintos, y en total se hacen seis llamadas, de las cuales tres son al procedimiento producto(), dos a división() y una a atangente(). El resultado obtenido sobre la distribución del tiempo de ejecución entre los procedimientos anteriores es la siguiente:

```
Flat profile:

Each sample counts as 0.01      seconds.
  %    cumulative    self             self      total
 time    seconds    seconds   calls  ms/call   ms/call   name
 49.92      6.36      6.36       2   3180.00   3180.00   división
 30.38     10.23      3.87       3   1290.00   1290.00   producto
 19.70     12.74      2.51       1   2510.00   2510.00   atangente
```

La instrumentación ha efectuado medidas cada 0,01 segundos. El tiempo total de ejecu- ción es de 12,74 segundos, según indica el último valor de la columna cumulative seconds. Por tanto, se han tomado un total de 12,74/0,01 = 1.274 muestras. Nótese que un número pequeño de muestras puede indicar una baja precisión en el proceso de monitorizacion.

De todo el tiempo de ejecución, el 49,92 % del tiempo total, es decir, 6,36 segundos, se utiliza en la ejecución del procedimiento división (columna self seconds). Este tiempo se ha consumido en dos llamadas al procedimiento, cada una de las cuales tarda 3.180 milisegundos (columna total ms/call). La columna self ms/call indica el tiempo que se dedica a ejecutar código propio del procedimiento, esto es, descontando las posibles llamadas a otros procedimientos dentro de éste (en este caso ambas columnas coinciden porque dentro de división no se llama a ningún otro procedimiento). Si se comparan estos valores con el resto podemos concluir que este procedimiento es más lento que los otros dos.

Por su lado, el procedimiento producto es el más rápido de todos, se llama tres veces, y cada llamada tarda 1.290 milisegundos en ejecutarse. La ejecución de éste representa el 30,38 % del tiempo total de programa. Finalmente, el procedimiento atangent es el que menos tiempo consume (el 19,70 % del total), y se ejecuta durante 2,51 segundos.

La información principal relativa al grafo de llamadas entre los procedimientos decla- rados en el programa bucles.c es la siguiente:

```
Call graph:

granularity: each sample hit covers 4 byte(s) for 0.08% of 12.74 seconds
```

| index | % time | self | children | called | name |
|---|---|---|---|---|---|
| | | | | | <spontaneous> |
| [1] | 100.0 | 0.00 | 12.74 | | main [1] |
| | | 6.36 | 0.00 | 2/2 | división [2] |
| | | 3.87 | 0.00 | 3/3 | producto [3] |
| | | 2.51 | 0.00 | 1/1 | atangente [4] |
| | | 6.36 | 0.00 | 2/2 | main [1] |
| [2] | 49.9 | 6.36 | 0.00 | 2 | división [2] |
| | | 3.87 | 0.00 | 3/3 | main [1] |
| [3] | 30.4 | 3.87 | 0.00 | 3 | producto [3] |
| | | 2.51 | 0.00 | 1/1 | main [1] |
| [4] | 19.7 | 2.51 | 0.00 | 1 | atangente [4] |

Esta información muestra la distribución del tiempo de ejecución de un procedimiento entre los procedimientos que se llaman desde él. En este caso únicamente ocurre para el procedimiento principal main, desde donde se hacen seis llamadas. Tal como se indica, la totalidad del tiempo de ejecución del programa, 12,74 segundos, corresponde a la ejecu- ción de este procedimiento (columna children). Este tiempo se distribuye entre las seis llamadas al resto de los procedimientos. La columna self representa el tiempo en que se ejecuta código propio del procedimiento, excluyendo posibles llamadas a otros procedi- mientos. Según se indica, todo el tiempo consumido por main es debido a la ejecución de los procedimientos que desde él se llaman, los cuales consumen todo su tiempo ejecutando código propio (sin ninguna llamada a otros procedimientos).

## 2.5. Problemas resueltos

**PROBLEMA 2.1** Considérese un sistema informático en el que la activación de un monitor software implica la ejecución de un total de 150 instrucciones máquina. Si el procesador del sistema tiene una velocidad de ejecución de 75 MIPS (*million of instructions per second* ), se pide:

1. Calcular el valor que ha de tener el periodo de muestreo si se quiere una sobrecarga (*overhead* ) del 5 %.
2. Con el periodo de muestreo obtenido, calcular el tamaño del fichero de resultados generado tras un periodo de medida de dos horas si en cada activación del monitor se graba una palabra de 32 bits.

**SOLuCıón:**

1. En primer lugar se puede estimar el tiempo que tarda en ejecutarse el monitor dividiendo el número de instrucciones ejecutadas entre la potencia de cálculo del procesador:

$$\frac{150 \text{ instrucciones}}{75 \text{ MIPS}} = \frac{150}{75 \times 10^{6} \text{ s}^{-1}} = 2 \times 10^{-6} \text{ s}$$

La sobrecarga introducida por la ejecución del monitor depende de la cadencia con que éste se ejecute para tomar las medidas. Si el periodo de muestreo es $T$, entonces la sobrecarga viene determinada por el cociente entre el tiempo que tarda el monitor en ejecutarse y este periodo. Por lo tanto, podemos escribir:

$$\text{Sobrecarga} = \frac{2 \times 10^{-6} \text{ s}}{T} = 0{,}05$$

Resolviendo la ecuación anterior se obtiene que, para una sobrecarga del 5 %, el periodo de muestreo $T$ ha de ser igual a 0,04 milisegundos.

2. Si se establece un periodo de muestreo de 0,04 milisegundos, en dos horas se tomarán:

$$\frac{2 \times 60 \times 60 \text{ s}}{0{,}04 \times 10^{-3} \text{ s}} = 18 \times 10^{7} \text{ muestras}$$

Dado que cada muestra ocupa una palabra de 32 bits, el volumen de información recogida será de $18 \times 10^{7} \times 32$ bits. Este mismo volumen expresado en MB será:

$$\frac{18 \times 10^{7} \times 32 \text{ bits}}{8 \times 1.024 \times 1.024} = 686{,}65 \text{ MB}$$

■

**PROBLEMA 2.2** El resultado de la ejecución de la orden vmstat 2 16 en un sistema informático que utiliza Linux como sistema operativo es el siguiente:

| procs | | | | memory | | | swap | | io | | system | | cpu | | |
|---|---|---|---|---|---|---|---|---|---|---|---|---|---|---|---|
| r | b | w | swpd | free | buff | cache | si | so | bi | bo | in | cs | us | sy | id |
| 0 | 0 | 0 | 0 | 26324 | 164928 | 19672 | 0 | 0 | 0 | 1 | 1 | 13 | 5 | 11 | 1 |
| 1 | 0 | 0 | 0 | 26324 | 164928 | 19672 | 0 | 0 | 0 | 0 | 113 | 97 | 12 | 29 | 59 |
| 5 | 0 | 0 | 0 | 26204 | 164928 | 19724 | 0 | 0 | 18 | 0 | 159 | 93 | 51 | 6 | 43 |
| 2 | 0 | 0 | 0 | 26136 | 164928 | 19728 | 0 | 0 | 4 | 0 | 118 | 127 | 68 | 0 | 32 |
| 2 | 0 | 0 | 0 | 26136 | 164928 | 19728 | 0 | 0 | 0 | 7 | 166 | 125 | 99 | 1 | 0 |
| 2 | 0 | 0 | 0 | 26136 | 164928 | 19728 | 0 | 0 | 0 | 0 | 121 | 160 | 99 | 1 | 0 |
| 2 | 0 | 0 | 0 | 26136 | 164928 | 19728 | 0 | 0 | 5 | 0 | 110 | 114 | 99 | 1 | 0 |
| 4 | 0 | 0 | 0 | 25604 | 164928 | 20048 | 0 | 0 | 2 | 0 | 440 | 231 | 81 | 19 | 0 |
| 2 | 0 | 0 | 0 | 25592 | 164928 | 20048 | 0 | 0 | 1 | 0 | 108 | 136 | 90 | 10 | 0 |
| 2 | 0 | 0 | 0 | 25792 | 164928 | 20048 | 0 | 0 | 0 | 1 | 120 | 182 | 98 | 2 | 0 |
| 3 | 0 | 0 | 0 | 25792 | 164928 | 20048 | 0 | 0 | 0 | 0 | 104 | 108 | 99 | 1 | 0 |
| 6 | 0 | 0 | 0 | 25792 | 164928 | 20048 | 0 | 0 | 32 | 0 | 115 | 137 | 97 | 3 | 0 |
| 8 | 0 | 0 | 0 | 25732 | 164928 | 20052 | 0 | 0 | 2 | 6 | 156 | 133 | 96 | 4 | 0 |
| 3 | 0 | 0 | 0 | 25732 | 164928 | 20052 | 0 | 0 | 0 | 0 | 103 | 81 | 78 | 22 | 0 |
| 3 | 0 | 0 | 0 | 25732 | 164928 | 20052 | 0 | 0 | 0 | 4 | 134 | 80 | 79 | 21 | 0 |
| 3 | 0 | 0 | 0 | 25732 | 164928 | 20052 | 0 | 0 | 0 | 0 | 111 | 100 | 76 | 24 | 0 |

1. ¿Cuál es la duración del periodo de medida?
2. ¿Tienen algún interés los datos ofrecidos en la primera línea? Razónese la respuesta.
3. ¿Cuál es el número medio de procesos que están esperando a ser ejecutados?
4. ¿Cuál es la utilización media del procesador en modo usuario?
5. Calcúlese la sobrecarga en el procesador producida por el sistema operativo.
6. ¿Ha habido actividad de intercambio de memoria (*swapping*) durante el periodo de medida? ¿Por qué?
7. ¿Cuál ha sido la actividad con los dispositivos de bloques?

**SOLUCIÓN:**

1. La orden vmstat 2 16 indica la toma de 16 medidas con un tiempo entre muestras de 2 segundos. Sin embargo, hay que tener en cuenta que la primera muestra se toma de forma inmediata, esto es, tan pronto como se ejecuta la orden. Por lo tanto, el periodo de medida será de 15 2 = 30 segundos, y no de 16 2 = 32 segundos, como se podría pensar inicialmente.
2. La primera línea no tiene ningún interés, ya que muestra valores medios de las medidas desde que el sistema se puso en marcha. Así, estas medidas carecen de valor significativo, y probablemente sean poco fiables. En consecuencia, la primera línea de datos que muestra la orden vmstat ha de descartarse. Esto implicará también que la simple ejecución de esta orden sin argumentos no tiene ninguna utilidad.
3. El número de procesos en espera para ser ejecutados, esto es, los que están preparados para ejecutarse, se indica por el parámetro r. Por tanto, y dejando de lado los datos ofre- cidos en la primera línea, se deberá calcular la media aritmética de la columna etiquetada con r para las 15 medidas válidas. Este valor será:

$$\bar{r} = \frac{1}{15}\sum_{i=1}^{15} r_i = \frac{48}{15} = 3{,}2 \text{ procesos}$$

Así, el número medio de procesos preparados para ejecutarse durante el periodo de medida contemplado ha sido de 3,2. Nótese que la columna b indica los procesos bloqueados en espera de algún recurso, como por ejemplo un dispositivo de entrada/salida, y la columna w muestra los procesos que, aunque se pueden ejecutar, han sido enviados al disco por intercambio de memoria (*swapping* ).

4. La utilización del procesador está expresada en las columnas us, sy y id. La columna us indica la utilización del procesador, en tanto por ciento, ejecutando código en modo usuario. Por tanto, la utilización media se calcula de la siguiente manera:

$$\overline{us} = \frac{1}{15}\sum_{i=1}^{15} us_i = \frac{1.222}{15} = 81{,}467$$

Así pues, la utilización media del procesador en modo usuario durante este periodo de medida ha sido del 81,467 %.

5. La sobrecarga producida por el sistema operativo se calcula como el porcentaje de tiem- po durante el cual el procesador ejecuta código del sistema operativo. Este parámetro, expresado en tanto por ciento, viene indicado en la columna etiquetada con sy, y su valor medio durante el periodo de medida será:

$$\overline{sy} = \frac{1}{15}\sum_{i=1}^{15} sy_i = \frac{144}{15} = 9{,}6$$

Por lo tanto, la sobrecarga debida al sistema operativo ha sido del 9,6 %.

6. La actividad de intercambio (*swapping* ), esto es, el transvase de procesos de memoria principal al disco, viene indicada por las columnas si y so. Estas columnas muestran la cantidad de memoria, expresada en KB/s, transvasada desde el disco hacia memoria principal, y al revés, respectivamente. Como estas columnas permanecen a cero durante el periodo de medida, podemos concluir que no ha habido actividad de esta clase.

7. La actividad con los dispositivos de entrada/salida de bloques está reflejada en las colum- nas bi y bo, las cuales indican la cantidad de bloques por segundo transferidos hacia los dispositivos de bloques, y desde éstos, respectivamente. Por tanto, según los datos reco- gidos, esta actividad ha sido más intensa desde los dispositivos de entrada/salida hacia memoria principal (columna bo) que a la inversa (columna bi). En concreto tendremos:

$$\overline{bi} = \frac{1}{15}\sum_{i=1}^{15} bi_i = \frac{64}{15} = 4{,}267 \text{ bloques/s}$$

$$\overline{bo} = \frac{1}{15}\sum_{i=1}^{15} bo_i = \frac{18}{15} = 1{,}2 \text{ bloques/s}$$

■

**PROBLEMA 2.3** Tras ejecutar la orden top en un sistema informático monoprocesador que utiliza Linux como sistema operativo se ha obtenido la siguiente información de acti- vidad:

```
   1:27pm    up 1 day,        1:11, 3 users, load average: 2.46, 0.80, 0.28
53 processes: 48 sleeping, 5 running, 0 zombie, 0 stopped
CPU states: 82.5% user,             0.5% system, 17.0% nice, 0.0% idle
Mem:   256464K  av,251672K used,        4792K free,0K shrd,         22792K  buff
Swap: 136512K    av,     1956K used,134556K free                     67484K  cached
```

| PID | USER | PRI | NI | SIZE | RSS | SHARE | STAT | LC | %CPU | %MEM | TIME | COMMAND |
|---|---|---|---|---|---|---|---|---|---|---|---|---|
| 6221 | pau | 0 | 0 | 600 | 600 | 484 | R | 0 | 27.5 | 0.2 | 0:43 | sieve |
| 6223 | pau | 0 | 0 | 596 | 596 | 480 | R | 0 | 27.3 | 0.2 | 0:28 | qnap |
| 6224 | pau | 0 | 0 | 596 | 596 | 480 | R | 0 | 27.3 | 0.2 | 0:26 | qnap |
| 6230 | pau | 0 | 10 | 584 | 562 | 468 | R N | 0 | 17.0 | 0.2 | 1:64 | trilog |
| 6231 | pau | 19 | 0 | 980 | 980 | 776 | R | 0 | 0.3 | 0.3 | 0:00 | top |
| 2715 | root | 20 | 0 | 23176 | 6212 | 2064 | S | 0 | 0.1 | 2.4 | 0:01 | X |
| 1 | root | 20 | 0 | 84 | 64 | 44 | S | 0 | 0.0 | 0.0 | 0:03 | init |
| 2 | root | 20 | 0 | 0 | 0 | 0 | SW | 0 | 0.0 | 0.0 | 0:00 | keventd |
| 3 | root | 20 | 0 | 0 | 0 | 0 | SW | 0 | 0.0 | 0.0 | 0:00 | kapmd |
| 4 | root | 20 | 19 | 0 | 0 | 0 | SWN | 0 | 0.0 | 0.0 | 0:00 | ksoftiq |
| 5 | root | 20 | 0 | 0 | 0 | 0 | SW | 0 | 0.0 | 0.0 | 0:11 | kswapd |
| 6 | root | 2 | 0 | 0 | 0 | 0 | SW | 0 | 0.0 | 0.0 | 0:00 | bdflush |
| 7 | root | 20 | 0 | 0 | 0 | 0 | SW | 0 | 0.0 | 0.0 | 0:00 | kupdatd |
| 8 | root | 19 | 0 | 0 | 0 | 0 | SW | 0 | 0.0 | 0.0 | 0:01 | kinoded |
| 366 | root | 20 | 0 | 600 | 596 | 488 | S | 0 | 0.0 | 0.2 | 0:00 | syslogd |
| 369 | root | 20 | 0 | 972 | 964 | 360 | S | 0 | 0.0 | 0.3 | 0:00 | klogd |
| 405 | root | 20 | 0 | 0 | 0 | 0 | SW | 0 | 0.0 | 0.0 | 0:00 | khubd |
| 487 | bin | 20 | 0 | 332 | 316 | 244 | S | 0 | 0.0 | 0.1 | 0:00 | portmap |
| 530 | at | 20 | 0 | 340 | 312 | 252 | S | 0 | 0.0 | 0.1 | 0:00 | atd |
| 554 | root | 14 | 0 | 900 | 816 | 684 | S | 0 | 0.0 | 0.3 | 0:02 | sshd |

Basándose en la información anterior, se pide:

1. ¿Cuánta memoria física tiene este computador?
2. ¿Qué porcentaje de la memoria física está usada actualmente?
3. ¿Cuál es la utilización media del procesador?
4. ¿Cuál es la carga media a lo largo de los últimos 15 minutos?
5. ¿Cómo es la evolución de la carga medida del sistema, ascendente o descendente?
6. ¿Hay algún proceso en ejecución con baja prioridad? ¿Cuál?
7. ¿Hay algún proceso residente en el disco debido a intercambio de memoria (*swapped out* )? ¿Cuál?

8. ¿Cuánta memoria física está ocupando el programa trilog?

9. ¿Cuánto tiempo lleva en ejecución el programa sieve?

**SOLuCIón:**

1. La memoria física tiene una capacidad de 256.464 KB.

2. Según indica el monitor, de toda la capacidad física de la memoria principal (256.464 KB), hay 251.672 KB usados, más de un 98 % del total. Sin embargo, esto no quiere decir que toda esta memoria principal, marcada como usada, esté realmente ocupada por algún programa.

3. La utilización media del procesador viene dada por la suma de las utilizaciones en modo usuario (user), sistema (system) y en ejecución de procesos con baja prioridad (nice). La suma de estos valores es:

   $$user + system + nice = 82{,}5 + 0{,}5 + 17{,}0 = 100$$

   Por tanto, el procesador se utiliza en un 100 %. Este valor también se podría haber calculado restando al 100 % del tiempo el porcentaje que está ocioso (idle): $100 - idle = 100 - 0 = 100$.

4. La carga medida de los últimos 15 minutos ha sido de 0,28 procesos.

5. La carga ha evolucionado de forma ascendente, porque hay un incremento del número de procesos en el núcleo del sistema operativo en el último minuto (2,46) respecto de los últimos 5 y 15 minutos (0,8 y 0,28, respectivamente).

6. Sí, porque hay dos procesos con el indicador N del parámetro STAT activado. Estos pro- cesos son trilog y ksoftiq. De estos dos procesos, el primero está en ejecución, y se puede observar que el porcentaje del tiempo de procesador que usa (17 %) es ligeramente menor que el ocupado por el resto de los procesos en ejecución (27 % aproximadamente).

7. Sí, porque hay varios procesos con el indicador W del parámetro STAT activado. Por ejemplo, keventd y kapmd.

8. La memoria física que ocupa un proceso viene indicada por el parámetro RSS (*resident set size*). En este caso, el proceso trilog ocupa 562 KB.

   Este parámetro no ha de confundirse con SIZE, que indica la memoria virtual ocupada por el proceso, dejando de lado la memoria no compartida (normalmente, el código ejecutable del proceso). Esta capacidad incluye las áreas de datos, tanto estáticas como dinámicas, del programa.

9. Según el monitor, el programa sieve lleva 43 segundos en ejecución.

■

**PROBLEMA 2.4** Un sistema informático que trabaja con el sistema operativo Linux tiene instalado el monitor de actividad sar (*system activity reporter* ). Este monitor se activa cada 20 minutos y tarda 450 milisegundos en ejecutarse por cada activación. En cada una de las activaciones se recoge información del sistema, se construye un registro de datos con esta información, y se añade al fichero histórico saDD del día DD correspondiente. Se pide:

1. Calcular la sobrecarga (*overhead* ) que genera este programa sobre el sistema informático.
2. Determinar el tamaño del directorio /var/log/sa a lo largo de dos semanas si el registro de datos generado por cada activación ocupa 3 KB.
3. Si el volumen máximo del directorio /var/log/sa es de 150 MB, ¿cuántos ficheros históricos saDD se pueden almacenar?
4. Describir qué efecto tiene la orden sar -d 2 30.
5. Indicar qué información histórica mostrará la siguiente orden: sar -A -s 9:00:00 -e 12:00:00 -f /var/log/sa/sa01.

**SOLuCIón:**

1. La sobrecarga se calcula por medio de la relación entre el tiempo que tarda cada activación del monitor y el intervalo de tiempo transcurrido entre dos activaciones consecutivas. En este caso tendremos:

$$\text{Sobrecarga} = \frac{450 \times 10^{-3}}{20 \times 60} = 0{,}000375$$

   Por tanto, la sobrecarga generada por el monitor sobre el sistema informático es del 0,0375 %.

2. Para dos semanas habremos de almacenar un total de 14 ficheros históricos saDD. En primer lugar, calculamos las muestras recogidas en un día:

$$\frac{24 \times 60}{20} = 72 \text{ muestras}$$

   El volumen de información recogida en 14 días que se almacenará en el directorio de ficheros históricos /var/log/sa será:

$$14 \times 72 \times 3 \text{ KB} = 3.024 \text{ KB}$$

3. El número máximo de ficheros se calcula dividiendo el espacio disponible entre el que ocupa cada fichero histórico:

$$\frac{150 \times 1.024\ KB}{72 \times 3\ KB} = \frac{153.600}{216} = 711{,}11$$

En consecuencia, en el directorio cabrán un total de 711 ficheros saDD enteros.

4. Recoge la información correspondiente al subsistema de disco (parámetro -d) en 30
muestras, con un intervalo de 2 segundos entre dos muestras consecutivas.

5. Muestra toda la información (parámetro -A) recogida el día 01 (fichero histórico sa01) entre las 9:00 y las 12:00 de la mañana.

■

**PROBLEMA 2.5** Respecto del monitor sar, ¿qué similitudes y qué diferencias podemos encontrar entre las dos órdenes siguientes?

1. sadc 3 10 datos_sadc
2. sar 3 10 > datos_sar

**SOLuCIón:** Las dos órdenes presentan el mismo patrón respecto a la toma de medidas: ambas efectúan 10 muestras, con 3 segundos entre muestras consecutivas. Sin embargo, las diferencias son más numerosas. La primera orden recoge toda la información disponible sobre la actividad del sistema, construye un registro binario y lo añade al fichero datos_sadc. Este fichero sólo puede ser leído mediante la orden sar -f datos_sadc, que puede incluir parámetros para seleccionar la información que se desee consultar. Por otra parte, la segunda orden sólo será capaz de obtener la información relativa al uso del procesador, previa llamada a sadc, y la grabará en formato texto en el fichero datos_sar. En resumen, el fichero datos_sadc contiene información binaria sobre toda la actividad del sistema recogida por el monitor, mientras que datos_sar contiene información en formato texto sobre la utilización del procesador del sistema. ■

**PROBLEMA 2.6** Un programa de cálculo numérico se ha instrumentado mediante la herramienta gprof y el resultado obtenido después de ejecutarlo ha sido el siguiente:

Flat profile:
Each sample counts as 0.02 seconds.

| % time | cumulative seconds | self seconds | calls | self ms/call | total ms/call | name |
|---|---|---|---|---|---|---|
| 51.44 | 170.52 | 170.52 | 12 | 14210.00 | 14210.00 | raíz |
| 26.34 | 257.83 | 87.31 | 2 | 43655.00 | 43655.00 | multi |
| 20.79 | 326.74 | 68.91 | 32 | 2153.44 | 2153.44 | tangente |
| 1.43 | 331.50 | 4.76 | 87 | 54.71 | 54.71 | suma |

Teniendo en cuenta estos datos se pide:

1. Determinar el número de muestras utilizada para generar los resultados.
2. ¿Cuál es el procedimiento más rápido? ¿Y el más lento? Justifíquese la respuesta.
3. Si el procedimiento más lento de los cuatro se sustituye por una nueva versión cuatro veces más rápida, ¿cuánto tiempo tardará en ejecutarse el programa? ¿Cuál es la mejora de rendimiento conseguida sobre el programa original con esta mejora?
4. Si los procedimientos raíz y multi se sustituyen por nuevas versiones tres y dos veces más rápidas, respectivamente, ¿qué mejora se obtendrá en el tiempo de ejecución del programa original?
5. ¿Cuánto se debería mejorar el procedimiento raíz si se quiere que el programa ori- ginal se ejecute en menos de 250 segundos?

**SOLuCIón:**

1. La cantidad de muestras se calcula dividiendo el tiempo que tarda el programa en ejecu- tarse entre el periodo de medida en que se recoge cada muestra:

$$\frac{331{,}50}{0{,}02} = 16.575 \text{ muestras}$$

2. El tiempo total que tarda en ejecutarse un procedimiento, esto es, desde que se llama hasta que acaba de ejecutarse la llamada, viene dado por la columna total ms call. Según los datos mostrados, el procedimiento más rápido es suma, que tarda 54,71 milisegundos, mientras que el más lento es multi, que tarda 43.655 milisegundos.
3. El nuevo tiempo de ejecución, después de hacer que el procedimiento multi se ejecute cuatro veces más rápidamente, será:

$$T = 170{,}52 + \frac{87{,}31}{4} + 68{,}91 + 4{,}76 = 266{,}0175 \text{ segundos}$$

La mejora de rendimiento conseguida se puede calcular dividiendo el tiempo de ejecución del programa original entre el nuevo tiempo:

$$A = \frac{331{,}50}{266{,}0175} = 1{,}2462$$

Así pues, el programa mejorado se ejecuta 1,2462 veces más rápidamente que el original, esto es, hay una mejora del 24,62 %.

4. Aplicamos la expresión de la ley de Amdahl teniendo en cuenta las fracciones de tiempo que se mejoran:

1 + 0,2079 + 0,0143 = 1,9034

0,2634

+

2

3

5. El procedimiento raíz consume el 51,44 % del tiempo de ejecución del programa. Así pues, si este procedimiento no se ejecutara, el programa tardaría 331,50 170,52 = 160,98. Para que el programa se ejecute en 250 segundos, el tiempo de ejecución del procedimiento raíz habría de pasar de 170,52 segundos a 250 160,98 = 89,02. En consecuencia, la mejora a efectuar sobre este procedimiento será:

$$\text{Mejora} = \frac{170{,}52}{89{,}02} = 1{,}9156$$

Esto es, hay que hacer que este procedimiento sea, aproximadamente, el doble de rápido; en concreto, la mejora será del 91,56 %.

Otra manera de resolver la cuestión puede ser la siguiente: se quiere reducir el tiempo de ejecución del programa en 331,50 250 = 81,5 segundos. Si solamente se mejora el procedimiento raíz, entonces su tiempo de ejecución se tendrá que rebajar en 170,52 81,5 = 89,02 segundos. Por tanto, la mejora necesaria que se deberá introducir en este procedimiento será 170,52/89,02 = 1,9156. ■

**PROBLEMA 2.7** Un programa escrito en lenguage C consta de una serie de llamadas a procedimiento. Dentro de la función main() se llama a proc1() y a proc2(); a su vez, desde proc1() se llama a proc11(), y desde proc2() se llama a proc21() y proc22(). El resultado de monitorizar la ejecución de este programa con gprof ha proporcionado la siguiente información:

Flat profile:

Each sample counts as 0.01 seconds.

| % time | cumulative seconds | self seconds | calls | self ms/call | total ms/call | name |
|---|---|---|---|---|---|---|
| 50.00 | 6.94 | 6.94 | 6 | 1156.67 | 1156.67 | proc11 |
| 22.19 | 10.02 | 3.08 | 20 | 154.00 | 154.00 | proc21 |
| 12.39 | 11.74 | 1.72 | 8 | 215.00 | 215.00 | proc22 |
| 10.16 | 13.15 | 1.41 | 4 | 352.50 | 1552.50 | proc2 |
| 4.54 | 13.78 | 0.63 | 2 | 315.00 | 3785.00 | proc1 |
| 0.72 | 13.88 | 0.10 | | | | main |

1. ¿Cuánto tiempo tarda en ejecutarse el programa?
2. Si hubiera que reducir significativamente el tiempo de ejecución del programa, ¿sobre qué procedimiento habría que hacer la mejora? ¿Por qué?
3. ¿En cuánto tiempo se ejecuta el procedimiento más rápido del programa?
4. ¿Cuánto tiempo transcurre por cada llamada al procedimiento proc1()?
5. ¿Cuánto tiempo tarda en ejecutarse el código propio del procedimiento proc1(), esto es, sin tener en cuenta las llamadas desde él a otros procedimientos?
6. Calcúlese el nuevo tiempo de ejecución del programa si la ejecución de los procedimientos proc21() y proc22() mejoran tres y siete veces, respectivamente.

**SOLuCIón:**

1. El programa tarda un total de 13,88 segundos en ejecutarse (último valor de la columna cumulative seconds).
2. La mejora habría que hacerla sobre el procedimiento proc11, ya que consume el 50 % del tiempo total de ejecución.
3. El procedimiento más rápido es proc21 (columna total ms/call), y tarda 154 milisegundos. Este procedimiento consume el 22,19 % del tiempo total de ejecución.
4. Cada llamada al procedimiento proc1() tarda 3.785 milisegundos (valor referido en la columna total ms/call).
5. El código propio contenido en el procedimiento proc1() tarda 315 milisegundos (columna self ms/call). La diferencia entre este tiempo y el anterior, esto es, 3.785 315 = 3.470 milisegundos, se emplea en ejecutar el procedimiento proc11(), que es llamado tres veces dentro de proc1(). Esto último se puede comprobar fácilmente porque proc1() se ejecuta dos veces y proc11() se ejecuta seis veces (véase la columna calls); como este último sólo se ejecuta desde el primero, se ejecutará necesariamente 6/2 = 3 veces.
6. El cálculo del nuevo tiempo de ejecución se puede llevar a cabo dividiendo directamente el tiempo consumido por ambos en el programa (columna self seconds) entre las mejoras correspondientes:

$$6{,}94 + \frac{3{,}08}{3} + \frac{1{,}72}{7} + 1{,}41 + 0{,}63 + 0{,}10 = 10{,}35 \text{ segundos}$$

En consecuencia, la ejecución del programa con los dos procedimientos mejorados se ha acelerado 13,88/10,35 = 1,34 veces. ■

**PROBLEMA 2.8** A continuación se muestra el resultado de la monitorización de la actividad de un programa escrito en C mediante la herramienta gprof. Sin embargo, debido a errores de tipo técnico, el informe obtenido carece de algunos datos que han sido marcados mediante el símbolo " x". Como información adicional, el grafo de llamadas que proporciona la herramienta de monitorización indica que todos los procedimientos son llamados úni- camente desde el programa principal main(), excepto ordena, que es llamado dentro del procedimiento procesa.

| % time | cumulative seconds | self seconds | calls | self ms/call | total ms/call | name |
|---|---|---|---|---|---|---|
| 94.77 | 26.11 | 26.11 | xxxx | 522.20 | 522.20 | ordena |
| 2.54 | 26.81 | 0.70 | 2 | 350.00 | 350.00 | invierte |
| 2.11 | 27.39 | 0.58 | 3 | 193.33 | xxxxx | normaliza |
| 0.58 | 27.55 | xxxx | 1 | 160.00 | 26270.00 | procesa |

Se pide resolver las siguientes cuestiones:

1. Tiempo que tarda en ejecutarse el procedimiento más rápido del programa.
2. Tiempo que tarda en ejecutarse el procedimiento más lento del programa.
3. ¿Cuántas llamadas se hacen a ordena desde el procedimiento procesa?
4. Tiempo que tarda en ejecutarse el código propio de procesa, esto es, sin tener en cuenta las llamadas realizadas al procedimiento ordena.
5. Indíquese qué alternativa, de las dos que se enumeran a continuación, permite dis- minuir más el tiempo de ejecución del programa completo: reducir a 20 el número de llamadas a ordena o rebajar su tiempo de ejecución a 300 ms por llamada.

**SOLuCIón:** De los datos mostrados en el informe podemos saber que el programa tarda 27,55 segundos en ejecutarse (último valor de la columna cumulative seconds). De todo este tiempo, el 94,77 % (esto es, 26,11 segundos) se consume en la ejecución del procedimiento ordena.

1. El tiempo total de ejecución de cada procedimiento viene especificado en la columna total ms/call. En esta columna falta por averiguar el tiempo que tarda el procedimiento normaliza. En particular, dado que no realiza ninguna llamada a otros procedimientos, el valor de la columna antes mencionada debe ser igual al valor correspondiente en la columna self ms/call, ya que todo el código ejecutado en este procedimiento es suyo propio. Así pues, el valor que falta por indicar es 193,33 ms. Teniendo este dato en cuenta, podemos concluir que el procedimiento más rápido es este mismo, normaliza.

2. El procedimiento más lento del programa, tras examinar la columna total ms/call, resulta procesa, que tarda 26.270 segundos. Nótese que este tiempo es el que transcurre desde que se comienza a ejecutar el procedimiento hasta que acaba, y en él influye tanto su código propio como las llamadas que haga a otros procedimientos (en este caso efectúa un número de llamadas a ordena que desconocemos por el momento).

3. Para saber las llamadas que se hacen al procedimiento ordena bastará con dividir el valor de la columna self seconds (total de tiempo que se ejecuta el código propio del procedimiento) entre el de la columna self ms/call (tiempo tardado en ejecutar el código propio por cada llamada), esto es, 26,11/0,5222 = 50.

4. El código propio de procesa tarda 160 ms en ejecutarse, según viene reflejado en la columna self ms/call. El resto del tiempo hasta completar los 26.270 ms es empleado en la ejecución del procedimiento ordena. En particular, podemos comprobar que este tiempo restante (26.270 160 = 26.110 ms) también se puede calcular como 50
522,20 = 26.110 ms.

5. Si el número de llamadas desde el procedimiento procesa a ordena se reduce de 50 a 20, el nuevo tiempo de ejecución del programa se puede calcular restándole al tiempo original la reducción conseguida. En particular, si nos ahorramos 30 llamadas a ordena, el tiempo que tardarían en ejecutarse será de 30 522,20 = 15.666 ms. Por tanto, el nuevo tiempo de ejecución del programa será de 27,55 15,666 = 11,884 segundos. La aceleración conseguida en el tiempo de ejecución del programa con esta mejora es de 27,55/11,884 = 2,318.

   Por el contrario, si se mantiene el número de llamadas al procedimiento ordena pero se reduce su tiempo de ejecución a 300 ms por llamada, entonces el tiempo de ejecución consumido por este procedimiento pasará de los 26,11 segundos originales a 50 0,3 = 15 segundos, es decir, se reduce el tiempo de ejecución en 26,11 15 = 11,11 segundos. Así pues, el tiempo de ejecución del programa completo será de 27,55 11,11 = 16,44 segundos. La aceleración conseguida en el tiempo de ejecución del programa con esta mejora es de 27,55/16,44 = 1,676.

   En definitiva, si se compara la aceleración conseguida por cada una de las dos alterna- tivas anteriores para reducir el tiempo de ejecución del programa completo, resulta más beneficioso emplear la primera de ellas. ■

## 2.6. Problemas con solución

**PROBLEMA 2.9** El monitor sar (*system activity reporter* ) de un sistema informático se activa cada 10 minutos. Este monitor tarda 300 ms en ejecutarse en cada activación. A partir de esta información se pide:

1. Estimar la sobrecarga (*overhead* ) que genera este programa sobre el sistema informático.
2. Calcular el tamaño del directorio /var/log/sa a lo largo de una semana si la infor- mación generada por activación ocupa 2 KB.
3. Describir el efecto de la orden sar -u 1 20.
4. Explicar el resultado que produce la ejecución de la orden sar -d -s 10:00:00 -e 14:00:00 -f /var/log/sa/sa23.

**SOLuCIón:**

1. La sobrecarga es del 0,05 %.
2. El volumen del directorio ocupará 2.016 KB.
3. Muestra la utilización del procesador un total de 20 veces a intervalos de 1 segundo.
4. Muestra la información referida a los discos desde las 10:00 hasta las 14:00 horas del día
23 del mes actual. ■

**PROBLEMA 2.10** A continuación se muestra el resultado obtenido tras ejecutar la orden top en un sistema informático que emplea Linux como sistema operativo:

2:52pm up 17 days, 3:41, 1 user, load average: 0.15,0.27,0.32
52 processes: 51 sleeping, 3 running, 0 zombie, 0 stopped
CPU states: 23.8% user, 14.0% system, 17.0% nice, 45.2% idle Mem: 257124K av,253052K used,4072K free,8960K shrd,182972K buff Swap: 261496K av,21396K used,240100K free,26344K cached

| PID | USER | PRI | NI | SIZE | RSS | SHARE | STAT | LIB | %CPU | %MEM | TIME | COMMAND |
|---|---|---|---|---|---|---|---|---|---|---|---|---|
| 807 | joan | 18 | 2 | 5708 | 5708 | 532 | R N | 0 | 23.0 | 2.2 | 6:16 | p_exec |
| 809 | joan | 14 | 2 | 5708 | 5708 | 532 | R N | 0 | 14.0 | 2.2 | 3:42 | p_exec |
| 185 | tomi | 1 | 0 | 824 | 824 | 632 | R | 0 | 0.5 | 0.3 | 0:00 | top |
| 201 | xp | 1 | 0 | 1272 | 1208 | 644 | S | 0 | 0.1 | 0.4 | 5:49 | xp_stat |
| 1 | root | 0 | 0 | 60 | 56 | 36 | S | 0 | 0.0 | 0.0 | 0:03 | init |
| 2 | root | 0 | 0 | 0 | 0 | 0 | SW | 0 | 0.0 | 0.0 | 0:00 | kflushd |
| 7 | root | 0 | 0 | 0 | 0 | 0 | SW | 0 | 0.0 | 0.0 | 0:00 | nfsiod |
| 194 | root | 0 | 0 | 72 | 4 | 4 | S | 0 | 0.0 | 0.0 | 0:00 | migetty |
| 195 | root | 0 | 0 | 68 | 0 | 0 | SW | 0 | 0.0 | 0.0 | 0:00 | migetty |
| 179 | root | 0 | 0 | 532 | 312 | 236 | S | 0 | 0.0 | 0.1 | 0:00 | sndmail |

1. ¿Cuánta memoria física tiene esta máquina?

2. ¿Qué porcentaje de la memoria física está usado según el monitor?
3. ¿Cuál es la utilización media del procesador?
4. ¿Cuál es la carga media del sistema a lo largo de los últimos 15 minutos?
5. ¿Cómo es la evolución de la carga media del sistema, ascendente o descendente?
6. ¿Hay algún proceso ejecutándose en baja prioridad?
7. ¿Hay algún proceso residente en el disco (*swapped out* )?
8. ¿Cuánta memoria física ocupa el monitor?

**SOLuCıón:**

1. La memoria física es de 257.124 KB.
2. El porcentaje ocupado de memoria física es del 98,4 %.
3. La utilización del procesador es del 54,8 %.
4. La carga media del sistema en los últimos 15 minutos es 0,32.
5. La evolución de la carga es descendente.
6. Sí, hay dos con el indicador N activado del parámetro STAT.
7. Sí, hay tres con el indicador W activado del parámetro STAT.
8. 824 KB. ■

**PROBLEMA 2.11** Tras ejecutar la orden vmstat 1 10 en un computador que utiliza Li- nux como sistema operativo se ha obtenido la siguiente información:

| procs | | | | memory | | | swap | | io | | system | | cpu | | |
|---|---|---|---|---|---|---|---|---|---|---|---|---|---|---|---|
| r | b | w | swpd | free | buff | cache | si | so | bi | bo | in | cs | us | sy | id |
| 3 | 0 | 0 | 10396 | 1120 | 396 | 50808 | 2 | 1 | 3 | 7 | 18 | 1 | 16 | 4 | 14 |
| 1 | 0 | 0 | 10396 | 1404 | 396 | 50696 | 0 | 0 | 0 | 4 | 275 | 130 | 77 | 23 | 0 |
| 1 | 0 | 0 | 10396 | 1404 | 396 | 50696 | 0 | 0 | 0 | 0 | 430 | 18 | 94 | 6 | 0 |
| 1 | 0 | 0 | 10396 | 1404 | 396 | 50696 | 0 | 0 | 0 | 0 | 328 | 18 | 91 | 9 | 0 |
| 1 | 0 | 0 | 10396 | 1404 | 396 | 50696 | 0 | 0 | 0 | 1 | 190 | 26 | 94 | 6 | 0 |
| 1 | 0 | 0 | 10396 | 1404 | 396 | 50696 | 0 | 0 | 0 | 0 | 202 | 18 | 95 | 5 | 0 |
| 1 | 0 | 0 | 10396 | 1404 | 396 | 50696 | 0 | 0 | 0 | 0 | 271 | 20 | 96 | 4 | 0 |
| 2 | 0 | 0 | 10396 | 1404 | 396 | 50696 | 0 | 0 | 0 | 0 | 177 | 16 | 95 | 5 | 0 |
| 2 | 0 | 0 | 10396 | 1404 | 396 | 50696 | 0 | 0 | 0 | 0 | 189 | 19 | 93 | 7 | 0 |
| 1 | 0 | 0 | 10396 | 1404 | 396 | 50696 | 0 | 0 | 0 | 18 | 200 | 21 | 95 | 5 | 0 |

1. Calcular el número medio de procesos preparados para ser ejecutados.
2. Calcular la utilización media del procesador ejecutando código en modo usuario.
3. Determinar la sobrecarga producida en el procesador por el sistema operativo.
4. ¿Ha habido alguna actividad de intercambio *swapping* durante el periodo de medida? ¿Por qué?

**SOLuCıón:**

1. El número medio de procesos preparados para ejecutares es 1,2.
2. El uso del procesador en modo usuario es del 92,22 % .
3. La sobrecarga debida a la utilización del procesador en modo supervisor es del 7,78 %.
4. No, porque los parámetros si y so están a cero. ■

**PROBLEMA 2.12** El monitor sar (*system activity reporter* ) de un computador se activa cada 5 minutos y tarda 400 ms en ejecutarse por cada activación. Se pide:

1. Calcular la sobrecarga (*overhead* ) que genera este monitor sobre el sistema informá- tico.
2. Si la información generada en cada activación ocupa 8.192 bytes, ¿cuántos ficheros históricos del tipo saDD se pueden almacenar en el directorio /var/log/sa si se dispone únicamente de 250 MB de capacidad libre?
3. Indicar los argumentos que se habrán de pasar a la orden sar si se desea mostrar la información referente a la utilización del procesador y la actividad de paginación del sistema el día 6 del mes actual desde las 15:00 hasta las 18:00 horas.

**SOLuCıón:**

1. La sobrecarga del monitor es del 0,13 %.
2. Se podrán almacenar 111 ficheros.
3. La orden es sar -uB -f /var/log/sa/sa06 -s 15:00:00 -e 18:00:00. ■

**PROBLEMA 2.13** Después de instrumentar un programa con la herramienta gprof el resultado obtenido ha sido el siguiente:

Flat profile:
Each sample counts as 0.01 seconds.

| % time | cumulative seconds | self seconds | calls | self ms/call | total ms/call | name |
|---|---|---|---|---|---|---|
| 60.28 | 70.32 | 70.32 | 10 | 7032.00 | 7032.00 | ordena |
| 28.96 | 104.10 | 33.78 | 8 | 4222.50 | 4222.50 | inicio |
| 10.76 | 116.66 | 12.56 | 34 | 369.41 | 369.41 | escala |

1. ¿Cuántas medidas se han tomado para obtener estos resultados?
2. Si el procedimiento más rápido de los tres se sustituye por otro tres veces más rápido, ¿cuánto tiempo tardará en ejecutarse el programa?
3. Si los tres procedimientos ordena, inicio y escala se sustituyen al mismo tiempo por nuevas versiones 2, 4 y 5 veces más rápidas, respectivamente, ¿qué mejora se obtendrá en el tiempo de ejecución sobre el programa original?

**SOLuCIón:**

1. Se han tomado 11.666 muestras.
2. El programa se ejecutará en 108,29 segundos.
3. La mejora será de 2,53. ■

---

**PROBLEMA 2.14** Un programa se ha instrumentado con la herramienta gprof del siste- ma operativo Linux y se ha obtenido el siguiente resultado después de ejecutarlo:

Flat profile:
Each sample counts as 0.01 seconds.

| % time | cumulative seconds | self seconds | calls | self ms/call | total ms/call | name |
|---|---|---|---|---|---|---|
| 70.10 | 94.36 | 94.36 | 15 | 6290.67 | 6290.67 | select |
| 29.90 | 134.60 | 40.24 | 43 | 935.81 | 935.81 | invert |

1. ¿Cuánto tiempo tardará en ejecutarse el programa si se hiciera el doble de llamadas a los dos procedimientos del mismo?
2. Si el procedimiento más rápido se sustituye por una nueva versión dos veces más rápida, ¿cuánto tiempo tardará en ejecutarse el programa?

3. Si los procedimientos select y invert se sustituyen al mismo tiempo por nuevas versiones siete y cuatro veces más rápidas, respectivamente, ¿qué mejora se obtendrá en el tiempo de ejecución del programa?

**SOLuCIón:**

1. El programa se ejecutará en 269,2 segundos.
2. El programa se ejecutará en 114,48 segundos.
3. La mejora será de 5,72. ■

## 2.7. Problemas sin resolver

**PROBLEMA 2.15** Un sistema informático dispone de una memoria cache de 8 KB externa al procesador y de una memoria principal de 64 KB. El tamaño de bloque de datos es de 8 bytes y el procesador tiene un espacio de direccionamiento de 128 KB. Indíquese qué información se necesitaría obtener si se quiere saber los bloques de cache accedidos por un programa si la correspondencia es:

1. Directa.
2. Asociativa por conjuntos de dos vías.

**PROBLEMA 2.16** En un computador se ha planificado la ejecución del monitor de actividad sar cada 10 minutos. Si cada activación de este monitor supone la ejecución de código en el procesador durante 40 ms, se pide calcular:

1. Sobrecarga que genera la ejecución del monitor sobre el computador.
2. Tamaño de un fichero de datos saDD, almacenado en el directorio /var/log/sa a lo largo de un día entero si la información recogida en cada activación ocupa un total de 6 KB.

**PROBLEMA 2.17** Indíquese qué orden (u órdenes), junto con los parámetros apropiados, se podría emplear para monitorizar los aspectos siguientes de la actividad en un computador que trabaja con el sistema operativo Linux:

1. Capacidad de memoria física ocupada por un proceso.
2. Número de cambios de contexto por segundo.

3. Carga media del sistema.
4. Número de interrupciones por segundo.
5. Capacidad libre de la unidad de disco magnético.
6. Usuarios conectados a la máquina.
7. Utilización del procesador en modo usuario.
8. Tiempo que lleva ejecutándose un proceso.
9. Tiempo que tarda un proceso en ejecutarse.

## 2.8. Actividades propuestas

**ACTIVIDAD 2.1** Analícese la información contenida en los ficheros siguientes del direc- torio /proc: cpuinfo, meminfo, ioports y stat.

**ACTIVIDAD 2.2** Mídase la actividad del sistema con la orden vmstat durante un periodo de 15 minutos. El periodo de muestreo será de tres segundos. Una vez recogida la infor- mación, represéntese gráficamente mediante algún programa gráfico como gnuplot. *Nota:* téngase en cuenta que la primera línea no contiene información relevante.

**ACTIVIDAD 2.3** Estúdiense a fondo las facilidades de la herramienta hdparm para ajus- tar los parámetros de configuración del disco de un computador. *Nota:* es necesario tener permisos de administrador del sistema (*root* ) para poder modificar los parámetros de fun- cionamiento del disco.

**ACTIVIDAD 2.4** Instálese el monitor sar en un computador con el sistema operativo Linux. Este programa está incluido en el paquete de utilidades de monitorización sysstat y se puede descargar de la página web http://perso.wanadoo.fr/sebastien.godard. *Nota:* es necesario tener permisos de administrador del sistema (*root* ).

**ACTIVIDAD 2.5** Ídem que la actividad anterior pero ahora con el monitor atsar (y el conjunto de utilidades adicionales que le acompañan) accesible en la dirección de Internet ftp.atcomputing.nl/pub/tools/linux. *Nota:* es necesario tener permisos de administra- dor del sistema (*root* ).

**ACTIVIDAD 2.6** Dado un sistema informático que dispone del monitor sar instalado, se pide lo siguiente:

1. ¿Cada cuánto tiempo se activa el monitor?

2. Determínese el tamaño de un fichero histórico correspondiente a un día entero. Te- niendo en cuenta el número de activaciones del monitor en un día, calcúlese el volumen que ocupa la información recogida en cada activación.
3. Planifíquese una sesión de medida de 5 minutos que recoja información sobre la actividad del disco. El intervalo de medida será de 12 segundos.
4. Compruébese la evolución de la actividad a lo largo de un día, con la ayuda de algún programa de visualización gráfica como gnuplot, referida a la utilización del procesador, el número de cambios de contexto y el uso del sistema de memoria.

**ACtiviDAD 2.7** Considérese el programa siguiente que calcula, un número considerable de veces, el factorial de un número aletatorio entre 0 y 31. Este cálculo se efectúa de dos maneras diferentes. La primera de ellas mediante un algoritmo recursivo, y la segunda mediante un algoritmo iterativo.

```
#include <stdlib.h> #define
VECES 3000000

long factorial_recursivo(long n); long
factorial_iterativo(long n);

void main()
{
   long a,b,i,j;

   for (i=1;i<=NVECES;i++)
   {
      j=1+(long) (31.0*rand()/(RAND_MAX+1.0));
      a = factorial_recursivo(j);
      b = factorial_iterativo(j);
   }
}

long factorial_recursivo(long n)
{
   if (n==0) return(1);
   return(n*factorial_recursivo(n-1));
}

long factorial_iterativo(long n)
{
   long i, fact = 1;

   for (i=1;i<=n;i++) fact = fact*i; return(fact);
}
```

Respecto del programa anterior se pide:

1. Determinar cuál de las versiones del algoritmo se ejecuta más rápidamente, y cuan- tificar la mejora en el tiempo de ejecución de la versión más rápida respecto de la más lenta.
2. Si un programa que tarda 45 minutos en ejecutarse utiliza la función implementada con la versión recursiva del algoritmo durante el 72 % del tiempo de ejecución, ¿cuánto tiempo tardará si esta función se sustituye por la versión iterativa? ¿Qué mejora del rendimiento se producirá?

**ACTIVIDAD 2.8** Un grupo de informáticos ha de diseñar una aplicación en lenguaje C que va a trabajar con matrices de grandes dimensiones de números reales. La aplicación realizará una serie de operaciones sobre estas matrices consistentes principalmente en el recorrido de sus elementos, para accesos de lectura y escritura. Con el objetivo de optimizar el diseño de la aplicación, el grupo de informáticos quiere averiguar, por un lado, si resulta más rápido acceder a los elementos de las matrices por filas o por columnas, y por otro, si es más costoso hacer una operación de lectura o escritura en memoria. Se pide diseñar un programa que ponga de manifiesto, si las hay, estas diferencias, y ayude a cuantificarlas. *Nota:* inténtese, en la medida de lo posible, reducir el efecto de la memoria cache del computador para que su existencia afecte lo menos posible a los resultados.

# Capítulo 3

## Análisis comparativo de rendimiento

Cualquier asunto que involucre la medida de prestaciones de un computador y una poste- rior comparación con otros sistemas provocará, inevitablemente, opiniones controvertidas. Partiendo de esta declaración de principios, es innegable que el análisis comparativo del rendimiento de varios computadores se ha ido perfeccionando con el paso del tiempo. Sin embargo, este proceso abarca tal variedad de factores, variables y, por qué no decirlo, inte- reses comerciales, que resulta muy complicado tener en cuenta todos los detalles al tiempo que se eliminan suspicacias.

La tendencia actual dentro del campo de la evaluación de prestaciones se orienta principalmente a utilizar índices que tienen en cuenta el tiempo de ejecución en un computador de un conjunto de programas de prueba o de evaluación (*benchmarks*). El carácter y na- turaleza de estos programas depende de qué se quiere evaluar de un sistema informático: desde un nivel más bajo ligado a componentes (por ejemplo, el procesador más el sistema de memoria más el compilador), hasta un nivel más elevado representado por el sistema completo (por ejemplo, un servidor web o un servidor de correo electrónico). Éste es el caso de corporaciones como SPEC (*Standard Performance Evaluation Corporation*), que propone índices denominados, con carácter general, SPECmarks, y cuyo significado depen- de del aspecto concreto que se esté evaluando. Algunos de estos índices, como por ejemplo SPECint o SPECfp, tienen en cuenta el tiempo de ejecución de un conjunto de progra- mas, y utilizan además algún tipo de normalización y cálculo de medias para reducir el rendimiento a un único indicador.

En este capítulo pondremos de manifiesto que el tiempo de ejecución es la única medida fiable para medir el rendimiento de un computador. Además, para que cualquier índice de prestaciones alternativo a éste resulte fiable (incluya o no el uso de algún tipo de media y normalización), también deberá reflejar este tiempo de ejecución.

En los apartados que vienen a continuación describiremos las principales características de algunas medidas de rendimiento, revisaremos las propiedades de los diferentes tipos de medias empleadas en el resumen de rendimientos y, finalmente, mostraremos algunas de las estrategias más sencillas para comparar el rendimiento de dos o más computadores.

## 3.1. Medidas de rendimiento

De entre todas la magnitudes medibles de un sistema informático susceptibles de ser utili- zadas como índices de prestaciones, el *tiempo* en llevar a cabo una actividad determinada representa la más intuitiva tanto para el analista como para el profano y, desde el punto de vista de la manipulación matemática, la menos susceptible de incorporar subterfugios. De hecho, ha sido ampliamente aceptado que el tiempo de ejecución de un programa re- presenta la medida exacta del rendimiento de un computador: aquel que ejecute la misma cantidad de trabajo en el menor tiempo posible será el más rápido.

Incluso teniendo en cuenta lo que acabamos de decir, existen otros índices de rendimien- to que, aun presentando algunos inconvenientes, se siguen utilizando, con mayor o menor acierto, en determinadas áreas de la informática. En particular, nos referimos a los índices MIPS, MFLOPS, MHz y CPI, todos ellos empleados con profusión por los diseñadores de procesadores.

Hasta hace bien poco, el procesador se ha venido considerando como la unidad funcional que representa de manera intrínseca el rendimiento del sistema completo. Tanto es así que, en numerosos textos sobre arquitectura de computadores, se suele equiparar el rendimiento del procesador con el rendimiento del computador en su conjunto. De hecho, la mayoría de los índices clásicos de prestaciones tienen como objetivo la medida del rendimiento del procesador. Por ejemplo, el índice MIPS (*million instructions per second* ) para un programa determinado se define como:

$$\frac{\text{Número de instrucciones}}{\text{Tiempo de ejecución} \times 10^6}$$

Aunque los MIPS han sido empleados tradicionalmente como índice para comparar las prestaciones de computadores con un mismo juego de instrucciones, presenta varios e importantes inconvenientes: no tiene en cuenta el tipo de instrucciones, puede variar según el programa ejecutado en un mismo computador y, más importante aún, el que un computador obtenga más MIPS que otro no implica necesariamente que proporcione un tiempo de ejecución menor. Como anécdota en tono jocoso, algunos autores han propuesto un significado alternativo para el acrónimo MIPS: *meaningless indicator of processor speed*.

Otro de los índices clásicos empleado para medir el rendimiento del procesador en el tratamiendo de números en coma flotante, especialmente en entornos de grandes com-putadores, es el MFLOPS (*million floating point operations per second* ), que computa el rendimiento en aritmética de coma flotante y se define como:

$$\frac{\text{Número de operaciones de coma flotante}}{\text{Tiempo de ejecución} \times 10^6}$$

El problema que plantea este índice es que el conjunto de operaciones de coma flotante suele variar de una arquitectura de procesador a otra (ocurre lo mismo con los MIPS cuando las máquinas a comparar no disponen del mismo juego o repertorio de instrucciones). Por otro lado, no es lo mismo, desde un punto de vista estrictamente temporal, hacer una suma que una división en coma flotante. Estos problemas se resuelven parcialmente con el establecimiento de los denominados MFLOPS *normalizados*, que permiten ponderar de una manera más justa la complejidad de cada operación en coma flotante.

Una métrica muy empleada en el ámbito comercial consiste en promocionar los computadores a partir de la frecuencia de reloj del procesador. Esta estrategia consigue captar la atención del comprador incauto porque resulta muy fácil asociar la idea de mayor fre- cuencia con la de menor tiempo de ejecución. Sin embargo, esta asociación resulta errónea en muchísimos casos porque deja de lado una gran cantidad de detalles tecnológicos y arquitectónicos tanto del procesador como del computador en su conjunto.

La mayoría de los textos académicos actuales que tratan sobre diseño del procesador tienen en cuenta varios factores que intervienen en el tiempo de ejecución de un programa. En particular, este tiempo se expresa en función del número $I$ de instrucciones ejecutadas, del número medio de ciclos de reloj necesarios para ejecutar una instrucción (CPI, *clock cycles per instruction*), y del tiempo de ciclo de reloj $t_c$:

$$\text{Tiempo de ejecución} = I \times \text{CPI} \times t_c$$

En la fórmula anterior intervienen tres factores con implicaciones muy diferentes. La variable $I$ depende tanto del repertorio de instrucciones como del compilador empleado en traducir el programa a lenguaje máquina. La variable CPI es un valor promedio y depende de la implementación que se haga del juego de instrucciones. Por último, el tiempo de ciclo o periodo de reloj $t_c$ depende de la frecuencia de reloj a la que funciona el procesador.

## 3.2. Resumen de rendimientos

Aunque el rendimiento de un computador tiene diversas y numerosas dimensiones, la mayo- ría de las personas involucradas en el diseño y evaluación del desempeño de computadores siguen reduciendo todo este espacio multidimensional a un único valor. El anhelo es que un simple número sea capaz de condensar toda la esencia del rendimiento de un sistema, cantidad que, por otro lado, se empleará para establecer comparaciones con otros compu- tadores. Esta reducción o unificación de valores se hace mediante las denominadas medidas de tendencia central: media, mediana y moda. De todas estas alternativas, el cálculo de medias o promedios suele ser el que con más asiduidad se emplea. De estas medias des- tacaremos tres: la media aritmética, la media armónica y la media geométrica, sin dejar de lado sus variantes ponderadas. Nótese que, en cualquier caso, la representatividad del valor medio calculado depende, en último extremo, de la variabilidad presente en los datos de partida.

Consideremos un conjunto de $n$ medidas de rendimiento $x_1, x_2, \ldots, x_n$. Las medias aritmética $\bar{x}_A$, armónica $\bar{x}_H$ y geométrica $\bar{x}_G$ de este conjunto de valores se definen, res- pectivamente, como:

$$\bar{x}_A = \frac{x_1 + x_2 + \cdots + x_n}{n} = \frac{1}{n}\sum_{i=1}^{n} x_i$$

$$\bar{x}_H = \frac{n}{\frac{1}{x_1} + \frac{1}{x_2} + \cdots + \frac{1}{x_n}} = \frac{n}{\sum_{i=1}^{n} \frac{1}{x_i}}$$

$$\bar{x}_G = \sqrt[n]{x_1 x_2 \cdots x_n} = \left(\prod_{i=1}^{n} x_i\right)^{\frac{1}{n}}$$

Las versiones ponderadas de las expresiones anteriores se emplean para dar mayor o me- nor relevancia a determinadas medidas dentro de todo el conjunto. Las medias ponderadas se calculan mediante las siguientes expresiones:

$$\bar{x}_{A,w} = w_1 x_1 + w_2 x_2 + \cdots + w_n x_n = \sum_{i=1}^{n} w_i x_i$$

$$\bar{x}_{H,w} = \frac{1}{\frac{w_1}{x_1} + \frac{w_2}{x_2} + \cdots + \frac{w_n}{x_n}} = \frac{1}{\sum_{i=1}^{n} \frac{w_i}{x_i}}$$

$$\bar{x}_{G,w} = x_1^{w_1} x_2^{w_2} \cdots x_n^{w_n} = \prod_{i=1}^{n} x_i^{w_i}$$

En las ecuaciones anteriores cada medida $x_i$ tiene asociado un peso $w_i$, con la restricción:

$$\sum_{i=1}^{n} w_i = 1$$

Cuando se utilizan los tiempos de ejecución de un conjunto de programas de prueba, el peso viene determinado por la frecuencia de uso de dichos tiempos (los programas que con mayor frecuencia se ejecuten serán los que tengan mayores pesos). Una opción alternativa, si no se conoce *a priori* el valor de los pesos, consiste en calcular su valor de manera que el tiempo total de ejecución se distribuya uniformemente entre todos ellos. En este caso los pesos se calculan mediante la expresión:

$$w_i = \frac{1}{T_i \sum_{j=1}^{n} \frac{1}{T_j}}$$

El uso de cada tipo de media depende del significado físico que represente cada medida de rendimiento $x_i$. Por ejemplo, si lo que se mide es el tiempo de ejecución de diversos programas, el valor medio calculado deberá ser proporcional a la suma de los tiempos de ejecución. Al contrario, si lo que se mide es el ratio entre el número de operaciones y el tiempo total de ejecución, el valor medio que se computa habrá de ser inversamente proporcional a la suma de los tiempos de ejecución.

Por ejemplo, la media aritmética es adecuada para promediar tiempos de ejecución y la media armónica para ratios, como por ejemplo MFLOPS. Sin embargo, ni la media aritmética sirve para promediar ratios ni la media armónica para promediar tiempos. Por su parte, la media geométrica no resulta apropiada para promediar tiempos ni ratios. Sin embargo, algunos autores han defendido su uso para resumir valores normalizados porque mantiene un orden consistente cuando se comparan las prestaciones de varios computadores (de hecho, se emplea en algunos índices de SPEC), aunque a veces este orden puede resultar erróneo si se compara con el establecido por la suma de los tiempos de ejecución. El adjetivo *consistente* significa en este contexto que el orden es el mismo, independientemente de la máquina que se tome como referencia para hacer la comparación.

En ocasiones, las medias anteriores se llevan a cabo sobre valores normalizados previamente (por ejemplo, dividiendo los tiempos de ejecución entre los obtenidos en una máquina de referencia). En cualquier caso, la mejor alternativa a seguir consiste en resu- mir el rendimiento y aplicar después la normalización al valor obtenido.

Para ilustrar el efecto del uso inadecuado de diferentes tipos de medias, consideremos los tiempos de ejecución (expresados en segundos) de 12 programas de prueba en cuatro computadores diferentes A, B, C y D mostrados en la Tabla 3.1. El computador R repre- senta la máquina que se toma como referencia (estos tiempos han sido publicados en la web de la organización SPEC y sirven para el cálculo del índice SPECint2000, el cual pretende medir el rendimiento del computador en aritmética entera).

La Figura 3.1 muestra gráficamente, para cada máquina, la media geométrica, la media armónica y la media aritmética de los tiempos de ejecución normalizados respecto de la máquina R (el caso de la media geométrica es similar a la forma de cálculo que emplea SPEC). Nótese que, tal como se ha señalado anteriormente, en estos tres casos el proce- dimiento que se sigue no es válido, ya que en todos ellos se están promediando ratios de tiempos de ejecución.

Para ilustrar los cálculos realizados consideraremos la máquina A. Esta máquina tarda un total de 1.831,1 segundos en ejecutar los 12 programas de prueba. Los valores normali-zados de los tiempos de ejecución obtenidos por esta máquina respecto de la máquina de referencia se calculan dividiendo los valores de la columna R entre los de la columna A. Así, estos valores son, para los 12 programas de prueba, los siguientes: 9,9, 9,1, 11,4, 6,6, 11,9, 10,1, 11,6, 11,9, 11,8, 14,3, 8,7 y 12,3. Así, tomando como base estos valores, la media geométrica resulta 10,61, la media armónica es 10,40, y finalmente, la media aritmética es 10,80.

| Programa | R | A | B | C | D |
|---|---|---|---|---|---|
| 1 | 1.400 | 141 | 170 | 136 | 134 |
| 2 | 1.400 | 154 | 166 | 215 | 301 |
| 3 | 1.100 | 96,8 | 94,2 | 146 | 201 |
| 4 | 1.800 | 271 | 283 | 428 | 523 |
| 5 | 1.000 | 83,8 | 90,1 | 77,4 | 81,2 |
| 6 | 1.800 | 179 | 189 | 199 | 245 |
| 7 | 1.300 | 112 | 131 | 87,7 | 75,5 |
| 8 | 1.800 | 151 | 158 | 138 | 142 |
| 9 | 1.100 | 93,5 | 122 | 104 | 118 |
| 10 | 1.900 | 133 | 173 | 118 | 142 |
| 11 | 1.500 | 173 | 170 | 179 | 240 |
| 12 | 3.000 | 243 | 264 | 338 | 508 |
| Suma | 19.100 | 1.831,1 | 2.010,3 | 2.166,1 | 2.710,7 |

**Tabla 3.1:** Tiempos de ejecución en diferentes máquinas.

Imaginemos que queremos comparar el rendimiento de los computadores B y C aten- diendo a los ratios de los tiempos de ejecución de cada programa. Si observamos la media geométrica, el rendimiento de ambas máquinas es similar (9,60 frente a 9,59). Sin embargo, si consideramos la media armónica, los dos computadores tienen rendimientos distintos: 9,45 para B y 8,94 para C, lo que proporciona una relación de medias de 9,45/8,94 = 1,06. Igual ocurre si consideramos la media aritmética, donde B obtiene un valor de 9,74 y C de 10,19: la relación de medias en este caso es de 10,19/9,74 = 1,05. Por tanto, el carácter cuantitativo de la conclusión varía según sea el tipo de media empleado.

La comparación anterior se puede llevar a cabo de forma satisfactoria si atendemos a los tiempos de ejecución (esto es, contemplando únicamente el tiempo total de ejecución del conjunto de programas). Así, los computadores B y C tienen rendimientos diferentes, ya que el primero tarda 2.010,3 segundos, mientras que el segundo tarda 2.166,1 segundos. En consecuencia, se puede ver de forma intuitiva que B es más rápido que C; en concreto, B resulta 2.166,1/2.010,3 = 1,08 veces más rápido que C ejecutando los programas de prueba.

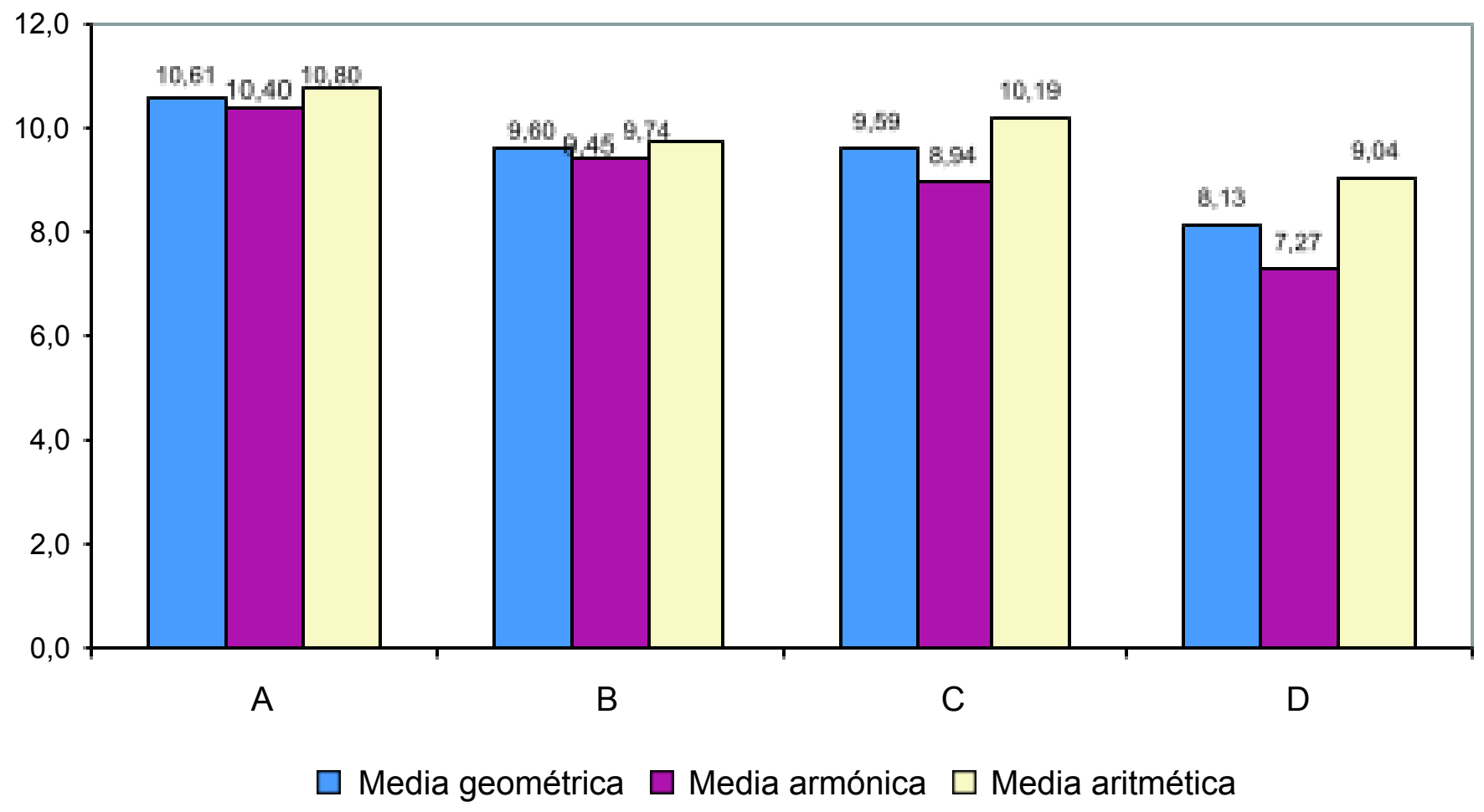

**Figura 3.1:** Comparativa entre medias de valores normalizados.

## 3.3. ¿Rendimientos significativamente diferentes?

Cuando se comparan las prestaciones de dos computadores tomando como base los tiempos de ejecución de un conjunto de $n$ programas de prueba puede resultar de utilidad deter- minar si las diferencias observadas tienen significancia estadística. Si consideramos que la variable $d_i$ representa la diferencia entre el tiempo de ejecución del programa $i$-ésimo en el primer computador y el tiempo de ejecución en el segundo, la expresión:

$$\bar{d} \pm t_{1-\frac{\alpha}{2},n-1} \times \frac{s}{\sqrt{n}}$$

establece el intervalo de confianza para la media $\bar{d}$ de las $n$ diferencias observadas. $t$ es la distribución de Student (con $n - 1$ grados de libertad), $s$ es la deviación típica, $\alpha$ representa el nivel de significación, y a $(1 - \alpha) \times 100$ se lo denomina el nivel de confianza. Habitualmente se utilizan niveles de confianza del 90 o 95 % ($\alpha = 0{,}10$ y $\alpha = 0{,}05$, respectivamente). Mientras que la fórmula anterior se emplea para valores de $n$ inferiores a 30, para valores superiores a esta cantidad se emplea la distribución normal:

$$\bar{d} \pm z_{1-\frac{\alpha}{2}} \times \frac{s}{\sqrt{n}}$$

Algunos valores de la distribución $t$ para un nivel de confianza del 95 % en función de los grados de libertad se muestran en la Tabla 3.2.

Nótese que la amplitud del intervalo depende del nivel de confianza, del número de programas de prueba utilizados y de la desviación típica. Cuando este intervalo incluye al 0 se puede afirmar que, con el nivel de confianza utilizado, los tiempos de ejecución no son significativamente diferentes. En caso contrario, se dice que hay diferencias significativas, y en consecuencia, los rendimientos de ambos computadores son significativamente distintos.

| $n-1$ | $t_{0{,}975,n-1}$ | $n-1$ | $t_{0{,}975,n-1}$ |
|---|---|---|---|
| 1 | 12,706 | 7 | 2,365 |
| 2 | 4,303 | 8 | 2,306 |
| 3 | 3,182 | 9 | 2,262 |
| 4 | 2,776 | 10 | 2,228 |
| 5 | 2,571 | 11 | 2,201 |
| 6 | 2,447 | 12 | 2,179 |

**Tabla 3.2:** Valores de la distribución *t* de Student para $\alpha$ = 0,05.

El uso de intervalos de confianza para comparar las prestaciones de dos computadores resulta un enfoque intuitivo y fácil de entender. Sin embargo, desde un punto de vista es- trictamente estadístico, esta aproximación es poco robusta. Una alternativa más completa para establecer comparaciones entre dos o más computadores consiste en emplear técnicas de análisis de la varianza (ANOVA, *analysis of variance*), que permiten separar la varia- ción total presente en los datos de partida en dos partes: la debida a factores puramente aleatorios y la correspondiente a las diferencias entre las alternativas.

## 3.4. Estrategias de comparación: ratios

El uso de ratios como índices de prestaciones ofrece un amplio abanico de posibilidades pa- ra comparar rendimientos. Un ratio consiste en un numerador y un denominador. Muchos de los índices utilizados en el ámbito informático son ratios (por ejemplo, CPI o MFLOPS) o bien se basan en el uso previo de éstos (por ejemplo, una media de tiempos de ejecución normalizados). Sin embargo, las comparaciones llevadas a cabo utilizando ratios pueden resultar confusas y, a menudo, afectadas de intereses partidistas. Para ilustrar los peli- gros que entraña el empleo de ratios, a continuación se muestran algunas peculiaridades derivadas de su uso para comparar rendimientos.

Imaginemos que queremos comparar el rendimiento de dos computadores, A y B. Si, por ejemplo, los tiempos de ejecución de A son siempre menores que los de B, el uso de ratios no puede ayudar a proclamar que B es más rápido que A. Sin embargo, si algunos programas se ejecutan más rápidamente en un computador que en otro, y viceversa, el ratio que define la normalización de tiempos puede ayudar a decantar la balanza hacia un lado. Supongamos que, para dos programas de prueba, los tiempos de ejecución obtenidos son los siguientes:

| Programa | A | B |
|---|---|---|
| 1 | 50 | 100 |
| 2 | 100 | 150 |
| Suma | 150 | 250 |
| Media aritmética | 75 | 125 |

Atendiendo a la suma de los dos tiempos de ejecución, la máquina A resulta 250/150 = 1,67 veces más rápida que B. A igual conclusión podemos llegar si comparamos las medias aritméticas: 125/75 = 1,67. Sin embargo, podemos utilizar ratios y alterar el resultado de la comparación normalizando los tiempos anteriores y tomando como máquina de referencia uno de los dos computadores en litigio. La normalización nos ofrece los siguientes valores (en paréntesis se denota la máquina que se toma como referencia):

| Programa | A (A) | B (A) | A (B) | B (B) |
|---|---|---|---|---|
| 1 | 1,00 | 2,00 | 0,50 | 1,00 |
| 2 | 1,00 | 1,50 | 0,67 | 1,00 |
| Suma | 2,00 | 3,50 | 1,17 | 2,00 |
| Media aritmética | 1,00 | 1,75 | 0,58 | 1,00 |

Por ejemplo, los dos primeros valores de la columna B(A) se obtienen tras dividir los tiempos de ejecución en B entre los tiempos correspondientes en A: 100/50 = 2,00 y 150/100 = 1,50. De igual forma, la columna A(B) se obtiene dividiendo los tiempos de ejecución en A entre los tiempos correspondientes en B: 50/100 = 0,50 y 100/150 = 0,67.

Tomemos, por ejemplo, la media aritmética de los tiempos de ejecución normalizados como índice de rendimiento (si se usara la suma el resultado sería idéntico). Si se emplea la máquina A como referencia se concluye que ésta es 1,75/1,00 = 1,75 veces más rápida que B, mientras que si se usa B como referencia entonces la mejora es ligeramente menor: 1,00/0,58 = 1,72. Obviamente, los diseñadores de la máquina A optarán por la primera opción a la hora de comparar.

Este ejemplo pone de manifiesto que los datos de partida pueden manipularse de di- ferentes maneras para obtener conclusiones más favorables. En cualquier caso, y como se ha mostrado en secciones anteriores, hay que tener en cuenta que promediar ratios no es una estrategia válida para comparar rendimientos, y menos aún cuando se emplea la media aritmética y una de las máquinas que se compara actúa como referencia en el proceso de normalización.

## 3.5. Problemas resueltos

**PROBLEMA 3.1** El rendimiento de un programa que implementa un conocido algoritmo numérico varía de acuerdo con las distintas secciones del código. En concreto, la generación de resultados del algoritmo se distribuye de acuerdo con los siguientes MFLOPS:

| Porcentaje de resultados | MFLOPS |
|---|---|
| 30 % | 1 |
| 20 % | 10 |
| 50 % | 100 |

Calcúlese el valor medio de los MFLOPS obtenidos por el algoritmo. ¿Cómo se distri- buye el tiempo de ejecución en función de los MFLOPS?

**SOLuCIón:** Dado que el índice de rendimiento es un ratio (operaciones dividido por tiempo) y éste difiere según el segmento de código ejecutado, el promedio se puede calcular mediante la media armónica ponderada:

$$\frac{1}{\frac{0,3}{1} + \frac{0,2}{10} + \frac{0,5}{100}} = 3,08 \text{ MFLOPS}$$

El valor anterior representa el rendimiento, expresado como ratio o tasa, del algoritmo.

Nótese la similitud de la expresión anterior con la ley de Amdahl.

Para calcular la distribución del tiempo de ejecución del algoritmo según los MFLOPS conseguidos podemos calcular la relación entre el porcentaje de resultados generados en cada fracción de código y la velocidad a la que éstos se generan. En particular, cuanto mayor sea el rendimiento del segmento, menor será el tiempo de ejecución consumido. Así pues, podemos escribir:

$$\frac{0,3}{1} + \frac{0,2}{10} + \frac{0,5}{100} = 0,3 + 0,02 + 0,005 = 0,325$$

Esta expresión indica que el algoritmo tarda en ejecutarse un 32,5 % del tiempo que tardaría si todos los resultados se obtuvieran con un rendimiento de 1 MFLOPS. Por tanto, la proporción de tiempo y la velocidad a la que se producen resultados se distribuyen de la siguiente manera: un 0,3/0,325 = 0,923 = 92,3 % del tiempo a 1 MFLOPS, un 0,02/0,325 = 0,062 = 6,2 % del tiempo a 10 MFLOPS, y finalmente, un 0,005/0,325 = 0,015 = 1,5 % del tiempo a 100 MFLOPS. En consecuencia, la mayor parte del tiempo de ejecución se emplea en producir resultados a la velocidad de 1 MFLOPS. Cualquier modificación que persiga obtener mejoras sustanciales en el rendimiento del algoritmo deberá recaer, necesariamente, en esta última porción de código. ■

---

**PROBLEMA 3.2** Una empresa de seguros está estudiando dos propuestas con el objetivo de actualizar los computadores personales de su instalación informática. El precio de cada computador es de 1.300 € y 1.450 €, para la propuesta A y B, respectivamente. Se estima que el número de computadores a reemplazar es de 75.

Los responsables informáticos de la empresa han ejecutado los ocho programas que utilizan habitualmente en un computador de cada propuesta, y han obtenido los tiempos de ejecución, expresados en segundos, que se muestran en la siguiente tabla:

| Programa | Modelo A | Modelo B |
|---|---|---|
| 1 | 23,6 | 24,0 |
| 2 | 33,7 | 41,6 |
| 3 | 10,1 | 8,7 |
| 4 | 12,9 | 13,5 |
| 5 | 67,8 | 66,4 |
| 6 | 9,3 | 15,2 |
| 7 | 47,4 | 50,5 |
| 8 | 54,9 | 52,3 |

Determínese si existen diferencias significativas en el rendimiento de los computadores personales de las dos propuestas.

**SOLUCIÓN:** Para saber si existen diferencias significativas en el rendimiento para las dos propuestas de adquisición construimos un intervalo de confianza para las diferencias observadas en los ocho tiempos de ejecución, a saber: 0,4, 7,9, 1,4, 0,6, 1,4, 5,9, 3,1 y 2,6. La media aritmética de estas diferencias es 1,56 y su desviación típica 3,75. El intervalo de confianza para un nivel de confianza del 95 % ($\alpha = 0,05$) será:

$$\bar{x} \pm t_{0,975,7} \times \frac{s}{\sqrt{n}} = -1,56 \pm 2,365 \times \frac{3,75}{\sqrt{8}} = -1,56 \pm 3,136$$

El valor 2,365 para la distribución $t$ se puede consultar en la Tabla 3.2. A raíz del resul- tado anterior, el intervalo de confianza es [ 4,696, 1,576]. Como este intervalo incluye al cero podemos afirmar que las diferencias observadas en los tiempos de ejecución de los programas no son significativas. En consecuencia, la mejor opción para actualizar los computadores de la empresa es la opción A, ya que resulta la menos costosa. En particular, la propuesta B resulta 1.450 €/1.300 € = 1,112 veces más cara que la A. ■

---

**PROBLEMA 3.3** Un estudio llevado a cabo mediante un monitor de ejecución de programas ha permitido cuantificar el tiempo medio de ejecución de las instrucciones que emplea una aplicación informática. Esta aplicación se ha ejecutado en dos procesadores, P y Q, con el mismo juego de instrucciones y se ha obtenido el siguiente resultado:

| Tipo de instrucción | Frecuencia de uso | Tiempo en P | Tiempo en Q |
|---|---|---|---|
| Memoria | 25 % | 70 ns | 72 ns |
| Comparación | 35 % | 32 ns | 27 ns |
| Salto | 25 % | 13 ns | 10 ns |
| Otras | 15 % | 18 ns | 12 ns |

Se pide:

1. Calcular el tiempo medio de ejecución de una instrucción en cada procesador y uti- lizarlo para cuantificar la mejora conseguida por el procesador más rápido.
2. Determinar el nuevo tiempo medio de ejecución de una instrucción en el procesador P si un nuevo diseño consigue que todas las instrucciones se ejecuten un 15 % más rápidamente.

**SOLuCıón:**

1. El tiempo medio de ejecución de una instrucción se calcula empleando la media aritmética ponderada, ya que conocemos el tiempo de ejecución de cada tipo y su frecuencia de uso por el programa. En consecuencia, podemos escribir:

$$T_P = 0{,}25 \times 70 + 0{,}35 \times 32 + 0{,}25 \times 13 + 0{,}15 \times 18 = 34{,}65 \text{ ns}$$
$$T_Q = 0{,}25 \times 72 + 0{,}35 \times 27 + 0{,}25 \times 10 + 0{,}15 \times 12 = 31{,}75 \text{ ns}$$

   Según estos resultados podemos concluir que el procesador Q ejecuta una instrucción, por término medio, 34,65/31,75 = 1,0913 veces más rápidamente.

2. Si todas las instrucciones se aceleran un 15 % el tiempo medio de ejecución también se verá mejorado en este porcentaje. En concreto, el nuevo tiempo medio de ejecución de una instrucción en el procesador P será de 34,65/1,15 = 30,13 nanosegundos.

**PROBLEMA 3.4** Considérese un programa de cálculo numérico que se ejecuta en dos minutos y hace las operaciones de coma flotante que se indican en la tabla adjunta. También se indican las operaciones normalizadas equivalentes.

| Operación | Cantidad | Operaciones normalizadas |
|---|---|---|
| ADD | $78 \times 10^6$ | 1 |
| SQRT | $29 \times 10^6$ | 3 |
| COS | $13 \times 10^6$ | 8 |
| EXP | $42 \times 10^6$ | 12 |

¿Cuál es el rendimiento conseguido por el computador con este programa de cálculo atendiendo a los MFLOPS? ¿Y si se mide en MFLOPS normalizados?

**SOLuCıón:** Los MFLOPS se calculan dividiendo el número de operaciones de coma flotante del programa entre el tiempo que se tarda en ejecutarlo:

$$\frac{(78 + 29 + 13 + 42) \times 10^6}{120 \times 10^6} = \frac{162}{120} = 1{,}35$$

Para calcular los MFLOPS normalizados repetimos el cálculo anterior teniendo en cuenta el número de operaciones normalizadas de cada tipo de operación de coma flotante:

$$\frac{(78 \times 1 + 29 \times 3 + 13 \times 8 + 42 \times 12) \times 10^6}{120 \times 10^6} = \frac{773}{120} = 6{,}442$$

■

---

**PROBLEMA 3.5** Un computador dispone de un procesador con un reloj que trabaja a 450 MHz. Este procesador estructura su juego de instrucciones en tres categorías: simples, normales y complejas. El número medio de ciclos por instrucción (CPI) para cada categoría se indica en la tabla adjunta.

| Tipo | CPI | Versión 1 | Versión 2 |
|---|---|---|---|
| Simple | 1 | $11 \times 10^6$ inst. | $12 \times 10^6$ inst. |
| Normal | 3 | $2 \times 10^6$ inst. | $2 \times 10^6$ inst. |
| Compleja | 5 | $3 \times 10^6$ inst. | $2 \times 10^6$ inst. |

El computador anterior se está utilizando para comparar el rendimiento de dos versiones de un compilador, V1 y V2. El número de instrucciones de cada categoría ejecutadas por un programa de prueba compilado con ambas versiones se indica también en la tabla anterior. Se pide calcular, para las dos versiones del compilador, el CPI medio y los MIPS conseguidos por el programa.

**SOLUCIÓN:** En primer lugar, nótese que ambas versiones del compilador producen el mismo número de instrucciones ejecutadas (la suma total de instrucciones es $16 \times 10^6$). El CPI medio de un programa se calcula dividiendo los ciclos de procesador utilizados en ejecutar el programa entre las instrucciones ejecutadas. Así pues, tendremos:

$$CPI_{V1} = \frac{(1 \times 11 + 3 \times 2 + 5 \times 3) \times 10^6 \text{ ciclos}}{(11 + 2 + 3) \times 10^6 \text{ instrucciones}} = \frac{32 \times 10^6}{16 \times 10^6} = 2$$

$$CPI_{V2} = \frac{(1 \times 12 + 3 \times 2 + 5 \times 2) \times 10^6 \text{ ciclos}}{(12 + 2 + 2) \times 10^6 \text{ instrucciones}} = \frac{28 \times 10^6}{16 \times 10^6} = 1{,}75$$

El tiempo de ejecución del programa se calcula dividiendo el número total de ciclos de procesador utilizados entre la frecuencia del reloj:

$$T_{V1} = \frac{32 \times 10^6}{450 \times 10^6} = 0{,}07111 \text{ s}$$

$$T_{V2} = \frac{28 \times 10^6}{450 \times 10^6} = 0{,}06222 \text{ s}$$

Si comparamos ambas versiones del compilador podemos concluir que la segunda de ellas permite que el programa se ejecute $0{,}07111/0{,}06222 = 1{,}14$ veces más rápidamente. Los

MIPS se calculan dividiendo el número total de instrucciones ejecutadas entre el tiempo total de ejecución del programa:

$$MIPS_{V1} = \frac{16 \times 10^6}{0{,}07111 \times 10^6} = 225$$

$$MIPS_{V2} = \frac{16 \times 10^6}{0{,}06222 \times 10^6} = 257{,}14$$

Este índice se podría haber calculado, también, a partir de la frecuencia del procesador y del CPI medio del programa:

$$MIPS = \frac{frecuencia}{CPI \times 10^6}$$

Aplicando esta fórmula obtenemos los mismos resultados:

$$MIPS_{V1} = \frac{450 \times 10^6}{2 \times 10^6} = 225$$

$$MIPS_{V2} = \frac{450 \times 10^6}{1{,}75 \times 10^6} = 257{,}14$$

■

**PROBLEMA 3.6** En la tabla mostrada a continuación se muestran los resultados obte- nidos (índices SPECint_base2000 y SPECint2000) por un determinado computador tras ejecutar el paquete de programas SPEC CPU2000 de aritmética entera y que han sido publicados en la página web de SPEC.

| Benchmarks | Base Ref Time | Base Run Time | Base Ratio | Peak Ref Time | Peak Run Time | Peak Ratio |
|---|---|---|---|---|---|---|
| 164.gzip | 1400 | 160 | 875 | 1400 | 159 | 880 |
| 175.vpr | 1400 | 317 | 442 | 1400 | 294 | 477 |
| 176.gcc | 1100 | 222 | 495 | 1100 | 190 | 578 |
| 181.mcf | 1800 | 517 | 348 | 1800 | 517 | 348 |
| 186.crafty | 1000 | 97.9 | 1022 | 1000 | 97.8 | 1022 |
| 197.parser | 1800 | 273 | 659 | 1800 | 272 | 661 |
| 252.eon | 1300 | 92.8 | 1401 | 1300 | 83.8 | 1551 |
| 253.perlbmk | 1800 | 170 | 1059 | 1800 | 170 | 1059 |
| 254.gap | 1100 | 134 | 819 | 1100 | 134 | 820 |
| 255.vortex | 1900 | 174 | 1089 | 1900 | 164 | 1157 |
| 256.bzip2 | 1500 | 258 | 582 | 1500 | 245 | 613 |
| 300.twolf | 3000 | 538 | 557 | 3000 | 529 | 567 |
| SPECint_base2000 | | | 720 | | | |
| SPECint2000 | | | | | | 749 |

Los tiempos que tardan en ejecutarse los programas en la máquina de referencia se reflejan en las columnas etiquetadas con Ref Time. Las dos columnas etiquetadas con Run Time corresponden a los tiempos obtenidos en la máquina a evaluar. Las columnas Base y Peak corresponden a los tiempos de ejecución obtenidos cuando el programa de prueba se compila sin y con parámetros de optimización, respectivamente.

Se pide calcular los índices SPECint_base2000 y SPECint2000 publicados en los datos anteriores. A partir de estos índices, ¿cuál es la mejora obtenida mediante las opciones de optimización en la compilación de los programas de prueba? ¿Coincide esta mejora con la obtenida si se emplea el tiempo total de ejecución como índice de prestaciones?

**SOLUCIÓN:** El índice SPECint_base2000 se calcula por medio de la media geométrica de los ratios entre los tiempos de ejecución en la máquina de referencia ($R_i$) y los tiempos obtenidos en la máquina objeto de estudio ($T_i$):

$$\text{SPECint_base2000} = \sqrt[12]{\prod_{i=1}^{12} \frac{R_i}{T_i}} \times 100 = 100 \times \sqrt[12]{\frac{1.400}{317} \times \cdots \times \frac{3.000}{538}} = 100 \times 7{,}1983 = 719{,}83$$

Dado que SPEC no emplea cifras decimales, el valor publicado es 720. Nótese que este valor está multiplicado por 100. De hecho, las columnas Base Ratio y Peak Ratio muestran los ratios entre los tiempos obtenidos por la máquina que se evalúa y los de la máquina de referencia (tiempos de ejecución normalizados), multiplicados por 100.

En el caso del índice SPECint2000 repetimos el cálculo anterior pero ahora con los tiempos obtenidos cuando se utilizan opciones de optimización en la compilación de los programas:

$$\text{SPECint2000} = \sqrt[12]{\prod_{i=1}^{12} \frac{R_i}{T_i}} \times 100 = 100 \times \sqrt[12]{\frac{1.400}{294} \times \cdots \times \frac{3.000}{529}} = 100 \times 7{,}4869 = 748{,}69$$

El valor de este índice publicado por SPEC es 749. Si comparamos los dos índices obtenidos podemos apreciar que hay un ligero incremento en el rendimiento cuando el código ejecutable de los programas se compila con opciones de optimización. En concreto, la mejora conseguida en este índice por estas optimizaciones es de 749/720 = 1,04, esto es, del 4 %. Si atendemos al tiempo total de ejecución, los doce programas tardan 2.953,7 y 2.855,6 segundos en ejecutarse, sin y con optimización en el proceso de compilación, respectivamente. Por tanto, la mejora conseguida es de 2.953,7/2.855,6 = 1,03, valor muy similar al obtenido utilizando los índices que propone SPEC. ■

**PROBLEMA 3.7** Un programa ejecuta un total de $120 \times 10^8$ instrucciones. De ellas, el 75 % se ejecutan en tres ciclos de reloj, mientras que el resto lo hace en cinco ciclos. Tras medir el tiempo de ejecución de este programa mediante la orden time del sistema operativo se ha obtenido la siguiente información:

```
real    0m84s
user    0m34s
sys     0m1s
```

Calcúlese el número medio de ciclos por instrucción (CPI) obtenidos por el programa, la frecuencia del procesador y los MIPS.

**SOLUCIÓN:** El programa ejecuta dos tipos de instrucciones: el primer tipo utiliza tres ciclos y representa el 75 % del total de instrucciones ejecutadas, mientras que el segundo emplea cinco ciclos y su contribución se reduce al 25 %. El número medio de ciclos por instrucción se puede calcular utilizando la media aritmética ponderada:

$$\text{CPI} = 0{,}75 \times 3 + 0{,}25 \times 5 = 3{,}5$$

Para calcular la frecuencia de reloj del procesador primero determinaremos el número de ciclos que tarda el programa en ejecutarse:

$$(120 \times 10^8) \times 3{,}5 = 420 \times 10^8 \text{ ciclos}$$

Según la información de la orden time, el programa se ejecuta en un total de 34 + 1 = 35 segundos (parámetros user más sys). Nótese que, aunque el usuario ha experimentado un tiempo de respuesta de 84 segundos, la diferencia entre este valor y los 35 segundos es tiempo que el programa está en espera compitiendo con otros programas por los recursos del computador. En consecuencia, la frecuencia de reloj que tiene el procesador será:

$$\frac{420 \times 10^8 \text{ ciclos}}{35 \text{ s}} = 12 \times 10^8 \text{ Hz} = 1{,}2 \text{ GHz}$$

Finalmente, los MIPS obtenidos por el programa se obtienen dividiendo las instrucciones ejecutadas entre el tiempo de ejecución:

$$\frac{120 \times 10^8 \text{ instrucciones}}{35 \text{ s} \times 10^6} = 342{,}86 \text{ MIPS}$$

■

**PROBLEMA 3.8** Dos empresas que diseñan procesadores han entrado en un litigio judicial a fin de demostrar cuál de ellas diseña el producto más rápido. Los dos procesadores que se están comparando son ATZUR y CIAN. En este contexto, el juez encargado del caso ha llamado a un informático en calidad de perito para dirimir el dilema. Como metodología de trabajo esta persona ha seleccionado un computador que soporta los dos procesadores y ha ejecutado un total de tres programas de prueba. Los tiempos de ejecución obtenidos en cada caso, expresados en minutos, han sido los siguientes:

| Programa | ATZUR | CIAN |
|---|---|---|
| 1 | 6 | 3 |
| 2 | 4 | 8 |
| 3 | 5 | 4 |

1. ¿Cuál sería, a grandes rasgos, el informe que recibiría el juez a fin de que emitiese una sentencia justa?
2. Dejando de lado la ética profesional, supongamos que la empresa que ha diseñado el procesador ATZUR hace un suculento ingreso en la cuenta corriente del perito. En este caso, ¿cómo podría el perito devolverle el favor a esta empresa (en términos de rendimiento, por supuesto)?
3. Ídem, pero suponiendo que es la empresa que fabrica CIAN quien hace el desintere- sado obsequio.

**SOLuCión:** Los únicos datos de partida vienen dados por los tiempos de ejecución obtenidos por los tres programas de prueba. Además, puesto que dependiendo del programa, un procesador resulta más rápido que el otro, tendremos opción de manipular los datos de alguna manera para que resulte vencedor uno de ellos.

1. Para emitir una sentencia justa nos podemos guiar, por ejemplo, por el tiempo total de ejecución. Aquel procesador que consiga ejecutar los tres programas de prueba en el menor tiempo será el que gane la comparativa. Sin embargo, en ambos casos el tiempo total de ejecución es de 15 minutos, por lo que se puede concluir que ambos presentan el mismo rendimiento.

   Otra posible manera de enfocar este informe es determinar si hay diferencias significativas en los tiempos de ejecución. En particular, las diferencias son 3, 4 y 1, cuya media aritmética es 0. Por tanto, e independientemente del ancho del intervalo de confianza, éste siempre incluirá el 0. Así pues, los tiempos de ejecución no presentan diferencias significativas.

2. Para beneficiar a una determinada opción se suele emplear algún tipo de normalización utilizando como máquina de referencia una de las máquinas que se comparan. En parti- cular, si se quiere beneficiar al procesador ATZUR dividiremos los tiempos de ejecución entre los de este procesador. La siguiente tabla muestra los valores normalizados, junto con el total y la media aritmética.

| Programa | ATZUR | CIAN |
|---|---|---|
| 1 | 1,0 | 0,5 |
| 2 | 1,0 | 2,0 |
| 3 | 1,0 | 0,8 |
| Suma | 3,0 | 3,3 |
| Media aritmética | 1,0 | 1,1 |

Como se puede apreciar, el procesador ATZUR obtiene una suma normalizada de 3,0, mientras que CIAN obtiene 3,3. Por lo tanto, si cuantificamos la mejora partiendo de esta normalización podemos concluir que ATZUR es 1,1 veces más rápido que CIAN. En términos porcentuales esta mejora es del 10 %.

Otra alternativa para beneficiar al procesador ATZUR consiste en utilizar una media aritmética ponderada, dando más peso a aquellos programas que tardan menos en este procesador. Por ejemplo, si asignamos arbitrariamente los pesos $w_1 = 0{,}1$, $w_2 = 0{,}8$ y $w_3 = 0{,}1$, las medias aritméticas ponderadas de los tiempos de ejecución son:

$$T_{ATZUR} = 0{,}1 \times 6 + 0{,}8 \times 4 + 0{,}1 \times 5 = 4{,}3 \text{ minutos}$$
$$T_{CIAN} = 0{,}1 \times 3 + 0{,}8 \times 8 + 0{,}1 \times 4 = 7{,}1 \text{ minutos}$$

Así pues, en este caso podemos decir que el procesador ATZUR obtiene un rendimiento $7{,}1/4{,}3 = 1{,}65$ veces superior al del procesador CIAN.

Finalmente, podríamos haber calculado los pesos relativos a cada programa de prueba para que el tiempo de ejecución se distribuya uniformemente entre todos ellos mediante la fórmula $w_i = 1/(T_i \times \sum_{j=1}^{n} \frac{1}{T_j})$. Usando los resultados obtenidos por el procesador ATZUR, los pesos calculados de esta manera son $w_1 = 0{,}27$, $w_2 = 0{,}41$ y $w_3 = 0{,}32$. Las medias ponderadas de los tiempos de ejecución obtenidas con estos pesos son:

$$T_{ATZUR} = 0{,}27 \times 6 + 0{,}41 \times 4 + 0{,}32 \times 5 = 4{,}86 \text{ minutos}$$
$$T_{CIAN} = 0{,}27 \times 3 + 0{,}41 \times 8 + 0{,}32 \times 4 = 5{,}37 \text{ minutos}$$

El resultado de esta estrategia permite concluir que ATZUR es $5{,}37/4{,}86 = 1{,}11$ veces más rápido que CIAN.

3. Repitiendo el procedimiento anterior de normalización, pero ahora respecto del procesador CIAN:

| Programa | ATZUR | CIAN |
|---|---|---|
| P1 | 2,0 | 1,0 |
| P2 | 0,5 | 1,0 |
| P3 | 1,25 | 1,0 |
| Suma | 3,75 | 3,0 |
| Media aritmética | 1,25 | 1,0 |

En virtud de estos valores, y empleando de nuevo el promedio normalizado, podemos concluir que el procesador CIAN es 1,25 veces más rápido que ATZUR (un 25 % en términos porcentuales).

Una conclusión similar se podría haber obtenido utilizando la media aritmética ponderada con unos pesos más favorables a CIAN, como, por ejemplo, $w_1 = 0{,}8$, $w_2 = 0{,}05$ y $w_3 = 0{,}15$. En este caso la media obtenida para ATZUR y CIAN es de 5,75 y 3,4 minutos, respectivamente, con lo que CIAN resulta 5,75/3,4 = 1,69 veces más rápido.

Finalmente, considerando que los tiempos de ejecución se distribuyen uniformemente para el procesador CIAN, los pesos asociados a cada uno de ellos son $w_1 = 0{,}47$, $w_2 = 0{,}18$ y $w_3 = 0{,}35$. El valor medio ponderado del tiempo total de ejecución con estos pesos para cada procesador resulta 5,29 y 4,25 minutos, para AZTUR y CIAN, respectivamente. Por tanto, la mejora de este último sobre el primero es de 5,29/4,25 = 1,25, esto es, un 25 % más rápido. ■

---

**PROBLEMA 3.9** Con el objetivo de evaluar el rendimiento en coma flotante de un computador se ha ejecutado un conjunto de cuatro programas de prueba. En la siguiente tabla se muestra, para cada programa, el tiempo de ejecución y el número de operaciones de coma flotante que lleva a cabo:

| Programa | Tiempo (s) | FLOP ($\times 10^9$) |
|---|---|---|
| P1 | 878 | 230 |
| P2 | 491 | 210 |
| P3 | 375 | 151 |
| P4 | 427 | 120 |

Calcúlese el número medio de MFLOPS que rinde este computador.

**SOLuCIón:** El número de MFLOPS para cada programa se calcula dividiendo el número de operaciones de coma flotante entre el tiempo de ejecución:

$$MFLOPS_{P1} = \frac{230 \times 10^9}{878 \times 10^6} = 262{,}0 \qquad MFLOPS_{P2} = \frac{210 \times 10^9}{491 \times 10^6} = 427{,}7$$

$$MFLOPS_{P3} = \frac{151 \times 10^9}{375 \times 10^6} = 402{,}7 \qquad MFLOPS_{P4} = \frac{120 \times 10^9}{427 \times 10^6} = 281{,}0$$

Para calcular el número medio de MFLOPS no se puede emplear la media aritmética, ya que este índice de prestaciones es un ratio con las operaciones en el numerador y el tiempo como denominador. En consecuencia, hay que utilizar la media armónica:

$$\frac{4}{\frac{1}{427,7} + \frac{1}{402,7} + \frac{1}{281,0}} = 328,0 \text{ MFLOPS}$$

Nótese que este valor también se puede obtener dividiendo la suma de las operaciones de coma flotante ejecutadas por los cuatro programas entre la suma de los tiempos de ejecución (nótese que hay una pequeña diferencia entre los dos valores debido a errores de redondeo):

$$\frac{(230 + 210 + 151 + 120) \times 10^9}{(878 + 491 + 375 + 427) \times 10^6} = \frac{711 \times 10^9}{2.171 \times 10^6} = 327,5 \text{ MFLOPS}$$

■

**PROBLEMA 3.10** Un estudio de evaluación de prestaciones pretende comparar el rendimiento obtenido por tres sistemas informáticos A, B y C. Para ello se han ejecutado dos programas de prueba, P1 y P2, cuyos tiempos de ejecución (en segundos) en cada sistema se reflejan en la tabla siguiente:

| Programa | A | B | C |
|---|---|---|---|
| P1 | 20 | 10 | 40 |
| P2 | 40 | 80 | 20 |

Se pide establecer una ordenación de los tres computadores, de mayor a menor rendi- miento, utilizando las siguientes alternativas para resumir y comparar prestaciones:

1. Suma del tiempo total de ejecución.
2. Suma normalizada del tiempo total de ejecución.
3. Media aritmética (sin y con normalización) de los tiempos de ejecución.
4. Suma y media aritmética de los tiempos de ejecución normalizados.
5. Media geométrica de los tiempos de ejecución.
6. Media geométrica de los tiempos de ejecución normalizados.

**SOLuCıón:** En este problema vamos a comprobar, una vez más, cómo las conclusiones de un estudio de comparación de rendimiento se pueden alterar empleando, de manera incorrecta, la aplicación de valores promedios y normalizaciones.

1. La medida más fiable para comparar rendimientos viene dada por el tiempo de ejecución de los programas de prueba. En este sentido la suma del tiempo de ejecución de estos programas será la que determine qué máquina es más rápida. Si se suman estos tiempos obtenemos:

| Programa | A | B | C |
|---|---|---|---|
| P1 | 20 | 10 | 40 |
| P2 | 40 | 80 | 20 |
| Suma | 60 | 90 | 60 |

En virtud de estos resultados las máquinas A y C presentan un rendimiento equivalente (tardan 60 segundos en ejecutar los dos programas), y ambas son 90/60 = 1,5 veces más rápidas que la máquina B.

2. El tiempo total de ejecución se puede normalizar respecto de cualquiera de las tres má- quinas:

| Programa | A | B | C |
|---|---|---|---|
| P1 | 20 | 10 | 40 |
| P2 | 40 | 80 | 20 |
| Suma | 60 | 90 | 60 |
| Respecto de A | 1,0 | 1,5 | 1,0 |
| Respecto de B | 0,67 | 1,0 | 0,67 |
| Respecto de C | 1,0 | 1,5 | 1,0 |

Si nos fijamos en las tres últimas filas podemos comprobar que la ordenación anterior se mantiene. En particular, las normalizaciones respecto de A y de C dan los mismos resultados: un tiempo de ejecución normalizado de 1,0 para estos dos computadores frente a 1,5 del computador B. La normalización respecto de C varía las cantidades de forma proporcional (A y C tardan el 67 % del tiempo de ejecución de B) y mantiene la misma ordenación.

3. La media aritmética se puede utilizar para promediar tiempos de ejecución porque man- tiene el significado físico de las cantidades que se resumen. La tabla siguiente muestra la media aritmética calculada sobre los tiempos de ejecución así como el valor normalizado de esta media respecto de los tres sistemas que se comparan:

| Programa | A | B | C |
|---|---|---|---|
| P1 | 20 | 10 | 40 |
| P2 | 40 | 80 | 20 |
| Media aritmética | 30 | 45 | 30 |
| Respecto de A | 1,0 | 1,5 | 1,0 |
| Respecto de B | 0,67 | 1,0 | 0,67 |
| Respecto de C | 1,0 | 1,5 | 1,0 |

Como se desprende de los valores mostrados en la tabla anterior, la media aritmética (normalizada o no) de los tiempos de ejecución mantiene las conclusiones de los apartados anteriores: los computadores A y C son, con un rendimiento equivalente, superiores al B. En concreto, ofrecen una mejora de 45/30 = 1,5 sobre el computador más lento.

4. La tabla siguiente muestra los resultados obtenidos tras calcular la suma y la media aritmética de los tiempos de ejecución normalizados respecto de los computadores A, B y C. Las columnas etiquetadas con (A), (B) y (C) muestran los valores normalizados de cada máquina en particular respecto de A, B y C, respectivamente.

| Prog. | A | (A) | (B) | (C) | B | (A) | (B) | (C) | C | (A) | (B) | (C) |
|---|---|---|---|---|---|---|---|---|---|---|---|---|
| P1 | 20 | 1,0 | 2,0 | 0,5 | 10 | 0,5 | 1,0 | 0,25 | 40 | 2,0 | 4,0 | 1,0 |
| P2 | 40 | 1,0 | 0,5 | 2,0 | 80 | 2,0 | 1,0 | 4,0 | 20 | 0,5 | 0,25 | 1,0 |
| Suma | 60 | 2,0 | 2,5 | 2,5 | 90 | 2,5 | 2,0 | 4,25 | 60 | 2,5 | 4,25 | 2,0 |
| Media | 30 | 1,0 | 1,25 | 1,25 | 45 | 1,25 | 1,0 | 2,125 | 30 | 1,25 | 2,13 | 1,0 |

De acuerdo con los valores anteriores, la normalización de los tiempos de ejecución pro- duce unos resultados diferentes a los ofrecidos por la suma de los tiempos de ejecución sin normalizar. Si se toma la máquina A como referencia, la suma de los tiempos nor- malizados ofrece los valores 2,0, 2,5 y 2,5 para A, B y C, respectivamente; esto es, la máquina A es la más rápida (mejora de 2,5/2 = 1,25) y las máquinas B y C tienen un rendimiento equivalente. En cambio, si tomamos B como referencia, los valores obtenidos son 2,5, 2,0 y 4,25; esto es, B es más rápida que A y que C, las que, por su parte, tienen rendimientos diferentes, siendo C la que presenta unas prestaciones mucho peores que el resto. Cuando se toma la máquina C como referencia, ésta sale muy beneficiada, ya que los valores obtenidos son 2,5, 4,25 y 2,0, y por tanto C resulta más rápida que A y B. Conclusiones similares obtenemos si analizamos la media aritmética de los tiempos normalizados. En consecuencia, podemos observar que el uso de la suma o la media arit- mética de los tiempos de ejecución normalizados mejora ostensiblemente la máquina que se toma como referencia y, peor aún, no mantiene el orden establecido por la suma de los tiempos de ejecución sin normalizar.

5. En la tabla siguiente podemos ver tanto la suma de los tiempos de ejecución como la media geométrica:

| Programa | A | B | C |
|---|---|---|---|
| P1 | 20 | 10 | 40 |
| P2 | 40 | 80 | 20 |
| Suma | 60 | 90 | 60 |
| Media geométrica | 28,3 | 28,3 | 28,3 |

En los tres casos la media geométrica ofrece el mismo resultado, y como consecuencia, la conclusión que se puede dirimir es que los tres computadores tienen un rendimiento equivalente, hecho que contrasta firmemente con la suma de los tiempos de ejecución.

6. Finalmente, la media geométrica de los tiempos de ejecución normalizados se presenta en la siguiente tabla:

| Prog. | A | (A) | (B) | (C) | B | (A) | (B) | (C) | C | (A) | (B) | (C) |
|---|---|---|---|---|---|---|---|---|---|---|---|---|
| P1 | 20 | 1,0 | 2,0 | 0,5 | 10 | 0,5 | 1,0 | 0,25 | 40 | 2,0 | 4,0 | 1,0 |
| P2 | 40 | 1,0 | 0,5 | 2,0 | 80 | 2,0 | 1,0 | 4,0 | 20 | 0,5 | 0,25 | 1,0 |
| Suma | 60 | 2,0 | 2,5 | 2,5 | 90 | 2,5 | 2,0 | 4,25 | 60 | 2,5 | 4,25 | 2,0 |
| Media | 28,3 | 1,0 | 1,0 | 1,0 | 28,3 | 1,0 | 1,0 | 1,0 | 28,3 | 1,0 | 1,0 | 1,0 |

Como se puede observar, la media geométrica establece el mismo orden entre los tres computadores independientemente de si se utilizan los tiempos de ejecución sin normalizar o normalizados (no importa qué máquina se tome como referencia). Ésta es la razón por la cual este tipo de media se emplea en el cálculo de índices como SPEC, donde los tiempos de ejecución se normalizan respecto de una máquina de referencia. Sin embargo, el orden que proporciona es erróneo, ya que según los valores que acabamos de calcular, las tres máquinas presentan rendimientos totalmente equivalentes, aunque sabemos que no tardan lo mismo en ejecutar los dos programas. Ésta es una de las razones por las que, en ocasiones, una máquina con un índice SPEC más elevado que otra puede tardar más tiempo en ejecutar un determinado conjunto de programas de prueba.

---

**PROBLEMA 3.11** Un equipo de programadores está diseñando una aplicación que trabaja fundamentalmente con vectores y matrices. Una gran parte del tiempo de ejecución de este programa es consumida por una subrutina de búsqueda. Los programadores han propuesto dos algoritmos distintos, A y B, para su implementación, pero no tienen claro cuál de ellos utilizar y han decidido llevar a cabo una serie de pruebas para comparar el rendimiento de ambos. La tabla siguiente muestra los tiempos de ejecución del programa (expresados en segundos) para diferentes datos de entrada:

| Algoritmo A | Algoritmo B |
|---|---|
| 125 | 132 |
| 140 | 139 |
| 133 | 127 |
| 128 | 131 |
| 138 | 141 |
| 126 | 123 |
| 136 | 135 |

¿Cuál de los dos algoritmos ofrece mejores prestaciones?

**SOLuCIóN:** Para comparar el rendimiento de los dos algoritmos primero calculamos la suma de los tiempos de ejecución. El algoritmo A tarda un total de 926 segundos, mientras que el B necesita 928 segundos. Así pues, ofrecen un tiempo total de ejecución muy similar, presentando el algoritmo A una mejora de 928/926 = 1,0022 sobre el B. Nótese que esta mejora es muy pequeña (no llega al 1 %).

Adicionalmente, podemos calcular un intervalo de confianza para las diferencias en los tiempos de ejecución y determinar si éstas son significativas. En particular, las diferencias observadas son: 7, 1, 6, 3, 3, 3 y 1. Su valor medio es 0,29 y la desviación típica 4,35. Como hay menos de 30 medidas utilizamos la distribución $t$. El intervalo de confianza para un nivel de confianza del 95 % ($\alpha = 0,05$) es:

$$\bar{x} \pm t_{0,975,6} \times \frac{s}{\sqrt{n}} = -0,29 \pm 2,447 \times \frac{4,35}{\sqrt{7}} = -0,29 \pm 4,02$$

El intervalo de confianza para la media de las diferencias es [−4,31, 3,74] y, dado que incluye el 0, podemos concluir que no hay diferencias significativas en el rendimiento de los dos algoritmos. Por tanto, el equipo de programadores puede utilizar cualquiera de los dos para implementar el programa. ■

**PROBLEMA 3.12** A continuación se muestran los tiempos de ejecución de seis programas de prueba en tres computadores distintos, A, B y C. Las cantidades están expresadas en segundos:

| Programa | A | B | C |
|---|---|---|---|
| 1 | 239 | 787 | 654 |
| 2 | 98 | 78 | 33 |
| 3 | 571 | 653 | 873 |
| 4 | 143 | 190 | 141 |
| 5 | 1.874 | 4.790 | 5.400 |
| 6 | 343 | 276 | 529 |

Se pide comparar el rendimiento de estos tres computadores utilizando la media geométrica de los tiempos de ejecución y de los tiempos de ejecución normalizados. ¿El orden obtenido con este tipo de media coincide con el establecido por el tiempo total de ejecución?

**SOLuCIón:** En primer lugar calculamos el tiempo total de ejecución obtenido por cada uno de los computadores: 3.268, 6.774 y 7.630 segundos, para A, B y C, respectivamente. Por lo tanto, el orden, desde el más rápido hasta el más lento, resulta: A, B y C. Este mismo orden se obtendría si, en vez de emplear la suma de los tiempos de ejecución, empleáramos la media aritmética: 544,7, 1.129,0 y 1.271,7.

En cambio, si calculamos la media geométrica de los tiempos de ejecución obtenemos 327,3, 464,7 y 443,3, lo que establece un orden distinto: A, C y B; es decir, según esta media el computador C obtiene mejor rendimiento que B. La Figura 3.2 muestra gráficamente estos valores.

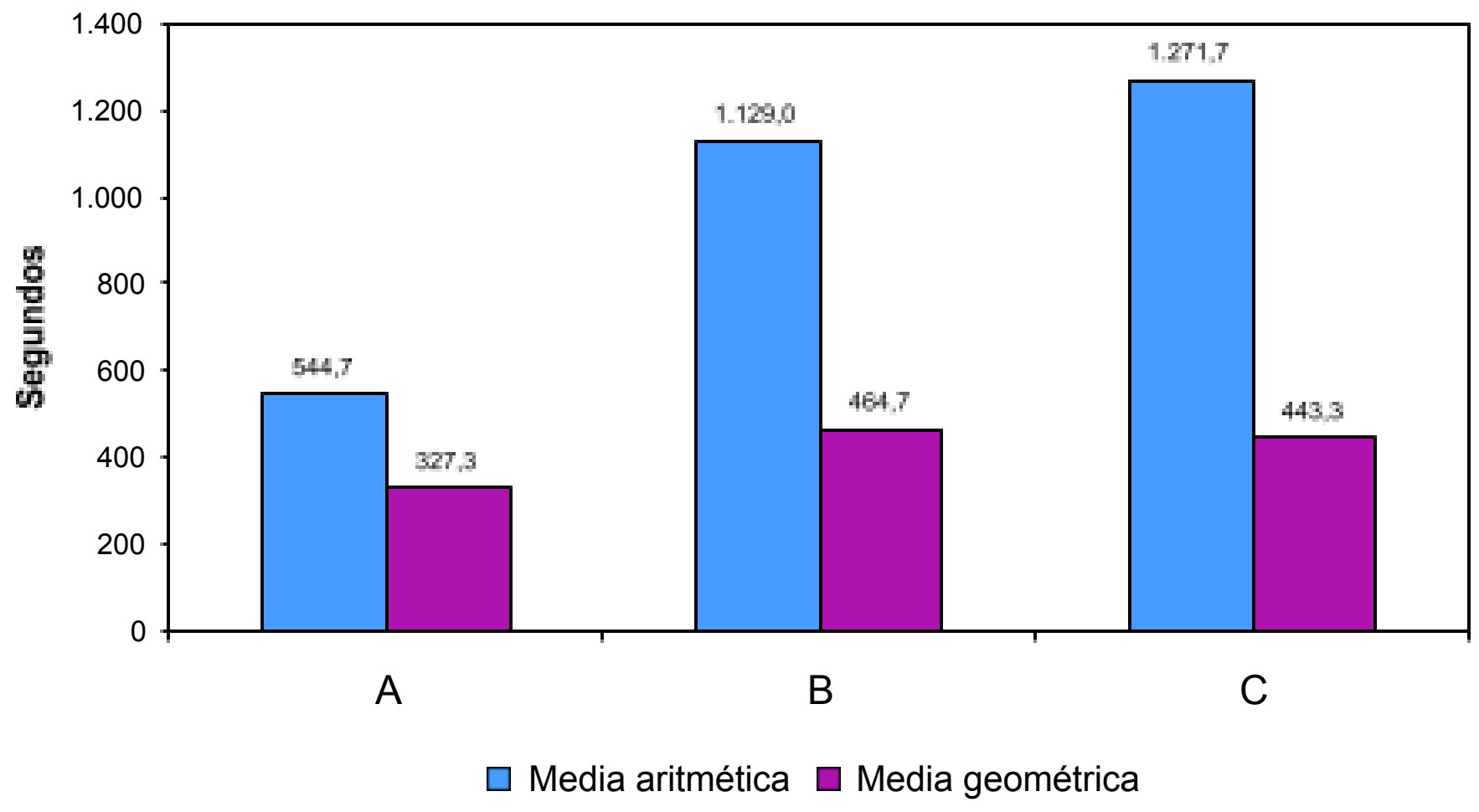

**Figura 3.2:** Comparación entre media aritmética y geométrica.

La siguiente tabla muestra los tiempos de ejecución normalizados (entre paréntesis se mues- tra el sistema que se ha tomado de referencia). La última fila muestra la media geométrica de los valores contenidos en cada columna.

| A (A) | B (A) | C (A) | A (B) | B (B) | C (B) | A (C) | B (C) | C (C) |
|---|---|---|---|---|---|---|---|---|
| 1,00 | 3,29 | 2,74 | 0,30 | 1,00 | 0,83 | 0,37 | 1,20 | 1,00 |
| 1,00 | 0,80 | 0,34 | 1,26 | 1,00 | 0,42 | 2,97 | 2,36 | 1,00 |
| 1,00 | 1,14 | 1,53 | 0,87 | 1,00 | 1,34 | 0,65 | 0,75 | 1,00 |
| 1,00 | 1,33 | 0,99 | 0,75 | 1,00 | 0,74 | 1,01 | 1,35 | 1,00 |
| 1,00 | 2,56 | 2,88 | 0,39 | 1,00 | 1,13 | 0,35 | 0,89 | 1,00 |
| 1,00 | 0,80 | 1,54 | 1,24 | 1,00 | 1,92 | 0,65 | 0,52 | 1,00 |
| 1,00 | 1,42 | 1,35 | 0,70 | 1,00 | 0,95 | 0,74 | 1,05 | 1,00 |

Si observamos los valores proporcionados por la media geométrica de los tiempos de ejecu- ción normalizados se puede comprobar que, para las tres normalizaciones, volvemos a obtener la misma ordenación: A, C y B. Así pues, la media geométrica ha dado siempre la misma orde- nación, independientemente de la máquina de referencia escogida, aunque no coincide con la obtenida por la suma ni la media aritmética de los tiempos de ejecución. ■

**PROBLEMA 3.13** La tabla siguiente muestra los tiempos de ejecución de tres programas de prueba en dos máquinas, A y B. Discútase si se puede aplicar alguna estrategia de análisis que concluya que la máquina A tiene un rendimiento superior al de la máquina B.

| Programa | A | B |
|---|---|---|
| 1 | 85 | 64 |
| 2 | 72 | 63 |
| 3 | 82 | 80 |

**SOLUCIÓN:** Dado que la máquina B ejecuta todos los programas en menos tiempo que la máquina A, no es posible concluir que esta última sea más rápida que la primera. En particular, si se toma en cuenta el tiempo total de ejecución de los programas, la máquina B resulta 239/207 = 1,16 veces más rápida que la máquina A.

Si se utilizase la media aritmética de los tiempos normalizados respecto de la máquina A, estrategia que, como sabemos, favorece al computador que actúa como referencia, los resultados serían los siguientes:

| Programa | A(A) | B(A) |
|---|---|---|
| 1 | 1,00 | 0,75 |
| 2 | 1,00 | 0,88 |
| 3 | 1,00 | 0,98 |
| Media aritmética | 1,00 | 0,87 |

A raíz de los datos mostrados en la tabla, la máquina A sigue siendo más lenta que la máquina B, ya que obtiene una media aritmética mayor. Incluso ponderando el peso de los valores que sirven para calcular esta media, sería imposible hacer que A mostrara mejores prestaciones que B. Iguales conclusiones podrían obtenerse para este caso de estudio si, en vez de emplear la media aritmética, se utilizase la media geométrica.

Adicionalmente, se puede calcular el intervalo de confianza de las diferencias observadas en los tiempos de ejecución. Por ejemplo, si restamos el tiempo que cada programa tarda en A menos el que tarda en B obtenemos las diferencias: 21, 9, y 2 segundos. El intervalo de confianza para estas diferencias ($n = 3$, $\alpha = 0,05$) es el siguiente:

$$\bar{d} \pm t_{1-\alpha/2,n-1} \times \frac{s}{\sqrt{n}} = 10,67 \pm 4,303 \times \frac{9,61}{\sqrt{3}}$$

es decir, [−13,21, 34,54], el cual incluye el cero y, por tanto, podemos concluir que no hay diferencias significativas en el rendimiento de las dos máquinas. Se obtendría la misma conclu- sión si las diferencias se calculasen al revés (valores negativos). En este sentido, el único hecho positivo respecto del rendimiento de la máquina A es que éste no resulta significativamente diferente del obtenido por la máquina B. ■

---

**PROBLEMA 3.14** El *benchmark* SYSmark2002 de la corporación BAPCO establece como índice de rendimiento de un computador una única cantidad denominada *SYSmark 2002 rating* obtenida como se describe a continuación de manera resumida. Los programas de prueba que integran este *benchmark* se organizan en dos grupos denominados *Internet Content Creation* y *Office Productivity*, y se calcula el tiempo total de ejecución de cada grupo. Estos dos tiempos obtenidos se normalizan respecto del tiempo tardado en una máquina de referencia denominada *Calibration Platform*, y finalmente cada uno de los dos ratios se multiplica por 100. Finalmente, los dos valores obtenidos se promedian utilizando la media geométrica.

En la página web de la corporación se han publicado los resultados correspondientes a dos computadores, que denominaremos A y B, y que se muestran a continuación:

| Máquina | *SYSmark 2002 rating* | *Internet Content Creation* | *Office Productivity* |
|---|---|---|---|
| A | 328 | 438 | 245 |
| B | 280 | 405 | 194 |

Sin embargo, en dicha página web no se encuentran disponibles los tiempos de ejecución obtenidos en cada computador para ambos grupos de programas de prueba (nótese que las dos últimas columnas de la tabla muestran ratios de tiempos de ejecución multiplicados por 100). Según los datos mostrados en la tabla anterior, la máquina A tiene un índice de rendimiento 328/280 = 1,17 veces más elevado que la máquina B.

Se pide realizar la comparación de los dos computadores utilizando como índice de rendimiento el tiempo de ejecución del *benchmark* completo. Supónganse para ello dos escenarios diferentes dados por los siguientes valores del tiempo de ejecución (expresados en segundos) en la máquina de referencia:

| Escenario | *Internet Content Creation* | *Office Productivity* |
|---|---|---|
| I | 1.000 | 1.000 |
| II | 10.000 | 80.000 |

**SOLuCIón:** En primer lugar es necesario estimar el tiempo de ejecución de los programas de prueba en cada una de las máquinas que se comparan a partir de la información publicada en la página web de BAPCO y reflejada en el enunciado del problema. Según la metodología empleada por SYSmark 2002, el índice 438 de la máquina A asociado al grupo de programas *Internet Content Creation* indica que, para este tipo de programas, esta máquina es 4,38 veces más rápida que la máquina tomada como referencia. Por ejemplo, si la máquina de referencia emplea 1.000 segundos, la máquina A tarda 1.000/4,38 = 228,3 segundos en ejecutar los mismos programas. Un razonamiento similar puede aplicarse al otro grupo de programas.

En particular, si consideramos el escenario I en el cual la máquina de referencia tarda 1.000 segundos en ejecutar cada grupo de programas, los tiempos que tardarán las máquinas A y B son los mostrados en la siguiente tabla (todos los tiempos se expresan en segundos):

| Máquina | Media aritmética | *Internet Content Creation* | *Office Productivity* |
|---|---|---|---|
| A | 318,24 | 228,31 | 408,16 |
| B | 381,19 | 246,91 | 515,46 |

Nótese que para resumir los tiempos de ejecución se ha utilizado la media aritmética, que es la adecuada en este caso (la media aritmética del tiempo de ejecución tiene significado físico). Si con estos datos queremos comparar el rendimiento de las máquinas A y B bastará con dividir el tiempo total de ejecución de la máquina más lenta entre la más rápida. Así, podemos concluir que A es aproximadamente 381,19/318,24 = 1,20 veces más rápida que B. Este resultado es, cuantitativamente hablando, diferente de la mejora de 1,17 obtenida con el índice nativo del *benchmark* que se indica en el enunciado.

Por otro lado, si suponemos el escenario II en el cual la máquina de referencia tarda 10.000 y 80.000 segundos en ejecutar el primer y segundo grupo de programas, respectivamente, los nuevos tiempos de ejecución que se obtienen en A y B son los siguientes:

| Máquina | Media aritmética | *Internet Content Creation* | *Office Productivity* |
|---|---|---|---|
| A | 17.468,08 | 2.283,11 | 32.653,06 |
| B | 21.853,12 | 2.469,14 | 41.237,11 |

En esta ocasión la mejora aproximada del computador A respecto del computador B es de 21.853,12/17.468,08 = 1,25, valor superior al obtenido en el primer escenario. Nótese que en el escenario II el segundo grupo de programas, que tiene un tiempo 8 veces mayor que el primero en la máquina de referencia, hace que su influencia en el tiempo total de ejecución sea mucho mayor.

Como se ha puesto de manifiesto, hemos podido plantear diferentes escenarios para com- parar dos máquinas A y B, determinados por la velocidad de la máquina de referencia, que proporcionan un mismo índice según SYSmark 2002, pero ofrecen distintas conclusiones en la comparación si utilizamos los tiempos de ejecución como índice de rendimiento. Así pues, el conocimiento del tiempo de ejecución es indispensable para cuantificar de forma concluyente y determinista la mejora de un computador respecto de otro.

---

**PROBLEMA 3.15** La tabla que se muestra a continuación refleja los tiempos de ejecución en segundos (Base run time) de los 14 programas de prueba que integran el *benchmark* SPEC CPU2000 empleados para el cálculo del índice SPECfp_base (rendimiento en coma flotante). En particular, los tiempos corresponden a la máquina de referencia R y a las dos máquinas, A y B, que se pretenden comparar.

| Programa | R | A | B |
|---|---|---|---|
| 1 | 1.600 | 159 | 150 |
| 2 | 3.100 | 180 | 168 |
| 3 | 1.800 | 185 | 182 |
| 4 | 2.100 | 223 | 580 |
| 5 | 1.400 | 104 | 98 |
| 6 | 2.900 | 193 | 180 |
| 7 | 2.600 | 208 | 207 |
| 8 | 1.300 | 127 | 128 |
| 9 | 1.900 | 152 | 149 |
| 10 | 2.200 | 232 | 230 |
| 11 | 2.000 | 167 | 170 |
| 12 | 2.100 | 195 | 201 |
| 13 | 1.100 | 250 | 247 |
| 14 | 2.600 | 275 | 230 |

Se pide comparar el rendimiento de los dos computadores A y B utilizando el índice SPECfp_base y el tiempo total de ejecución.

**SOLUCIÓN:** El cálculo del índice SPECfp_base se basa en la media geométrica de los ratios de los tiempos de ejecución respecto de la máquina de referencia multiplicados por 100 (Base ratio). Estos ratios multiplicados por 100 están reflejados en la siguiente tabla:

| Programa | A | B |
|---|---|---|
| 1 | 1.006 | 1.067 |
| 2 | 1.722 | 1.845 |
| 3 | 973 | 989 |
| 4 | 942 | 362 |
| 5 | 1.346 | 1.429 |
| 6 | 1.503 | 1.611 |
| 7 | 1.250 | 1.256 |
| 8 | 1.024 | 1.016 |
| 9 | 1.250 | 1.275 |
| 10 | 948 | 957 |
| 11 | 1.198 | 1.176 |
| 12 | 1.077 | 1.045 |
| 13 | 440 | 445 |
| 14 | 945 | 1.130 |

Por ejemplo, el valor 1.006 asociado a la máquina A para el primer programa se calcula dividiendo el tiempo que tarda la máquina de referencia (1.600 segundos) entre el tiempo que tarda la máquina A (159 segundos) y el resultado se multiplica por 100: $(1.600/159) \times 100 =$ 1.006,29. Nótese que en la tabla no se han reflejado los valores decimales, tal como hace SPEC en su página web.

Finalmente, el índice SPECfp_base se obtiene mediante la media geométrica de los valores mostrados en la tabla anterior. Estos índices resultan 1.071 y 1.032 para A y B, respectivamente. Así pues, podemos concluir que la máquina A tiene un índice SPECfp_base 1.071/1.032 = 1,04 veces mayor que la máquina B.

Por otro lado, si la comparación se hace en base a los tiempos de ejecución bastará con sumar los tiempos de cada programa en ambas máquinas. En particular, sumando los tiempos mostrados en la tabla del enunciado obtenemos que A tarda 2.650 segundos en ejecutar todos los programas, mietras que B tarda 2.920 segundos. En consecuencia, la máquina A resulta 2.920/2.650 = 1,10 veces más rápida que la B. ■

## 3.6. Problemas con solución

**PROBLEMA 3.16** Un estudio que pretende comparar el rendimiento de dos computadores A y B ha utilizado un conjunto de cuatro programas de prueba. Los tiempos de ejecución de estos programas, expresados en segundos, se reflejan en la siguiente tabla:

| Programa | Máquina A | Máquina B |
|---|---|---|
| P1 | 16,9 | 14,3 |
| P2 | 33,8 | 31,7 |
| P3 | 23,6 | 25,8 |
| P4 | 67,9 | 75,4 |

1. Compárense los rendimientos de ambas máquinas utilizando el tiempo total de ejecución.
2. Utilícese una estrategia de análisis que permita concluir que la máquina B es más rápida que la máquina A.
3. ¿Son significativas las diferencias observadas en los tiempos de ejecución de ambas máquinas?

**SOLUCIÓN:**

1. La máquina A resulta 1,04 veces más rápida que B.
2. Si se normalizan los tiempos de ejecución tomando como referencia la máquina B, ésta resulta 1,02 veces más rápida que A.
3. No, porque el intervalo de confianza incluye el 0. ■

**PROBLEMA 3.17** Un monitor de ejecución de programas ha analizado el comportamiento de una aplicación dedicada al cálculo numérico en dos máquinas, A y B, con la misma arquitectura de procesador. La información aportada ha permitido determinar la frecuencia de ejecución de cada tipo de instrucción y su tiempo de ejecución (expresado en nanose- gundos):

| Tipo de instrucción | Frecuencia | Tiempo en A | Tiempo en B |
|---|---|---|---|
| Acceso a memoria | 18 % | 100 | 110 |
| Aritméticas | 47 % | 70 | 80 |
| Trigonométricas | 23 % | 275 | 150 |
| Otras | 12 % | 10 | 9 |

Calcúlese el tiempo medio de ejecución de una instrucción en cada computador y em- pléese este tiempo para comparar el rendimiento de ambas máquinas.

**SOLuCión:** El tiempo medio de ejecución de una instrucción en A y B es de 115,35 y 92,98 ns, respectivamente. En consecuencia, la máquina B es, por término medio, 1,24 veces más rápida que A en la ejecución de una instrucción. ■

**PROBLEMA 3.18** Un computador dispone de un procesador que funciona con un reloj a 100 MHz. Este procesador dispone de tres tipos de instrucciones con un número medio de ciclos acorde a su complejidad de ejecución. En particular, los tipos de instrucción A, B y C tardan una media de 1, 2 y 3 ciclos en ejecutarse, respectivamente.

Este computador se utiliza para comparar el rendimiento de dos compiladores distintos, S y K. Para ello se ha monitorizado la ejecución de un programa de prueba compilado con ambos compiladores. El número de instrucciones ejecutadas de cada tipo se muestra en la siguiente tabla:

| Compilador | Tipo A | Tipo B | Tipo C |
|---|---|---|---|
| S | $5 \times 10^6$ | $1 \times 10^6$ | $1 \times 10^6$ |
| K | $10 \times 10^6$ | $1 \times 10^6$ | $1 \times 10^6$ |

Calcúlese, para cada compilador, el tiempo de ejecución del programa de prueba y los MIPS conseguidos. ¿Hay alguna contradicción en los resultados obtenidos?

**SOLuCión:** El programa compilado con S tarda 0,1 segundos y consigue alcanzar 70 MIPS, mientras que el compilado con K tarda 0,15 segundos y obtiene 80 MIPS. Este ejemplo pone de manifiesto que los MIPS no sirven para predecir el tiempo de ejecución. En particular, la mejora obtenida por S es de 0,15/0,1 = 1,5; sin embargo, utilizando los MIPS como índice de prestaciones, el compilador K permite obtener una mejora de 80/70 = 1,14. ■

**PROBLEMA 3.19** En la tabla adjunta se indican los tiempos de ejecución en dos computadores, A y B, de un conjunto de programas de prueba para aritmética entera (gcc, latex, gzip, bzip2) y aritmética de coma flotante (float, trilog, savage, linpack, whetstone). Los tiempos se expresan en segundos.

| Programa | Tiempo en A | Tiempo en B |
|---|---|---|
| gcc | 58,2 | 43,6 |
| latex | 32,1 | 21,3 |
| gzip | 42,9 | 32,8 |
| bzip2 | 11,0 | 8,2 |
| float | 54,2 | 40,2 |
| trilog | 46,6 | 45,1 |
| savage | 49,3 | 46,3 |
| linpack | 25,8 | 32,8 |
| whetstone | 52,0 | 58,3 |

Compárese el rendimiento de las dos máquinas mediante el tiempo total de ejecución teniendo en cuenta todos los programas del conjunto. Determínese si hay diferencias significativas entre las dos máquinas cuando ejecutan, por un lado, programas de aritmética entera, y por otro, de coma flotante.

**SOLUCIÓN:** Empleando el tiempo total de ejecución de todos los programas podemos concluir que la máquina B es 1,13 veces más rápida. En aritmética entera las diferencias son significativas, resultanto la máquina B 1,36 veces más rápida que A. En aritmética de coma flotante no hay diferencias significativas, resultando la máquina B ligeramente más rápida. ■

---

**PROBLEMA 3.20** Un programa que se ejecuta en 87 segundos hace las operaciones de coma flotante reflejadas en la tabla siguiente. También se indican las operaciones nor- malizadas equivalentes a cada tipo de operación matemática. Exprese el rendimiento del computador tanto en MFLOPS como en MFLOPS normalizados.

| Operación | Cantidad | Operaciones normalizadas |
|---|---|---|
| ADD, SUB | $54 \times 10^6$ | 1 |
| DIV | $36 \times 10^6$ | 2 |
| SQRT | $13 \times 10^6$ | 4 |
| SIN, ATAN | $21 \times 10^6$ | 6 |
| LOG | $35 \times 10^6$ | 10 |

**SOLUCIÓN:** El computador ejecuta 1,83 MFLOPS y 7,52 MFLOPS normalizados. ■

---

**PROBLEMA 3.21** A continuación se muestran los tiempos de ejecución experimentados en tres computadores, A, B y R, para un conjunto de cinco programas de prueba:

| Programa | A | B | R |
|---|---|---|---|
| 1 | 96,2 | 95,3 | 103,9 |
| 2 | 13,1 | 10,2 | 53,8 |
| 3 | 79,6 | 67,4 | 156,3 |
| 4 | 45,2 | 51,8 | 98,1 |
| 5 | 88,3 | 89,3 | 238,5 |

Calcúlese el índice de prestaciones de las máquinas A y B a la manera de SPEC, toman- do como referencia la máquina R. Compárese el rendimiento de estas máquinas atendiendo tanto a este índice como al tiempo total de ejecución. ¿Hay diferencias significativas en los rendimientos de A y B?

**SOLuCIón:** La media geométrica de los tiempos de ejecución normalizados respecto de la máquina R son 2,20 y 2,32 para A y B, respectivamente. Según SPEC, estos índices serían 220 y 232, respectivamente. En cualquier caso, estos datos permiten concluir que B rinde 1,06 veces más que A. En cambio, la suma de los tiempos de ejecución son 322,4 y 314,0, respectivamente, lo que rebaja la mejora conseguida a 1,03. Respecto a las diferencias, éstas no son significativas porque el intervalo de confianza incluye el 0. ■

## 3.7. Problemas sin resolver

**PROBLEMA 3.22** La siguiente tabla muestra los tiempos de ejecución de una serie de programas de prueba en tres sistemas informáticos distintos. La última columna muestra el número de instrucciones ejecutadas por cada programa.

| Programa | S1 | S2 | S3 | Instrucciones |
|---|---|---|---|---|
| 1 | 43 | 54 | 35 | $3{,}32 \times 10^{10}$ |
| 2 | 98 | 97 | 216 | $8{,}31 \times 10^{12}$ |
| 3 | 38 | 43 | 62 | $2{,}49 \times 10^{9}$ |
| 4 | 27 | 36 | 41 | $1{,}77 \times 10^{10}$ |
| 5 | 59 | 68 | 23 | $3{,}92 \times 10^{10}$ |
| 6 | 32 | 57 | 37 | $4{,}23 \times 10^{10}$ |

Suponiendo que todos los programas tienen la misma importancia en este estudio de evaluación, compárense las prestaciones de estos tres sistemas en base a:

1. Media aritmética de los tiempos de ejecución.
2. Media aritmética de los tiempos de ejecución normalizados respecto del primer sis- tema.
3. MIPS.

**PROBLEMA 3.23** Repítase el problema anterior suponiendo que los pesos atribuidos a cada programa de prueba son, respectivamente: 0,10, 0,35, 0,25, 0,10, 0,05 y 0,15.

**PROBLEMA 3.24** La siguiente tabla muestra los tiempos de ejecución de tres programas de prueba en tres sistemas distintos, A, B y C. Utilícense diversas estrategias de compara- ción para demostrar la superioridad de cada uno de ellos respecto de los otros dos.

| Programa | A | B | C |
|---|---|---|---|
| 1 | 25 | 50 | 75 |
| 2 | 50 | 75 | 25 |
| 3 | 75 | 25 | 50 |

**PROBLEMA 3.25** Un procesador con arquitectura RISC (*Reduced Instruction Set Computer* ) es capaz de ejecutar $50 \times 10^6$ instrucciones por segundo, mientras que otro procesador con arquitectura CISC (*Complex Instruction Set Computer* ) ejecuta $45 \times 10^6$ instrucciones por segundo. Razónese cuál de estos dos procesadores presenta un mejor rendimiento.

**PROBLEMA 3.26** Con el objetivo de analizar el efecto de la memoria cache en el rendimiento de un computador se ha medido el tiempo de ejecución de una serie de programas de prueba. Estos tiempos, expresados en segundos, son los siguientes:

| Programa | Sin cache | Con cache |
|---|---:|---:|
| 1 | 128 | 82 |
| 2 | 113 | 81 |
| 3 | 98 | 76 |
| 4 | 139 | 122 |
| 5 | 117 | 105 |

Determínese si las diferencias observadas son significativas y, en caso afirmativo, calcú- lese la mejora conseguida en el rendimiento debido al uso de la memoria cache.

## 3.8. Actividades propuestas

**ACTIVIDAD 3.1** Consúltense las páginas web de las empresas Intel (www.intel.com) y AMD (www.amd.com) para saber cuáles son los mecanismos que utilizan estas empresas para medir el rendimiento de sus procesadores.

**ACTIVIDAD 3.2** La organización TPC (*Transaction Processing Performance Council* ) es una corporación de empresas cuyo objetivo es evaluar el rendimiento de sistemas infor- máticos en aplicaciones transaccionales y de bases de datos (OLTP, *on-line transaction processing*). Consúltese la página web www.tpc.org para analizar qué tipos de programas de prueba se proponen para evaluar el rendimiento de estos sistemas y qué índices de prestaciones se utilizan.

**ACTIVIDAD 3.3** Repítase la actividad anterior para el caso de la corporación SPEC (*Standard Performance Evaluation Corporation*) cuya página web es www.spec.org.

**ACTIVIDAD 3.4** Repítase la actividad anterior para el caso de la corporación BAPCo (*Business Applications Performance Corporation*) cuya página web es www.bapco.com, y que desarrolla diversos conjuntos de programas de prueba como SYSmark, MobileMark o WebMark.

**ACTIVIDAD 3.5** Considérense los tiempos de ejecución obtenidos por dos máquinas, A y B, en la ejecución de dos programas de prueba:

| Programa | A | B |
|---|---|---|
| 1 | 10 | 1 |
| 2 | 100 | 1.000 |

Se quiere comparar el rendimiento de ambos computadores mediante la media armónica de los tiempos de ejecución normalizados respecto a una máquina de referencia R. Esta normalización se lleva a cabo dividiendo el tiempo obtenido en R entre el obtenido en la máquina que se compara. La comparación se ha de hacer para cada uno de los escenarios siguientes definidos por diferentes tiempos de ejecución en la máquina de referencia:

| Programa | Escenario I | Escenario II | Escenario III |
|---|---|---|---|
| 1 | 1 | 1 | 1 |
| 2 | 100 | 10 | 1.000 |

¿Son conclusiones similares a las que se obtienen utilizando el tiempo total de ejecución de los dos programas en cada máquina?

**ACTIVIDAD 3.6** Considérese la misma situación que la actividad anterior pero con las siguientes variantes: los tiempos reflejados en la primera tabla son ahora el número de ciclos que tardan en ejecutarse los programas de prueba, y los tiempos que definen los tres escenarios son ahora el número de instrucciones ejecutadas por cada programa (no hay máquina de referencia). Nótese que los tiempos de ejecución en las dos máquinas A y B se mantienen constantes para los tres escenarios.

En este caso la comparación del rendimiento de los computadores A y B se hará me- diante la media aritmética del número medio de ciclos por instrucción (CPI) obtenidos por cada programa. Las conclusiones obtenidas, ¿son similares a las que se obtienen utilizando el número de ciclos que tardan en ejecutarse los dos programas en cada máquina?

# Capítulo 4

## Introducción al análisis operacional

El análisis operacional forma parte de una serie de técnicas, denominadas analíticas, em- pleadas en la estimación del rendimiento de los sistemas informáticos. Estas técnicas hacen uso de un modelo de comportamiento del computador y su carga, y calculan los índices de prestaciones a partir de este modelo.

Desde un punto de vista amplio el computador puede concebirse como un conjunto de dispositivos físicos relacionados entre sí y una serie de trabajos que hacen uso de ellos. Los dispositivos comprenden, por ejemplo, el procesador, los discos y la memoria, mientras que los trabajos representan los programas que se ejecutan en la máquina.

Uno de los paradigmas que más éxito han tenido para modelar el comportamiento de los sistemas informáticos es el basado en redes de colas de espera (*queueing networks*) introducido por Jackson en la década de 1950. Aunque hayan pasado más de cincuenta años desde su aparición, este tipo de modelos sigue gozando de plena vigencia. El objetivo que se persigue con el diseño de estos modelos es, principalmente, estimar el tiempo que un trabajo necesita para que sea procesado por el sistema informático.

Dentro del marco que establecen los modelos basados en redes de colas de espera, el análisis operacional, presentado por Buzen y Denning a finales de la década de 1970, aborda la construcción de modelos de colas dejando de lado hipótesis estadísticas de difícil verificación. En este sentido, las técnicas que provee el análisis operacional comprenden una serie de relaciones muy sencillas entre variables directamente observables del sistema informático. En este contexto, la palabra *operacional* equivale a “directamente medible”. Así, una hipótesis operacionalmente comprobable es una hipótesis que puede ser verificada por medida. Por ejemplo, una hipótesis cuya veracidad es fácilmente comprobable usando técnicas de medida puede ser la siguiente: el número de llegadas de peticiones a un sistema es igual al número de finalizaciones, en un tiempo suficientemente largo.

## 4.1. Estaciones de servicio

Una estación de servicio (*service station*, *queue*) es un objeto abstracto compuesto por un servidor y una cola de espera. El servidor representa al recurso físico del computador, mientras que la cola de espera modela la cola de trabajos que esperan recibir servicio (esto es, aguardan a utilizar el recurso físico).

Los parámetros temporales más importantes de una estación de servicio desde el punto de vista del rendimiento son dos: el tiempo de servicio y el tiempo de respuesta. El primero es el tiempo que transcurre desde que un trabajo empieza a utilizar el recurso hasta que lo deja libre. El segundo incluye este tiempo de servicio más el tiempo que el trabajo pasa aguardando en la cola de espera. Cuando se puede atender a más de un trabajo en paralelo, las estaciones de servicio incluyen más de un servidor. La Figura 4.1 muestra gráficamente tres tipos de estaciones de servicio: con un único servidor y una cola de espera (a), con dos servidores y una cola de espera (b), y con infinitos servidores (c), la cual no tiene cola de espera porque los trabajos que llegan siempre encuentran un servidor disponible.

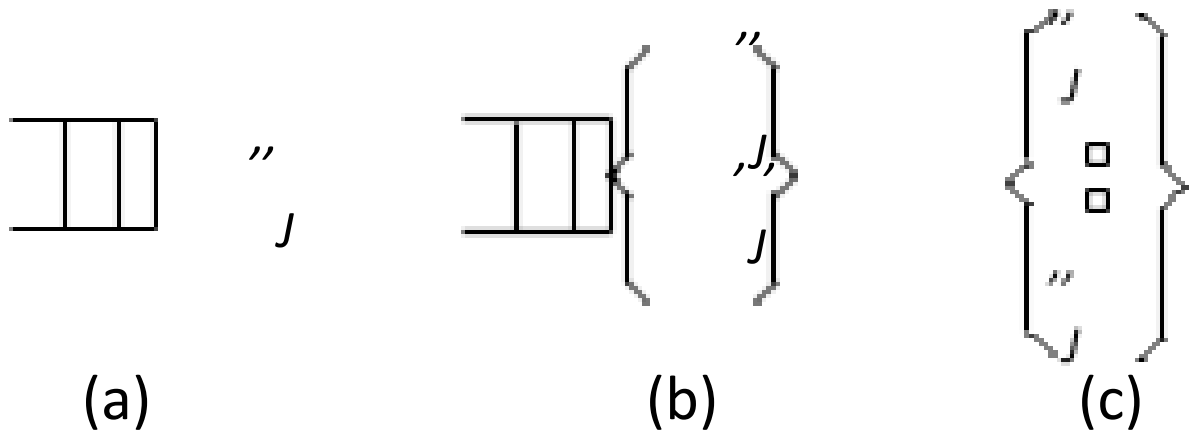

**Figura 4.1:** Diferentes estaciones de servicio.

Cuando una estación tiene infinitos servidores se dice que es de tipo *retardo*, ya que los clientes no esperan para adquirir un servidor. Cuando esto no ocurre, esto es, la estación tiene un número finito de servidores, se dice que la estación es de tipo *cola*, y los clientes pueden sufrir demoras debidas a la espera hasta conseguir un servidor libre.

## 4.2. Redes de colas de espera

El comportamiento de la mayoría de los sistemas informáticos está caracterizado por la presencia de varios puntos de congestión originados por la compartición de recursos. En estos casos es difícil, y a veces demasiado restrictivo, representar el comportamiento del sistema mediante una única estación de servicio como las que acabamos de describir. En su lugar, resulta más adecuado modelar explícitamente los diferentes puntos de congestión del sistema. El modelo resultante es una red de colas, es decir, un conjunto de estaciones de servicio interconectadas a través de las cuales circulan los trabajos siguiendo un patrón determinista o probabilista.

Formalmente podemos definir una red de colas como un grafo dirigido cuyos nodos son las estaciones de servicio. Los arcos entre estos nodos indican las transiciones posibles entre las estaciones. Los trabajos que circulan a través de la red pueden ser de clases diferentes. A su vez, los trabajos de clases diferentes pueden seguir recorridos distintos a través de la red.

Las redes pueden ser clasificadas según los tipos de trabajos que circulan por sus estaciones. Si en todas las estaciones de la red los trabajos tienen el mismo comportamiento, tanto en lo que se refiere a los tiempos de servicio como en lo que atañe al camino que siguen, decimos que la red tiene una única clase de trabajos (monoclase). En una red con varias clases de trabajos (multiclase) los trabajos de una misma clase siguen un patrón idéntico de comportamiento, probablemente diferente del resto de las clases, en lo que se refiere a tiempos de servicio y/o encaminamiento.

Las redes también pueden ser clasificadas según la topología del grafo subyacente. Las redes abiertas se caracterizan por la existencia de, al menos, una fuente de trabajos y uno o más sumideros que absorben los trabajos que salen del sistema y, así mismo, la posibilidad de encontrar al menos un camino que, a partir de cada nodo, lleve a un sumidero. Las redes abiertas se emplean para modelar el comportamiento de sistemas que soportan cargas transaccionales. En una red de este tipo el número de trabajos que hay en el sistema varía con el tiempo. La productividad de una red abierta suele ser un dato conocido porque su valor es igual a la tasa de entrada al sistema. Los índices que interesan de este tipo de redes son el tiempo de respuesta y el número de trabajos dentro del sistema. En la Figura 4.2 se muestra una red de colas abierta que comprende tres estaciones de servicio; en ella también se pueden apreciar sendos símbolos que identifican la fuente de trabajos y el sumidero de la red.

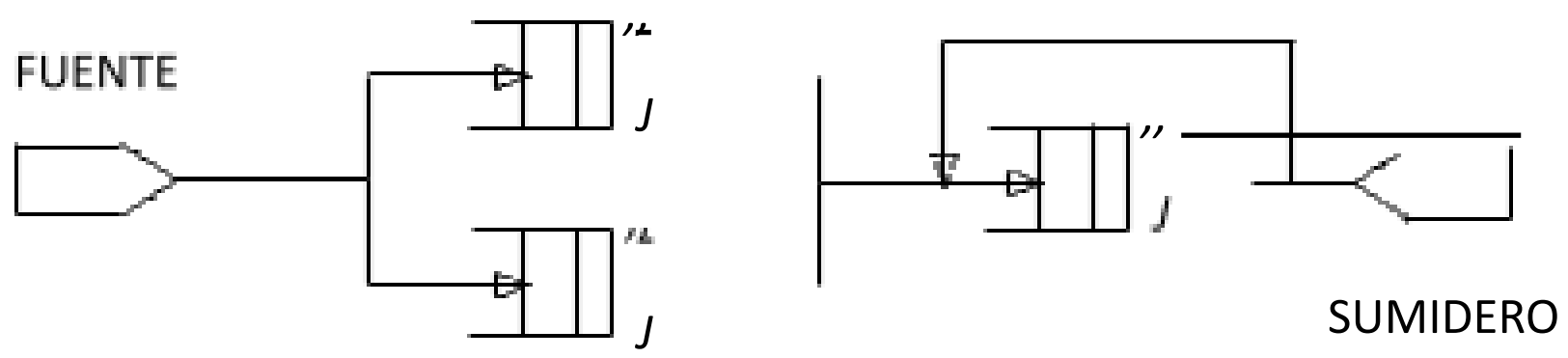

**Figura 4.2:** Ejemplo de red de colas abierta.

Las redes cerradas son redes en las cuales los trabajos ni entran ni salen y, por tanto, su número permanece constante. En algunos casos interesa contemplar un modelo de colas cerrado como un sistema en el que la salida está unida a la entrada, de manera que los tra- bajos que "salen" del mismo, inmediatamente, "regresan" a él. Con esta visión del sistema, el flujo de trabajos a través del enlace entre la salida y la entrada define la productivi- dad de la red. En estas redes resulta de gran interés conocer el tiempo de respuesta y la productividad. Los sistemas con carga de tipo interactivo y con carga por lotes (*batch*) se modelan mediante redes cerradas. La Figura 4.3 muestra un ejemplo de este tipo de redes. Nótese que la estación con infinitos servidores se suele utilizar para representar el tiempo transcurrido entre la finalización de una petición al sistema y el comienzo de una nueva, y que, como tal, no modela ningún dispositivo físico del sistema informático. En caso de que la carga fuera por lotes la estación con infinitos servidores desaparecería del modelo y se uniría la salida con la entrada de esta estación.

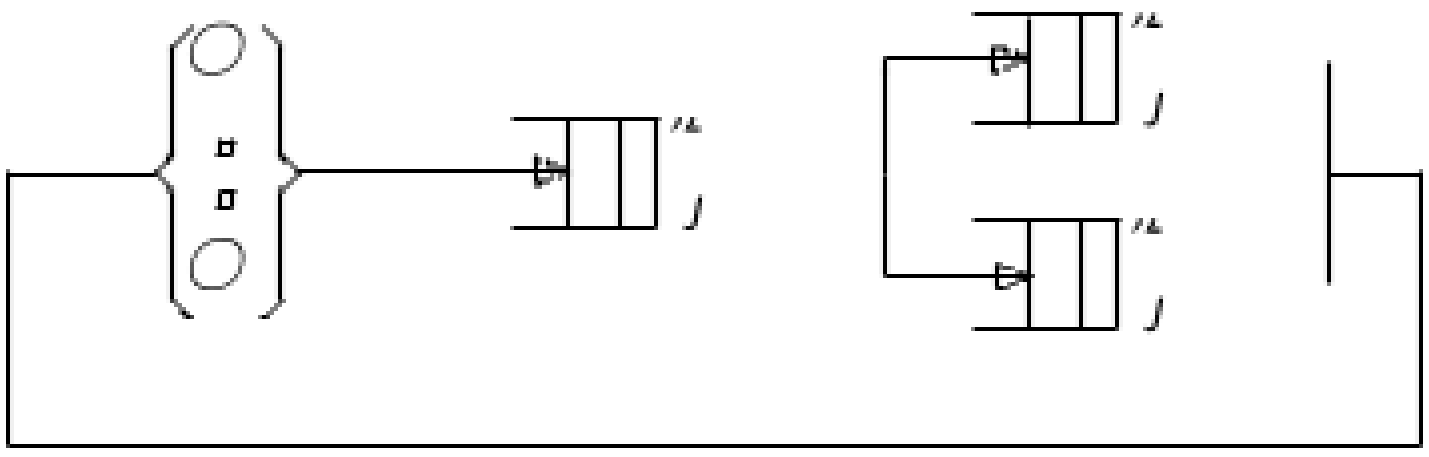

**Figura 4.3:** Ejemplo de red de colas cerrada (carga interactiva).

Finalmente, en una red con múltiples clases de trabajos es posible que la red sea abierta para un tipo de trabajos y cerrada para otro. En este caso se dice que la red es mixta.

Dentro del conjunto de patrones de interconexión que se pueden establecer entre las estaciones de servicio de un modelo hay algunos que tienen más utilidad que otros. Entre los primeros cabe destacar por su importancia el denominado modelo del servidor cen- tral (*central server model* ), que fue introducido por Buzen en 1973. Este modelo intenta reproducir el comportamiento de los programas cuando se ejecutan en un computador. De acuerdo con este paradigma, un programa que se va a ejecutar en un computador es enviado a la cola de procesos del procesador. Una vez que comienza a ejecutarse, el pro- grama realiza una petición de acceso a un dispositivo de entrada/salida (por ejemplo, un disco magnético). Una vez satisfecha esta petición de entrada/salida, el programa vuelve a ser planificado para su ejecución en el procesador, y así sucesivamente hasta que acaba y abandona el procesador. Nótese que este modelo no considera de manera explícita la memoria principal del computador; únicamente tiene en cuenta el procesador y las uni- dades de almacenamiento. La Figura 4.4 muestra el modelo de servidor central en el caso de un sistema con un procesador y dos dispositivos de almacenamiento, aplicado a una carga interactiva (a) y a una carga transaccional (b). Finalmente, la salida del procesador por parte de un trabajo tiene una componente probabilística, ya que tiene dos alternativas básicas: bien volver de nuevo a los discos o bien empezar un nuevo tiempo de reflexión (modelo cerrado) o salir del sistema (modelo abierto).

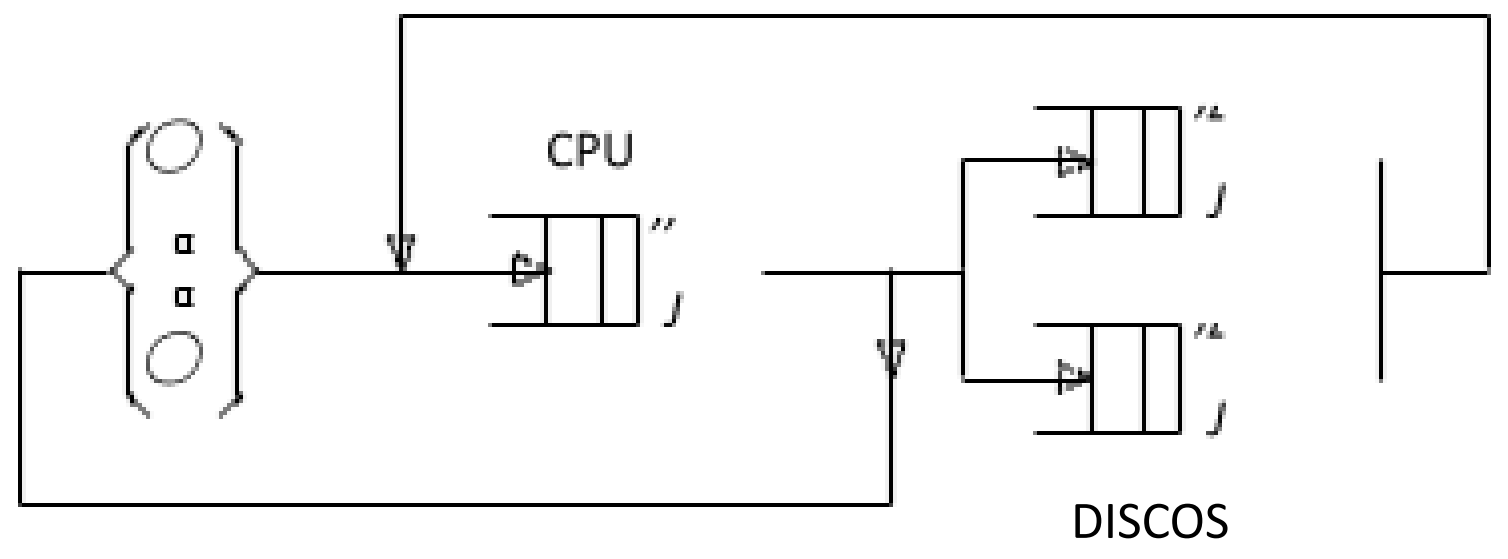

(a) Interactivo

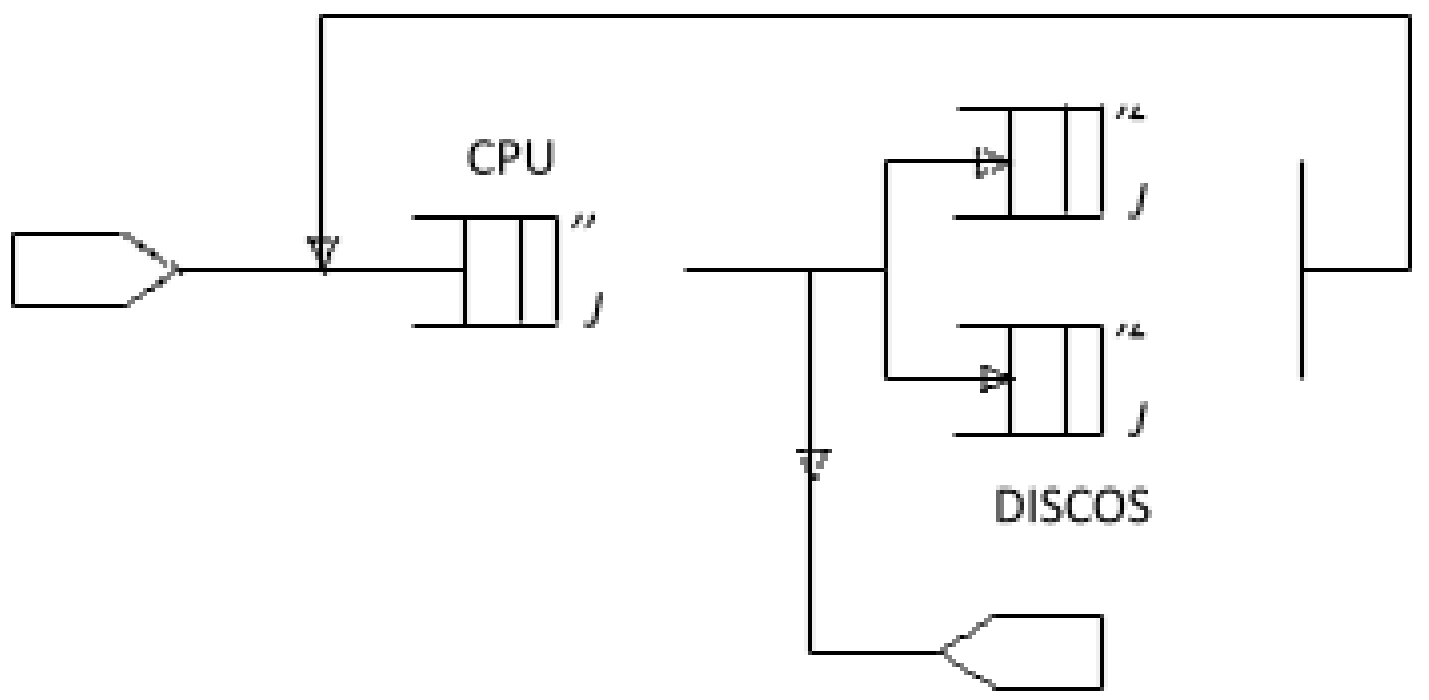

(b) Transaccional

**Figura 4.4:** El modelo del servidor central.

## 4.3. Leyes operacionales

Las variables operacionales son cantidades directamente medibles durante un periodo de observación finito. Si observamos un dispositivo cualquiera $i$ de un sistema informático como una caja negra durante un periodo de tiempo $T$ , obtenemos las siguientes medidas: número de llegadas ($A_i$), número de salidas o terminaciones ($C_i$), y tiempo total en que el dispositivo está ocupado ($B_i$). De estas magnitudes operacionales podemos derivar las siguientes:

Tasa de llegada: $\lambda_i = A_i/T$ trabajos por unidad de tiempo
Productividad: $X_i = C_i/T$ trabajos por unidad de tiempo
Utilización: $U_i = B_i/T$ (sin unidades)
Tiempo de servicio: $S_i = B_i/C_i$ unidades de tiempo por trabajo

El inverso de la tasa de llegadas se denomina tiempo entre llegadas ($T/A_i$), mientras que el inverso del tiempo de servicio se denomina tasa de servicio ($\mu_i = C_i/B_i$).

Las variables operacionales pueden cambiar de valor de un periodo de observación a otro. Sin embargo, hay relaciones que se mantienen para cualquier periodo de observación.

Estas relaciones se denominan leyes operacionales, y no dependen de hipótesis sobre la distribución estadística que siga los tiempos de servicio y los tiempos entre llegadas.

Como punto de partida supondremos que durante el periodo de observación $T$ el siste- ma se encuentra operando en un estado estable o de equilibrio, lo que recibe el nombre de hipótesis del flujo equilibrado de trabajos. En este estado se cumple que el número de traba- jos que entra es igual al número de trabajos que sale ($A_i = C_i, \forall i$). Tomando un periodo de observación $T$ suficientemente grande se consigue, generalmente, que la diferencia $A_i - C_i$ sea muy pequeña comparada con $C_i$ y, por tanto, que la hipótesis sea aproximadamente correcta. Nótese que $A_i = C_i$ implica $\lambda_i = X_i$. Esta hipótesis operacional, comprobable por medida, nos será muy útil de ahora en adelante.

A continuación se describirán las leyes operacionales que establecen relaciones entre las distintas variables de carácter operacional.

## Ley de la utilización

La utilización de un dispositivo se puede expresar en función del número de terminaciones mediante la siguiente fórmula:

$$U_i = \frac{B_i}{T} = \frac{C_i}{T} \times \frac{B_i}{C_i} = X_i \times S_i \qquad (4.1)$$

Esta expresión permite, por tanto, relacionar la productividad de un dispositivo con su tiempo de servicio. Si además se cumple la hipótesis del flujo equilibrado de trabajos obtendremos una expresión equivalente a la anterior en función de la tasa de llegada:

$$U_i = \lambda_i \times S_i$$

## Ley del flujo forzado

Esta ley es de gran importancia y relaciona la productividad del sistema $X_0$ con la pro- ductividad de un dispositivo individual $X_i$. En un modelo abierto la productividad está definida por el número de trabajos que abandona el sistema por unidad de tiempo. En cambio, en un modelo cerrado, ningún trabajo abandona el sistema. Aun así, los trabajos que atraviesan el enlace que une la salida con la entrada se comportan como si abando- naran el sistema e inmediatamente reentraran en él. La productividad del sistema en este último caso viene dada por el número de trabajos que atraviesan este enlace por unidad de tiempo.

Supongamos que cada trabajo realiza $V_i$ peticiones o visitas al dispositivo $i$. Si el flujo de trabajos está equilibrado, el número de trabajos que sale del sistema $C_0$ (o atraviesa el enlace exterior) y el número de trabajos que atraviesa el dispositivo $i$ están relacionados por la expresión:

$$C_i = C_0 \times V_i \qquad \text{y, por tanto,} \qquad V_i = \frac{C_i}{C_0}$$

La variable $V_i$ recibe el nombre de razón de visitas al dispositivo $i$. La productividad total del sistema $X_0$ durante el periodo de observación es:

$$X_0 = \frac{C_0}{T}$$

mientras que la productividad del dispositivo $i$ es:

$$X_i = \frac{C_i}{T} = \frac{C_i}{C_0} \times \frac{C_0}{T}$$

En consecuencia, podemos obtener una expresión de $X_i$ en función de las variables $V_i$ y $X_0$:

$$X_i = X_0 \times V_i \qquad (4.2)$$

que representa la expresión de la ley del flujo forzado. Esta ley establece que el flujo a través de un determinado dispositivo de la red determina el flujo en cualquier otro dispositivo. La ley del flujo forzado es válida si lo es la hipótesis del flujo equilibrado de trabajos. Combinando los resultados de la ley del flujo forzado (ecuación 4.2) y de la ley de la utilización (ecuación 4.1) podemos obtener la siguiente expresión para el valor de la utilización del dispositivo:

$$U_i = X_i \times S_i = X_0 \times V_i \times S_i = X_0 \times D_i$$

donde $D_i = V_i S_i$ recibe el nombre de demanda de servicio sobre el dispositivo $i$ en todas las visitas que un trabajo realiza al mismo. La relación anterior establece que la utilización de cada dispositivo del sistema es proporcional a su demanda de servicio.

Las razones de visita son una forma de especificar el encaminamiento de los trabajos a través de la red. Otra descripción equivalente se puede realizar mediante la proporción de trabajos, también denominada probabilidad de encaminamiento o de transición. Así, las probabilidades de encaminamiento, $p_{ij}$, indican la proporción de trabajos que cuando salen de la estación $i$ se dirigen a la estación $j$, o de forma equivalente, indican la probabilidad de que un trabajo pase a la estación $j$ después de terminar su servicio en la estación $i$. En este sentido se tendrá que $p_{ij} = C_{ij}/C_i$, y en particular, $p_{0j} = A_{0j}/A_0$ y $p_{i0} = C_{i0}/C_0$.

Razones de visita y probabilidades de encaminamiento son equivalentes en el sentido de que a partir de una se obtienen las otras. En un sistema con $K$ estaciones de servicio en que se cumple la hipótesis del flujo equilibrado de trabajos se tiene:

$$C_j = \sum_{i=0}^{K} C_i \quad p_{ij}$$

donde el subíndice 0 representa el exterior del sistema y $p_{i0}$ es la proporción de trabajos que, después de recibir servicio en la estación $i$, abandonen la red. Dividiendo ambos lados de la igualdad por $C_0$ obtenemos:

$$V_j = \sum_{i=0}^{K} V_i \quad p_{ij}$$

que representan las denominadas ecuaciones de las razones de visita. Como cada visita al mundo exterior corresponde a una terminación de un trabajo, tendremos que siempre se cumplirá $V_0 = 1$.

## Ley de Little

La ley de Little, que data de principios de la década de 1960, también es una ley operacional. La única hipótesis requerida para su aplicación es la del flujo equilibrado de trabajos. Si llamamos $N_i$ al número de trabajos y $R_i$ al tiempo de respuesta de la estación de servicio $i$, la ley de Little establece que:

$$N_i = \lambda_i \times R_i \tag{4.3}$$

y como hemos exigido que se cumpla la hipótesis del flujo equilibrado de trabajos podemos sustituir $\lambda_i$ por $X_i$:

$$N_i = X_i \times R_i \tag{4.4}$$

Esta ley es de gran interés en el estudio de los modelos de colas, ya que combina índices de suma importancia en los estudios de rendimiento: tiempo de respuesta y productividad. Además, a esto hay que unir el hecho de que se puede aplicar a cualquier parte del modelo con la única condición de que se cumpla el flujo equilibrado de trabajos.

## Ley general del tiempo de respuesta

El número de trabajos en una red de colas formada por $K$ estaciones se puede expresar como $N = N_1 + N_2 + \cdots + N_K$. Si sustituimos los valores de $N_i$ de acuerdo con la ley de Little tendremos:

$$X_0 \times R = X_1 \times R_1 + X_2 \times R_2 + \cdots + X_K \times R_K = \sum_{i=1}^{K} X_i \times R_i$$

Dividiendo ambos miembros de la igualdad por $X_0$ y aplicando la ley del flujo forzado quedará la expresión:

$$R = V_1 \times R_1 + V_2 \times R_2 + \cdots + V_K \times R_K = \sum_{i=1}^{K} V_i \times R_i \qquad (4.5)$$

Esta expresión recibe el nombre de ley general del tiempo de respuesta, y permite ver claramente que el tiempo de permanencia de un trabajo en un sistema depende del número de visitas que realiza a cada dispositivo y del tiempo de respuesta que experimenta en él por cada una de las visitas.

## Ley del tiempo de respuesta interactivo

Todos los modelos de sistemas con carga interactiva pueden dividirse conceptualmente en dos partes: una que modela el tiempo de reflexión (subsistema de terminales) y otra que contiene los dispositivos físicos del computador contemplados por el modelo (subsistema central). El tiempo de reflexión (*think time*), identificado habitualmente mediante la va- riable $Z$, es el tiempo que transcurre desde que un trabajo abandona el subsistema central hasta que entra de nuevo en él; para sistemas interactivos se tiene que $Z > 0$, mientras que en sistemas por lotes el valor de $Z$ es cero. El tiempo de respuesta del sistema, $R$, corresponderá al tiempo que un trabajo pasa en el subsistema central.

El funcionamiento del sistema es el que sigue: los usuarios generan peticiones desde los terminales que se sirven en el subsistema central, y vuelven posteriormente a los terminales. Estos terminales están modelados por una estación con infinitos servidores (no hay tiempo de espera en cola). Transcurrido el tiempo de reflexión los usuarios generan la siguiente petición. La Figura 4.5 muestra gráficamente un esquema de red de colas para una carga interactiva.

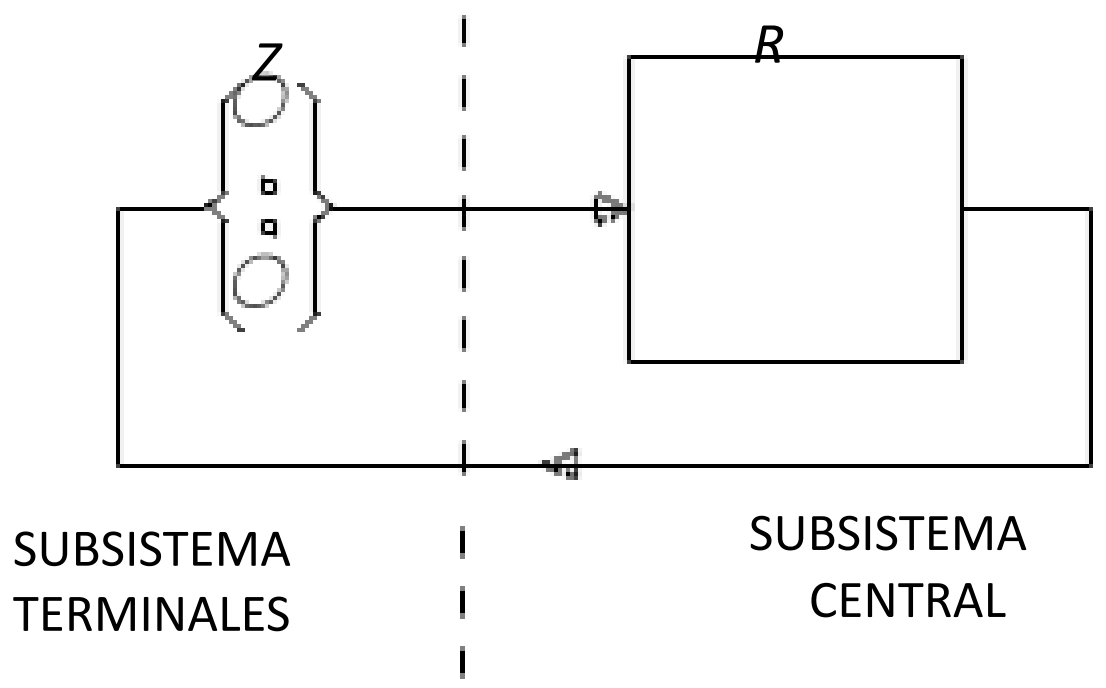

**Figura 4.5:** Ejemplo de modelo para un sistema interactivo.

Podemos aplicar la ley de Little al conjunto de los dos subsistemas. Este conjunto incluye el subsistema central y el subsistema de terminales. El número de trabajos en el conjunto es $N$. El tiempo medio que permanece un trabajo en el conjunto es igual al tiempo que permanece en los terminales, $Z$, más el tiempo que permanece en el subsistema central, $R$. Por tanto, aplicando la ley de Little, se puede escribir:

$$N = (Z + R) \times X_0$$

y despejando la variable $R$ obtenemos la expresión para la ley del tiempo de respuesta interactivo:

$$R = \frac{N}{X_0} - Z \quad (4.6)$$

Nótese que el número de trabajos en los terminales viene dado, empleando la ley de Little, por $Z \times X_0$ y el número de trabajos dentro del sistema que compiten por los recursos es $R \times X_0$.

## 4.4. Problemas resueltos

**PROBLEMA 4.1** El disco de un computador se ha monitorizado durante un periodo de medida de 30 segundos. Durante este tiempo han llegado 11 peticiones y han acabado 12. Se sabe que el disco ha estado vacío durante 2,5 segundos, y se ha podido medir el tiempo de respuesta de 9 peticiones. Estos tiempos, expresados en segundos, son: 8,2, 9,1, 2,3, 5,9, 2,0, 6,2, 4,1, 6,5 y 7,3. Se pide calcular:

1. La exactitud con que se cumple la hipótesis del flujo equilibrado de trabajos.
2. La tasa de llegadas de peticiones al disco y el tiempo entre llegadas.
3. La productividad del disco.
4. El tiempo de respuesta del disco.
5. La utilización del disco.
6. El tiempo de servicio del disco.

**SOLUCIÓN:**

1. El número de llegadas y el de salidas en el disco no coinciden durante el periodo de medida, aunque tienen valores similares. En particular, la relación entre los dos valores es $A/C = 11/12 = 0{,}917$. El error relativo que se comete al suponer que hay flujo equilibrado de trabajos es 1 − 0,917 = 0,083, valor que también se puede calcular de la siguiente manera:

$$\frac{A - C}{C} = \frac{11 - 12}{12} = 0,083 = 8,3\ \%$$

Por tanto, la hipótesis del flujo equilibrado de trabajos se cumple aproximadamente en un 91,7 %.

2. La tasa de llegadas de peticiones al disco se calcula dividiendo el número de peticiones que llegan entre la duración del periodo de medida:

$$\lambda = \frac{A}{T} = \frac{11}{30} = 0,367 \text{ peticiones/s}$$

En consecuencia, el tiempo medio entre llegadas es de 30/11 = 2,73 segundos.

3. La productividad del disco se calcula dividiendo el número de peticiones que acaban entre la duración del periodo de medida:

$$X = \frac{C}{T} = \frac{12}{30} = 0,4 \text{ peticiones/s}$$

4. El tiempo medio de respuesta $R$ del disco se puede obtener a partir de la media aritmética de los tiempos de respuesta $r_i$ observados para las 9 peticiones completadas por el disco:

$$R = \frac{\sum_{i=1}^{i=9} r_i}{9} = \frac{8,2 + 9,1 + 2,3 + \cdots + 7,3}{9} = 5,73 \text{ s}$$

5. La utilización se obtiene como la relación entre el tiempo en que el disco está ocupado y la longitud del periodo de medida:

$$U = \frac{B}{T} = \frac{T - 2,5}{T} = \frac{27,5}{30} = 0,917$$

6. El tiempo de servicio se calcula a partir de la ley de la utilización, la cual expresa la utilización como producto de la productividad y el tiempo de servicio ($U = X\ S$).

Así pues:

$$\frac{0,917}{0,4} = 2,29 \text{ s}$$

Nótese que la diferencia entre el tiempo de respuesta $R$ y el tiempo de servicio $S$ repre- senta el tiempo de espera en cola: $R$ $S$ = 5,73 2,29 = 3,44. Esta cantidad representa el tiempo medio que un trabajo que llega al disco espera hasta comenzar a recibir servicio, o de forma equivalente, el tiempo que emplea el disco para atender a los trabajos que se encuentran por delante de él en la cola de espera. ■

**PROBLEMA 4.2** Un segmento de red local transmite 3.000 paquetes por segundo. Cada paquete tiene un tiempo medio de transmisión de 0,1 ms. Determínese la utilización de este segmento de red.

**SOLUCIÓN:** Si aplicamos la ley de la utilización, que relaciona la productividad del dispositivo y su tiempo de servicio, obtenemos:

$$U = X \times S = 3.000 \times 0{,}0001 = 0{,}3$$

Por tanto, la utilización del segmento de red es del 30 %. ■

**PROBLEMA 4.3** Considérese una red Ethernet con un ancho de banda de 10 Mbps. Las peticiones emitidas desde un cliente a un servidor a través de esta red constan de tres paquetes de 1.518 bytes cada uno, mientras que las respuestas desde el servidor requieren el envío de nueve paquetes de idéntico tamaño hacia el cliente. Calcúlese la demanda de servicio de la red que provoca cada transacción entre el cliente y el servidor.

**SOLUCIÓN:** La demanda de servicio $D$ que cada transacción efectúa sobre la red es equivalente al tiempo total de transmisión; es decir, hay que considerar tanto el tiempo de servicio de la petición como el tiempo de servicio de la respuesta. Esto representa un número de 3 + 9 = 12 paquetes enviados a través de la red. Así pues, podremos escribir:

$$D = \frac{8 \times 1.518 \times (3 + 9)}{10 \times 10^6} = 0{,}0146 \text{ s}$$

Por tanto, cada interacción entre el cliente y el servidor requiere un total de 14,6 ms de uso efectivo de la red. ■

**PROBLEMA 4.4** En un sistema cliente–servidor se considera que las transacciones usan 4 ms de procesador en el cliente, 6 ms de procesador en el servidor y leen 12 bloques de 1.024 bytes del disco del servidor. De las características técnicas del disco se sabe que el tiempo medio de posicionamiento es de 8 ms, la latencia media es 3,6 ms y el ratio de transferencia es 24 MB/s. Se pide calcular:

1. Las demandas de servicio de las transacciones en los procesadores del cliente y del servidor, expresadas en segundos.

2. El tiempo medio de servicio del disco.
3. La demanda de servicio del disco del servidor suponiendo que los bloques están grabados en pistas diferentes o, en el mejor de los casos, situados de forma consecutiva.
4. ¿Qué componentes del tiempo de servicio del disco influyen más en el rendimiento?

**SOLUCIÓN:**

1. Los valores correspondientes a las demandas de servicio de los procesadores vienen dados por el enunciado del problema: en el cliente la demanda del procesador es de 0,004 segundos, mientras que en el servidor es de 0,006 segundos.
2. El tiempo de servicio $S$ de un disco está integrado por la suma de tres componentes básicos: el retardo de búsqueda o posicionamiento, la latencia o retardo rotacional y el tiempo de transferencia de los datos. Así pues, el tiempo que el disco tarda en acceder a un bloque de datos de 1.024 bytes es:

$$S = 0{,}008 + 0{,}0036 + \frac{1.024}{24 \times 10^6} = 0{,}01164 \text{ s}$$

Nótese que para calcular el tiempo de transferencia se ha dividido por 24 $10^6$, ya que la velocidad de transferencia se expresa en unidades del sistema internacional (M = $10^6$) y no en unidades de capacidad de almacenamiento (M = $2^{20}$).

3. La demanda de servicio $D$ del disco depende del número de accesos que hace cada transacción y del tiempo de servicio que requiere cada acceso. Según el enunciado, cada transacción visita la unidad de disco un total de 12 veces. El tiempo de servicio de un acceso depende de la ubicación de los datos en la superficie del mismo. Si, en el peor de los casos, los bloques leídos se distribuyen de forma aleatoria, cada acceso tarda en servirse una media de 0,01164 segundos, como hemos visto en el punto anterior. Por tanto, tendremos:

$$D = V \times S = 12 \times 0{,}01164 = 0{,}13968 \text{ s}$$

Si, por el contrario, los bloques se sitúan en la misma pista del disco en posiciones consecutivas, solamente hay que contabilizar un tiempo de posicionamiento y un retardo rotacional para el primer acceso; esto es, la demanda de servicio requerida para acceder a 12 bloques de 1.024 bytes cada uno quedará reducida a:

$$D = 0{,}008 + 0{,}0036 + 12 \times \frac{1.024}{24 \times 10^6} = 0{,}01209 \text{ s}$$

Nótese que la mejora conseguida al disponer de todos los bloques dentro de la misma pista es de 0,13968/0,01209 = 11,55.

4. El tiempo de posicionamiento (8 ms) y la latencia rotacional (3,6 ms) son los dos com- ponentes que más retrasan el servicio de las transacciones, mientras que el tiempo de transferencia es mucho menor. Por ejemplo, el tiempo de transferencia de un bloque de 1.024 bytes es:

$$\frac{1.024}{24 \times 10^6} = 4{,}26667 \times 10^{-5} \text{ s} = 0{,}04267 \text{ ms}$$

De aquí que la demanda de servicio pueda descender casi hasta el valor del tiempo de servicio del disco si disponemos todos los bloques sobre la misma pista, porque el cabezal de lectura del brazo del disco ha de hacer un único movimiento de posicionamiento para situarse sobre la pista. ■

**PROBLEMA 4.5** Considérese una unidad de disco duro cuya controladora dispone de una cierta cantidad de memoria cache. El tiempo medio de acceso a la controladora es 0,1 ms, el tiempo medio de posicionamiento del disco es de 5 ms, la latencia rotacional media es de 6 ms y el tiempo medio de transferencia es de 0,3 ms. Si la memoria cache tiene una probabilidad de acierto del 95 %, determínese el tiempo medio de servicio de la unidad de disco. Adicionalmente, calcúlese esta última variable en caso de que la probabilidad de acierto sea del 90, 80 y 70 %.

**SOLUCIÓN:** El tiempo servicio del disco incluirá, independientemente de la situación física de la información (en disco o en memoria cache), el retardo introducido por la controladora (gestión de la memoria cache y operaciones de lectura y/o escritura en esta memoria). Adicionalmente, el retardo propio del disco duro habrá que considerarlo sólo en el caso de que los datos no se encuentren en la memoria cache y haya que acceder a la información contenida en él. El tiempo de servicio de la unidad de disco, para una probabilidad de acierto del 95 %, se puede calcular como:

$$S_{(0,95)} = 0{,}1 + (1 - 0{,}95) \times (5 + 6 + 0{,}3) = 0{,}655 \text{ ms}$$

De forma análoga, el tiempo de servicio para cada valor de la probabilidad de acierto será:

$$S_{(0,90)} = 0{,}1 + (1 - 0{,}90) \times (5 + 6 + 0{,}3) = 1{,}23 \text{ ms}$$

$$S_{(0,80)} = 0{,}1 + (1 - 0{,}80) \times (5 + 6 + 0{,}3) = 2{,}36 \text{ ms}$$

$$S_{(0,70)} = 0{,}1 + (1 - 0{,}70) \times (5 + 6 + 0{,}3) = 3{,}49 \text{ ms}$$

Nótese cómo pequeñas variaciones en la tasa de aciertos de la memoria cache repercuten en un incremento considerable del tiempo medio de servicio de la unidad de disco. ■

**PROBLEMA 4.6** Un servidor web tiene un tiempo medio de respuesta de 12 milisegundos y recibe una media de 500 peticiones por segundo. Calcúlese el número medio de peticiones que hay en este servidor.

**SOLUCIÓN:** El número medio de peticiones en el servidor web se puede calcular mediante la ley de Little, la cual relaciona el tiempo de permanencia en el sistema y su productividad:

$$N = X \times R = 500 \times 12 \times 10^{-3} = 6 \text{ peticiones}$$

■

---

**PROBLEMA 4.7** Un procesador recibe una media de dos programas por segundo. Cada programa experimenta un tiempo medio de ejecución de 0,4 segundos y un tiempo medio de respuesta de 2 segundos. Se pide calcular:

1. Utilización media del procesador.
2. Tiempo medio de espera en la cola del procesador.
3. Número medio de programas en la cola de espera del procesador.

**SOLUCIÓN:**

1. Si se aplica la ley de la utilización, y además se tiene en cuenta el cumplimiento de la hipótesis del flujo equilibrado de trabajos, podemos obtener:

   $$U = X \times S = \lambda \times S = 2 \times 0{,}4 = 0{,}8$$

   Así pues, la utilización del procesador es del 80 %.
2. El tiempo de espera en cola corresponde al tiempo que un programa pasa, una vez ha llegado al procesador, sin recibir servicio. Por tanto, este tiempo de espera se puede calcular restando el tiempo de servicio del tiempo de respuesta: $2 - 0{,}4 = 1{,}6$ segundos.
3. Se puede aplicar la ley de Little a la cola de espera del procesador. Como la productividad de la cola es la misma que la del procesador y, además, se conoce el tiempo de espera en esta cola, los programas que por término medio están esperando a conseguir el procesador para ejecutarse son $2 \times 1{,}6 = 3{,}2$. ■

---

**PROBLEMA 4.8** Después de monitorizar el procesador de un servidor web durante un periodo de 30 segundos se sabe que ha sido utilizado durante 27 segundos. Así mismo, se han contabilizado 74 llegadas y 72 salidas de peticiones.

1. ¿Cuál es la tasa de llegadas al procesador?
2. ¿Cuál es la productividad del procesador?
3. Determínese la utilización del procesador.
4. Si cada trabajo hace una media de cuatro visitas al procesador, ¿cuál es la produc- tividad del servidor web?

**SOLuCIón:** Según los datos del problema, existe una ligera discrepancia entre las llegadas y salidas del procesador durante el periodo de medida, por lo que la hipótesis del flujo equilibrado de trabajos no es totalmente válida; en concreto, el error que cometemos suponiéndola cierta es:

$$\frac{A - C}{C} = \frac{74 - 72}{74} = 0{,}027 = 2{,}7\ \%$$

1. La tasa de llegadas se calcula dividiendo el número de llegadas contabilizadas entre la duración del periodo de medida:

$$\lambda = \frac{A}{T} = \frac{74}{30} = 2{,}467 \text{ peticiones/s}$$

2. La productividad del procesador se calcula de manera análoga a la tasa de llegadas pero utilizando el número de salidas:

$$X = \frac{C}{T} = \frac{72}{30} = 2{,}400 \text{ peticiones/s}$$

   Nótese que el valor de la productividad y la tasa de llegadas no son exactamente iguales, ya que el número de llegadas y el de salidas difieren ligeramente.

3. La utilización del procesador es una variable adimensional que se calcula dividiendo el tiempo que ha estado ocupado entre el tiempo total que dura el periodo de medida:

$$U = \frac{B}{T} = \frac{27}{30} = 0{,}9$$

4. La productividad $X_0$ del servidor web puede calcularse aplicando la ley del flujo forzado al procesador. Si $X$ y $V$ representan la productividad y la razón de visita al procesador, respectivamente, podemos escribir:

$= V$

$$= \frac{2,4}{4} = 0,6 \text{ peticiones/s}$$

Nótese que la productividad del servidor web, $X_0$, es bastante menor que la productividad del procesador, $X$. ■

**PROBLEMA 4.9** Una red FDDI que actúa como red central o *backbone* interconecta varias redes de área local a través de distintos encaminadores. Un monitor de redes ha desvelado que el encaminador de una red Ethernet tiene una latencia de 0,2 ms por cada paquete transmitido y que el tráfico en esta red es de 20.000 paquetes por segundo. Así mismo, los datos también han desvelado que el encaminador que conecta una red Fast Ethernet a la red FDDI contiene una media de dos paquetes en tránsito y que el tráfico en la misma es de 2.000 paquetes por segundo. Calcúlese el número medio de paquetes en tránsito entre la red central y la red Ethernet, así como el tiempo medio que tarda en gestionar el envío de un paquete el encaminador que enlaza la red Fast Ethernet y la red central.

**SOLuCIón:** En primer lugar, el número medio de paquetes en tránsito entre la red central y la red Ethernet se calcula aplicando la ley de Little, ya que se conoce el tráfico en esta última red (que pasa necesariamente por el encaminador) y el tiempo medio que tarda el encaminador en enviar cada paquete. Por tanto, podremos escribir:

$$N = X \times R = 20.000 \times 0{,}0002 = 4 \text{ paquetes}$$

Por otro lado, también se puede aplicar la misma ley para calcular el tiempo que tarda el encaminador de la red Fast Ethernet en transmitir un paquete:

$$\frac{4}{2.000} = 0{,}002 \text{ s}$$

■

---

**PROBLEMA 4.10** Un servidor de ficheros fue monitorizado durante una hora en la que se produjeron 32.400 operaciones de entrada/salida. Asimismo, se pudo comprobar que el número medio de peticiones activas en el servidor fue 9. Determínese el tiempo de respuesta medio de las peticiones al servidor de ficheros.

**SOLuCIón:** La productividad del servidor de ficheros se puede calcular dividiendo el número de peticiones servidas entre el periodo de medida:

$$X = \frac{32.400}{3.600} = 9 \text{ peticiones/s}$$

Dado que se conoce el número medio de peticiones activas dentro del servidor, podemos aplicar la ley de Little para calcular el tiempo medio de respuesta:

$$R = \frac{N}{X} = \frac{9}{9} = 1 \text{ s}$$

**PROBLEMA 4.11** En un sistema interactivo la utilización del disco es del 50 %, la razón de visita 20 y el tiempo de servicio 0,025 segundos. El sistema tiene actualmente conectados a un total de 25 usuarios con un tiempo medio de reflexión de 18 segundos. ¿Cuál es el tiempo medio de respuesta del sistema que experimentan estos usuarios?

**SOLuCıón:** En primer lugar podemos utilizar los datos referidos al disco para calcular la productividad del sistema completo. Si $U$ , $V$ y $S$ denotan la utilización, la razón de visita y el tiempo de servicio del disco, respectivamente, podemos escribir la ley de la utilización como $U = X \times S$. En esta expresión podemos aplicar la ley del flujo forzado y sustituir la variable $X$ por $X_0 \times V$ . En consecuencia tenemos:

$$U = X \times S = X_0 \times V \times S$$

A partir de la expresión anterior la productividad $X_0$ del sistema completo se calcula como:

$$X_0 = \frac{U}{V \times S} = \frac{0,5}{20 \times 0,025} = 1 \text{ trabajo/s}$$

Ahora ya estamos en condiciones de calcular el tiempo de respuesta $R$ del sistema empleando la ley del tiempo de respuesta de los sistemas interactivos:

$$R = \frac{N}{X_0} - Z = \frac{25}{1} - 18 = 7 \text{ s}$$

Por tanto, cada usuario conectado al sistema experimentará un tiempo medio de respuesta de 7 segundos. ■

**PROBLEMA 4.12** Las transacciones a una base de datos realizan una media de cinco operaciones de entrada/salida al servidor que la contiene. Este servidor ha sido monitorizado durante dos horas en las que se ejecutaron 28.800 transacciones. Determínese:

1. La productividad media del servidor que almacena la base de datos.
2. La utilización del disco si cada acceso de entrada/salida al mismo tarda una media de 25 ms.
3. La demanda de servicio del disco.

**SOLuCıón:**

1. La productividad del servidor se calcula dividiendo el número de transacciones servidas entre la duración del periodo de medida:

$$X_0 = \frac{28.800}{7.200} = 4 \text{ transacciones/s}$$

2. Dado que el número medio de visitas $V$ al disco es conocida, se puede aplicar la ley del flujo forzado y obtener así la productividad $X$ del disco:

$$X = X_0 \times V = 4 \times 5 = 20 \text{ accesos/s}$$

Ahora podemos aplicar la ley de la utilización al disco:

$$U = X \times S = 20 \times 0{,}025 = 0{,}5$$

3. La demanda de servicio del disco $D$ se calcula multiplicando la razón de visitas por el tiempo de servicio:

$$D = V \times S = 5 \times 0{,}025 = 0{,}125 \text{ s}$$

■

**PROBLEMA 4.13** El sitio web de una librería virtual recibe una media de 25 visitas por segundo. La mayoría de estas visitas se dedican a hojear el catálogo virtual de libros. Sólo una de cada cinco visitas se emplea para hacer un pedido de libros. Cada orden de pedido provoca la activación de un programa CGI que se ejecuta en el servidor web consumiendo 100 ms de tiempo de procesamiento. Determínese la utilización del procesador debida a la ejecución de los programas CGI. ¿Cuál sería la utilización del procesador si los programas CGI fueran rediseñados y tardasen un 25 % menos de tiempo en ejecutarse?

**SOLUCIÓN:** De las 25 visitas por segundo que recibe el servidor web, sólo la quinta parte genera la ejecución de un programa CGI, esto es, 5 visitas por segundo. Si se considera que existe flujo equilibrado de trabajos podemos calcular la utilización del procesador debida a este tipo de visitas a partir de la siguiente expresión:

$$U = X \times S = 5 \times 0{,}1 = 0{,}5$$

Esto significa que los programas CGI provocan una utilización del 50 % en el procesador. Si este tipo de programas se rediseñan para que se ejecuten en un 25 % menos de tiempo, esto es, en $0{,}1\ (1\ 0{,}25) = 0{,}075$ segundos, tendremos que la utilización del procesador se reducirá hasta:

$$U = X \times S = 5 \times 0{,}075 = 0{,}375$$

Téngase en cuenta que la utilización del procesador puede ser superior a los valores que acabamos de calcular, ya que únicamente estamos considerando la proporción debida a la ejecución de los programas CGI. ■

**PROBLEMA 4.14** El sitio web de una compañía aérea fue monitorizado durante una hora en la que se produjeron 21.600 peticiones. El servidor recibe peticiones a cuatro tipos de objetos almacenados en él: páginas HTML, imágenes, documentos formateados y vídeos. Se ha observado que la distribución de la carga de peticiones es la siguiente:

| Tipo | Identificador | Proporción ( %) | Tamaño medio (KB) |
|---|---|---|---|
| 1 | Páginas | 25 | 15 |
| 2 | Imágenes | 50 | 25 |
| 3 | Documentos | 15 | 2 |
| 4 | Vídeos | 10 | 500 |

Determínese el ancho de banda, expresado en Kbps, del servidor web a partir de los datos recogidos por las herramientas de monitorización.

**SOLuCIón:** En primer lugar calculamos el ancho de banda o productividad del servidor, expresado en bits por segundo (bps), atendiendo a cada tipo de petición:

$$X_1 = 21.600 \times 0,25 \times 15 \times \frac{1.024 \times 8}{3.600} = 18.4320 \text{ bps}$$

$$X_2 = 21.600 \times 0,50 \times 25 \times \frac{1.024 \times 8}{3.600} = 614.400 \text{ bps}$$

$$X_3 = 21.600 \times 0,15 \times 2 \times \frac{1.024 \times 8}{3.600} = 14.745,6 \text{ bps}$$

$$X_4 = 21.600 \times 0,10 \times 500 \times \frac{1.024 \times 8}{3.600} = 2.457.600 \text{ bps}$$

Por tanto, el tráfico soportado por el servidor web durante el periodo de medida se calcula como la suma del tráfico de todos los tipos de peticiones, que en este caso es de 3.271.065,6 bps, o de forma equivalente, 3.271,07 Kbps o 3,27 Mbps. Nótese que en las unidades que expresan la productividad, las constantes K y M vienen representadas por $10^3$ y $10^6$, respectivamente, ya que están referidas a parámetros temporales y no a capacidad de almacenamiento. ■

**PROBLEMA 4.15** Un fabricante de coches usa una intranet para compartir los diseños de prototipos nuevos. Esta intranet es utilizada por los departamentos de ingeniería y de producción. Los departamentos tienen 100 y 1.000 empleados, respectivamente, y se considera que, en el primero, el 95 % de los ingenieros usan la intranet, mientras que en el segun- do este porcentaje se reduce al 75 %. Durante las horas de trabajo los ingenieros realizan 75 operaciones por hora que solicitan una media de 5 ficheros, mientras que los emplea- dos de producción hacen 50 operaciones por hora que solicitan una media de 3 ficheros. El tamaño medio de los ficheros de diseño es de 125 KB. Determínese el ancho de banda mínimo para la intranet de manera que pueda soportar el tráfico de información requerido.

Así mismo, determínese si la intranet se podría implementar físicamente mediante una red Ethernet (velocidad de transmisión de 10 Mbps) o una red Fast Ethernet (velocidad de transmisión de 100 Mbps).

**SOLUCIÓN:** Para saber la cantidad de información por unidad de tiempo generada en la intranet hay que considerar el número de usuarios involucrados, las operaciones que genera cada uno de ellos y la información transmitida en cada operación. Así pues, el ancho de banda o productividad $X$ expresada en millones de bits por segundo (Mbps) generado en la intranet será:

$$X = (0{,}95 \times 100) \times \frac{75}{3.600} \times (5 \times 125 \times 1.024 \times 8) + (0{,}75 \times 1.000) \times \frac{50}{3.600} \times (3 \times 125 \times 1.024 \times 8)$$

$$= 10{,}13 + 32 = 42{,}133 \text{ Mbps}$$

En consecuencia, la red que se utilizase debería tener, como mínimo, un ancho de banda de 42,133 Mbps. Por tanto, la intranet ha de implementarse, al menos, con una red de tipo Fast Ethernet, que tiene una velocidad de transmisión de 100 Mbps. Una red Ethernet, que funciona a 10 Mbps, sería insuficiente para satisfacer las necesidades del departamento de ingeniería. ■

---

**PROBLEMA 4.16** El monitor de un servidor dedicado al comercio electrónico ha esti- mado la razón de visita y el tiempo de respuesta de los tres dispositivos afectados por las peticiones. Esta información se refleja en la tabla adjunta. Calcúlese el tiempo de respuesta que experimenta una transacción en este servidor.

| Dispositivo | Razón de visita | Tiempo de respuesta (ms) |
|---|---|---|
| 1 | 1,6 | 1,0 |
| 2 | 2,0 | 5,8 |
| 3 | 5,9 | 1,1 |

**SOLUCIÓN:** Dado que se conocen las razones de visita y los tiempos de respuesta de todos los dispositivos del servidor se puede aplicar la ley general del tiempo de respuesta de un sistema:

$$R = \sum_{i=1}^{3} V_i \times R_i = 1{,}6 \times 1{,}0 + 2{,}0 \times 5{,}8 + 5{,}9 \times 1{,}1 = 19{,}69 \text{ ms}$$

Así pues, cada transacción con el servidor experimenta un tiempo medio de respuesta de 19,69 milisegundos. ■

**PROBLEMA 4.17** Consideremos un servidor web con un procesador y un disco. Este sistema recibe una media de $\lambda$ = 3 peticiones por segundo. Las peticiones siguen el modelo de comportamiento del servidor central. Los tiempos de servicio y las razones de visita a cada dispositivo se indican en la siguiente tabla:

| Dispositivo | Razón de visita | Tiempo de servicio (s) |
|---|---|---|
| Procesador (1) | 5 | 0,02 |
| Disco (2) | 4 | 0,05 |

Se pide calcular:

1. La demanda de servicio de cada dispositivo.
2. Si el tiempo de respuesta del procesador y del disco es 0,0286 y 0,1250 segundos, respectivamente, calcúlese el tiempo de respuesta del servidor web.
3. El número medio de peticiones en el sistema.
4. La productividad y la utilización de cada dispositivo.

## SOLUCIÓN:

1. La demanda de servicio de cada dispositivo es:

$$D_1 = V_1 \times S_1 = 5 \times 0{,}02 = 0{,}1 \text{ s}$$
$$D_2 = V_2 \times S_2 = 4 \times 0{,}05 = 0{,}2 \text{ s}$$

2. El tiempo de respuesta del sistema se puede calcular a partir de los tiempos de respuesta de cada dispositivo y de la razón de visita utilizando la ley general del tiempo de respuesta:

$$R = \sum_{i=1}^{2} V_i \times R_i = 0{,}0286 \times 5 + 0{,}1250 \times 4 = 0{,}643 \text{ s}$$

3. El número de peticiones dentro del servidor web se calcula utilizando la ley de Little. Si $X_0$ denota la productividad del servidor web, y se cumple la hipótesis del flujo equilibrado de trabajos, tendremos:

$$N = X_0 \times R = \lambda \times R = 3 \times 0{,}643 = 1{,}929 \text{ peticiones}$$

Este número de peticiones, 1,929, son las que, por término medio, están utilizando el procesador y el disco.

4. La productividad de los dispositivos puede calcularse a partir de su razón de visitas y la productividad del servidor web utilizando la ley del flujo forzado:

$$X_1 = X_0 \times V_1 = 3 \times 5 = 15 \text{ peticiones/s}$$
$$X_2 = X_0 \times V_2 = 3 \times 4 = 12 \text{ peticiones/s}$$

El cálculo de esta productividad también se podría haber llevado a cabo a partir de las probabilidades de encaminamiento. Por ejemplo, la productividad del sistema es una fracción de la productividad del procesador:

$$\lambda = X_0 = X_1 \times p_{1,0}$$

donde $p_{1,0}$ es la proporción de peticiones que salen del procesador y abandonan el servidor web, o de forma equivalente, la probabilidad de que una petición que sale del procesador salga del servidor web. Esta proporción o probabilidad se puede calcular a partir de las razones de visita:

$$p_{1,0} = \frac{V_0}{V_1} = \frac{1}{5} = 0{,}2$$

Por tanto, la productividad del procesador será:

$$X_1 = \frac{\lambda}{p_{1,0}} = \frac{3}{0{,}2} = 15 \text{ peticiones/s}$$

Por otra parte, la productividad del disco también es una fracción de la productividad del procesador. En concreto, esta fracción viene determinada por la probabilidad de encami- namiento del procesador hacia el disco, que se calcula como:

$$p_{1,2} = \frac{V_2}{V_1} = \frac{4}{5} = 0{,}8$$

En consecuencia, la productividad del disco será:

$$X_2 = X_1 \times p_{1,2} = 15 \times 0{,}8 = 12 \text{ peticiones/s}$$

Finalmente, las utilizaciones de los dos dispositivos se calculan por medio de la ley de Little:

$$U_1 = X_1 \times S_1 = 15 \times 0{,}02 = 0{,}3$$
$$U_2 = X_2 \times S_2 = 12 \times 0{,}05 = 0{,}6$$

■

**PROBLEMA 4.18** Consideremos un sistema informático interactivo con un procesador, una unidad de disco y una unidad de cinta. Los tiempos de servicio y razones de visita de estos dispositivos se muestran en la siguiente tabla:

| Dispositivo | Razón de visita | Tiempo de servicio (s) |
|---|---|---|
| Procesador (1) | 9 | 0,005 |
| Disco (2) | 7 | 0,02 |
| Cinta (3) | 1 | 0,3 |

Los trabajos que sirve el sistema siguen el modelo de comportamiento del servido central. El número de usuarios conectados es de 20 y su tiempo medio de reflexión de 8 segundos.

1. Calcúlense las demandas de servicio de cada dispositivo.
2. Determínense las probabilidades de encaminamiento al disco y a la cinta para los trabajos que salen del procesador.
3. Si la productividad del sistema informático es de 2,2241 trabajos por segundo, ¿cuál es el número medio de trabajos que están en reflexión?
4. ¿Cuántos trabajos hay en el sistema?
5. ¿Cuál es el tiempo de respuesta del sistema informático?
6. Calcúlese la productividad y la utilización de cada dispositivo.

## SOLUCIÓN:

1. La demanda de servicio de los tres dispositivos considerados es:

$$D_1 = V_1 \times S_1 = 9 \times 0{,}005 = 0{,}045 \text{ s}$$
$$D_2 = V_2 \times S_2 = 7 \times 0{,}02 = 0{,}14 \text{ s}$$
$$D_3 = V_3 \times S_2 = 1 \times 0{,}3 = 0{,}3 \text{ s}$$

2. Las probabilidades de encaminamiento se calculan a partir de las razones de visita:

$$p_{1,2} = \frac{V_2}{V_1} = \frac{7}{9} = 0{,}7778$$
$$p_{1,3} = \frac{V_3}{V_1} = \frac{1}{9} = 0{,}1111$$

Por otro lado, la probabilidad de encaminamiento hacia los terminales se determina como:

$$p_{1,0} = \frac{V_0}{V_1} = \frac{1}{9} = 0{,}1111$$

3. El número de trabajos en reflexión se puede calcular aplicando la ley de Little a los terminales:

$$X \times Z = 2{,}2241 \times 8 = 17{,}7928 \text{ trabajos}$$

Así pues, de los 20 usuarios que consideramos en el modelo, una media de 17,7928 están en reflexión. El resto está compitiendo por los tres recursos del sistema.

4. El número de trabajos dentro del sistema informático se calcula restando aquellos que están en reflexión del número total: $20 - 17{,}7928 = 2{,}2071$.

5. El tiempo de respuesta del sistema se calcula fácilmente por medio de la ley del tiempo de respuesta de un sistema interactivo:

$$R = \frac{N}{X_0} - Z = \frac{20}{2{,}2241} - 8 = 0{,}9924 \text{ s}$$

6. La productividad de cada dispositivo se determina a partir de la productividad del sistema y de las probabilidades de encaminamiento:

$$X_1 = \frac{X_0}{p_{1,0}} = \frac{2{,}2241}{0{,}1111} = 20{,}02 \text{ peticiones/s}$$

$$X_2 = X_1 \times p_{1,2} = 20{,}02 \times 0{,}7778 = 15{,}57 \text{ peticiones/s}$$

$$X_3 = X_1 \times p_{1,3} = 20{,}02 \times 0{,}1111 = 2{,}2241 \text{ peticiones/s}$$

Nótese que los valores anteriores también se pueden calcular de manera más directa utilizando la ley del flujo forzado:

$$X_1 = X_0 \times V_1 = 2{,}2241 \times 9 = 20{,}02 \text{ peticiones/s}$$

$$X_2 = X_0 \times V_2 = 2{,}2241 \times 7 = 15{,}57 \text{ peticiones/s}$$

$$X_3 = X_0 \times V_3 = 2{,}2241 \times 1 = 2{,}2241 \text{ peticiones/s}$$

Finalmente, las utilizaciones se calculan aplicando la ley de la utilización:

$$U_1 = X_1 \times S_1 = 20{,}02 \times 0{,}005 = 0{,}1001$$

$$U_2 = X_2 \times S_2 = 15{,}57 \times 0{,}02 = 0{,}3114$$

$$U_3 = X_3 \times S_3 = 2{,}2242 \times 0{,}3 = 0{,}6672$$

■

**PROBLEMA 4.19** Un modelo de sistema informático con cuatro dispositivos presenta los parámetros que se indican a continuación:

| Dispositivo | Razón de visita | Tiempo de servicio (ms) |
|---|---|---|
| 1 | 8 | 8,3 |
| 2 | 7 | 2,1 |
| 3 | 1 | 5,4 |
| 4 | 6 | 6,0 |

Si el sistema ha servido un total de 120 trabajos en medio minuto, calcúlense las productividades y las utilizaciones de los dispositivos 1 y 4, así como el tiempo de respuesta del sistema.

**SOLUCIÓN:** La productividad del sistema informático se calcula dividiendo el número de trabajos servidos entre el tiempo de medida:

$$X_0 = \frac{125}{30} = 4 \text{ trabajos/s}$$

Suponiendo que se cumple la hipótesis del flujo equilibrado de trabajos, la productividad de los dispositivos puede determinarse mediante la ley del flujo forzado:

$$X_1 = X_0 \times V_1 = 4 \times 8 = 32 \text{ peticiones/s}$$

$$X_4 = X_0 \times V_4 = 4 \times 6 = 24 \text{ peticiones/s}$$

Por su parte, la utilización de cada dispositivo se puede calcular mediante la ley de la utilización:

$$U_1 = X_1 \times S_1 = 32 \times 8,3 \times 10^{-3} = 0,2656$$

$$U_4 = X_4 \times S_4 = 24 \times 6,0 \times 10^{-3} = 0,1440$$

Finalmente, el tiempo medio de respuesta del sistema se calcula aplicando la ley general del tiempo de respuesta. Nótese que en su expresión intervienen todos los dispositivos del sistema:

$$R = \sum_{i=1}^{4} V_i \times R_i = 8 \times 8,3 + 4 \times 2,1 + 1 \times 5,4 + 6 \times 6,0 = 116,2 \text{ ms}$$

■

## 4.5. Problemas con solución

**PROBLEMA 4.20** Considérese una red de área local Gigabit Ethernet con un ancho de banda de 1 Gbps. Las peticiones desde un cliente al servidor constan de 4 paquetes de 1.518 bytes cada uno, mientras que las respuestas desde el servidor requieren el envío de 12 paquetes de este mismo tamaño. Calcúlese el tiempo de transmisión total de la transacción a través de la red.

**SOLuCIón:** El tiempo requerido de transmisión es de 0,1943 ms. ■

**PROBLEMA 4.21** Considérese una red Fast Ethernet que interconecta varias redes de área local a través de distintos encaminadores (*routers*) actuando a modo de red central (*backbone*). En concreto, una red Token Ring está conectada a la red central a través de un encaminador que posee una latencia de 0,3 ms por paquete. La red central transmite a la Token Ring una media de 10.000 paquetes por segundo. Adicionalmente, una red Ethernet también está conectada a la red central a través de un repetidor (*hub*). El número medio de paquetes en tránsito desde la Ethernet a la red central es de 3 y la Ethernet transmite 1.000 paquetes por segundo. Calcúlese:

1. El número medio de paquetes en tránsito de la red central a la Token Ring.
2. El tiempo de tránsito del repetidor que enlaza la red Ethernet y la central.

**SOLuCIón:** El número de paquetes en tránsito desde la red central a la red Token Ring es de 3, mientras que el repetidor que conecta la red Ethernet a la central tiene un tiempo de respuesta de 0,003 segundos. ■

**PROBLEMA 4.22** El responsable del sistema informático de un banco quiere reemplazar el programa de gestión de la base de datos. La publicidad que el proveedor ha suminis- trado sobre un nuevo gestor afirma que éste resulta un 40 % más rápido que el actual. El responsable de la instalación ha utilizado una implementación del *benchmark* TPC-C para evaluar el rendimiento tanto del gestor actual como del nuevo, obteniendo 23.000 y 35.000 tpm (transacciones por minuto), respectivamente. Estímese la productividad que permitirá alcanzar el nuevo gestor de la base de datos en el sistema informático si el actual realiza 500 tps (transacciones por segundo).

**SOLuCIón:** La productividad que permitirá alcanzar el nuevo gestor será de 760,8 tps ■

**PROBLEMA 4.23** Consideremos un sistema interactivo con 24 usuarios conectados. Calcúlese la productividad de este sistema si el tiempo de reflexión es de 30 segundos y el tiempo de respuesta del sistema experimentado por cada usuario es de 2 segundos.

**SOLuCIón:** La productividad del sistema es de 0,75 trabajos/s. ■

**PROBLEMA 4.24** La tarjeta de red de un computador recibe 125 paquetes por segundo y tarda una media de 2 milisegundos en enviar cada paquete. ¿Cuál es la utilización de esta tarjeta?

**SOLuCIón:** La utilización de la tarjeta es 0,25. ■

**PROBLEMA 4.25** Un monitor indica que la utilización del procesador de un sistema interactivo es 0,75 y su demanda de servicio de 3 segundos. El tiempo de respuesta del sistema es 15 segundos y hay un total de 10 usuarios conectados. Calcúlese el tiempo medio de reflexión de los usuarios.

**SOLUCIÓN:** El tiempo medio de reflexión es de 25 segundos. ■

**PROBLEMA 4.26** La productividad de un computador es de 3 programas ejecutados por segundo. Si cada programa requiere una media de 4 accesos al disco, ¿cuál es la producti- vidad del disco?

**SOLUCIÓN:** La productividad del disco es de 12 accesos/s. ■

**PROBLEMA 4.27** Un sistema informático está compuesto por un procesador y un disco. Las razones de visita a estos dos dispositivos son 4 y 3, respectivamente. El tiempo medio de respuesta del sistema experimentado por un trabajo es de 20 segundos. Si se sabe que el tiempo de respuesta del disco es dos veces más grande que el del procesador, calcúlese el tiempo de respuesta de estos dos dispositivos.

**SOLUCIÓN:** El tiempo de respuesta del procesador es de 2 segundos, mientras que el del disco es de 4 segundos. ■

## 4.6. Problemas sin resolver

**PROBLEMA 4.28** Después de utilizar un monitor durante un periodo de medida de 5 segundos se ha obtenido la siguiente información sobre los instantes de llegada y de salida de diferentes trabajos al disco de un sistema informático:

| Trabajo | Llegada | Salida |
|---|---|---|
| 1 | – | 0,98 |
| 2 | – | 1,82 |
| 3 | – | 2,10 |
| 4 | 0,34 | 2,58 |
| 5 | 1,76 | 2,95 |
| 6 | 2,21 | 3,12 |
| 7 | 3,84 | 4,24 |
| 8 | 4,39 | 4,66 |
| 9 | 4,77 | – |
| 10 | 4,90 | – |

1. ¿Con qué precisión se cumple la ley del flujo equilibrado de trabajos?
2. ¿Cuál es la productividad del disco?
3. Determínese, por cálculo directo, el tiempo medio de respuesta del disco.
4. Calcúlese mediante la ley de Little el número medio de trabajos en el disco.
5. ¿Cuál es la utilización del disco? ¿Y su tiempo de servicio?

---

**PROBLEMA 4.29** El sitio web de una empresa dedicada a productos de cosmética recibe una media de 30 peticiones por segundo. La mayoría de estas peticiones se usan para hojear el catálogo virtual de productos. Solamente una de cada 10 peticiones realiza un pedido de cosméticos. Cada una de estas órdenes de pedido provoca la ejecución de un programa en Java (*servlet* ) en el procesador del servidor web y una demanda de 100 ms al disco del servidor. Determínese la utilización del disco debido a las peticiones de pedido.

---

**PROBLEMA 4.30** El administrador de un sistema informático tiene que elegir entre dos servidores para implementar un sitio web dedicado al comercio electrónico. Para ello ha ejecutado una implementación del *benchmark* TPC-W en ambos servidores, y uno de ellos resulta un 30 % más rápido en el servicio de las transacciones del test. Determínese el tiempo de respuesta de ambos servidores si el más lento alcanza una productividad de 25 tps (transacciones por segundo) cuando hay 3 transacciones de media residentes en el servidor.

---

**PROBLEMA 4.31** Un servidor web compuesto por tres dispositivos recibe peticiones de dos clases, A y B. Los parámetros relativos a cada clase se muestran en la siguiente tabla, donde los tiempos se expresan en milisegundos:

| Dispositivo | $V_{i,A}$ | $S_{i,A}$ | $R_{i,A}$ | $V_{i,B}$ | $S_{i,B}$ | $R_{i,B}$ |
|---|---|---|---|---|---|---|
| 1 | 3 | 2,5 | 9,3 | 2 | 1,5 | 5,5 |
| 2 | 2 | 7,0 | 18,0 | 1 | 3,0 | 4,2 |
| 3 | 4 | 1,0 | 4,8 | 8 | 0,4 | 4,7 |

Calcúlese la demanda de servicio a cada dispositivo y el tiempo medio de respuesta del sistema para cada clase de peticiones.

---

## 4.7. Actividades propuestas

**ACTIVIDAD 4.1** En esta actividad se propone un trabajo de campo. Para ello elíjase un establecimiento comercial (por ejemplo, una peluquería, una panadería, etc.). Se pide monitorizar los instantes de entrada y de salida de los clientes del establecimiento durante un tiempo razonable, que dependerá de la actividad (por ejemplo, una hora), anotando el instante de entrada y el de salida, así como el instante en que reciben servicio. Compruébese con qué exactitud se cumple la hipótesis del flujo equilibrado de trabajos, y calcúlense los siguientes índices de prestaciones del establecimiento: número medio de clientes, tiempo medio de servicio, tiempo medio de respuesta y número medio de clientes en espera.

# Capítulo 5

## Aplicaciones del análisis operacional

El capítulo anterior ha tratado los aspectos fundamentales del análisis operacional como herramienta que permite establecer relaciones sencillas entre variables relacionadas con el rendimiento de los sistemas informáticos a partir de un modelo. En este capítulo presen- taremos la aplicación de esta técnica para estimar el rendimiento de un computador y, a su vez, para cuantificar el efecto en las prestaciones de mejoras en el sistema.

En particular, se van a presentar dos algoritmos de resolución con el fin de obtener una estimación del tiempo de respuesta y de la productividad de un sistema informático modelado mediante una red de colas de espera. En particular, las redes que consideraremos serán muy sencillas: habrá una única clase de trabajos y supondremos que tanto los tiempos de servicio de las estaciones como los tiempos entre llegadas de trabajos en los modelos abiertos se distribuyen de forma exponencial. Esta simplificación de la realidad que se modela resulta, en muchas ocasiones, satisfactoria para los tiempos entre llegadas, aunque no tanto para los tiempos de servicio. Sin embargo, en conjunto podemos considerarla como una aproximación aceptable de la realidad, aunque en última instancia esto dependerá del tipo de sistema que se esté considerando.

Junto con la estimación del rendimiento por medio de un algoritmo de resolución a un modelo de colas de espera, también contemplamos la alternativa de calcular los valores más optimistas del rendimiento que podemos esperar de un sistema informático. Estos límites, que reciben el adjetivo de asintóticos, se pueden establecer de una manera muy sencilla sin necesidad de emplear ningún algoritmo de resolución, y nos proporcionarán una cota superior tanto del tiempo de respuesta como de la productividad alcanzable por el sistema.

Finalmente, hablaremos brevemente sobre las alternativas de que disponemos para ac-tuar sobre un sistema informático con el fin de mejorar su rendimiento. Veremos que, mediante el análisis de un modelo de colas, bien sea aplicando los algoritmos de resolu- ción o los límites asintóticos, resulta sencillo evaluar cuantitativamente el efecto que una determinada terapia tiene sobre las prestaciones del sistema.

## 5.1. Estimación del rendimiento

En este apartado vamos a presentar de forma resumida dos algoritmos clásicos para resolver modelos de colas sencillos y estimar así el rendimiento del sistema, el cual vendrá dado por el tiempo de respuesta y la productividad. Para ello partiremos de dos hipótesis. La primera establece que si un trabajo está sirviéndose en una estación, el tiempo que le falta para abandonar el servidor es independiente del tiempo que ya lleva en servicio. La segunda hipótesis es la siguiente: en un sistema abierto, el tiempo que transcurre hasta la próxima llegada es independiente del instante en que se produjo la última. Estas dos hipótesis equivalen a suponer que tanto la distribución del tiempo de servicio de las estaciones como la distribución del tiempo de llegadas en un modelo abierto son exponenciales. Debido a sus propiedades estadísticas, se dice que esta distribución carece de memoria (*memoryless property*).

Antes de plantear los algoritmos de resolución introduciremos una expresión para cal- cular el tiempo de respuesta de una estación de servicio *i* de tipo cola. Esta expresión adquiere la forma:

$$R_i = (N_i + 1) \times S_i \tag{5.1}$$

y puede justificarse de la siguiente forma: un trabajo que llega a la estación *i* encuentra $N_i$ trabajos en ella y esperará $N_i$ $S_i$ unidades de tiempo a que se sirvan, más $S_i$ para recibir su propio servicio. Nótese que se está utilizando la propiedad de que el tiempo de servicio se distribuye exponencialmente (carece de memoria) y, por tanto, no es necesario tener en cuenta el tiempo de servicio ya recibido por el cliente que está en el servidor cuando se produce la llegada. Esta propiedad de carencia de memoria no puede ser comprobada operacionalmente, y por lo tanto la ecuación anterior no constituye una ley operacional. Nótese que, operando sobre la expresión anterior y sustituyendo $N_i$ por $X_i$ $R_i$ (ley de Little), podemos relacionar el tiempo de respuesta de una estación *i* con su tiempo de servicio $S_i$ y su utilización $U_i$:

$$R_i = (X_i \times R_i + 1) \times S_i \Rightarrow R_i = \frac{S_i}{1 - X_i \times S_i} = \frac{S_i}{1 - U_i} \tag{5.2}$$

Mediante esta última relación que permite calcular fácilmente $R_i$ y las leyes operacio- nales que hemos visto en el capítulo anterior podemos plantear sendos algoritmos para resolver redes abiertas y cerradas, respectivamente.

## Algoritmo para redes abiertas

Como punto de partida supondremos conocidos la razón de visita $V_i$ y el tiempo de servicio $S_i$ de las $K$ estaciones de la red (sean de tipo cola o de tipo retardo). Como se ha dicho antes, tanto los tiempos de servicio como los tiempos entre llegadas se suponen distribuidos exponencialmente. Así mismo, se supondrá conocida la tasa de llegadas $\lambda$ al sistema, la cual será igual a la productividad del sistema, ya que suponemos válida la hipótesis del flujo equilibrado de trabajos. El objetivo del algoritmo es calcular las variables: $X_i$, $N_i$, $R_i$ y $U_i$ para cada estación, y $R$ y $N$ para toda la red.

Así pues, en primer lugar podemos calcular la demanda de servicio de cada estación haciendo $D_i = V_i \times S_i$. Las utilizaciones $U_i$ se obtienen mediante la expresión $U_i = \lambda \times D_i$, y las productividades $X_i$ haciendo $X_i = \lambda \times V_i$. En estos momentos ya podemos calcular $R_i$, que será igual a $S_i/(1 - U_i)$ si la estación es de tipo cola o bien $S_i$ si es de tipo retardo. El número de trabajos en cada estación se obtiene aplicando la ley de Little $N_i = X_i \times R_i$.

Finalmente, el tiempo de respuesta del sistema se obtiene a partir de los $R_i$ y $V_i$ aplicando la ley general del tiempo de respuesta: $R = \sum_{i=1}^{i=K} V_i R_i$, y el número de trabajos en el mismo se calcula sumando los trabajos contenidos en todas las estaciones del modelo: $N = \sum_{i=1}^{i=K} N_i$, o bien aplicando la ley de Little al sistema completo.

Como ejemplo sencillo de aplicación de este algoritmo, supongamos una red de colas abierta que recibe una tasa de llegadas $\lambda$ de dos trabajos por segundo, con dos dispositivos, 1 y 2, que tienen los tiempos de servicio y razones de visita expresados en la siguiente tabla:

| Dispositivo | Razón de visita | Tiempo de servicio (s) |
|---|---|---|
| 1 | 6 | 0,01 |
| 2 | 7 | 0,02 |

En primer lugar se pueden calcular las utilizaciones:

$$U_1 = \lambda \times D_1 = \lambda \times V_1 \times S_1 = 2 \times 6 \times 0{,}01 = 0{,}12$$
$$U_2 = \lambda \times D_2 = \lambda \times V_2 \times S_2 = 2 \times 7 \times 0{,}02 = 0{,}28$$

En este momento ya se pueden calcular los tiempos de respuesta de cada estación utilizando la ecuación 5.2:

$$R_1 = \frac{S_1}{1 - U_1} = \frac{0{,}01}{1 - 0{,}12} = 0{,}0114 \text{ s}$$
$$R_2 = \frac{S_2}{1 - U_2} = \frac{0{,}02}{1 - 0{,}28} = 0{,}0278 \text{ s}$$

Finalmente, el tiempo de respuesta del sistema y el número de trabajos contenidos en él se calculan utilizando las relaciones:

$$R = V_1 \times R_1 + V_2 \times R_2 = 6 \times 0{,}0114 + 7 \times 0{,}0278 = 0{,}263 \text{ s}$$
$$N = \lambda \times R = 2 \times 0{,}263 = 0{,}526 \text{ trabajos}$$

## Algoritmo para redes cerradas

Este algoritmo también se denomina *análisis del valor medio*. Igual que en el caso anterior, supondremos conocidos $V_i$ y $S_i$ para todas las estaciones del modelo, además del tiempo de reflexión $Z$ (que será nulo si se trata de un sistema por lotes). Las variables a calcular son similares al caso anterior, y la diferencia estriba en que ahora no se conoce la productividad del sistema, sino que se ha de estimar; en cambio, al tratarse de un modelo cerrado, sí se sabe el número de trabajos $N$ en el sistema

Antes de proponer el algoritmo plantearemos la ecuación que permite estimar $R_i$ para las estaciones de tipo cola teniendo en cuenta que ahora su valor dependerá del número de trabajos $N$ en el sistema:

$$R_i(N) = [N_i(N - 1) + 1] \times S_i$$

donde $N_i(N - 1)$ es el número de trabajos en la estación $i$ cuando en la red cerrada hay $N - 1$ trabajos. Esta relación establece que el estado de la red visto por un trabajo que está en tránsito de una estación a otra (el trabajo ha abandonado una estación, pero aún no se ha incorporado a la siguiente), tiene la misma distribución que el estado que vería un observador aleatorio si el número total de trabajos en la red fuese $N - 1$. Esta afirmación es bastante intuitiva, ya que un trabajo en tránsito no puede observarse a sí mismo en ninguna estación.

La ecuación anterior relaciona dos índices de prestaciones, uno para $N$ y otro para $N-1$, dando lugar a un procedimiento de cálculo iterativo. Los valores para la primera iteración son fáciles de establecer: para $N = 0$ se cumple $N_i = 0$ y por tanto $R_i(1) = S_i$, $i = 1, \ldots, K$. Para las estaciones de tipo retardo se cumple, además, que $R_i(N) = S_i$, $\forall N$. Así pues, el algoritmo de resolución tendrá la siguiente forma:

Para $n$ desde 1 hasta $N$ hacer:

$$R_i(n) = (N_i(n - 1) + 1) \times S_i, \text{ con } N_i(0) = 0$$

$$R(n) = \sum_{i=1}^{K} V_i \times R_i(n), \qquad X(n) = \frac{n}{Z + R(n)}$$

$$N_i(n) = X(n) \times V_i \times R_i(n) \qquad X_i(n) = X(n) \times V_i$$

$$U_i(n) = X(n) \times V_i \times S_i$$

Como ejemplo sencillo de aplicación de este algoritmo, supongamos una red de colas cerrada con tres trabajos y dos dispositivos, 1 y 2, que tienen los tiempos de servicio y razones de visita expresados en la siguiente tabla:

| Dispositivo | Razón de visita | Tiempo de servicio (s) |
|---|---|---|
| 1 | 15 | 0,03 |
| 2 | 14 | 0,5 |

Supondremos que la carga es interactiva con un tiempo de reflexión $Z = 5$ segundos. Para aplicar el algoritmo habrá que hacer un total de 3 iteraciones, una por cada trabajo presente en el sistema. Para cada iteración calcularemos, en primer lugar, el tiempo de res- puesta de cada estación; a continuación se calculará el tiempo de respuesta del sistema y su productividad, y finalmente se podrán determinar el número de trabajos, la productividad y la utilización de cada estación del modelo. La Tabla 5.1 muestra los datos obtenidos en cada iteración del algoritmo (no se muestra ni las utilizaciones ni la productividades individuales):

| Trabajos | $R_1$ | $R_2$ | $R$ | $X_0$ | $N_1$ | $N_2$ |
|---|---|---|---|---|---|---|
| 1 | 0,0300 | 0,5000 | 7,4500 | 0,0803 | 0,0361 | 0,5622 |
| 2 | 0,0311 | 0,7811 | 11,4920 | 0,1219 | 0,0569 | 1,3335 |
| 3 | 0,0317 | 1,1667 | 16,8090 | 0,1376 | 0,0654 | 2,2468 |

**Tabla 5.1:** Ejemplo de aplicación del algoritmo del valor medio

El tiempo de respuesta del sistema, a partir de los datos presentados en la tabla, es de 16,8090 segundos, mientras que la productividad es de 0,1376 trabajos por segundo.

## 5.2. Límites asintóticos

Como ya se ha señalado, una consecuencia de la ley del flujo forzado es que las utilizaciones de los dispositivos son proporcionales a las demandas totales de servicio. Por tanto, aquel con mayor demanda de servicio tendrá la mayor utilización; tal dispositivo se denomina cuello de botella, y su papel resulta determinante en las prestaciones del sistema completo. El cuello de botella puede estar localizado en varios dispositivos cuando sus demandas de servicio sean iguales y, además, sean las más altas.

Así mismo, cuando la carga del sistema incrementa su magnitud, el dispositivo que tien- de a congestionarse en primer lugar es este cuello de botella. Cuando la utilización de este dispositivo presenta valores cercanos a 1 se dice que está *saturado*. Por esta razón resulta muy interesante que las utilizaciones o demandas de los dispositivos de un sistema sean lo más parecidas posible. Cuando esto ocurre se dice que el sistema está equilibrado (*balanced system*). La mejora del comportamiento del dispositivo cuello de botella redundará en un incremento significativo del rendimiento del sistema completo. En cambio, este incremento será marginal cuando la mejora se haga en cualquiera de los restantes dispositivos.

Los límites asintóticos de las prestaciones representan una técnica de aplicación muy sencilla para acotar, desde un punto de vista optimista, los mejores valores de la producti- vidad y el tiempo de respuesta de un sistema informático. Al dispositivo cuello de botella del sistema informático lo denotaremos mediante el subíndice $b$. Una vez localizado este dispositivo se cumplirán las siguientes igualdades:

$$D_b = \max\{D_1, D_2, \ldots, D_K\} = V_b \times S_b \quad U_b = X_0 \times D_b$$

Como primera aproximación al establecimiento de estos límites asintóticos optimistas consideraremos inicialmente un modelo de colas abierto. El valor máximo de la tasa de llegadas que el sistema es capaz de soportar será aquel que sature completamente el dis- positivo cuello de botella, esto es, que provoque $U_b = 1$. Como se cumple la hipótesis del flujo equilibrado de trabajos, podremos escribir:

$$U_b = X_b \times S_b = X_0 \times V_b \times S_b = X_0 \times D_b = \lambda D_b$$

Sea $\lambda_{opt}$ el valor más alto de la tasa de llegadas que el sistema puede aceptar, la cual será equivalente a la productividad del sistema que denotaremos por $X_{opt}$. Particularizando la expresión anterior para $U_b = 1$ tendremos:

$$\text{Si } U_b = 1 \Rightarrow X_{opt} \times D_b = 1 \Rightarrow X_{opt} = \frac{1}{D}$$

Cuando la tasa de llegadas al sistema toma el valor $\lambda = X_{opt}$ el sistema satura el cuello de botella y el número de trabajos en el sistema crece de forma indefinida, haciendo que éste se vuelva inestable.

Si tomamos en cuenta el tiempo de respuesta, el valor optimista del mismo $R_{opt}$ viene dado cuando el trabajo que llega al sistema lo encuentra vacío; en consecuencia, no habrá de esperar en ningún dispositivo, ya que tan pronto como llegue a él recibirá servicio. En este caso, su valor será equivalente a la suma las demandas de servicio que haga a los diferentes dispositivos del sistema. Si el modelo tiene $K$ estaciones de servicio tendremos:

$$R_{opt} = \sum_{i=1}^{K} D_i = D$$

Resumiendo los resultados obtenidos para el modelo abierto obtenemos las siguientes expresiones para los límites asintóticos:

$$X_{opt} = \frac{1}{D_b} \qquad R_{opt} = D = \sum_{i=1}^{K} D_i$$

En el caso de un modelo cerrado, los límites asintóticos para la productividad y el tiempo de respuesta se establecen de un modo similar. Nos centraremos en el estudio de una red de colas cerrada que modela un sistema interactivo (el caso de un sistema por lotes es similar haciendo el tiempo de reflexión igual a cero). Dado que la carga del sistema viene establecida por $N$ , consideraremos dos situaciones: carga muy baja (sistema vacío: $N = 0$) y carga muy alta (sistema saturado: valores de $N$ suficientemente grandes para saturar el cuello de botella).

Supongamos en primer lugar que el sistema no tiene ningún dispositivo saturado. El valor más optimista para el tiempo de respuesta, $R_{opt}$, es aquel que experimenta un trabajo que no tiene que esperar por utilizar los dispositivos. De modo análogo al caso abierto tenemos:

$$R_{opt} = \sum_{i=1}^{K} D_i = D$$

La particularización de la ley del tiempo de respuesta interactivo para este valor $R_{opt}$ nos permite obtener una expresión optimista para la productividad:

$$\text{Si } R_{opt} = \frac{N}{X_{opt}} - Z \Rightarrow X_{opt} = \frac{N}{D + Z}$$

Si ahora consideramos un escenario distinto en el que la carga es tal que el cuello de botella está saturado ($U_b = 1$), el valor más alto que cabría esperar ahora para la productividad sería:

$$\text{Si } U_b = 1 \Rightarrow X_b \times S_b = X_{opt} \times V_b \times S_b = 1 \Rightarrow X_{opt} = \frac{1}{D}$$

Particularizando de nuevo la expresión de la ley del tiempo de respuesta interactivo para este valor de la productividad tendremos:

$$R_{opt} = \frac{N}{X_{opt}} - Z \Rightarrow R_{opt} = N \times D_b - Z$$

Si ahora resumimos los resultados obtenidos para el modelo cerrado teniendo en cuenta las dos situaciones establecidas por la carga (valores bajo y alto) obtenemos las siguientes expresiones para los límites asintóticos:

$$R_{opt} = \max\{D, D_b \times N - Z\}$$

$$X_{opt} = \min\left\{\frac{N}{D + Z}, \frac{1}{D_b}\right\}$$

El punto de cruce de las dos rectas en cualquiera de las dos expresiones anteriores puede calcularse, por ejemplo, como:

$$\frac{N}{D+Z} = \frac{1}{D_b} \Rightarrow N = \frac{D+Z}{D_b}$$

Al valor anterior de $N$ se lo conoce como punto teórico de saturación, ya que con él se consigue, desde un punto de vista optimista y asintótico, la productividad teórica más alta alcanzable por el sistema. Si se considera el tiempo de respuesta, a partir de este valor no se puede garantizar el tiempo mínimo establecido por $D$ porque los trabajos en el sistema experimentan esperas en los dispositivos, al menos en el cuello de botella. Como el número de trabajos en el sistema viene dado por un número entero, en la práctica el valor anterior se suele expresar como un valor entero:

$$N^* = \frac{D+Z}{D_b}$$

Consideremos un ejemplo sencillo de modelo basado en el servidor central con una carga de tipo interactivo. Supongamos que los trabajos tienen un tiempo medio de reflexión de 6 segundos y que la red de colas tiene tres dispositivos: un procesador y dos discos, cuyas razones de visita y tiempos de servicio se indican en la siguiente tabla:

| Dispositivo | Razón de visita | Tiempo de servicio (s) |
|---|---|---|
| Procesador (1) | 32 | 0,0375 |
| Disco (2) | 25 | 0,02 |
| Disco (3) | 6 | 0,05 |

A partir de la información anterior se puede calcular la demanda de servicio de cada dispositivo:

$$D_1 = V_1 \times S_1 = 32 \times 0{,}0375 = 1{,}2 \text{ s}$$
$$D_2 = V_2 \times S_2 = 25 \times 0{,}02 = 0{,}5 \text{ s}$$
$$D_3 = V_3 \times S_3 = 6 \times 0{,}05 = 0{,}3 \text{ s}$$

A partir de las demandas de servicio podemos concluir que el cuello de botella del sistema es el procesador. Además, su demanda de servicio es muy superior a las demandas de los discos. Por otro lado, el valor del punto teórico de saturación es:

$$N^* = \frac{D+Z}{D_b} = \frac{2+6}{1{,}2} = [6{,}67| = 7 \text{ trabajos}$$

Si se calculan los límites asintóticos del rendimiento obtenemos las siguientes expresio- nes:

$$R_{opt} = \text{máx}\{D, D_b \times N - Z\} = \text{máx}\{2, 1{,}2 \times N - 6\}$$

$$X_{opt} = \text{mín}\left\{\frac{N}{D+Z}, \frac{1}{D_b}\right\} = \text{mín}\left\{\frac{N}{8}, 0{,}833\right\}$$

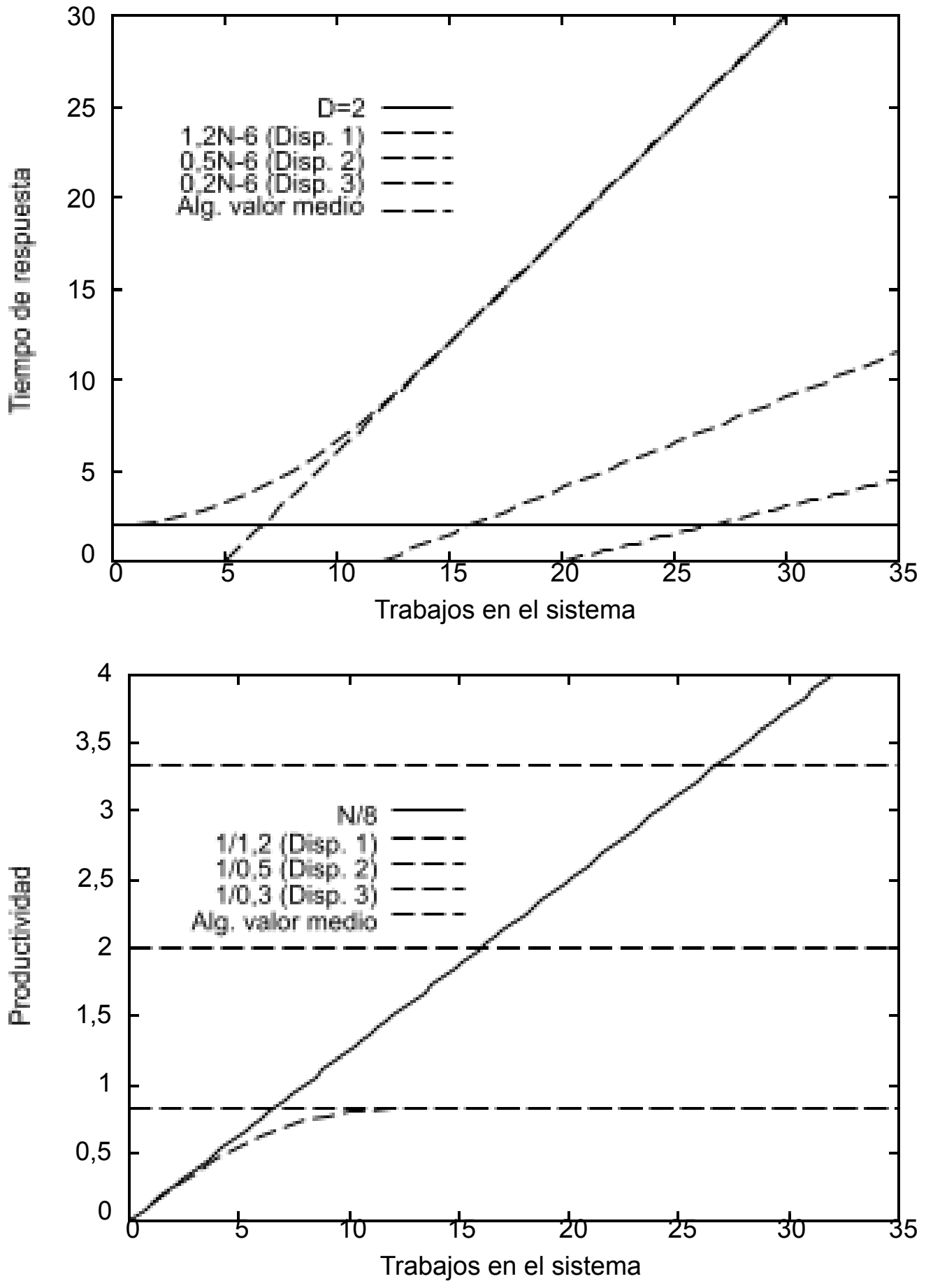

**Figura 5.1:** Límites asintóticos del tiempo de respuesta y de la productividad.

Estos límites asintóticos, junto con la estimación del rendimiento calculada a partir del algoritmo del valor medio, están reflejados en la Figura 5.1. Nótese que en ambas gráficas se han incluido las rectas inducidas por todos los dispositivos así como la estimación de rendimiento que provee el algoritmo del valor medio. De esta manera se aprecia con más claridad la importancia del dispositivo cuello de botella como limitador del rendimiento del sistema completo.

Por ejemplo, si se examina la evolución del tiempo de respuesta se puede apreciar cómo, cuando hay un único trabajo en el sistema, la asíntota proporciona el mismo valor que el algoritmo. Sin embargo, a medida que aumenta el número de trabajos, la previsión que hace la asíntota es inferior a la estimación del algoritmo porque supone la existencia de un único trabajo en el sistema. Hay que esperar hasta aproximadamente los 15 trabajos en el sistema para que la previsión optimista del límite asintótico coincida con el resultado que aporta el algoritmo. Si se analiza la gráfica relativa a la productividad se constata un comportamiento similar. En cualquiera de los dos casos es interesante poner de manifiesto que la asíntota que determina el comportamiento del sistema es la del cuello de botella. Sin embargo, no se ha tenido en cuenta la sobrecarga producida por el sistema operativo, que aumenta con el número de trabajos en el sistema.

## 5.3. Mejora del rendimiento

La mejora del rendimiento de un sistema informático no es una tarea trivial, ya que hay multitud de factores que influyen en él, como por ejemplo los componentes físicos del com- putador y el comportamiento de los programas que se ejecutan en él, desde el sistema operativo hasta las aplicaciones de los usuarios. En cualquier caso, la mejora del rendi- miento implicará la localización del cuello de botella del sistema (que podrá encontrarse en el hardware o en el software), responsable último de las prestaciones globales, y actuar sobre él.

Dejando de lado las particularidades de cada sistema, hay dos aproximaciones de ca- rácter general que suelen emplearse para mejorar el rendimiento de un computador. La primera de ellas consiste en actuar sobre los componentes físicos del mismo mejorándolos o aumentando su número (*upgrading techniques*). Aquí podemos contemplar tanto el reem- plazo de los dispositivos existentes por versiones más modernas y rápidas (menor tiempo de servicio); por ejemplo, añadir nuevas unidades iguales o similares a las existentes con el objetivo de realizar varias tareas en paralelo. Aunque estas acciones terapéuticas pueden parecer sencillas de llevar a cabo, en la práctica no suelen resultar así. En primer lugar, la adición de nuevos componentes supone una inversión nada desdeñable desde un punto de vista económico y, en segundo lugar, hay determinados componentes físicos que son difíciles de reemplazar de una manera aséptica. Por ejemplo, la sustitución de un procesador o de la memoria principal de un computador personal por componentes más modernos requiere que la placa base sea capaz de asumirlos, hecho poco probable si ha pasado mucho tiempo desde la adquisición del equipo original. Algo parecido ocurre si se quiere añadir un nuevo procesador manteniendo el existente: es necesario que la placa base provea tal posibilidad. Si hablamos, por otro lado, de ampliar el número de discos duros del computador, depen- deremos tanto de las posibilidades que ofrezca la interficie (por ejemplo, IDE, SCSI o Fibre Channel) como del espacio físico disponible para ubicarlos.

La segunda técnica de mejora, que recibe el nombre de ajuste o sintonización (*tun- ning techniques*), engloba todas aquellas acciones sobre los programas que se ejecutan en un computador con el objetivo de mejorar el uso que hacen de los dispositivos físicos. Aquí podemos considerar, por ejemplo, los parámetros de configuración tanto del sistema operativo como de las aplicaciones de los usuarios. La aplicación de esta técnica depende fundamentalmente del grado de conocimiento tanto del programa a modificar como del comportamiento e interacción del mismo con los dispositivos físicos del sistema. A esto habría que añadir, además, la propia facilidad y disponibilidad del programa para ser modificado. Un ejemplo típico de este tipo de técnicas en sistemas con varios procesadores o unidades de disco magnético consiste en equilibrar las demandas de servicio que los programas requieren de estos dispositivos.

Como ejemplo sencillo de aplicación consideremos el sistema cuyo rendimiento está reflejado en la Figura 5.1. Para mejorar las prestaciones de este sistema no queda más remedio que actuar sobre el cuello de botella, que en este caso concreto es el procesador.
¿Qué pasará si, por ejemplo, se sustituye esta unidad por una dos veces más rápida? Con esta terapia se está reduciendo su tiempo de servicio de 0,0375 a 0,01875 segundos, por lo que su nueva demanda de servicio es $D_1 = V_1 \times S_1 = 32 \times 0{,}01875 = 0{,}6$ segundos. Este valor, aunque se ha reducido a la mitad, es todavía el más alto del sistema, por lo que este dispositivo seguirá siendo el cuello de botella. Los nuevos límites asintóticos serán:

$$R_{opt} = \max\{D, D_b \times N - Z\} = \max\{1{,}4,\ 0{,}6 \times N - 6\}$$

$$X_{opt} = \min\left\{\frac{N}{D+Z}, \frac{1}{D_b}\right\} = \min\left\{\frac{N}{7{,}4}, 1{,}667\right\}$$

Así pues, vemos que tanto el valor del tiempo de respuesta mínimo como el de la productividad máxima han mejorado, y las asíntotas del cuello de botella han sufrido un desplazamiento, que puede apreciarse en la Figura 5.2. En particular, si se examina la gráfica del tiempo de respuesta la asíntota $D$ del sistema original se ha desplazado ligeramente hacia abajo (pasa de 2 a 1,4 segundos). Sin embargo, la diferencia más notable se aprecia en la asíntota del cuello de botella, que pasa de $1{,}2 \times N - 6$ a $0{,}6 \times N - 6$ (la pendiente pasa a valer la mitad). Así mismo, la repercusión de la mejora puede apreciarse también en el resultado obtenido por el algoritmo del valor medio. A similares conclusiones se puede llegar si se analiza la evolución de la productividad, cuyo valor máximo ha crecido desde $1/1{,}2 = 0{,}833$ trabajos por segundo en el sistema original hasta justamente el doble: $1/0{,}6 = 1{,}667$ trabajos por segundo. En cualquier caso, las asíntotas que establecen los discos (dispositivos 2 y 3) han permanecido iguales.

Asimismo, también puede comprobarse que el punto teórico de saturación ha mejorado sensiblemente pasando de 7 trabajos en el sistema original a:

$$N^* = \frac{D+Z}{D_b} = \frac{1{,}4+6}{0{,}6} = \lceil 12{,}33 \rceil = 13 \text{ trabajos}$$

Cabe destacar, por último, que si la mejora del procesador hubiera sido mayor, por ejemplo incorporando uno nuevo 3 veces más rápido que el original, entonces su demanda de servicio hubiera pasado a valer $D_1 = V_1 \times S_1 = 32 \times 0{,}0125 = 0{,}4$ segundos, valor por debajo de la demanda del primer disco (dispositivo 2). En consecuencia, este disco sería el nuevo cuello de botella y sus asíntotas serían las que determinarían el rendimiento del sistema global.

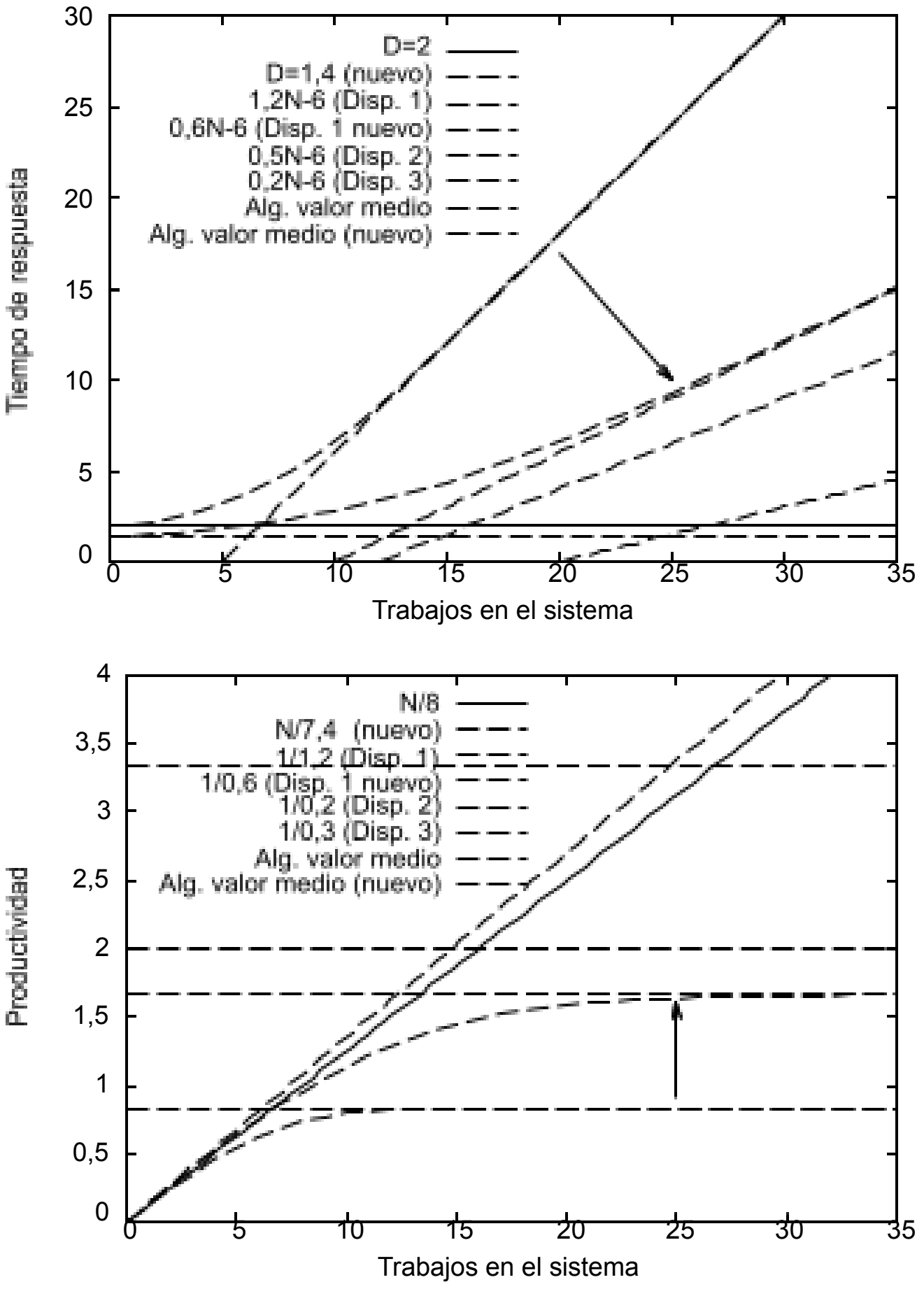

**Figura 5.2:** Nuevos límites asintóticos del tiempo de respuesta y de la productividad.

## 5.4. Problemas resueltos

**PROBLEMA 5.1** Consideremos un modelo de sistema informático interactivo con un procesador y un disco. El sistema tiene 10 usuarios con un tiempo de reflexión de 8 segundos. El procesador tiene un tiempo de servicio de 0,03 segundos y el disco de 0,1 segundos. Las razones de visita del procesador y del disco son 8 y 7, respectivamente. Aplíquese el algoritmo de resolución para redes cerradas para estimar el tiempo medio de respuesta del sistema informático.

**SOLUCIÓN:** El algoritmo del análisis del valor medio se aplicará de forma iterativa desde 1 hasta los 10 trabajos que hay en el sistema. Si el procesador y el disco vienen indicados con los subíndices 1 y 2, respectivamente, el resultado de cada iteración viene dado en la Tabla 5.2. Si observamos los resultados de la última fila, el tiempo medio de respuesta del sistema informático

| Trabajos | $R_1$ | $R_2$ | $R$ | $X_0$ | $N_1$ | $N_2$ |
|---|---|---|---|---|---|---|
| 1 | 0,0300 | 0,1000 | 0,9400 | 0,1119 | 0,0268 | 0,0783 |
| 2 | 0,0308 | 0,1078 | 1,0013 | 0,2222 | 0,0548 | 0,1677 |
| 3 | 0,0316 | 0,1168 | 1,0705 | 0,3307 | 0,0837 | 0,2703 |
| 4 | 0,0325 | 0,1270 | 1,1493 | 0,4372 | 0,1137 | 0,3888 |
| 5 | 0,0334 | 0,1389 | 1,2394 | 0,5412 | 0,1446 | 0,5261 |
| 6 | 0,0343 | 0,1526 | 1,3430 | 0,6422 | 0,1764 | 0,6860 |
| 7 | 0,0353 | 0,1686 | 1,4626 | 0,7398 | 0,2089 | 0,8731 |
| 8 | 0,0363 | 0,1873 | 1,6013 | 0,8332 | 0,2417 | 1,0925 |
| 9 | 0,0373 | 0,2092 | 1,7628 | 0,9219 | 0,2747 | 1,3503 |
| 10 | 0,0382 | 0,2350 | 1,9511 | 1,0049 | 0,3074 | 1,6533 |

**Tabla 5.2:** Aplicación del algoritmo del valor medio

con 10 trabajos es de 1,9511 segundos, y su productividad de 1,0049 trabajos por segundo.

■

**PROBLEMA 5.2** Consideremos un modelo abierto de sistema informático compuesto por un procesador y dos discos distintos. Los tiempos de servicio, expresados en segundos, y las razones de visita de cada dispositivo se indican en la siguiente tabla:

| Dispositivo | Razón de visita | Tiempo de servicio (s) |
|---|---|---|
| Procesador (1) | 16 | 0,01 |
| Disco (2) | 7 | 0,02 |
| Disco (3) | 8 | 0,03 |

Calcúlense las asíntotas optimistas del tiempo de respuesta y la productividad de este sistema. Asimismo, determínense las utilizaciones de los dispositivos del modelo si la tasa de llegadas es $\lambda = 0{,}002$ trabajos/s.

**SOLuCión:** Primeramente calcularemos la demanda de servicio a cada dispositivo:

$$D_1 = V_1 \times S_1 = 16 \times 0{,}01 = 0{,}16 \text{ s}$$
$$D_2 = V_2 \times S_2 = 7 \times 0{,}02 = 0{,}14 \text{ s}$$
$$D_3 = V_3 \times S_3 = 8 \times 0{,}03 = 0{,}24 \text{ s}$$

El valor más pequeño que podemos esperar del tiempo de respuesta, $D$, viene dado por la suma de las demandas de servicio de todos los dispositivos:

$$R_{opt} = D = 0{,}16 + 0{,}14 + 0{,}24 = 0{,}54 \text{ s}$$

El valor máximo de la productividad vendrá dado por la tasa de llegadas que satura el cuello de botella (dispositivo 3). En este caso tendremos:

$$X_{opt} = \frac{1}{D_b} = \frac{1}{D_3} = \frac{1}{0,24} = 4,167 \text{ trabajos/s}$$

Finalmente, si $\lambda = 0,002$ trabajos/s las utilizaciones de los diferentes dispositivos se calculan del siguiente modo:

$$U_1 = \lambda \times D_1 = 0,002 \times 0,16 = 0,32$$
$$U_2 = \lambda \times D_2 = 0,002 \times 0,14 = 0,28$$
$$U_3 = \lambda \times D_3 = 0,002 \times 0,24 = 0,48$$

Nótese que el dispositivo cuello de botella (dispositivo número 3) presenta los valores más elevados tanto de la demanda de servicio como de la utilización. Cualquier acción cuyo objetivo fuese mejorar el rendimiento del sistema habría de afectar a este dispositivo.

---

**PROBLEMA 5.3** Consideremos un modelo cerrado de sistema informático compuesto por un procesador y dos discos. El tiempo de reflexión es de 18 segundos. Las demandas de servicio de cada dispositivo, expresadas en segundos, se indican en la siguiente tabla:

| Dispositivo | Demanda de servicio (s) |
|---|---|
| Procesador (1) | 10 |
| Disco (2) | 12 |
| Disco (3) | 8 |

Calcúlense las asíntotas optimistas del tiempo de respuesta y la productividad de este sistema, así como el punto teórico de saturación. ¿Cuál sería el número máximo de trabajos que permitiría obtener un tiempo de respuesta inferior a 60 segundos?

**SOLUCIÓN:** Si observamos las demandas de servicio podemos concluir que el cuello de botella es el primer disco (dispositivo 2) por tener el mayor valor de este índice. El valor más pequeño que podemos esperar del tiempo de respuesta, $D$, viene dado por la suma de las demandas de servicio de todos los dispositivos:

$$R_{opt} = D = 10 + 12 + 8 = 20 \text{ s}$$

El valor más alto de la productividad vendrá dado por la tasa de llegadas que satura el cuello de botella:

$$X_{opt} = \frac{1}{D_b} = \frac{1}{D_2} = \frac{1}{12} = 0,0833 \text{ trabajos/s}$$

Con estos dos valores podemos expresar los límites asintóticos como:

$$R_{opt} = \max\{D, D_b \times N - Z\} = \max\{20, 12 \times N - 18\}$$

$$X_{opt} = \min\left\{\frac{N}{D+Z}, \frac{1}{D_b}\right\} = \min\left\{\frac{N}{38}, 0{,}0833\right\}$$

El punto teórico de saturación será:

$$N^* = \frac{D+Z}{D_b} = \frac{38}{12} = \lceil 3{,}1667 \rceil = 4 \text{ trabajos}$$

Asintóticamente hablando, a partir de 4 trabajos en el sistema ya no es posible considerar, desde un punto de vista optimista, un tiempo de respuesta inferior a $D = 20$ segundos. Por tanto, a la derecha del punto de saturación el valor optimista del tiempo de respuesta viene dado por la recta $12\ N\ 18$. Para saber hasta cuántos trabajos puede haber en el sistema manteniendo un tiempo de respuesta optimista por debajo de los 60 segundos habremos de resolver la inecuación:

$$12 \times N - 18 < 60 \Rightarrow N < 6{,}5$$

En consecuencia, el sistema habría de tener un máximo de 6 trabajos en el sistema para garantizar un tiempo de respuesta optimista por debajo de los 60 segundos. ■

**PROBLEMA 5.4** Un usuario percibe que el 15 % de sus peticiones a un servidor web son servidas por la cache de su computador, con un tiempo de respuesta medio de 0,5 segundos. En cambio, el tiempo medio de respuesta de las peticiones no encontradas en la cache es de 5 segundos. La instalación de un nuevo navegador aumenta la probabilidad de acierto de la cache local hasta el 50 % de las peticiones. Determínese el tiempo de respuesta percibido por el usuario antes y después de la actualización del navegador. Adicionalmente, calcúlese la mejora en el tiempo de respuesta y compárese al incremento de la probabilidad de acierto de la cache.

**SOLUCIÓN:** El tiempo medio de respuesta $t$ percibido por el usuario se puede expresar en función de la probabilidad de acierto $p_c$ en la memoria cache de su computador. Si $t_c$ denota el tiempo de respuesta de una petición cuando hay acierto en la cache y $t_w$ denota el tiempo de respuesta de una petición que es atendida por el servidor web podemos escribir:

$$t = t_c \times p_c + t_w \times (1 - p_c)$$

En consecuencia, el tiempo de respuesta $t_1$ experimentado usando el navegador original será:

$$t_1 = 0{,}5 \times 0{,}15 + 5 \times (1 - 0{,}15) = 4{,}325 \text{ s}$$

Después de actualizar el navegador el tiempo de respuesta $t_2$ será:

$$t_2 = 0{,}5 \times 0{,}5 + 5 \times (1 - 0{,}5) = 2{,}75 \text{ s}$$

En consecuencia, la mejora en el tiempo de respuesta es de $t_1/t_2 = 4{,}325/2{,}75 = 1{,}57$ veces, mientras que la mejora en la tasa de aciertos es de 0,5/0,15 = 3,33. Como se puede ver, la mejora en el tiempo de respuesta de las peticiones es menor que el incremento en la tasa de acierto. Esto es así porque el tiempo de acceso al servidor web incluye tanto el acierto como el fallo en el acceso a la memoria cache. ■

**PROBLEMA 5.5** Un usuario consulta páginas web de un tamaño medio de 300 KB. El tiempo de transmisión a través de la red de una página de 300 KB es de 4 segundos, el tiempo medio de residencia en el servidor web en el que se encuentra la página es de 3 segundos y el tiempo que el navegador emplea en formatear y mostrar la página en el computador del usuario es de 0,5 segundos. Insatisfecho con este tiempo de respuesta, ha decidido sustituir el procesador de su computador por uno nuevo 1,5 veces más rápido.

Calcúlese el nuevo tiempo de respuesta experimentado por el usuario después de la actualización suponiendo que el tiempo respuesta del navegador es totalmente dependiente de la velocidad del procesador. ¿Cuál es la mejora de rendimiento obtenida?

**SOLuCIóN:** El tiempo de respuesta experimentado por el usuario cuando accede a una página web está compuesto por los tres componentes indicados: acceso al servidor, transferencia por la red y presentación por parte del navegador. Así pues, este tiempo tiene un valor original de:

$$t_1 = 3 + 4 + 0{,}5 = 7{,}5 \text{ s}$$

Tras reemplazar el procesador por uno 1,5 veces más rápido el nuevo tiempo de respuesta que obtendrá es:

$$t_2 = 3 + 4 + \frac{0{,}5}{1{,}5} = 7{,}333 \text{ s}$$

En consecuencia, tras mejorar 1,5 veces la velocidad del procesador, la mejora experimentada por el usuario en el tiempo de respuesta es de $t_1/t_2 = 7{,}5/7{,}333 = 1{,}022$. Nótese que el rendimiento global se ve afectado muy poco porque la mejora se aplica al componente del tiempo de respuesta que menos afecta al tiempo total. ■

**PROBLEMA 5.6** Consideremos un modelo abierto de sistema informático compuesto por un procesador y dos discos idénticos. El tiempo medio entre llegadas de clientes es de 0,6 segundos, los cuales se comportan de acuerdo con el modelo del servidor central. Los tiempos de servicio, expresados en segundos, y las razones de visita de cada dispositivo se indican en la siguiente tabla:

| Dispositivo | Razón de visita | Tiempo de servicio (s) |
|---|---|---|
| Procesador (1) | 17 | 0,03 |
| Disco (2) | 6 | 0,04 |
| Disco (3) | 10 | 0,04 |

Se pide calcular:

1. La tasa de llegadas al sistema.
2. Las demandas de servicio de los dispositivos.
3. El dispositivo cuello de botella.
4. Las probabilidades de encaminamiento a los discos.
5. El tiempo mínimo de respuesta del sistema informático.
6. La productividad de los dispositivos del sistema.
7. El valor máximo de la tasa de llegadas que soporta el sistema.
8. El tiempo de respuesta de cada dispositivo.
9. El tiempo de respuesta del sistema informático.
10. El número de trabajos que hay en el sistema.

**SOLuCıón:**

1. La tasa de llegadas de trabajos al sistema se calcula como la inversa del tiempo medio entre llegadas:

$$\lambda = \frac{1}{0,6} = 1,667 \text{ trabajos/s}$$

2. Las demandas de servicio se determinan a partir de la razón de visita y de los tiempos de servicio:

$$D_1 = V_1 \times S_1 = (V_2 + V_3 + 1) \times S_1$$
$$= 17 \times 0,03 = 0,51 \text{ s}$$
$$D_2 = V_2 \times S_2 = 6 \times 0,04 = 0,24 \text{ s}$$
$$D_3 = V_3 \times S_3 = 10 \times 0,04 = 0,4 \text{ s}$$

3. El dispositivo cuello de botella es aquel que tiene la demanda de servicio más elevada. Por tanto, en este caso el cuello de botella es el procesador del sistema.

4. Las probabilidades de encaminamiento a los discos se calculan a partir de las razones de visita. Si el subíndice 0 representa el exterior del sistema, podemos escribir:

$$p_{1,0} = \frac{V_0}{V_1} = \frac{1}{17} = 0{,}05882$$

$$p_{1,2} = \frac{V_2}{V_1} = \frac{6}{17} = 0{,}35294$$

$$p_{1,3} = \frac{V_3}{V_1} = \frac{10}{17} = 0{,}58824$$

Así pues, se puede comprobar que:

$$p_{1,2} + p_{1,3} + p_{1,0} = 1$$

5. El tiempo de respuesta mínimo, $D$, es la suma de las demandas de servicio de todos los dispositivos representados en el modelo:

$$D = D_1 + D_2 + D_3 = 0{,}51 + 0{,}24 + 0{,}4 = 1{,}15 \text{ s}$$

6. Las productividades de los dispositivos se pueden calcular a partir de la tasa de llegadas al sistema y de las probabilidades de encaminamiento:

$$X_1 = \frac{\lambda}{p_{1,0}} = 28{,}333 \text{ trabajos/s}$$

$$X_2 = X_1 \times p_{1,2} = 10 \text{ trabajos/s}$$

$$X_3 = X_1 \times p_{1,3} = 16{,}666 \text{ trabajos/s}$$

7. El valor máximo de la tasa de llegadas que soporta el sistema viene determinado por la demanda de servicio del cuello de botella:

$$\lambda_{máx} = \frac{1}{D_b} = \frac{1}{0{,}51} = 1{,}961 \text{ trabajos/s}$$

8. El tiempo medio de respuesta de un dispositivo cualquiera $i$ se calcula a partir de la expresión no operacional $R_i = (N_i + 1) \times S_i$. Si se aplica la ley de Little y se sustituye la variable $N_i$ por $X_i R_i$, entonces obtenemos la expresión conocida del tiempo de respuesta:

$$R_i = \frac{S_i}{1 - X_i \times S_i} = \frac{S_i}{1 - U_i}$$

Si aplicamos esta expresión a los distintos dispositivos que conforman el modelo tendre- mos:

$$R_1 = \frac{0,03}{1 - 28,333 \times 0,03} = \frac{0,03}{1 - 0,85} = 0,2 \text{ s}$$

$$R_2 = \frac{0,04}{1 - 10 \times 0,04} = \frac{0,04}{1 - 0,4} = 0,0667 \text{ s}$$

$$R_3 = \frac{0,04}{1 - 16,666 \times 0,04} = \frac{0,04}{1 - 0,667} = 0,12 \text{ s}$$

9. El tiempo medio de respuesta del sistema informático se calcula por medio de la ley general del tiempo de respuesta, que contempla la razón de visita a cada dispositivo y su tiempo medio de respuesta:

$$R = V_1 \times R_1 + V_2 \times R_2 + V_3 \times R_3 = 5 \text{ s}$$

10. Conociendo la tasa de llegadas al sistema informático y su tiempo medio de respuesta, el número medio de trabajos que hay en él se determina fácilmente aplicando la ley de Little a todo el sistema:

$$N = \lambda \times R = 1,667 \times 5 = 8,333 \text{ trabajos}$$

■

**PROBLEMA 5.7** El modelo de un sistema informático interactivo incluye un procesador, un disco y una cinta. Los 10 terminales conectados al sistema se comportan de acuerdo con el modelo de servidor central y tienen un tiempo medio de reflexión de 5 segundos. Los tiempos de servicio, expresados en segundos, y las razones de visita de cada dispositivo se indican en la siguiente tabla:

| Dispositivo | Razón de visita | Tiempo de servicio (s) |
|---|---|---|
| Procesador (1) | 11 | 0,02 |
| Disco (2) | 8 | 0,04 |
| Cinta (3) | 2 | 0,1 |

Basándose en este modelo de sistema informático, se pide determinar:

1. Las demandas de servicio de los dispositivos.
2. El dispositivo cuello de botella.
3. Las probabilidades de encaminamiento al disco y la cinta.

4. El tiempo de respuesta mínimo del sistema informático.
5. La productividad máxima que puede conseguir el sistema.
6. El valor del punto teórico de saturación $N^*$.
7. Las asíntotas optimistas del tiempo de respuesta y de la productividad del sistema, $R_{opt}$ y $X_{opt}$.
8. El tiempo medio de respuesta del sistema informático si se conoce que, por término medio, hay 8,107 trabajos en reflexión.
9. El número de trabajos en el sistema informático que compiten por los recursos del mismo.

**SOLUCIÓN:**

1. La demanda de servicio de un dispositivo se calcula a partir de la razón de visita y el tiempo de servicio:

$$D_1 = V_1 \times S_1 = 11 \times 0{,}02 = 0{,}22 \text{ s}$$
$$D_2 = V_2 \times S_2 = 8 \times 0{,}04 = 0{,}32 \text{ s}$$
$$D_3 = V_3 \times S_3 = 2 \times 0{,}1 = 0{,}2 \text{ s}$$

2. El dispositivo cuello de botella de este sistema es el disco, ya que tiene la demanda de servicio más alta de los tres.

3. Las probabilidades de encaminamiento desde el procesador al disco y a la cinta se calculan a partir de las razones de visita. Si los terminales (trabajos en reflexión) vienen referidos con el subíndice 0, entonces tendremos:

$$p_{1,0} = \frac{V_0}{V_1} = \frac{1}{11} = 0{,}0909$$
$$p_{1,2} = \frac{V_2}{V_1} = \frac{8}{11} = 0{,}7273$$
$$p_{1,3} = \frac{V_3}{V_1} = \frac{2}{11} = 0{,}1818$$

Se puede comprobar, adicionalmente, que la suma de estas tres probabilidades de enca- minamiento es igual a la unidad:

$$p_{1,0} + p_{1,2} + p_{1,3} = 1$$

4. El tiempo de respuesta mínimo, $D$, se calcula como la suma de las demandas de servicio de todos los dispositivos del sistema visitados por los trabajos:

$$D = D_1 + D_2 + D_3 = 0{,}22 + 0{,}32 + 0{,}2 = 0{,}74 \text{ s}$$

5. La productividad máxima del sistema informático está determinada por el dispositivo cuello de botella:

$$\frac{1}{D_b} = \frac{1}{0{,}32} = 3{,}125 \text{ trabajos/s}$$

6. El punto teórico de saturación se determina como:

$$N^* = \frac{D + Z}{D_b} = \frac{0{,}74 + 5}{0{,}32} = \lceil 17{,}94 \rceil = 18 \text{ trabajos}$$

   Dado que únicamente hay 10 trabajos en el sistema podemos concluir que el nivel de carga del sistema es relativamente bajo, ya que su valor se encuentra bastante alejado de este punto.

7. Las asíntotas optimista para el tiempo de respuesta y la productividad del sistema, en función del número $N$ de trabajos, son las siguientes:

$$R_{opt} = \max\{D, D_b \times N - Z\} = \max\{0{,}74,\ 0{,}32 \times N - 5\}$$

$$X_{opt} = \min\left\{\frac{N}{D + Z}, \frac{1}{D_b}\right\} = \min\left\{3{,}125, \frac{N}{5{,}74}\right\}$$

8. Si hay $N_Z = 8{,}107$ trabajos por término medio en reflexión, entonces podemos calcular la productividad del sistema aplicando la ley de Little a los terminales:

$$X = \frac{N_Z}{Z} = \frac{8{,}107}{5} = 1{,}6214 \text{ trabajos/s}$$

   Ahora podemos aplicar la ley de Little al modelo completo del sistema y calcular así su tiempo medio de respuesta:

$$R = \frac{N}{X} - Z = \frac{10}{1{,}6214} - 5 = 1{,}1675 \text{ s}$$

9. El número de trabajos que compiten por los recursos del sistema se calculará restando del número total aquellos que están en reflexión:

$$N - N_Z = 10 - 8{,}107 = 1{,}893 \text{ trabajos}$$

■

**PROBLEMA 5.8** El modelo cerrado de un sistema informático consta de dos dispositivos, procesador y cinta, cuyos tiempos de servicio (expresados en segundos) y razones de visita se reflejan en la siguiente tabla:

| Dispositivo | Razón de visita | Tiempo de servicio (s) |
|---|---|---|
| Procesador (1) | 8 | 0,1 |
| Cinta (2) | 7 | 0,2 |

Si el tiempo medio de reflexión es de 2 segundos y hay un total de 5 trabajos en el modelo, se pide calcular las asíntotas optimistas del tiempo de respuesta y productividad del sistema, $R_{opt}$ y $X_{opt}$, así como el punto teórico de saturación.

**SOLUCIÓN:** Para conocer el cuello de botella hay que determinar el dispositivo con mayor demanda de servicio. Si calculamos estas variables para los dos que hay en el sistema tendremos:

$$D_1 = V_1 \times S_1 = 8 \times 0{,}1 = 0{,}8 \text{ s}$$
$$D_2 = V_2 \times S_2 = 7 \times 0{,}2 = 1{,}4 \text{ s}$$

Por tanto, el cuello de botella del sistema es la cinta. Asimismo, el tiempo mínimo de respuesta del sistema, $D$, será:

$$D = D_1 + D_2 = 0{,}8 + 1{,}4 = 2{,}2 \text{ s}$$

La productividad máxima se calcula como el inverso de la demanda de servicio del dispositivo cuello de botella:

$$\frac{1}{D_b} = \frac{1}{1{,}4} = 0{,}7143 \text{ trabajos/s}$$

Las asíntotas optimistas del sistema son:

$$R_{opt} = \max\{D, D_b \times N - Z\} = \max\{2{,}2,\ 1{,}4 \times N - 2\}$$
$$X_{opt} = \min\left\{\frac{N}{D+Z}, \frac{1}{D_b}\right\} = \min\left\{0{,}7143, \frac{N}{4{,}2}\right\}$$

Finalmente, el punto teórico de saturación viene determinado por la expresión:

$$N^* = \frac{D+Z}{D_b} = \frac{2{,}2+2}{1{,}4} = \lceil 3 \rceil = 3 \text{ trabajos}$$

■

**PROBLEMA 5.9** A continuación se muestran los tiempos de servicio y razones de visita del procesador y disco de un sistema informático dedicado a servir las páginas web de una empresa:

| Dispositivo | Razón de visita | Tiempo de servicio (s) |
|---|---|---|
| Procesador (1) | 11 | 0,25 |
| Disco (2) | 10 | 0,08 |

El tiempo medio entre llegadas de peticiones al sistema es de 3 segundos. Los administradores del sistema han recibido numerosas quejas de los usuarios y han decidido situar el tiempo medio de respuesta del servidor web por debajo de los 15 segundos. Para conseguir este objetivo disponen de tres opciones distintas, las cuales se indican a continuación:

1. Reemplazar el procesador por otro el doble de rápido que el actual.
2. Reemplazar el disco por uno nuevo que reduce en un tercio el tiempo de servicio del disco original.
3. Optimizar y recompilar el código que ejecuta el servidor web, con lo que se consi- gue rebajar el tiempo de servicio del procesador de los 0,25 segundos originales a 0,20 segundos.

Se pide determinar cuáles de las opciones anteriores permite conseguir el objetivo mar- cado por los administradores del servidor web.

**SOLuCión:** Antes de analizar las propuestas de mejora analizaremos el sistema informático original. En primer lugar calculamos las demandas de servicio para localizar el cuello de botella:

$$D_1 = V_1 \times S_1 = 11 \times 0{,}25 = 2{,}75 \text{ s}$$
$$D_2 = V_2 \times S_2 = 10 \times 0{,}08 = 0{,}8 \text{ s}$$

El cuello de botella es el procesador, el cual presenta, además, un valor muy elevado de la demanda de servicio si se compara con la del disco. En consecuencia, únicamente resultarán eficaces aquellas propuestas de mejora que afecten directamente al procesador.

Para calcular el tiempo medio de respuesta que experimentan los usuarios del servidor web aplicamos el algoritmo de resolución de redes abiertas. Primero calcularemos las productividades de los dispositivos. Si el exterior del sistema viene indicado por el subíndice 0, tendremos:

$$X_1 = \frac{\lambda}{p_{1,0}} = \frac{1/3}{1/11} = 3{,}667 \text{ trabajos/s}$$

$$X_2 = X_1 \times p_{1,2} = 3{,}667 \times \frac{10}{11} = 3{,}333 \text{ trabajos/s}$$

Ahora podemos calcular el tiempo medio de respuesta de cada dispositivo:

$$R_1 = \frac{0,25}{1 \quad 3,667 \quad 0,25} = \frac{0,25}{1 \quad 0,917} = 3,012 \text{ s}$$

0,08 0,04

$$R_2 = \frac{}{1 - 3,333 \times 0,08} = \frac{}{1 - 0,267} = 0,109 \text{ s}$$

El tiempo medio de respuesta del servidor web será, aplicando la ley general del tiempo de respuesta:

$$R = V_1 \times R_1 + V_2 \times R_2 = 34,22 \text{ s}$$

Por lo tanto, cada petición enviada por un usuario tarda, por término medio, poco más de medio minuto. A continuación se analiza, por separado, cada una de las propuestas de mejora para rebajar este tiempo por debajo de los 15 segundos.

1. Si se sustituye el procesador por otro doble de rápido, la demanda de servicio de este dispositivo se verá reducida a la mitad de la original:

   $$D_1 = \frac{2,75}{2} = 1,375 \text{ s}$$

   Nótese que la razón de visita al procesador no cambia por esta sustitución. Si comparamos este valor con la demanda de servicio del disco, podemos concluir que el procesador sigue siendo el cuello de botella. Si aplicamos de nuevo el algoritmo de resolución de redes abier- tas obtenemos un tiempo medio de respuesta del sistema informático de 3,63 segundos, y por tanto, se consigue el objetivo de rebajar este índice por debajo de los 15 segundos.

2. Con esta actualización el tiempo de servicio del nuevo disco se calcula como sigue:

   $$S_2 = \frac{2}{3} \times 0,08 = 0,0533 \text{ s}$$

   Dado que las razones de visita se mantienen constantes, el cuello de botella del sistema seguirá estando localizado en el procesador y, además, las prestaciones del sistema no se verán afectadas significativamente. De hecho, si se calcula el tiempo de respuesta del sistema obtenemos un valor de 33,65 segundos, esto es, sólo se ha conseguido rebajar el tiempo medio de servicio del servidor web en aproximadamente medio segundo. En resumen, esta terapia no mejora de forma significativa el rendimiento del sistema.

3. Con la optimización y recompilación del código de la aplicación se consigue disminuir el tiempo de servicio del procesador y, por ende, su demanda de servicio:

$$D_1 = V_1 \times S_1 = 11 \times 0{,}2 = 2{,}2 \text{ s}$$

Si se compara este valor con la demanda del disco, podemos observar que el procesador continúa siendo el cuello de botella. Sin embargo, su demanda de servicio ha mejorado un 25 %. El nuevo tiempo de respuesta de cada petición al servidor web, aplicando el algoritmo de resolución de redes abiertas, es de 9,34 segundos, el cual se sitúa bastante por debajo de los 15 segundos.

En resumen, podemos concluir que las dos terapias que actúan sobre el cuello de botella han sido las más efectivas en la reducción del tiempo de respuesta del sistema. Este tiempo de respuesta mejora 34,09/3,63 = 9,34 veces si se sustituye el procesador por otro el doble de rápido, mientras que si se optimiza y recompila la aplicación informática la mejora obtenida es 34,09/9,34 = 3,65. ■

---

**PROBLEMA 5.10** Consideremos un sistema informático interactivo con 15 usuarios y un tiempo de reflexión de 6 segundos. El sistema consta de un procesador, un disco magnético y un CD-ROM. Los tiempos de servicio y las razones de visita se muestran en la siguiente tabla:

| Dispositivo | Razón de visita | Tiempo de servicio (s) |
|---|---|---|
| Procesador (1) | 13 | 0,01 |
| Disco (2) | 10 | 0,3 |
| CD-ROM(3) | 2 | 0,1 |

El administrador pretende rebajar el tiempo medio de respuesta del sistema informático por debajo de los 5 segundos. Discútase cuál de las siguientes alternativas permite conseguir este objetivo:

1. Equilibrar las demandas de servicio de los dispositivos de almacenamiento.
2. Sustituir el disco magnético por dos que sean doble de rápidos, repartiendo entre los dos las 10 visitas que cada trabajo hace al disco original.
3. Reemplazar el disco magnético por uno nuevo con un tiempo de servicio de 0,07 segundos.

**SOLUCIÓN:** En primer lugar analizaremos el modelo del sistema informático original. Para determinar el cuello de botella del sistema calcularemos las razones de visita a los dispositivos del modelo:

$$D_1 = V_1 \times S_1 = 13 \times 0{,}01 = 0{,}13 \text{ s}$$
$$D_2 = V_2 \times S_2 = 10 \times 0{,}3 = 3 \text{ s}$$
$$D_3 = V_3 \times S_3 = 2 \times 0{,}1 = 0{,}2 \text{ s}$$

Según estos valores, el cuello de botella del sistema es el disco magnético. Si se resuelve el modelo aplicando el algoritmo del valor medio, el tiempo de respuesta del sistema informático es de 39 segundos, valor muy por encima de los 5 segundos que pretende conseguir el administrador. A continuación veremos el efecto de las diferentes terapias propuestas sobre este índice de rendimiento.

1. El equilibrado de las demandas de servicio de los dispositivos de almacenamiento requiere un ajuste de sus razones de visita. En concreto, se habrán de calcular las nuevas razones de visita $V_2'$ y $V_3'$ que satisfacen el siguiente sistema de ecuaciones:

$$V_2' + V_3' = 12$$
$$V_2' \times S_2 = V_3' \times S_3$$

   Los valores que resuelven el sistema anterior son $V_2' = 3$ y $V_3' = 9$. En este caso el cuello de botella viene determinado por los dos dispositivos de almacenamiento, ya que tienen una nueva demanda de 0,9 segundos, valor más alto que la demanda del procesador. El tiempo de respuesta del sistema informático calculado mediante el algoritmo de valor medio es de 9,1496 segundos. Por tanto, la terapia aplicada ha tenido un efecto significativo sobre el sistema, produciendo una mejora del tiempo de respuesta de 39/9,1496 = 4,26. Sin embargo, el objetivo de rebajar este tiempo por debajo de 5 segundos no se ha conseguido.

2. La sustitución del disco original por dos el doble de rápidos permite obtener una demanda de servicio de cada uno de los discos muy inferior a la del disco original. Esto se debe a que la razón de visita a cada uno de ellos pasa de 10 a 5, y su tiempo de servicio de 0,3 a 0,15. En concreto, la nueva demanda de estos discos es de $5 \times 0,15 = 0,75$ segundos.

   Dado que este valor de la demanda es el más alto, los dos discos nuevos representan ahora el cuello de botella del sistema. La resolución del modelo mediante el algoritmo del valor medio da un tiempo de respuesta del sistema de 6,9646 segundos, y por tanto, la mejora de este índice es de 39/6,9646 = 5,60. Aunque la mejora es superior a la conseguida por la primera terapia, tampoco se consigue rebajar, aunque por poco, el tiempo de respuesta por debajo de los 5 segundos pretendidos.

3. El disco nuevo que se utiliza en el reemplazo tiene una demanda de servicio de $10 \times 0,07 = 0,7$ segundos. Aunque este valor indica que el disco sigue siendo el cuello de botella, la mejora de este dispositivo es considerable. En concreto, la resolución del modelo mediante el algoritmo del valor medio permite estimar el tiempo medio de respuesta del sistema en 4,7497 segundos, lo que permite, por muy poco, alcanzar el objetivo deseado. La mejora conseguida en este índice de prestaciones es de 39/4,7497 = 8,21 respecto del sistema original. ■

**PROBLEMA 5.11** Un equipo de informáticos está diseñando un servidor web para una gran empresa dedicada a la venta por catálogo. Se sabe que este servidor deberá atender una media de dos peticiones de página web por segundo. El tiempo medio de servicio del procesador es de 0,02 segundos, su razón de visita de 13 y cada petición al servidor web realizará, por término medio, 12 visitas al sistema de almacenamiento. Los diseñadores del sistema están discutiendo la viabilidad de dos posibles configuraciones para este sistema de almacenamiento:

- Configuración A: un único disco magnético con un tiempo medio de servicio de 0,035 segundos.
- Configuración B: dos discos magnéticos con un tiempo medio de servicio de 0,07 segundos cada uno de ellos (ambos reciben la misma carga).

Calcúlese, para cada configuración, el tiempo mínimo de respuesta de una petición así como la productividad máxima que puede alcanzar el servidor web. Cuantifíquese la mejora de la mejor opción respecto del tiempo medio de respuesta del servidor web.

**SOLuCIón:** El tiempo mínimo de respuesta de un sistema $D$ se calcula sumando las deman- das de servicio de los dispositivos visitados por los trabajos. De manera particular para cada configuración tenemos:

$$D_A = 13 \times 0{,}02 + 12 \times 0{,}035 = 0{,}68 \text{ s}$$
$$D_B = 13 \times 0{,}02 + 6 \times 0{,}07 + 6 \times 0{,}07 = 1{,}1 \text{ s}$$

Nótese que, para la configuración B, las visitas al sistema de almacenamiento se han dis- tribuido de forma equitativa entre los dos discos.

La productividad máxima alcanzada por el sistema con cada configuración depende de la situación del dispositivo cuello de botella. En ambas configuraciones la demanda de servicio del procesador es la misma: $13 \times 0{,}02 = 0{,}26$ segundos. En la configuración A, la demanda de servicio del disco es de $12 \times 0{,}035 = 0{,}42$ segundos; por su parte, en la configuración B, la demanda de servicio de cada disco es la misma: $6 \times 0{,}070 = 0{,}42$ segundos. En consecuencia, el cuello de botella en ambas configuraciones se sitúa en los discos del sistema de almacenamiento, y las dos alcanzan la misma productividad máxima:

$$\frac{1}{D_b} = \frac{1}{0{,}42} = 2{,}38 \text{ peticiones/s}$$

Según se aprecia, aunque ambas configuraciones permiten alcanzar la misma productividad máxima, la configuración A es $1{,}1/0{,}68 = 1{,}62$ veces más rápida considerando únicamente el tiempo mínimo de respuesta, esto es, en un contexto de baja carga.

El tiempo medio de respuesta del servidor web se calcula aplicando el algoritmo de reso- lución para redes abiertas. En particular, el tiempo de respuesta del sistema cuando se usa

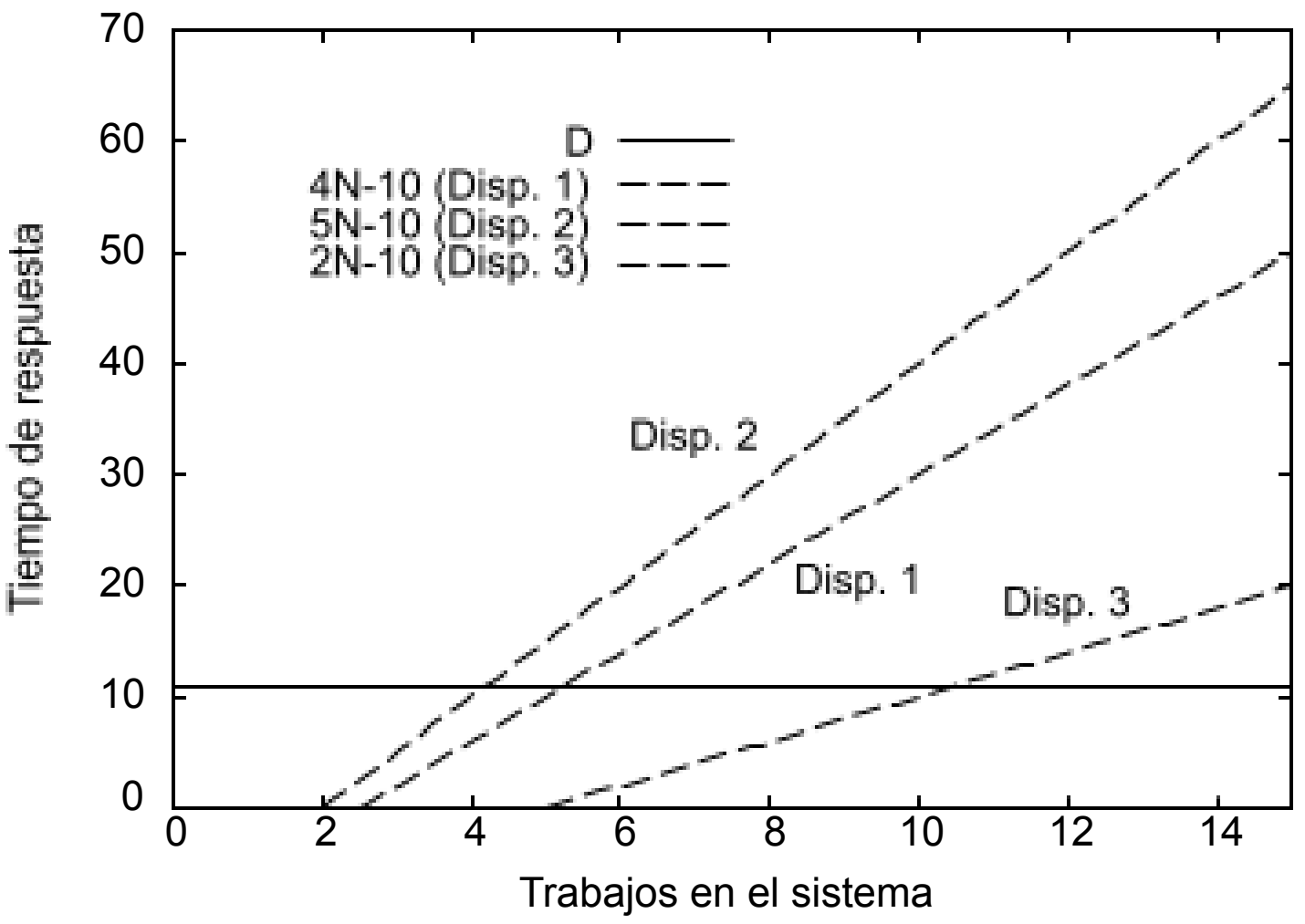

**Figura 5.3:** Asíntotas del tiempo de respuesta.

la configuración A se puede calcular utilizando la expresión de la ley general del tiempo de respuesta:

$$13 \times \frac{0{,}02}{1-2\times 13\times 0{,}02} + 12 \times \frac{0{,}035}{1-2\times 12\times 0{,}035} = 3{,}167 \text{ s}$$

De manera similar, el tiempo de respuesta del sistema cuando el subsistema de discos se configura con la opción B es:

$$13 \times \frac{0{,}02}{1-2\times 13\times 0{,}02} + 2\times 6 \times \frac{0{,}07}{1-2\times 6\times 0{,}07} = 5{,}792 \text{ s}$$

En consecuencia, si se emplea el tiempo medio de respuesta como índice de rendimiento, la configuración A resulta 5,792/3,167 = 1,83 veces más rápida que la B. ■

---

**PROBLEMA 5.12** Las asíntotas optimistas del tiempo de respuesta, expresado en segun- dos, de un sistema interactivo con tres dispositivos se muestran en la Figura 5.3. ¿Cuál es el cuello de botella de este sistema? ¿Por qué? Indíquese alguna acción de mejora sobre el sistema para disminuir el tiempo de respuesta y justifíquese numéricamente.

**SOLuCIón:** La recta asociada a un dispositivo cualquiera *i* tiene la forma $D_i \times N - Z$, donde $D_i$ es el valor de la demanda de servicio. El cuello de botella del sistema es el dispositivo número 2, ya que tiene la demanda más alta y, por tanto, la pendiente más elevada, con un valor de 5 segundos. El tiempo mínimo de respuesta *D* es de 11 segundos, que corresponde a la suma de las demandas de servicio de los tres dispositivos.

La mejora del tiempo de respuesta se podría llevar a cabo, por ejemplo, incorporando otro dispositivo similar al número 2, y equilibrando sus demandas de servicio, que pasarían a valer 2,5 segundos para cada uno de los dos. Con esta acción el cuello de botella pasaría a ser el dispositivo número 1 con una demanda de servicio de 4 segundos. Cualquier acción posterior que quisiera reducir el tiempo de respuesta habría de hacerse, necesariamente, sobre este último dispositivo. ■

---

**PROBLEMA 5.13** El administrador de un sitio web ha monitorizado la actividad del sistema para calcular la demanda de servicio y la razón de visita de los dispositivos que más afectan al rendimiento. En particular, la información obtenida se muestra a continuación:

| Dispositivo | Demanda de servicio (s) | Razón de visita |
|---|---|---|
| Procesador (1) | 2 | 10 |
| Disco (2) | 5 | 12 |
| CD-ROM (3) | 6 | 2 |

El sitio web recibe una media de 0,1 peticiones por segundo durante las horas del día de máxima actividad. Determínese:

1. La tasa máxima de llegadas capaz de soportar el sitio web.
2. El tiempo mínimo de respuesta de una petición.
3. El tiempo medio de respuesta de una petición.
4. Una acción sobre el sistema (sintonización, actualización, o ambas) que permita me- jorar, como mínimo, 1,3 veces el tiempo medio de respuesta.

**SOLuCIón:**

1. La máxima tasa de llegadas capaz de soportar el sitio web viene dada por el inverso de la demanda más alta, en particular, $1/6 = 0{,}1667$ peticiones/s.
2. El tiempo mínimo de respuesta $D$ que experimenta una petición se calcula sumando las demandas de servicio que efectúa una petición sobre los recursos del sistema. En este caso se tendrá:

$$D = 2 + 5 + 6 = 13 \text{ s}$$

3. Para poder estimar el valor medio del tiempo de respuesta es necesario aplicar el algoritmo de resolución de redes abiertas, el cual se basa en el cálculo del tiempo de respuesta de una estación, $R_i$, mediante la expresión:

$$R_i = \frac{S_i}{1 - X_i \times S_i} = \frac{S_i}{1 - U_i} = \frac{S_i}{1 - \lambda \times D_i}$$

Previamente debemos calcular los tiempos medios de servicio de cada dispositivo $S_i$ a partir de las demandas de servicio y las razones de visita mediante la expresión $S_i = D_i/V_i$. En particular, se tendrá: $S_1 = 2/10 = 0{,}2$, $S_2 = 5/12 = 0{,}417$ y $S_3 = 6/2 = 3$ segundos.

Teniendo en cuenta que el sistema recibe una tasa de llegadas de $\lambda = 0{,}1$ peticiones/s, el tiempo de respuesta de cada dispositivo será:

$$R_1 = \frac{S_1}{1 - \lambda \times D_1} = \frac{0{,}2}{1 - 0{,}1 \times 2} = 0{,}25 \text{ s}$$

$$R_2 = \frac{S_2}{1 - \lambda \times D_2} = \frac{0{,}417}{1 - 0{,}1 \times 5} = 0{,}834 \text{ s}$$

$$R_3 = \frac{S_3}{1 - \lambda \times D_3} = \frac{3}{1 - 0{,}1 \times 6} = 7{,}5 \text{ s}$$

Con los valores anteriores calculados ya podemos aplicar la ley general del tiempo de respuesta:

$$R = V_1 \times R_1 + V_2 \times R_2 + V_3 \times R_3$$
$$= 10 \times 0{,}25 + 12 \times 0{,}834 + 2 \times 7{,}5 = 27{,}508 \text{ s}$$

4. Como hemos visto en el apartado anterior, el tiempo medio de respuesta que experimenta una petición en las horas de máxima actividad es de 27,508 segundos. Este valor está muy afectado por el tiempo de respuesta del CD-ROM, esto es, $R_3$, dispositivo que representa el cuello de botella del sistema, ya que tiene la demanda de servicio más alta ($D_3 = 6$ s). En este sentido, cualquier acción sobre el sistema que persiga disminuir el tiempo medio de respuesta debe incidir directamente sobre el funcionamiento de este dispositivo.

   Imaginemos que podemos llevar a cabo una acción tal que el número de accesos a la unidad de CD-ROM se reduzca a la mitad, es decir, de 2 visitas se pasa a 1. Esto po- dría conseguirse, por ejemplo, ampliando adecuadamente la memoria cache del sistema operativo para las operaciones de entrada/salida involucradas con este dispositivo. Conse- cuentemente, en esta nueva situación la nueva demanda de servicio del CD-ROM pasará a valer la mitad, esto es, $D_3 = 6/2 = 3$ segundos. Su tiempo medio de respuesta $R_3$ será:

$$S_3$$

$$R_3 = \frac{3}{1 - \lambda \times D_3} = \frac{3}{1 - 0,1 \times 3} = 4,286 \text{ s}$$

Finalmente, el nuevo tiempo medio de respuesta del sistema tras la modificación del sistema será:

$$\begin{aligned} R &= V_1 \times R_1 + V_2 \times R_2 + V_3 \times R_3 \\ &= 10 \times 0{,}25 + 12 \times 0{,}384 + 1 \times 4{,}29 = 16{,}798 \text{ s} \end{aligned}$$

En consecuencia, la mejora del tiempo medio de respuesta obtenida tras la modificación introducida respecto del sistema original será de 27,508/16,798 = 1,64, valor superior a la mejora de 1,3 planteada en el enunciado del problema.

Nótese que, una vez llevada a cabo la sintonización del sistema, el nuevo cuello de botella es el disco, dado que su demanda de servicio se convierte en la más alta. Ulteriores acciones sobre el sistema que pretendan mejorar el rendimiento deberán ir encaminadas a modificar el comportamiento de este dispositivo. ■

## 5.5. Problemas con solución

**PROBLEMA 5.14** Un usuario consulta páginas web de un tamaño medio de 200 KB. El tiempo de transmisión a través de la red de una página de 200 KB es de 5 segundos, el tiempo medio de residencia en el servidor web en el que se encuentra la página es de 4 segundos y la productividad del navegador en formatear y mostrar la página en el computador del usuario es de 25.600 bytes/segundo. Insatisfecho con este tiempo de respuesta, ha decidido sustituir el procesador de su computador por uno nuevo 2 veces más rápido.

Calcúlese el nuevo tiempo de respuesta experimentado por el usuario después de la actualización suponiendo que el tiempo de respuesta del navegador es totalmente dependiente de la velocidad del procesador. ¿Cuál es la mejora de rendimiento obtenida?

**SOLUCIÓN:** El tiempo de respuesta obtenido tras la actualización es de 13 segundos, y la mejora de 1,3. ■

**PROBLEMA 5.15** Un sistema interactivo está compuesto por un procesador y dos discos. Las razones de visita son 25, 15 y 9, mientras que los tiempos de servicio son 0,2, 0,8 y 0,6 segundos, respectivamente. El tiempo medio de reflexión es de 18 segundos. Determínese:

1. El dispositivo cuello de botella.
2. Las asíntotas $R_{opt}$ y $X_{opt}$.
3. El punto teórico de saturación $N^*$,

4. El número de trabajos que, como máximo, admite el sistema si se ha de respetar un tiempo de respuesta máximo de 100 segundos.

**SOLuCıón:**

1. El cuello de botella es el primer disco.
2. $R_{opt} = \max\{22{,}4,\ 12 \times N - 18\}$ y $X_{opt} = \min\{N/40{,}4,\ 0{,}08\}$.
3. Cuatro trabajos.
4. Nueve trabajos. ■

---

**PROBLEMA 5.16** El modelo de un sistema transaccional consta de un procesador, un disco magnético y un CD-ROM. Sus tiempos de servicio son 20, 30 y 10 milisegundos, y las razones de visita 16, 7 y 8, respectivamente. Determínese:

1. El dispositivo cuello de botella.
2. El tiempo mínimo de respuesta del sistema.
3. Valor mínimo de la tasa de llegadas que satura el sistema.
4. Valor máximo que puede alcanzar la utilización del disco magnético.

**SOLuCıón:**

1. El cuello de botella es el procesador.
2. El tiempo mínimo de respuesta es 0,61 segundos.
3. La tasa mínima que satura el sistema es 3,125 trabajos por segundo.
4. La utilización máxima del disco es 0,6563. ■

---

**PROBLEMA 5.17** El modelo de colas de un sistema informático incluye un procesador, un disco magnético y una unidad de cinta. Las razones de visita y los tiempos de servicio se detallan en la siguiente tabla:

| Dispositivo | Razón de visita | Tiempo de servicio (s) |
|---|---|---|
| Procesador (1) | 18 | 0,005 |
| Disco (2) | 15 | 0,02 |
| Cinta (3) | 2 | 0,1 |

El sistema está sometido a una carga transaccional con una tasa media de llegadas de tres trabajos por segundo, y se sabe que la productividad del procesador es de 54 trabajos por segundo. Los trabajos se comportan de acuerdo con el modelo del servidor central. Se pide calcular:

1. El dispositivo cuello de botella.
2. Las probabilidades de encaminamiento a los dispositivos de almacenamiento.
3. Tiempo de respuesta mínimo del sistema.
4. Productividad máxima del sistema.
5. Tiempo medio de respuesta del sistema.
6. Número medio de trabajos en el sistema.

**SOLuCıón:**

1. El cuello de botella es el disco magnético.
2. Las probabilidades de encaminamiento son $p_{1,2} = 0{,}833$, $p_{1,3} = 0{,}111$.
3. El tiempo de respuesta mínimo es 0,59 segundos.
4. La productividad máxima alcanzable es 3,33 trabajos por segundo.
5. El tiempo de respuesta del sistema es 3,62 segundos.
6. En el sistema hay 10,87 trabajos. ■

---

**PROBLEMA 5.18** Consideremos el sistema informático del problema anterior, pero ahora sometido a una carga interactiva. En este caso hay 25 usuarios en el sistema, y su tiempo medio de reflexión es de 5 segundos. Mediante un monitor de actividad se ha podido comprobar que la productividad del sistema es de 3,2072 trabajos por segundo. Se pide calcular:

1. El tiempo de respuesta del sistema informático.
2. El punto teórico de saturación $N^*$.
3. Las asíntotas optimistas del tiempo de respuesta y productividad.
4. El nuevo punto teórico de saturación si se equilibran las demandas a los dos disposi- tivos de almacenamiento.

**SOLuCIón:**

1. El tiempo de respuesta es 2,795 segundos.
2. El valor de $N^*$ es 19 trabajos.
3. Las asíntotas son $R$opt = máx{0,59, 0,3 × $N$ – 5} y $X$opt = mín{$N$/5,59, 3,33}
4. El nuevo valor de $N^*$ es 20 trabajos. ■

## 5.6. Problemas sin resolver

**PROBLEMA 5.19** Una empresa dedicada al comercio por Internet dispone de un servidor web compuesto por un procesador y tres unidades de CD-ROM idénticas. Los tiempos de servicio de estos dispositivos son 50 y 200 milisegundos, para cada tipo, respectivamente. Cada transacción que sirve el sistema provoca 10 visitas a cada disco y un total de 31 visitas al procesador. El sistema recibe una media de 0,3 transacciones por segundo. Se pide calcular:

1. Dispositivo cuello de botella.
2. Tiempo mínimo de respuesta de una transacción.
3. Tasa máxima de llegadas que soporta el sistema.
4. Utilización de cada dispositivo.
5. Tiempo medio de respuesta del sistema.
6. Número medio de transacciones que compiten por los recursos.
7. Tiempo medio de respuesta del sistema si las tres unidades de CD-ROM se sustituyen por una única unidad con un tiempo de servicio de 100 milisegundos y se mantiene el mismo número de visitas al subsistema de almacenamiento.

**PROBLEMA 5.20** Considemos el modelo de un sistema informático que consta de un procesador y dos unidades de disco magnético idénticas. Los trabajos del sistema siguen el comportamiento del servidor central. Las razones de visita y los tiempos de servicio se detallan en la siguiente tabla:

| Dispositivo | Razón de visita | Tiempo de servicio (s) |
|---|---|---|
| Procesador (1) | 25 | 0,01 |
| Disco (2) | 12 | 0,04 |
| Disco (3) | 12 | 0,04 |

1. Supóngase que el sistema es sometido a una carga transaccional, con una tasa de llegadas de dos trabajos por segundo. En este supuesto se pide calcular:

   a) El tiempo de respuesta del sistema informático.
   b) ¿Cuál es la productividad máxima que soporta este sistema?
   c) El nuevo tiempo de respuesta del sistema si un disco soporta el doble de accesos que el otro.
   d) ¿Qué pasaría si se reemplazasen los dos discos por uno nuevo con un tiempo de servicio de 0,03 segundos?

2. Supóngase ahora que el sistema soporta una carga interactiva con 20 usuarios y un tiempo medio de reflexión de 3 segundos. En este supuesto se pide calcular:

   a) El tiempo de respuesta del sistema informático.
   b) ¿Cuál es la productividad máxima que soporta este sistema?
   c) Las asíntotas optimistas del tiempo de respuesta y de la productividad.
   d) ¿Qué pasaría si se reemplazasen los dos discos por uno nuevo con un tiempo de servicio de 0,03 segundos? Compárese este resultado con el caso transaccional y justifíquese la diferencia obtenida.

## 5.7. Actividades propuestas

**ACTIVIDAD 5.1** Diséñese un programa en un lenguaje de alto nivel denominado solred que implemente los algoritmos de resolución de redes de colas estudiados en este capítulo. El programa se ejecutará desde la línea de órdenes y admitirá los siguientes parámetros de entrada:

- Tipo de red (0: abierta, 1: cerrada).
- Tasa de llegadas si se trata de un modelo abierto o número de trabajos y tiempo de reflexión si es cerrado.
- Número de dispositivos a considerar en el modelo.
- Razón de visita y tiempo de servicio de cada dispositivo.

La llamada a este programa para los dos ejemplos de resolución mostrados en la introducción teórica de este capítulo será:

- Modelo abierto: solred 0 2 2 6 0.01 7 0.02

- Modelo cerrado: solred 1 3 5 2 15 0.03 14 0.05

La salida del programa aplicada al ejemplo resuelto en la presentación del algoritmo de resolución de modelos abiertos será la siguiente:

```
      ***************************************************************
    *   Vi    *   Si    *   Di    *   Ui    *   Ni    *   Ri    *   Xi    *
    ******************************************************************

*    *         *         *         *         *         *
* 1  *  6.0000*   0.0100*   0.0600*   0.1200*   0.1364*
*    *         *         *         *         *         *
* 2  *  7.0000*   0.0200*   0.1400*   0.2800*   0.3889*
                                                              *         *
                                                        0.0114* 12.0000*
                                                              *         *
                                                        0.0278* 14.0000*
*    *         *         *         *         *         *         *         *
    ******************************************************************

***********************************************
*  TRABAJOS EN EL SISTEMA                *    0.5253*
*                                        *          *
*  TIEMPO DE RESPUESTA                   *    0.2626*
*  TIEMPO DE RESPUESTA MÍNIMO (D)        *    0.2000*
*                                        *          *
*  PRODUCTIVIDAD                         *    2.0000*
*  PRODUCTIVIDAD MÁXIMA                  *    7.1429*
***********************************************

************************************
*            LÍMITES ASINTÓTICOS          *
************************************
* Ropt = 0.2000                           *
* Xopt = 7.1429                           *
************************************
```

Por su parte, la salida del programa cuando éste se aplica al caso de estudio descrito en la explicación del algoritmo del valor medio tendrá el siguiente aspecto:

```
      ***************************************************************
    *   Vi    *   Si    *   Di    *   Ui    *   Ni    *   Ri    *   Xi    *
*****************************************************************
*    *         *         *         *         *         *         *         *
* 1   * 15.0000*   0.0300*   0.4500*   0.2111*   0.2437*   0.0346*   7.0382*
*    *         *         *         *         *         *         *         *
* 2   * 14.0000*   0.0500*   0.7000*   0.3284*   0.4102*   0.0624*   6.5690*
*    *         *         *         *         *         *         *         *
*****************************************************************
```

```
********************************************
*  TRABAJOS EN EL SISTEMA                *          3*
*  TRABAJOS EN LOS DISPOSITIVOS          *     0.6539*
*  TRABAJOS EN REFLEXIÓN                 *     2.3461*
*  PUNTO DE SATURACIÓN (N*)              *          9*
*                                        *           *
*  TIEMPO DE RESPUESTA                   *     1.3937*
*  TIEMPO DE RESPUESTA MÍNIMO (D)        *     1.1500*
*                                        *           *
*  PRODUCTIVIDAD                         *     0.4692*
*  PRODUCTIVIDAD MÁXIMA                  *     1.4286*
********************************************

***********************************
*          LÍMITES ASINTÓTICOS           *
***********************************
* Ropt = máx { 1.15,        0.70*N- 5.00} *
* Xopt = mín {N/ 6.15,        1.43}       *
***********************************
```

**ACTIVIDAD 5.2** Respecto al programa implementado en la actividad anterior, inclúyase la posibilidad de que se muestren los resultados para diferentes valores de la tasa de entrada (modelos abiertos) o del número de trabajos en el sistema (modelos cerrados). Estos valores de la carga se pueden especificar con tres parámetros: valor inicial, valor final e incremento. El formato de los datos de salida puede ser el especificado anteriormente, que ofrece un gran nivel de detalle, o bien mostrar la productividad y el tiempo de respuesta del sistema para cada valor de la carga; este último formato es adecuado para poder representar los valores de forma gráfica, por ejemplo haciendo uso del programa gnuplot o una hoja de cálculo.

**ACTIVIDAD 5.3** Los límites asintóticos optimistas establecidos en este capítulo son aplicables en condiciones muy generales. Sin embargo, no sirven para establecer cotas superiores del tiempo de respuesta ni cotas inferiores de la productividad. En un trabajo publicado en el año 1982 por Zahorjan et al. [48] se propone un conjunto de límites asintóticos que establecen cotas, tanto inferior como superior, y muy próximas entre sí, para un modelo cerrado (interactivo o por lotes). Estos límites se establecen suponiendo que el sistema es- tá equilibrado (*balanced system*), esto es, todos los dispositivos tienen la misma demanda de servicio. Además, como hipótesis de partida se supone que todos los dispositivos del sistema tienen un único servidor (excepto la estación que modela el tiempo de reflexión). En particular, las cotas para el tiempo de respuesta son:

$$\text{máx}\left\{ N \times D_b - Z,\ D + (N-1) \times D_m \times \frac{D}{D+Z} \right\} \le R_{opt} \le D + (N-1) \times D_b \times \frac{(N-1) \times D}{(N-1) \times D + Z}$$

y las cotas para la productividad son:

$$\frac{N}{Z + D + (N-1) \times D_b \times \frac{(N-1)\times D}{(N-1)\times D + Z}} \leq X_{\text{opt}} \leq \text{mín}\left(\frac{1}{D_b}, \frac{N}{Z + D + (N-1) \times D_m \times \frac{D}{D+Z}}\right)$$

donde $D_m = D/K$, siendo $K$ el número de dispositivos en el sistema.

Dibújense las cotas anteriores para el rendimiento de los sistemas analizados en las Figuras 5.1 y 5.2.

# Capítulo 6

## Caracterización de la carga

Las prestaciones de un sistema informático dependen del tipo de acciones o trabajos que lleva a cabo. Así, por ejemplo, los procesos de cálculo intensivo, utilizados en la investiga- ción científica para la resolución de ecuaciones complejas, tienen distintas características que el proceso transaccional típico de un banco provocado por sus cajeros automáticos, distintas también de las de un servidor web de suministro de películas bajo demanda. Mientras que el proceso científico exige una alta capacidad de cómputo, y, por tanto, de procesador, el proceso bancario necesita mucha más entrada/salida de bloques pequeños de información y apenas requiere cálculos intermedios; a lo sumo, la resta de una canti- dad de otra para calcular el nuevo saldo en una cuenta corriente, una vez realizada una salida de dinero por un cajero automático. Por su parte, la tarea del servidor de vídeo es enviar por la red telefónica ficheros muy voluminosos. El proceso científico no necesita normalmente ninguna utilización de red informática externa porque los cálculos se realizan en el mismo edificio donde se encuentra el científico, mientras que en el proceso bancario tiene una influencia muy importante la velocidad de la línea que conecte los 10.000 cajeros automáticos al servidor central de base de datos del banco. El envío de vídeo por la red telefónica requiere de un ancho de banda grande para que la transmisión de las películas se complete en un tiempo razonable. Esto es, dependiendo del tipo de trabajo a realizar, del tipo de carga, se demandarán más unos recursos informáticos que otros (procesador, entrada/salida y red).

El analista de prestaciones desea conocer cómo se van a comportar uno o varios sistemas informáticos ante una determinada carga de trabajo. Normalmente no se dispone de la carga a que van a someter esos sistemas, pero sí se pueden utilizar modelos de la misma con características similares a la carga futura, lo que se ha denominado anteriormente como carga de prueba. Por este motivo, los estudios de evaluación de prestaciones se basan normalmente en modelos de carga, modelos que se extraen tras una caracterización previa de la misma. Una vez que el modelo está disponible, se pueden estudiar los efectos de los cambios en la carga y en el sistema de una manera controlada simplemente cambiando los parámetros de dicho modelo.

En la parte teórica de este capítulo se va a describir en primer lugar cómo se puede representar la carga de un sistema mediante un modelo de carga, y dados varios modelos, determinar cuál de ellos representa mejor la carga del sistema en estudio. Posteriormente se explicará la técnica más utilizada para la caracterización de la carga, que es la técnica de agrupamiento de carga, por ser la más empleada, la más intuitiva y haber paquetes estadís- ticos disponibles para ello. Se termina la parte teórica con el estudio del comportamiento de los usuarios que acceden a un servidor web, introduciendo los gráficos del modelo de comportamiento del usuario, que nos ayudarán a caracterizar este tipo específico de carga.

Se denomina *carga* (*workload* ) a todas las demandas que realizan los usuarios a un sistema informático (solicitudes a una página web, actualizaciones y consultas a una base de datos, consultas a un almacén de datos, etc.) durante un intervalo de tiempo. Normalmente es difícil tener acceso a la carga real para realizar estudios de evaluación de prestaciones, y por este motivo se utilizan modelos de carga. Además, los modelos de carga permiten la portabilidad y la reproducción de las necesidades en distintos entornos, y el coste de su uso es menor que trabajar con la carga real.

Se llama *caracterización de la carga* al proceso por el cual se define un modelo de carga que reproduce lo mejor posible las características de la carga real. El modelo de carga se establecerá en función de los parámetros que puedan afectar al comportamiento del sistema y los objetivos que se persiguen. La caracterización de la carga consiste en la descripción de la carga por medio de *parámetros cuantitativos* y funciones; el objetivo es generar un modelo de carga que sea capaz de captar y reproducir el comportamiento de la carga y de sus características más importantes.

Uno de los objetivos de la evaluación del rendimiento o de las prestaciones es adaptar la configuración de un sistema informático a la carga existente o propuesta. Por lo tanto, es necesario tener la capacidad de describir la carga de tal forma que se pueda determinar el hardware necesario para soportarla con el nivel de servicio acordado.

Algunos de los parámetros cuantitativos que se utilizan en la caracterización de la carga son los siguientes: número y tipo de usuarios (consulta sencilla o compleja a una página web, actualización sencilla o compleja en una base de datos, etc.), peticiones a páginas de un servidor web, peticiones a un servidor de comercio electrónico, número de termi- nales activos, tasa de llegada de transacciones, tiempo total de procesador, tiempo medio de procesador entre dos operaciones de entrada/salida y sus distribuciones, capacidad de memoria principal utilizada, número de operaciones de entrada/salida, sus distribuciones y sus características (aleatorias, secuenciales, en ráfagas, etc.), número y características de los ficheros en disco y en cinta, número de páginas impresas, número de procesos por lotes lanzados en paralelo, etc.

También se utilizan algunos *parámetros cualitativos*, como el lenguaje de programación y los tipos de actividad realizados (edición, compilación, ejecución), para combinar una descripción funcional de los modelos con su parte cuantitativa.

## 6.1. Representatividad de un modelo de carga

Los modelos de carga, por la misma razón de ser modelos, son siempre aproximaciones y, por tanto, representan una abstracción de la realidad que se pretende representar. Así, cuando se utiliza un modelo para representar la carga de un sistema informático, éste debe describir, de la forma más fidedigna posible, aquellos aspectos más importantes de esa carga. Es decir, el modelo de carga debe considerar qué detalles de ésta realzar a la vista del estudio de rendimiento que se lleva a cabo y del funcionamiento del sistema que se analiza. La ***representatividad de la carga*** es una medida de la similitud entre el modelo y la carga real. Una vez que el modelo de carga está disponible se pueden estudiar los efectos de cambios en la carga de una manera controlada simplemente cambiando el peso de los parámetros en el modelo.

Una problemática comúnmente planteada es decidir, ante dos ($W'$ y $W''$) o más mode- los, cuál de ellos es más representativo de la carga real ($W$). Estos problemas se resuelven de la siguiente manera:

1. Se normalizan los parámetros de todos los modelos y de la carga real, dando un valor en el intervalo $(0, 1)$. Sea

   - $v_j$, el valor del parámetro $j$ para un componente de la carga.
   - $v_{ij}$, el valor medio del parámetro $j$ para los componentes de clase $i$. $v_{jmín}$,
   - el valor mínimo de $v_j$ para todos los componentes.
   - $v_{jmáx}$, el valor máximo de $v_j$ para todos los componentes.

   El valor normalizado $v'_{ij}$ será:

$$v'_{ij} = \frac{v_{ij} - v_{jmín}}{v_{jmáx} - v_{jmín}} \quad (6.1)$$

2. Se calcula la distancia entre los valores normalizados de los modelos y el valor normalizado de la carga real como el valor absoluto de la diferencia de los valores normalizados, esto es, $|v'_{ij} - v_{ij}|$ y $|v''_{ij} - v_{ij}|$, o bien sus correspondientes distancias cuadráticas.

3. Se pondera doblemente la distancia entre los parámetros del modelo y los de la carga que se quiere modelizar, teniendo en cuenta:

- $q_i$: el porcentaje de programas de clase $i$ en la carga.
- $w_j$: el peso asociado al parámetro $j$. Estos pesos pueden tomar cualquier valor.

Esta distancia se calcula mediante la siguiente ecuación:

$$D' = \sum_{i} q_i \sum_{j=1}^{k} w_j \times |v_{ij} - v'_{ij}|$$

$$D'' = \sum_{i=1} q_i \sum_{j=1} w_j \times |v''_{ij} - v_{ij}|$$

Los pesos asociados a los parámetros ($w_j$) tienen unos valores u otros dependiendo del objetivo respecto al cual se quiera medir la representatividad. Por ejemplo, si se quiere evaluar la bondad de un modelo respecto a la actividad de entrada/salida (E/S), se dará un peso mayor a los $w_j$ que estén relacionados con operaciones de E/S.

4. El modelo que diste menos de la carga real será el más representativo para ese objetivo específico.

Vamos a aplicar lo descrito anteriormente para calcular la distancia que existe entre una carga real $W$ y un par de modelos de carga $W'$ y $W''$ que se caracterizan por los parámetros indicados en la siguiente tabla, para el caso del objetivo 1 de dar mayor importancia al procesador ($w_{11} = 2$, $w_{21} = 0{,}5$) y el caso del objetivo 2 de dar mayor importancia a la entrada/salida ($w_{21} = 0{,}5$, $w_{22} = 2$),

| Parámetro | $W$ | $W'$ | $W''$ |
|---|---|---|---|
| Tiempo procesador (s) | 3 | 2,5 | 2,8 |
| Tiempo de E/S (s) | 2 | 2,1 | 1,7 |
| Tiempo procesador (s) | 5 | 1 | 4 |
| Tiempo de E/S (s) | 4 | 1 | 3 |

y los rangos de dichos parámetros vienen dados por la siguiente tabla: Parámetro

Valor máximo Valor mínimo Rango

Primero normalizamos los parámetros de todos los modelos, resultando la tabla normalizada siguiente:

| Parámetro | $W$ | $W'$ | $W''$ |
|---|---|---|---|
| Tiempo procesador (s) | 0,5 | 0,375 | 0,45 |
| Tiempo de E/S (s) | 0,33 | 0,36 | 0,23 |

En segundo lugar, se calculan las distancias entre los valores normalizados, resultando la siguiente tabla de distancias:

| Parámetro | $D'$ | $D''$ |
|---|---|---|
| Tiempo procesador (s) | 0,125 | 0,05 |
| Tiempo de E/S (s) | 0,03 | 0,1 |

Para terminar, se pondera para el objetivo 1 y para el objetivo 2 la distancia entre los parámetros del modelo y los de la carga que se quiere modelizar, resultando:

$$D'_1 = 2 \times 0{,}125 + 0{,}5 \times 0{,}03 = 0{,}265$$

$$D''_1 = 2 \times 0{,}05 + 0{,}5 \times 0{,}1 = 0{,}15$$

$$D'_2 = 0{,}5 \times 0{,}125 + 2 \times 0{,}03 = 0{,}0925$$

$$D''_2 = 0{,}5 \times 2{,}1 + 2 \times 0{,}1 = 0{,}2025$$

Por lo tanto, el modelo $W''$ modeliza mejor el procesador (objetivo 1) mientras que el modelo $W'$ modeliza mejor la entrada/salida (objetivo 2).

## 6.2. Técnica de agrupamiento (clustering)

El análisis mediante agrupamiento es una técnica matemática para agrupar medidas, su- cesos o trabajos individuales que son similares en algún aspecto o de alguna manera. Estos trabajos se describen mediante los valores numéricos de un conjunto de parámetros, tales como el tiempo de procesador utilizado, el número de operaciones de entrada/salida reali- zadas a los diferentes periféricos, la memoria necesaria para la ejecución, etc. La elección de los parámetros utilizados para caracterizar los trabajos es importante y depende del propósito para el que se haya construido el modelo. Ya que el objetivo es agrupar los tra- bajos que son lo más similares posible, es necesario cuantificar el concepto de grados de similitud. Una medida de la similitud es la distancia entre dos trabajos que se representan como puntos en un espacio multidimensional con ejes correspondientes a los parámetros utilizados. Esta distancia se obtiene mediante la fórmula tradicional euclídea de la raíz cuadrada de la suma de los cuadrados de la diferencia entre los valores de los parámetros para dos trabajos. Ya que solamente se está interesado en comparar las dos distancias, no es necesario el cálculo de la raíz cuadrada; es suficiente comparar los cuadrados de las distancias. Hay algunos programas estadísticos, como el SPSS, el SAS o el Statgraphics, que realizan automáticamente el análisis de agrupamientos.

A continuación se va a describir un ejemplo detallando los pasos que se dan en la aplicación de este método. La Figura 6.1 muestra las demandas de procesador y de entra- da/salida de 25 trabajos. Los trabajos se pueden clasificar en cinco grupos, como se indica en la figura. Por lo tanto, en lugar de utilizar 25 trabajos para cada análisis, se pueden utilizar cinco trabajos para representar las demandas de recursos promedio de cada grupo.

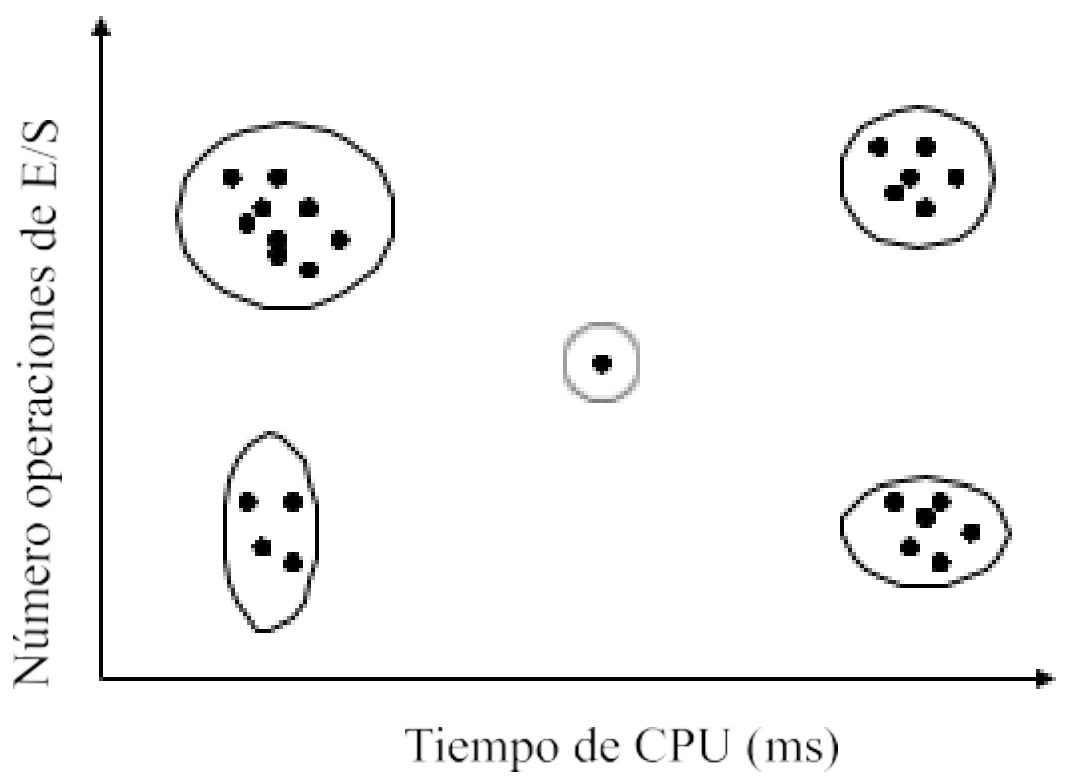

**Figura 6.1:** Ejemplo de agrupamiento de 25 trabajos en cinco clases.

Para caracterizar los datos de la carga medidos utilizando agrupamiento, los pasos son los siguientes:

1. *Tomar una muestra, esto es, un subconjunto de los componentes de la carga.* Normalmente el número de componentes medidos es demasiado grande para que se pueda utilizar en un análisis de agrupamiento. Por lo tanto, es necesario seleccionar un subconjunto pequeño al que llamaremos muestra.

2. *Seleccionar los parámetros de la carga.* Cada componente tiene asociado un gran número de parámetros (demandas de recursos). Algunos de estos parámetros son muy importantes bien porque pertenecen al recurso cuello de botella o al recurso más caro. Los parámetros menos importantes se pueden omitir del análisis de agrupamiento, reduciendo el coste del análisis. Los dos criterios clave para seleccionar los parámetros son su impacto en el rendimiento y en la varianza.

3. *Transformar los parámetros si es necesario.* Si la distribución de un parámetro está muy sesgada, se debería considerar la posibilidad de sustituir el parámetro por una función de él. Por ejemplo, en un estudio se puede utilizar el tiempo de procesador en forma logarítmica porque el analista puede indicar que dos programas que consumen 1 y 2 segundos de procesador son tan diferentes como aquellos que consumen 1 o 2 milisegundos. En este caso, el cociente entre el tiempo de procesador (2) se ha considerado más importante que su diferencia (1 segundo y 1 milisegundo).

4. *Quitar los valores extremos.* Son los puntos que tienen unos valores de los parámetros extremos, cayendo lejos de la mayoría de los otros puntos. Puede haber efectos que pudieran obligar a considerar estos valores extremos en algunas ocasiones.

5. *Poner en una escala adecuada todas las observaciones.* Se recomienda que los valores de los parámetros se representen en una escala de tal forma que los valores relativos y los rangos sean aproximadamente iguales. Hay cuatro técnicas de escalado que se usan:

a) *Normalizar a cero la media y a uno la varianza (valores normalizados centra- dos y reducidos).* Sea $\{x_{1k}, x_{2k}, \ldots, x_{mk}\}$ los valores del parámetro $k$. El valor escalado $x'_{ik}$ también citado como $Z$, que corresponde a $x_{ik}$, viene dado por:

$$Z = x'_{ik} = \frac{x_{ik} - \overline{x_k}}{s_k} \tag{6.2}$$

Aquí $\overline{x_k}$ y $s_k$ son la media aritmética y la desviación típica del parámetro $k$, respectivamente.

Ésta es la técnica de escalado más utilizada, porque es una medida libre de unidades, que permite evitar los problemas que surgen al utilizar parámetros con valores relativos y rangos muy diferentes. Esta medida se llama también *medida estándar* o valoración $Z$ (*Z score*).

Los beneficios que aporta esta técnica con respecto a la de la distancia euclídea se van a presentar por medio de un ejemplo. La diferencia entre el tiempo de ejecución de procesador de 1 s y de 1,1 s es de una distancia euclídea de 0,1 s. La diferencia entre el tiempo de ejecución de procesador de 1.000 ms y de 1.100 ms es de una distancia euclídea de 100 ms. Estos dos conjuntos de cantidades son equivalentes proporcionalmente, pero una tiene una distancia euclídea 100 veces mayor que la otra. Para evitar tener que seleccionar una unidad, se utiliza la medida estándar. En nuestro ejemplo supóngase que el tiempo de ejecución medio del procesador es de 0,8 s o 800 ms y que la desviación típica es de 0,5 s o 500 ms. Los valores $Z$ de ambos tiempos de ejecución de procesador son:

$$Z_1 = \frac{1{,}0 - 0{,}8}{0{,}5} = \frac{1.000 - 800}{500} = 0{,}4$$

$$Z_2 = \frac{1{,}1 - 0{,}8}{0{,}5} = \frac{1.100 - 800}{500} = 0{,}6$$

Esto es, la medida estándar de los dos tiempos de ejecución equivalentes de procesador resultan en el mismo valor numérico. Habiendo convertido los valores de los parámetros a $Z$, el cuadrado de las distancias de las mediciones en el periodo de observación, se puede encontrar con la siguiente ecuación:

$$D_{ij} = \sum_{k=1}^{n} (Z_{ik} - Z_{jk})^2 \tag{6.3}$$

siendo $D_{ij}$ el cuadrado de la distancia entre las mediciones $i$ y $j$, y $Z_{ik}$ el valor $Z$ para el parámetro $k$ de la medida $i$.

b) *Pesos.* La ecuación de normalización mediante los pesos viene dada por la siguiente ecuación:

$$x'_{ik} = w_k \times x_{ik} \tag{6.4}$$

El peso $w_k$ se puede asignar dependiendo de la importancia relativa del pará- metro o es inversamente proporcional a la desviación típica de los valores del parámetro.

c) *Normalización del rango.* El rango se cambia de $[x_{mín,k}, x_{máx,k}]$ a $[0, 1]$. La fórmula del escalado es:

$$x'_{ik} = \frac{x_{ik} - x_{mín,k}}{x_{máx,k} - x_{mín,k}} \tag{6.5}$$

Aquí $x_{mín,k}$ y $x_{máx,k}$ son los valores mínimo y máximo, respectivamente, del parámetro $k$.

d) *Normalización con percentiles.* Los datos son escalados de tal forma que el 95 % de los valores caigan entre 0 y 1:

$$x'_{ik} = \tag{6.6}$$

Aquí $x_{2,5,k}$ y $x_{97,5,k}$ son los percentiles 2,5 y 97,5, respectivamente, para el parámetro $k$.

6. *Seleccionar una medida (métrica) de la distancia.* La cercanía de dos puntos se mide definiendo una medida de la distancia. Se suele emplear uno de los tres métodos siguientes:

a) *Distancia euclídea.* La distancia $d$ existente entre los dos componentes de la carga denotados como $\{x_{i1}, x_{i2}, \ldots, x_{in}\}$ y $\{x_{j1}, x_{j2}, \ldots, x_{jn}\}$ se define como:

$$d = \sqrt{\sum (x_{ik} - x_{jk})^2} \tag{6.7}$$

b) *Distancia euclídea ponderada.* En este caso, la distancia entre los dos compo- nentes viene dada por:

$$d = \sqrt{\sum_{k=1} a_k (x_{ik} - x_{jk})^2} \qquad (6.8)$$

siendo $a_k$ para $k = 1, 2, \ldots, n$ los pesos elegidos convenientemente para los $n$ parámetros.

c) *Distancia chi-cuadrado.* En este último caso la distancia entre los dos compo- nentes viene dada por la expresión:

$$d = \sum_{k=1} \left( \frac{(x_{ik} - x_{jk})^2}{x_{ik}} \right) \qquad (6.9)$$

La distancia euclídea es la métrica de distancia más utilizada. La distancia euclídea ponderada se emplea si los parámetros no se han escalado o si los parámetros tienen niveles de importancia muy diferentes. La distancia chi-cuadrado se utiliza cuando se ponen de manifiesto proporciones entre componentes.

7. *Realizar el agrupamiento.* El objetivo es conseguir una agrupación en clases de forma que las varianzas dentro de cada clase sean las menores posibles y que las varianzas entre las clases sean las mayores posibles. Hay muchos métodos para realizar esta fase, pero solamente se van a explicar con cierto detalle dos de ellos.

   Los primeros métodos son los *métodos globales no jerárquicos*. Se basan en realizar una subdivisión inicial del espacio en $k$ clases e ir mejorando iterativamente el agru- pamiento de $m$ elementos de $n$ componentes desplazándolos de una clase a otra. El agrupamiento óptimo se alcanza cuando ya no se pueden desplazar elementos de una clase a otra porque no se aumenta la distancia euclídea entre clases.

   El otro método es el *método del árbol de extensión mínima* (*Minimum Spanning Tree*, MST). Se trata de un método jerárquico ascendente de agrupamiento, que comienza con $n$ clases de un componente cada una y sucesivamente se unen las clases más cercanas, según el siguiente esquema:

   a) Comenzar con $k = n$ clases, es decir, todo componente es una clase.
   b) Encontrar el centroide de la clase $i$, $i = 1, 2, \ldots, k$. El centroide tiene los valores de los parámetros igual a la media de todos los puntos de la clase.
   c) Calcular la matriz de distancia entre las clases. El elemento $(i, j)$ de la matriz es la distancia entre los centroides de las clases $i$ y $j$. Se puede utilizar cualquiera de las distancias explicadas previamente en el apartado de métricas de distancias.
   d) Encontrar el elemento más pequeño no nulo de la matriz de distancias. Supón- gase que $d_{lm}$ es la distancia entre las clases $l$ y $m$, que es la mínima. Se juntan las clases $l$ y $m$. También se juntan todos aquellos pares de clases que tengan la misma distancia.
   e) Repetir los pasos $b$ a $d$ hasta que todos los componentes sean parte de una clase.

8. *Interpretar los resultados.* Una vez se ha decidido el número de grupos adecuado, se pueden eliminar aquellos grupos cuya demanda de recursos afecte poco al rendimiento del sistema. El siguiente paso es interpretar las clases en términos funcionales. Si una serie de componentes en un grupo pertenecen a un único tipo de aplicación, ayuda mucho el etiquetarlo con un nombre apropiado que nos indique a qué tipo de aplicación corresponden. Generalmente será posible etiquetar los grupos mediante las demandas de recursos que realizan: mucho o poco procesador, mucha o poca E/S, etc.

9. *Cambiar los parámetros, o el número de grupos, y repetir los pasos del 3 al 7.*

10. *Seleccionar los componentes representativos de cada grupo.*

A continuación se va a exponer un ejemplo de agrupamiento para facilitar su compren- sión. Considérese una carga con cuatro componentes, los programas Mixto (M), Lectura (L), Escritura (E) y Cálculo (C), y dos parámetros, el tiempo de procesador y el número de operaciones de E/S. Los valores de los parámetros son los que se muestran en la tabla siguiente:

| Programa | Tiempo procesador | Operaciones de E/S |
|---|---|---|
| M | 2 | 3 |
| L | 1 | 5 |
| E | 1 | 6 |
| C | 4 | 1 |

Vamos a calcular el centroide de esta muestra así como la distancia utilizando la métrica de la distancia euclídea. Aplicamos los pasos descritos anteriormente con referencia al método del árbol de extensión mínima, que son los siguientes:

- Paso 1: se consideran cuatro clases coincidiendo cada una de las clases con cada uno de los cuatro programas.
- Paso 2: los centroides son {2,3}, {1,5}, {1,6} y {4,1}.
- Paso 3: utilizando la métrica de la distancia euclídea, la matriz de distancias resulta ser la siguiente:

| Programa | M | L | E | C |
|---|---|---|---|---|
| M | 0 | 2,24 | 3,16 | 2,83 |
| L | | 0 | 1 | 5 |
| E | | | 0 | 5,83 |
| C | | | | 0 |

- Paso 4: la distancia mínima entre las clases es de 1 entre L y E. Estas dos clases se juntan en otra nueva clase. De nuevo volvemos ahora al paso 2 para calcular los nuevos centroides de las tres clases existentes. Los pasos propios del agrupamiento son del 2 al 4 y se van a repetir cada vez que se haga un nuevo agrupamiento. Se van a repetir estos pasos hasta que todos los componentes formen parte de una sola clase.

- Paso 2: el centroide de la clase LE es $\left(\frac{1+1}{2}, \frac{5+6}{2}\right)$, esto es, {1, 5,5}. Los otros centroides son los mismos que antes.
- Paso 3: ahora hay tres clases, y la matriz de la distancia euclídea resulta ser la siguiente:

| Programa | M | LE | C |
|---|---|---|---|
| M | 0 | 2,69 | 2,83 |
| LE | | 0 | 6,26 |
| C | | | 0 |

- Paso 4: la distancia mínima entre clases es de 2,69 entre M y LE. Por lo tanto, estas dos clases se unen.
- Paso 2: el centroide de la clase MLE es $\left(\frac{2+1}{2}, \frac{3+5,5}{2}\right)$, esto es, {1,5, 4,25}.

Hay dos clases con dos centroides, según se representan en la Figura 6.2. El segundo centroide coincide con el punto C.

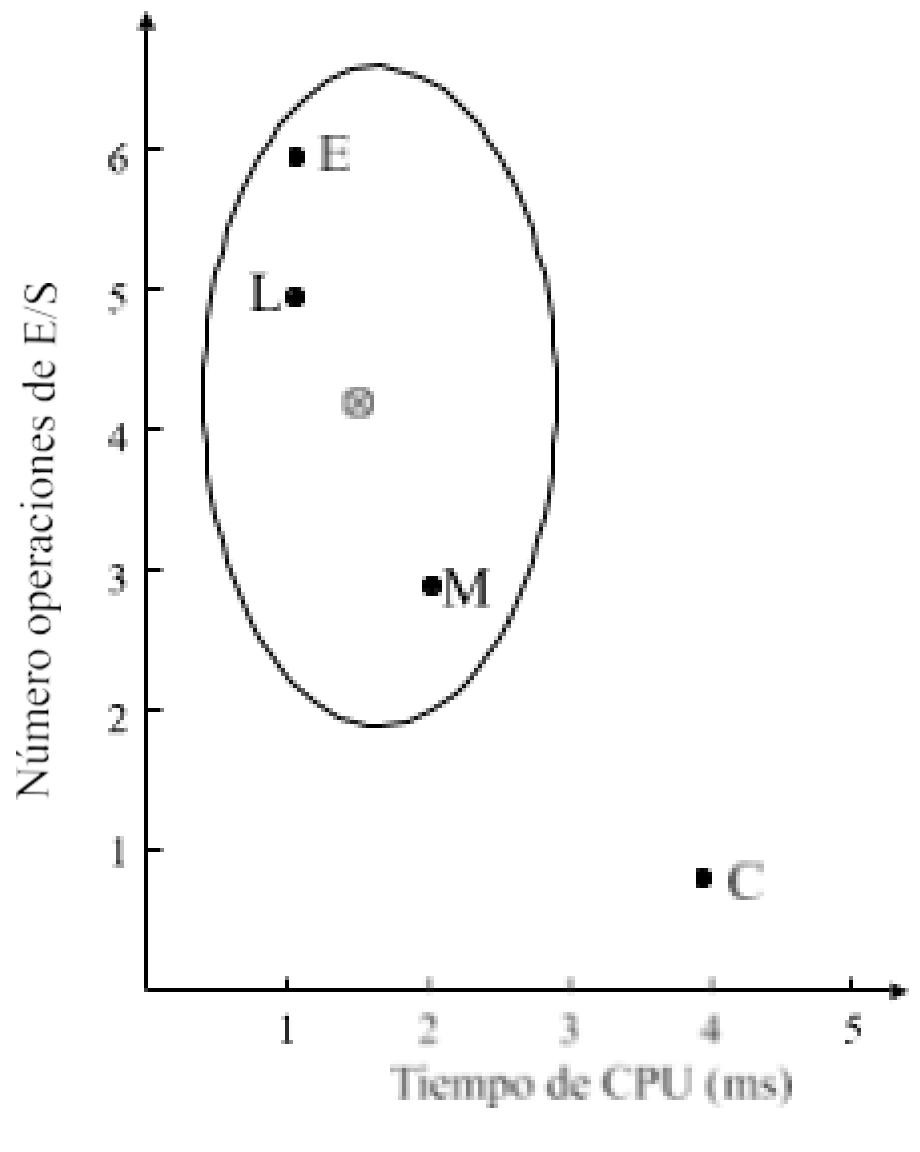

**Figura 6.2:** Agrupamiento realizado en el ejemplo descrito.

- Paso 3: la matriz de distancias resulta ser ahora la siguiente:

| Programa | MLE | C |
|---|---|---|
| MLE | 0 | 4,10 |
| C | | 0 |

- Paso 4: la distancia mínima entre clases es de 4,10. La unión entre la clase MLE y C resulta en una sola clase MLEC, cuyo centroide es $\left\{\frac{1,5+4}{2}, \frac{4,25+1}{2}\right\}$, esto es, {2,75, 2,63}.

Los resultados del proceso de agrupamiento se pueden representar como un árbol extendido llamado *dendograma*. Cada rama del árbol representa una clase y se dibuja en vertical hasta una altura a la que la clase se une con las clases vecinas. El árbol extendido para el ejemplo desarrollado anteriormente se muestra en la Figura 6.3. Los componentes L y E se unen a la altura de 1. Los componentes M y LE se unen a la altura 2,69. La última unión ocurre a la altura de 4,10 cuando se une MLE y C.

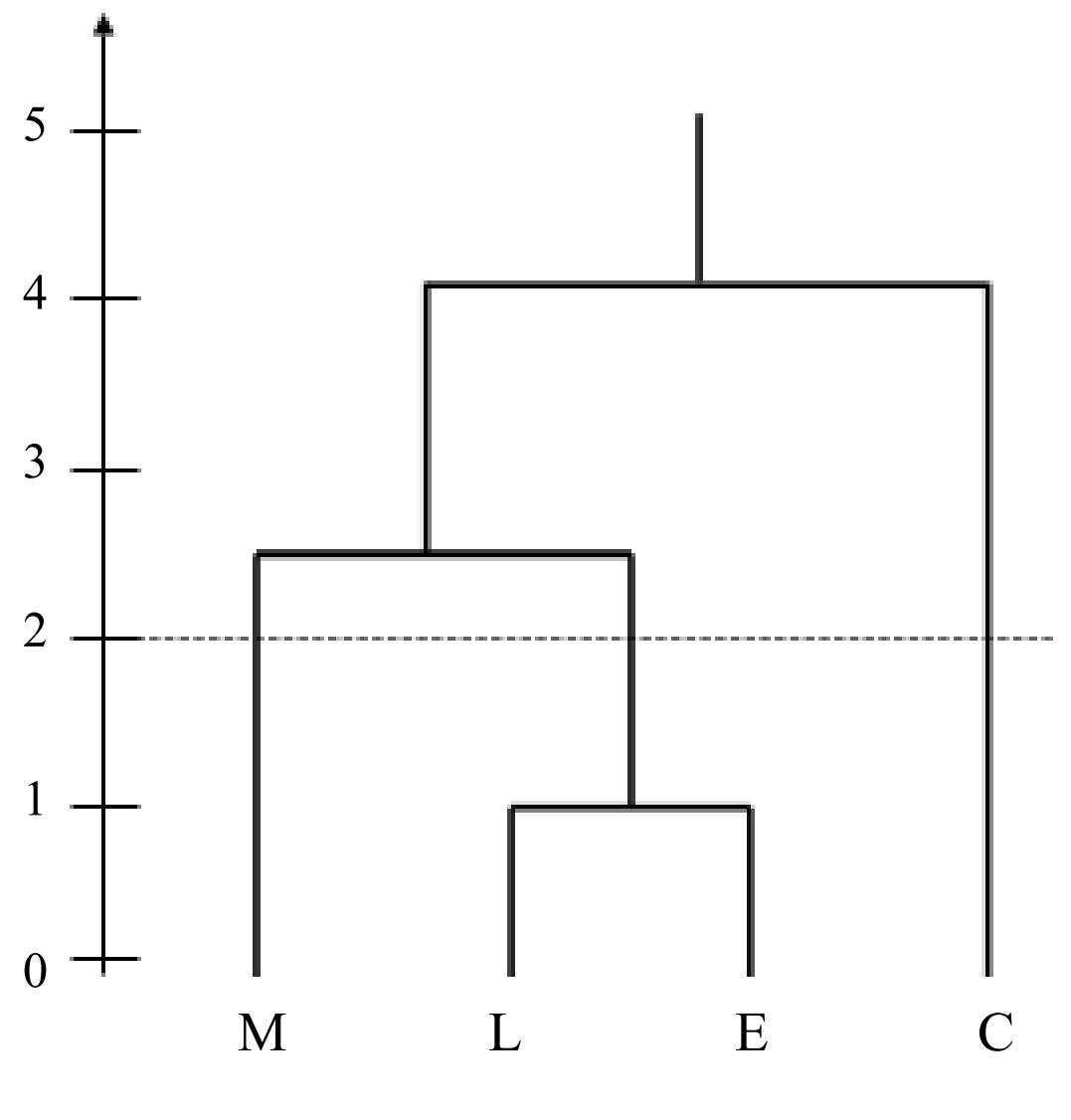

**Figura 6.3:** Dendograma asociado al ejemplo descrito.

El propósito de dibujar el árbol extendido es obtener las clases dentro de las cuales hay una distancia menor de la permitida por un cierto valor. Por ejemplo, si la distancia permitida máxima dentro de una clase es de 2, se traza una línea horizontal a esta altura (mostrada como línea discontínua en la Figura 6.3), que corta tres ramas del árbol que representan las clases M, LE y C, indicando que dentro de estas clases la distancia entre sus componentes (distancia intraclase) es menor de 2.

Una pregunta frecuente cuando se trabaja en técnicas de agrupamiento es cuántas clases representan con exactitud la carga. Esta pregunta se puede contestar analizando la variación de dos medidas: la distancia media entre los puntos de una clase y su centroide (distancia intraclase), y la distancia media entre centroides (distancia entre clases). Esta variación se puede caracterizar mediante el coeficiente de variación, esto es, el cociente entre la desviación típica y la media. Se ha de elegir un número relativamente pequeño de clases tal que el coeficiente de la varianza dentro de la clase (intraclase) sea pequeña y el coeficiente de la varianza entre clases sea grande. El cociente entre el coeficiente de variación intraclase y entre clases, indicado por $\beta_{cv}$ , es una guía útil para determinar la calidad del proceso de agrupamiento. Se debe elegir un número de clases para el cual el valor de $\beta_{cv}$ no disminuya al aumentar el número de clases.

## 6.3. Otros criterios de agrupamiento

En este apartado se desarrolla un planteamiento más cualitativo y menos cuantitativo que los anteriores. Las técnicas de particionamiento dividen la carga en clases tales que sus ele- mentos (poblaciones) están formados por componentes bastante homogéneos. El objetivo es agrupar componentes que son similares de alguna forma. Esto es, esta técnica indica que es mejor representar la carga mediante varias clases o grupos sin intentar representarla mediante una sola clase. El motivo de particionar la carga es doble: mejorar la repre- sentatividad de la caracterización y aumentar la capacidad de predicción del modelo. A continuación se citan algunos atributos utilizados para establecer particiones de la carga en clases de componentes similares.

- *Según el tipo de aplicaciones.* Una carga puede tener agrupados sus componentes de acuerdo con la aplicación a la que pertenecen. Supóngase que se quiere caracterizar el tipo de tráfico que fluye a través de la conexión a Internet de una empresa. Debido a la disparidad en tamaños y la duración entre las transacciones más comunes (por ejemplo, HTTP, telnet) y nuevas aplicaciones multimedia (por ejemplo, Mbone), es crítico dividir el tráfico de red en distintos tipos de aplicaciones. La Tabla 6.1 muestra un ejemplo de los tipos de tráfico y la cantidad de bytes transmitidos durante el periodo de observación.
- *Según los objetos.* Se puede dividir la carga de acuerdo con el tipo de objetos mane- jados por las aplicaciones. En un servidor web, una carga se puede particionar por los tipos de documentos a los que se accede desde los clientes. Como ejemplo, la Tabla 6.2 presenta una carga agrupada según el tipo de documentos.
- *Según la utilización de los recursos de los distintos tipos de aplicaciones.* La demanda de recursos por componente también se puede emplear para dividir la carga en clases o grupos. Éste es el tipo de agrupamiento más utilizado, empleando normalmente estos dos parámetros: el tiempo de procesador y el tiempo de E/S. La Tabla 6.3 muestra un ejemplo de clases de órdenes de un sistema informático de tipo transaccional. En este ejemplo, el tiempo máximo de procesador y el número máximo de operaciones de E/S solicitadas por una transacción se consideran los elementos críticos del sistema. Por ejemplo, la transacción de tipo ligera incluye a las transacciones que demandan más de 6 ms y menos de 30 ms de tiempo de procesador, y un tiempo de E/S que varía entre 50 y 200 ms.

| Tipo de aplicación | KB transmitidos |
|---|---|
| HTTP | 4.500 |
| Mbone | 600 |
| ftp | 400 |
| telnet | 98 |
| Otros | 65 |

**Tabla 6.1:** Particionamiento de carga basado en tipo de aplicación Internet.

| Tipo de documento | Acceso ( %) |
|---|---|
| Páginas web (html) | 40 |
| Imágenes (gif, jpeg) | 20 |
| Formateados (doc, ps, dvi) | 15 |
| Sonido (wav) | 10 |
| Vídeo (mpeg, avi) | 5 |
| Otros | 10 |

**Tabla 6.2:** Particionamiento de carga basado en tipo de documento.

| Transacción | Proporción | Máx. procesador (ms) | Máx E/S (ms) |
|---|---|---|---|
| Muy ligera | 40 % | 6 | 50 |
| Ligera | 30 % | 30 | 200 |
| Media | 20 % | 120 | 600 |
| Pesada | 10 % | 800 | 1.000 |

**Tabla 6.3:** Particionamiento de carga basado en la utilización de recursos.

## 6.4. Acceso a servidores web

Se está comenzando a utilizar los gráficos de comportamiento para caracterizar o describir las elecciones realizadas por los usuarios de una página web cuando acceden a ella y, por lo tanto, caracterizar la carga provocada por esos accesos. Los gráficos del modelo del comportamiento del usuario (*Customer Behavior Model Graph*, CBMG) son una herramienta habitual de representación de los patrones de navegación de un grupo de clientes o usuarios en un servidor web. El CBMG tiene *n* estados, donde el primero siempre es el de entrada y el último es el de salida. La salida del sistema se puede realizar desde varios estados del sistema y está representada mediante una flecha con su porcentaje correspondiente. Los arcos, con sus probabilidades asociadas, representan las secuencias de las órdenes tal co- mo son ejecutadas por los usuarios en su comportamiento habitual. El número de estados (funciones de negocio) depende del tipo de servicio que se dé en esa web. Éste es un caso particular de los modelos markovianos aplicado a los servidores web. Un CBMG típico de una tienda en línea, como puede ser un supermercado, una librería tipo Amazon, etc., es el presentado en la Figura 6.4.

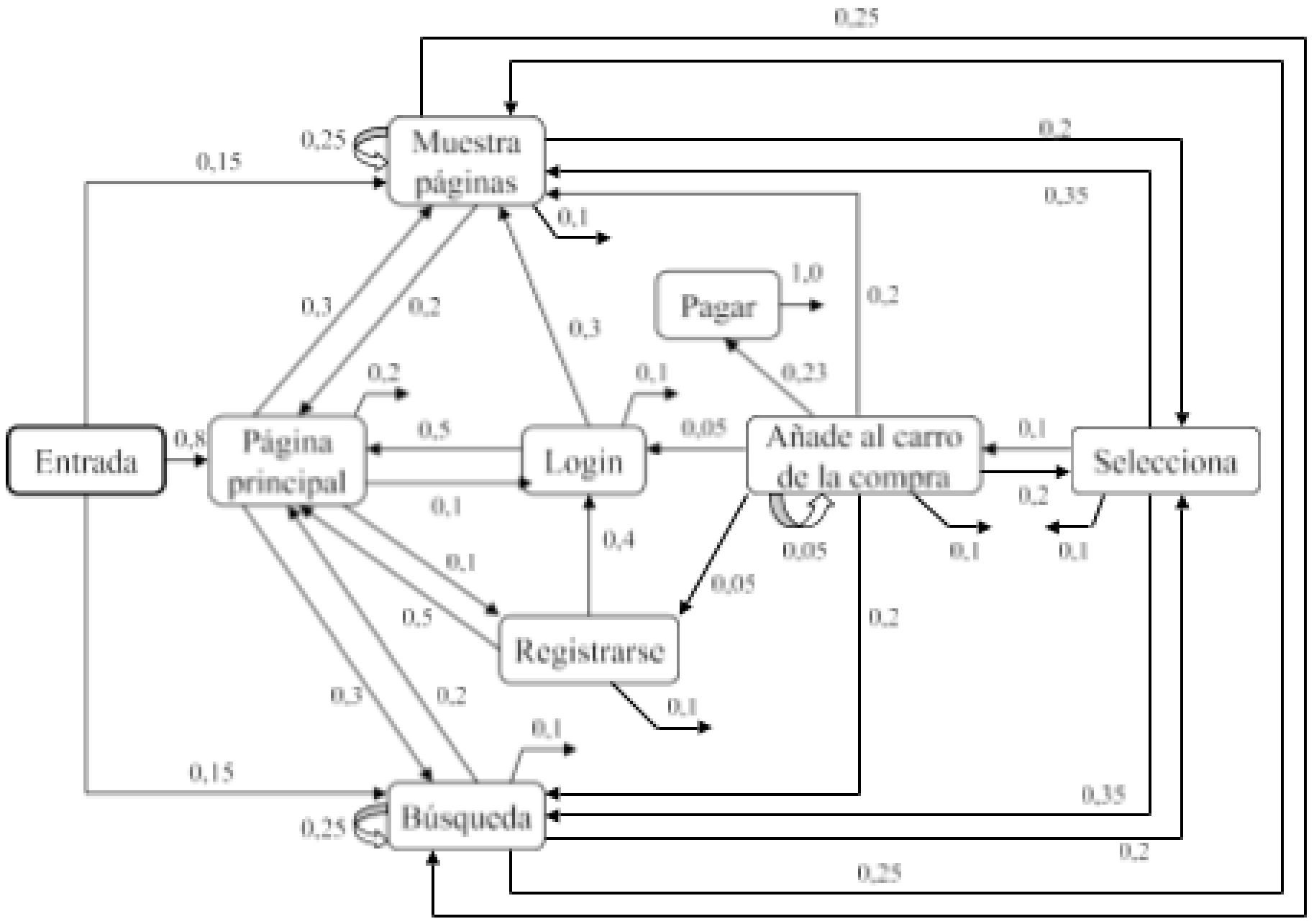

**Figura 6.4:** Gráfico del modelo de comportamiento del usuario de compras en línea.

Algunos estados o funciones que son comunes en cualquier servidor web, y que aparecen en casi todos los CBMG, son los siguientes:

- *Entrada*: conexión al sitio. Un usuario se mueve a este estado después de solicitar la conexión al sitio.
- *Página principal* : ésta es la página principal de entrada a este sitio. Éste es el estado

del usuario una vez que ha seleccionado la URL del sitio y ha entrado.

- *Registrarse*: registro de un nuevo usuario. Para poder utilizar esta aplicación de compra electrónica en línea el cliente o usuario debe haberse registrado previamente, y utilizar una palabra clave para su acceso.
- *Búsqueda*: búsqueda en la base de datos del sitio. Un usuario va a este estado después de solicitar una búsqueda.
- *Selecciona*: ver uno de los resultados de la búsqueda. Un usuario accede a este estado como resultado de la búsqueda o de mostrar las páginas del sitio.
- *Muestra páginas*: sigue las conexiones dentro del sitio. Éste es el estado alcanzado una vez que el usuario ha seleccionado una de las conexiones disponibles en la página principal para ver cualquiera de las páginas del sitio.

Los otros estados específicos para el CBMG de la tienda en línea son los siguientes:

- *Aŭade al carro de la compra*. Añade el artículo seleccionado al carro de la compra.
- *Pagar*. Paga los artículos que han sido seleccionados en el carro de la compra y sale de la tienda.

Con el objetivo de calcular las tasas de visita a cada uno de los estados se ha simplificado el gráfico anterior a otro más sencillo que se presenta en la Figura 6.5.

Desarrollemos este ejemplo del modelo de comportamiento de los usuarios en su acceso a un servidor web. Este CBMG simplificado tiene siete estados; el estado salida no está indicado específicamente. Llamemos $V_j$ al número medio de veces que se visita el estado $j$ por cada vez que se visita el sitio de comercio electrónico. El número de visitas al estado “Añade al carro de la compra” es igual al número de visitas al estado “Selecciona” multipli- cado por la probabilidad de que el usuario pase del estado “Selecciona” al estado “Añade al carro de la compra”. Por lo tanto, podemos escribir la siguiente relación:

$$V_{\text{Añade al carro de la compra}} = V_{\text{Selecciona}} \times 0,2$$

El número de visitas al estado $j$ del CBMG es igual a la suma del número de visitas a todos los estados del CBMG multiplicado por la probabilidad de transición de cada uno de estos estados al estado $j$. Esto es, para cualquier estado $j$ ($j = 2, \ldots, n \; 1$) de ~~un~~ CBMG, se puede escribir la ecuación:

$$V_j = \sum_{k=1}^{n-1} V_k \; p_{k,j}$$

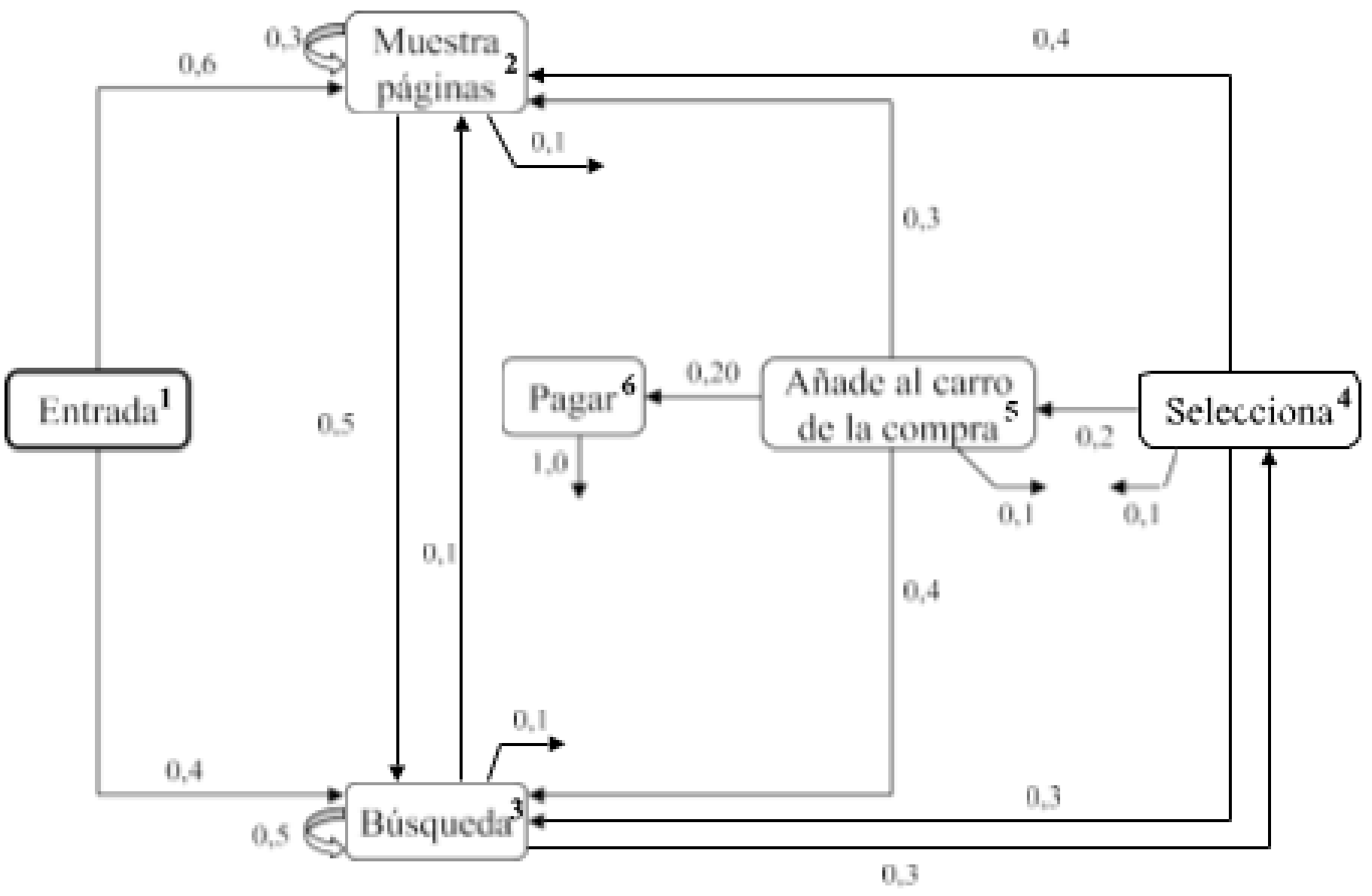

Figura 6.5: Gráfico simplificado del modelo de comportamiento del usuario de compras en línea.

siendo $p_{k,j}$ la probabilidad de que un usuario pase del estado $k$ al estado $j$. Dado que $V_1 = 1$ (porque el estado 1 es el estado de entrada), se puede encontrar el número medio de visitas $V_j$ resolviendo el sistema de ecuaciones lineal siguiente:

$$V_1 = 1$$

$$V_j = \sum_{k=1}^{n-1} V_k \; p_{k,j} \text{ siendo } j = 2, \ldots, n \qquad (1)$$

Los resultados obtenidos tras resolver el sistema de ecuaciones anteriores son los siguientes: $V_2 = 2{,}91$, $V_3 = 4{,}80$, $V_4 = 1{,}44$, $V_5 = 0{,}29$ y $V_6 = 0{,}06$.

Se pueden obtener medidas muy útiles de los CBMG, como la longitud media de la sesión, que es la suma de las tasas de visita, o el ratio de compra. En el caso que nos ocupa, la longitud media es la suma de las tasas de visita de los estados 2 al 6 que es igual a 9,467, y el ratio de compra por visita viene dado por $V_{pagar}$ que es de un 6 %.

De los archivos de bitácora de HTTP se pueden deducir los valores más importantes que caracterizan un determinado CBMG. Consideremos el caso de un conjunto de servidores web de un sitio de comercio electrónico. Se pueden juntar todos los archivos de bitácora de estos servidores utilizando la señal temporal del mismo. Cada línea de este archivo de bitácora podría tener las siguientes entradas:

- Identificación del usuario ($u$): identificación del usuario que ha realizado la solicitud. Se pueden utilizar herramientas para identificar las solicitudes que proceden del mismo usuario durante una sesión.

- Tipo de solicitud ($s$): indica el tipo de solicitud. Puede ser una solicitud para realizar una búsqueda, una selección de algunos de los resultados de la búsqueda, una solicitud de añadir un producto al carro de la compra, etc.
- Momento de la solicitud ($t$): instante en que la solicitud llega al sitio web. ▪

Tiempo de ejecución ($x$): tiempo de ejecución de la solicitud.

Posteriormente se utilizaría un algoritmo de aglutinación de las solicitudes de cada usuario agrupándolas por sesiones suponiendo que una sesión tiene una duración máxima, por ejemplo, de 30 minutos.

Pongamos un ejemplo. Supóngase que se ha analizado un archivo de bitácora (*log HTTP*) de un sitio de comercio electrónico que tiene un CBMG igual al presentado en la Figura 6.5. El algoritmo de aglutinación de solicitudes ha generado 20.000 sesiones con 340.000 líneas del archivo de bitácora de solicitudes. Después de ejecutar los algoritmos de agrupamiento en el archivo de bitácora de las sesiones utilizando seis grupos, se obtienen los que se presentan en la tabla siguiente:

| Grupo | 1 | 2 | 3 | 4 | 5 | 6 |
|---|---|---|---|---|---|---|
| Porcentaje de las sesiones | 46 % | 27 % | 11 % | 8 % | 6 % | 2 % |
| Tasa de compra por visita | 6 | 5 | 4 | 3 | 2 | 1,5 |
| Longitud de la sesión | 6 | 15 | 25 | 30 | 50 | 80 |
| $V_{\text{Añade al carro}}$ | 0,06 | 0,2 | 0,4 | 0,9 | 2 | 4 |
| $V_{\text{Muestra}} + V_{\text{Búsqueda}}$ | 4 | 12 | 20 | 25 | 40 | 70 |

La primera línea muestra el porcentaje de las sesiones que entran en cada grupo. Por ejemplo, el grupo 1 representa casi la mitad del total. La línea segunda muestra la tasa de compra por visitas realizadas, que representa el porcentaje de clientes que compran de la web. La longitud de la sesión indica el número medio de solicitudes efectuadas por cada usuario al sitio web. La línea cuarta muestra la tasa de añadir al carro de la compra, que representa el número medio de veces por sesión que un usuario añade un artículo al carro de la compra. Sin embargo esta operación no implica una compra, como se puede deducir de la comparación entre este valor y los de compra. La última línea de la tabla indica el número de operaciones de muestra y búsqueda asociadas con los clientes de cada grupo.

Del análisis de la tabla anterior se pueden deducir varios patrones de comportamiento muy distintos. El grupo 1, que representa la mayoría de las sesiones (46 %), tiene una longitud de la sesión muy corta (6) y el porcentaje más alto de clientes que compran de la página web. En el otro extremo se tiene el grupo 6 que representa la menor proporción de clientes que tiene la longitud de las sesiones mayor y el porcentaje menor de compras de la tienda. La tasa de compra por visita disminuye con la longitud de la sesión.

Hay otra manera distinta de calcular las tasas de visita a partir del archivo de bitácora (peticiones HTTP) que se va a explicar a continuación. En este caso se utiliza el modelo de visitas del cliente (*Customer Visit Model*, CVM) en lugar del CBMG. En la tabla siguiente se presenta la tasa de visitas realizada en diez sesiones.

| Sesión | $V_{Muestra}$ | $V_{Búsqueda}$ | $V_{Añade\ al\ carro}$ | $V_{Selecciona}$ | $V_{Paga}$ |
|---|---|---|---|---|---|
| 1 | 5 | 12 | 2 | 5 | 1 |
| 2 | 10 | 15 | 1 | 14 | 0 |
| 3 | 4 | 7 | 2 | 4 | 1 |
| 4 | 18 | 20 | 3 | 15 | 0 |
| 5 | 4 | 12 | 2 | 7 | 1 |
| 6 | 6 | 11 | 3 | 7 | 1 |
| 7 | 7 | 12 | 2 | 7 | 1 |
| 8 | 5 | 4 | 1 | 2 | 1 |
| 9 | 7 | 10 | 1 | 8 | 1 |
| 10 | 15 | 20 | 1 | 18 | 0 |

Necesitamos agrupar las sesiones en grupos más pequeños y representativos. Se aplican técnicas de agrupamiento tomando como medida de la distancia la distancia entre dos vectores de razones de visita. Considérense las sesiones $a$ y $b$ caracterizadas por los vectores tasas de visita $V_a = (V_2^a, \cdots, V_{n-1}^a)$ y $V_b = (V_2^b, \cdots, V_{n-1}^b)$. La distancia entre las sesiones $a$ y $b$ es la siguiente:

$$d_{V_a,V_b} = \sqrt{\sum_{i=2}^{n-1} \left(V_i^a - V_i^b\right)^2}$$

Los resultados de aplicar el algoritmo de agrupamiento hasta dos grupos a los valores de la tabla anterior se presentan en la tabla siguiente. El cálculo de cada uno de los pasos de este agrupamiento se detalla en el Problema 6.7.

| Grupo | $V_{Muestra}$ | $V_{Búsqueda}$ | $V_{Añade}$ | $V_{Selecciona}$ | $V_{Paga}$ |
|---|---|---|---|---|---|
| 6715938 | 5,38 | 8,19 | 1,56 | 5,13 | 1 |
| 4102 | 13,25 | 17,5 | 1,5 | 15,25 | 0 |

Se puede ver que el primer grupo (identificador 6715938) recoge a todos los clientes que compran y el segundo grupo aglutina a todos los usuarios que no compran. Se deduce que las características del comportamiento de los compradores entre sí y de los no compradores entre sí son muy semejantes porque se han agrupado en dos grupos separados, compradores y no compradores.

Para poder realizar la planificación de la capacidad y los estudios de dimensionamiento de un sitio de comercio electrónico se necesita trasladar los resultados de la caracterización de la carga de los CBMG al consumo de recursos informáticos. Los *diagramas de interacción cliente/servidor* (*Client/Server Interaction Diagram*, CSID) se utilizan para convertir los estados especificados en los CBMG en demandas de recursos de procesador, disco y red. De esta forma podemos caracterizar la carga a nivel de recurso. Por ejemplo, el CSID de la función de búsqueda en una página web se representa en la Figura 6.6. Esto es, la función de búsqueda se lanza desde un cliente, va primero al servidor web, de aquí pasa al servidor de aplicaciones que lanza la solicitud al servidor de base de datos, que realiza la búsqueda devolviéndola al servidor de aplicaciones, el cual lo devuelve al servidor web, que a su vez lo devuelve al cliente que lo había solicitado. En cada uno de los servidores se consume un tiempo de los recursos procesador y disco. En el envío de mensajes e información entre el cliente y los servidores, y los servidores entre sí, se consumirá un tiempo del recurso red.

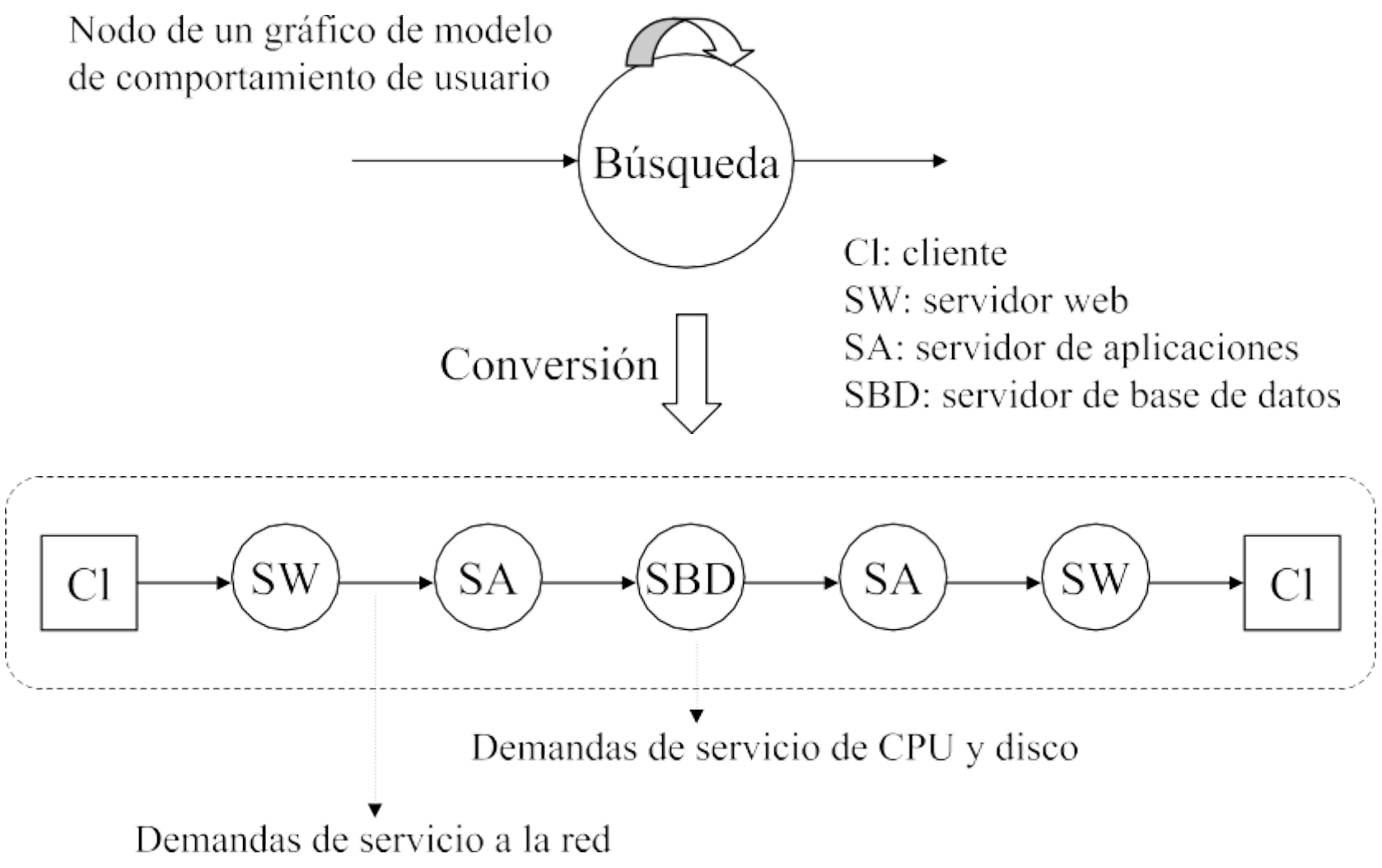

**Figura 6.6:** Diagrama de interacción cliente/servidor de la función de comercio electrónico de búsqueda.

Desarrollemos un ejemplo; supóngase que en el caso que nos ocupa el servidor de la base de datos tiene un procesador y dos discos con tiempos de demanda de servicio de 8, 18 y 16 ms, respectivamente, para una ejecución de la función búsqueda. La tabla siguiente muestra la tasa de llegada $\lambda$, en sesiones por segundo, la tasa de visita a la función búsqueda, y las demandas de servicio debidas a esta función, para cada grupo de clientes del ejemplo anterior (compradores y no compradores).

| Grupo | $\lambda$ | $V_{Búsqueda}$ | $D_{proc}$ (s) | $D_{d1}$ (s) | $D_{d2}$ (s) |
|---|---|---|---|---|---|
| Compradores | 0,07 | 8,18 | 0,0704 | 0,15858 | 0,13088 |
| No compradores | 0,03 | 17,5 | 0,14 | 0,315 | 0,28 |

Las demandas de servicio de los recursos procesador (p), disco 1 (d1) y disco 2 (d2) debido a la función de búsqueda realizada por los compradores (c) y los no compradores (nc), resultan del producto del número de búsquedas que se realizan por segundo por la demanda de recursos que realiza cada una de estas búsquedas. Esto es:

$$D_{p,c} = 8{,}18 \times 0{,}008 = 0{,}0704 \text{ ms}$$
$$D_{d1,c} = 8{,}18 \times 0{,}018 = 0{,}1586 \text{ ms}$$
$$D_{d2,c} = 8{,}18 \times 0{,}016 = 0{,}1301 \text{ ms}$$
$$D_{p,nc} = 17{,}5 \times 0{,}008 = 0{,}14 \text{ ms}$$
$$D_{d1,nc} = 17{,}5 \times 0{,}018 = 0{,}315 \text{ ms}$$
$$D_{d2,nc} = 17{,}5 \times 0{,}016 = 0{,}28 \text{ ms}$$

La utilización de los recursos es igual al producto de la demanda de servicio en ese recurso por la tasa de llegada. Esto es, la utilización de los recursos debido a la función de búsqueda es:

$$\lambda_c = 0{,}07 \times 8{,}18 = 0{,}05726 \text{ búsquedas/s}$$
$$U_{p,c} = 0{,}0704 \times 0{,}05726 = 0{,}0403 = 4{,}03\ \%$$
$$U_{d1,c} = 0{,}1586 \times 0{,}05726 = 0{,}0908 = 9{,}08\ \%$$
$$U_{d2,c} = 0{,}1301 \times 0{,}05726 = 0{,}0749 = 7{,}49\ \%$$
$$\lambda_{nc} = 0{,}03 \times 17{,}5 = 0{,}525 \text{ búsquedas/s}$$
$$U_{p,nc} = 1{,}4 \times 0{,}525 = 0{,}0735 = 7{,}35\ \%$$
$$U_{d1,nc} = 0{,}315 \times 0{,}525 = 0{,}1653 = 16{,}53\ \%$$
$$U_{d2,nc} = 0{,}28 \times 0{,}525 = 0{,}147 = 14{,}7\ \%$$

Lo que se ha hecho para la función de búsqueda se podría llevar a cabo para el resto de las funciones del servidor web, obteniendo el total de la demanda de los recursos procesador, disco 1 y disco 2, así como el total de su utilización.

## 6.5. Problemas resueltos

**PROBLEMA 6.1** En un sistema informático se procesa una carga real $W$, generándose a partir de ella dos modelos, $W_1$ y $W_2$. Hay tres tipos de aplicaciones, una de cálculo ($C$), otra de lectura ($L$) y una tercera intermedia ($I$), según se especifica en la tabla anexa. En ella se indica el tiempo de procesador, el tiempo de presencia y el número de operaciones de E/S lógicas. Los tiempos vienen expresados en segundos.

| Parámetros | $WC$ | $WL$ | $WI$ | $W_1C$ | $W_1L$ | $W_1I$ | $W_2C$ | $W_2L$ | $W_2I$ |
|---|---|---|---|---|---|---|---|---|---|
| Procesador | 80 | 30 | 40 | 90 | 25 | 42 | 190 | 15 | 80 |
| Presencia | 1.000 | 400 | 500 | 1.100 | 350 | 530 | 2.000 | 200 | 250 |
| Operaciones E/S | 300 | 5.000 | 1.000 | 600 | 2.500 | 500 | 280 | 4.000 | 900 |

Los rangos asignados a cada parámetro de la carga se indican a continuación:

| Parámetros | mín | máx | Rango (máx − mín) |
|---|---|---|---|
| Procesador | 20 | 220 | 200 |
| Presencia | 200 | 2.200 | 2.000 |
| Operaciones E/S | 100 | 3.100 | 3.000 |

Dadas las características de ambos modelos, se pide analizar cuál de los dos explica mejor la carga real $W$ , de acuerdo con los objetivos y criterios de ponderación que se exponen a continuación:

1. Objetivo 1: dar más importancia a la actividad del procesador y al tiempo de pre- sencia ($w_{11}$ = 2, $w_{21}$ = 2, $w_{31}$ = 0).
2. Objetivo 2: dar más importancia a la actividad de E/S lógicas ($w_{12}$ = 0, $w_{22}$ = 1, $w_{32}$ = 2).

La variable $w_{ij}$ representa el peso del parámetro $i$ en el objetivo $j$. Los porcentajes de programas existentes de cada clase son: $q_1$ = 40 %, $q_2$ = 40 % y $q_3$ = 20 %.

**SOLuCIón:** Se van a seguir los pasos indicados en la parte teórica de este capítulo especifi- cados en el apartado de medida de la representatitividad de un modelo de carga. Por lo tanto, debemos dar los siguientes pasos:

1. Se normalizan los parámetros de todos los modelos y de la carga real. Por ejemplo, el valor normalizado del tiempo de procesador, del tiempo de presencia y de las operaciones lógicas de E/S para los programas de cálculo de la carga real se calculan así:

$$T_{\text{Procesador}} = \frac{80-20}{220-20} = 0,3$$

$$= \frac{1.000-200}{2.200-200} = 0,4$$

$$T_{\text{Presencia}} = \frac{300-100}{3.100-100} = 0,67$$

$\text{Operaciones}_{E/S}$

Procediendo de igual manera con todos los programas, la tabla normalizada resultante es la siguiente:

| Parámetros | $WC$ | $WL$ | $WI$ | $W_1C$ | $W_1L$ | $W_1I$ | $W_2C$ | $W_2L$ | $W_2I$ |
|---|---|---|---|---|---|---|---|---|---|
| Procesador | 0,3 | 0,05 | 0,1 | 0,35 | 0,025 | 0,11 | 0,85 | 0 | 0,3 |
| Presencia | 0,4 | 0,1 | 0,15 | 0,45 | 0,075 | 0,165 | 0,9 | 0 | 0,025 |
| Operaciones E/S | 0,67 | 1,63 | 0,3 | 0,17 | 0,8 | 0,13 | 0,06 | 1,3 | 0,27 |

2. Se calcula la distancia entre los valores normalizados de los modelos y el valor normalizado de la carga real como el valor absoluto de la diferencia de los valores normalizados. La tabla de distancias obtenida es la siguiente:

| Parámetros | $D'C$ | $D'L$ | $D'I$ | $D''C$ | $D''I$ | $D''I$ |
|---|---|---|---|---|---|---|
| Procesador | 0,05 | 0,025 | 0,01 | 0,55 | 0,05 | 0,2 |
| Presencia | 0,05 | 0,025 | 0,015 | 0,5 | 0,1 | 0,125 |
| Operaciones E/S | 0,1 | 0,83 | 0,17 | 0,006 | 0,33 | 0,03 |

3. Se pondera para el objetivo 1 doblemente la distancia entre los parámetros del modelo y los de la carga que se quieren modelizar, resultando:

$$D'_1 = 0,4 \times 0,2 + 0,4 \times 0,1 + 0,2 \times 0,05 = 0,13$$
$$D''_1 = 0,4 \times 2,1 + 0,4 \times 0,3 + 0,2 \times 0,65 = 8,65$$

ya que:

$$
\begin{aligned}
0,05 \times 2 + 0,05 \times 2 + 0,1 \times 0 &= 0,2 \\
0,025 \times 2 + 0,025 \times 2 + 0,83 \times 0 &= 0,1 \\
0,01 \times 2 + 0,015 \times 2 + 0,17 \times 0 &= 0,05 \\
0,55 \times 2 + 0,5 \times 2 + 0,006 \times 0 &= 2,1 \\
0,05 \times 2 + 0,1 \times 2 + 0,33 \times 0 &= 0,3 \\
0,2 \times 2 + 0,125 \times 2 + 0,03 \times 0 &= 0,65
\end{aligned}
$$

Los pesos asociados a los parámetros $w_{11}$ y $w_{21}$ son de 2 para el objetivo 1, porque en este objetivo se mide la representatividad del modelo de carga respecto de la actividad del procesador y el tiempo de presencia, que son específicamente los parámetros 1 y 2. Para este objetivo 1 el peso de las actividades de E/S lógicas es nulo, y por este motivo $w_{31} = 0$.

El modelo que dista menos de la carga real es el más representativo para este objetivo primero, esto es, el modelo $W_1$.

4. Se pondera para el objetivo 2 doblemente la distancia entre los parámetros del modelo y los de la carga que se quieren modelizar, resultando:

$$D_2' = 0{,}4 \times 0{,}25 + 0{,}4 \times 1{,}685 + 0{,}2 \times 0{,}345 = 0{,}843$$
$$D_2 = 0{,}4 \times 0{,}512 + 0{,}4 \times 0{,}76 + 0{,}2 \times 0{,}185 = 0{,}7788$$

ya que:

$$\begin{aligned}
0{,}05 \times 0 + 0{,}05 \times 1 + 0{,}1 \times 2 &= 0{,}25 \\
0{,}025 \times 0 + 0{,}025 \times 1 + 0{,}83 \times 2 &= 1{,}685 \\
0{,}01 \times 0 + 0{,}015 \times 1 + 0{,}17 \times 2 &= 0{,}345 \\
0{,}55 \times 0 + 0{,}5 \times 1 + 0{,}006 \times 2 &= 0{,}512 \\
0{,}05 \times 0 + 0{,}1 \times 1 + 0{,}33 \times 2 &= 0{,}76 \\
0{,}2 \times 0 + 0{,}125 \times 1 + 0{,}03 \times 2 &= 0{,}185
\end{aligned}$$

El peso asociado al parámetro $w_{32}$ es de 2 para el objetivo 2 porque en este objetivo se mide la representatividad del modelo de carga respecto de la actividad de E/S lógicas, que es específicamente el parámetro 3. Para este objetivo 2 el peso de la actividad del procesador es nulo, y por este motivo, $w_{12} = 0$. Sin embargo, el peso asociado al parámetro $w_{22}$ es de 1 porque en este objetivo influye en algo el tiempo de presencia que es el parámetro 2.

El modelo que dista menos de la carga real es el más representativo para este segundo objetivo, esto es, el modelo $W_2$. ■

**PROBLEMA 6.2** La tabla siguiente contiene los valores de los parámetros que caracterizan la carga real $W$ de un sistema informático y dos modelos $W_1$ y $W_2$ de representación de la misma.

| Parámetro | $WC1$ | $WC2$ | $WC3$ |
|---|---|---|---|
| % de la carga total | 50 % | 30 % | 20 % |
| Tiempo de procesador (s) | 100 | 50 | 75 |
| Operaciones E/S | 200 | 1.000 | 700 |
| Memoria real(MB) | 300 | 550 | 400 |
| Ficheros de disco | 10 | 6 | 7 |
| Páginas impresas | 2.000 | 3.000 | 2.300 |
| Parámetro | $W_1C1$ | $W_1C2$ | $W_1C3$ |
| Tiempo de procesador (s) | 100 | 50 | 80 |
| Operaciones E/S | 240 | 1.350 | 950 |
| Memoria real (MB) | 300 | 560 | 410 |
| Ficheros de disco | 11 | 10 | 12 |
| Páginas impresas | 2.050 | 3.100 | 2.200 |
| Parámetro | $W_2C1$ | $W_2C2$ | $W_2C3$ |
| Tiempo de procesador (s) | 140 | 70 | 95 |
| Operaciones E/S | 200 | 1.000 | 800 |
| Memoria real (MB) | 330 | 580 | 410 |
| Ficheros de disco | 13 | 8 | 10 |
| Páginas impresas | 2.100 | 3.200 | 2.400 |

Los rangos se indican en la tabla siguiente:

| Parámetro | Rango (máximo − mínimo) |
|---|---|
| Tiempo de procesador (s) | 200 − 0 |
| Operaciones E/S | 2.100 − 100 |
| Memoria real (MB) | 1.000 − 100 |
| Ficheros de disco | 20 − 0 |
| Páginas impresas | 4.100 − 100 |

Se pide calcular cuál de los dos modelos caracteriza mejor la carga real para cada uno de los siguientes objetivos:

- Objetivo 1: con respecto a la actividad del procesador, dando los siguientes pesos a cada uno de los cinco parámetros: $w_1 = 2$, $w_2 = 1$, $w_3 = 1$, $w_4 = 1$ y $w_5 = 0{,}5$.
- Objetivo 2: con respecto a la actividad de operaciones de E/S, dando los siguientes pesos a cada uno de los cinco parámetros: $w_1 = 1$, $w_2 = 2$, $w_3 = 1$, $w_4 = 2$ y $w_5 = 0{,}5$.
- Objetivo 3: con respecto a la actividad conjunta de E/S, procesador y memoria principal, dando los siguientes pesos a cada uno de los cinco parámetros: $w_1 = 1$, $w_2 = 1$, $w_3 = 1$, $w_4 = 1$ y $w_5 = 1$.

**SOLUCIÓN:** Se van a seguir los siguientes pasos:

1. Se normalizan los parámetros de todos los modelos y de la carga real. La tabla normalizada resultante es la siguiente:

| Parámetros | $WC1$ | $WC2$ | $WC3$ |
|---|---|---|---|
| Procesador | 0,5 | 0,25 | 0,375 |
| Op. E/S | 0,05 | 0,45 | 0,3 |
| Memoria | 0,2 | 0,45 | 0,3 |
| Ficheros | 0,5 | 0,3 | 0,35 |
| Páginas | 0,475 | 0,725 | 0,55 |

| Parámetros | $W_1C1$ | $W_1C2$ | $W_1C3$ | $W_2C1$ | $W_2C2$ | $W_2C3$ |
|---|---|---|---|---|---|---|
| Procesador | 0,5 | 0,25 | 0,4 | 0,7 | 0,35 | 0,475 |
| Op. E/S | 0,07 | 0,625 | 0,425 | 0,05 | 0,45 | 0,35 |
| Memoria | 0,2 | 0,46 | 0,31 | 0,23 | 0,48 | 0,31 |
| Ficheros | 0,55 | 0,4 | 0,6 | 0,65 | 0,4 | 0,5 |
| Páginas | 0,488 | 0,75 | 0,525 | 0,5 | 0,775 | 0,575 |

2. Se calcula la distancia entre los valores normalizados de los modelos y el valor normalizado de la carga real como el valor absoluto de la diferencia de los valores normalizados. La tabla de distancias que se obtiene es:

| Parámetros | $D_{C1}$ | $D_{C2}$ | $D_{C3}$ | $D_{C1}$ | $D_{C2}$ | $D_{C3}$ |
|---|---|---|---|---|---|---|
| Tiempo de procesador (s) | 0 | 0 | 0,025 | 0,2 | 0,1 | 0,1 |
| Operaciones de E/S | 0,02 | 0,175 | 0,125 | 0 | 0 | 0,05 |
| Memoria real (MB) | 0 | 0,01 | 0,01 | 0,03 | 0,03 | 0,01 |
| Ficheros de disco | 0,05 | 0,1 | 0,25 | 0,05 | 0,05 | 0,15 |
| Páginas impresas | 0,013 | 0,025 | 0,025 | 0,025 | 0,05 | 0,025 |

3. Se pondera para el objetivo 1 doblemente la distancia entre los parámetros del modelo y los de la carga que se quieren modelizar:

$$D'_1 = 0,5 \times 0,076 + 0,3 \times 0,31 + 0,2 \times 0,477 = 0,216$$

$$D_1 = 0,5 \times 0,1325 + 0,3 \times 0,305 + 0,2 \times 0,4225 = 0,242$$

ya que:

$$0 \times 2 + 0,02 \times 1 + 0 \times 1 + 0,05 \times 1 + 0,013 \times 0,5 = 0,076$$

$$0 \times 2 + 0,175 \times 1 + 0,01 \times 1 + 0,1 \times 1 + 0,025 \times 0,5 = 0,31$$

$$0,025 \times 2 + 0,125 \times 1 + 0,01 \times 1 + 0,25 \times 1 + 0,025 \times 0,5 = 0,477$$

$$0,2 \times 2 + 0 \times 1 + 0,03 \times 1 + 0,05 \times 1 + 0,025 \times 0,5 = 0,1325$$

$$0,1 \times 2 + 0 \times 1 + 0,03 \times 1 + 0,05 \times 1 + 0,05 \times 0,5 = 0,305$$

$$0,1 \times 2 + 0,05 \times 1 + 0,01 \times 1 + 0,15 \times 1 + 0,025 \times 0,5 = 0,4225$$

El peso asociado (la importancia) al parámetro $w_{11}$ es de 2 para el objetivo 1 porque en este objetivo se mide la representatividad del modelo de carga respecto de la actividad de procesador, que es específicamente el parámetro 1. Para este objetivo 1 el peso de las operaciones de E/S en disco, la memoria real utilizada y el número de ficheros utilizados es menor que el del tiempo del procesador, exactamente la mitad, igual a 1; y por este motivo, $w_{21} = w_{31} = w_{41} = 1$. Para este objetivo 1 el peso del número de líneas impresas es muy pequeño, exactamente una cuarta parte, y por tanto, se tiene $w_{51} = 0,5$. Cabe destacar que en este ejercicio todos los parámetros influyen con distinto peso en el objetivo, mientras que en el ejercicio anterior algunos de los pesos se igualaban a cero. Esto puede hacer que las diferencias entre los modelos sea menor.

El modelo que dista menos de la carga real es el más representativo para este primer objetivo, esto es, el modelo $W_1$. Cabe destacar que la diferencia de las distancias entre los dos modelos y la carga real es muy pequeña.

4. Se pondera para el objetivo 2 doblemente la distancia entre los parámetros del modelo y los de la carga que se quieren modelar:

$$_2 = 0,5 \times 0,1465 + 0,3 \times 0,5725 + 0,2 \times 0,7975 = 0,4045$$

$$_2 = 0,5 \times 0,3425 + 0,3 \times 0,255 + 0,2 \times 0,612 = 0,260$$

ya que:

$$0 \times 1 + 0,02 \times 2 + 0 \times 1 + 0,05 \times 2 + 0,013 \times 0,5 = 0,1465$$

$$0 \times 1 + 0,175 \times 2 + 0,01 \times 1 + 0,1 \times 2 + 0,025 \times 0,5 = 0,5725$$

$$0,025 \times 1 + 0,125 \times 2 + 0,01 \times 1 + 0,25 \times 2 + 0,025 \times 0,5 = 0,7975$$

$$0,2 \times 1 + 0 \times 2 + 0,03 \times 1 + 0,05 \times 2 + 0,025 \times 0,5 = 0,3425$$

$$0,1 \times 1 + 0 \times 2 + 0,03 \times 1 + 0,05 \times 2 + 0,05 \times 0,5 = 0,255$$

$$0,1 \times 1 + 0,05 \times 2 + 0,01 \times 1 + 0,15 \times 2 + 0,025 \times 0,5 = 0,612$$

El peso asociado (la importancia) al parámetro $w_{22}$ y $w_{42}$ es de 2 para el objetivo 2, porque en este objetivo se mide la representatividad del modelo de carga respecto de la actividad de las operaciones de E/S en disco, que es específicamente el parámetro 2, con que está íntimamente relacionado también el parámetro 4. Para este objetivo 2 el peso del tiempo de procesador y la memoria real utilizada es menor que los dos anteriores, exactamente la mitad, igual a 1; y por este motivo, $w_{12} = w_{32} = 1$. Para este objetivo 2 el peso (la importancia) del número de líneas impresas es muy pequeño, exactamente una cuarta parte, y en consecuencia, $w_{52} = 0,5$.

El modelo que dista menos de la carga real es el más representativo para este segundo objetivo, esto es, el modelo $W_2$.

5. *Se pondera para el objetivo* 3 *doblemente la distancia entre los parámetros del modelo y los de la carga que se quieren modelizar* :

$$D'_3 = 0{,}5 \times 0{,}083 + 0{,}3 \times 0{,}310 + 0{,}2 \times 0{,}425 = 0{,}2195$$

$$D_3 = 0{,}5 \times 0{,}305 + 0{,}3 \times 0{,}23 + 0{,}2 \times 0{,}335 = 0{,}289$$

ya que:

$$0 + 0{,}02 + 0 + 0{,}05 + 0{,}013 = 0{,}083$$
$$0 + 0{,}175 + 0{,}01 + 0{,}1 + 0{,}025 = 0{,}310$$
$$0{,}025 + 0{,}125 + 0{,}01 + 0{,}25 + 0{,}025 = 0{,}425$$
$$0{,}2 + 0 + 0{,}03 + 0{,}05 + 0{,}025 = 0{,}305$$
$$0{,}1 + 0 + 0{,}03 + 0{,}05 + 0{,}05 = 0{,}23$$
$$0{,}1 + 0{,}05 + 0{,}01 + 0{,}15 + 0{,}025 = 0{,}335$$

El peso asociado (la importancia) a cada uno de los cinco parámetros es igual para darle igual importancia a cada uno de ellos.

El modelo que dista menos de la carga real es el más representativo para este tercer objetivo, esto es, el modelo $W_1$. ■

**PROBLEMA 6.3** Un servidor de base de datos de un servidor web se monitorizó durante un periodo de pico de treinta minutos. Durante este periodo, el tiempo de procesador utilizado y el número de operaciones de E/S de cada una de las 1.200 transacciones se grabaron. Estas 1.200 transacciones se agruparon en las doce clases que se recogen en la tabla siguiente.

| Clase | Tiempo de procesador (ms) | Operaciones de E/S |
|---|---|---|
| 1 | 20 | 7 |
| 2 | 50 | 25 |
| 3 | 50 | 200 |
| 4 | 60 | 180 |
| 5 | 12 | 15 |
| 6 | 45 | 20 |
| 7 | 50 | 25 |
| 8 | 55 | 200 |
| 9 | 12 | 30 |
| 10 | 9 | 25 |
| 11 | 22 | 48 |
| 12 | 52 | 20 |

Se pide hacer una valoración razonada de si es aconsejable realizar un agrupamiento hasta llegar a una sola clase o es mejor agrupar la carga en varias clases.

**SOLuCIón:** El primer paso es ordenar la tabla de datos para ver si existe una cierta agrupación natural de estos datos. La representación gráfica, presentada en la Figura 6.7, nos ayuda a ver que existen tres clases o grupos naturales bien diferenciados, que son los siguientes:

1. Transacciones que necesitan poco tiempo de procesador y pocas operaciones de E/S.
2. Transacciones que demandan mucho tiempo de procesador y muchas operaciones de E/S.
3. Transacciones que demandan mucho tiempo de procesador pero pocas operaciones de E/S.

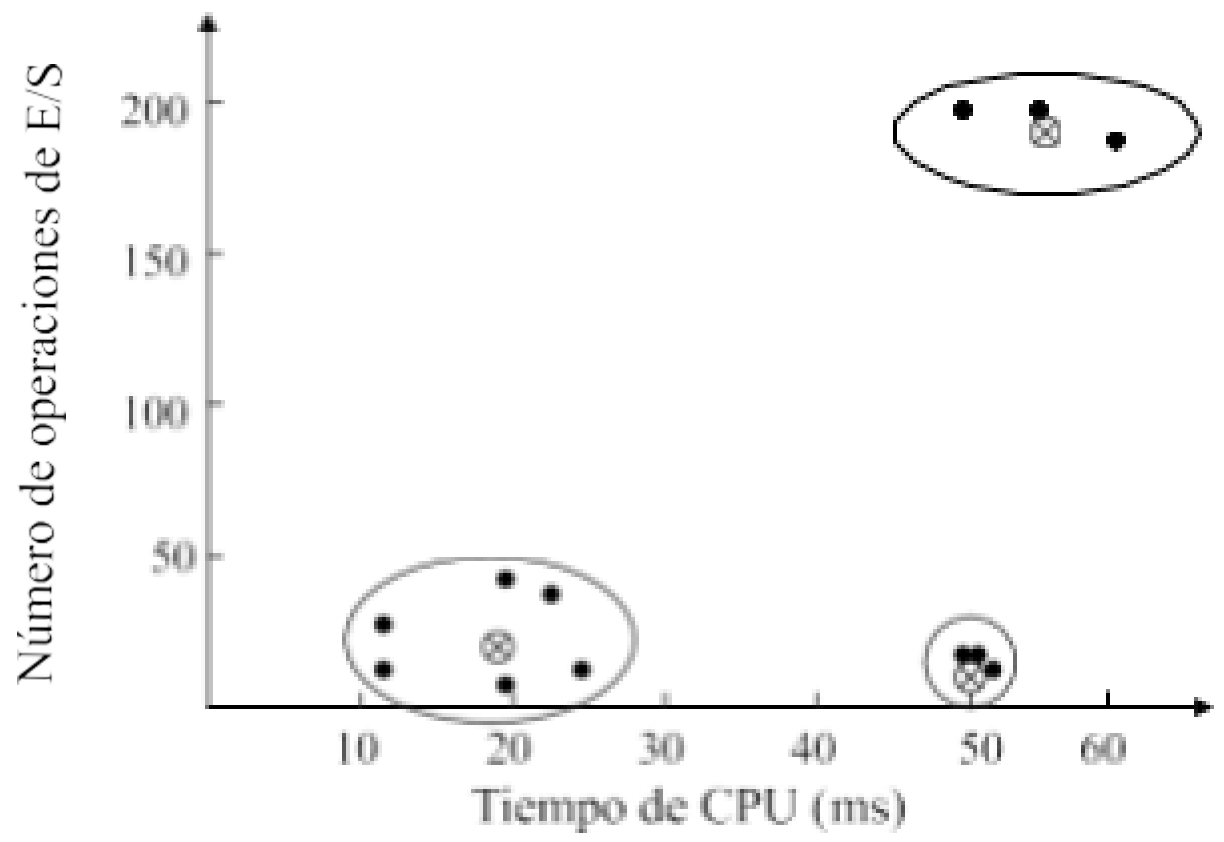

**Figura 6.7:** Agrupamiento realizado en el Problema 6.3.

Un modelo de carga compuesto por las coordenadas de las tres clases suministra una repre- sentación más compacta del consumo de los recursos de la carga que aquella obtenida por la unión de las tres clases en una sola clase. La Figura 6.7 también presenta el centroide de cada grupo natural. Está claro que si se representaran todos los puntos por sólo un punto, el caso de una sola clase, se obtendría una representación con mucho menos conocimiento y sentido de la carga global que aquella suministrada por las tres clases. Por lo tanto, el número de clases elegido para representar la carga afecta a la exactitud y riqueza de información del modelo de carga.

---

**PROBLEMA 6.4** Se tiene un servidor web de documentos. La tabla siguiente presenta los tiempos de procesador, de entrada/salida y de presencia en el sistema de los diez tipos de solicitudes de documentos a ese servidor web. Evidentemente, esto sólo es un ejemplo para realizar un problema de cálculo, se necesitaría un número mayor de muestras para que fuese significativo.

| Solicitud | Procesador (ms) | E/S (ms) | Ejecución (ms) |
|---|---|---|---|
| 1 | 9,4 | 39 | 70 |
| 2 | 12,9 | 10 | 144 |
| 3 | 15,3 | 11,9 | 154 |
| 4 | 8,7 | 39 | 64 |
| 5 | 11 | 90 | 112 |
| 6 | 17 | 140 | 162 |
| 7 | 216 | 1.200 | 4.400 |
| 8 | 130 | 120 | 152 |
| 9 | 90 | 50 | 150 |
| 10 | 170 | 140 | 190 |

El tiempo de ejecución corresponde al tiempo total que el servidor necesita para ser- vir la solicitud. Se solicita hacer una caracterización de la carga siguiendo un criterio de alta representatividad, teniendo en cuenta que el tiempo de ejecución es directamente pro- porcional al tamaño del documento y utilizando la distancia euclídea como métrica de agrupamiento.

**SOLuCıón:** Suponiendo que todos los tipos de solicitudes aparezcan con la misma frecuencia, la primera idea que se nos viene a la mente es agrupar todos los tipos de solicitudes en un solo tipo, denominándola solicitud tipo de la web, calculando la media aritmética de los valores de los recursos utilizados. Los parámetros de caracterización son el tiempo de procesador y el tiempo de E/S. La tabla siguiente muestra el par de parámetros de una solicitud media y el total del tiempo debido a los diez tipos de solicitudes que caracteriza a la carga de la tabla de datos.

| Solicitud | Procesador (ms) | E/S (ms) | Número de componentes |
|---|---|---|---|
| Única | 68 | 184 | 10 |
| Totales | 680,3 | 1.840 | 10 |

La carga real está representada ahora por un modelo compuesto por diez solicitudes carac- terizadas por el par de parámetros $\{68, 184\}$, que coloca en el servidor las mismas demandas de procesador y E/S que las solicitudes originales.

Pero el enunciado del problema nos pide la caracterización de la carga siguiendo un *criterio de alta representatividad*, esto es, mediante varias clases si esto es posible para conseguir una alta representatividad. Un análisis más detallado de los parámetros nos anima a hacer una agrupación en tres clases distintas para conseguir una representación más compacta. Además, cada clase se puede asociar con el tiempo de ejecución o el tamaño del documento. Una clase incluye todos aquellos componentes que son similares entre sí en el uso de recursos. Por lo tanto, se definen tres clases de documentos: pequeños, medianos y grandes, según el siguiente orden:

- La primera clase, llamada de *documentos pequeños*, incluye aquellas solicitudes en que los tiempos de ejecución es menor de 120, que recogen las solicitudes 1, 4 y 5.
- La segunda clase, llamada de *documentos medianos*, consiste en todas aquellas solicitudes con tiempo de ejecución entre 120 y 2.000, que recogen las solicitudes 2, 3, 6, 8, 9 y 10.
- La tercera clase, llamada de *documentos grandes*, que recoge la solicitud que tiene tiempo de ejecución mayor de 2.000, esto es, la solicitud 7.

La tabla siguiente presenta los tres pares de parámetros medios que representan ahora la carga original.

| Tipo de solicitud | Procesador (ms) | E/S (ms) | Componentes |
|---|---|---|---|
| Documentos pequeños | 9,7 | 56 | 3 |
| Documentos medianos | 72,5 | 78,6 | 6 |
| Documentos grandes | 216 | 1.200 | 1 |
| Total | 68 | 2.000 | 10 |

La agrupación de las solicitudes en tres clases guarda algunas características importantes de la carga real y hace más clara la distinción entre las solicitudes que demandan cantidades diferentes de tiempo de recursos.

La resolución de este problema se ha realizado mediante una agrupación subjetiva y utili- zando la media como técnica de agrupamiento y no siguiendo un criterio más científico como el que se podría haber utilizado empleando las técnicas de agrupamiento explicadas en la parte teórica. Además, se ha realizado una agrupación manual en función del tiempo de ejecución para simplificar la resolución del problema. Habitualmente realizaremos el agrupamiento utilizando dos parámetros, que suelen ser el tiempo de procesador y el tiempo o el número de operaciones de E/S. Por este motivo hemos agrupado manualmente en este caso en función del tiempo de ejecución y, posteriormente, hemos realizado un agrupamiento más científico y metodológico utilizando la técnica de agrupamiento sobre el tiempo de procesador y de E/S. ■

---

**PROBLEMA 6.5** Supóngase que se tiene un servidor web de documentos que está dedicado a entregar documentos HTML bajo demanda. Se grabaron en los archivos de bitácora el comportamiento de la ejecución de todas las solicitudes. Se seleccionó una muestra aleatoria de siete solicitudes en un periodo específico con el objeto de tomar una muestra pequeña. La tabla siguiente presenta los parámetros que caracterizan la carga para el propósito de este estudio.

| Documento | Tamaño (KB) | Accesos |
|---|---|---|
| 1 | 40 | 70 |
| 2 | 4 | 260 |
| 3 | 11 | 300 |
| 4 | 100 | 10 |
| 5 | 5 | 280 |
| 6 | 30 | 100 |
| 7 | 90 | 25 |

Se pide reducir a tres clases la carga de la muestra seleccionada.

**SOLUCIÓN:** Si se analizan los datos de la carga se deduce que se trata de valores de magnitud muy diferente, y por lo tanto es aconsejable hacer un cambio de escala que los aproxime un poco más. Vamos a presentar dos técnicas para cumplir con este objetivo. La primera efectúa el escalado mediante la normalización a cero de la media y la varianza, esto es, presenta los valores normalizados centrados y reducidos. La segunda utiliza la conversión a valores logarítmicos. Esta última será la empleada en la resolución completa del problema, mientras que la primera se emplea únicamente para ilustrar un ejemplo de su uso.

La tabla siguiente presenta los valores transformados mediante la normalización a cero de la media y la varianza. El valor medio del tamaño y del número de accesos es 40 y 149,29, respectivamente. Por otro lado, la desviación típica del tamaño es 39,92 y la del número de accesos es 126,24. Las dos columnas de la derecha muestran el valor calculado como $(x\ m)/\sigma$, donde $x$ es el valor medido de la tabla de datos, $m$ la media y $\sigma$ la desviación típica de la columna de datos respectiva.

| Documento | $\frac{x-m}{\sigma}$ del tamaño | $\frac{x-m}{\sigma}$ de accesos |
|---|---|---|
| 1 | 0,00 | 0,63 |
| 2 | 0,90 | 0,88 |
| 3 | 0,73 | 1,19 |
| 4 | 1,50 | 1,10 |
| 5 | 0,88 | 1,06 |
| 6 | 0,25 | 0,39 |
| 7 | 1,25 | 0,98 |

La tabla siguiente muestra los valores transformados por la función $\log_{10} x$, siendo $x$

el valor de la tabla de datos.

| Documento | $\log_{10} x$ del tamaño | $\log_{10} x$ de accesos |
|---|---|---|
| 1 | 1,60 | 1,85 |
| 2 | 0,60 | 2,41 |
| 3 | 1,04 | 2,48 |
| 4 | 2,00 | 1,00 |
| 5 | 0,70 | 2,45 |
| 6 | 1,48 | 2,00 |
| 7 | 1,95 | 1,40 |

Se va a utilizar la distancia euclídea como la métrica para evaluar las distancias entre las clases. El número de clases inicial es igual al número de datos de partida, y los valores de los parámetros de los centroides coinciden con los valores de los parámetros de los componentes, ya representados en la tabla anterior.

Se calculan las distancias entre las clases. La variable $d_{CiCj}$ representa la distancia euclídea entre la clase $i$ y la clase $j$. Solamente indicamos dos de ellas para que se vea cómo se calculan:

$$d_{C1C2} = \sqrt{(1,60-0,60)^2 + (1,85-2,41)^2} = 1,15$$

$$d_{C1C3} = \sqrt{(1,60-1,04)^2 + (1,85-2,48)^2} = 0,84$$

Los resultados de estos cálculos se presentan en la primera matriz de distancias entre clases de este ejercicio, en la tabla siguiente:

| Clase | C1 | C2 | C3 | C4 | C5 | C6 | C7 |
|---|---|---|---|---|---|---|---|
| C1 | 0 | 1,15 | 0,84 | 0,93 | 1,09 | 0,20 | 0,57 |
| C2 | | 0 | 0,44 | 1,99 | 0,10 | 0,97 | 1,69 |
| C3 | | | 0 | 1,76 | 0,34 | 0,65 | 1,41 |
| C4 | | | | 0 | 1,95 | 1,13 | 0,40 |
| C5 | | | | | 0 | 0,90 | 1,64 |
| C6 | | | | | | 0 | 0,77 |
| C7 | | | | | | | 0 |

La distancia mínima entre las clases de la tabla anterior es la que se produce entre las clases C2 y C5, esto es, 0,10. Las coordenadas del centroide de esta nueva clase son $\frac{0,60+0,70}{2} = 0,65$ y $\frac{2,41+2,45}{2} = 2,43$.

Ahora se tienen seis clases. Se recalcula la matriz de distancias para el nuevo conjunto de clases aplicando otra vez la ecuación de la distancia euclídea. Cabe tener en cuenta que las únicas distancias nuevas que hemos de calcular son las de la nueva clase con las clases restantes. Los resultados se presentan en la segunda matriz de distancias entre clases de este ejercicio, en la tabla siguiente:

| Clase | C1 | C25 | C3 | C4 | C6 | C7 |
|---|---|---|---|---|---|---|
| C1 | 0 | 1,12 | 0,84 | 0,93 | 0,20 | 0,57 |
| C25 | | 0 | 0,39 | 1,97 | 0,93 | 1,66 |
| C3 | | | 0 | 1,76 | 0,65 | 1,41 |
| C4 | | | | 0 | 1,13 | 0,40 |
| C6 | | | | | 0 | 0,77 |
| C7 | | | | | | 0 |

La distancia mínima en esta última tabla es la que se produce entre las clases C1 y C6, esto es, 0,20. Las coordenadas del centroide de esta nueva clase son $\frac{1,60+1,48}{2} = 1,54$ y $\frac{1,84+2}{2} = 1,92$.

Tenemos ahora cinco clases. Recalculamos la matriz de distancias para el nuevo conjunto de clases aplicando otra vez la ecuación de la distancia euclídea. Las únicas distancias nuevas que hemos de calcular son las de la nueva clase con las clases restantes. Los resultados los presentamos en la tercera matriz de distancias entre clases de este ejercicio, en la tabla siguiente:

| Clase | C16 | C25 | C3 | C4 | C7 |
|---|---|---|---|---|---|
| C16 | 0 | 1,02 | 0,75 | 1,03 | 0,67 |
| C25 | | 0 | 0,39 | 1,97 | 1,66 |
| C3 | | | 0 | 1,76 | 1,41 |
| C4 | | | | 0 | 0,40 |
| C7 | | | | | 0 |

La nueva distancia mínima entre las clases de la tabla anterior se produce entre las clases C25 y C3, esto es, 0,39. Las coordenadas del centroide de esta nueva clase son $\frac{0,65+1,04}{2} = 0,85$

y $\frac{2,43+2,48}{2} = 2,45$.

Realizamos el mismo proceso que en las anteriores ocasiones presentando la cuarta matriz de distancias entre clases de este ejercicio, en la tabla siguiente:

| Clase | C16 | C253 | C4 | C7 |
|---|---|---|---|---|
| C16 | 0 | 0,87 | 1,03 | 0,67 |
| C253 | | 0 | 1,86 | 1,53 |
| C4 | | | 0 | 0,40 |
| C7 | | | | 0 |

La nueva distancia mínima se produce entre las clases C4 y C7, esto es, 0,40. Las coorde- nadas del centroide de esta nueva clase C47 son $\frac{2+1,95}{2} = 1,97$ y $\frac{1+1,40}{2} = 1,20$.

Ya se tienen las tres clases que solicitaba el enunciado del problema, que son la C16, la C235 y la C47. Se recalcula la matriz de distancias para el nuevo conjunto de clases. Los resultados se presentan en la quinta y última matriz de distancias entre clases de este ejercicio, en la tabla siguiente:

| Clase | C16 | C253 | C47 |
|---|---|---|---|
| C16 | 0 | 0,87 | 0,85 |
| C253 | | 0 | 1,69 |
| C47 | | | 0 |

La carga original dada en la tabla de datos del enunciado del problema se ha agrupado en tres clases con componentes de características similares. Cada clase está representada por valores de los parámetros iguales a la media de los valores de los parámetros de todos los componentes de esa clase. La tabla siguiente presenta la salida del proceso de agrupamiento, contiene la descripción de los centroides que representan cada clase y el número de componentes (número de accesos). Cada clase corresponde a una clase de la carga. Se han ordenado de acuerdo con el tamaño del documento. Los valores se han pasado a la escala original, es decir, se calculan como la inversa de $\log x$, esto es, como $10^x$, siendo $x$ el valor del parámetro de la clase en su centroide. Por ejemplo, $10^{0,85} = 7,08$, $10^{1,54} = 34,67$, $10^{1,98} = 95,50$, $10^{2,45} = 281,84$,
$10^{1,92} = 83,18$ y $10^{1,20} = 15,85$.

| Tipo de documento | Clase | Tamaño (KB) | Accesos | Componentes |
|---|---|---|---|---|
| Pequeño | C253 | 7,08 | 281,84 | 3 |
| Intermedio | C16 | 34,67 | 83,18 | 2 |
| Grande | C47 | 95,50 | 15,85 | 2 |

Analizando estas tres clases se puede deducir lo siguiente:

- La clase C253 está compuesta por documentos de tamaño pequeño y suponen la mayor parte del número de accesos.
- La clase C16 se compone de documentos de tamaño intermedio y suponen una parte del número de accesos.
- La clase C47 está compuesta de documentos de gran tamaño y suponen una parte muy pequeña del número de accesos. ■

**PROBLEMA 6.6** En la Tabla 6.8 se indican las mediciones del tiempo de respuesta ($t_r$) y del tiempo de servicio ($t_s$) realizadas en más de doscientos subsistemas de disco magnético con distintas cantidades de memoria cache entre los años 1991 y 2003 como parte de un trabajo de investigación en España.

Los datos del problema también se presentan en la Figura 6.9. No se ha representado el punto de los subsistemas de disco sin memoria cache porque se sale de la figura por su valor muy alto.

Se pide agrupar las clases arriba indicadas en tres grupos aplicando la técnica de las distancias euclídeas, así como dibujar el dendograma que se produzca.

| Clase | Tamaño cache (GB) | $t_r$ (ms) | $t_s$ (ms) |
|---|---|---|---|
| 1 | 0 | 45,3 | 31,9 |
| 2 | 1 | 11,4 | 9 |
| 3 | 2 | 8,5 | 7,2 |
| 4 | 3 | 5,6 | 4,6 |
| 5 | 4 | 5 | 4,2 |
| 6 | 8 | 4 | 3,6 |
| 7 | 16 | 3 | 2,7 |

**Figura 6.8:** Parámetros de subsistemas de disco magnético con memoria cache para el Problema 6.6.

**SOLUCIÓN:** Como se puede ver en la figura, hay una correlación considerable entre las dos variables, tiempo de respuesta y tiempo de servicio. Calculamos la matriz de distancias utilizando la técnica de las distancias euclídeas sobre los parámetros del tiempo de respuesta y el tiempo de servicio.

Se calculan las distancias entre las clases. $d_{C_iC_j}$ representa la distancia euclídea entre la clase *i* y la clase *j*, que se calcula utilizando la ecuación de la distancia euclídea. Solamente indicamos dos de ellas para que se vea cómo se calculan:

$$d_{C1C2} = \sqrt{(45,3-11,4)^2 + (31,9-9)^2} = 40,9$$

$$d_{C1C3} = \sqrt{(45,3-8,5)^2 + (31,9-7,2)^2} = 44,3$$

Los resultados de estos cálculos se presentan en la primera matriz de distancias entre clases de este ejercicio, en la tabla siguiente:

| Clase | C1 | C2 | C3 | C4 | C5 | C6 | C7 |
|---|---|---|---|---|---|---|---|
| C1 | 0 | 40,9 | 44,3 | 48,2 | 48,9 | 50,1 | 51,4 |
| C2 | | 0 | 3,4 | 7,3 | 8,0 | 9,2 | 10,5 |
| C3 | | | 0 | 3,9 | 4,6 | 5,8 | 7,1 |
| C4 | | | | 0 | 0,7 | 1,9 | 3,2 |
| C5 | | | | | 0 | 1,2 | 2,5 |
| C6 | | | | | | 0 | 1,3 |
| C7 | | | | | | | 0 |

La distancia mínima entre las clases de la tabla anterior es la que se produce entre las clases C4 y C5, esto es, 0,7. Las coordenadas del centroide de esta nueva clase son $\frac{5.6+5.0}{2} = 5,3$ y $\frac{4.6+4.2}{2} = 4,4$.

Se calculan las distancias de esta clase C45 nueva al resto de las clases, resultando la siguiente matriz de distancias:

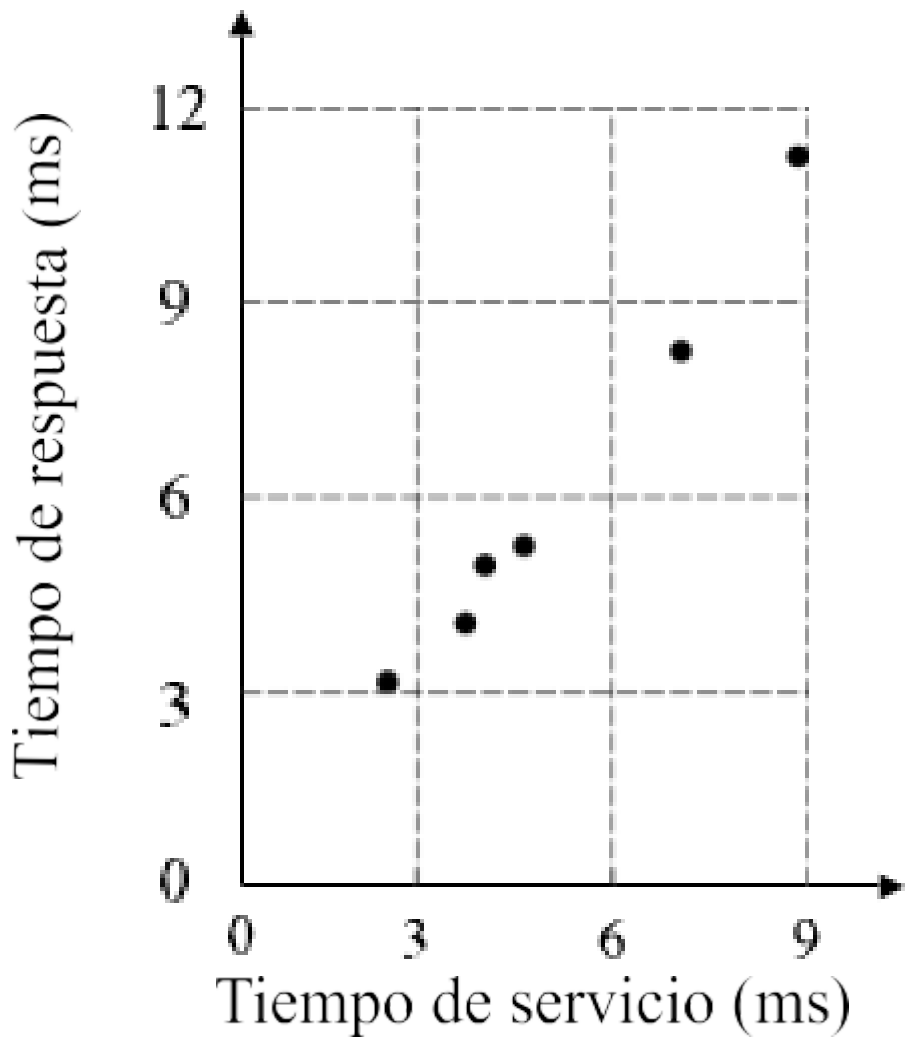

**Figura 6.9:** Histograma de subsistemas de disco magnético con memoria cache para el Problema 6.6.

| Clase | C1 | C2 | C3 | C45 | C6 | C7 |
|---|---|---|---|---|---|---|
| C1 | 0 | 40,9 | 44,3 | 48,5 | 50,1 | 51,4 |
| C2 | | 0 | 3,4 | 7,6 | 9,2 | 10,5 |
| C3 | | | 0 | 4,3 | 5,8 | 7,1 |
| C45 | | | | 0 | 1,5 | 2,9 |
| C6 | | | | | 0 | 1,3 |
| C7 | | | | | | 0 |

La distancia mínima entre las clases de la tabla anterior es 1,3, que se produce entre las clases C6 y C7. Las coordenadas del centroide de esta nueva clase son $\frac{4+3}{2}$ = 3,5 y $\frac{3,6+2,7}{2}$ = 3,2.

Se calculan las distancias de esta clase C67 nueva al resto de las clases, resultando la siguiente matriz de distancias:

| Clase | C1 | C2 | C3 | C45 | C67 |
|---|---|---|---|---|---|
| C1 | 0 | 40,9 | 44,3 | 48,5 | 50,7 |
| C2 | | 0 | 3,4 | 7,6 | 9,8 |
| C3 | | | 0 | 4,3 | 6,4 |
| C45 | | | | 0 | 2,2 |
| C67 | | | | | 0 |

La distancia mínima entre las clases de la tabla anterior es 2,2, que se produce entre las clases C45 y C67. Las coordenadas del centroide de esta nueva clase son $\frac{5,3+3,5}{2}$ = 4,4 y $\frac{4,4+3,2}{2}$ = 3,8.

Se calculan las distancias de esta clase C4567 nueva al resto de las clases, resultando la siguiente matriz de distancias:

| Clase | C1 | C2 | C3 | C4567 |
|---|---|---|---|---|
| C1 | 0 | 40,9 | 44,3 | 47,5 |
| C2 | | 0 | 3,4 | 6,6 |
| C3 | | | 0 | 3,2 |
| C4567 | | | | 0 |

La distancia mínima entre las clases de la tabla anterior es 3,2, que se produce entre las clases C4567 y C3. Las coordenadas del centroide de esta nueva clase son 7,3 y 6,2.

Se calculan las distancias de esta clase C4567 nueva al resto de las clases, resultando la siguiente matriz de distancias:

| Clase | C1 | C2 | C45673 |
|---|---|---|---|
| C1 | 0 | 40,9 | 45,9 |
| C2 | | 0 | 5,0 |
| C45673 | | | 0 |

Hemos llegado a la agrupación en tres clases, que son la clase C1 de subsistemas de disco sin memoria cache, la clase C2 con 1 GB de memoria cache, y el resto de las clases se han agrupado en una tercera clase C45673 que recoge a todos los subsistemas con memoria cache de 2 GB o más. Esta agrupación es razonable, ya que los subsistemas sin memoria cache tienen unos tiempos de respuesta y de servicio muy elevados (45,3 ms y 31,9 ms). En cuanto se introduce memoria cache (1 GB) los tiempos de respuesta y de servicio bajan drásticamente (11,4 ms y 9 ms). Los incrementos adicionales de memoria cache mejoran esos tiempos (varían entre los 8,5 y los 3 ms para el tiempo de respuesta, y entre 7,2 y 2,7 ms para el tiempo de servicio), pero en una proporción menor que el simple hecho de introducir la memoria cache. Es interesante destacar también que las distancias más cercanas se producen entre los subsistemas que tienen más memoria cache, las clases C4, C5, C6 y C7, para posteriormente ir añadiendo a este tercer grupo también las que tienen menor memoria cache, la clase C3, y la siguiente, si hubiéramos seguido agrupando, la clase C2. El dendograma de este problema se presenta en la Figura 6.10.

■

**PROBLEMA 6.7** En la tabla siguiente se presenta un pequeño conjunto de muestras de la tasa de visitas correspondiente a diez sesiones de acceso a un servidor web.

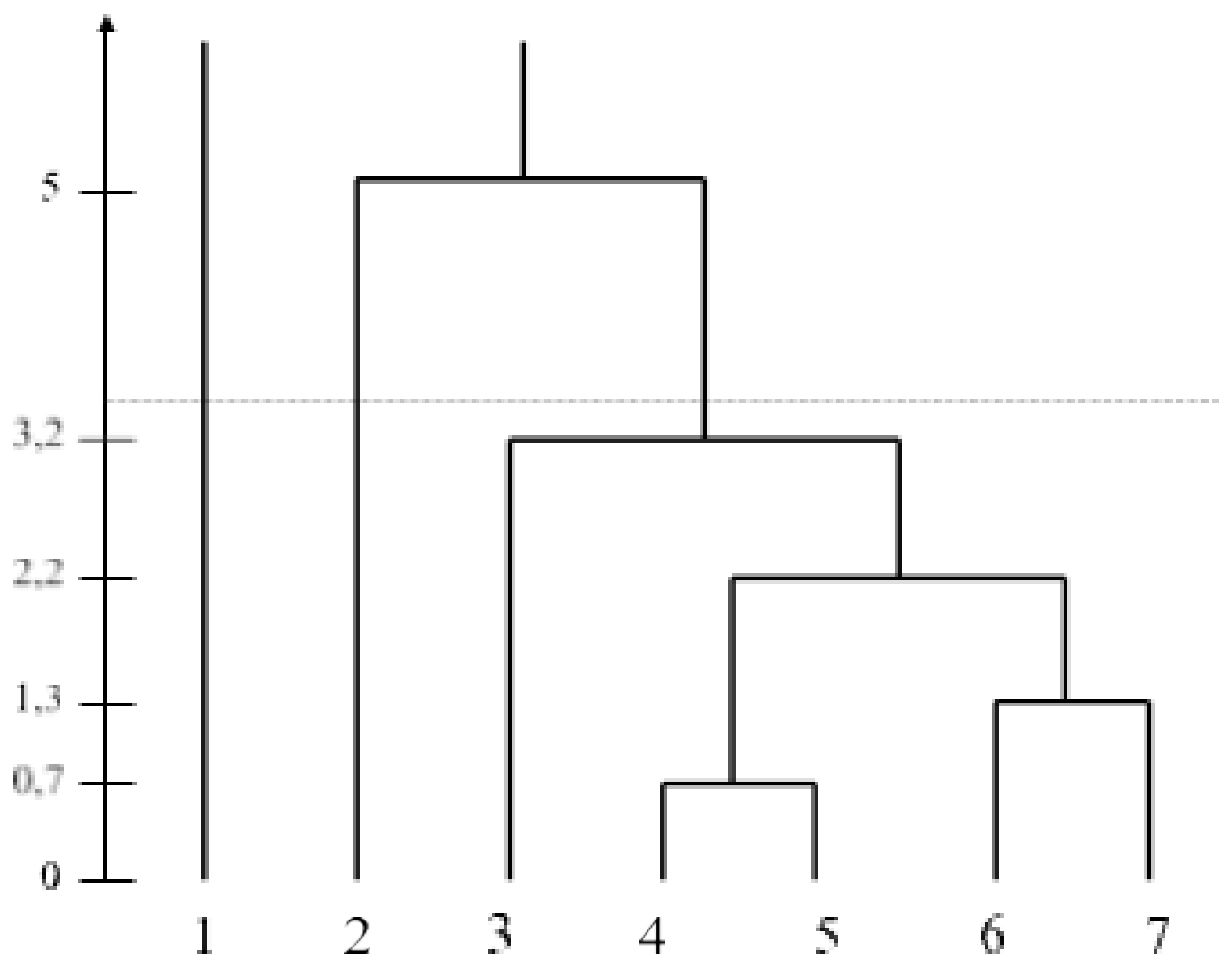

**Figura 6.10:** Dendograma correspondiente al Problema 6.6.

| Sesión | $V_{Muestra}$ | $V_{Búsqueda}$ | $V_{Añade\ al\ carro}$ | $V_{Selecciona}$ | $V_{Paga}$ |
|---|---|---|---|---|---|
| 1 | 5 | 12 | 2 | 5 | 1 |
| 2 | 10 | 15 | 1 | 14 | 0 |
| 3 | 4 | 7 | 2 | 4 | 1 |
| 4 | 18 | 20 | 3 | 15 | 0 |
| 5 | 4 | 12 | 2 | 7 | 1 |
| 6 | 6 | 11 | 3 | 7 | 1 |
| 7 | 7 | 12 | 2 | 7 | 1 |
| 8 | 5 | 4 | 1 | 2 | 1 |
| 9 | 7 | 10 | 1 | 8 | 1 |
| 10 | 15 | 20 | 1 | 18 | 0 |

Se pide agrupar hasta un nivel que parezca razonable.

**SOLuCIóN:** Se utiliza la distancia euclídea como la métrica para evaluar las distancias entre las clases. Se calculan las distancias entre las clases iniciales que se presentan en la matriz de distancias en la siguiente tabla:

| Clase | C1 | C2 | C3 | C4 | C5 | C6 | C7 | C8 | C9 | C10 |
|---|---|---|---|---|---|---|---|---|---|---|
| C1 | 0 | 10,82 | 5,20 | 18,30 | 2,24 | 2,65 | 2,83 | 8,60 | 4,24 | 18,30 |
| C2 | | 0 | 14,21 | 9,70 | 9,80 | 9,27 | 8,31 | 17,06 | 8,43 | 8,12 |
| C3 | | | 0 | 22,10 | 5,83 | 5,48 | 6,56 | 3,87 | 5,92 | 22,10 |
| C4 | | | | 0 | 18,06 | 17,03 | 15,84 | 24,47 | 16,58 | 4,69 |
| C5 | | | | | 0 | 2,45 | 3,00 | 9,54 | 3,87 | 17,55 |
| C6 | | | | | | 0 | 1,73 | 8,89 | 2,65 | 16,97 |
| C7 | | | | | | | 0 | 9,70 | 2,45 | 15,84 |
| C8 | | | | | | | | 0 | 8,72 | 24,76 |
| C9 | | | | | | | | | 0 | 16,28 |
| C10 | | | | | | | | | | 0 |

Podemos fijarnos en todas aquellas distancias menores de 3, que son: 2,24, 2,65, 2,83, 2,45, 1,73, 2,65 y 2,45, leyendo de izquierda a derecha y de arriba abajo. La distancia menor es 1,73 entre las clases C6 y C7. El centroide de esta nueva clase o grupo es 6,5, 11,5, 2,5, 7 y 1.

En este problema vamos a intentar no presentar todas las ocasiones la nueva matriz o tabla de distancias completa que nos llevaría a ocupar mucho espacio. Para calcular la nueva matriz de distancias basta con calcular la distancia de esta nueva clase al resto de las antiguas clases, resultando la siguiente nueva fila de la tabla:

| Clase | C1 | C2 | C3 | C4 | C5 | C8 | C9 | C10 |
|---|---|---|---|---|---|---|---|---|
| C67 | 2,60 | 8,76 | 5,98 | 16,42 | 2,60 | 9,24 | 2,39 | 16,39 |

La única distancia menor de 3 de esta nueva tabla es 2,60. Comparando este valor con 2,24, que es la única distancia que queda menor de 3 al eliminar las filas y columnas en las que aparecían las antiguas clases C6 y C7, se deduce que la distancia nueva menor es ahora 2,24, entre las clases C1 y C5. El centroide de esta nueva clase es 4, 5, 12, 2, 6 y 1. Como en el caso anterior, calculamos solamente las distancias de esta nueva clase al resto de las clases, resultando la siguiente nueva fila de la matriz de distancias:

| Clase | C2 | C3 | C4 | C67 | C8 | C9 | C10 |
|---|---|---|---|---|---|---|---|
| C15 | 10,26 | 5,41 | 18,15 | 2,34 | 9,01 | 3,90 | 17,90 |

La única distancia menor de 3 de esta nueva fila es 2,34. Comparando el resto de las distancias que son todas mayores de 3, se deduce que la distancia nueva menor es ahora precisamente 2,34, entre las clases 67 y 15. La distancia de 2,60 no es aplicable ya que la C5 ha desaparecido como clase independiente. El centroide de esta nueva clase C6715 es 5,5, 11,75, 2,25, 6,5 y 1. Como en el caso anterior, calculamos solamente las distancias de esta nueva clase al resto de las clases, resultando la siguiente nueva fila de la matriz de distancias:

| Clase | C2 | C3 | C4 | C8 | C9 | C10 |
|---|---|---|---|---|---|---|
| C6715 | 9,[illegible] | 5,58 | 17,27 | 9,06 | 3,02 | 17,12 |

La distancia mínima nueva es 3,02, entre las clases C6715 y C9. El centroide de esta nueva clase C67159 es 6,25, 10,88, 1,63, 7,25 y 1. A continuación presentamos la nueva matriz de distancias completa:

| Clase | C2 | C3 | C4 | C67159 | C8 | C10 |
|---|---|---|---|---|---|---|
| C2 | 0 | 14,21 | 9,70 | 8,83 | 17,06 | 8,12 |
| C3 | | 0 | 22,10 | 5,55 | 3,87 | 22,10 |
| C4 | | | 0 | 16,86 | 24,47 | 4,69 |
| C67159 | | | | 0 | 8,76 | 16,63 |
| C8 | | | | | 0 | 24,76 |
| C10 | | | | | | 0 |

La distancia mínima nueva es 3,87, entre las clases C3 y C8. El centroide de la nueva clase C38 es 4,5, 5,5, 1,5, 3 y 1. A continuación presentamos la nueva matriz de distancias:

| Clase | C2 | C38 | C4 | C67159 | C10 |
|---|---|---|---|---|---|
| C2 | 0 | 15,58 | 9,70 | 8,83 | 8,12 |
| C38 | | 0 | 23,23 | 7,07 | 23,38 |
| C4 | | | 0 | 16,86 | 4,69 |
| C67159 | | | | 0 | 16,63 |
| C10 | | | | | 0 |

La distancia mínima nueva es 4,69, entre las clases C4 y C10. El centroide de la nueva clase C410 es 16,5, 20, 2, 16,53 y 0. A continuación presentamos la nueva matriz de distancias:

| Clase | C2 | C38 | C410 | C67159 |
|---|---|---|---|---|
| C2 | 0 | 15,58 | 8,63 | 8,83 |
| C38 | | 0 | 23,19 | 7,07 |
| C410 | | | 0 | 16,86 |
| C67159 | | | | 0 |

La distancia mínima nueva es 7,07, entre las clases C67159 y C38. El centroide de la nueva clase C6715938 es 5,38, 8,19, 1,56, 5,13 y 1. A continuación presentamos la nueva matriz de distancias:

| Clase | C2 | C410 | C6715938 |
|---|---|---|---|
| C2 | 0 | 8,63 | 12,16 |
| C410 | | 0 | 19,85 |
| C6715938 | | | 0 |

La distancia mínima nueva es 8,63, entre las clases C410 y C28. El centroide de la nueva clase C4102 es 13,25, 17,5, 1,5, 15,25 y 0.

Resultan dos clases o grupos. La clase C6715938 se caracteriza por aquellas visitas que han pagado en su visita, mientras que la clase C4102 recoge aquellas visitas donde el cliente se ha ido sin pagar. Este agrupamiento final es bastante llamativo y lleva a la consideración de que la actuación de los compradores entre sí es bastante similar, mientras que la de los no compradores entre sí también lo es, diferenciándose mucho el comportamiento entre los compradores y los no compradores.

---

**PROBLEMA 6.8** Se tiene una red de almacenamiento (*Storage Area Network*, SAN), compuesta por un sistema de discos magnéticos compartido por varios servidores de proceso, cuya carga de trabajo de entrada/salida se presenta en la tabla siguiente:

| Tipo de carga | E/S por s | Acceso | % lectura/escritura | Bloque |
|---|---|---|---|---|
| Transaccional | 10.000 | Aleatorio | 75 % lectura | 4 KB |
| FPSW | 10.000 | Aleatorio | 100 % lectura | 512 B |
| FGSW | 50 | Aleatorio | 100 % lectura | 1 MB |
| S ficheros | 200 | Aleatorio | 80 % lectura | 32 KB |
| S vídeo | 0,1 | Secuencial | 100 % lectura | 2 GB |
| SEVS | 100 | Secuencial | 100 % escritura | 128 KB |

Los acrónimos de la tabla indican: FPSW: ficheros pequeños en servidor web; FGSW: ficheros grandes en servidor web; S: servidor de; SEVS: servidor de edición de vídeo y salvado.

El rendimiento de los sistemas de almacenamiento se mide por el número de operaciones de entrada/salida y el tiempo de respuesta asociado en los procesos aleatorios, y por la velocidad de transferencia medida en MB/s en los procesos secuenciales. Se pide:

1. Explicar las características o parámetros de cada uno de los tipos de carga especifi- cados, indicando algún ejemplo para facilitar su comprensión.

2. Calcular la carga total de entrada/salida medida en número de operaciones de entrada/salida para accesos aleatorios y en MB/s en accesos secuenciales.

**SOLUCIÓN:**

1. Primero se explicarán la características de cada uno de los tipos de carga indicados.

   - *El tipo de carga transaccional* representa a servidores de base de datos que realizan transacciones. El acceso a la base de datos es normalmente aleatorio, ya que la transacción lee o actualiza un registro determinado. Por ejemplo, una consulta del saldo en mi cuenta corriente bancaria desde una cajero automático lleva al siste- ma a leer solamente unos campos específicos de mi registro de la base de datos.

Si posteriormente se retiró dinero por dicho cajero automático, se producirá una actualización sobre el campo saldo del registro que incluye los datos de mi cuenta bancaria. Para autorizar esta salida de dinero tendré que proporcionar una clave de autentificación por el teclado del cajero automático, que será comparada con la clave que se encuentra en un campo del registro de mi cuenta bancaria. En este sencillo proceso se han realizado varias lecturas y solamente una escritura. Por este motivo es razonable que el enunciado indique un 75 % de lecturas y un 25 % de escrituras, que en otros casos es especificado como una tasa de lectura:escritura de 3:1. El porcentaje de lectura sobre escritura de estos servidores depende mucho del tipo de transacciones que en él se realicen. Un proceso transaccional intensivo en escritura puede tener una tasa de lectura:escritura de 2:1. Un proceso transaccional ligero en escritura o intensivo en lectura puede tener una tasa de lectura:escritura de 5:1.

La información en las bases de datos normalmente se encuentra fraccionada en bloques de 4 KB, porque al querer recuperar normalmente uno o varios campos de un sólo registro, se optimiza el rendimiento al hacer la transferencia de información en bloques lo más pequeños posible. Con bloques pequeños se conseguirá que la transferencia de la información del disco al procesador sea lo más rápida posible y además se utilizarán los recursos el menor tiempo posible.

- *El tipo de carga en un servidor web* se ha clasificado en el enunciado del problema en dos clases (FPSW y FGSW), dependiendo del tamaño del fichero que se va a servir desde el servidor web. Los usuarios accederán a la información y datos de un servidor web de una forma aleatoria y en la mayoría de los casos para leer información. El enunciado supone que este servidor web es sólo informativo, ya que no permite la escritura de información por parte del usuario. Hay otros servidores web, que pueden incluir dentro de sus posibles funciones la realización de transacciones, como la retirada de dinero de un cajero automático, o la compra del derecho de ver un fichero de vídeo o audio, etc. Se supone que este tipo de carga transaccional se ha incluido dentro de la primera carga del enunciado del problema, la transaccional, y que el resto de la carga contra el servidor web va a ser de sólo lectura.

  Los tamaños de las solicitudes pueden cambiar desde un bloque pequeño de 512 bytes hasta un bloque grande de 1 MB, produciéndose la mayoría de las solicitudes para obtener bloques de información pequeños.

- *El tipo de carga de servidor de ficheros* (S ficheros) es aquella que sirve ficheros a los usuarios que los solicitan. La mayoría de los accesos a estos ficheros son de lectura (80 %), aunque también se producen actualizaciones sobre los mismos. Piénsese en el disco de un servidor ofimático que da servicio de documentación, hojas de cálculo, procesador de textos y presentaciones a un departamento de cien personas. Algunos de estos documentos serán formularios estándar que cada usuario se baja a su computador personal para adaptarlos convenientemente (contratos comerciales) o simplemente leerlos (prácticas comerciales). Estos últimos formularios son de sólo lectura, mientras que las hojas de cálculo y textos generados por un usuario y que permanecen en el servidor del departamento sufren lecturas y escrituras.

El acceso a estos ficheros es aleatorio. El tamaño de los ficheros típicamente está entre 4 y 64 KB, y por este motivo se ha elegido en el enunciado 32 KB.

- *El tipo de carga de servidor de vídeo* (S vídeo) es un caso específico del tipo de carga de servidor de ficheros, que se caracteriza por que en él se efectuán procesos de lectura (100 % lectura) secuenciales (100 % secuencial) de ficheros muy grandes, de unos 2 GB. Estos servidores de vídeo se están comenzando a utilizar mucho más en la actualidad y necesitan de un nivel alto de continuidad en la calidad de servicio, para evitar lo que se llaman *blancos en la emisión*, esto es, la falta de un suministro sostenido adecuado de transferencia de vídeo (MB/s) al emisor de vídeo en directo, que puede llevar a que la emisión de vídeo en el punto de destino se degrade, o incluso, llegue a interrumpirse.
- *El tipo de carga de servidor de edición de vídeo y salvado* (SEVS) se caracteriza por operaciones de escritura (100 % escritura) de procesos secuenciales (100 % se- cuencial) en chorro (*streaming* ) de ficheros grandes (de más de 64 KB). Un caso especial es el de salvado de información, que consiste en realizar una copia de in- formación para poder recuperarla en el caso de avería del disco. Se suele realizar sobre cinta magnética. También se puede realizar una segunda copia sobre disco, normalmente llamada *copia remota de disco*, que se genera leyendo de los *discos primarios* y escribiendo en chorro su información en los *discos secundarios*. Estos discos secundarios normalmente se encuentran en otro edificio, ya sea a distancia de *campus* o a distancias mayores, para poder trabajar con ellos en el caso de que ocurra un desastre en el centro primario.

2. En segundo lugar se calculará la carga total de entrada/salida medida en número de ope- raciones de entrada/salida para accesos aleatorios y en MB/s para accesos secuenciales. Suponiendo que todas las cargas se produzcan con la misma frecuencia, el número total de operaciones de entrada/salida por segundo es el siguiente:

$$10.000 + 10.000 + 50 + 200 + 0{,}1 + 100 = 20.350{,}1$$

Se puede ver que las cargas de ficheros grandes en servidor web, servidor de vídeo, y servidor de edición de vídeo y salvado influyen muy poco, solamente en un 1 %, en el número de operaciones de entrada/salida total. Por lo tanto, se podrían haber despreciado en el cálculo. El número total de MB/s transferidos en forma secuencial es el siguiente:

$$\frac{0{,}1\ \text{pet}}{\text{s}} \times \frac{128\ \text{KB}}{\text{pet}} + \frac{100\ \text{pet}}{\text{s}} \times \frac{1\ \text{MB}}{\text{pet}} +$$

50 pet / s = 262,8 MB/s

Se puede comprobar que las cargas de tipo transaccional, ficheros pequeños de servidor web y servidor de ficheros, influyen poco en el número de MB/s, y, por este motivo, no se han tenido en cuenta en el cálculo. Serían 4 MB/s, 0,512 MB/s y 6,4 MB/s, respectivamente. ■

## 6.6. Problemas con solución

**PROBLEMA 6.9** El tiempo de procesador y el número de operaciones de E/S de cinco de las siete clases de programas que trabajan contra una base de datos relacional y unos programas de edición de textos se muestran en la tabla siguiente:

| Clase | Nombre del programa | Procesador (ms) | E/S por s |
|---|---|---|---|
| C3 | Actualizaciones pesadas | 8 | 27 |
| C4 | Actualizaciones medianas | 6 | 27 |
| C5 | Actualizaciones ligeras | 6 | 12 |
| C6 | Consulta pesada | 4 | 91 |
| C7 | Editor de textos ligero | 1 | 33 |

Se pide preparar un dendograma utilizando un algoritmo de árbol de extensión mínima para el análisis de agrupamiento.

**SOLuCIóN:** El dendograma se presenta en la Figura 6.11.

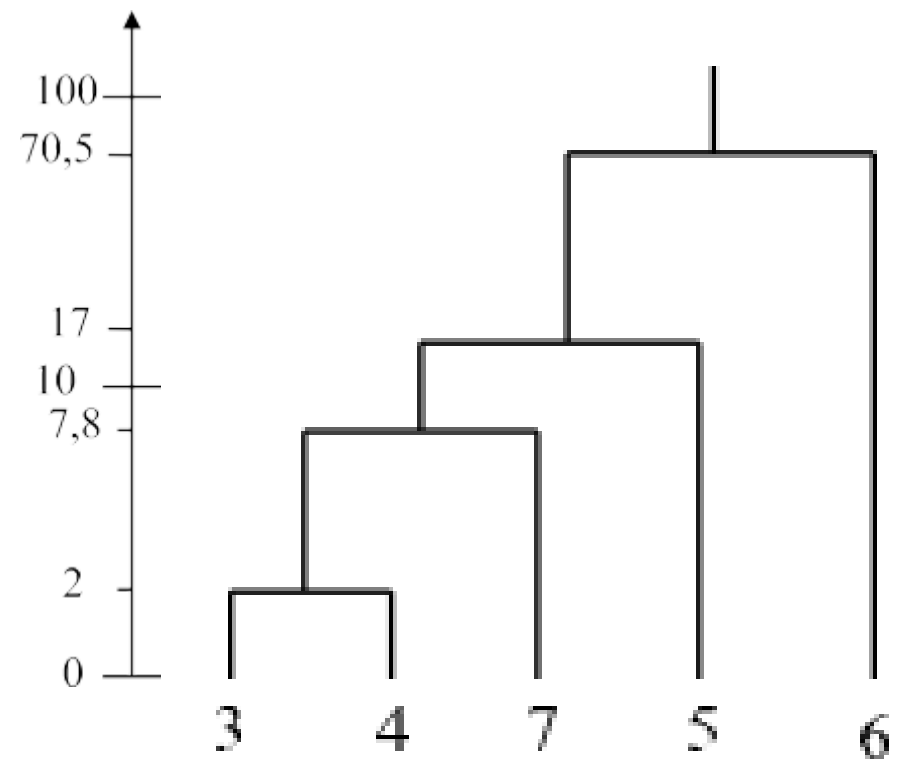

**Figura 6.11:** Dendograma realizado con los datos del Problema 6.9.

■

**PROBLEMA 6.10** Dibújese un gráfico del modelo de comportamiento del usuario de la página web de un ayuntamiento que ofrece entre sus servicios los tres siguientes:

- Descarga de un fichero de vídeo, de audio o de un documento impreso.
- Ver un vídeo en tiempo real.
- Suscribirse para recibir envíos.

Indíquese las probabilidades de utilización de cada una de las funciones de la página web.

**SOLUCIÓN:** Un gráfico del modelo del comportamiento del usuario posible de una página web de ese ayuntamiento se representa en la Figura 6.12. Los usuarios normalmente se descargarán la información en su ordenador personal (70 %) y en muy contadas ocasiones verán un vídeo en línea (10 %).

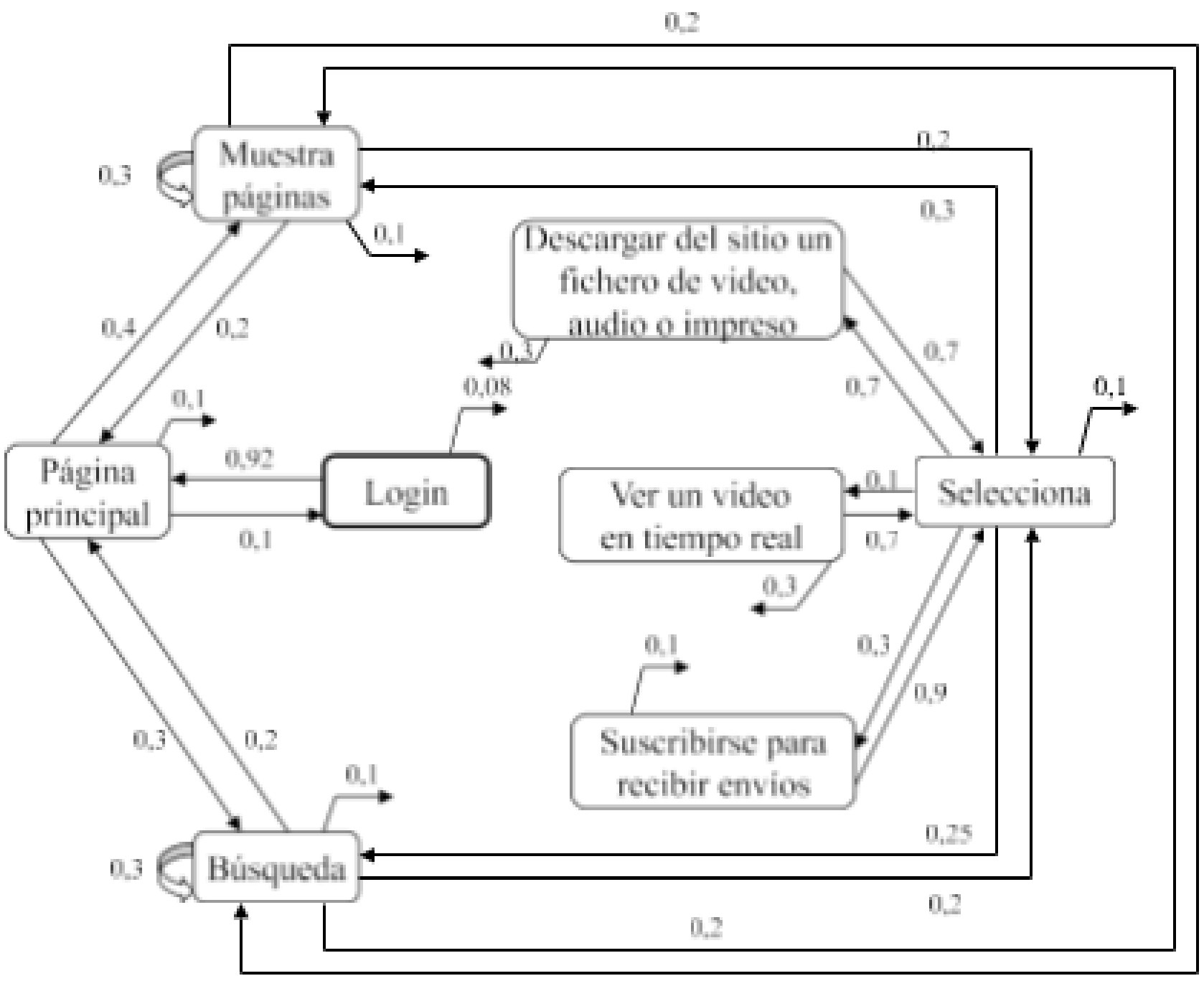

**Figura 6.12:** Gráfico del modelo de comportamiento del usuario de una página web de un ayuntamiento.

■

**PROBLEMA 6.11** Dibújese un gráfico del modelo de comportamiento del usuario de la página web de un banco que ofrece entre sus servicios los habituales de consulta del estado de la cuenta corriente, de órdenes de transferencia y de pagos de facturas, etc. Indíquese unas probabilidades razonables de utilización de cada una de las funciones de la página web.

**SOLuCión:** Un gráfico del modelo del comportamiento del usuario de una página web de un banco se representa en la Figura 6.13.

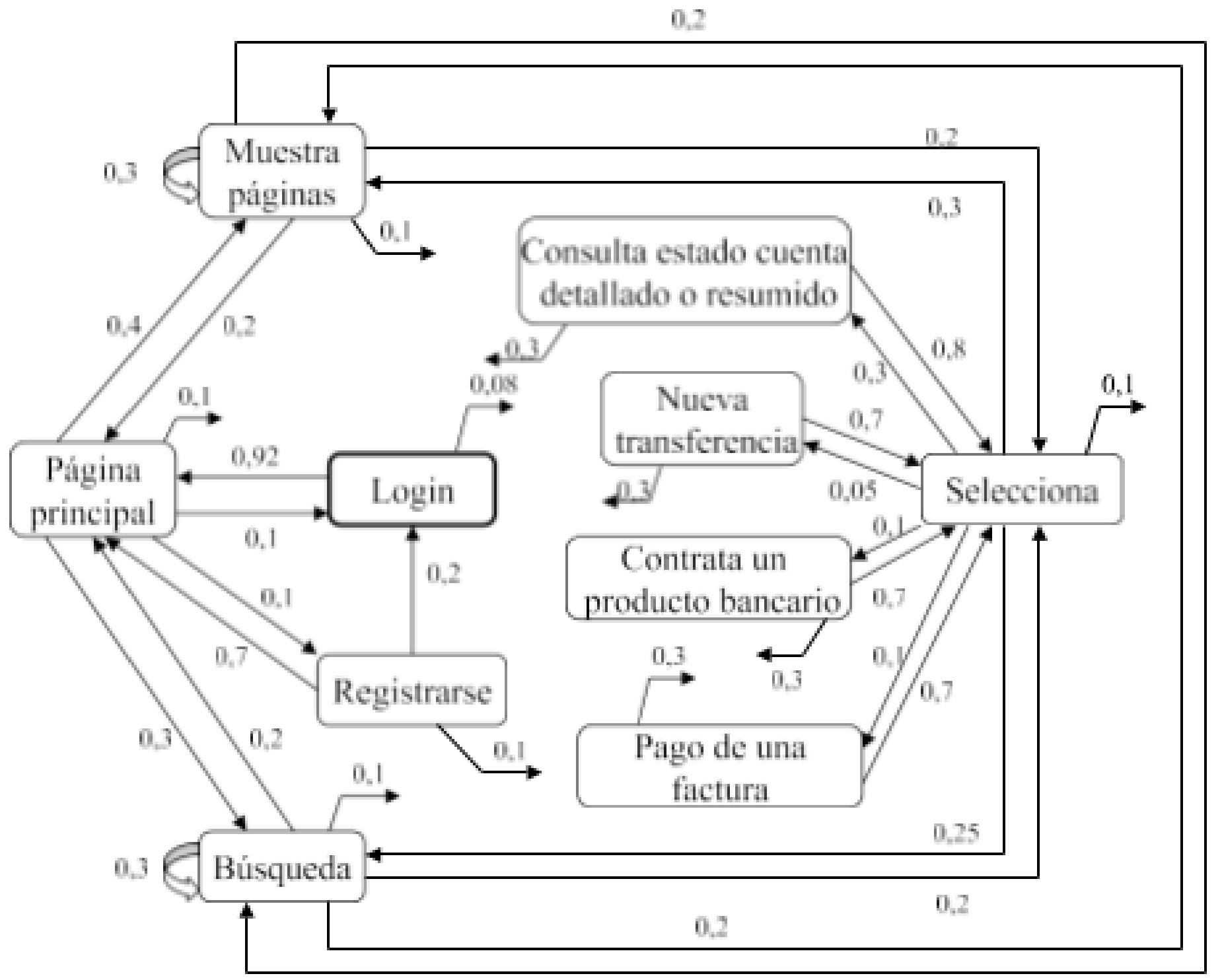

**Figura 6.13:** Gráfico del modelo de comportamiento del usuario de una página web de un banco.

Respecto a las probabilidades señaladas en dicho gráfico cabe indicar que la selección más frecuente va a ser la consulta del estado de la cuenta corriente (30 %), alguna vez se realiza el pago de alguna factura o la contratación de algún producto bancario, y en muy contadas ocasiones se ordena una transferencia bancaria. ■

## 6.7. Problemas sin resolver

**PROBLEMA 6.12** Un servidor está sometido a dos cargas de características distintas: la carga 1 (C1) es típica de un sistema de cálculo intensivo, mientras que la carga 2 (C2) es típica de un servidor de web de vídeo bajo demanda. Las tablas siguientes contienen los valores de los parámetros que caracterizan esta carga real $W$ de un sistema informático y dos modelos $W_1$ y $W_2$ de representación de la misma.

| Parámetro | $WC1$ | $WC2$ |
|---|---|---|
| % de la carga total | 50 % | 50 % |
| Tiempo de procesador (s) | 100 | 20 |
| Tiempo de E/S (s) | 20 | 100 |
| Tiempo de red (s) | 10 | 80 |

| Parámetro | $W_1C1$ | $W_1C2$ | $W_2C1$ | $W_2C2$ |
|---|---|---|---|---|
| Tiempo de procesador (s) | 95 | 18 | 90 | 15 |
| Tiempo de E/S (s) | 15 | 90 | 18 | 95 |
| Tiempo de red (s) | 8 | 75 | 7 | 70 |

Los rangos se indican en la tabla siguiente:

| Parámetro | Rango (máximo − mínimo) |
|---|---|
| Tiempo de procesador (s) | 100 − 5 |
| Tiempo de E/S (s) | 100 − 5 |
| Tiempo de red (s) | 90 − 5 |

Se pide calcular cuál de los dos modelos caracteriza mejor la carga real para cada uno de los siguientes objetivos:

- Objetivo 1: con respecto a la actividad del procesador, dando los siguientes pesos a cada uno de los tres parámetros: $w_1 = 2$, $w_2 = 0{,}5$, $w_3 = 0{,}5$.
- Objetivo 2: con respecto a la actividad de E/S, dando los siguientes pesos a cada uno de los tres parámetros: $w_1 = 0{,}5$, $w_2 = 2$, $w_3 = 0{,}5$.
- Objetivo 3: con respecto a la actividad conjunta de procesador, E/S y red, dando los siguientes pesos a cada uno de los tres parámetros: $w_1 = 1$, $w_2 = 1$, $w_3 = 1$.

Se pide también que se expliquen las diferencias de los consumos de recursos por cada una de las dos cargas C1 y C2.

---

**PROBLEMA 6.13** Se ha realizado una serie de mediciones durante los días laborables de una semana en la hora punta del proceso de una transacción tipo que se ejecuta en un computador de la central de reservas aéreas y se han detectado los consumos medios de los recursos según se presentan en la siguiente tabla. Se pide realizar un agrupamiento hasta llegar a dos clases.

| Solicitud | Procesador (ms) | E/S (ms) | Red (ms) |
|---|---|---|---|
| Lunes | 18 | 25 | 9 |
| Martes | 17 | 28 | 8 |
| Miércoles | 20 | 30 | 10 |
| Jueves | 19 | 29 | 9 |
| Viernes | 22 | 33 | 8 |

**PROBLEMA 6.14** Dibújese un gráfico del comportamiento del usuario de una página web de venta de computadores en línea, que incluya, además de las funciones habituales de cualquier página web, aquéllas que permitan ofertas especiales, la opción de contratar configuraciones estándar o la de construirse su propia configuración, etc. Indíquense unas probabilidades razonadas de la utilización de cada una de las funciones de la página web.

## 6.8. Actividades propuestas

**ACTIVIDAD 6.1** Aplíquese el modelo de caracterización de la carga de servidor web des- crito en la teoría de este capítulo al servidor web que se tenga más cercano: el de la universidad o el de la empresa de un amigo.

**ACTIVIDAD 6.2** Léase el artículo *Workload Characterization Issues and Methodologies* de M. Calzarossa, L. Massari y D. Tessera, citado en la bibliografía, y dedúzcanse algunos de los distintos tipos de carga que pueden existir: multimedia, entornos cliente/servidor y sistemas paralelos, así como el modo en que se puede hacer la caracterización de cada uno de estos tipos de carga.

**ACTIVIDAD 6.3** Consúltese el artículo *I/O Characterization in Open Environments* de W. W. Hsu y A. J. Smith, citado en la bibliografía, para deducir el tipo de carga de entrada/salida que se produce en los servidores Unix, Linux y NT en distintos entornos y realícense agrupamientos de las mismas utilizando criterios de similitud.

# Capítulo 7

## Planificación de la capacidad

Actualmente las empresas y los organismos públicos quieren ofrecer nuevos servicios a sus clientes y ciudadanos con un nivel de satisfacción garantizado. Pensemos, por ejemplo, en la nueva forma de realizar la declaración de la renta que actualmente ofrece la Agencia Española de Administración Tributaria (AEAT) vía web. Una de las actividades que debe realizar el analista de prestaciones de sistemas informáticos es prever la carga que estos servicios ofrecidos van a producir y dimensionar los recursos necesarios de hardware para proporcionar el nivel de servicio esperado. El analista de prestaciones de la AEAT para este servicio debe estimar el número de personas que harán simultáneamente la declaración, el tipo de visita que realizarán, la carga producida, y los recursos necesarios para que el ciudadano tenga un tiempo de respuesta suficientemente bueno para que quede satisfecho. A este proceso se lo denomina planificación de la capacidad.

La planificación de la capacidad observa las necesidades de negocio que se deben satisfacer, entendiendo y analizando las cargas de trabajo que se van a ejecutar y el servicio (tiempo de respuesta) que se quiere dar, y detalla los recursos físicos (capacidad) necesarios.

La planificación de la capacidad es el proceso de identificar la configuración de un sistema para suministrar el rendimiento satisfactorio para las cargas de trabajo futuras proyectadas. También se puede definir como el proceso de predecir cuándo los niveles de la carga futura saturarán el sistema y determinar el modo más efectivo, en coste, de retrasar la saturación del sistema todo lo posible. Por tanto, se entiende capacidad como la productividad máxima del sistema. Los recursos sobre los que habitualmente se fija la atención son: procesador, memoria, disco, redes de área local o de almacenamiento, y redes de comunicaciones.

Una metodología básica y sencilla en el proceso de planificación de un sistema infor- mático se puede realizar siguiendo los pasos siguientes:

1. Dotar de instrumentación al sistema.
2. Monitorizar la utilización del sistema.
3. Caracterizar la carga.
4. Predecir el rendimiento bajo diferentes alternativas.
5. Seleccionar la alternativa con el menor coste y/o el mayor rendimiento.

Una buena planificación de la capacidad, revisada con la periodicidad conveniente, puede, al menos, minimizar el ajuste de rendimiento en un sistema. El plan de capacidad predice los dispositivos que podrían convertirse en el cuello de botella del sistema, avan- zando posibles soluciones antes de que se produzca su saturación. Los estudios relacionados con la planificación de la capacidad dependen de una caracterización correcta de la carga de los entornos existentes en dicho sistema informático.

En el presente capítulo se comenzará describiendo el concepto de la capacidad adecuada de un sistema informático para satisfacer las cargas de trabajo a las que va a estar some- tido, suministrando una calidad de servicio determinada o un acuerdo de nivel de servicio fijado con los usuarios de dicho sistema, todo ello utilizando una arquitectura y aplicativos determinados, y dentro de unos costes limitados. Posteriormente se indican los niveles de gestión y planificación de la capacidad, desde un nivel inicial en el que se carece de esta gestión, hasta el nivel superior, en el que se utilizan los criterios de las aplicaciones de negocio para predecir los niveles de servicio de una manera automática. Después se expli- can los métodos de predicción más comúnmente utilizados, que son los de regresión lineal, medias móviles y suavizado exponencial, que se utilizarán en la mayoría de los problemas planteados en este capítulo. A continuación, se introduce el concepto de unidad de predic- ción natural, que es una variable de negocio cuyo valor está directamente relacionado con los recursos utilizados por una aplicación en el sistema. Se han incluido varios problemas sobre la unidad de predicción natural para familiarizarse con su uso. Posteriormente se explica el nuevo concepto de capacidad bajo demanda que automatiza la asignación de la capacidad necesaria a una carga del sistema para cumplir un nivel de servicio determina- do. Esto permite una utilización más versátil de los recursos y una mejora en la calidad de servicio. Por último se dan algunas indicaciones de cómo realizar la planificación en escenarios donde la carga de trabajo varía mucho.

## 7.1. Capacidad adecuada

Un asunto de suma importancia cuando se planifica la capacidad de una instalación es determinar cuál es la capacidad adecuada para dar soporte a la operación de la misma. La capacidad adecuada es función de los tres elementos que destacamos a continuación.

En primer lugar, los acuerdos de nivel de servicio (*Service Level Agreements*, SLA). Son umbrales de productividad, rendimiento y de disponibilidad exigidos y pactados en- tre los grupos de usuarios y la organización de soporte de los sistemas informáticos de la instalación. Algunos ejemplos de acuerdos de nivel de servicio podrían expresarse de mo- do informal como: “disponibilidad de servicio de un 99,9 %”, “el tiempo de respuesta para consultas sencillas a bases de datos no debería exceder los dos segundos”, “el tiempo de res- puesta para búsquedas en bases de datos realizadas a través de la web no deberían exceder el 20 % del tiempo de respuesta en Internet”, “el sistema ha de realizar al menos 30.000 transacciones de unas determinadas características al día”, etc. En todos estos acuerdos, aunque de modo impreciso, se tienen incluso en cuenta factores como la intensidad de los usuarios finales, que corresponde al cociente entre el tiempo medio de respuesta del siste- ma y el tiempo medio de reflexión de esos usuarios cuando interaccionan con el sistema, la satisfacción en el uso del sistema, o incluso el tiempo de respuesta máximo soportable por los usuarios en acceso remoto antes de la desconexión (“clientes de un solo clic”).

Un término llamado calidad de servicio (*Quality of Service*, QoS) que se venía emplean- do anteriormente sólo para las redes de comunicaciones, tiene una cierta semejanza con el de acuerdo de nivel de servicio, pero este último es más general. Calidad y servicio tienen diferentes significados según el contexto: servicio se puede definir como el comportamiento esperado o resultado de un sistema; por lo tanto, calidad de servicio es el grado en que se consigue este resultado. Una forma de cuantificar la calidad de servicio es calcular el ratio de la deficiencia de la calidad de servicio ante un nivel deseado de la misma, tal como se indica en la siguiente ecuación:

$$\text{Desviación QoS} = \frac{\text{QoS conseguida} - \text{QoS deseada}}{\text{QoS deseada}}$$

Desgraciadamente, es muy difícil resumir en un solo valor las calidades de servicio conseguida y deseada, respectivamente.

En segundo lugar, la capacidad adecuada para alcanzar un nivel de servicio acordado se puede conseguir utilizando dispositivos y subsistemas de tipos distintos. La arquitectura seleccionada puede depender de las exigencias del aplicativo que se ejecuta en el sistema, del grado de experiencia y madurez en la explotación del sistema, de la facilidad de admi- nistración o de otros motivos, que no tienen por qué estar directamente relacionados con el rendimiento.

En tercer y último lugar, hay que abordar los planes de sistemas y destacar la limitación de su coste. Todas las organizaciones tienen presupuestos que restringen la capacidad de la que se puede disponer. Para valorar las distintas alternativas, se debería utilizar lo que se llama el coste total de propiedad, que tiene en cuenta todos los costes asociados al arranque (*startup*) y a la explotación u operación del sistema durante un tiempo determinado. Los costes de arranque incluyen los gastos de compra de hardware y software, los costes de instalación e implantación, los costes de personal y los de formación inicial. Los costes de operación incluyen los costes de mantenimiento de hardware y software, los costes de personal para mantener el sistema y los costes ambientales (consumo eléctrico, aire acondicionado y espacio, principalmente).

Se podría decir que un sistema informático tiene una capacidad adecuada si los niveles de servicio se cumplen continuamente para una tecnología y estándares especificados, y si los servicios se suministran dentro de los límites de coste acordados.

## 7.2. Niveles de gestión y planificación

El analista de prestaciones tiene que tener en cuenta varios parámetros a la hora de elegir la herramienta de planificación de la capacidad más adecuada, que son los siguientes: el tiempo con que quiere conocer cuándo las métricas de rendimiento van a alcanzar su umbral, la sofisticación de la herramienta, el hardware, el sistema operativo, el software básico y las aplicaciones que se van a utilizar, la integración de esta herramienta con otras herramientas de gestión del sistema, y por último, si va a desarrollar una herramienta propia o va a comprar una ya existente.

La planificación de la capacidad se está comenzando a gestionar automáticamente. A continuación describimos con mayor detalle los seis niveles en la sofisticación de la gestión y planificación de la capacidad que existen en la actualidad:

**Nivel** 0 No hay un programa de gestión de la capacidad. La gestión de la capacidad se realiza ocasionalmente.

**Nivel** 1 Se realizan medidas de tendencia y predicción de la utilización en periodos de pico. Se planifica la capacidad de los recursos con unas revisiones periódicas.

**Nivel** 2 Se conoce exactamente las utilizaciones de cada uno de los recursos debidas a las cargas de trabajo significativas.

**Nivel** 3 Existe un sistema automático de análisis y predicción de la carga de trabajo.

**Nivel** 4 Se predicen automáticamente los niveles de servicio a partir de las predicciones de capacidad.

**Nivel** 5 Se utilizan los criterios de las aplicaciones de negocio a través de un modelo que sirve para predecir los niveles de servicio.

No todas las organizaciones están en el mismo nivel de gestión, aunque se hace necesario conocer, al menos, qué métodos y técnicas se emplean para predecir la carga futura y, por tanto, la capacidad del sistema.

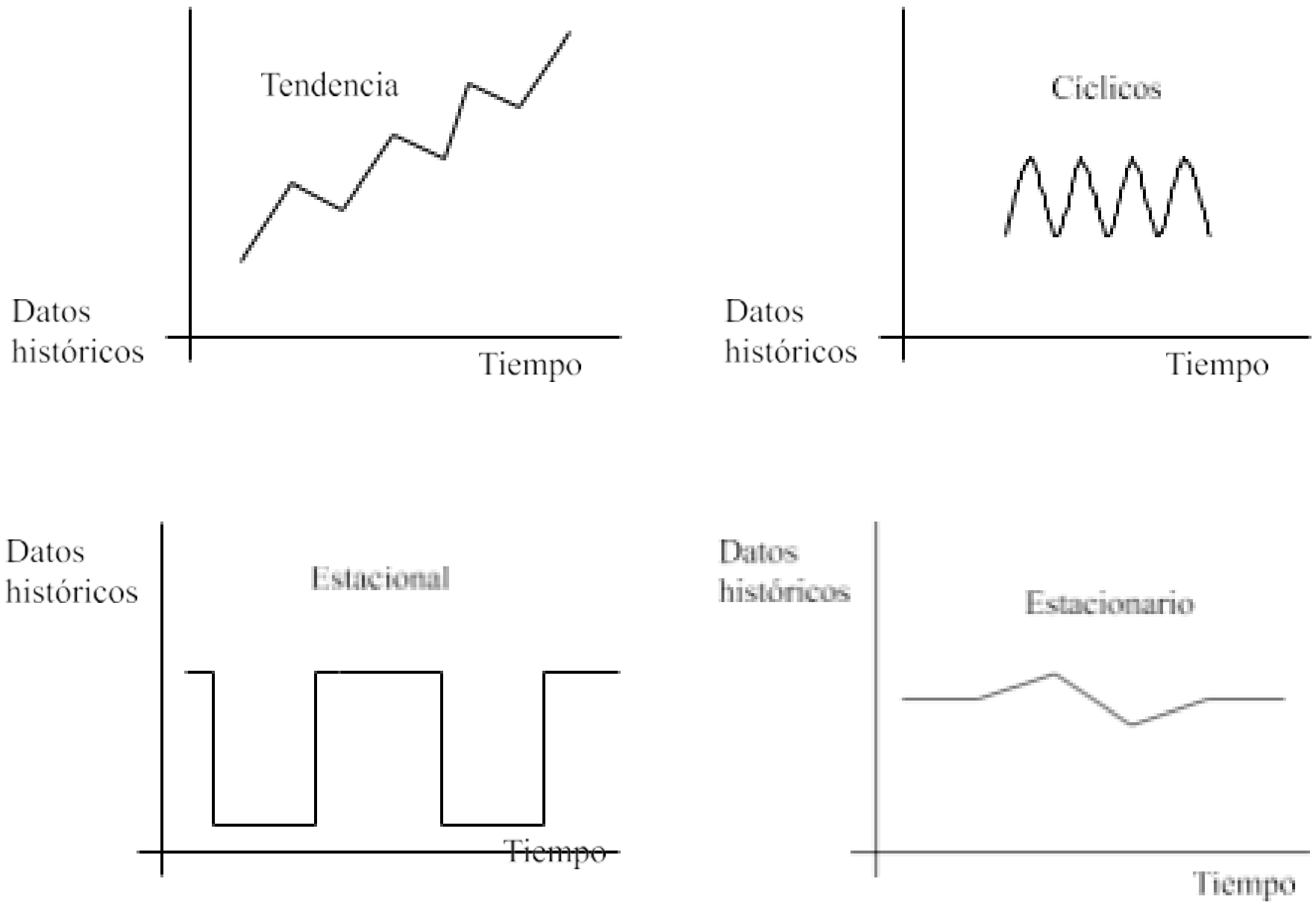

**Figura 7.1:** Patrones de datos históricos.

## 7.3. Los métodos de predicción

Los métodos de predicción suelen dividirse en dos tipos, llamados cuantitativos y cualita- tivos. Los métodos cuantitativos se basan en la existencia de datos históricos para estimar los valores futuros de los parámetros de la carga de trabajo. Los métodos cualitativos son un proceso subjetivo basado en el análisis y la intuición sobre un mercado considerado, así como los planes de negocio, las opiniones de expertos, las analogías históricas y cualquier otra información relevante del escenario tecnológico del sistema. Los métodos cuantitati- vos hacen uso de técnicas estadísticas de predicción. Los valores obtenidos con los métodos cuantitativos se pueden ajustar a los obtenidos con los métodos cualitativos, o bien con otras técnicas de evaluación cuantitativa como la monitorización, el *benchmarking* y el análisis operacional de modelos de sistemas.

A la hora de seleccionar la técnica de predicción cuantitativa más adecuada se han de considerar los factores siguientes:

- La disponibilidad y la fiabilidad de los datos históricos.
- La exactitud y el horizonte de planificación.
- El patrón encontrado en los datos históricos.

Como se puede ver en la Figura 7.1, es posible identificar cuatro patrones de datos his- tóricos: tendencia, cíclico, estacional y estacionario. Mientras que el patrón de tendencia refleja una carga de trabajo que tiende claramente a aumentar o a disminuir, el patrón estacionario no muestra ningún signo de variación sistemática debido a que presenta una media constante. Los patrones estacionales y cíclicos son similares con respecto a la pre- sencia de fluctuaciones. La diferencia es la periodicidad de las fluctuaciones que ocurren en el patrón estacional, por ejemplo, los servidores web para estudiantes en edad universitaria muestran un decrecimiento significativo del tráfico durante los meses de verano. La dife- rencia con los patrones cíclicos es la periodicidad de las fluctuaciones que aparecen siempre en los estacionales.

A continuación se expondrán tres técnicas cuantitativas, típicamente estadísticas, para la predicción de la carga de trabajo: regresión lineal, medias móviles y suavización expo- nencial.

1. *Regresión lineal.* Los modelos de regresión se utilizan para estimar el valor de una variable como una función de otras variables. La variable que hay que predecir se llama variable dependiente y las variables utilizadas para predecir su valor se lla- man variables independientes. La relación matemática establecida entre las variables puede tomar muchas formas, tales como curvas polinómicas (por ejemplo, lineales o cuadráticas) u otras. En el caso de la regresión lineal, la relación utilizada supone que la variable dependiente es una función lineal de las variables independientes. Las técnicas de regresión son apropiadas para trabajar con datos no estacionales que muestran una tendencia. En concreto, la regresión lineal simple supone que los da- tos históricos muestran un patrón de evolución lineal. Suponiendo que sólo hay una variable independiente, la ecuación general para calcular la línea de regresión viene dada por:

$$y = a + b \times x \qquad (7.1)$$

donde $y$ es la variable dependiente, $x$ es la variable independiente, $a$ es el corte con el eje de ordenadas y $b$ es la pendiente de la línea de regresión que representa la relación entre las dos variables. El método de los mínimos cuadrados determina los valores de $a$ y $b$ que minimizan la suma de los cuadrados del error de la predicción. Por lo tanto,

$$b = \frac{\sum_{i=1}^{n} x_i y_i - n \times \bar{x} \times \bar{y}}{\sum_{i=1}^{n} x_i^2 - n \times \bar{x}^2} \qquad (7.2)$$

$$a = \bar{y} - b \times \bar{x} \qquad (7.3)$$

donde $(x_i, y_i)$, $(i = 1, \ldots n)$, son las coordenadas de los $n$ puntos de datos observados, $\bar{y} = \frac{1}{n}\sum_{i=1}^{n} y_i$ es la media de los $y_i$, y finalmente, $\bar{x} = \frac{1}{n}\sum_{i=1}^{n} x_i$ es la media de los $x_i$.

No es objeto de este texto profundizar en la regresión con un mayor número de variables independientes u otros tipos de regresión.

2. *Medias móviles*. Ésta es una técnica de predicción simple que hace que el valor predi- cho para el siguiente periodo sea la media de observaciones previas. Cuando se aplica a datos casi estacionarios, la exactitud alcanzada por la técnica es normalmente alta. Una serie temporal se considera estacionaria cuando no hay cambio sistemático ni en la media ni en la varianza. Esta técnica es apropiada para predicciones a corto plazo. El valor predicho viene dado por la ecuación:

$$f_{t+1} = \frac{y_t + y_{t-1} + \cdots + y_{t-n+1}}{n}, \qquad (7.4)$$

donde $f_{t+1}$ es el valor de la predicción, $y_t$ es el valor observado hasta el instante $t$, y $n$ es el número de observaciones utilizadas para calcular $f_{t+1}$. Se debería seleccionar un valor de $n$ que minimice el error de predicción, que es definido mediante el cuadrado de la diferencia entre el valor predicho y el valor actual. El error cuadrático medio viene dado por el valor:

$$\frac{1}{n}\sum_{t=1}^{n}(y_t - f_t)^2$$

Se pueden probar diferentes valores de $n$ para encontrar el que da un menor error cuadrático medio.

3. *Suavizado exponencial*. Las tendencias históricas se pueden analizar utilizando suavizado exponencial. Esta técnica se debería utilizar para datos no estacionales que no muestran una tendencia sistemática. El suavizado exponencial realiza una media ponderada de las observaciones pasadas y la presente para predecir un valor. Este valor se puede interpretar como el valor esperado en el futuro. Es similar a la técnica de medias móviles con respecto a la forma en que ambas técnicas calculan el valor de predicción. La diferencia está en que el suavizado exponencial coloca más peso en las observaciones recientes. El motivo de utilizar diferentes pesos recae en la hipótesis de que las últimas observaciones dan una indicación mejor del futuro cercano. Para mantener la coherencia en la nomenclatura de las variables con el apartado anterior, el valor que hay que predecir se calcula como:

$$f_{t+1} = (1 - \alpha)\, f_t + \alpha\, (y_{t+1}) \qquad (7.5)$$

Es decir, el valor que se obtiene en la predicción es la suma del valor de la predicción en el periodo anterior más el valor observado en la actualidad, ponderados según una probabilidad que puede ser fija o variable. Operando sobre la expresión anterior se obtiene:

$$f_{t+1} = f_t + \alpha\,(y_{t+1} - f_t) \qquad (7.6)$$

donde $f_{t+1}$ es el valor esperado del periodo $t + 1$, $y_{t+1}$ es el valor observado en el instante $t + 1$, $f_t$ es el valor estimado en el instante $t$ y $\alpha$ es el peso que se le otorga al valor observado más reciente ($0 < \alpha < 1$).

Resultado de la aplicación de esta técnica sobre una serie de datos con gran variedad aleatoria es la producción de una serie de estimaciones más suaves (de ahí la denominación) haciendo intervenir valores anteriores según un peso elegido en función del conocimiento de la serie temporal.

En la aproximación de Tustin, se puede tomar más de un valor estimado anterior (si se quiere dar mayor relevancia a lo estimado), o bien se puede tomar más de un valor observado (si se pretende dar mayor relevancia a varias observaciones consecutivas). Por ejemplo, se podría expresar el valor estimado como:

$$f_{t+1} = (1 - \alpha)\,f_t + \frac{\alpha\,(y_{t+1} + y_t)}{2} \qquad (7.7)$$

donde $f_{t+1}$ es el valor estimado del periodo $t + 1$, $y_{t+1}$ es el valor observado en el instante $t + 1$ y $y_t$ es el valor observado en el instante $t$. De este modo, se toma la media de los dos últimos valores observados para la estimación.

El peso habitualmente es fijo, pero puede hacerse variable con el curso de las observaciones, de tal forma que cuanto más larga es la serie temporal, mayor peso tiene $\alpha$. Puesto que la suma de la probabilidad total es uno, el peso se determina con una función cuya serie converge a uno. Una función de peso típica es

$$\alpha = \frac{2n - 1}{2n + 1} \qquad (7.8)$$

donde $n$ es el número de orden en la serie temporal de observaciones. Así, con esta función comienza con $\alpha = 0{,}5$ y tiende a ser uno, a medida que aumenta $n$. De forma general, esta función puede ser:

$$\alpha = \frac{m \times n - 1}{m \times n + 1}$$

donde $m > 2$ actúa como multiplicador del orden de la serie, provocando un mayor peso para las observaciones recientes desde el principio del histórico considerado.

Antes de realizar la predicción, la técnica seleccionada se debe validar con los datos disponibles. Esto se puede hacer utilizando parte de los datos históricos antiguos para ejercitar el modelo. Los datos restantes, que corresponden a los valores más actuales, se pueden entonces comparar con los valores predichos para valorar la exactitud del método. También se pueden hacer pruebas para valorar el error cuadrático medio de cada técnica en estudio de tal forma que se seleccione la que da menor error cuadrático medio.

Un ejemplo de aplicación de las técnicas de predicción podría ser estimar la demanda del procesador en una aplicación de catálogo en línea en la web. En función del número de artículos existentes en el catálogo (que van variando con el tiempo) y los accesos que se realizan al servidor web durante un periodo de observación, se podrían utilizar las tres técnicas de predicción con los datos del histórico de monitorización del procesador. Así, según que los datos presenten una tendencia, sean estacionarios o presenten gran variedad de proceso, se utilizan unas técnicas u otras para conocer la capacidad futura.

## 7.4. Unidades de predicción natural

En la planificación de la capacidad, la predicción de la carga de trabajo futura es de capital importancia. Una parte muy grande de esta predicción es la tasa de cambio de la carga de trabajo debido a las aplicaciones actuales. Existe gran cantidad de información respecto a los cambios en la carga de trabajo disponible en los departamentos de planificación y gestión de la empresa, como, por ejemplo, los crecimientos que prevé tener en sus departamentos, productos, clientes, proveedores, áreas de negocio, etc. Una unidad de predicción natural (*Natural Forecast Unit*, NFU) es una variable de negocio cuyo valor está directamente relacionado con los recursos consumidos por una o varias aplicaciones. Por ejemplo, la paga de un empleado en una empresa puede generar una transacción de nómina una vez al mes. A su vez, las medidas o cálculos realizados con los monitores pueden determinar la utilización media de procesador, el número de operaciones de entrada/salida, etc., para cada tipo de transacción, y, en particular, para las transacciones de nómina. Por lo tanto, una predicción del número de empleados dentro de un año puede transformarse en una buena estimación de la utilización de los recursos por la aplicación de nómina dentro de un año. Es decir, el número de empleados es la unidad de predicción natural para esta aplicación. Otras unidades de predicción natural típicas podrían ser el número de artículos para un sistema de inventario; el número de pólizas de seguros para el sistema principal de una compañía de seguros; etc. Los componentes esenciales de la caracterización de la carga utilizando las unidades de predicción natural para los propósitos de estimación del rendimiento son los siguientes:

- Medidas del trabajo solicitado al sistema utilizando variables de negocio como mé- trica, que son propiamente las unidades de predicción natural.
- Las relaciones que muestren el número de transacciones generadas para el periodo caracterizado para cada unidad de predicción natural.
- Las relaciones que indiquen los recursos del sistema que consume cada transacción.

En algunos casos no es fácil identificar las unidades de predicción natural. Una vez identificadas, es necesario asociar contadores de transacciones y de necesidades de recursos a las mismas. El resultado de este análisis es un conjunto de coeficientes que describen el número de transacciones de cada tipo para la hora pico por unidad de predicción natural, y el requerimiento medio de recursos para cada transacción, preferiblemente en unidades independientes de la productividad de los dispositivos. El modelo de carga de trabajo resultante puede servir para varios propósitos. En principio, puede servir para preparar un sistema de cargo económico por uso del sistema para cada departamento o función de la empresa. Pero, por supuesto, su función más importante es hacer disponibles los coeficientes para traducir las previsiones de NFU en necesidades de recursos del sistema. Para poder utilizar las NFU, los programas aplicativos y el sistema operativo se deben instrumentalizar para que los monitores permitan la recogida de datos necesarios y también las tasas de transacciones correspondientes. Los analistas de sistemas pueden realizar entrevistas a los usuarios para determinar varios tipos de datos concernientes a las NFU. Algunos ejemplos son los siguientes:

- ¿Cuáles son las principales variables NFU para el usuario de este departamento? Las variables que se utilizan actualmente ¿son todavía las más apropiadas?
- ¿Cúales son las tasas de crecimiento más probables para estas NFU? ¿Representa esto cambios importantes de las tasas de crecimiento históricas?
- ¿Es probable que haya cambios entre las NFU y el consumo de recursos del sistema? Por ejemplo, si una transacción se complica para aumentar la realización de un mayor número de tareas, esto provocará probablemente un mayor consumo de recursos.
- ¿Cuáles son las aplicaciones más probables que solicitará este usuario? ¿Qué prioridad asignará a cada una de ellas? ¿Se cubrirán estas aplicaciones con las NFU actuales o se necesitarán unas nuevas?

## 7.5. Capacidad bajo demanda

Una aplicación interesante de la planificación de la capacidad es la determinación del rendimiento de los sistemas durante su ejecución (gestión de nivel 3 o superior). De esta forma, el sistema de análisis y predicción:

1. Analiza las medidas de rendimiento que se van realizando en tiempo real.
2. Modela automáticamente sus diferentes facetas.
3. Inician acciones automáticamente para mantener los tiempos de servicio acordados y la utilización de los recursos.

La capacidad bajo demanda automatiza la asignación de la capacidad necesaria a una carga del sistema para cumplir un nivel de servicio determinado. Sus beneficios son varios:

1. Ahorra el personal necesario que realizaba manualmente estas tareas antes. El per- sonal podría estar dedicado a otras tareas menos repetitivas.
2. Difiere la inversión en la substitución (reemplazo) de máquinas o contratación de nuevos servicios, consolidando aplicaciones y utilizando un gestor de recursos com- partidos. Tradicionalmente las aplicaciones se han aislado ejecutándose en recursos distintos por razones de seguridad, aislamiento y rendimiento. Pero esta rigidez lleva a situaciones en que una aplicación necesita recursos adicionales mientras que otras aplicaciones están infrautilizando sus servidores, como se muestra en la Figura 7.2.

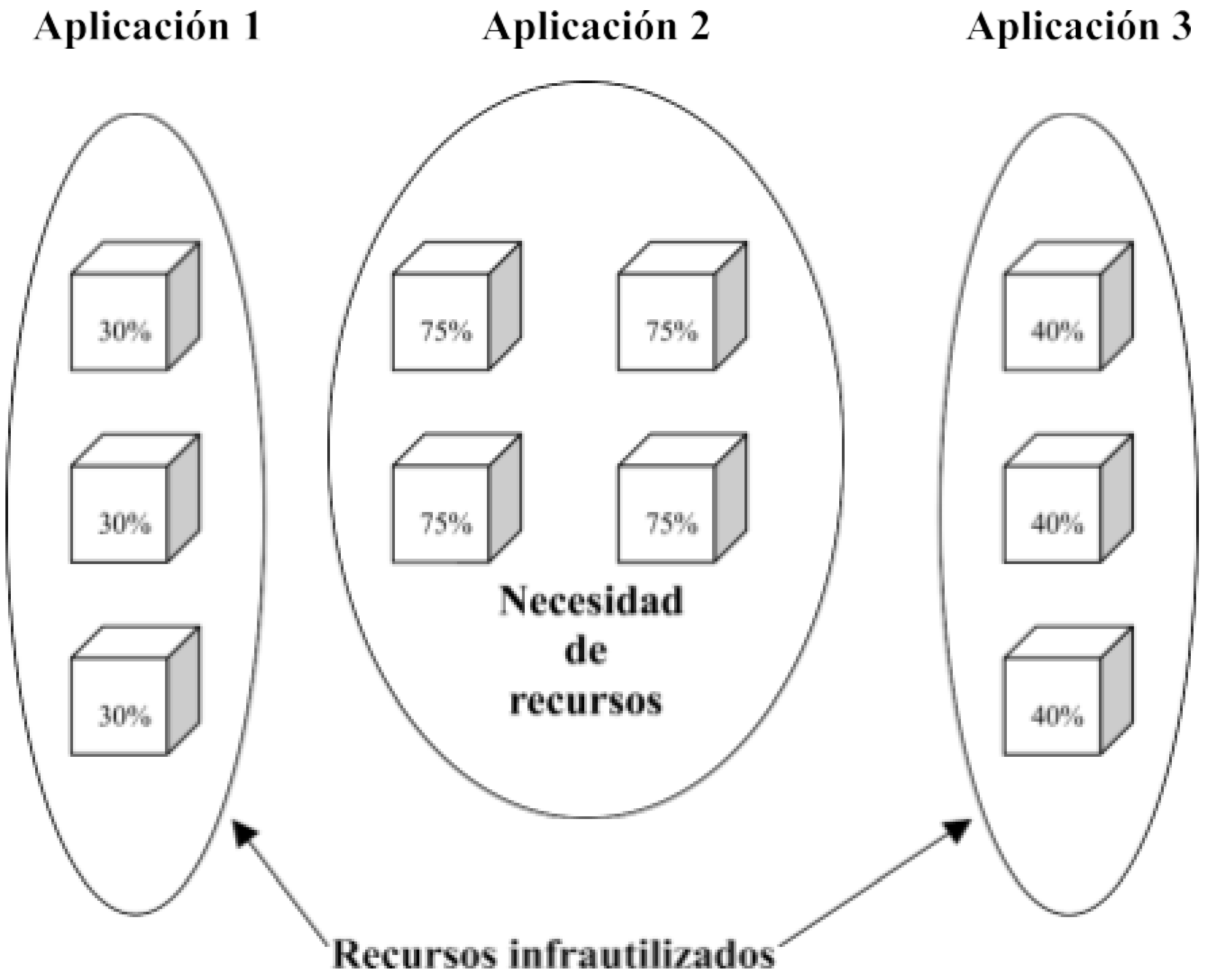

**Figura 7.2:** Asignación de recursos rígida a las aplicaciones.

Supóngase un suministrador de servicios de Internet (*Internet Service Provider*, ISP) de varias aplicaciones: venta de libros en línea, reserva de billetes de avión, noticias deportivas, etc. La utilización de capacidad bajo demanda empleando una gestión de recursos compartidos permite que si se tiene un pico de demanda en la aplicación de venta de libros porque está teniendo lugar la feria del libro, se puede asignar la capacidad sobrante de otras aplicaciones a esta aplicación. Una gestión de este tipo se muestra en la Figura 7.3, donde se muestra la asignación de recursos a la aplicación 2, que es la que está más cargada, provenientes de las aplicaciones 1 y 3, dejando un entorno de utilización de los recursos más equilibrado entre las tres aplicaciones.

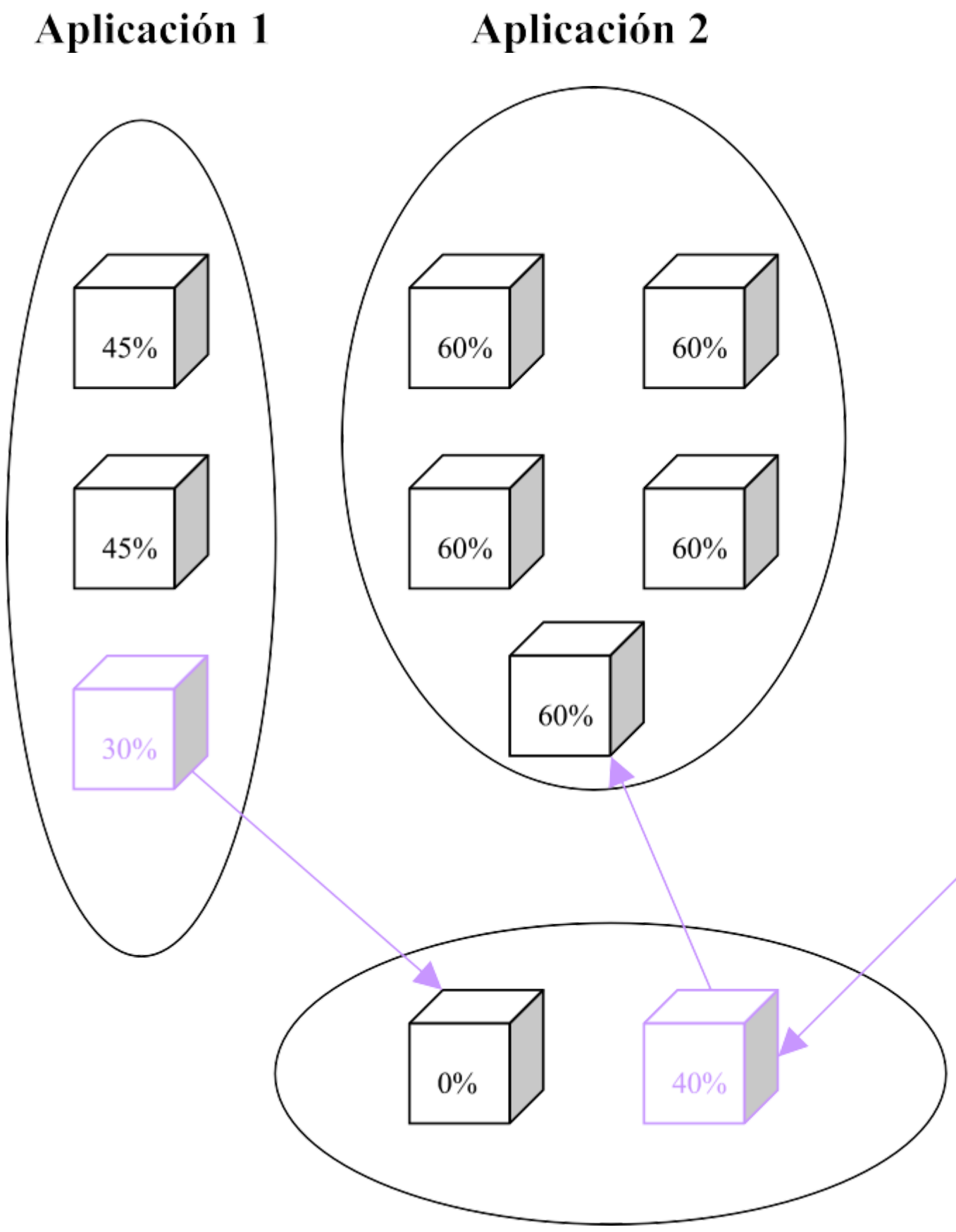

**Figura 7.3:** Asignación de recursos flexible bajo demanda a las aplicaciones.

En la asignación de recursos inicial, de carácter más rígido, la aplicación 2 tenía sus recursos, típicamente procesadores, cargados en un 75 %, mientras que la aplicación 3 solamente los tenía cargados en un 40 %, y la aplicación 1 incluso en menos del 30 %. Suponiendo que todos los recursos tienen la misma potencia, se puede asignar un procesador dedicado anteriormente a la aplicación 3 a la aplicación 2, y dejar uno de la aplicación 1 como reserva (sin carga) para que pueda entrar en caliente en caso de que cualquiera de los otros procesadores se averíe. La carga final sería del 45, 60 y 60 %, respectivamente, en los procesadores dedicados a cada una de las aplicaciones.

Otro ejemplo muy interesante puede ser las agrupaciones de servidores web (*web clusters*), que contienen la misma información, donde un conmutador (*web switch*) se encarga de encaminar las transacciones HTTP entrantes en función de la carga actual de los servidores. De este modo, dependiendo de la política de servicio implementada en el conmutador y de la carga de los servidores, se pueden establecer subgrupos de servidores de tamaño variable para las distintas transacciones. Así, las transacciones prioritarias, si las hubiera, podrían adquirir más servidores disponibles en su grupo con el correspondiente detrimento de las no prioritarias, si el acuerdo de servicio requerido así lo dispusiera.

3. Facturación de hardware y software bajo demanda. Esto supone el pago por los recursos que se están utilizando tal como actualmente se factura el consumo de agua o de energía eléctrica en una población: tanto se consume, tanto se paga.

4. Aceleración de la implantación de la externalización de la informática de las empresas (*outsourcing*). La garantía de los acuerdos de servicio requeridos por los usuarios, mediante el uso de una planificación de la capacidad en función de la demanda, provoca una mayor aceptación de que esos servicios se realicen externamente a la organización que los demanda, es decir, que los sistemas no pertenezcan propiamente a las empresas que los necesitan.

## 7.6. Planificación en escenarios muy variables

Se ha comprobado que algunos sistemas distribuidos de gran envergadura, como Internet, tienen cargas con ráfagas de tráfico (*burstiness*) independientemente de la escala temporal en que se midan. Esto es, la cantidad de información transmitida tiene picos altísimos y caídas pronunciadas, que se pueden observar tomando escalas temporales diferentes. Se denomina *ratio de tráfico en pico* al cociente entre la frecuencia media de tráfico durante el pico y la frecuencia media de tráfico en una instalación.

Este tipo de distribución se denomina autosimilitud en la literatura (*self-similarity*). En particular, si se representa la carga de trabajo que llega durante un cierto periodo de observación, en trabajos por unidad de tiempo, las gráficas que muestran el número de trabajos por segundo, tienen formas muy parecidas a las del número de trabajo por minuto, por hora, por día, por mes o incluso por año. Es decir, representando el número de trabajos en el eje de ordenadas y las unidades de tiempo en el de abcisas, el comportamiento a ráfagas se presenta similar en todas las escalas de tiempo.

Por otra parte, los elementos de información que se almacenan en los servidores web, habitualmente archivos o aplicaciones dinámicas, tienen normalmente tamaños que van de lo ínfimo a lo gigantesco, con una distribución de cola pesada (*heavy-tailed* ) donde pocos elementos de gran tamaño tienen mucha influencia en el rendimiento del servidor.

Consideremos que se desea estimar el tiempo de demanda de una petición de acceso a un archivo en un servidor web, que ya se está ejecutando desde hace algún tiempo y además se sabe el tiempo medio de demanda de los accesos al servidor, entonces:

1. Si la distribución del tiempo de demanda es de cola corta (*short-tailed* ), cuanto más tiempo haya pasado desde el principio de la ejecución, menor es el tiempo adicional que se tendrá que esperar hasta la finalización del acceso solicitado. Ello es debido a que la distribución tiene una cola media menor que la demanda media, como por ejemplo en la distribución uniforme.

2. Si la distribución no tiene memoria, el tiempo adicional que se tendrá que esperar hasta la finalización del acceso es independiente del tiempo que ya haya pasado. Es decir, la distribución tiene una cola media igual a la demanda media, como por ejemplo la distribución exponencial.

3. Sin embargo, si la distribución es de cola larga, el tiempo adicional que se tendrá que esperar hasta la finalización del acceso al servidor web crece con el tiempo que ya se ha esperado. Es decir, la distribución tiene una cola media mayor que la demanda media del servidor, como por ejemplo la distribución de Pareto.

En los servidores web se dan los tres fenómenos anteriormente citados, el tráfico se produce a ráfagas, la carga es autosimilar y la información reside en objetos con tamaños (archivos o transacciones dinámicas) que siguen una distribución de Pareto. Aunque es difícil predecir las demandas de uso de los sistemas de este tipo, los administradores deben prepararse para ello.

En muchos casos, es posible identificar las raíces de los cambios repentinos en la deman- da de los usuarios en este tipo de escenarios y también es posible categorizar los fenómenos que producen las ráfagas de tráfico a un sitio web. Comencemos por clasificar los tipos de eventos:

- Eventos nuevos predecibles. Surgen en sitios web específicos a causa de nuevos suce- sos predecibles. Tras hechos impredecibles, como los desastres naturales (huracanes, terremotos y tormentas), sucede que un ingente número de personas buscan informa- ción en servidores especializados. Los sitios relacionados, por ejemplo, con el tiempo meteorológico suelen contener la predicción de estos sucesos mediante modelos que predicen estos desastres naturales y deben estar preparados para recibir avalanchas de consultas.

- Anuncios de productos o servicios. Un servicio de la preparación de liquidación de impuestos en línea se colapsó en la primera semana de operación. La compañía es- peraba 500.000 clientes en un periodo de doce semanas. Sin embargo, el sitio recibió 220.000 clientes en la primera semana y el sistema se colapsó. Se puede concluir de esta experiencia que un negocio virtual electrónico puede establecer la fecha de anun- cio de un producto o servicio, pero es muy difícil que se pueda predecir el éxito que éste tendrá. Por lo tanto, se debe planificar una infraestructura flexible de tal forma que los recursos se puedan añadir o retirar según las necesidades del momento.

- Eventos especiales. Las navidades, los eventos deportivos o cinematográficos, unas elecciones, etc., provocan siempre un crecimiento en el tráfico de Internet. Este incremento en el número de usuarios en línea es claramente previsible con varios meses de antelación. La pregunta que de nuevo podemos formularnos es si es posible predecir la intensidad de la avalancha. La contestación es clara: mediante la planificación de la capacidad del sitio web.

- Eventos nuevos impredecibles. Los usuarios de Internet siguen las noticias nuevas rompedoras y lo mismo hace el tráfico generado por sus consultas. Siempre que ocurre un suceso importante (guerras, escándalos, accidentes, desplomes de la bolsa, etc.), el tráfico se dispara en los sitios de noticias escritas y televisadas. Por ejemplo, durante una crisis del mercado de valores cuando el índice industrial Dow Jones se desplomó, se negoció un volumen récord de acciones. Los negocios de agencias de bolsa se inundaron con visitantes buscando cotizaciones y realizando pedidos de compra y/o venta, lo que generó picos de tráfico de siete a diez veces la media usual del volumen de órdenes.

Por otra parte, aunque el tratamiento de la llegada a ráfagas del tráfico y de la autosimilaridad de éste se escapa del ámbito de este texto, se puede disminuir el efec- to de la distribución de cola pesada en los tamaños de los archivos web. De forma práctica, se puede esperar que un gran porcentaje de las peticiones HTTP que llegan a un servidor web, soliciten documentos de pequeño tamaño, mientras que un pequeño porcentaje de las mismas soliciten documentos que son de uno o dos órdenes de magnitud más grandes que los primeros. Debido a la gran variabilidad de los tamaños de los documentos, tomar tamaños medios para toda la población de peticiones tendría poca significación estadística. Categorizar las peticiones en un número finito de clases, definidas por rangos de tamaños, mejora el significado y la exactitud de las métricas de rendimiento. De este modo, la in- tensidad de la carga de trabajo por clase se reflejará en la frecuencia media de peticiones por clase que lleguen al servidor web.

La popularidad de la información contenida en los servidores web es un fenómeno conocido que a veces se ha tratado a través de la ley de Zipf. Esta ley fue aplicada originalmente para relacionar la popularidad de las palabras de un idioma y su frecuencia de uso, pero también se puede utilizar para caracterizar la frecuencia de acceso a los archivos web. La ley de Zipf indica que unas pocas palabras son muy utilizadas, o unos pocos ficheros son muy accedidos, y que muchas palabras son muy poco empleadas, o un número muy elevado de ficheros son muy poco accedidos. La aplicación de la ley de Zipf a Internet enuncia que si se establece un orden de popularidad de los ficheros contenidos en un servicio web (ser-vidor o grupo de servidores), la frecuencia de acceso a esos ficheros es aproximadamente proporcional al inverso de su popularidad. O lo que es lo mismo, el número de referencias a un fichero web ($P$) es inversamente proporcional ($k$ es una constante positiva) al orden en la lista de popularidad de ese fichero ($r$) según la ecuación siguiente:

$$P = \frac{k}{r} \tag{7.9}$$

Es decir, si se ordenan los ficheros web según su popularidad ($\rho$) empezando por el número uno, el $i$-ésimo fichero más popular es accedido dos veces más que el $2i$-ésimo en la lista. Cuanto menor sea el orden en la lista de popularidad de un fichero, esto es, cuanto más arriba esté el fichero en la lista de popularidad, mayor será el número de referencias que se hagan a él. Este resultado puede ser muy interesante para realizar una primera caracterización de la carga cuando se conoce la frecuencia de acceso total a un servidor web con eventos como los que se han descrito anteriormente.

## 7.7. Problemas resueltos

**PROBLEMA 7.1** El administrador de sistemas del departamento de Tecnología de la Información de una compañía que posee una división de comercio electrónico para sus productos, monitoriza el tráfico del sitio web cada treinta minutos y calcula la tasa media de llegada de las transacciones durante el periodo. Al final del día, el sistema suministra el ratio de tráfico en pico del día. También, al final de la semana, el director recibe un informe de rendimiento semanal, en el que se resalta el momento con mayor demanda. Del análisis de los gráficos del tráfico, el administrador deduce que los picos ocurren en la mitad de la semana. Para anticipar posibles problemas de colapso, el director solicita al administrador sobre cuál será el ratio estimado de la demanda pico de la semana próxima, basándose en la tabla de ratios de tráfico en pico indicados en la Tabla 7.1.

| Semana | Ratio tráfico en pico |
|---|---|
| 1 | 33,5 |
| 2 | 26,3 |
| 3 | 29,9 |
| 4 | 24,8 |
| 5 | 22,6 |
| 6 | 23,2 |
| 7 | 27,1 |
| 8 | 25,7 |

**Tabla 7.1:** Estadística de clientes semanales.

**SOLuCIón:** La tabla de datos asociada a este problema muestra el ratio de tráfico en pico durante las ocho semanas últimas. Para estimar el ratio de tráfico en pico para la siguiente semana, el administrador decidió utilizar la técnica de medias móviles con los tres valores más recientes en la tabla. Por lo tanto, el ratio de tráfico en el pico para la siguiente semana, empleando la ecuación (7.4), resulta ser de:

$$f_9 = \frac{23{,}2 + 27{,}1 + 25{,}7}{3} = 25{,}3$$

Evidentemente, si hubiera tomado todas las semanas el ratio hubiera sido de 26,6■

**PROBLEMA 7.2** Una compañía de investigación de Internet monitoriza los clientes, esto es, visitantes y compradores, de una tienda de juguetes en línea. El número de clientes (base de clientes) es una información clave para la planificación de recursos e infraestructura para la tienda. La Tabla 7.2 muestra el tamaño de la base de clientes de los seis últimos meses del año.

| Mes | Número de clientes |
|---|---|
| Julio | 708.000 |
| Agosto | 654.000 |
| Septiembre | 636.000 |
| Octubre | 712.000 |
| Noviembre | 608.000 |
| Diciembre | 704.000 |

**Tabla 7.2:** Evolución de la base de clientes del enunciado del Problema 7.2.

Además de estos datos, la dirección tiene la información siguiente derivada del comportamiento del cliente:

- El número de visitas mensuales de un cliente al servidor web es de 2,7.
- El ratio medio de compra de una visita es de 1,87 %.
- La carga media de trabajo en una sesión de cliente es de 4,7 transacciones.

El director del departamento de Informática solicita calcular el número de trasacciones estimadas para el mes de enero del año próximo. Se pide que se utilice la técnica de predicción de suavizado exponencial para resolver este ejercicio con un peso fijo del 60 % para las medidas de los meses anteriores.

| Mes | Clientes | Previsión ($\alpha$ = 0,6) |
|---|---|---|
| Julio | 708.000 | 708.000 |
| Agosto | 654.000 | 675.600 |
| Septiembre | 636.000 | 651.840 |
| Octubre | 712.000 | 687.936 |
| Noviembre | 608.000 | 639.974 |
| Diciembre | 704.000 | 678.390 |

**Tabla 7.3:** Evolución de la base de clientes utilizando la técnica exponencial suave en el Problema 7.2.

**SOLuCIón:** Se aplica la técnica de suavizado exponencial a los datos de la tabla del enunciado, con lo que se obtienen los valores de predicción en la tercera columna de la Tabla 7.3.

En particular, el tamaño estimado de la base de clientes en el mes de Diciembre, calculado a partir de los datos medidos en diciembre y de la previsión anterior, se obtiene mediante la ecuación (7.6):

$$f = 639.974 + 0{,}60 \times (704.000 - 639.974) = 678.390$$

Si se interpreta este valor como el previsto para enero del siguiente año, el número estimado de transacciones mensuales se calcula como sigue. En primer lugar calculamos el número total de visitas multiplicando la media de las visitas mensuales por cliente y la base de clientes:

$$2{,}7 \times 678.390 = 1.831.653$$

Las transacciones por mes vendrán dadas por el producto entre la media de transacciones por visita y el número total de visitas:

$$4{,}7 \times 1.831.653 = 8.698.769$$

Por lo tanto, el número de transacciones que se estima para enero es de 8.698.769■

**PROBLEMA 7.3** Considérese un servidor web que da visibilidad a los anuncios de una empresa. La actividad del servidor se mide por el número de páginas y gráficos solicitados por los visitantes, por la duración de la visita y por el tiempo entre visitas. Supóngase que se conoce el número de visitas realizadas durante los doce últimos meses tal como se representa en la Tabla 7.4.

Se pide calcular el número de visitantes al servidor web durante las próximas navidades, esto es, en el próximo diciembre, para así poder suministrar un servicio de alta calidad. Se solicita que se utilice la técnica de predicción de regresión lineal.

| Mes | Visitas históricas |
|---|---|
| Abril | 65.110 |
| Mayo | 73.333 |
| Junio | 75.345 |
| Julio | 68.235 |
| Agosto | 58.011 |
| Septiembre | 75.644 |
| Octubre | 65.856 |
| Noviembre | 80.456 |
| Diciembre | 90.322 |
| Enero | 65.100 |
| Febrero | 74.201 |
| Marzo | 76.212 |

**Tabla 7.4:** Evolución mensual de la base de clientes durante los doce últimos meses.

**SOLUCIÓN:** Se utiliza el método de la regresión lineal para predecir el número de visitas. Se puede representar el número de visitas con $y$, la variable dependiente, y el número del mes con la variable independiente $x$. De la Tabla 7.4 podemos deducir los valores intermedios siguientes:

$$n = 12$$

$$\bar{x} = \frac{1+12}{2} = 6{,}5$$

$$\bar{y} = \frac{65.110 + 73.333 + \cdots + 76.212}{12} = 72.479{,}1$$

$$\sum_{i=1}^{12} x_i^2 = 650$$

$$\sum_{i=1}^{12} x_i \times y_i = 5.753.963$$

$$n \times \bar{x}^2 = 507$$

$$n \times \bar{x} \times \bar{y} = 5.653.370{,}7$$

Se calculan los valores de $b$ y $a$ con las ecuaciones (7.2) y (7.3), obteniendo los siguientes valores:

$$b = 703{,}4$$

$$a = 67.906{,}7$$

En consecuencia, la relación lineal entre el mes ($x$) y el número de visitas ($y$) se puede escribir empleando la ecuación (7.1) como:

$$y = 67.906{,}7 + 703{,}4 \times x$$

Por lo tanto, se puede predecir el número de visitas al servidor web para nuevos valores de $x$. Por ejemplo, el siguiente diciembre es el mes 21, y por lo tanto el número de visitas que se predicen es:

$$y = 67.906{,}7 + 703{,}4 \times 21 = 82.678{,}1$$

■

---

**PROBLEMA 7.4** En muchas organizaciones el correo electrónico se considera una aplicación de misión crítica porque de su funcionamiento depende la velocidad en las comunicaciones de la empresa. Para mantener los niveles de servicio esperados, se mide constantemente el volumen de correo electrónico por medio del número de mensajes. La columna segunda de la Tabla 7.5 muestra el número mensual de mensajes procesados por el servidor de correo electrónico en los últimos diecisiete meses.

Para evitar problemas de rendimiento, el administrador del sistema quiere conocer anticipadamente cuál será el número de mensajes procesados el mes de septiembre aplicando las tres técnicas de predicción estudiadas: regresión lineal, medias móviles y suavizado exponencial.

**SOLUCIÓN:**

1. *Técnica de regresión lineal.* La primera técnica que se utiliza para predecir el número de mensajes es el método de la regresión lineal con las ecuaciones correspondientes. Se puede representar el número de mensajes con $y$, como variable dependiente, y el número del mes con la variable independiente $x$. De la Tabla 7.5 se pueden deducir los valores intermedios siguientes:

$$n = 17$$

$$\bar{x} = \frac{1 + 17}{2} = 9$$

$$\bar{y} = \frac{646.498 + 783.485 + \cdots + 840.558}{17} = 776.424{,}4$$

$$\sum_{i=1}^{17} x_i^2 = 1.785$$

$$\sum_{i=1}^{17} x_i\, y_i = 126.470.664$$

| Mes | Número actual de mensajes |
|---|---|
| Abril | 646.498 |
| Mayo | 783.485 |
| Junio | 498.583 |
| Julio | 471.315 |
| Agosto | 494.311 |
| Septiembre | 549.204 |
| Octubre | 974.004 |
| Noviembre | 1.001.598 |
| Diciembre | 706.086 |
| Enero | 835.888 |
| Febrero | 1.149.200 |
| Marzo | 1.066.325 |
| Abril | 984.593 |
| Mayo | 715.774 |
| Junio | 690.877 |
| Julio | 790.916 |
| Agosto | 840.558 |

**Tabla 7.5:** Evolución del número de mensajes de correo electrónico durante los últimos diecisiete meses.

$$n \times \bar{x}^2 = 1.377$$
$$n \times \bar{x} \times \bar{y} = 118.792.935,2$$

Se calculan los valores de $b$ y $a$ sustituyendo los valores anteriores en las ecuaciones (7.2) y (7.3) de regresión obteniendo $b = 18.817,97$ y $a = 607.062,67$. Se obtiene la ecuación de la recta de tendencia, resultando la expresión siguiente para la ecuación (7.1):

$$y = 607.062,67 + 18.817,97 \times x$$

Las representaciones de los datos medidos y de los resultados obtendidos con las tres técnicas de predicción se muestran en la Figura 7.4.

Por lo tanto, se puede predecir el número de mensajes al servidor web para nuevos valores de $x$. Por ejemplo, el siguiente septiembre es el mes 18, que es lo que se nos pide en el enunciado del problema, y por lo tanto el número de mensajes predichos es:

$$y = 607.062,67 + 18.827,97 \times 18 = 945.786,13$$

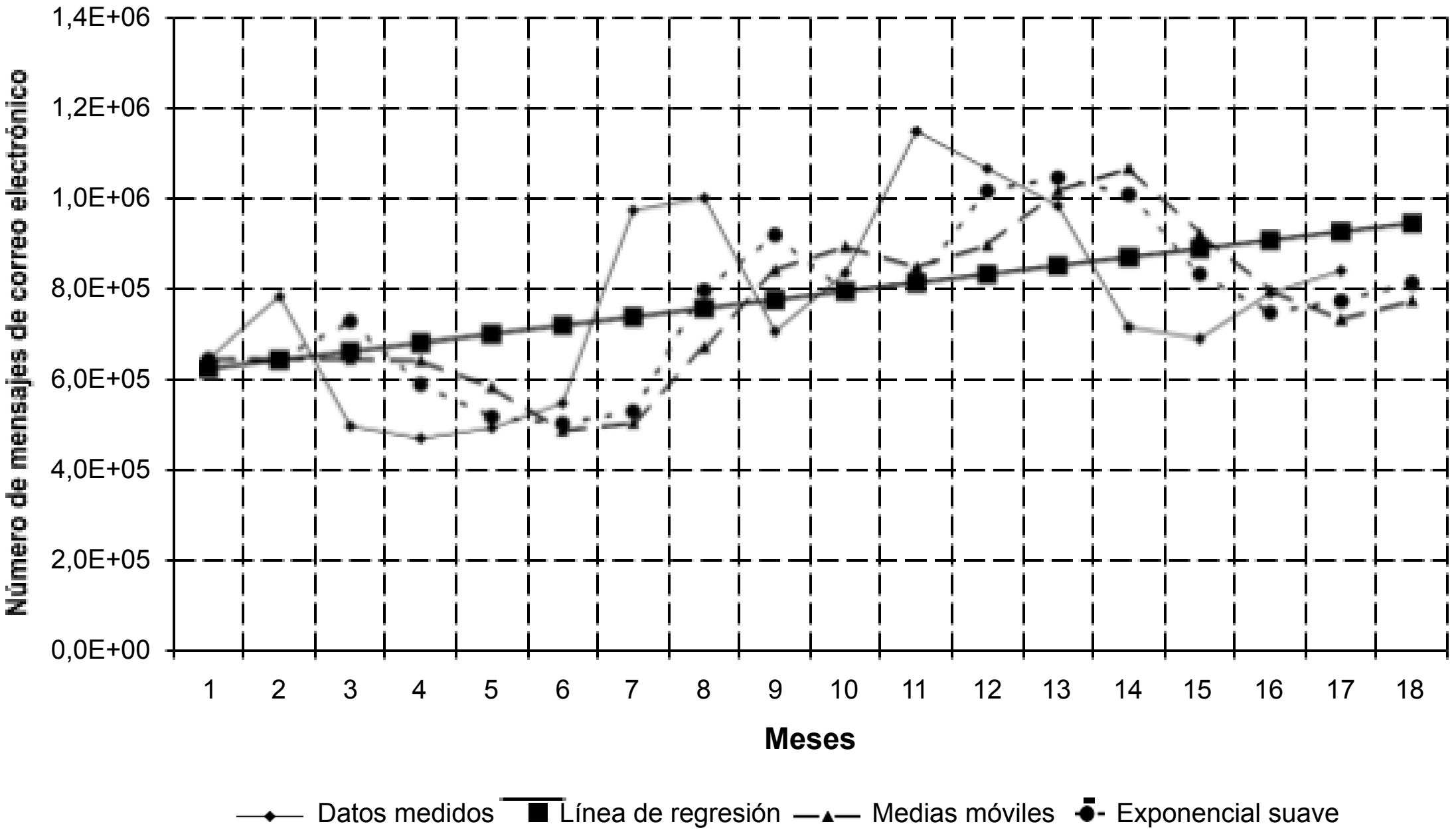

**Figura 7.4:** Datos y previsión utilizando las tres técnicas de predicción.

2. *Técnica de las medias móviles.* La segunda técnica que se aplica es la de medias móviles. Para ello se eligen los tres últimos valores de *n* de los datos existentes en la Tabla 7.5. El uso de tres observaciones da el menor error cuadrático medio. Por lo tanto, el número de mensajes para septiembre, utilizando la ecuación (7.4), será:

$$f_{\text{septiembre}} = \frac{690.877 + 790.916 + 840.558}{3} = 774.117$$

3. *Técnica del suavizado exponencial.* El problema consiste en conocer el peso de los datos históricos de las diecisiete últimas observaciones. Supongamos que se otorga un valor $\alpha = 0{,}6$ hasta agosto. Aplicando el cálculo suavizado expresado por la ecuación (7.6), se obtiene la tercera columna de la Tabla 7.6. Nótese que el valor predicho la primera vez (para el mes de mayo) es igual al valor observado (en abril), ya que no hay observaciones previas.

   En particular, el número de mensajes estimados para septiembre es el calculado en agosto:

$$f\text{agosto} = 773.673 + 0{,}6 \times (840.558 - 773.673) = 813.804$$

■

| Mes | Número actual de mensajes | Predicción |
|---|---|---|
| Abril | 646.498 | 646.498 |
| Mayo | 783.485 | 728.690 |
| Junio | 498.583 | 590.625 |
| Julio | 471.315 | 519.039 |
| Agosto | 494.311 | 504.202 |
| Septiembre | 549.204 | 531.203 |
| Octubre | 974.004 | 796.883 |
| Noviembre | 1.001.598 | 919.712 |
| Diciembre | 706.086 | 791.536 |
| Enero | 835.888 | 818.147 |
| Febrero | 1.149.200 | 1.116.779 |
| Marzo | 1.066.325 | 1.046.507 |
| Abril | 984.593 | 1.009.358 |
| Mayo | 715.774 | 833.207 |
| Junio | 690.877 | 747.809 |
| Julio | 790.916 | 773.673 |
| Agosto | 840.558 | 813.804 |

**Tabla 7.6:** Evolución del número de mensajes de correo electrónico utilizando la técnica de predicción exponencial suave.

**PROBLEMA 7.5** Una compañía productora de música vende los éxitos archivados en sus servidores a sus clientes conectados a través de Internet. A través de los históricos pasados, la compañía espera recibir un millón de visitas diarias a partir del mes siguiente debido a que se lanzará el nuevo disco de un grupo muy popular, que se supone que alcanzará el primer puesto en la lista de ventas. Se le pide al analista de prestaciones que estime el número de visitas a los cinco discos más vendidos, incluido el nuevo lanzamiento, para poder prever el almacenamiento de los archivos musicales en unidades de E/S de alta velocidad. Desgraciadamente, el analista no dispone de ninguna información adicional para la previsión. ¿Cómo se podría cuantificar la carga con tan pocos datos?

**SOLuCión:** Si se aplica la ley de Zipf, expresada en la ecuación (7.9), la frecuencia de acceso a los archivos musicales es proporcional a su popularidad. Consideremos que todos los accesos previstos para el mes siguiente acceden a los archivos del nuevo éxito; por tanto, su frecuencia es uno y recibirá un millón de visitas diarias. Por otra parte, el número de visitas a los archivos de los siguientes cuatro éxitos musicales se obtiene reduciendo la frecuencia de acceso a medida que su popularidad esperada disminuye (menor orden en la lista de ventas). Es decir, se aplica la ley para calcular el número de visitas del éxito *i*-ésimo dividiendo el número de visitas total entre el orden en que figura este éxito *i*-ésimo.

De este modo, el número de visitas de los cinco discos más populares serán los correspon-dientes a 1.000.000 visitas para el nuevo éxito, 500.000 para el segundo de la lista de éxitos y
333.333, 250.000 y 200.000 para los restantes. ■

**PROBLEMA 7.6** Tres divisiones de una empresa tienen cargas de trabajo que consumen recursos en la hora de pico de carga que corresponden a diciembre del año 2003, como se indica en la Tabla 7.7.

| Departamento | Consumo de procesador (s) |
|---|---|
| A | 300 |
| B | 530 |
| C | 250 |

**Tabla 7.7:** Consumo de procesador para el Problema 7.6.

El resto de la carga se puede considerar que consume 150 segundos de procesador en la hora punta. Se pide:

1. Calcular las contribuciones relativas de la carga de trabajo como porcentajes de la carga en la hora punta.
2. ¿Cuáles serán los componentes relativos de la carga de trabajo en diciembre de dentro de dos años si las mejores estimaciones del crecimiento de la carga son del 40, 10 y 20 % anual, para los departamentos A, B y C, respectivamente, y para el resto de los departamentos de un 15 %?

**SOLuCIón:**

1. El consumo total de procesador viene dado por:

$$300 + 530 + 250 + 150 = 1.230 \text{ s}$$

Las contribuciones relativas de carga de trabajo como porcentajes de la carga en la hora pico son las siguientes:

$$\text{Departamento A} = \frac{300}{1.230} \times 100 = 24,4\ \%$$

$$\text{Departamento B} = \frac{530}{1.230} \times 100 = 43,1\ \%$$

$$\text{Departamento C} = \frac{250}{1.230} \times 100 = 20,3\ \%$$

$$\text{Resto departamentos} = \frac{150}{1.230} \times 100 = 12,2\ \%$$

2. El crecimiento del consumo del procesador para los años siguientes se calcula, para los tres departamentos, a continuación:

Dpto. A en un año = $300 \times 1{,}4 = 420$ Dpto.
A en dos años = $300 \times 1{,}4^2 = 588$
Dpto. B en un año = $530 \times 1{,}1 = 583$
Dpto. B en dos años = $530 \times 1{,}1^2 = 641{,}3$
Dpto. C en un año = $250 \times 1{,}2 = 300$
Dpto. C en dos años = $250 \times 1{,}2^2 = 360$
Resto en un año = $150 \times 1{,}15 = 172{,}5$
Resto en dos años = $150 \times 1{,}15^2 = 198{,}4$
Consumo total en 2005 = 1.787,7

Las contribuciones relativas por cada departamento dentro de dos años (2005) serán las siguientes:

Depto. A = $\frac{588}{1.787{,}7} \times 100 = 32{,}9\ \%$

Depto. B = $\frac{641{,}3}{1.787{,}7} \times 100 = 35{,}9\ \%$

Depto. C = $\frac{360}{1.787{,}7} \times 100 = 20{,}1\ \%$

Resto = $\frac{198{,}4}{1.787{,}7} \times 100 = 11{,}1\ \%$

Si se comparan las contribuciones relativas pasados dos años (2005) con las iniciales (2003) se puede ver el gran incremento que se ha producido en la contribución relativa del departa- mento A, que ha quitado contribución fundamentalmente al departamento B, que de todas formas sigue teniendo la mayor contribución. Si se sigue esta tendencia, en tan sólo un año el departamento A pasará a ser el de mayor contribución relativa. La contribución relativa del departamento C y del resto de los departamentos varía muy poco durante estos dos años.

---

**PROBLEMA 7.7** Una compañía posee una Intranet de 100 Mbps que utiliza para las co- municaciones de su sistema distribuido. Se sabe que el ancho de banda disponible ha sido suficiente para mantener el aplicativo cliente-servidor de la organización. Sin embargo, desde que se han adquirido ciertos equipos de edición y producción de vídeo en el de- partamento de publicidad, se han producido grandes retrasos en las comunicaciones que posiblemente se deban a la transferencia de multimedia compartiendo el ancho de banda con el aplicativo tradicional. Se ha monitorizado el tráfico de red durante cinco días labo- rables, con lo que se ha obtenido la Tabla 7.8, que muestra los porcentajes de ancho de banda ocupado durante el pico de carga.

| Día | Porcentaje ocupación red |
|---|---|
| 1 | 64,0 |
| 2 | 78,5 |
| 3 | 49,8 |
| 4 | 97,4 |
| 5 | 99,0 |

**Tabla 7.8:** Porcentaje de ocupación de la red.

Se solicita la realización de una previsión de la ocupación de red durante el pico de trabajo diario mediante suavizado exponencial dando un peso del 90 % a la media de los dos últimos días monitorizados. Compárese la previsión con un peso variable mayor o igual que el 90 % para el valor monitorizado diario.

**SOLuCión:** Se aplica la aproximación de Tustin en el primer caso, con lo que las predicciones diarias se obtienen a partir de la ecuación (7.7) como:

$$f_{t+1} = (1-\alpha) f_t + \alpha \frac{(y_{t+1} + y_t)}{2}$$

mientras que si se aplica un peso variable mínimo del 90 % se calculan mediante la ecua- ción (7.5):

$$f_{t+1} = (1-\alpha) f_t + \alpha (y_{t+1}),$$

donde, según la ecuación (7.8):

$$\alpha = \frac{2n-1}{2n+1}$$

y $n$ es el número de orden en la serie diaria de observaciones. Así, con esta función comienza con $\alpha = 0{,}9$ y tiende a ser uno, a medida que aumenta el valor de $n$. La Tabla 7.9 muestra los porcentajes según el tipo de peso utilizado.

Como se puede ver por ambas predicciones, la ocupación de red para el día siguiente es altísima y posiblemente provocará colapsos debido al nuevo tráfico de vídeo en la Intranet.

| Día | Monitor | Peso fijo | Peso variable |
|---|---|---|---|
| 1 | 64,0 | 64,0 | 64,0 |
| 2 | 78,5 | 70,5 | 77,2 |
| 3 | 49,8 | 71,2 | 52,1 |
| 4 | 97,4 | 80,7 | 93,9 |
| 5 | 99,0 | 96,4 | 98,6 |

**Tabla 7.9:** Porcentaje obtenido con las predicciones de peso fijo y peso variable.

**PROBLEMA 7.8** La Intranet de una empresa de publicidad tiene por cuello de botella la propia red Ethernet. Para mejorar la velocidad de transferencia se ha planificado reempla- zar Ethernet por una Fast Ethernet o una Gigabit Ethernet. En particular, la transferencia de archivos multimedia sobre la Intranet es la que provoca largas esperas y colisiones de transmisión. Mientras no se reemplaza la red, se intenta minimizar el efecto del cuello de botella mediante el uso de protocolos de transmisión sensibles al ancho de banda disponi- ble. Así, una vez codificados los datos multimedia, los protocolos de transmisión adaptan la velocidad de transferencia en el emisor en función de la velocidad de reconocimiento que retorna del receptor. De este modo, no absorben todo el ancho de banda, a costa de almacenar en memorias secundarias en emisión y recepción.

Debido a otros requerimientos tecnológicos, el analista de prestaciones posee dos proto- colos multimedia adaptativos que podrían servir para tales fines: MobilMeanTR y SoftEx- Protocol. MobilMeanTR analiza la velocidad de transmisión de ciertas tramas durante la transferencia y mediante la técnica de medias móviles predice el ancho de banda remanen- te. Por el contrario, SoftExProtocol calcula en tiempo real el ancho de banda remanente a través de la aplicación de la suavización exponencial de la velocidad de las tramas muestra. Se propone al analista de prestaciones que elija un protocolo, teniendo en cuenta que, en función de la previsión del ancho de banda remanente, el protocolo transmite a la máxima calidad. En la Tabla 7.10 se ha tomado una serie de monitorizaciones del ancho de banda real disponible para multimedia.

Además, disponemos de la siguiente información acerca de los dos protocolos mencio- nados:

- MobilMeanTR prácticamente no consume recursos de red, pero SoftExProtocol, de- bido a su naturaleza de tiempo real, tiene un 1 % de sobrecarga en ancho de banda.
- Por el contrario, mientras que SoftExProtocol puede transmitir con cinco niveles de calidad: 1, 10, 50, 100 y 500 KBps, MobilMeanTR sólo puede transmitir con tres: 1, 50 y 500 KBps.
- Si el ancho de banda remanente real es menor que el previsto, ambos protocolos descienden automáticamente a calidades inferiores, pero de modo distinto: Mobil- MeanTR baja al mínimo de 1 KBps hasta el siguiente periodo de estimación, puesto que no analiza en tiempo real, mientras que SoftExProtocol baja a la calidad de transmisión inmediatamente inferior a la prevista hasta que desaparecen los errores de transmisión debidos a pérdidas de tramas de reconocimiento.

| Muestra | Ancho de banda (Mbps) |
|---|---|
| 1 | 4,0 |
| 2 | 8,5 |
| 3 | 9,8 |
| 4 | 7,4 |
| 5 | 9,0 |
| 6 | 2,0 |
| 7 | 5,5 |
| 8 | 9,8 |
| 9 | 1,0 |
| 10 | 5,0 |

**Tabla 7.10:** Ancho de banda disponible.

- MobilMeanTR calcula el ancho de banda remanente a partir de las tres últimas muestras. SoftExProtocol otorga un peso del 80 % a la última monitorización.

A partir de la información aportada, se pide:

1. Calcular el ancho de banda disponible previsto por ambos protocolos a partir de las muestras de la tabla.
2. Calcular la calidad de transmision en KBps de ambos protocolos en función de las previsiones. ¿Qué protocolo envía teóricamente una mayor calidad de información? ¿Qué protocolo disminuye la tasa de error?

**SOLuCIón:**

1. Se calcula la previsión del ancho de banda disponible de ambos protocolos con sus res- pectivas técnicas estadísticas; los resultados se presentan en la Tabla 7.11.

   Por ejemplo, el valor de MobilMeanTR para la muestra 9 resulta al aplicar la ecuación (7.4) para las tres últimas muestras de la siguiente forma:

$$f_9 = \frac{1 + 9{,}8 + 3{,}5}{3} = 4{,}8$$

   Por ejemplo, el valor de SoftExProtocol para la muestra 2 resulta al aplicar la ecuación (7.6) con un valor de $\alpha = 0{,}8$ de la siguiente forma:

$$f_2 = 4 + 0{,}80 \times (8{,}5 - 4) = 7{,}6$$

| Muestra | Disponible | MobilMeanTR | SoftExProtocol |
|---|---|---|---|
| 1 | 4,0 | 4,0 | 4,0 |
| 2 | 8,5 | 6,2 | 7,6 |
| 3 | 9,8 | 7,4 | 9,3 |
| 4 | 7,4 | 8,6 | 7,8 |
| 5 | 9,0 | 8,7 | 8,7 |
| 6 | 2,0 | 6,1 | 3,3 |
| 7 | 3,5 | 4,8 | 3,5 |
| 8 | 9,8 | 5,1 | 8,5 |
| 9 | 1,0 | 4,8 | 2,5 |
| 10 | 5,0 | 5,3 | 4,5 |

**Tabla 7.11:** Previsión del ancho de banda disponible para ambos protocolos.

2. Se toma la calidad máxima de transmisión del protocolo sabiendo que el ancho de banda se ha medido en Mbps y las velocidades de transmisión están en KBps. Así, por ejemplo, 4 Mbps equivalen a la calidad máxima de 500 KBps, ya que 1 byte está compuesto por 8 bits, con lo que ambos protocolos no ocupan nunca más del 40 % del ancho de banda máximo de la red Ethernet de 10 Mbps.

   Obsérvese que MobilMeanTR transmite siempre a la máxima calidad, puesto que estima que siempre hay un ancho de banda remanente igual o superior a 4 Mbps. Por el contrario, SoftExProtocol transmite a 100 KBps durante tres de los periodos observados, periodos 3, 7 y 9, porque en esos tres periodos el ancho de banda remanente es inferior a 4 Mbps. Con estos resultados, la elección se decantaría por MobilMeanTR.

   Sin embargo, si se compara el ancho de banda disponible medido y el previsto, Mobil- MeanTR es demasiado optimista y calcula la media móvil muy por encima de la realidad. Por ejemplo, el error entre la medida 9 y su estimación por MobilMeanTR es mucho mayor que el error producido por SoftExProtocol. Por tanto, muy previsiblemente se pro- ducirán pérdidas de tramas de reconocimiento y colisiones debido a la congestión de la comunicación, con lo que finalmente MobilMeanTR tendrá que transmitir a la calidad mínima de 1 KBps, como se ha indicado en las notas adicionales del enunciado. Ello es debido a que MobilMeanTR no analiza el tráfico en tiempo real, mientras que SoftEx- Protocol se adaptará al ancho de banda disponible, bajando a la calidad de transmisión inmediatamente inferior de 100 KBps.

   En el peor de los casos, el ancho de banda disponible monitorizado es de 1,0 Mbps, pero eso permite hasta 125 KBps, de tal forma que SoftExProtocol no tiene que disminuir de calidad de transmisión por errores de congestión, puesto que utiliza sólo 100 KBps en la transferencia multimedia. Por el contrario, MobilMeanTR cae a 1 KBps por errores de comunicación por sobreestimación de ancho de banda remanente.

   En la Tabla 7.12 se representan las velocidades de transmisión disponibles y las que podrá transmitir cada protocolo tras el tratamiento de errores. No se tiene en cuenta la sobrecarga del 1 % de SoftExProtocol.

| Muestra | Disponible (Mbps) | MobilMeanTR (KBps) | SoftExProtocol (KBps) |
|---|---|---|---|
| 1 | 4,0 | 500 | 500 |
| 2 | 8,5 | 500 | 500 |
| 3 | 9,8 | 500 | 500 |
| 4 | 7,4 | 500 | 500 |
| 5 | 9,0 | 500 | 500 |
| 6 | 2,0 | 1 | 100 |
| 7 | 3,5 | 1 | 100 |
| 8 | 9,8 | 500 | 500 |
| 9 | 1,0 | 1 | 100 |
| 10 | 5,0 | 500 | 500 |

**Tabla 7.12:** Velocidades de transmisión previstas para ambos protocolos.

El resultado es que SoftExProtocol transmite a 380,0 KBps de media durante el periodo de observación menos un 1 % de sobrecarga, lo que resulta en 376,2 KBps, mientras que MobilMeanTR trasmite a 350,3 KBps de media, lo que no supone gran diferencia.

**PROBLEMA 7.9** Una compañía multinacional fabrica y vende ascensores. Parte impor- tante de su negocio son los contratos de mantenimiento de los ascensores instalados. Las dos cargas de trabajo interactivas más importantes en su computador central son la admi- nistración de ventas y administración de mantenimiento. Estas dos suponen un 80 % de la carga en la hora punta. El sistema de contabilidad general es una tercera carga importante que se ejecuta actualmente en un sistema batch, pero se espera que pase a ser interactiva en un plazo de dos años.

Se han identificado unidades de predicción naturales (NFU) para estas tres cargas de trabajo. Para la administración de ventas es el número de entradas en el fichero de clientes, para la administración de mantenimiento es el número de contratos de mantenimiento, mientras que para la contabilidad general es el número de millones de euros de ventas.

Se muestran las predicciones de crecimiento de las unidades de predicción naturales en la Tabla 7.13, así como las mediciones realizadas en el año 2003. Las variables mostradas en esta tabla son: $P$ : clientes en miles; $C$: número de contratos de mantenimiento; $V$ : millones de euros de ventas; $p$: pequeño; $m$: mediano, y $g$: grande.

Las estadísticas del total de transacciones en hora pico, clasificadas en transacciones cortas o ligeras, medianas, y largas o pesadas, para las dos primeras cargas de trabajo se midieron en diciembre del año 2003 para dar los datos que se presentan en la Tabla 7.14.

| Año | $P_p$ | $P_m$ | $P_a$ | $C_p$ | $C_m$ | $C_a$ | $V_p$ | $V_m$ | $V_a$ |
|---|---|---|---|---|---|---|---|---|---|
| 2003 | – | 300 | – | – | 100 | – | – | 15 | – |
| 2004 | 300 | 330 | 360 | 105 | 110 | 120 | 16 | 17 | 18 |
| 2005 | 310 | 360 | 420 | 105 | 120 | 135 | 17 | 19 | 22 |
| 2006 | 320 | 390 | 500 | 125 | 140 | 150 | 18 | 21 | 25 |

**Tabla 7.13:** Previsiones de las unidades de predicción naturales para el Problema 7.9.

| | Cortas | Medianas | Largas |
|---|---|---|---|
| Transacciones de ventas | 400 | 80 | 10 |
| Transacciones de mantenimiento | 500 | 100 | 15 |
| Segundos de procesador por transacción | 0,2 | 0,8 | 7,5 |
| Entrada/salida a disco por transacción | 4 | 30 | 400 |

**Tabla 7.14:** Número de transacciones totales y de cada tipo por hora en una hora pico para el Problema 7.9.

Diciembre suele ser el mes con mayor carga porque se cierra el año fiscal de la mayoría de las empresas. El contador de transacciones indica el número de transacciones en la hora pico.

Para la carga de contabilidad general, la productividad en mayo del año 2003 se midió como una media de 500 registros por cada ejecución nocturna. Mayo se seleccionó como un mes promedio, ya que diciembre es el final del año fiscal. Las mejores estimaciones actuales prevén que, cuando el sistema pase a en línea en el año 2006, el 80 % de las actualizaciones serán vía el sistema interactivo, y que por cada transacción de actualización habrá dos transacciones de consulta. Las necesidades de recursos para estas consultas serán similares a las de las transacciones ligeras en el sistema de administración de ventas, mientras que las actualizaciones serán parecidas a las transacciones medianas. Además se estima que el 25 % de las transacciones en línea ocurrirán en la hora de pico.

Las tendencias anteriores indican que los recursos utilizados por transacción crecerán a una tasa anual del 8 % para el uso del procesador, y del 11 % para el uso del disco. No hay datos suficientes para diferenciar entre las tasas de crecimiento en uso para los diferentes tipos de transacciones. Los cambios en los números de las transacciones por NFU se han determinado mediante el análisis de tendencias de datos históricos y estimaciones de los usuarios y son de un 6 % para ventas y de un 8 % para mantenimiento. No hay datos para la contabilidad general; se supone que es de un 6 %, en línea con la administración de ventas.

Basándose en los datos aportados se pide calcular:

1. Las necesidades de procesador medidas en segundos en los años 2004, 2005 y 2006 representadas gráficamente, utilizando las hipótesis de crecimiento bajo, medio y alto.
2. Las necesidades de entrada/salida medidas en número de operaciones en los años

2004, 2005 y 2006 representadas gráficamente, utilizando las hipótesis de crecimiento bajo, medio y alto.

**SOLUCIÓN:** Los cálculos se han realizado utilizando una hoja de cálculo. La predicción de crecimiento bajo para la necesidad del recurso procesador para al año 2004 se obtiene como la suma de las predicciones basadas en las estimaciones de clientes de ventas y contratos de mantenimiento dando los siguientes pasos:

- Cálculo de la necesidad de procesador para la hora pico del año 2004 debido al número de transacciones cortas ($t_c$), medias ($t_m$) y largas ($t_l$) de *ventas*. Para ello se multiplica el número de transacciones de venta por el tiempo consumido de procesador para cada uno de los tipos de transacciones y se suman. Esto es:

$$400\ t_c \times 0{,}2 + 80\ t_m \times 0{,}8 + 10\ t_l \times 7{,}5 = 219\ \text{s}$$

- Cálculo de la necesidad de procesador para la hora pico del año 2004 debido a las *tran- sacciones de mantenimiento*. Repetiremos el mismo proceso que en el caso anterior:

$$500\ t_c \times 0{,}2 + 100\ t_m \times 0{,}8 + 15\ t_l \times 7{,}5 = 292{,}5\ \text{s}$$

Por lo tanto, el total de demanda de procesador necesaria en diciembre de 2003 ha sido de 511,5 segundos.

- Cálculo de la necesidad de operaciones de E/S para la hora pico del año 2004 por las
*transacciones de ventas*. Aplicamos de nuevo el mismo procedimiento:

$$400\ t_c \times 4 + 80\ t_m \times 30 + 10\ t_l \times 400 = 8.000\ \text{E/S}$$

- Cálculo de la necesidad de operaciones de E/S para la hora pico del año 2004 por las *transacciones de mantenimiento*. Se multiplica el número de transacciones de manteni- miento por el número de operaciones de E/S para cada uno de los tipos de transacciones (cortas, medias y largas) y se suman. Esto es:

$$500\ t_c 4 + 100\ t_m \times 30 + 15\ t_l \times 400 = 11.000\ \text{E/S}$$

Por lo tanto, el total de operaciones de E/S realizadas en diciembre de 2003 ha sido de
19.000 operaciones de E/S.

- Cálculo de la necesidad de procesador para la hora pico del año 2004, utilizando la unidad de predicción natural pequeña. En este caso se ha de realizar el cálculo como la suma de la demanda de procesador necesaria para las transacciones de ventas y la demanda de procesador necesaria para las transacciones de mantenimiento.

$$219\ s \times \frac{300}{300} \times (1{,}08) \times (1{,}06) + 292{,}5\ s \times \frac{105}{100} \times (1{,}08) \times (1{,}08) = 608{,}9\ s$$

- Cálculo de la necesidad de procesador para la hora pico del año 2004, utilizando la unidad de predicción natural media. Se ha de realizar el cálculo como la suma de la demanda de procesador necesaria para las transacciones de ventas y la necesaria para las transacciones de mantenimiento.

$$219\ s \times \frac{330}{300} \times (1{,}08) \times (1{,}06) + 292{,}5\ s \times \frac{110}{100} \times (1{,}08) \times (1{,}08) = 651{,}1\ s$$

De la misma forma se calcularían la necesidad de procesador para el año 2004 en el caso de previsión de crecimiento alto, y los valores de operaciones de E/S para los casos de previsiones de crecimiento baja, media y alta para el año 2004.

- Cálculo de demanda de procesador debida a la contabilidad general en el año 2006.

Las transacciones de contabilidad general se añaden desde 2006 en adelante. Su contri- bución se estima suponiendo que, si la aplicación estuviera ejecutándose en el año×2003, habría 500 0,8 transacciones por día de actualizaciones, haciendo 500 0,8 0,25 = 100 actualizaciones por hora en pico; y, por lo tanto, 200 consultas por hora en pico, ya que por cada actualización se producen dos consultas. Cada consulta consume lo mismo que una transacción corta de ventas y cada actualización consume lo mismo que una tran- sacción media de ventas. Por lo tanto, el consumo en el año 2006 debido a la aplicación de contabilidad para una previsión de crecimiento baja es la siguiente:

$$200\ t \times 0{,}2\ \frac{s}{t} + 100\ t \times 0{,}8\ \frac{s}{t} \times \frac{}{15} \times (1{,}07)^3 \times (1{,}05)^3 = 216{,}0\ s$$

$$200\ t \times 4\ \frac{E/S}{t} + 100\ t \times 0{,}8\ \frac{E/S}{t} \times \frac{}{15} \times (1{,}1)^3 \times (1{,}05)^3 = 7.427{,}6\ E/S$$

18

Los valores resultantes de los cálculos realizados para obtener el consumo de procesador según el tipo de crecimiento se presentan en la Tabla 7.15. Por su parte, el consumo de la E/S se muestra en la Tabla 7.16. La representación gráfica de estas dos tablas lleva a dos gráficas: la Figura 7.5 muestra las tres curvas con las necesidades de procesador previstas en segundos, mientras que la Figura 7.6 muestra los resultados para las necesidades de operaciones entrada/salida en disco.

Hay varios aspectos de este problema que se pasan a discutir a continuación:

| Año | Pequeño | Medio | Grande |
|---|---|---|---|
| 2003 | – | 511,5 | – |
| 2004 | 608,9 | 651,1 | 710,2 |
| 2005 | 714,4 | 821,9 | 939,0 |
| 2006 | 1.146,7 | 1.270,4 | 1.508,6 |

**Tabla 7.15:** Consumo previsto de procesador según el tipo de crecimiento.

| Año | Pequeño | Medio | Grande |
|---|---|---|---|
| 2003 | – | 19.000 | – |
| 2004 | 23.259 | 24.860 | 27.120 |
| 2005 | 28.043 | 32.260 | 36.846 |
| 2006 | 45.016 | 52.137 | 60.461 |

**Tabla 7.16:** Consumo previsto de operaciones de E/S en disco según el tipo de crecimiento.

- El primero es los efectos de los "multiplicadores modestos". Las actividades de la compañía crecen en un factor de, aproximadamente, un 50 %, pero las necesidades de procesador y de disco crecen en un factor de entre un 205 y un 300 %. La composición de las necesidades de recursos aumentadas por transacción y las transacciones aumentadas por NFU producen las necesidades adicionales de crecimiento. Sin embargo, el resultado está en consonancia con afirmaciones que se encuentran en la literatura; en ellas se asevera que, mientras una organización crezca entre un 5 y un 10 % al año, las necesidades de cálculo pueden crecer entre un 30 y un 40 % al año.

- Este ejemplo es muy sencillo comparado con los casos de la vida real. Se han considerado tres tipos de carga de trabajo de interés; éste es un número pequeño. Se han dejado de lado al menos dos variables del sistema importantes: el espacio en disco y la capacidad de transmisión de la red. Asimismo, se han utilizado reglas particularmente sencillas para estimar el crecimiento en el uso, y no se justifica la utilización de reglas más complejas sin tener datos suficientemente precisos.

- Finalmente, las formas de las curvas indican que traer la carga de trabajo de la contabilidad general en línea no produce mucho cambio en la carga, aunque sí un cambio en la pendiente de crecimiento. ■

**PROBLEMA 7.10** Una empresa telefónica europea quiere realizar una planificación de la capacidad para decidir el espacio en disco que necesita una nueva aplicación de inteligencia de negocio que utiliza una base de datos DB2. La longitud media de los registros y el número de registros en enero de cada tipo de datos vienen especificados en la Tabla 7.17.

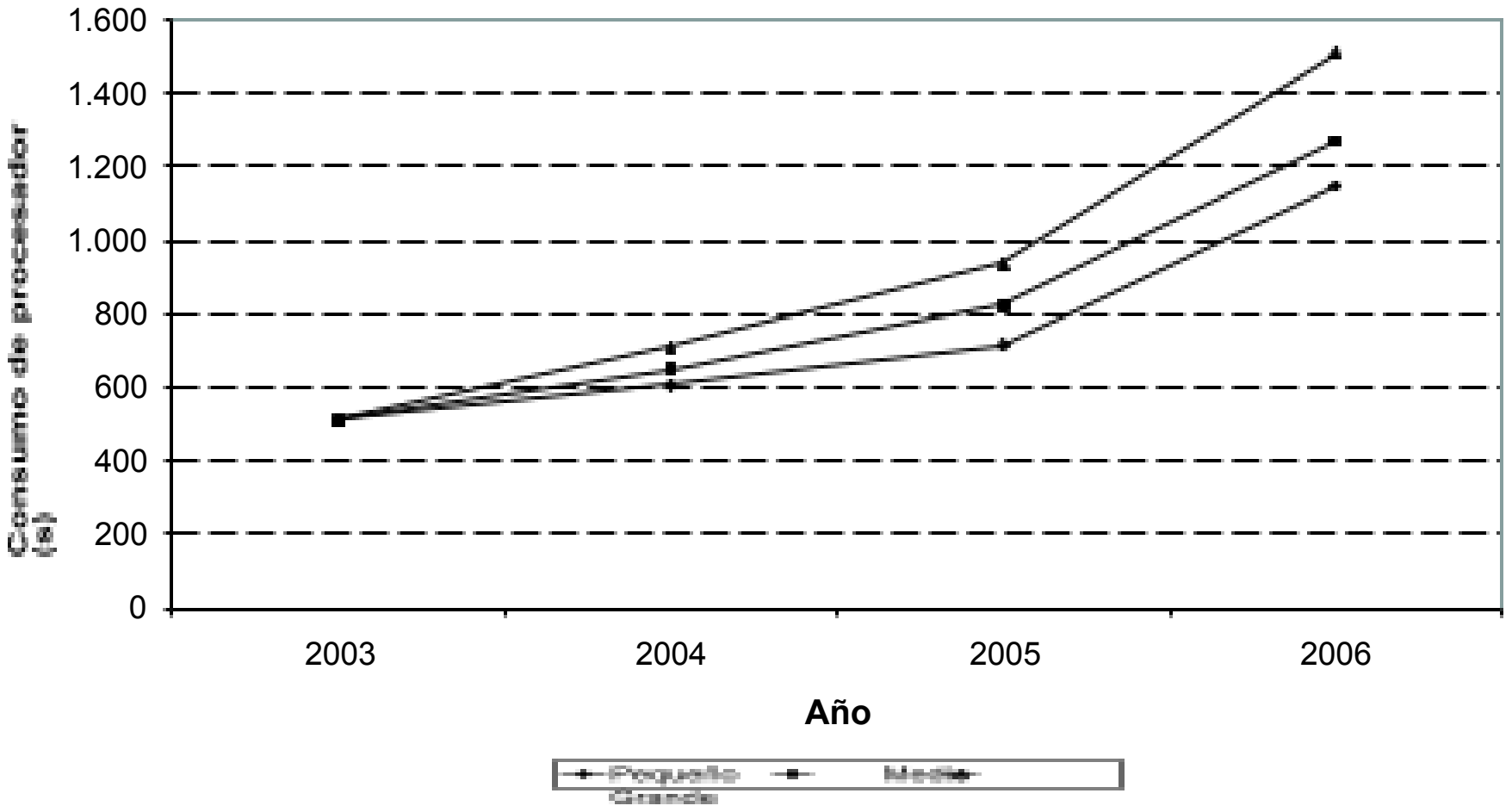

**Figura 7.5:** Necesidades de procesador previstas.

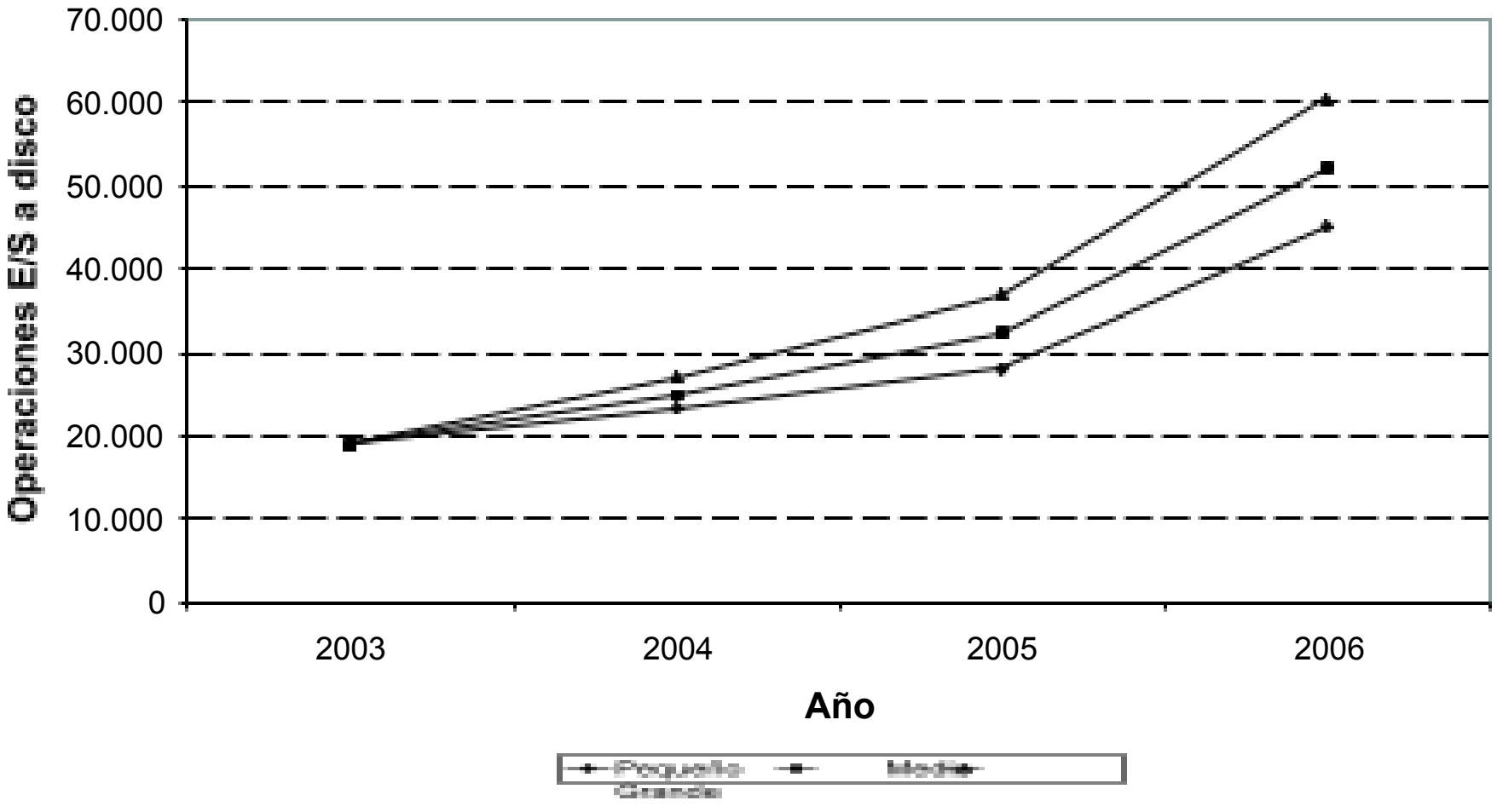

**Figura 7.6:** Necesidades de operaciones de E/S a disco previstas.

Al número de registros de enero de 2004 indicados en la Tabla 7.17 hay que sumarle un incremento mensual del 5 % para calcular los de febrero de 2004, que son datos totalmente nuevos respecto de los de enero. Si enero tiene 100 registros nuevos, febrero tendrá 105 registros nuevos, marzo tendrá 110 registros nuevos, etc. Por lo tanto, para calcular el total de registros en el año 2004 hay que sumar los generados en los doce meses. Existe una regla empírica que indica que el espacio necesario en disco para una base de datos es tres veces superior al que necesitan los datos brutos.

Se pide calcular el espacio en disco necesario para almacenar la información de los doce meses del año 2004 suponiendo dos casos: el primero en el que la configuración de discos esté en RAID 1; el segundo en el que la configuración esté en RAID 5 con un tamaño de grupo de ocho discos, dedicando el espacio equivalente a uno de ellos a la paridad y el equivalente a los otros siete a datos, pero con los datos y la paridad distribuida entre los ocho discos. En ambos casos se utilizan discos de 146 GB y se tiene un disco de reserva por cada 40 discos físicos.

| Tipo | Longitud media (bytes) | Registros |
|---|---|---|
| 1 | 312 | 5.850.000 |
| 2 | 272 | 3.540.000 |
| 3 | 312 | 65.000 |
| 4 | 487 | 140 |
| 5 | 324 | 14.000 |
| Total | | 9.469.140 |

**Tabla 7.17:** Características de los registros de la base de datos.

**SOLuCIón:** Para solucionar este problema primero se va a calcular el número de GB necesarios para almacenar toda la información neta incluida en los registros detallados en el enunciado, teniendo en cuenta el incremento mensual del 5 %. Posteriormente se multiplicará por tres para obtener el espacio necesario en disco (regla empírica). Después se calculará el número de discos totales como la suma de los necesarios para mantener esa capacidad en un determinado tipo de paridad (RAID 1 ó RAID 5) y los discos de reserva.

Para calcular el tamaño bruto estimado para los meses de enero y febrero del año 2004 basta con multiplicar la longitud media por el número de registros:

Para enero: $312 \times 5.850.000 = 1.825.200.000$ bytes

Para febrero: $1.825.200.000 \times 1{,}05 = 1.916.460.000$ bytes

El factor por el que hay que multiplicar para calcular la suma de todos los meses del año 2004, teniendo en cuenta que cada mes recoge registros de llamadas distintas, es el siguiente:

$$1 + 1{,}05 + 1{,}05^2 + 1{,}05^3 + \cdots + 1{,}05^9 + 1{,}05^{10} + 1{,}05^{11} = 15{,}917$$

La Tabla 7.18 presenta los resultados de los cálculos realizados.

Aplicando la regla empírica indicada, y considerando, por un lado, que en una arquitectura de redundancia RAID 1 se necesita el doble de disco y, por otro, que se utilizan discos de 146 GB, se puede deducir que el número de discos necesarios para el caso de RAID 1 es el siguiente:

$$\text{Discos dedicados al RAID 1} = \frac{44.774 \text{ GB} \times 3 \times 2}{146 \text{ GB}} = 1.840$$

| Datos | Longitud media | Registros | Tamaño enero | Tamaño 2004 |
|---|---|---|---|---|
| 1 | 312 bytes | 5.850.000 | 1.825,2 | 29,052 |
| 2 | 272 bytes | 3.540.000 | 962,88 | 15,326 |
| 3 | 312 bytes | 65.000 | 20,28 | 0,322 |
| 4 | 487 bytes | 140 | 0,068 | 0,001 |
| 5 | 324 bytes | 14.000 | 4,536 | 0,072 |
| Total | | 9.469.140 | 2.812,964 GB | 44,774 TB |

**Tabla 7.18:** Cálculo del tamaño del fichero de enero de 2004 (GB) y del tamaño total de todos los meses del año 2004 (TB).

$$\text{Discos de reserva} = \frac{1.840}{40} = 46$$

Total de discos necesarios con RAID 1 = 1.840 + 46 = 1.886

Para el caso de la configuración en RAID 5 hacen falta muchos menos discos:

$$\text{Discos dedicados al RAID 5} = \frac{44.774 \text{ GB} \times 3 \times \frac{8}{7}}{146} = 1.052$$

$$\text{Discos de reserva} = \frac{1.052}{40} = 27$$

Total de discos necesarios con RAID 5 = 1.052 + 27 = 1.079

Los platos de disco se incluyen en subsistemas de disco. Ahora se debería preguntar a los distintos fabricantes qué subsistema de disco le aconsejan para este aplicativo y ajustar sus necesidades de disco a un número entero de subsistemas de disco.

**PROBLEMA 7.11** La empresa telefónica del problema anterior quiere ahora dimensionar el procesador necesario para el aplicativo de consultas y además para atender la carga transaccional durante su periodo en línea de ocho horas laborables.

Para el dimensionamiento debido a la carga del aplicativo de consultas se aconseja em- plear el índice de rendimiento inteligente relativo de negocio (*Relative Business Intelligent Performance*, RBI), que mide el número de veces que un modelo de computador determi- nado es más potente para el aplicativo inteligencia de negocio que el modelo de referencia (MR). El RBI se calcula como la media entre los valores ROLTP y Specrate95. El modelo ROLTP (*Relative OnLine Transaction Processing*) tiene 1 ROLTP y 15,18 Specrate95; esto es, para calcular el número de RBI debido a los Specrate95 hay que dividir por 15,18 el valor dado en Spec. El fabricante de estos computadores, que utilizan el sistema operativo UNIX, proporciona ROLTP y Specrate95 para la mayoría de sus modelos. El fabricante ha calculado que el modelo

M1 de 24 vías tiene 377 RBI y es capaz de leer 500 MB/s sostenidos en este tipo de aplicativo. Por lo tanto, se puede deducir que los MB/s por RBI son 1,33.

La carga de trabajo del aplicativo de consultas de la empresa telefónica está compuesta por un 40 % de consultas ligeras con 2 MB de transferencia por consulta, un 55 % de consultas medianas sobre un 1,5 % del total de datos de la base de datos, y un 5 % de consultas pesadas sobre un 10 % del total de datos de la base de datos. Se supone que hay 30 usuarios realizando 17 consultas al día, y que solamente la tasa de consultas durante el periodo del horario laboral se reduce al 33 % del total de consultas.

Por otro lado, para el dimensionamiento debido a la carga transaccional se aconseja utilizar el *benchmark* TPC-C, que se utiliza para carga transaccional. En pico llegan 42 ficheros por minuto, y cada fichero tiene aproximadamente 5.000 registros de transacciones. Según el *benchmark* TPC, 1 ROLTP equivale a 12 transacciones/s. Un nodo de ancho de 375 MHz de este tipo de computador proporciona una capacidad de 80 ROLTP.

Esta familia de máquinas puede tener una serie de nodos dentro de cada computador. Se pide calcular el número de nodos necesarios de 375 MHz de ancho de este tipo de computador teniendo en cuenta que tiene 80 ROLTP y 875 Specrate95 por nodo y un factor de acoplamiento de 1,23 para el número de nodos resultante; de 375 MHz high teniendo en cuenta que tiene 319,3 ROLTP y 3.352 Specrate95 por nodo y un factor de acoplamiento de 1,09 para el número de nodos resultante; de p680 24w teniendo en cuenta que tiene 736 ROLTP y 5.184 Specrate95 por nodo y un factor de acoplamiento de 1,02 para el número de nodos resultante.

**SOLuCıón:** Primero se calculan los RBI necesarios para el aplicativo de consultas. Posterior- mente se hace la conversión de los RBI en el número de nodos necesarios para soportar esta carga. En segundo lugar se calcula el número de transacciones por segundo debido al transac- cional. Posteriormente se calcula el número de nodos necesarios para soportar esta carga. Por último, se debería calcular el número de nodos totales necesarios teniendo en cuenta el factor de acoplamiento.

Vayamos dando los pasos indicados. Primero se calculan los RBI necesarios para el aplicativo de consultas. El número de consultas por hora durante la jornada laboral provocado por los 30 usuarios es el siguiente:

$$30 \text{ usuarios} \times 17 \frac{\text{consultas}}{\text{día x usuario}} \times \frac{1 \text{ día}}{24 \text{ horas}} \times 0{,}33 = 7 \frac{\text{consultas}}{\text{hora}}$$

La media de MB/consulta viene dada por la suma de las provocadas por las consultas ligeras y de las provocadas por las consultas medianas y pesadas, teniendo en cuenta que el tamaño total de la base de datos, calculado en el ejercicio anterior, es de 44.774 GB:

$$0{,}4 \times \frac{2 \text{ MB}}{\text{consulta}} + (0{,}55 \times 0{,}015 + 0{,}05 \times 0{,}1) \times \frac{44.774.000 \text{ MB}}{\text{consulta}} = 593.260 \frac{\text{MB}}{\text{consulta}}$$

El ancho de banda necesario, expresado en MB/s, se calcula de la siguiente manera:

$$593.260\ \frac{\text{MB}}{\text{consulta}} \times 7\ \frac{\text{consultas}}{\text{hora}} \times \frac{1\ \text{hora}}{3.600\ \text{s}} = 1.154\ \text{MB/s}$$

Los RBI necesarios vienen dados por:

$$1.154\ \frac{\text{MB}}{\text{s}} \times \frac{1\ \text{RBI}}{1{,}33\ \frac{\text{MB}}{\text{s}}} = 867\ \text{RBI}$$

Para indicar qué máquina concreta hay que instalar se deberán convertir los RBI en el número de nodos necesarios para soportar esta carga. El número de RBI de un nodo viene dado por la media aritmética entre los ROLTP y Specrate95, teniendo en cuenta que el número de RBI equivalentes por Spec viene dado por la división de Specrate95 y 15,18, según el modelo ROLTP. Los RBI de cada uno de los nodos de 375 MHz wide, 375 MHz high y p680 24w son los siguientes, respectivamente:

$$= 68{,}82\ \frac{\text{RBI}}{\text{sistema}}$$

$$319{,}3 + \ldots = 270{,}05\ \frac{\text{RBI}}{\text{sistema}}$$

$$= 538\ \frac{\text{RBI}}{\text{sistema}}$$

El número de nodos necesario en cada uno de los tres casos viene dado por el número entero superior a los cocientes siguientes:

$$\frac{867 \times 1,23}{68,82} = 16$$

$$\frac{867 \times 1,09}{270,05} = 4$$

$$\frac{867 \times 1,02}{538} = 2$$

En segundo lugar se calcula el número de transacciones por segundo debido al transaccional, que viene dado por:

$$42\ \frac{\text{ficheros}}{\text{min}} \times 5.000\ \frac{\text{transacciones}}{\text{fichero}} \times \frac{1\ \text{min}}{60\ \text{s}} = 4.166$$

Se calcula el número de nodos de ancho 375 MHz para la carga transaccional, que viene dado por:

trans accio nes 1 ROLTP 1 nodo

s

4.166 s × 12 transacciones × 80 ROLTP = 5

La configuración para soportar simultáneamente las dos cargas ha de sumar el número de nodos necesarios para cada una de ellas, teniendo en cuenta el factor de acoplamiento entre nodos. ■

## 7.8. Problemas con solución

**PROBLEMA 7.12** Se pide analizar algunas predicciones de aumento de cargas de traba- jo. Las entrevistas con los departamentos de usuarios han producido las predicciones de crecimiento anual expresadas en porcentaje indicadas en la Tabla 7.19. NFU actuales quie- re decir unidades de predicción natural actuales en el año 2004, que indican la carga de trabajo sobre el recurso analizado en el año actual 2004.

| Departamento | | Ventas | Financiero | Personal | I+D |
|---|---|---|---|---|---|
| NFU actuales | | 100 | 85 | 54 | 20 |
| Predicción 1 | Alta | 30 | 10 | 45 | 20 |
| | Media | 12 | 8 | 30 | 10 |
| | Baja | 0 | 5 | 25 | 0 |
| Predicción 2 | Alta | 40 | 40 | 40 | 40 |
| | Media | 25 | 25 | 25 | 25 |
| | Baja | 6 | 6 | 6 | 6 |
| Predicción 3 | Alta | 20 | 15 | 40 | 20 |
| | Media | 15 | 10 | 25 | 12 |
| | Baja | 5 | 5 | 15 | 8 |

**Tabla 7.19:** Estadísticas de transacciones para el Problema 7.12.

Se pide determinar las estimaciones de las tasas de crecimiento alta, media y baja para las cuatro NFU basándose en las medias simples de las estimaciones.

**SOLUCIÓN:** Se calcula la media simple de la predicción alta, media y baja para cada uno de los departamentos según se indica a continuación:

na

nci

en

er

tas

o

Fi

30
+
40
+
20
=

= 30
3
10
+
40
+
15
=

= 21,7
3

$$\text{Personal}_{\text{alta}} = \frac{45 + 40 + 40}{3} = 41{,}7$$

$$\text{I+D}_{\text{alta}} = \frac{20 + 40 + 20}{3} = 26{,}7$$

$$\text{Ventas}_{\text{media}} = \frac{12 + 25 + 15}{3} = 17{,}3$$

$$\text{Financiero}_{\text{media}} = \frac{8 + 25 + 10}{3} = 14{,}3$$

$$\text{Personal}_{\text{alta}} = \frac{30 + 25 + 25}{3} = 26{,}7$$

$$\text{I+D}_{\text{alta}} = \frac{10 + 25 + 12}{3} = 15{,}7$$

$$\text{Ventas}_{\text{baja}} = \frac{0 + 6 + 5}{3} = 3{,}7$$

$$\text{Financiero}_{\text{baja}} = \frac{5 + 6 + 5}{3} = 5{,}3$$

$$\text{Personal}_{\text{baja}} = \frac{25 + 6 + 15}{3} = 15{,}3$$

$$\text{I+D}_{\text{baja}} = \frac{0 + 6 + 8}{3} = 4{,}7$$

Esto es, el cálculo de las predicciones de crecimiento no afecta ni se ve afectado por las unidades de predicción naturales. ■

---

**PROBLEMA 7.13** Se pide utilizar todas las predicciones del Problema 7.12 pero dando un peso distinto a cada una de ellas de tal modo que aquellas que se juzguen más fiables hagan una mayor contribución a los resultados. En particular:

1. Propóngase un conjunto de pesos para las previsiones. Explíquese por qué se han elegido estos pesos.
2. Calcúlense las previsiones de medias ponderadas para una predicción de crecimiento alta, media y baja. Compárense los resultados con los obtenidos en el Problema 7.12.
3. Descríbase una situación en la que el uso de medias ponderadas proporcione resulta- dos sustancialmente diferentes a aquellos obtenidos con la media simple.

**SOLuCIón:**

1. Se propone dar un peso del 40 % a cada una de las predicciones que parecen estar más elaboradas, que son la 1 y la 3, mientras que se da un peso de un 20 % a la predicción 2 porque parece estar menos elaborada, al suponer un crecimiento igual para cada uno de los departamentos.

2. Se calculan las medias ponderadas con los pesos fijados en la solución del subapartado anterior para cada crecimiento alto, medio y bajo, según se especifica a continuación.

$$\text{Ventas}_{\text{alta}} = 30 \times 0{,}4 + 40 \times 0{,}2 + 20 \times 0{,}4 = 28$$

$$\text{Financiero}_{\text{alta}} = 10 \times 0{,}4 + 40 \times 0{,}2 + 15 \times 0{,}4 = 22$$

$$\text{Personal}_{\text{alta}} = 45 \times 0{,}4 + 40 \times 0{,}2 + 40 \times 0{,}4 = 42$$

$$\text{I+D}_{\text{alta}} = 20 \times 0{,}4 + 40 \times 0{,}2 + 20 \times 0{,}4 = 24$$

$$\text{Ventas}_{\text{media}} = 12 \times 0{,}4 + 25 \times 0{,}2 + 15 \times 0{,}4 = 15{,}8$$

$$\text{Financiero}_{\text{media}} = 8 \times 0{,}4 + 25 \times 0{,}2 + 10 \times 0{,}4 = 12{,}2$$

$$\text{Personal}_{\text{media}} = 30 \times 0{,}4 + 25 \times 0{,}2 + 25 \times 0{,}4 = 27$$

$$\text{I+D}_{\text{media}} = 10 \times 0{,}4 + 25 \times 0{,}2 + 12 \times 0{,}4 = 13{,}8$$

$$\text{Ventas}_{\text{baja}} = 0 \times 0{,}4 + 6 \times 0{,}2 + 5 \times 0{,}4 = 3{,}2$$

$$\text{Financiero}_{\text{baja}} = 5 \times 0{,}4 + 6 \times 0{,}2 + 5 \times 0{,}4 = 5{,}2$$

$$\text{Personal}_{\text{baja}} = 25 \times 0{,}4 + 6 \times 0{,}2 + 15 \times 0{,}4 = 17{,}2$$

$$\text{I+D}_{\text{baja}} = 0 \times 0{,}4 + 6 \times 0{,}2 + 8 \times 0{,}4 = 4{,}4$$

Al comparar estos resultados con los del Problema 7.12 se puede deducir que son muy parecidos, siendo la diferencia entre los valores en casi todos los casos menor del 10 %.

3. Una de estas situaciones se da cuando una de las predicciones ha recibido un mayor número de declaraciones por parte de las personas entrevistadas en todos los departamentos de la empresa. Supóngase, por ejemplo, que de las veinte entrevistas y estudios técnicos realizados se deduce que la interrelación del crecimiento entre los distintos departamentos es tan grande que el peso de la predicción 2 ha de ser del 80 %. En este caso, va a resultar una gran diferencia entre las medias simples y las ponderadas, ya que en este último caso el crecimiento alto se acercará mucho al 40 %, el medio al 25 % y el bajo al 6 %.

---

**PROBLEMA 7.14** El sistema transaccional *Ludwig* recibía 53.110 transacciones por segundo (tps) de carga media hace 12 meses en el centro de procesos de datos del Consorcio de Agencias de Viaje. Estaba prevista la adquisición por parte del consorcio de un grupo de agencias independientes que incrementarían la carga un 10 % mensual sobre la carga de enero, hasta su completa inserción en dos años. Ahora que ha pasado la mitad del periodo de inserción, el comportamiento ha sido sensiblemente distinto al esperado, tal como se muestra en la Tabla 7.20, que recoge la evolución del número de transacciones por segundo del pasado año.

Se pide:

| Mes | tps |
|---|---|
| Enero | 53.110 |
| Febrero | 61.222 |
| Marzo | 62.345 |
| Abril | 57.312 |
| Mayo | 57.897 |
| Junio | 63.544 |
| Julio | 52.856 |
| Agosto | 68.932 |
| Septiembre | 78.932 |
| Octubre | 66.280 |
| Noviembre | 68.932 |
| Diciembre | 63.002 |

**Tabla 7.20:** Evolución del número de transacciones por segundo del pasado año en el sistema *Ludvig*.

1. Determinar las expresiones de las dos rectas de tendencia según se tome un incre- mento lineal a partir del mes de enero pasado o la recta de regresión lineal.
2. ¿Cuál es tasa de transacciones por segundo con la previsión del incremento de un 10 % mensual para el mes siguiente y para dentro de otro año? ¿Cuál es la previsión para el mes siguiente y para dentro de otro año, aplicando la técnica de regresión lineal?
3. Se vuelve a realizar una previsión a la baja, pasando por alto la recta de tendencia, y se otorga una previsión de un 2 % sobre el número de transacciones por segundo de diciembre. ¿Cuál es la previsión para enero del año siguiente?

**SOLuCıón:**

1. Las rectas de tendencia se indican a continuación. La primera es la recta de tendencia con incremento lineal del 10 %, mientras que la segunda se corresponde con la recta de regresión lineal.

$$y = 53.110{,}00 + 5.311{,}00 \times x \quad y = 55.221{,}30 + 1.175{,}74 \times x$$

2. En enero la predicción del aumento lineal del 10 % es francamente pesimista, en compa- ración con la regresión lineal: $y = 116.842$ (incremento lineal del 10 %) e $y = 69.330{,}18$ (regresión lineal). Si la previsión es dentro de 12 meses, la diferencia se hace abismal: $y = 180.574$ e $y = 83.439{,}06$, para un incremento lineal del 10 % y regresión lineal, respectivamente.

3. Con la correción a la baja de la previsión basándose en el mes de diciembre, la recta de tendencia empírica sería:

$$y = 63.002{,}00 + 1.260{,}04 \times x$$

de donde resulta una previsión para 13 meses más tarde, cercana a la regresión lineal efectuada con los meses tabulados, pero sigue siendo una aproximación. La previsión será $y = 79.382{,}52$. ■

## 7.9. Problemas sin resolver

**PROBLEMA 7.15** Una empresa posee un agrupamiento de servidores web (*web cluster* ) que poseen los mismos contenidos, y un conmutador (*web switch*) se encarga de seleccio- nar el destino de las peticiones de información que provienen de transacciones HTTP. El conmutador ejecuta un programa de planificación que elige el servidor destino en función de la productividad observada y prevista a través de la técnica de suavizado exponencial. En la Tabla 7.21 se muestran la productividad de dos servidores del agrupamiento durante el periodo de monitorización que realiza el conmutador.

| Muestra | Servidor 1 | Servidor 2 |
|---|---|---|
| 1 | 70 | 70 |
| 2 | 65 | 67 |
| 3 | 63 | 66 |
| 4 | 71 | 65 |
| 5 | 60 | 63 |
| 6 | 72 | 60 |

**Tabla 7.21:** Transacciones por segundo (tps) de los dos servidores.

1. Calcúlense las productividades previstas de los dos servidores si se asigna un peso fijo del 90 % a las monitorizaciones suministradas.
2. Calcúlense las rectas de tendencia de la productividad mediante regresión lineal de ambos servidores. Dibújese un gráfico que muestre la tendencia.
3. ¿Cuál es el servidor más productivo? ¿Cuál será el servidor más productivo en el futuro?

**PROBLEMA 7.16** Una agencia de viajes maneja un sistema transaccional de reserva y venta de billetes de avión. En la Tabla 7.22 se presenta el número de transacciones realizadas en los seis últimos meses. En esta agencia trabajan tres agentes de ventas, que utilizan el sistema cinco horas diarias durante 20 días laborables en promedio. Para poder mantener un tiempo de respuesta aceptable para los usuarios, el sistema no puede recibir más de 25 transacciones por minuto (tpm). Calcúlese la carga prevista para el siguiente mes en transacciones por minuto a partir de la media de todos los meses anteriores y con suavizado exponencial con un peso fijo para las mediciones del 80 %. ¿Podrá mantenerse el acuerdo de nivel de servicio?

| Mes | Número total de transacciones |
|---|---:|
| Febrero | 490.200 |
| Marzo | 266.325 |
| Abril | 415.493 |
| Mayo | 135.774 |
| Junio | 690.777 |
| Julio | 909.016 |

**Tabla 7.22:** Transacciones del sistema transaccional de reserva de billetes.

---

**PROBLEMA 7.17** Una compañía de seguros tiene un sistema de entrada en línea para los pagos de las primas. Actualmente los pagos se procesan como una operación secundaria (*back room*), con una tasa de entrada esencialmente estacionaria durante el turno de trabajo principal de 3.000 transacciones por hora durante las ocho horas del turno de trabajo.

La compañía propone mejorar el servicio a sus clientes colocando terminales en los mostradores de los cajeros en sus sucursales, y actualizar sus registros cuando los clientes paguen. Se anticipa que el 30 % de los clientes dispondrán ellos mismos de este servicio, y que la mitad de esos clientes aparecerán durante la hora punta de mediodía de las 13:00 horas. Se pide:

1. ¿Cuál es la la nueva tasa de transacciones en la hora punta estimada para esta carga de trabajo?
2. Se anticipa que durante los próximos cinco años el número de pólizas conseguidas aumentará en un 8 % compuesto y la proporción de los poseedores de pólizas que pagarán en persona aumentará linealmente en un 35 %, mientras que el ratio de pico a promedio de éstas se mantendrá constante. Dibújese un gráfico que muestre los cambios durante el periodo de las transacciones para la hora punta.

**PROBLEMA 7.18** Un club de fans del grupo musical *Fatal Performance* posee un servi- dor web que contiene información referente a su actividad como asociación, calendario de conciertos, videoclips y canciones del grupo, ventas, etc. El próximo mes de marzo se lanza al mercado discográfico *Overhead is Devil*, el nuevo trabajo del grupo. Se espera que el número de accesos mensuales suba diez veces con respecto a la previsión media sin tener en cuenta este nuevo evento. En la Tabla 7.23 se presenta el número de accesos al servidor web durante los últimos meses monitorizados, que corresponden al periodo de abril del pasado año hasta agosto del presente.

| Mes | Número de accesos |
|---|---|
| Abril | 46.498 |
| Mayo | 83.485 |
| Junio | 98.583 |
| Julio | 71.315 |
| Agosto | 94.311 |
| Septiembre | 49.204 |
| Octubre | 74.004 |
| Noviembre | 61.598 |
| Diciembre | 06.086 |
| Enero | 35.888 |
| Febrero | 49.200 |
| Marzo | 66.325 |
| Abril | 84.593 |
| Mayo | 15.774 |
| Junio | 69.777 |
| Julio | 90.916 |
| Agosto | 40.558 |

**Tabla 7.23:** Accesos al servidor web del club de fans del grupo musical *Fatal Performance*.

1. Calcúlese el número de accesos previsto a partir de la muestra temporal mediante regresión lineal en febrero y marzo del año que viene, tras el lanzamiento.

2. El servidor incorpora actualmente el archivo multimedia del video-clip *Device's delay* correspondiente a un trabajo anterior, que es el archivo más popular del grupo con la mitad de los accesos totales al servidor. Calcúlese el número aproximado de accesos a este archivo, cuando quede desplazado en marzo al décimo lugar, tras los nueve videoclips del nuevo disco.

## 7.10. Actividades propuestas

**ACTIVIDAD 7.1** Búsquese en Internet información sobre el índice de rendimiento *Relative Business Intelligent*, indicando la capacidad máxima de RBI de procesadores del mercado actual.

**ACTIVIDAD 7.2** Establézcase contacto con algún proveedor de hardware y pregúntese por las herramientas que esta empresa suele utilizar para realizar planificación de la capacidad. Por ejemplo, IBM utiliza la herramienta CP2000 para sus sistemas *mainframe*.

**ACTIVIDAD 7.3** Realícese una entrevista a los usuarios del sistema de información de una empresa para pedirles que definan cuáles son los acuerdos de nivel de servicio a los que les gustaría llegar con el Departamento de Sistemas de Información de la empresa. Asimismo, determínese el grado de sensibilidad de dichos niveles en el Departamento de Sistemas de Información.

**ACTIVIDAD 7.4** Contáctese con departamentos de Sistemas de Información y determí- nese qué nivel de gestión y planificación de la capacidad realizan.

**ACTIVIDAD 7.5** Léase el artículo *System Performance Management and Capacity Plan- ning Tutorial* de B. Domanski, citado en la bibliografía, para obtener una comprensión más completa y enriquecida de este capítulo.

# Capítulo 8

# Rendimiento de un servidor web

En este capítulo se propondrá un caso de estudio que trata de ilustrar la utilización de algunas de las técnicas expuestas en este texto. Aunque se realizan ciertas simplificaciones por cuestiones de espacio y complejidad, el problema que se propone es suficientemente completo. Se trata de realizar un estudio sobre el rendimiento de un servidor web en una Intranet corporativa. La resolución por pasos de este caso permite avanzar en el cono- cimiento de la evaluación cuantitativa del servidor, desde prácticamente la ausencia de cualquier información acerca de su rendimiento, hasta la planificación futura del mismo. Se sugiere acceder a este capítulo una vez que se conocen las técnicas de rendimiento de- sarrolladas en los anteriores, bien si se lee por secciones o en su totalidad. Es interesante también comparar el uso de las diversas técnicas en problemas sencillos y autocontenidos, frente a otros de envergadura mayor como el caso que se propone.

## 8.1. Planteamiento y metodología de resolución

Una compañía fabricante de componentes de la industria del automóvil posee aproximadamente $1{,}500$ empleados, entre los que se incluyen ingenieros, operarios, técnicos, gestores, administrativos y otro personal. Se podría decir que casi la totalidad de los empleados de la organización tiene, de algún modo, acceso a un computador personal o a una estación de trabajo. La compañía ha implementado una Intranet con varias subredes departamentales. La Intranet sirve para muchos de los cometidos en la gestión y comunicación corporativa, como, por ejemplo, la formación de los empleados, el soporte de ayuda a usuarios, el manejo de memorandos internos y formularios, etc. Concretamente, el departamento de ingeniería, que engloba unos $100$ empleados, posee un servidor web, en su propia subred, que sirve para diseminar noticias e información entre sus usuarios. De este modo, el servidor web es utilizado principalmente por el personal de ingeniería para depositar la información que consideran oportuna. Todo el personal de la compañía es susceptible de ser usuario de dicha información que reside en las páginas web de los ingenieros y otros archivos que no se encuentran jerarquizados o estructurados bajo ninguna página de presentación; sin embargo, se puede considerar que la mayoría de los accesos provienen del propio departamento y de su subred.

El administrador del sistema del departamento de ingeniería desea conocer el rendi- miento actual de su servidor web, así como los factores que pueden limitar su capacidad futura. Por otra parte, debido a que el hardware del servidor web es un poco obsoleto, el administrador no conoce cuál de los dispositivos es el cuello de botella. Tampoco sabe si será necesario reemplazar todo o parte del servidor web, si sólo se necesitará reconfigurarlo o esperar a que se llegue cerca del límite de productividad para hacerlo; límite que, por cierto, también desconoce. Éstas y otras cuestiones obligan a realizar, al menos, un estudio de rendimiento y planificación de la capacidad del servidor web.

La Figura 8.1 ilustra las principales fases en la metodología de resolución de un pro- blema de rendimiento. La metodología descansa en dos modelos: el modelo de carga y el modelo de rendimiento. A estos dos modelos podría añadírseles el modelo de costes que englobaría la evaluación económica del rendimiento, pero esta materia está fuera del al- cance de este texto. El modelo de carga captura la demanda de recursos y la intensidad de la carga de trabajo durante un tiempo de observación representativo. El modelo de rendi- miento se usa para conocer, predecir y sintonizar el rendimiento del sistema en función de los parámetros de la carga y la descripción simplificada del propio sistema. La resolución de este caso de estudio seguirá los pasos descritos por esta metodología de trabajo utilizando muchas de las técnicas presentadas en los capítulos anteriores.

En este caso de estudio la resolución propuesta se realiza a partir del modelo de car- ga, sobre el cual se realizan experimentos de *benchmarking* controlados. La información recolectada por el proceso de *benchmarking* y los distintos monitores del sistema operativo servirán como valores de entrada de las variables de las leyes operacionales. Estas leyes operacionales permitirán obtener otros parámetros de interés, con objeto de realizar un modelo de rendimiento del servidor web. El modelado del servidor a través de un formalis- mo matemático nos permitirá realizar distintos experimentos con el modelo abstracto y no con el servidor real como hasta el momento. Estas pruebas con un modelo abstracto serán de gran utilidad para, por ejemplo, predecir la carga futura del servidor web del departa- mento de ingeniería o bien para ajustar, mejorar o sintonizar el sistema. A lo largo de los apartados siguientes se pondrá de manifiesto el uso de las diversas técnicas que permitan estos objetivos en este caso de ejemplo.

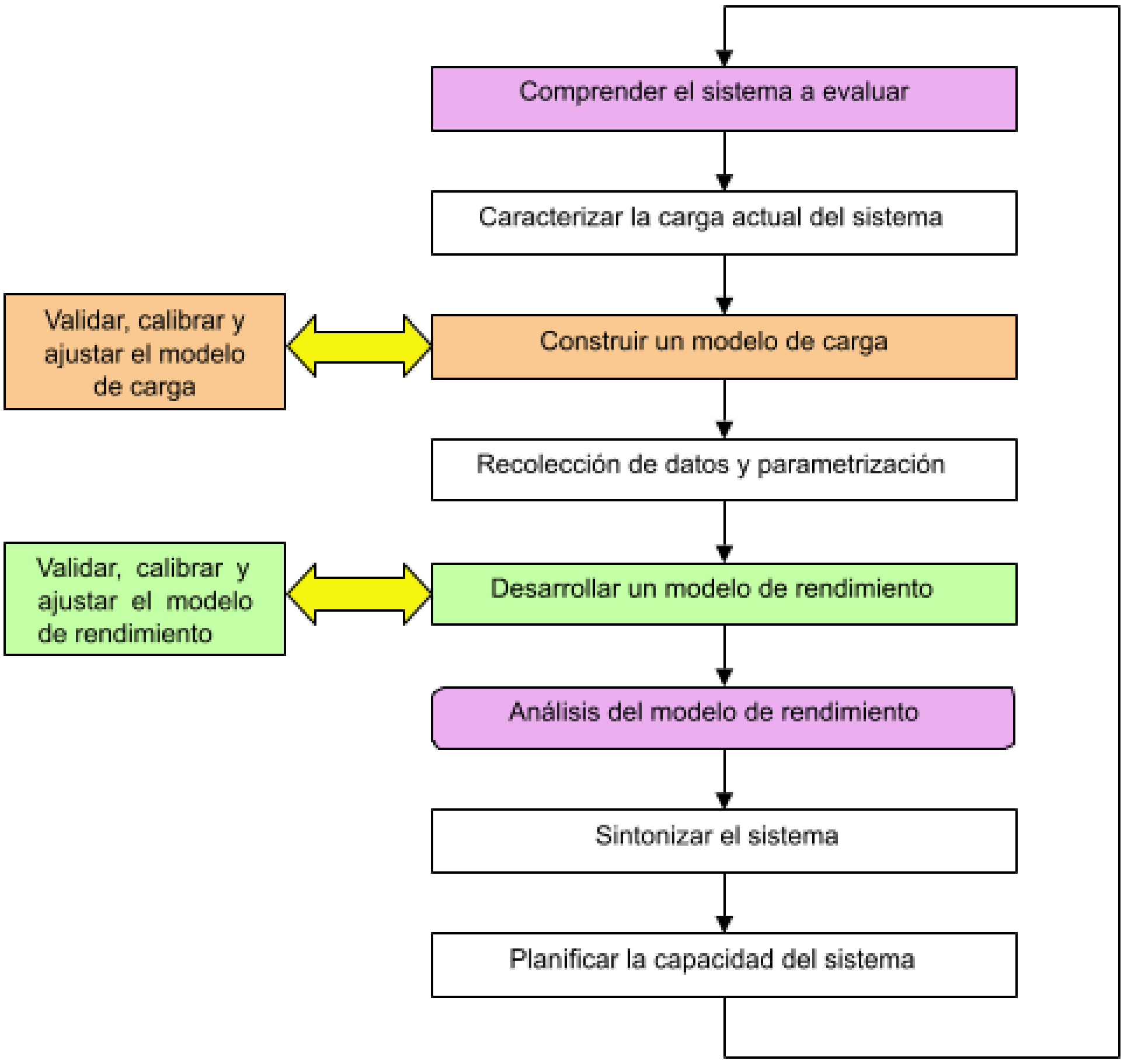

**Figura 8.1:** Una metodología para el estudio del rendimiento.

## 8.2. Escenario de explotación del servidor web

La fase inicial de la metodología propuesta para el caso de estudio consiste en conocer el conjunto de dispositivos hardware, software y protocolos que se presentan en el escenario de explotación.

El estudio del sistema se ha realizado sobre un escenario abierto, al cual pueden acceder usuarios desde la red de área local de la compañía y, en menor grado debido a medidas de seguridad, unos pocos usuarios externos pueden acceder desde cualquier localización vía Internet. Este último punto se ha obviado en el caso de estudio, pero se volverá a él más adelante. La Intranet corresponde a una Fast Ethernet a 100 Mbps. Aunque se conoce que los usuarios accederán con las herramientas propias de su estación de trabajo al servidor web, por ejemplo el navegador, se desconoce *a priori* el entramado de puentes y encaminadores (*routers*) de la red y menos aún, en el caso externo, el modo de conexión externa a través de ISPs. Durante el desarrollo de este capítulo, se propondrán soluciones sencillas para conocer el efecto de estas posibilidades.

| | |
|---|---|
| Procesador | Intel Pentium II 233 MHz |
| Disco 1 | Quantum FIREBALL SE4.34 4,3 GB |
| Disco 2 | Seagate ST340810A 40 GB |
| Memoria RAM | 192 MB |

**Tabla 8.1:** Configuración del servidor web.

### 8.2.1. Configuración del servidor web

El servidor web del caso de estudio es un computador personal, con dos discos duros (controladora incluida), uno de los cuales se utiliza como almacenamiento de bibliotecas de sistema y otros archivos no activos para el servicio de web (véase la Tabla 8.1). En concreto, es el disco 2 el que contiene la información de interés para el estudio.

El servidor web dispone del software *Apache Web Server*. Apache es uno de los servido- res web más utilizados en Internet. Su popularidad se debe a una serie de características, como son: el tener código fuente abierto, mantener una evolución rápida y continuada de versiones, poder ser utilizado por desarrolladores de cualquier plataforma y además ser gratuito.

Como opción predeterminada, Apache provee de dos archivos de registro o bitácoras de navegación (*log files*); uno de ellos corresponde al uso del servidor web y el otro a los errores. El primero de ellos va a ser de gran utilidad para el inicio de la resolución de las cuestiones a las que se enfrenta el administrador de este sistema.

### 8.2.2. Información disponible

El administrador del servidor web, a pesar de que posee estadísticas de éste, no conoce el rendimiento actual del servicio que provee, y menos aún la previsión del rendimiento que podrá dar en un futuro. Las estadísticas semanales y mensuales que realiza son las habituales en este tipo de instalaciones y muestran gran cantidad de datos globales (nú- mero de páginas accedidas, errores, etc.) y parciales (páginas más populares, página con mayor información trasferida, etc.). Esta información provista por servicios que se pueden encontrar en Internet, aunque muy interesante para el control cualitativo de la utilización del servidor, puede no serlo tanto para realizar un estudio cuantitativo del servicio que se ofrece a los usuarios (tiempo de respuesta, productividad, latencia, etc.). Del mismo modo, el rendimiento actual puede ser de gran interés, pero también lo es conocer el que tendría en determinadas situaciones (cuello de botella, número máximo de transacciones permitidas, etc.) y qué ocurriría si se pudiera cambiar el hardware (procesador, disco duro, etc.) por otro potencialmente más rápido. Por otro lado, en este servidor web, se producía la parado- ja de que en periodos de mayor actividad, los usuarios decían obtener mejores prestaciones que en periodos teóricamente menos cargados. Todas estas cuestiones no tienen respuesta en esta instalación, puesto que nunca se ha planteado el rendimiento como variable de capacidad futura, y el estudio pretende, al menos, orientarnos acerca del rendimiento del servidor web.

## 8.3. Estudio de rendimiento del servidor web

### 8.3.1. Construcción de un modelo de carga

Recordemos que el término de caracterización de la carga hace referencia al proceso de construir un modelo que describa cuantitativamente la carga que se ejecuta en el sistema a analizar. La carga del servidor web se puede definir como el conjunto de peticiones de HTTP procesadas en el sistema durante un intervalo de observación. Por tanto, durante esta primera actividad, se deberían recoger las peticiones HTTP durante un cierto tiempo de interés y posteriormente realizar el modelo de carga.

#### Almacenamiento de datos de carga

Cuando un usuario se conecta a un servidor web se produce un intercambio de información. La mayor parte de la información de control sobre la conversación HTTP que se establece entre los usuarios y el servidor web, se registra en la bitácora de acceso y la bitácora de error. Estas bitácoras son ficheros que se encuentran en el servidor, accesibles por el administrador del sistema. A continuación se muestra con detalle el contenido de una entrada de este fichero:

```
212.166.131.69 [16/Feb/2002:12:55:58 +0100]
"GET /imatges/foto.html HTTP/1.1" 200 28870
"HTTP://deping.compa.com/" "Mozilla/4.0"
```

Si se observa la entrada del ejemplo mostrado se obtienen los siguientes parámetros:

- Identificación del computador solicitante, que en la línea del ejemplo que nos ocupa es 212.166.131.69, que corresponde a la dirección IP de la máquina del usuario que efectuó una invocación a un método de HTTP.
- Fecha y hora en la que ocurrió la solicitud (*timestamp*). Este membrete incluye tam- bién la información de la zona horaria (+0100 en la entrada del ejemplo).
- Transacción HTTP de la solicitud que se hizo al servidor. Normalmente corresponde con la invocación de alguno de los métodos HEAD, GET y POST HTTP, seguido de la URL (o URI) solicitada por el computador del usuario (cliente) y de la versión del protocolo HTTP en la que se esperaba la respuesta del servidor. En la línea del ejemplo el usuario solicita la página foto.html, ubicada dentro del directorio *imatges*, y espera la respuesta en la versión 1.1 del protocolo HTTP.

- Código de respuesta, que indica si la solicitud ha tenido éxito o bien muestra el tipo de error que se ha producido. En el caso de la línea del ejemplo el código es el 200, que indica que la operación fue satisfactoria.
- Bytes tranferidos al computador cliente, sin incluir los encabezados propios del protocolo HTTP, es decir, el tamaño del fichero.
- Dirección URL completa.
- Datos del sistema y del navegador del cliente.

Una vez conocida la información que nos proporciona la bitácora de acceso, se debe tomar una muestra de ésta. El primer problema es determinar el tamaño de la muestra. Se podría estudiar estadísticamente la bitácora guardándola en un fichero y luego intentar encontrar el periodo de observación suficiente. Otra solución que no excluye la anterior es guiarse por los conocimientos del administrador del sistema o la experiencia del analista de prestaciones. Así, se puede establecer el intervalo de uso característico o el que le interesa al responsable del sistema. Normalmente existen programas que analizan los datos de acceso de los servidores web, que utilizan los responsables de los sistemas y permiten establecer ciertos patrones básicos de uso. Más adelante se comentarán estas posibilidades.

En el caso concreto de este servidor web se consideró que el intervalo de una semana laboral era el más interesante, puesto que coincide con el horario de actividad de los usuarios. Así se tomaron los datos de la bitácora de acceso del servidor de varias semanas laborables que, una vez analizadas con los paquetes estadísticos oportunos, se consideró que producían la misma carga. De hecho, hasta cada día era muy parecido al anterior. Sin embargo, lo que finalmente se supuso como carga de trabajo fue una semana de cinco días laborales, de entre las recogidas en la bitácora. Este fichero de carga real, tras una serie de procedimientos que se explicarán a continuación, produce el modelo de carga característico.

### Filtrado y preprocesado de los datos de carga

Una vez conseguido el fichero de carga real, se observa en el mismo que el porcentaje de operaciones GET es muy superior al de otras transacciones HTTP. Recordemos que la orden GET se utiliza para recoger cualquier tipo de información de un servidor web. Se utiliza siempre que el usuario hace un clic sobre un enlace o se teclea directamente una dirección URL. Como resultado, el servidor HTTP envía el documento correspondiente a la URL seleccionada, o bien activa un módulo CGI, que generará a su vez la información de retorno. Consultado este fenómeno con el administrador del sistema y verificado con las estadísticas de otras semanas y meses, se concluye que se trata de un servidor estático con información de recuperación y consulta para los usuarios.

Así se produce un primer filtrado de los datos contenidos en los ficheros de acceso y se preprocesan eliminando las escasas líneas de la bitácora correspondientes a operaciones distintas de GET. El resultado del filtrado inicial es un fichero de traza con los accesos de tipo HTTP GET correspondiente a un día laborable de 8 horas. Así, durante el periodo de observación, que a partir de ahora denominaremos $T$ y que corresponde a $5 \times 8 \times 60 \times 60 =$ 144.000 segundos, se produjeron 55.910 accesos GET.

## Caracterización de la carga

Para poder establecer la caracterización de la carga y disminuir el impacto de la diferencia de magnitudes en la misma, se necesitan criterios de clasificación según la naturaleza del fichero de traza, es decir, dividir en clases de acceso distintas a partir del conocimiento que se tiene de la carga. Cada uno de los criterios de clasificación produce un número de clases diferente sobre los mismos datos de carga.

Debido a que la carga real no es más que una colección de componentes heterogéneas, esta partición se hace basándose en los atributos del fichero del que se dispone. Para ilustrar dos clasificaciones con criterios diferentes, se toman típicamente dos atributos: el tamaño y el tipo de los ficheros consultados. Así, como ejemplo ilustrativo de este caso de estudio, cada uno de los ficheros de acceso con operaciones GET se clasificará por el tamaño del fichero solicitado por la transacción HTTP.

Por tanto, el fichero de acceso recogido durante el periodo de observación, de una sema- na laborable ($T$), se clasificará por el tamaño del archivo que se solicita en la transacción. Así tomaremos $T$ = 144.000 segundos en los que se produjeron 55.910 transacciones. Por tanto, se producen 0,3882 transacciones por segundo (tps), equivalentes a algo más de 23 transacciones por minuto (tpm).

## Modelo de carga del servidor web

Como el objetivo final de caracterizar la carga es conocer cómo rinde y rendirá el servidor web, se necesita una representación compacta de esa carga, y ésa es la razón de construir un modelo. Para la construcción de un modelo de carga del servidor web, se debe filtrar de nuevo el fichero de carga real para que sólo permanezcan los valores del criterio de clasificación o atributos de carga. El resultado de estos filtros será un fichero de traza con los tamaños de los archivos accedidos. Para realizar este segundo filtrado se pueden construir programas *ad hoc*, por ejemplo en lenguaje Perl, o bien utilizar paquetes estadísticos que permitan filtrar los datos del fichero a procesar, por ejemplo SPSS. En cualquier caso, se debe tener en cuenta que la cantidad de datos que se producen en el intervalo de observación puede ser tan elevada que se necesite partir y luego ensamblar los resultados. En el caso que nos ocupa el fichero de carga con transacciones HTTP GET antes de clasificarlo era de unas decenas de MB. Si fuese necesaria tal partición previa para el análisis de clasificación, es bastante natural hacerlo por días, horas e incluso decenas de minutos y luego ensamblar fácilmente los datos resultantes.

A partir del fichero de traza, correspondiente al periodo de observación $T$ , se obtiene el conjunto de archivos solicitados al servidor web, el tamaño de cada uno de ellos y, por último, su número de accesos. Llegado este punto, se puede realizar una primera clasifica- ción según el tamaño del archivo consultado esté en un intervalo medido en bytes. Estos intervalos son arbitrarios pero deben ser suficientemente representativos. Así se obtienen diferentes grupos, cada uno formado por el conjunto de accesos a archivos que tengan un tamaño que esté dentro del intervalo preestablecido para cada grupo. Por ejemplo, se po- dría realizar una primera clasificación de los accesos a archivos en clases según ese intervalo arbitrario: el primer grupo engloba accesos a archivos de hasta K bytes; luego se agrupa de K en K bytes hasta el tamaño de fichero mayor recogido en la bitácora. Pero se podría establecer cualquier otro criterio de clasificación inicial. El resultado de la apreciación an- terior distingue $n$ grupos diferentes con accesos durante el periodo $T$ . De cada grupo, se calcula el tamaño medio de los archivos que comprende, así como el número de accesos que recibe cada grupo.

El siguiente paso construye una representación de la carga más compacta, que se de- nomina modelo de la carga. Para obtener dicho modelo de la carga se aplican técnicas estadísticas de agrupamiento (*clustering*). En particular, se ha utilizado la técnica del ár- bol de extensión mínima (MST, *Minimal Spanning Tree*) (véase el Capítulo 6). En el caso de modelo de carga que nos ocupa se comienza con 39 grupos donde el fichero de ma- yor tamaño es del orden de $10^7$ bytes. El proceso de agrupamiento se detiene al llegar a 5 clusters, en que las distancias entre centroides se consideran demasiado alejadas para fusionarlas. Por razones de espacio y simplicidad, se han obviado estas manipulaciones del fichero de bitácora inicial en este caso de estudio.

Resumiendo, la carga original de trabajo recogida en el fichero de bitácora ha sido filtrada eliminando todas las transacciones distintas del tipo GET, y después ha sido pro- cesada y filtrada de nuevo eliminando la información distinta de la del tamaño del fichero solicitado; a continuación, ha sido agrupada en 39 clases representando el tamaño medio del archivo y el número medio de accesos recibidos de cada grupo y tras el agrupamiento a través de la técnica MST se han quedado en cinco clases de peticiones HTTP. Cada clase está representada por unos valores, que son la media de los valores de los componentes que forman la clase.

La Tabla 8.2 representa el modelo de carga al que estuvo sometido el servidor durante el tiempo de observación, establecido anteriormente, de cinco días laborables ($T$). Esta carga se ha clasificado en cinco tipos de peticiones HTTP diferentes dependiendo del tamaño del archivo solicitado. Estas clases de peticiones, según el tamaño de archivo, se han identificado como tipos “muy pequeño”, “pequeño”, “mediano”, “grande” y “muy grande”. De este modo, por ejemplo, las peticiones de tipo “muy pequeño” han realizado una media de 4.141 accesos y tienen un tamaño medio de 113,79 bytes; las de tipo “pequeño” han realizado una media de 2.992 accesos y su tamaño medio es de 597,12 bytes, y así sucesivamente hasta llegar a las de mayor tamaño, a las que se ha asignado el identificador "muy grande", con un tamaño superior a 1 MB pero con escasamente 15 accesos durante toda la semana. Puesto que se han producido 55.910 accesos durante el intervalo $T$, el número total de accesos se recalcula a partir del porcentaje de la clase. Por ejemplo, la clase "muy pequeño" representa el 44,64 % de todos los accesos de la traza, que corresponderían a algo más de 24.958 transacciones.

| Tipo | Cluster | TM (bytes) | NA | % accesos | Total accesos |
|---|---|---|---|---|---|
| Muy pequeño | 1 | 113,79 | 4.141,00 | 44,64 | 24.958,22 |
| Pequeño | 2 | 597,12 | 2.992,00 | 32,26 | 18.036,56 |
| Mediano | 3 | 13.794,71 | 1.919,33 | 20,69 | 11.567,78 |
| Grande | 4 | 140.788,16 | 209,38 | 2,26 | 1.263,56 |
| Muy grande | 5 | 1.246.479,87 | 14,88 | 0,15 | 83,86 |

**Tabla 8.2:** Modelo de carga del servidor web durante el periodo de observación.

### 8.3.2. Recolección de datos de parametrización

En un escenario ideal, los valores de los parámetros necesarios para la construcción de un modelo de rendimiento estarían dados directamente por las propias facilidades del sistema a modelar. Sin embargo, en algunos casos puede que el sistema todavía no exista, y en la mayoría el analista de prestaciones sólo dispone de medidas muy generales del rendimiento no orientadas al modelo de carga concreto de su instalación. En los servidores web, las medidas necesarias para parametrizar el modelo de rendimiento se reducen a conocer las demandas de procesador, E/S y red que establecen las peticiones HTTP; siendo éstos los dispositivos hardware que más intervienen en la consecución del proceso solicitado. Los va- lores de dichas demandas específicas para el modelo de carga del servidor web se obtendrán de la monitorización de experimentos controlados de *benchmarking* y su posterior análisis operacional. Dichos experimentos estarán diseñados en función de la disponibilidad de *ben- chmarks* y monitores adecuados para tal fin, en el escenario considerado. En la Figura 8.2 se puede observar el entorno controlado de *benchmarking* para el caso de estudio. Mientras el servidor web recibe la carga sintética generada por el *benchmark* desde una estación cliente de la Intranet, los monitores residentes en el servidor se ejecutan y almacenan datos de la actividad de las transacciones HTTP. Ambas actividades han sido invocadas por el analista de prestaciones, durante periodos de inactividad de los usuarios de la Intranet.

#### *Benchmarking*

Un modo muy útil para estudiar el rendimiento de un sistema dado es someterlo a una parte de la carga actual y medir los valores de interés del mismo. En contrapartida, la carga real es difícil de medir, por lo que en la mayoría de los casos no suele ser efectivo.

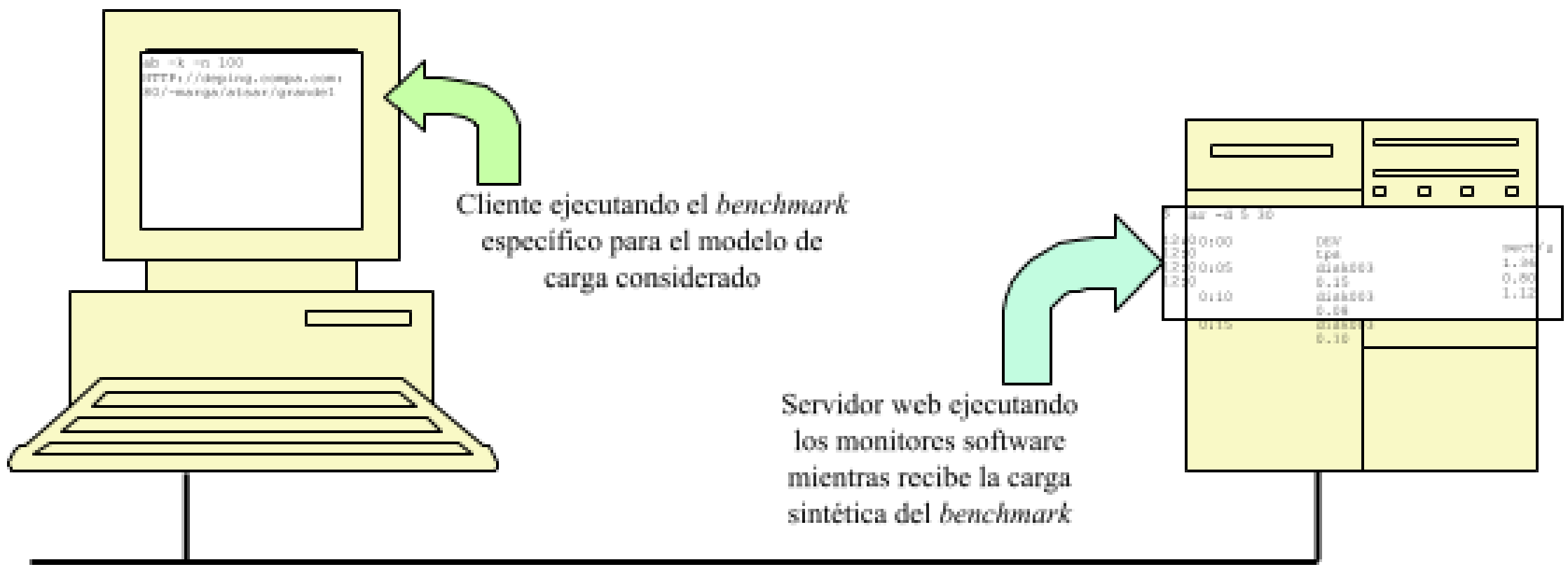

**Figura 8.2:** Entorno controlado para la monitorización del servidor web.

La alternativa general es hacer uso de los programas de prueba (*benchmark* ) y utilizar sus resultados para comprender el rendimiento de los diferentes sistemas.

Ya que el propio servidor web proporciona un programa *benchmark*, parece obvio utilizarlo en este caso de estudio. Por lo que se ha utilizado el *benchmark* que proporciona el servidor Apache, que recibe el nombre de ab (*Apache Benchmark* ). Se debe recordar que el servidor está en funcionamiento al menos 8 horas al día y, por tanto, puede haber interferencia con los usuarios habituales. Si las medidas del *benchmark* se toman durante las horas de utilización, se tiene que discernir qué carga es artificial y qué carga es real, lo que no parece deseable. Por el contrario, se pueden hacer los experimentos de *benchmar- king* cuando no haya prácticamente interferencia con un uso controlado y restringido. En el desarrollo de este apartado se justificará cuál de los momentos de experimentación es el más deseable.

## Programa de prueba ab

El *benchmark* ab es una herramienta parametrizable que puede obtenerse libremente del suministrador del servidor. Dentro de las diferentes opciones que presenta, la ejecución del *benchmark* se realiza desde una estación cliente contra el servidor, por ejemplo en modo orden:

ab [k] [ n request] [HTTP://hostname:port]/path]

donde las opciones son las siguientes:

- k habilita la característica KeepAlive del servidor; es decir, el servidor es capaz de ejecutar múltiples peticiones en una sesión HTTP, característica que posee el servidor

Apache que se está analizando. Los resultados del *benchmark* son dependientes de este parámetro. Si el servidor está configurado con esta opción para las transacciones HTTP en uso habitual, se debe activar para que la carga artificial se asemeje al máximo a la carga real.

- n request indica el número de peticiones a ejecutar, es decir, la cantidad de carga artificial a someter al servidor. El fichero seleccionado se solicita, de forma secuencial, un total de *n* veces consecutivas.
- hostname es el servidor al que se le envían las peticiones HTTP; en el caso de estudio presente el servidor web del departamento de ingeniería.
- port representa el puerto en el que está escuchando el servidor; en este caso, al ser un servidor web es el puerto número 80.
- path indica el fichero al que se quiere acceder, con la ruta completa de su ubicación en el servidor. Este *benchmark* realiza la carga siempre contra un archivo característico o muestra. Esta característica del *benchmark* ab es común a otros *benchmarks* debido a la facilidad de implementación del programa. Sin embargo, como comprobaremos después, obliga a ciertas simplificaciones y a un trabajo adicional al analista de prestaciones que realiza el estudio de este caso.

El principal resultado de la ejecución del *benchmark* ab es el throughput o productividad desde la estación cliente al servidor (ver Capítulo 4). Es decir, cuántas peticiones por unidad de tiempo puede servir el sistema. A continuación se muestra una ejecución del *benchmark* ab junto con la información que proporciona:

```
ab k n 100 HTTP://deping.compa.com:80/~marga/atsar/grande1 Benchmarking

deping.compa.com (be patient)...

Server Software: Apache/1.3.27 Server
Hostname: deping.compa.com Server Port: 80

Document Path: /~marga/atsar/grande1 Document
Length : 181 bytes

Concurrency Level : 1
Time taken for test: 0.203 seconds Complete
request: 100
Failed request: 0
Total transferred: 50692 bytes Html
transferred: 18100 bytes Request per
second: 492.61
```

Transfer Rate: 249.71 kb/s received

Connection times (ms)

| | | Min | avg | max |
|---|---|---|---|---|
| Connect: | | 0 | 1 | 15 |
| Processing: | 0 | 1 | 0 | |
| Total: | 0 | 1 | 15 | |

En este ejemplo se ha efectuado un total de 100 peticiones a la página web ubicada en la dirección deping.compa.com/ marga/atsar/grande1. La opción KeepAlive ha sido invocada puesto que el servidor web también la tiene habilitada, permitiendo varias pe- ticiones HTTP en la misma sesión. La presentación de los resultados de la ejecución se interpreta como sigue:

- Server Software, versión del software del servidor web al que se accede.
- Server Hostname, nombre del servidor web al que se accede.
- Server Port, puerto en el que está el servidor recibiendo.
- Document Path, ruta del directorio del archivo muestra.
- Document Length, tamaño en bytes del archivo muestra.
- Concurrency Level, en este caso de estudio, al no indicarle que se deseaban peticio- nes concurrentes, su valor es 1.
- Time taken for test, tiempo que ha durado la ejecución del experimento.
- Complete request, peticiones que se han realizado con éxito.
- Failed request, peticiones que han fallado.
- Total transferred, número de bytes transferidos, incluyendo las cabeceras de los distintos protocolos que intervienen.
- Html tranferred, número de bytes HTML transferidos; únicamente se considera el tamaño del archivo muestra.
- Request per second, productividad medida en peticiones por segundo.
- Transfer rate, productividad medida en KB por segundo.
- Connection times, tiempo de conexión (mínimo, medio y máximo) en milisegundos.

Como se puede observar, el conjunto de resultados que proporciona el *benchmark* ab, aunque muy simples, pueden ser de gran ayuda para tener una idea aproximada del comportamiento cuantitativo del servidor visto desde un cliente. Por lo tanto, una vez conocido el modelo de carga, se puede ejecutar el *benchmark* ab con la carga caracterizada a través de archivos muestra bien elegidos.

### *Benchmarking* para el modelo de carga del servidor

En este modelo de carga se dispone de cinco clases de transacciones que demandan archi- vos de tamaños medios diferentes (véase la Tabla 8.2). Para adaptar el modelo de carga al *benchmark* ab, se debe encontrar, para cada clase, un archivo muestra que realmente se encuentre en el servidor y que tenga un tamaño igual o similar al tamaño medio caracterís- tico. Otros *benchmarks* permiten elegir un conjunto de archivos muestra y no sólo uno. En cualquier caso, la tarea del analista de prestaciones es seleccionar qué archivos muestra del servidor representarían artificialmente el modelo de carga de los usuarios reales. En el caso que nos ocupa, se puede realizar una rutina de búsqueda en el fichero filtrado de la carga inicial o bien utilizar paquetes estadísticos estándar para localizar los archivos muestra. El resultado de la búsqueda proporciona unos archivos muestra como los de la Tabla 8.3 (todos los tamaños se expresan en bytes).

| Clase | TM (bytes) | NA | Documento | Tamaño |
|---|---|---|---|---|
| Muy pequeño | 113,79 | 4.141,00 | fon_r5_c1.gif | 114 |
| Pequeño | 597,12 | 2.992,00 | men_r03_c1.gif | 617 |
| Mediano | 13.794,71 | 1.919,33 | cache.jpe | 14.486 |
| Grande | 140.788,16 | 209,38 | assembler.html | 153.845 |
| Muy grande | 1.246.479,87 | 14,88 | vrml2000.pdf | 1.319.793 |

**Tabla 8.3:** Archivos de muestra para la carga sintética del *benchmark* ab.

Una vez localizados los archivos muestra de cada clase en el servidor, se ejecuta el *benchmark* con cada uno de ellos, con un número de repeticiones suficiente para que la estimación de la medida sea correcta. Evidentemente, lo más preciso sería estudiar el número de ejecuciones necesarias para garantizar la corrección de los resultados del *benchmarking*, pero a efectos prácticos basta un número de repeticiones elevado. Por otra parte, si se pre- tende que no haya interferencia de carga real durante la experimentación y garantizar las condiciones de la misma, el *benchmark* debería ejecutarse de modo controlado. En este ca- so, las mediciones se realizaron durante los fines de semana, cuando el número de usuarios era nulo, desactivando incluso el acceso desde el exterior.

## Análisis de los resultados del *benchmarking*

Entre los datos más relevantes que nos proporciona el *benchmark* ab, el que puede servir *a posteriori* para modelar el comportamiento del servidor web es, sin duda, el tiempo de respuesta de las peticiones HTTP. Dicho tiempo de respuesta es el que transcurre desde que se inicia con el envío de las peticiones HTTP desde el computador cliente, hasta que llegan de vuelta los servicios demandados en esas peticiones, al cliente. Este *benchmark* envía una petición cada vez que recibe la respuesta de la anterior, mediante la secuencia: petición, respuesta, petición, . . . , respuesta. De este modo, si se divide el tiempo de test del *benchmark* entre el número de peticiones generadas en ese tiempo, se obtiene el tiempo de respuesta de cada una de esas peticiones en el servidor. Este tiempo de respuesta corresponde al valor medio de un ciclo petición–respuesta en el computador desde el cual se efectúan los experimentos de *benchmarking*. Invirtiendo el tiempo de respuesta medio calculado por el *benchmark* para una sola petición HTTP se obtiene la productividad media del servidor web para ese tipo de transacción muestra en las condiciones del test (véase el Capítulo 4).

Puesto que el modelo de carga consiste en cinco grupos de transacciones características, se ha realizado una cantidad de experimentos múltiplo del número caracterizado de accesos. Así, por ejemplo, como en el modelo de carga se determinó que las transacciones del grupo "grande" realizaban 209 accesos al servidor, se ha ejecutado el *benchmark* 100 veces con 2090 accesos por experimento a la página característica correspondiente. La duración media de las 100 ejecuciones fue de 32,49 segundos, lo que supone una productividad máxima del servidor de algo más de 64 tps. Para cada experimento de 100 bloques de ejecución se emplearon los archivos muestra correspondientes, considerándose que tal número de bloques era suficientemente representativo. En la Tabla 8.4 se presentan los valores medios de la productividad máxima de las 100 pruebas con un número de accesos 10 veces superior a los referidos en el modelo de carga.

| Clase | Duración test(s) | Número peticiones | Productividad (tps) |
|---|---:|---:|---:|
| Muy pequeño | 62,03 | 41.410 | 667,580 |
| Pequeño | 52,81 | 29.920 | 566,559 |
| Mediano | 42,06 | 19.190 | 456,253 |
| Grande | 32,49 | 209 | 64,327 |
| Muy grande | 18,74 | 150 | 8,004 |

**Tabla 8.4:** Productividad máxima alcanzada durante los experimentos de *benchmarking*.

Se debe considerar que los resultados obtenidos son sólo representativos de forma par- cial, debido a que se han obtenido a partir de un archivo de muestra que existe en el servidor web, pero no es la carga real observada durante un cierto periodo. Muy al contrario, se ha realizado una carga artificial o de prueba, para determinar aproximadamente el tiempo de respuesta de las transacciones HTTP en el servidor cuando hay una única transacción sirviéndose.

Por otra parte, este tiempo engloba el tiempo de conexión, el tiempo de transmisión de los archivos, el tiempo de resolución del nombre del servidor en el DNS, el tiempo en que el servidor atiende la petición y la transmisión de la respuesta. En todos estos tiempos influyen decisivamente la velocidad de los dispositivos del servidor (procesador, disco, cache. . .) y la capacidad de transferencia de Internet y la red local. Más adelante se tendrá que discernir la efectividad de la cache de E/S para no llevarse a engaño en cuanto al rendimiento.

Evidentemente, un *benchmark* orientado a peticiones HTTP no proporciona informa- ción acerca del uso de los dispositivos hardware, lo cual es lógico, puesto que tampoco se puede reconocer esa información en el propio protocolo. Será necesario el uso de otras técnicas de evaluación del rendimiento para determinar el uso de procesador, las visitas a los discos, la utilización de la cache, la latencia de la red, etc., que requiere cada petición HTTP. Es decir, se necesitará monitorizar el servidor, mientras recibe las peticiones HTTP de una carga de prueba. De este modo combinaremos el proceso de *benchmarking* con la monitorización del sistema.

### Monitorización

A la hora de realizar cualquier tipo de estudio sobre un sistema informático es necesario tener información real acerca de lo que está ocurriendo en ese sistema. Para ello es de vital importancia disponer de herramientas que permitan obtener esa información. Un monitor es una herramienta utilizada para observar la actividad de un sistema informático mientras es utilizado por sus usuarios. En general, observa el comportamiento del sistema, recoge datos estadísticos de la ejecución de los programas, analiza los datos recogidos y presenta los resultados. Durante el desarrollo de este apartado, se deben confrontar dos problemas: el periodo de monitorización y el conocimiento de la carga sometida; es decir, resolver las cuestiones de la interferencia con la carga real actual y discernir la carga a monitorizar en el sistema.

En este caso de estudio, se han utilizado varios de los monitores software incluidos en el servidor web, cuyo sistema operativo es Linux. Recordemos que todo monitor que se ejecute en el sistema produce inherentemente una sobrecarga (*overhead* ), es decir, un consumo de sistema imputable a los procesos de medida. Aun así, nos vemos obligados a conocer cuántos recursos consume la carga que soporta el servidor web, por lo que se va a ejecutar una serie de monitores y herramientas del sistema operativo. Las herramientas utilizadas en este apartado han sido descritas en el Capítulo 2.

## Procesador

El monitor vmstat permite monitorizar la actividad de la memoria virtual del sistema. Con él, además de conocer el espacio de memoria virtual usado y libre podremos ver los fallos de página que se producen, así como el porcentaje de uso del procesador de los distintos procesos ejecutables. En este último punto radica nuestro interés en el caso de estudio.

Sin embargo, hay que señalar que las peticiones HTTP que recibe el servidor provocan una serie de procesos en el sistema que habrá que identificar *a priori*. Es decir, el protocolo HTTP y los procesos de ejecución en el sistema operativo provocados por el protocolo no tienen una relación intrínseca directa. Por ello, cuando tenemos suficiente privilegio para poder utilizar el monitor vmstat, como administradores, se debe controlar su ejecución para discernir sin lugar a dudas qué procesos pertenecen a la carga HTTP. De nuevo, las técnicas de rendimiento se utilizan conjuntamente; así combinaremos el uso del *benchmark* ab y del monitor vmstat.

Para eliminar la posible interferencia de otros usuarios en las pruebas experimentales, se elige un momento sin uso del servidor o bien se aísla durante los momentos de prueba. Después se ejecuta el *benchmark* ab con las cargas de muestra que se aplicaron en el aparta- do anterior mientras el monitor vmstat está arrancado (de hecho, se efectuará esta misma operación con otros monitores). Finalmente, se recogen los resultados de las mediciones del monitor en un fichero contable. Posiblemente, habrá que construir ciertos scripts de Linux para realizar estas operaciones de arranque, almacenamiento y parada del monitor en cuestión.

De este modo, mientras se produce la carga de prueba con los archivos muestra de cada clase en el modelo de carga, el monitor vmstat anota la utilización de los recursos del sistema. Cuando termina la carga del *benchmark*, se puede detener la monitorización y se observan los periodos de actividad e inactividad anotados. Los periodos activos corres- ponden al momento en que el sistema está prácticamente ocioso y aislado para el resto de los usuarios, aunque ejecutará los procesos que se produzcan al recibir la carga de prueba HTTP.

A continuación se presenta un segmento de las soluciones que proporciona el programa vmstat:

```
$ vmstat 1 4
procs                  memory      swap       io      system          cpu
r b w    swpd free   buff   cache   si  so    bi bo    in      cs   us  sy  id
0 0 0   18408 5428  25212   80456   0   0    10  9     1       8    3   0  90
0 0 0   18408 5428  25212   80456   0   0     0  0   151      50    0   0 100
0 0 0   18408 5428  25212   80456   0   0     0  0   139      50    0   0 100
0 0 0   18408 5420  25220   80456   0   0     0 20   148      54    0   1  99
```

Puesto que se conoce aproximadamente la duración global de las pruebas de test a través del *benchmark* ab cuando se somete el servidor web, aislado y controlado, a las transaccio-nes muestra, se puede también conocer a priori la duración deseada de la monitorización de vmstat. Así, se puede especificar un tiempo entre muestras de un solo segundo, por ejemplo, y observar la monitorización durante ráfagas repetidas de carga de test del *ben- chmark*. Evidentemente, todo monitor produce cierta sobrecarga que consideraremos casi despreciable en comparación al consumo de recursos de las transacciones HTTP.

Como se puede observar, son varios los sujetos de medida que proporciona el monitor vmstat pero, para nuestro caso de estudio, los datos más relevantes están en las tres últimas columnas, las cuales dan información acerca del uso del procesador. En particular, la columna us indica el porcentaje utilizado en modo usuario, la columna sy la utilización por el sistema operativo y, por último, la columna id, el porcentaje que se encuentra libre.

Con los datos que suministra el monitor durante los periodos de carga con el *benchmark*, se puede conocer la utilización que sufre el procesador por parte de la carga, por ejemplo, de los 1.000 accesos de cada archivo muestra en el modelo de carga caracterizado.

Sin embargo, una de las características de la medida debe ser su repetibilidad, esto es, la posibilidad de realizar una medición diferentes veces, y que el resultado de este proceso de medida sea siempre el mismo. En caso de que no sea así, las sucesivas medidas se usan para reducir los errores intrínsecos del proceso de medición. Por tanto, esta batería de pruebas se realiza un número de veces elevado para garantizar las medidas. En este caso se realizaron 100 baterías de pruebas de *NA* accesos con cada archivo muestra. Ahora bien, en la monitorización, el resultado de una medición será algo distinto unas veces de otras, ya que, normalmente, no es posible repetir exactamente las mismas condiciones de carga, en los mismos instantes. Por ello, se suele hablar de monitorización y no de medición, ya que lo que estrictamente se efectúa es un seguimiento de la actividad realizada.

En resumen, se ejecutaron, simultáneamente, el *benchmark* ab para cada clase de peti- ción HTTP del modelo de carga, y el monitor vmstat, con el suficiente número de muestras para englobar la duración de las fases de test. Este proceso se repitió una serie de veces, en concreto 100, para baterías de un número igual de accesos al fichero prototipo (véase la Tabla 8.3), para así tener la característica de repetibilidad anteriormente mencionada. Finalmente, con los ficheros contables generados con los scripts de Linux, se calculó la media de utilización del procesador para cada tipo de petición HTTP. Los resultados se presentan en la Tabla 8.5.

Como se aprecia en la Tabla 8.5, la suma de los porcentajes medios no completa el 100 % de uso del procesador, lo que indica un remanente ocioso de este dispositivo durante el test. Ello es debido a que durante la carga de prueba, el procesador posiblemente esté gestionando los dispositivos de E/S, red, etc. De hecho, ya se puede observar que, cuanto mayor es el tamaño del fichero solicitado al servidor, menor es la utilización media del procesador. Por otra parte, también se observa que los porcentajes de usuario y sistema por clase no difieren mucho, aunque, para la clase “muy grande”, el de sistema es más del doble que el de usuario. Este último fenómeno parece confirmar que cuanto mayor es el tamaño del fichero solicitado mayor gestión se necesita por parte del sistema operativo.

| Clase | % usuario | % sistema |
|---|---|---|
| Muy pequeño | 54,66 | 43,33 |
| Pequeño | 49,38 | 43,72 |
| Mediano | 21,48 | 23,75 |
| Grande | 5,75 | 5,40 |
| Muy grande | 1,33 | 3,38 |

**Tabla 8.5:** Resumen de la monitorización del procesador mediante vmstat.

Así pues, mediante el monitor vmstat se ha conseguido conocer, de forma aproximada, qué porcentaje de uso de procesador requiere cada una de las peticiones HTTP del modelo de carga. Esta medida está representada por el valor medio de la utilización del procesador durante la ejecución de la carga de prueba representada por el *benchmark*, prescindiendo de la sobrecarga en el servidor provocada por el monitor software.

Una vez conocidos los datos sobre el procesador, sería muy interesante saber cuál es el rendimiento del disco; por ejemplo, conocer la cantidad de tiempo que dedica a cada petición HTTP, así como cuántas veces es accedido en el servicio de cada una de éstas.

### Entrada/Salida

El monitor sar es una herramienta del sistema operativo Linux que proporciona datos estadísticos de los diferentes dispositivos del sistema. A continuación se puede observar una muestra de sus resultados:

```
$ sar -d 5      30

12:00:00        DEV         tps     sect/s
12:00:05        disk003     0.15    1.36
12:00:10        disk003     0.08    0.80
12:00:15        disk003     0.10    1.12
```

La columna etiquetada con DEV (*device*) indica el disco que se está analizando; en este caso se trata del dispositivo en que se encuentran los datos de interés para este caso de estudio. La columna tps indica el número de transferencias por segundo del dispositivo lógico; sin embargo, se debe tener en cuenta que varias transferencias lógicas pueden com- binarse en una sola petición de E/S física. El mismo programa, pero con otro modificador, sar –b, muestra en la misma columna tps el número de transferencias por segundo al disco físico. De esta manera se puede discernir, cuando las transacciones HTTP consultan ficheros del servidor, si hay actividad de disco o de cache de E/S.

Al igual que se hizo con el monitor vmstat, se ejecutará el monitor sar a la vez que se accede al archivo muestra de cada clase del modelo de carga, a través del *benchmark* ab, en una monitorización sin interferencias de otros usuarios. De este modo, se puede confirmar que los datos que proporciona sar son debidos a las peticiones generadas por el *benchmark* en un entorno controlado. Como en el caso del monitor vmstat, el periodo de activación y el número de muestras se seleccionan según la duración del test del *benchmark* ab.

Se hace necesario mencionar que, para que se produzcan realmente lecturas en el disco, la cache de E/S debe permanecer ociosa, puesto que en caso contrario no se realiza ninguna visita al disco, como demostraría el monitor sar. Para que la cache no actúe, o bien se desactiva durante los periodos de test, lo cual puede no ser fácil, o bien se generan ficheros nuevos idénticos para accederlos una vez, a fin de que la tasa de aciertos (*hit ratio*) de la cache sea nula durante un breve instante. Se elige esta segunda opción, se crea un fichero copia del fichero prototipo con otro nombre antes de la monitorización, se monitoriza su único acceso, se borra y se vuelve a crear. Este procedimiento se automatiza mediante un procedimiento y se aíslan los tiempos de monitorización de los de creación y borrado de ficheros intermedios que no serían propios del uso habitual del servidor.

En la Tabla 8.6 se presentan los resultados deducidos a partir de la ejecución del monitor sar durante la carga repetida de los archivos muestra elegidos. A partir de un número de muestras elevado, midiendo la productividad del dispositivo conocido como disk003 por el sistema operativo, dividiendo la media de transferencias por segundo del monitor sar por las transacciones por segundo del *benchmark* ab se obtiene el número medio de visitas al disco físico. Otros monitores de Linux, como atsar, proporcionan, entre otros datos, el número de lecturas por segundo, que serviría también para este cálculo aproximado.

| Clase | Razón de visita |
|---|---|
| Muy pequeño | 1,0 |
| Pequeño | 1,0 |
| Mediano | 1,0 |
| Grande | 6,0 |
| Muy grande | 12,0 |

**Tabla 8.6:** Razón de visita al disco deducida de la monitorización del servidor mediante sar.

Analizando la salida que proporciona el monitor sar, se comprueba que una vez que un archivo determinado es solicitado, la próxima vez que se consulta no hay actividad en el disco. Se deduce de ello, fenómeno esperable, que los archivos muestra de cada clase, una vez solicitados, son guardados en la memoria cache del disco durante un cierto periodo de tiempo. Ello nos lleva a intentar conocer el rendimiento no sólo del disco, del cual únicamente conocemos las visitas por transacción, sino también de la memoria cache de E/S. Por otra parte, se comprueba que el uso del procesador no es igual cuando los ficheros son solicitados por primera vez que cuando se supone que están almacenados en la cache de E/S. Sobre este punto se volverá más adelante.

En cuanto al rendimiento del disco, siempre se pueden usar los datos del fabricante sobre el tiempo de posicionamiento del brazo (*seek* ), la latencia rotacional y el tiempo de transferencia teórica, o bien utilizar otro monitor para determinar la velocidad de acceso. Debido a que también deseamos conocer el rendimiento de la memoria cache, podemos aprovechar la utilidad hdparm. En este caso, casi es obligado utilizar una utilidad del sistema que tendrá en cuenta la ocupación y la fragmentación de la información, y no sólo los cálculos basados en las características descritas por el fabricante.

La utilidad hdparm es una herramienta del sistema operativo Linux que proporciona todo tipo de datos relacionados con los discos duros y memorias cache de éstos que tenga el sistema. Ejecutando dicha utilidad varias veces (se recomiendan dos o tres con el sistema ocioso) se obtienen los resultados de productividad del disco duro y la cache, concretamente en nuestro caso de estudio, hdparm muestra:

```
$ hdparm -Tt /dev/hda
/dev/hda:

Timing buffer-cache reads: 128 MB in 1.63 seconds= 77.58 MB/s Timing Buffered
disk reads: 64 MB in 4.06 seconds= 9.98 MB/s
```

Como se observa, el monitor software nos proporciona la productividad (*throughput* ) de los discos y la cache expresada en MB/s. Ambos valores se obtienen realizando lecturas secuenciales de la información sin retraso añadido por el sistema de ficheros. En cuanto a la etiqueta del fabricante del disco, la información de la que disponemos en la siguiente:

```
Main Specifications
Product Description: Seagate U6 hard drive - 40 GB - ATA-100 Type: Hard drive -
standard
Form Factor: 3.5" x 1/3H internal
Dimensions (WxDxH): 10.2 cm x 14.7 cm x 2.6 cm Weight: 0.7
kg
Formatted Capacity: 40 GB
Interface Type: DMA/ATA-100 (Ultra) Data
Transfer Rate: 100 MBps Average Seek Time:
8.9 ms
Spindle Speed: 5400 rpm
Cylinders: 16383
Heads (Physical): 2 Sectors
per Track: 63
Recoverable Errors (Non-Recoverable Errors ): ( 1 per 1013 ) Cache / Buffer Size: 2
MB
Warranty: 1 year warranty
...
Performance
Drive Transfer Rate: 100 MBps (external) / 54.5 MBps (internal)
```

Seek Time: 8.9 ms (average) / 22 ms (max)
Track-to-Track Seek Time: 1.2 ms
Average Latency: 5.6 ms
Spindle Speed: 5400 rpm

Como se puede observar, el ratio de transferencia que aparece en las especificaciones del disco duro es diez veces más alto que el de la monitorización con hdparm, pero no tiene en cuenta que el bus puede reducir esa velocidad de transferencia. La diferencia entre las velocidades de transferencia interna y externa seguramente podría reducirse a través del cambio de configuración del dispositivo en el sistema operativo, como veremos más adelante. Sin embargo, se supondrá, de momento, que esta monitorización responde a la realidad actual y que el sistema no se va a reconfigurar.

En segundo lugar, aunque el resultado de hdparm teóricamente incluye el tiempo de posicionamiento y la latencia rotacional, no se tiene constancia de en qué condiciones se realizan las pruebas de monitorización, excepto que son lecturas secuenciales. Por tanto, podemos concluir que es una prueba bastante optimista. Para ser un poco más realista, se tomará la productividad de la monitorización de hdparm como la velocidad de transferencia del disco. Es decir, que para cada acceso al disco, se asignará su servicio medio como la suma del tiempo medio para el posicionamiento del brazo en la pista correspondiente más la latencia rotacional media para situarse al comienzo de la pista de lectura; ambos datos provenientes de la etiqueta de características del fabricante. A esta suma se le añadirá el tiempo medio de transferencia de la cantidad de información, que se calcula a partir de la monitorización. Por tanto,

- Si se supone que la probabilidad de que la información solicitada en una transacción está distribuida uniformemente entre todos los sectores que conforman las pistas del disco, el tiempo medio de latencia tarda la mitad de un giro del disco. Puesto que el disco gira a 5.400 rpm, tarda 11,111 ms en dar una vuelta, lo que supone 5,555 ms de latencia rotacional.

- El tiempo medio de posicionamiento del brazo no sigue una distribución uniforme debido a su tecnología electromecánica. También influyen en su cálculo la sectorización del disco y, sobre todo, la fragmentación de la información. Con tiempo de posiciona- miento del brazo mínimo, que es teóricamente nulo, el máximo de 22 ms y el tiempo de posicionamiento entre pistas de 1,2 ms, posiblemente podríamos construir una función parabólica del comportamiento de este retraso. Sin embargo, por cuestiones de simplicidad tomaremos el valor medio especificado de 8,9 ms, que corresponde a la etiqueta del fabricante del disco. Este valor es muy pesimista, una regla empírica que se suele utilizar es sólo considerar una tercera o incluso una cuarta parte de este valor. Sin embargo, tomaremos el valor medio de desplazamiento de brazo, debido a la gran cantidad de ficheros de menor tamaño que se solicitan durante la carga de prueba y que posiblemente estén almacenados de forma fragmentada.

- La velocidad media de transferencia del disco es de 9,98 MB/s y la de la cache de E/S de 77,58 MB/s, tomando los valores de la monitorización. Por lo que a partir del volumen de información a transferir, se puede calcular el tiempo que emplea en realizar la operación de E/S.

- Obsérvese que las velocidades de transferencia que suministra el fabricante son sensiblemente diferentes. Posteriormente, intentaremos optimizar la velocidad monitori- zada para acercarla a estos valores.

- Se puede considerar que sólo se realiza una visita por transacción a la cache de E/S puesto que tiene un tamaño de 2 MB, superior al tamaño medio de cualquier transacción HTTP. Este fenómeno se comprueba con el monitor sar -b.

## Red

En cuanto a la red de acceso al servidor y la correspondiente latencia debida a la conexión debemos hacer varias simplificaciones. La primera será suponer que los clientes de la Intra- net están conectados directamente a la red Ethernet de 100 Mbps. Esto es prácticamente real desde el conocimiento que se tiene de los usuarios internos, a través de la información del administrador del sistema y de las pruebas que se han realizado con el mismo monitor que se va utilizar para varias localizaciones. Y la segunda simplificación se refiere preci- samente a los usuarios de Internet que provienen del exterior. Puesto que es un número abismalmente inferior y acceden a la misma información que los usuarios de Intranet, se calcula el porcentaje de tráfico exterior a partir de los ficheros estadísticos de acceso. Tras la observación de los ficheros involucrados en la carga del sistema, detallada en secciones anteriores, se puede establecer que menos de un 0,0001 % de los accesos provienen del exterior, con lo que no se considerarán en este caso estudio.

## Análisis operacional

El objetivo del análisis operacional es llegar a establecer relaciones entre las variables que caracterizan la carga real o sintética y las que miden el comportamiento. Está formado por una serie de leyes que son las que relacionan las variables operacionales entre sí (véase el Capítulo 4).

Para construir el modelo de rendimiento del servidor web, se necesita conocer el tiempo de servicio de cada uno de los dispositivos que forman el sistema. Gracias a los monitores software hemos conocido algunos de esos valores; sin embargo, no todos los necesarios. Para ello se usarán las leyes operacionales y los datos obtenidos por las diferentes técnicas que se han expuesto hasta ahora.

## Entrada/Salida

La monitorización del disco y la ley de la demanda permiten parametrizar el comporta- miento del disco duro:

- El número de visitas por transacción fue obtenido a través de la monitorización del disco con sar (véase la Tabla 8.6). Por ejemplo, las transacciones de clase "grande" realizan una media de seis accesos (visitas) al disco duro.
- El tiempo de servicio medio por transacción se obtiene de la suma de los valores medios de posicionamiento, latencia y transferencia. Todos los accesos sufren en pro- medio un posicionamiento de 8,9 ms y una latencia rotacional de 5,555 ms. En el caso de las transacciones de clase "grande", por ejemplo, hay que añadir la sexta parte de 15,415 ms que corresponde al tiempo de transferencia de los 153.845 bytes en seis accesos a 9,98 MB/s.
- La demanda de servicio se puede conocer multiplicando el número medio de accesos por el tiempo de servicio que solicitan. Así, por ejemplo, las transacciones "grandes" realizan seis visitas al disco en las que solicitan un servicio medio de 17,204 ms, con lo que la demanda media por transacción de clase "grande" es de 103,204 ms.

Análogamente para todos los tipos de transacciones HTTP muestra, se obtienen los datos operacionales del disco duro que se muestran en la Tabla 8.7. El tamaño se expresa en bytes, y tanto el tiempo de servicio como la demanda de servicio vienen indicados en milisegundos.

| Clase | Visitas | Tamaño | Servicio | Demanda |
|---|---|---|---|---|
| Muy pequeño | 1,0 | 114 | 14,466 | 14,466 |
| Pequeño | 1,0 | 617 | 14,516 | 14,516 |
| Mediano | 1,0 | 14.486 | 15,906 | 15,906 |
| Grande | 6,0 | 153.845 | 17,024 | 103,224 |
| Muy grande | 12,0 | 1.319.793 | 25,475 | 305,703 |

**Tabla 8.7:** Demanda y servicio medio por clase de transacción del disco duro.

Un fenómeno observable en este primer análisis es que, tal como se esperaba, la demanda media de las transacciones de mayor tamaño se ve muy incrementada debido al número de accesos al disco necesarios para completarlas. Sin embargo, puesto que la cache tiene un tamaño de 2 MB (véanse los datos del disco aportados por el fabricante), vamos suponer que caben todas las transacciones en la memoria. Por lo tanto, la tasa de visitas a la cache por transacción es única y el servicio medio por clase equivale a la demanda media.

Lógicamente, no consideraremos el tiempo de posicionamiento medio ni el tiempo rotacional en el cómputo del servicio medio de la cache, tan sólo la transferencia media de 77,58 MB/s. En la Tabla 8.8 se presentan los datos operacionales de la cache de E/S.

| Clase | Visitas | Tamaño | Servicio | Demanda |
|---|---|---|---|---|
| Muy pequeño | 1,0 | 114 | 0,0015 | 0,0015 |
| Pequeño | 1,0 | 617 | 0,0079 | 0,0079 |
| Mediano | 1,0 | 14.486 | 0,1867 | 0,1867 |
| Grande | 1,0 | 153.845 | 1,9830 | 1,9830 |
| Muy grande | 1,0 | 1.319.793 | 17,012 | 17,012 |

**Tabla 8.8:** Demanda y servicio medio por clase de transacción de la memoria cache de E/S.

La demanda media y el servicio medio de la cache son equivalentes, puesto que sólo se realiza una visita por acceso. Este hecho junto con la gran velocidad de transferencia hacen que las diferencias de rendimiento entre el disco y la cache sean muy significativas.

## Procesador

Podemos continuar con los datos del procesador, tomando los resultados de la ejecución del *benchmark* ab y el monitor vmstat. Recordemos que la ejecución del *benchmark* produce la frecuencia de saturación del servidor, es decir, que se utiliza aproximadamente un 100 % durante la descarga de las transacciones muestra. Por tanto, la productividad asociada a la frecuencia de saturación es la máxima y corresponde a los valores de la Tabla 8.3, cuando se realizan pruebas con un 100 % de transacciones de un único tipo.

Por otro lado, gracias a la ley de la utilización se pueden calcular las demandas medias de procesador por cada clase de transacción. Por ejemplo, el *benchmark* produce 4.141 transacciones de tipo “muy pequeña”, y el servidor responde a 667,58 tps. Si consume una media de 54,66 % de procesador en modo usuario durante la experimentación, la demanda correspondiente es el cociente de la utilización y la productividad, es decir, 0,6490 ms. El resto de los cálculos se realizan análogamente y se reflejan en la Tabla 8.9 (la productividad máxima se expresa en transacciones por segundo).

Los datos confirman que el manejo de las transacciones de menor tamaño consumen casi la misma cantidad de proceso. Sin embargo, a medida que se necesita realizar mayor transferencia de información, la demanda de procesador se hace mucho mayor.

Aunque se desconoce el número medio de visitas que se realizan al procesador (*threads*), por cada transacción HTTP que recibe el servidor web, se puede aproximar a partir de la E/S. Se puede decir que el número medio de visitas al procesador es superior en una unidad a las visitas al disco, tal como se considera en el modelo del servidor central. Este número medio de visitas es equivalente al número de *threads*. Así, mediante la ley de la demanda, se calculan los tiempos medios de servicio por *thread*. Por ejemplo, si la demanda en modo usuario de las transacciones de la clase de tipo "pequeño" es de 0,6421 ms y cada transacción realiza dos visitas al procesador, el tiempo de servicio medio de cada acceso es de la mitad. Esta aproximación permite calcular más datos operacionales de este dispositivo, que se exponen en la Tabla 8.10 (los tiempos están expresados en milisegundos).

| Clase | Uso procesador usuario | sistema | Prod. máx. | Demanda usuario | sistema | global |
|---|---|---|---|---|---|---|
| Muy pequeño | 54,66 | 43,33 | 667,580 | 0,8187 | 0,6490 | 1,4678 |
| Pequeño | 49,38 | 43,72 | 566,559 | 0,8715 | 0,7727 | 1,6443 |
| Mediano | 21,48 | 23,75 | 456,253 | 0,4707 | 0,5206 | 0,9913 |
| Grande | 5,75 | 5,40 | 64,327 | 0,8939 | 0,8503 | 1,7442 |
| Muy grande | 1,33 | 3,38 | 8,004 | 1,6616 | 4,2228 | 5,8845 |

**Tabla 8.9:** Demanda de usuario y sistema por clase de transacción en el procesador.

| Clase | Demanda | Visitas | Servicio |
|---|---|---|---|
| Muy pequeño | 1,4678 | 2 | 0,7339 |
| Pequeño | 1,6443 | 2 | 0,8221 |
| Mediano | 0,9913 | 2 | 0,4956 |
| Grande | 1,7442 | 7 | 0,2491 |
| Muy grande | 5,8845 | 13 | 0,4526 |

**Tabla 8.10:** Demanda y servicio medio por clase de transacción en el procesador.

Llegados a este punto, ya se han obtenido los valores suficientes para proceder a realizar el modelado de rendimiento del servidor web, aplicando un modelo de redes de colas. A partir de este modelo se estimarán otros parámetros de rendimiento.

### 8.3.3. Construcción de un modelo de rendimiento

En los modelos orientados a colas se suele utilizar un conjunto de estaciones de servicio relacionadas entre sí, que constituyen lo que se denomina red de colas. Así, en el modelo que nos ocupa, las estaciones de servicio se corresponden con los recursos del hardware, es decir, el procesador, el disco duro, etc., mientras que las colas representan las facili- dades del sistema hardware y/o software del sistema para gestionar la espera por dichos dispositivos. Los clientes de esas colas son los usuarios de esos recursos, en este caso las transacciones HTTP que fluyen en el sistema. Como el número de clientes no se puede

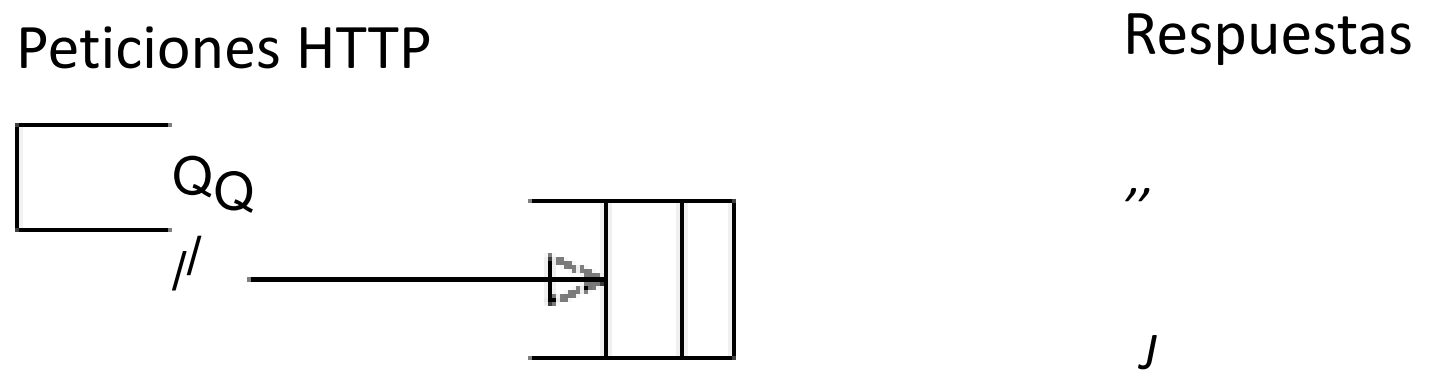

**Figura 8.3:** Modelo de rendimiento de estación única del servidor web.

considerar constante, se opta por un modelo abierto y orientado al servidor.

Para llegar a un modelo de rendimiento del servidor web que se considere suficiente para el estudio de rendimiento, se va a pasar por dos etapas de refinamiento. En cada etapa se irá especificando un poco más el modelo, es decir, añadiendo elementos y utilizando parámetros más específicos, con el objetivo final de obtener una representación abstracta pero simplificada de la realidad.

### Modelo simple

Antes de enfrentarnos a la tarea de modelar los principales dispositivos que componen el servidor web, se puede realizar un modelo de “caja negra”, opaco en cuanto a la distribución de la carga de trabajo interna. En este modelo, tan simple, consideramos el servidor como un todo, que recibe peticiones HTTP, las procesa, y expulsa las transacciones HTTP de respuesta. Dicho modelo representa todo el comportamiento del sistema. El modelo enton- ces es sólo una estación con un tiempo de servicio distinto para cada tipo de petición (véase la Figura 8.3). Este modelo simple es un resumen del comportamiento del sistema, pero aun así permite conocer la carga que soporta y el servicio global que obtiene esa carga.

La carga del sistema se representará mediante una estación fuente o manantial, que genera las llegadas de transacciones al servidor según los porcentajes de carga establecidos. Dicha estación fuente necesita un parámetro que le indique con qué periodo debe generar un cliente, o lo que es lo mismo, cada cuánto tiempo tiene que generar una transacción HTTP de una cierta clase. Este parámetro se obtiene de la caracterización de la carga. Por ejemplo, la clase “muy pequeña” representa el 44,64 % de la carga, lo que supone que de las 55.910 transacciones totales llegarían algo más de 24.958 de esta clase al servidor. Puesto que el periodo de observación $T$ es una semana laborable de 144.000 segundos, el periodo medio entre llegadas de transacciones de esta clase es de 5,7696 segundos. Este resultado no es más que el inverso de la productividad media calculada, ya que se supone el equilibrio entre el flujo de entrada y de salida del sistema. En la Tabla 8.11 se representan el número de peticiones calculado a partir de la distribución de porcentajes del modelo de carga, la frecuencia de llegada en transacciones por minuto (tpm) y el periodo medio entre llegadas consecutivas para cada una de las clases de transacciones consideradas durante el periodo de medida $T$.

El modelo de la Figura 8.3 sólo es una representación reducida a través del formalismo de las redes de colas, pero el interés residirá en su evaluación mediante análisis opera-cional. En consecuencia, todos los tiempos medios de servicio, demandas, utilizaciones y demás variables del modelo se han obtenido mediante cálculo operacional basándose en la monitorización durante la ejecución del *benchmark*. Finalmente, se ha supuesto que la sobrecarga que pudieran producir los monitores software utilizados forma parte del modelo de rendimiento del sistema.

| Clase | Número de peticiones | Porcentaje peticiones | Frecuencia de llegadas (tpm) | Tiempo entre llegadas (s) |
|---|---|---|---|---|
| Muy pequeño | 24.958,08 | 44,64 | 10,3992 | 5,7696 |
| Pequeño | 18.036,56 | 32,26 | 7,5152 | 7,9838 |
| Mediano | 11.567,78 | 20,69 | 4,8199 | 12,4484 |
| Grande | 1.263,56 | 2,26 | 0,5264 | 113,9817 |
| Muy grande | 83,86 | 0,15 | 0,0349 | 1.717,1977 |
| Total | 55.910 | 100 | 23,2958 | 2,5755 |

**Tabla 8.11:** Frecuencia y periodo entre llegadas de transacciones durante el periodo de medida $T$.

Para poder conocer la demanda media por transacción HTTP del modelo simple, se tiene que calcular anteriormente el valor de la utilización media del servidor web, cuando soporta la carga del periodo de medida $T$. Las consideraciones a tener en cuenta son la siguientes:

- Mediante las pruebas de *benchmarking* se ha podido establecer la productividad máxima del servidor (véase la Tabla 8.4) recibiendo transacciones de las cinco clases por separado. En estas cargas de prueba, el *benchmark* lleva, al menos teóricamente, al servidor hasta el punto de saturación de un sistema transaccional. Por tanto, la productividad máxima por clase reflejada en esta tabla es la frecuencia de saturación por clase de transacción del servidor, es decir, aquella frecuencia que hace que el servidor tenga una utilización del 100 %.

- Debido a la ausencia de concurrencia en las pruebas de *benchmarking*, cada vez que llega la respuesta de una petición HTTP, se genera inmediatamente otra, con lo que el número de peticiones activas dirigidas al servidor web es como máximo de una. De hecho, para las pruebas, incluso se podría considerar un modelo de colas cerrado, *batch* puro, es decir, con un tiempo de reflexión nulo. Por lo tanto, la determinación de la demanda máxima a través del *benchmarking* sería equivalente bien consideremos un sistema abierto o cerrado.

- En las pruebas de *benchmarking*, debido a la repetición de la misma transacción muestra, la cache de E/S tiene almacenados los datos a transferir, por lo que se puede considerar que el disco no tiene actividad durante las pruebas.

- Aplicando la ley de la utilización se conocen los valores del servicio medio obtenido por las transacciones en la prueba. Por ejemplo, la frecuencia de saturación del ser- vidor mientras es sometido a 4.141 transacciones “muy pequeñas” es de 667,58 tps. De este modo, puesto que la utilización es máxima, la demanda es la inversa de la frecuencia de saturación. Se puede concluir entonces que, cuando el servidor web es utilizado un 100 % por las transacciones de clase “muy pequeña”, la demanda máxima por transacción es de 1,4979 ms en promedio.

- Debido a que la distribución de la carga caracterizada por clases suma el 100 % de la carga con el reparto según la Tabla 8.12, las frecuencias máximas se ven reducidas por los porcentajes correspondientes. Como puede verse en esta tabla, a medida que las transacciones aumentan de tamaño, la demanda máxima que se efectúa al servidor, en consonancia con la suma de las demandas parciales del procesador (Tabla 8.10) y de la cache de E/S (Tabla 8.8), también aumenta.

| Clase | Frecuencia saturación | Porcentaje peticiones | Frec. saturac. por clase (tps) | Demanda máx. (ms) |
|---|---|---|---|---|
| Muy pequeño | 667,580 | 44,64 | 298,007 | 1,4979 |
| Pequeño | 566,559 | 32,26 | 182,772 | 1,7650 |
| Mediano | 456,253 | 20,69 | 94,398 | 2,1917 |
| Grande | 64,327 | 2,26 | 1,672 | 15,5455 |
| Muy grande | 8,004 | 0,15 | 0,012 | 124,9375 |

**Tabla 8.12:** Frecuencia de saturación y demanda media por clase de transacción.

Para conocer el rendimiento del servidor web, a través de la carga caracterizada durante el periodo de observación, se debe calcular primero la utilización del servidor, es decir, la utilización de la estación de servicio en el modelo simple por parte de los clientes recibidos. En este sentido consideraremos:

- Puesto que se conoce la demanda media por clase en la frecuencia de saturación, se puede considerar que éste es todo el servicio que otorga el servidor a los clientes cuando se utiliza un 100 %.

- Debido a que la carga caracterizada considera cinco tipos diferentes de transacciones, se puede ponderar la utilización del servidor en la frecuencia máxima por clase, tal como se ha realizado en la Tabla 8.12.

- Gracias a la ley de la utilización, se puede calcular la correspondiente al servidor cuando las frecuencias de carga por clase son inferiores a la de la saturación, tal como las que se producen durante el periodo de medida. Por ejemplo, si llegan al servidor transacciones de clase “mediana” a razón de 4,8199 tpm o 0,8033 tps durante el periodo de observación *T* demandando en media 0,4534 ms, la utilización que sufre el servidor web debido esa carga es del 0,0036 %.

| Clase | Demanda máx. (clase) | Frecuencia llegadas (tpm) | Utilización del servidor web ( %) |
|---|---|---|---|
| Muy pequeño | 1,4979 | 10,3992 | 0,026 |
| Pequeño | 1,7650 | 7,5152 | 0,022 |
| Mediano | 2,1917 | 4,8199 | 0,017 |
| Grande | 15,5455 | 0,5264 | 0,006 |
| Muy grande | 124,9375 | 0,0349 | 0,007 |

**Tabla 8.13:** Utilización media del servidor web durante el periodo de observación.

Como se puede observar a partir de los datos reflejados en la Tabla 8.13, la utilización total del servidor web durante el periodo de observación *T* es inferior al 0,08 %. Obsérvese que llegan unas 23 tpm y el servidor podría soportar más de 576 tps, o de forma equiva- lente, más de 34.560 tpm si sumamos las frecuencias de saturación ponderadas según la distribución de clases de esta tabla.

Si se quiere conocer el tiempo de respuesta del servidor web, se puede estimar el rendimiento medio a través de la ley de Little. Para el modelo simple, todo el servidor constituye una sola estación en un modelo abierto. Aunque se conoce el número máximo de usuarios, no se sabe cuál es el número de transacciones generadas por sesión. Sin embargo, el ser- vicio medio de un cliente en la estación es equivalente a su demanda media, puesto que consideramos una única visita por transacción. En la Tabla 8.14 se presenta un resumen de las demandas de servicio máximas, las utilizaciones del servidor por clase de transacción y el tiempo de respuesta del servidor (expresado en milisegundos) calculado según la ley de Little.

### Modelo final

Una vez que conocemos el rendimiento del servidor web, a través del modelado de cola única, se pueden modelar los dispositivos que intervienen durante el *benchmarking* y la monitorización conjunta. Este modelo considera los dispositivos principales que conforman el servidor. Éste era el principal obstáculo al que se enfrentaba el caso de estudio, puesto que en base a la información de la bitácora del servidor no se podía inferir el rendimiento de sus dispositivos a partir de la carga transaccional recibida. Este modelo presenta una serie de cambios con respecto al anterior. A continuación se detallan la naturaleza de dichos cambios y la causa de las modificaciones (véase la representación gráfica del modelo en la Figura 8.4).

| Clase | Porcentaje peticiones | Utilización servidor ( %) | Demanda (ms) | Tiempo de respuesta (ms) |
|---|---|---|---|---|
| Muy pequeño | 44,64 | 0,026 | 1,4979 | 1,5378 |
| Pequeño | 32,26 | 0,022 | 1,7650 | 1,8047 |
| Mediano | 20,69 | 0,017 | 2,1917 | 2,2296 |
| Grande | 2,26 | 0,006 | 15,5455 | 15,6393 |
| Muy grande | 0,15 | 0,007 | 124,9375 | 125,8182 |
| Servidor | 100,00 | 0,078 | 2,2302 | 2,4188 |

**Tabla 8.14:** Demanda, utilización y tiempo de respuesta medio del modelo simple.

Las diferentes estaciones, red, procesador y E/S tienen unos tiempos de servicio cal- culados en apartados anteriores. Así, las estaciones generadoras de clientes o fuentes, una para cada clase de petición, tienen un tiempo de servicio inverso a la frecuencia con la que deben generar las transacciones HTTP. Estas estaciones fuente encaminan a los clientes hacia la estación Ethernet representando que las transacciones HTTP se introducen en la Intranet del servidor web, tal como se estableció en el modelo simple (véase la Tabla 8.12).

Los parámetros de rendimiento del procesador y de la cache fueron calculados en el apartado de recolección de datos y parametrización, sobre todo a partir del análisis ope- racional de los datos de monitorización. No obstante, falta aún conocer cómo modelar el comportamiento de la red Ethernet y el disco duro.

## Red

Para expresar el tiempo que representa la latencia en la conectividad de la Intranet, se ha utilizado un procedimiento de cómputo asimétrico que se ejecuta dependiendo de la clase de cliente, diferenciando peticiones y respuestas de la solicitud HTTP. Hay que tener en cuenta que, a un mensaje de red que transporte una transacción HTTP, se le van añadiendo una serie de bytes de cabecera y de control para ser gestionado por los niveles inferiores del protocolo TCP/IP. Estos bytes adicionales producirán cierta sobrecarga en la transmisión final del mensaje. Así, debemos considerar diferentes fuentes del retraso adicional:

- Sobrecarga producida por el protocolo TCP: 20 bytes.
- Sobrecarga producida por el protocolo IP: 20 bytes.
- Sobrecarga debida a la latencia de la red sobre la que se transmite. En nuestro caso es una Ethernet; para esta red, el tamaño máximo de la unidad de transmisión es de 1.500 bytes. Por lo tanto, cuando los paquetes exceden de este tamaño se fragmentan de forma adecuada para que puedan circular por la Ethernet. Se deberá añadir a cada paquete transmitido por la Ethernet, una sobrecarga de 18 bytes.

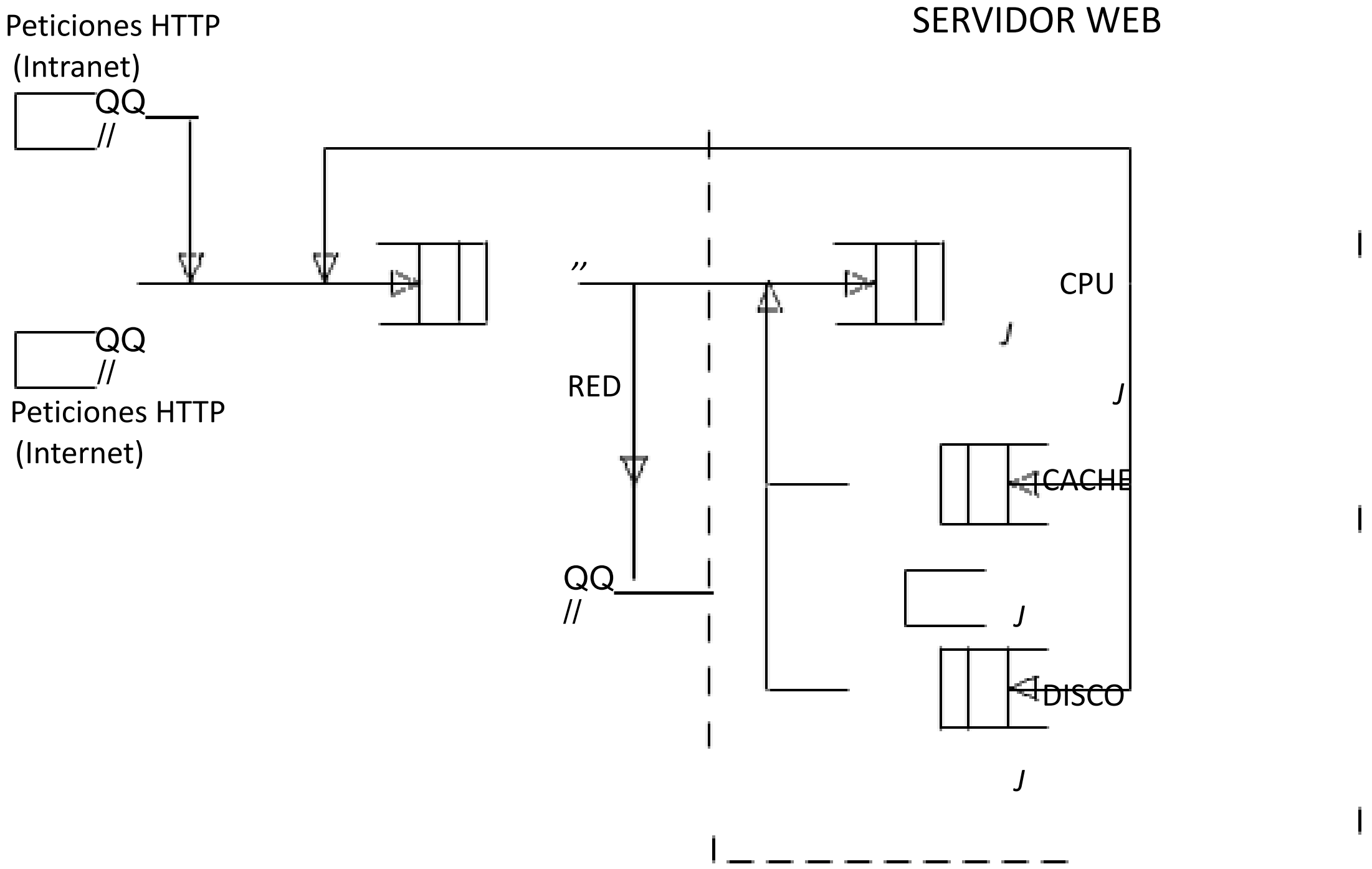

**Figura 8.4:** Modelo de rendimiento final del servidor web.

Todas estas sobrecargas se han expresado en bytes, pero lógicamente su transmisión produce el correspondiente retraso debido al incremento de información de control de las transacciones HTTP. Evidentemente, otro dato importante a precisar es el ancho de banda de la red Fast Ethernet, que en el caso que nos ocupa es de 100 Mbps.

Por tanto, ya que los modelos de carga están caracterizados por distintas clases de clientes, hay que distinguir que los clientes del modelo simple incluían las peticiones de servicio y sus correspondientes respuestas. Así pues, si el cliente es de tipo petición, la trama HTTP tendrá una longitud media de 170 bytes más 58 bytes de sobrecarga, que es un tamaño común en el protocolo HTTP. Por el contrario, si el cliente es de tipo respuesta, la longitud de la trama será igual al tamaño en bytes del fichero solicitado más 40 bytes de sobrecarga. Recordemos que si el tamaño de la respuesta excede los 1.500 bytes, se divide en subtramas y a cada una se le debe añadir la sobrecarga de red, que es de 18 bytes.

Conocidos los tamaños medios de las peticiones y sus respuestas para cada clase de cliente y el ancho de banda de la red, es decir, su velocidad de transferencia, podemos calcular el tiempo medio de demanda de la estación Ethernet.

En la Tabla 8.15 están agrupados los distintos componentes de latencia de la red, debido a la transmisión. Esto evidentemente es una simplificación que no incluye, por ejemplo, las colisiones CSMA/CD en las distintas subredes, ni los retardos debidos a la conmutación de las mismas. Aunque ambos fenómenos influyen en la eficiencia de la comunicación, se de- bería conocer con detalle la arquitectura de la red para establecerlos de forma aproximada. Aunque no son parte de la demanda de red que realizan las peticiones HTTP en el servidor, su fragmentación en tramas provoca colisiones que podrían modelarse del siguiente modo.

| Clase | Bytes a transferir |
|---|---|
| Muy pequeño | 228 + 114 + 40 = 382 |
| Pequeño | 228 + 617 + 40 = 885 |
| Mediano | 228 + 10 × 18 + 14.486 + 40 = 14.934 |
| Grande | 228 + 103 × 18 + 153.845 + 40 = 155.967 |
| Muy grande | 228 + 880 × 18 + 1.319.793 + 40 = 1.335.901 |

| Clase | Porcentaje | Visitas | Demanda transferencia |
|---|---|---|---|
| Muy pequeño | 44,64 | 1 + 1 | 0,0305 |
| Pequeño | 32,26 | 1 + 1 | 0,0708 |
| Mediano | 20,69 | 1 + 10 | 1,1947 |
| Grande | 2,26 | 1 + 103 | 12,4773 |
| Muy grande | 0,15 | 1 + 880 | 106,8720 |

**Tabla 8.15:** Demanda media de red por transacción.

Se podría incrementar el retraso en la comunicación en un factor que dependa del número de tramas en la red. Así, a la demanda media de transferencia por trama se le puede añadir la latencia CSMA/CD:

$$D_{\text{CSMA/CD}} = \text{Número de tramas} \times S \times C(k)$$

donde $S$ es el tiempo de detección de la colisión (*slot time*), que en una red de tipo Fast Ethernet corresponde a $5{,}12\ \mu$s, $C(k)$ es el número medio de colisiones por petición que corresponde a la expresión:

$$C(k) = \frac{1 - A(k)}{A(k)}$$

donde $A(k)$ es la probabilidad de una transmisión sin colisiones:

$$A(k) = \left(1 - \frac{1}{k}\right)^{k-1}$$

y finalmente $k$ es el número de tramas que utilizan la red. En el caso de que la trama sea única el retardo adicional es $S \times C(1)$, que anula el término. $C(k)$ representa el número medio de colisiones por petición cuando hay $k$ tramas que se está intentando transmi- tir simultáneamente. En este caso, se producen múltiples colisiones hasta que una de las estaciones consigue transmitir su trama de forma satisfactoria. Por ejemplo, las transac- ciones “grandes” utilizan una trama de petición de 228 bytes, cuya respuesta se fracciona en 103 tramas, debido al tamaño del fichero a transferir, ocupando el canal un mínimo de 12,4773 ms. La petición de ida es una única trama, con menor probabilidad de colisionar que su respuesta de 103 tramas.

Sin embargo, el retraso por colisión depende también del número de estaciones que intentasen acceder a la red. No se conoce *a priori* el número de estaciones que desea transmitir simultáneamente, aunque se conozca el número de usuarios. Obsérvese que en el caso hipotético de que todo el departamento de ingeniería intentase enviar una trama simultáneamente $C(k)$, sería inferior a 1 ms y, en el peor de los casos, si todos los empleados transmitieran simultáneamente una trama sería del orden de 10 ms. Pero eso significaría que todas las estaciones comparten el medio sin concentradores ni subdivisiones de red, que precisamente es lo que no ocurre en este caso en el que el servidor web se encuentra en la subred de ingeniería. Puesto que no se tiene información acerca de la congestión de la subred en la que se encuentra el servidor ni de su arquitectura, en el modelo no se pueden aventurar estas cantidades, que dependen de la configuración de los concentradores y la tecnología de red empleada. Estos retrasos por colisión se podrían modelar con mayor información sobre la red corporativa, pero no son imputables a la demanda de servicio de las peticiones en el servidor. Por otra parte, dado que se comprobó que la carga de usuarios de Internet era similar a la de la Intranet, y además su porcentaje era irrelevante sobre el total, este tipo de tráfico no se ha considerado en el modelo.

### Disco duro

Sin lugar a dudas, la modificación más importante a tener en cuenta es que se ha añadido la estación de servicio que modela el disco duro. Para su parametrización se tiene que investigar la tasa de acierto de la cache, es decir, con qué probabilidad el servidor accede a la cache de E/S o directamente al disco físico, para acceder al fichero solicitado. En la mayoría de los subsistemas de discos medianamente eficientes, la probabilidad de acceso a la cache ronda entre un 0,8 (80 %) y un 0,9 (90 %). En el proceso de carga con el *benchmark*, es del 100 % ya que el proceso de experimentación se basa en la repetición de las transacciones.

En la Tabla 8.16 se recuerdan los valores de demanda de la cache de E/S y el disco duro cuando la tasa de acierto es el 100 % o el 0 %, es decir, se accede a un dispositivo o al otro. Todos los tiempos se expresan en milisegundos, y el tamaño en bytes.

En la Tabla 8.17 se muestran las demandas de cada uno de los dispositivos de E/S del sistema, variando la tasa de acierto de la memoria cache. Los resultados que se presentan se han obtenido a partir de los valores de calculados mediante análisis operacional. Por ejemplo, la demanda de disco y de cache de cada una de las columnas de la Tabla 8.16 se obtiene de la suma ponderada de las demandas por clase de transacción. Posteriormente, se multiplica esa demanda ponderada por la probabilidad de la tasa de acierto, en el caso de la memoria cache, y por el suceso opuesto, en el caso del disco duro.

| Clase | Visitas | Porcentaje | Tamaño | Demanda cache | Demanda disco |
|---|---|---|---|---|---|
| Muy pequeño | 1 | 44,64 | 114 | 0,0015 | 14,466 |
| Pequeño | 1 | 32,26 | 617 | 0,0079 | 14,516 |
| Mediano | 1 | 20,69 | 14.486 | 0,1867 | 15,906 |
| Grande | 6 | 2,26 | 153.845 | 1,9830 | 103,224 |
| Muy grande | 12 | 0,15 | 1.319.793 | 17,0120 | 305,703 |
| Sistema E/S | 1,13 | 100,00 | 8.703 | 0,1121 | 17,222 |

**Tabla 8.16:** Demanda media comparada de la E/S del servidor web.

| Clase | Demanda cache E/S (*TA* = 1,0) | (*TA* = 0,9) | (*TA* = 0,8) | Demanda disco |
|---|---|---|---|---|
| Muy pequeño | 0,0015 | 1,4478 | 2,8943 | 14,466 |
| Pequeño | 0,0079 | 1,4587 | 2,9095 | 14,516 |
| Mediano | 0,1867 | 1,7586 | 3,3305 | 15,906 |
| Grande | 1,9830 | 12,1071 | 22,2312 | 103,224 |
| Muy grande | 17,0120 | 45,8811 | 74,7502 | 305,703 |

**Tabla 8.17:** Demandas de la cache de E/S en función de la tasa de acierto (*TA*) en el servidor web.

En las tablas anteriores se puede apreciar que la variación de la tasa de acierto provoca grandes modificaciones a los tiempos de demanda de los dispositivos (la demanda se expresa en milisegundos). Al disminuir la tasa de acierto, se aumenta la probabilidad de acceso al disco duro, con lo que la demanda de E/S crece rápidamente con respecto al modelo simple. Así, por ejemplo, cuando el 90 % de los accesos se realizan en la cache, la demanda global de E/S es más de once veces superior a la misma demanda del modelo sin disco del apartado anterior. Peor es el caso si la tasa de acierto baja al 80 %, donde la duración total de las transferencias de E/S para las transacciones más grandes se dispara. El efecto de esta variación es el aumento traumático en la demanda media del servidor web, como se puede apreciar en la Tabla 8.18, donde se asume una tasa de acierto del 80 % para la memoria cache de E/S.

Puesto que la demanda por clase se ha visto incrementada con respecto a una tasa de acierto del 100 % durante el proceso de *benchmarking*, lógicamente la utilización del servidor se verá incrementada también y con ello la demanda media. En la Tabla 8.19 se han dispuesto los incrementos de demanda media por clase debido al descenso de la tasa de acierto considerada y el tiempo de respuesta correspondiente aproximado a través de la ley de Little.

| Clase | Demanda procesador | Demanda cache E/S | Demanda de red | Demanda acumulada |
|---|---|---|---|---|
| Muy pequeño | 1,4678 | 2,8943 | 0,0305 | 4,3924 |
| Pequeño | 1,6443 | 2,9095 | 0,0708 | 4,6246 |
| Mediano | 0,9913 | 3,3305 | 1,1947 | 5,5165 |
| Grande | 1,7442 | 22,2312 | 12,4773 | 36,4527 |
| Muy grande | 5,8845 | 74,7502 | 106,8720 | 187,5067 |
| Servidor | 1,4390 | 3,5336 | 0,7282 | 5,7008 |

**Tabla 8.18:** Demanda media por dispositivo y clase de transacción para el modelo final.

| Clase | Frecuencia llegadas (tpm) | Demanda acumulada | Utilización servidor ( %) | Tiempo de respuesta |
|---|---|---|---|---|
| Muy pequeño | 10,3992 | 4,3924 | 0,076 | 4,7536 |
| Pequeño | 7,5152 | 4,6246 | 0,058 | 4,9093 |
| Mediano | 4,8199 | 5,5165 | 0,044 | 5,8611 |
| Grande | 0,5264 | 36,4527 | 0,032 | 37,6577 |
| Muy grande | 0,0349 | 187,5067 | 0,011 | 189,5922 |
| Servidor | 23,295 | 5,7008 | 0,221 | 7,3181 |

**Tabla 8.19:** Demanda, utilización y tiempo de respuesta del modelo final.

Como se puede ver en la Tabla 8.19, las transacciones que demandan ficheros mayores, aunque incrementan la demanda media, no son las que más contribuyen a la utilización global del servidor. Sin embargo, se aprecia un descenso notable de la frecuencia máxima por clase de transacción que podría soportar el servidor web, debido principalmente al disco duro. Las frecuencias máximas del modelo final se obtienen gracias a la aplicación de la ley de la demanda y los porcentajes de cada clase de transacción HTTP (véase la Tabla 8.20).

Cuando se ha propuesto el modelo final y se han obtenido los principales parámetros de rendimiento del servidor web, a través de la caracterización de la carga, el *benchmarking*, la monitorización del sistema y la aplicación del análisis operacional, se va a proceder a la detección y propuesta de eliminación de los cuellos de botella del sistema. Este análisis de los cuellos de botella permitirá señalar mejoras posibles en el rendimiento para posteriormente

| Clase | Modelo simple | | Modelo final | |
|---|---|---|---|---|
| | Demanda | Frec. saturación | Demanda | Frec. saturación |
| Muy pequeño | 1,4979 | 298,007 | 4,3924 | 101,630 |
| Pequeño | 1,7650 | 182,772 | 4,6246 | 69,757 |
| Mediano | 2,1917 | 94,398 | 5,5165 | 37,505 |
| Grande | 15,5455 | 1,672 | 36,4527 | 0,620 |
| Muy grande | 124,9375 | 0,012 | 187,5067 | 0,008 |
| Servidor | 2,2302 | 576,861 | 5,7008 | 209,520 |

**Tabla 8.20:** Comparación de la frecuencia de saturación (tps) y demanda medias de los modelos simple y final.

calcular la capacidad futura del sistema.

### 8.3.4. Mejora del rendimiento

A partir del modelo final, más detallado, se puede ajustar el rendimiento del sistema, por ejemplo mediante la detección y eliminación del cuello de botella del servidor web. Se entiende por cuello de botella el dispositivo del sistema que recibe una mayor demanda por trabajo y, por tanto, el que se saturará primero por sobrecarga (véase el Capítulo 5). Para determinar los dispositivos susceptibles de ser cuellos de botella en el servidor web, se deben calcular las demandas globales y por clase de cliente. El cuello de botella será aquel dispositivo que tenga la demanda más elevada para la carga de trabajo característica del momento, y que llegaría primero al 100 % de utilización, con un incremento de la carga del mismo perfil.

#### Entrada/Salida

Si se examina de nuevo la Tabla 8.18 se puede observar que una vez que se ha incluido el disco en el modelo de rendimiento, es la E/S, cache más disco, el dispositivo que presenta una demanda mayor, tanto para cada clase de transacción como globalmente. Por lo que podemos concluir que la E/S es el cuello de botella del servidor web. Con una tasa de acierto del 80 %, las frecuencias de saturación globales por clase de transacción se reducen en comparación con las del modelo cache, como se presentó en la Tabla 8.20. Ambos con- juntos de frecuencias de saturación suponen una utilización del 100 % del servidor web. Así, mientras que la demanda ponderada de E/S para el modelo intermedio era de 2,3256 ms, en el modelo final con una tasa de acierto del 80 % subía a 5,7471 ms. Invirtiendo es- tos valores y aplicando la segunda ley de la utilización se determinan las frecuencias de saturación medias. De este modo, en el modelo intermedio, que virtualmente no emplea el disco, el servidor podría soportar algo más de 576 tps. Sin embargo, con una tasa de acierto del 80 %, la intervención del disco duro en la transferencia de información hace que el servidor soporte menos de 209 tps. Afortunadamente, estas frecuencias distan mucho de las que recibe actualmente el servidor, pero podemos intentar mejorar el rendimiento del sistema a través de la mejora de su cuello de botella.

| Clase | Modelo simple Frec. sat. (tps) | Modelo final Frec. sat. (tps) | Frecuencia llegadas (tpm) |
|---|---|---|---|
| Muy pequeño | 298,007 | 101,630 | 10,399 |
| Pequeño | 182,772 | 69,757 | 7,515 |
| Mediano | 94,398 | 36,963 | 4,812 |
| Grande | 1,672 | 0,605 | 0,526 |
| Muy grande | 0,012 | 0,007 | 0,035 |
| Servidor | 576,861 | 208,962 | 23,295 |

**Tabla 8.21:** Frecuencias de saturación y de llegada durante el periodo de observación.

Evidentemente, si se eleva la tasa de acierto de la cache, la demanda baja considerablemente, como se señaló en la Tabla 8.17. Así, por ejemplo, con valores de la tasa de acierto de la cache próximos al 100 %, el cuello de botella pasaría a ser el procesador, puesto que tiene la demanda ponderada más alta para la carga caracterizada. Si observamos indivi- dualmente las distintas clases de transacciones, vemos que realmente la red es el cuello de botella de las transacciones “mediana”, “grande” y “muy grande”. En cualquier caso, puede ser casi imposible mantener una tasa de éxito en la E/S tan alta, puesto que ni con un 90 % sería suficiente para que el cuello de botella fuese otro dispositivo diferente al disco duro.

Otra solución más factible para mejorar el comportamiento de la E/S pasa por reconfi- gurar el disco duro a través de utilidades del sistema operativo. La utilidad hdparm puede servir no sólo para detectar la velocidad de transferencia, sino para cambiar los parámetros de configuración que afectan al funcionamiento del disco (véase el Capítulo 2). Recordemos que el disco estaba transfiriendo a unos 9,98 MB/s. Aunque la velocidad de transferencia es sólo una parte del servicio del disco (el posicionamiento medio es de 8,9 ms y la latencia media de 5,555 ms), es una medida de rendimiento decisiva en todas las transacciones que realiza el servidor web. Si se demanda al sistema operativo acerca de la configuración del dispositivo de E/S, se obtiene lo siguiente:

```
$ hdparm    /dev/hda
/dev/hda:

multcount      = 0 (off)
I/O support    = 0 (default 16 bit)
Unmaskirq      = 0 (off)
Using_dma      = 0 (off)
```

```
Keepsettings = 0 (off) nowerr
                    = 0 (off)
readonly            = 0 (off)
readahead           = 8 (on)
...
```

Ésta es la configuración predeterminada, pero no es necesariamente óptima (aunque puede que la más segura), puesto que funcionaría casi con cualquier disco físico. Algunos parámetros interesantes que se podrían modificar son:

- multcount: controla cuántos sectores son traídos (*fetch*) desde el disco en una interrupción de E/S. Habilitando esta opción, se reduce típicamente la sobrecarga del sistema operativo por el manejo del disco físico.
- I/O support: controla cómo se pasan los datos del bus a la controladora del disco. El valor típico para controladoras modernas es 3, que se refiere a modo 32 bits.
- Unmaskirq: permite desenmascarar otras interrupciones mientras procesa interrup- ciones de disco. Por ejemplo, deja que Linux atienda a otras tareas, como el tráfico de red, mientras espera que el disco retorne los datos solicitados.
- Using_dma: utiliza asignación dinámica de memoria.

Aplicando los modificadores correspondientes se puede cambiar la configuración del manejo del disco por parte del sistema operativo, pero antes habría que asegurarse de que no es peligroso para los datos:

```
$ hdparm    -X66 d1 u1 m16 c3 /dev/had

/dev/had:
setting 32-bit I/O support flag to 3 setting
multcount to 16
setting unmaskirq to 1 (on) setting
using_dma to 1 (on)
setting xfermode to 66 (UltraDMA mode2)
...

$ hdparm -Tt /dev/hda
/dev/hda:

Timing buffer-cache reads: 128 MB in 1.75 seconds= 73.14 MB/s Timing Buffered
disk reads: 64 MB in 1.89 seconds= 33.86 MB/s
```

Como se puede ver tras el cambio de configuración, la velocidad de transferencia de la cache ha bajado a 73,14 MB/s, pero la ganancia en el disco ha sido espectacular: de 9,98 MB/s se ha pasado a 33,86 MB/s. De todos modos, puesto que el test que realiza hdparm lo hace secuencialmente, la mejora de los accesos al disco no sería tan importante cuando se considera la carga real con los ficheros de menor tamaño que son los que más se consultan, sólo sería realmente importante en los de gran tamaño. Sin embargo, la mejora no impide que la E/S siga siendo el cuello de botella, como se aprecia si se comparan los valores mostrados en las Tablas 8.22 y 8.23 con los de las Tablas 8.7 y 8.8, respectivamente. Esto es debido a que no se reduce ni el tiempo de posicionamiento medio ni tampoco la latencia rotacional.

| Clase | Visitas | Tamaño | Servicio | Demanda |
|---|---|---|---|---|
| Muy pequeño | 1 | 114 | 14,458 | 14,458 |
| Pequeño | 1 | 617 | 14,473 | 14,473 |
| Mediano | 1 | 14.486 | 14,882 | 14,882 |
| Grande | 6 | 15.3845 | 15,212 | 91,273 |
| Muy grande | 12 | 1.319.793 | 17,303 | 212,437 |

**Tabla 8.22:** Demanda y servicio medio del disco después de llevar a cabo la sintonización.

| Clase | Visitas | Tamaño | Servicio | Demanda |
|---|---|---|---|---|
| Muy pequeño | 1 | 114 | 0,0015 | 0,0015 |
| Pequeño | 1 | 617 | 0,0084 | 0,0084 |
| Mediano | 1 | 14.486 | 0,1980 | 0,1980 |
| Grande | 1 | 153.845 | 2,1034 | 2,1034 |
| Muy grande | 1 | 1.319.793 | 18,0447 | 18,0447 |

**Tabla 8.23:** Demanda y servicio medio de la memoria cache de E/S después de llevar a cabo la sintonización.

Aunque la velocidad de transferencia de la cache ha descendido algo, la mejora en la del disco es sensible, pero sólo permite que la demanda de E/S mejore en las transacciones mayores, de menor peso en la caracterización de la carga. La Tabla 8.24 compara la demanda de la E/S antes y después de la sintonización (los resultados se expresan en milisegundos).

La posibilidad de reducir el desplazamiento del brazo y la latencia media exigiría reorganizar el contenido del disco, pero dada la naturaleza de la información contenida en un servidor web, no parece una solución viable. Una forma de mejorar el rendimiento de E/S, más evidente que las anteriores, consiste en sustituir el disco duro por otro con un servicio medio más rápido, sobre todo en posicionamiento medio y latencia rotacional, pero en este

| Clase | Demanda antes | Demanda después |
|---|---|---|
| Muy pequeño | 2,8943 | 2,8929 |
| Pequeño | 2,9095 | 2,9013 |
| Mediano | 3,3305 | 3,1348 |
| Grande | 22,2312 | 19,9373 |
| Muy grande | 74,7502 | 56,9231 |
| Servidor | 3,5336 | 3,4119 |

**Tabla 8.24:** Comparativa de la demanda de E/S media actual y sintonizada para una tasa de acierto de 0,8 en la memoria cache de E/S.

caso habría que reemplazar el dispositivo con el coste añadido correspondiente.

Recapitulando, para reducir el efecto del cuello de botella que suponen los accesos al disco duro, la demanda de E/S debería ser como máximo la del procesador. Este requeri- miento surge de resolver la ecuación:

$$1{,}4390 = TA \times 0{,}1121 + (1 - TA) \times 17{,}222$$

donde la variable TA representa la tasa de acierto en la memoria cache de E/S. Si se resuelve la ecuación anterior obtenemos una tasa de acierto superior al 92 %. Al ser este porcentaje muy difícil de alcanzar, en la práctica se puede optar por disminuir la demanda de E/S por reconfiguración del dispositivo en el sistema o reemplazándolo a un coste superior. En el primer caso, la mejora se apreciará en los accesos de gran tamaño, puesto que se reduce sólo la velocidad de transferencia global. De este modo, con la misma tasa de acierto, la demanda de la clase “muy grande” pasa a ser prácticamente la mitad de la correspondiente a la red. En el caso de adquirir un nuevo sistema de E/S, se debería encontrar alguno cuya latencia media y desplazamiento medio del brazo fuesen muy reducidos, aun manteniendo una tasa de acierto del 80 % y la misma velocidad de transferencia reconfigurada. Así, se podría llegar a una demanda máxima para la clase “grande” de 13,37 ms, que corresponde a la de la transmisión de la red Fast Ethernet. De nuevo, para la tecnología actual, la forma de llegar tener demandas medias de E/S inferiores en las clases de tamaño menor exige una tasa de acierto de prácticamente un 100 %. Estas conclusiones se pueden extraer de los datos mostrados en la Tabla 8.25, donde se observa que la ganancia media de la sintonización no es apreciable. No obstante, si se produjese un cambio en el perfil de las transacciones a favor de las de mayor tamaño, la ganancia sería más importante.

## Procesador

A pesar de tener una utilización muy alta, la demanda media de este dispositivo no es la mayor de las consideradas y, por tanto, no parece objetivo primario en la eliminación de cuellos de botella. Sin embargo, tiene la segunda demanda media más alta, después de la E/S. Otro argumento diferente es la obsolescencia del procesador, que bien podría susti- tuirse por alguno de mayor velocidad. Por ejemplo, se podría sustituir el Pentium II por un Pentium 4 a 3,06 GHz. Este procesador, según el fabricante, es seis veces más eficiente utilizado, precisamente, como un servidor web. De nuevo, la mejora en la demanda media de procesador, que pasaría de 1,490 ms a 0,2333 ms, produciría problemas de rendimien- to de E/S, en primer término, y de red, en las clases de mayor tamaño, puesto que se producirían antes las interrupciones de ambos dispositivos con un procesador más rápido.

| Clase | Demanda procesador | Demanda red | Demanda E/S anterior | Demanda E/S sintonizada |
|---|---|---|---|---|
| Muy pequeño | 1,4678 | 0,0305 | 2,8943 | 2,8929 |
| Pequeño | 1,6443 | 0,0708 | 2,9095 | 2,9013 |
| Mediano | 0,9913 | 1,1947 | 3,3305 | 3,1348 |
| Grande | 1,7442 | 12,4773 | 22,2312 | 19,9373 |
| Muy grande | 5,8845 | 106,8720 | 74,7502 | 56,9231 |
| Servidor | 1,4390 | 0, 7282 | 3,5336 | 3,4119 |

**Tabla 8.25:** Demanda media de los dispositivos con una E/S sintonizada.

### Red

Si se consiguiera una mejora sustancial en la E/S, y que el procesador fuese más rápido, el siguiente paso sería disminuir la demanda media de la red. Este dispositivo se presenta como susceptible de mejora puesto que es también muy sensible al tamaño de las transacciones de respuesta (véase, por ejemplo, la Tabla 8.15). De hecho, para la clase “muy grande” y una tasa de acierto del 80 % en el disco original sin reconfigurar, la red ya era el cuello de botella. Sintonizando el disco la diferencia de demandas medias para esa clase aumenta, como se comprueba en la Tabla 8.23.

Una solución costosa, pero efectiva, sería sustituir la red Fast Ethernet por una Gigabit Ethernet que aumentaría la velocidad de transmisión de 100 Mbps a 1 Gbps, es decir, disminuiría la demanda media por transacción unas diez veces aproximadamente. Debe tenerse en cuenta que la mejora de la red sin mejorar el disco provocaría un colapso en el servidor debido a que las transacciones de petición, que poseen siempre un menor tamaño, ahora llegarían con mayor velocidad al servidor, provocando mayores esperas en el acceso a la E/S y en segundo término un mayor castigo del procesador original.

## 8.3.5. Planificación de la capacidad

La caracterización de la carga produjo un perfil que ha servido para estudiar el rendimiento actual del servidor web en este caso. Incluso nos hemos permitido calcular las frecuencias de saturación con el cuello de botella actual (la E/S), mejorando el cuello de botella y también proponer el reemplazo de otros dispositivos menos susceptibles de ser sustituidos. Pero ¿qué ocurriría si cambiasen los porcentajes por clase? Evidentemente, se debe prever si aumenta la demanda global del sistema con la caracterización actual y a su vez estudiar las consecuencias de rendimiento que tendría un cambio de perfil de los usuarios (véase el Capítulo 7).

Supongamos que lo que se produce es un cambio de perfil en los usuarios del servidor y se demandan las transacciones más peligrosas para los dispositivos más lentos: E/S y red. A partir de la ley operacional de la demanda, se pueden establecer las nuevas frecuencias de saturación por clase de transacción (véase la Tabla 8.19). En la Tabla 8.26 se plantean distintos escenarios a partir de la carga observada actual.

| Clase | Frec. saturac. Modelo final | ( %) | Perfil 2 Frec. sat. | ( %) | Perfil 3 Frec. sat. |
|---|---|---|---|---|---|
| Muy pequeño | 101,630 | 45 | 102,449 | 40 | 91,066 |
| Pequeño | 69,757 | 32 | 69,195 | 30 | 64,870 |
| Mediano | 37,505 | 20 | 36,254 | 20 | 36,254 |
| Grande | 0,620 | 2 | 0,548 | 5 | 1,371 |
| Muy grande | 0,008 | 1 | 0,053 | 5 | 0,266 |
| Servidor | 209,520 | 100 | 208,499 | 100 | 193,827 |

| Clase | ( %) | Perfil 4 Frec. sat. | ( %) | Perfil 5 Frec. sat. |
|---|---|---|---|---|
| Muy pequeño | 30 | 68,299 | 5 | 11,383 |
| Pequeño | 20 | 43,246 | 10 | 21,623 |
| Mediano | 10 | 18,127 | 15 | 27,190 |
| Grande | 20 | 5,484 | 25 | 6,580 |
| Muy grande | 20 | 1,064 | 45 | 2,394 |
| Servidor | 100 | 136,220 | 100 | 69,170 |

**Tabla 8.26:** Frecuencias de saturación expresadas en tps de diversos perfiles de carga.

Si el perfil de los usuarios, accediendo a la información actual, cambiase hasta una distribución de probabilidades casi inversa a la actual, es decir, con un 45 % de accesos a ficheros de gran tamaño, la frecuencia de saturación podría llegar a ser inferior a 70 tps. Y en el caso más desfavorable, si el 100 % de las transacciones fuesen “muy grandes”, el servidor web sólo podría soportar 8 tps, como se recalcó previamente en la Tabla 8.4.

Por tanto, si la información del servidor se mantuviese, es decir, los tamaños de ficheros medios siguiesen siendo parecidos, pero las frecuencias de acceso cambiasen de perfil, el cuello de botella del sistema determinaría el crecimiento vegetativo de la instalación. En cualquier caso habría que caracterizar la carga de nuevo.

Veamos qué podría ocurrir de producirse la variación del rendimiento de la Tabla 8.26. Supongamos que los cambios de perfil observados corresponden a valores mensuales. Así, el perfil original corresponde al caso de estudio si se hubiera realizado hace cuatro meses, el denominado “perfil 2” es la medida que se tomó hace tres meses y así sucesivamente hasta el perfil 5, que corresponde a la medida actual del rendimiento. Si se utiliza la técnica de las medias móviles, podemos predecir la frecuencia de saturación del perfil 6, $F_S(6)$, esto es, la máxima productividad alcanzable del mes siguiente, a través de la expresión:

$$F_S(6) = \frac{F_S(1) + F_S(2) + \cdots + F_S(5)}{5} = 162{,}952 \text{ tps}$$

lo que resulta, evidentemente, un poco extraño puesto que sería un cambio de tendencia al alza. La técnica de medias móviles sirve para datos muy estacionarios, y la muestra del perfil 6 es muy diferente a las anteriores. Por tanto, no parece una técnica apropiada para los perfiles observados.

Si se usa la regresión lineal con el método de los mínimos cuadráticos, se podría predecir el valor según la recta de tendencia. En este caso puede parecer bastante adecuado, puesto que en las tablas anteriores se observa una tendencia clara al descenso de rendimiento. Sin embargo, esta predicción sería muy optimista porque los dos últimos meses acusan un descenso muy fuerte de rendimiento, y la pendiente arrastra el peso de las muestras anteriores que eran más suaves.

Con estos perfiles podría utilizarse la técnica de suavización exponencial, con un peso muy alto para las observaciones recientes. Si utilizamos un peso fijo del 90 %, la fórmula a aplicar para obtener la frecuencia estimada futura en el mes $i$-ésimo, $F_S(i)$, será:

$$F_S(i + 1) = F_S(i) + 0{,}9 \times (F_S(i + 1) - F_S(i))$$

De nuevo, supongamos que la Tabla 8.26 refleja las frecuencias de saturación de los cinco meses considerados. Debido a que se necesita una frecuencia estimada de inicialización, se toma la frecuencia máxima del modelo de carga estudiado.

| Frecuencia | Mes 1 | Mes 2 | Mes 3 | Mes 4 | Mes 5 |
|---|---|---|---|---|---|
| Medida | 209,520 | 208,499 | 193,827 | 136,220 | 69,170 |
| Estimada | 209,520 | 208,601 | 195,304 | 142,128 | 76,465 |

**Tabla 8.27:** Frecuencias estimadas mediante suavización exponencial de peso fijo.

Si se observa la Tabla 8.27 con detalle, la diferencia entre la media exponencial suavizada y el valor real se va incrementando con el número de muestras. Ello es debido a que la productividad del servidor es cada vez menor, pero se arrastra un peso del 10 % del histórico que hace aumentar el valor estimado con respecto al valor de productividad máxima medido. Esta previsión supondría un pequeño aumento de productividad con respecto a la realidad pero no tan alto como la previsión por medias móviles, y mucho menor que con la regresión lineal. Aun así, parece que se está rompiendo lo que podría ser una tendencia de peor rendimiento por cada mes que pasa.

También se puede tomar un peso variable incremental en función del número de mues- tras, es decir, se da un peso mayor a medida que aumenta el número de muestras. Por ejemplo, el peso se puede calcular como:

$$\text{Peso} = \frac{M + N - 1}{M + N + 1}$$

donde $N$ es el número de muestras, que en este caso son mensuales, y $M$ es un desplaza- miento. Este desplazamiento $M$ se puede fijar para que el peso, en el inicio, sea un valor muy bajo (se da mayor peso al valor histórico), o muy alto (se da mayor peso a las obser- vaciones recientes); de cualquier forma, irá creciendo con el número de muestras hasta el límite del 100 % en las observaciones recientes si el número de muestras fuera infinito. Así, por ejemplo, se podría comenzar con un peso del 90 % en la primera muestra, tal como se había tomado en la suavización fija anterior: para obtener un peso igual a 0,9, $M$ debe ser igual a 10 cuando $N = 1$. A medida que $N$ crece, el peso de los valores recientes aumenta también, tal como se observa en los resultados de la Tabla 8.28. Evidentemente, la elección de cualquier otra función creciente para el cálculo del peso de las observaciones recientes cuyo límite sea 1 produciría valores distintos.

| Frecuencia | Mes 1 | Mes 2 | Mes 3 | Mes 4 | Mes 5 |
|---|---|---|---|---|---|
| Medida | 209,520 | 208,499 | 193,827 | 136,220 | 69,170 |
| Estimada | 209,520 | 208,591 | 195,067 | 140,751 | 74,323 |
| Peso | 0,900 | 0,909 | 0,916 | 0,923 | 0,928 |

**Tabla 8.28:** Frecuencias estimadas por suavización exponencial de peso variable.

Otras aproximaciones, como la de Tustin, pueden combinar la media móvil con la sua- vización exponencial. Ya que el error entre dos meses consecutivos es menor que cualquier otro superior, podría proponerse que el histórico tuviese en cuenta no sólo el mes anterior, sino la media de las dos medidas anteriores. Evidentemente, con esta aproximación se pue- de ajustar con el desplazamiento de la Tabla 8.28, pero siempre resultará una previsión muy optimista debido al valor de arrastre adicional.

Con estos perfiles con un descenso de productividad tan pronunciado, sobre todo en el último mes, la técnica de predicción que posiblemente se ajuste mejor es la suavización exponencial con un peso elevado en las últimas observaciones. El peso puede ser fijo o variable, pero en cualquier caso nuevas observaciones determinarían estos detalles.

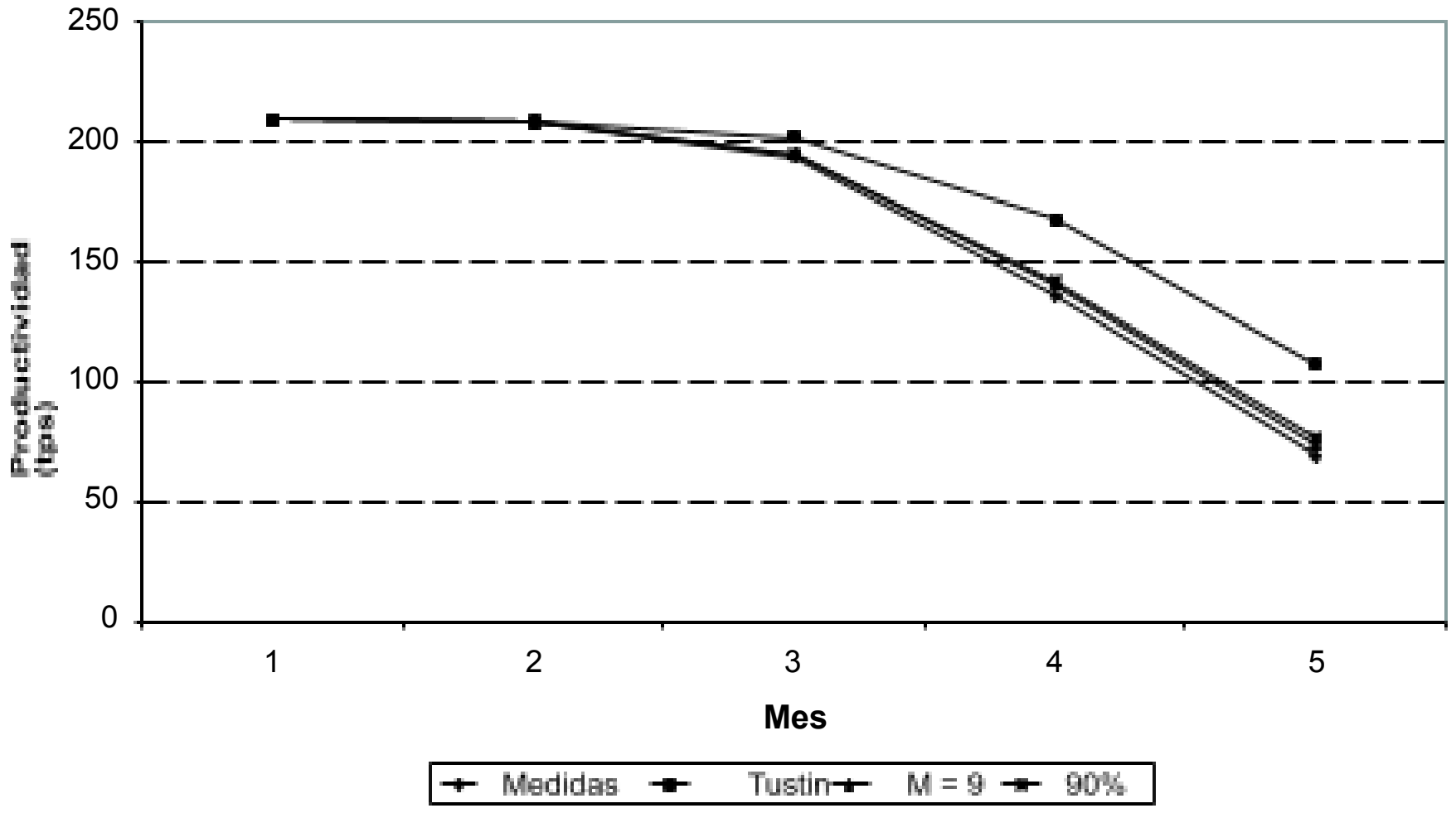

**Figura 8.5:** Productividad máxima estimada por suavización exponencial.

Finalmente, en la Figura 8.5 se representan los distintos valores de la frecuencia estima- da y la real para los distintos procedimientos de suavización exponencial y en la Figura 8.6 un comparativo de las tres técnicas en este caso de estudio.

### 8.3.6. Conclusiones

En este capítulo se ha tratado de ilustrar el uso de algunas de las técnicas expuestas en los capítulos anteriores mediante un caso de estudio práctico y actual. A partir de la ausencia de información sobre las prestaciones de un servidor web en una Intranet corporativa y con las herramientas mínimas, se ha planteado una metodología aproximada del análisis del rendimiento. Esta metodología descansa sobre dos modelos: el modelo de carga y el modelo de rendimiento. Por cuestiones de espacio y simplicidad, a partir de un modelo de carga ya clasificado por tamaño de archivo solicitado, se han aproximado los valores suficientes para construir un modelo de rendimiento. Puesto que la naturaleza de las transacciones HTTP, que son fuente de carga del servidor web, no permite establecer directamente el consumo de recursos de los dispositivos, se han usado las técnicas básicas del análisis de prestaciones para establecer dichos valores. De este modo, se han monitorizado algunos dispositivos esenciales del servidor mientras era sometido a la carga artificial de un *bench- mark* comercial con los mismos perfiles del modelo de carga y se han relacionado los valores obtenidos con las leyes operacionales. Con los valores recolectados de la utilización de los dispositivos por parte de las transacciones HTTP del modelo de carga se han establecido dos modelos de prestaciones del servidor web para conocer su rendimiento, uno más simple considerando al servidor como una estación de servicio única y otro más refinado considerando sus dispositivos principales: procesador, E/S y red. Del análisis del modelo refinado de rendimiento del servidor web, se ha alcanzado un grado mayor en el conocimiento de la demanda que las transacciones HTTP hacen de los dispositivos, del tiempo de respuesta actual en función de la utilización del servidor web y del cuello de botella de la instalación. El modelado también ha permitido establecer pautas de mejora y sintonización del servidor web, así como la planificación de la capacidad futura en función de los perfiles de carga que pudiesen establecerse.

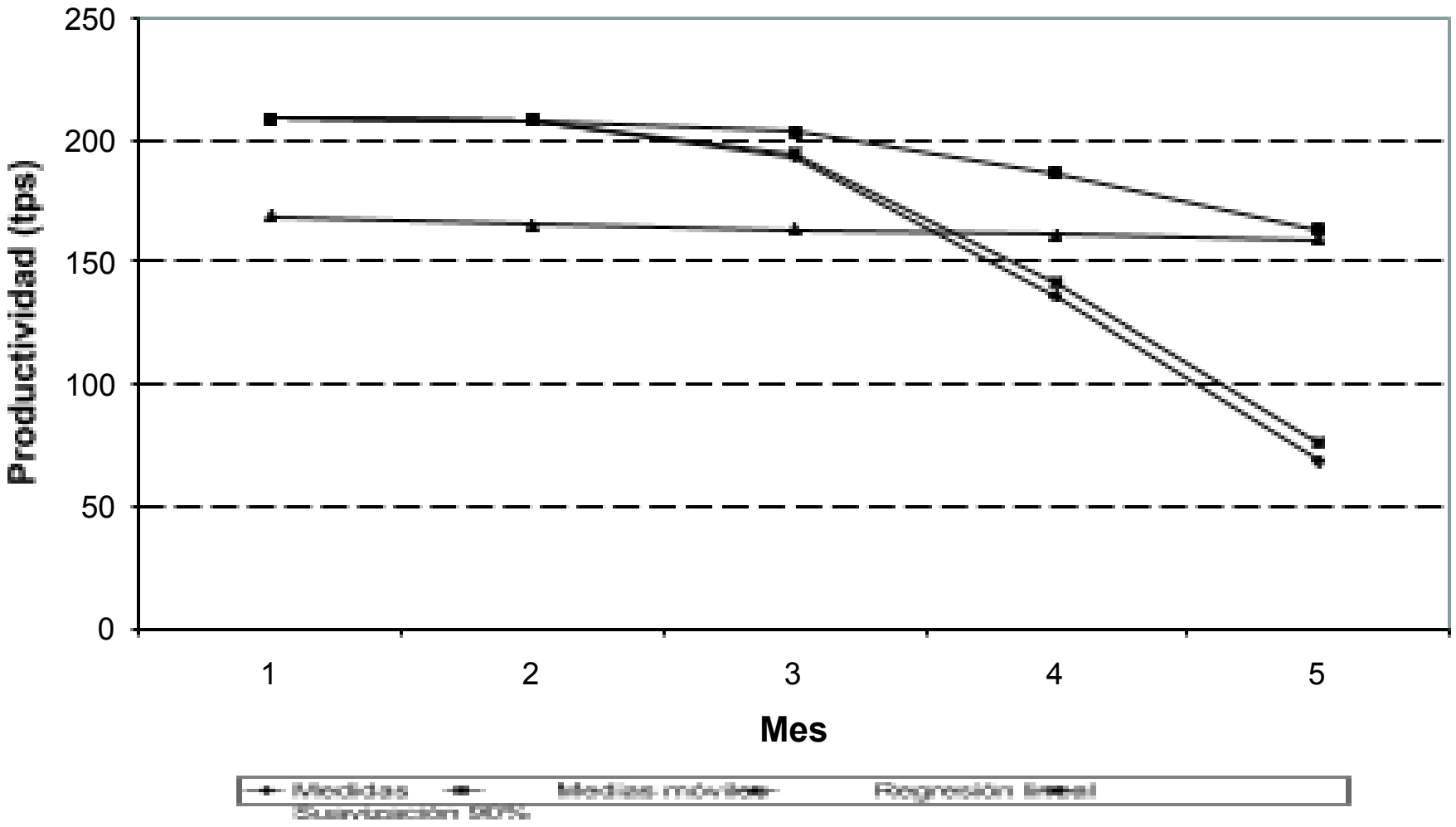

**Figura 8.6:** Productividad máxima estimada por diversas técnicas.

Evidentemente, la solución propuesta en este caso de estudio no es única; se podrían haber realizado otros modelos de carga y otros modelos de rendimiento. También debe señalarse que, aunque se hubieran seguido exactamente los mismos pasos que en la solu- ción adoptada aquí, los modelos podrían haber sido más simples o, por el contrario, más refinados. Ésta es una característica esencial en el análisis del rendimiento de los sistemas y particularmente de las metodologías basadas en modelos, que son intrínsecamente par- ciales en doble sentido. Son parciales porque descansan en la agrupación, síntesis y realce de detalles en comparación con la realidad, y también son parciales porque los modelos son una idea abstracta que el analista de prestaciones tiene de esa realidad, tanto si es observable como si sólo es un proyecto de futuro. Ése es el interés y la grandeza que posee el análisis del rendimiento de los sistemas y que esperamos haber trasladado durante el desarrollo de este libro.

## 8.4. Problemas sin resolver

**PROBLEMA 8.1** La caracterización de la carga en el caso de estudio se ha realizado a partir del tamaño de los ficheros solicitados por las peticiones HTTP. Sin embargo, es posible caracterizar la carga a través de los tipos de fichero solicitados. Si ésta hubiera sido la caracterización elegida, ¿cómo se agruparían los tipos de ficheros en clases? ¿Sería necesario aplicar la técnica de *clustering*? ¿Sobre qué variables de monitorización se aplicarían los agrupamientos?

---

**PROBLEMA 8.2** ¿Qué ocurriría si el *benchmark* utilizara la memoria cache del computador cliente que suministra la carga? ¿Por qué no se realiza el *benchmarking* con accesos desde un navegador?

---

**PROBLEMA 8.3** ¿Cómo cambiarían los modelos si el *benchmark* ab usara transacciones almacenadas en la memoria cache del servidor? ¿Qué ocurriría si las pruebas se redireccionasen a un servidor espejo?

---

**PROBLEMA 8.4** La ejecución del *benchmark* desde una estación cliente implica que los resultados del experimento son dependientes parcialmente de la localidad, es decir, que las facilidades de comunicación entre el cliente y el servidor influirán en las mediciones. ¿Cómo se puede obviar o reducir este efecto?

---

**PROBLEMA 8.5** Los modelos de redes de colas que se han utilizado en este estudio son orientados al servidor. ¿Qué información sería necesaria si se hubiera efectuado un mode- lado orientado al cliente?

---

**PROBLEMA 8.6** ¿Cómo podría haberse resuelto el modelo de colas del servidor web si se hubiera supuesto que los tamaños de los archivos solicitados seguían una distribución gamma o de Pareto con esos valores medios?

---

**PROBLEMA 8.7** Concentrar mucho la carga en ciertas momentos produciría más ráfagas (*burstiness*), con lo que asumir valores medios típicos distribuidos exponencialmente, como se hace habitualmente en el análisis operacional, sería un error. ¿Cómo afectaría esto al modelado de la carga a partir de la construcción de los ficheros de bitácora?

---

**PROBLEMA 8.8** Para modelar el comportamiento del disco duro en el caso de estudio, se ha supuesto que la velocidad de transferencia media es la que el monitor hdparm muestra. Sin embargo, el test que realiza este monitor incluye al menos el tiempo de posicionamiento del brazo en la pista de test y varios desplazamientos a pistas colindantes puesto que se trata de lecturas secuenciales. También incluirá al menos la latencia rotacional media. ¿Cómo afectaría al modelo restar estos tiempos del test? ¿Cuál es la influencia de no haberlos considerado? Tómese en cuenta la precisión de la monitorización y el tipo de chequeo del monitor en concreto para contestar a las cuestiones anteriores.

**PROBLEMA 8.9** La ley de Zipf predice que la popularidad de un documento almacenado en un servidor web es aproximadamente inversa a su frecuencia de uso. ¿Podría haberse caracterizado la carga con estadísticas cualitativas sobre los documentos más populares del servidor web? ¿Qué ahorro hubiera significado en este caso de estudio?

## 8.5. Actividades propuestas

**ACTIVIDAD 8.1** Búsquense en Internet programas que manejen estadísticas de servido- res a través de la carga de un fichero bitácora con los accesos HTTP. Identifíquense qué cuestiones se resuelven a partir del fichero bitácora, el formato de los ficheros de entrada y las gráficas de salida, etc.

**ACTIVIDAD 8.2** Búsquense otros algoritmos de agrupamiento (*clustering*) diferentes del árbol de extensión mínima (MST, *Minimum Spanning Tree*). ¿Qué diferencias aportan con respecto al empleado en este caso de estudio?

**ACTIVIDAD 8.3** Búsquense en Internet otros *benchmarks* para servidores web, como por ejemplo *httperf* o *Webstone*. Descárguese alguno de ellos y ejecútese contra algún servidor al que se tenga acceso. ¿Qué diferencias de uso se observan? ¿Son los resultados de la ejecución de estos *benchmarks* sensiblemente distintos para cargas similares? ¿Qué ventajas ofrecen frente al *benchmark* ab? ¿Cómo presentan los datos de la experimentación?

**ACTIVIDAD 8.4** El *benchmark* TPC-W no es realmente un ejecutable como los ejemplos de la actividad anterior. TPC-W es un modelo de construcción de *benchmark* para co- mercio electrónico, que contiene todas las fases diseño excepto la propia programación del *benchmark*. Consúltese la documentación del TPC-W en Internet. ¿Qué tipo de empresa comercial propone la especificación como sujeto de experimentación? ¿Qué transacciones propone como modelo de carga artificial?

**ACTIVIDAD 8.5** Una de las mayores dificultades a las que puede enfrentarse un analis- ta de prestaciones es la de intentar discernir las unidades de trabajo en que trabajan los distintos servicios. Consúltese el formato de las transacciones HTTP y considérese la posi- bilidad de conocer el gasto de recursos a través de los procesos de sistema operativo, que evidentemente no aparecerán ni en el mismo número, ni serán de la misma naturaleza que las transacciones observadas del protocolo HTTP.

**ACTIVIDAD 8.6** Accédase a la orden man de un sistema operativo Linux o Unix, o búsquense en Internet las características de los monitores y utilidades software siguientes:

bonnie, hdparm, memstat, uax, pmap y free. Algunos de ellos pertenecen a ambos sistemas operativos, otros sólo a uno, y otros no se incorporan de manera predeterminada en el sistema.

**ACTIVIDAD 8.7** A pesar de que la información que nos proporcione el administrador del sistema sea fiable, existen herramientas sencillas para reconocer la vía de conexión que sufren las transacciones desde un cliente al servidor, incluso desde el exterior. Búsquese un trazador, como por ejemplo NeoTrace o traceroute, que siga la ruta TCP/IP desde la estación de trabajo en la que se ejecuta hasta un servidor. Analícense los resultados de la misma y determínense los retrasos de conectividad debido a las redes telefónicas, las redes de datos, los encaminadores (*routers*), etc.

**ACTIVIDAD 8.8** Las cargas relacionadas con servidores, sobre todo en Internet, exhiben comportamientos de gran variabilidad. Lo que se conoce como tráfico a ráfagas (*burstiness*) y distribuciones de cola pesada (*heavy-tailed* ). El primer fenómeno está relacionado con que la carga no se distribuye igual en el tiempo, sino que se sufre de mucha congestión en momentos muy concretos, lo que produce cuellos de botella transitorios o intermitentes. El segundo fenómeno trata con que los tamaños de los ficheros consultados en la web sigan distribuciones de tipo Pareto y Logarítmico Normal. ¿Cómo se puede averiguar si la carga corresponde a esa congestión? ¿Cómo determinar la distribución que más se asemeja a los tamaños de ficheros consultados? ¿Cree que clasificar la carga ha disminuido el efecto de este último fenómeno?

**ACTIVIDAD 8.9** El método de los mínimos cuadrados es una resolución de la regresión lineal simple. Sin embargo, existen otras formas de generalizar la regresión, como, por ejemplo, la no lineal y la múltiple. Razónese cómo emplear estas técnicas estadísticas para la planificación de la capacidad del servidor web del caso de estudio.

# Bibliografía